\documentclass[a4paper,11pt,fleqn]{book}
\usepackage[utf8]{inputenc}

\title{Book template}
\author{Stefano Lanza}
\date{September 2019}

\usepackage[backend=bibtex, maxbibnames=9, sorting=none, style=numeric-comp]{biblatex}
\if{}

\DeclareFieldFormat*{title}{\mkbibemph{#1}}
\DeclareFieldFormat*{citetitle}{\mkbibemph{#1}}
\DeclareFieldFormat{journaltitle}{#1}

\renewbibmacro*{in:}{%
	\ifentrytype{article}
	{}
	{\printtext{\bibstring{in}\intitlepunct}}}

\newbibmacro*{pubinstorg+location+date}[1]{%
	\printlist{#1}%
	\newunit
	\printlist{location}%
	\newunit
	\usebibmacro{date}%
	\newunit}

\renewbibmacro*{publisher+location+date}{\usebibmacro{pubinstorg+location+date}{publisher}}
\renewbibmacro*{institution+location+date}{\usebibmacro{pubinstorg+location+date}{institution}}
\renewbibmacro*{organization+location+date}{\usebibmacro{pubinstorg+location+date}{organization}}

\fi{}

\usepackage{bibentry}

\usepackage{graphicx}
\usepackage{wrapfig}

\usepackage[T1]{fontenc}
\usepackage[utf8]{inputenc}
\usepackage[french,german,english]{babel}

\usepackage{amsmath}
\usepackage{multirow}
\usepackage{array}
\newcolumntype{L}{>{\centering\arraybackslash}m{3cm}}

\usepackage{lipsum}

\usepackage{pifont}
\newcommand{\cmark}{\ding{51}}
\newcommand{\xmark}{\ding{55}}


\usepackage{graphicx}
\usepackage[table]{xcolor}

\definecolor{darkred}{RGB}{145,6,6}
\definecolor{darkblue}{RGB}{4,11,111}
\definecolor{bluegreen}{RGB}{0,153,153}
\definecolor{ochre}{RGB}{250,231,152}

\usepackage{subfig}
\usepackage{booktabs}
\usepackage{lipsum}
\usepackage{microtype}
\usepackage{url}
\usepackage[final]{pdfpages}

\usepackage{fancyhdr}

\fancypagestyle{plain}{
	\fancyhf{}

	\fancyfoot[EL,OR]{\thepage}}
\fancypagestyle{addpagenumbersforpdfimports}{
	\fancyhead{}
	
	\fancyfoot{}
	\fancyfoot[RO,LE]{\thepage}
}

\usepackage{listings}
\usepackage{bm}

\makeatletter
\def\cleardoublepage{\clearpage\if@twoside \ifodd\c@page\else
    \hbox{}
    \thispagestyle{empty}
    \newpage
    \if@twocolumn\hbox{}\newpage\fi\fi\fi}
\makeatother \clearpage{\pagestyle{plain}\cleardoublepage}

\usepackage{color}
\usepackage{tikz}
\usepackage[explicit]{titlesec}
\newcommand*\chapterlabel{}
\titleformat{\chapter}[display]  
	{\normalfont\Huge} 
	{\gdef\chapterlabel{\thechapter\ }}     
 	{0pt} 
 	  {\begin{tikzpicture}[remember picture,overlay]
    \node[yshift=-8cm] at (current page.north west)
      {\begin{tikzpicture}[remember picture, overlay]
        \draw[fill=darkblue,darkblue] (-0.2,-0.2) rectangle(34.5mm,12mm);
        \node[anchor=north east,yshift=-7.2cm,xshift=33mm,minimum height=30mm,inner sep=0mm] at (current page.north west)
        {\parbox[top][30mm][t]{15mm}{\raggedleft $\phantom{\textrm{l}}${\fontfamily{ppl}\selectfont\color{white}\chapterlabel}}};  
        \node[anchor=north west,yshift=-7.2cm,xshift=37mm,text width=\textwidth,minimum height=30mm,inner sep=0mm] at (current page.north west)
              {\parbox[top][30mm][t]{\textwidth}{\fontfamily{ppl}\selectfont\color{darkblue}#1}};
       \end{tikzpicture}
      };
   \end{tikzpicture}
   \gdef\chapterlabel{}
  } 

\titleformat{\section}
  {\normalfont\Large\bfseries\sffamily}
  {\llap{\colorbox{darkblue}{\makebox[10mm][r]  {\color{white}\thesection}}\hspace{1em}}}
  {0em}{\color{darkblue}{#1}}

\titleformat{\subsection}
  {\normalfont\large\sffamily}
  {\llap{\colorbox{ochre}{\makebox[10mm][r]  {\color{darkblue}\thesubsection}}\hspace{1em}}}
  {0em}{\color{darkblue}{#1}}

\titleformat{\subsubsection}
{\normalfont\sffamily}
{}
  {0em}
  {\begin{tikz}
  			\fill[rotate=45,darkred] (0,0) rectangle (0.2,0.2);
  	\end{tikz} \hspace{0.25em}\color{darkred}{#1}}

\titlespacing*{\chapter}{0pt}{50pt}{20mm}
\titlespacing*{\section}{0pt}{13.2pt}{3mm}  
\titlespacing*{\subsection}{0pt}{13.2pt}{2mm}
\titlespacing*{\subsubsection}{0pt}{13.2pt}{0mm}

\newcounter{myparts}
\newcommand*\partlabel{}
\titleformat{\part}[display]  
	{\thispagestyle{empty}\normalfont\Huge} 
	{\gdef\partlabel{\thepart\ }}     
 	{0pt} 
 	  {\setlength{\unitlength}{20mm}
	  \addtocounter{myparts}{1}
	  \begin{tikzpicture}[remember picture,overlay]
    \node[anchor=north west,xshift=-65mm,yshift=-6.9cm-\value{myparts}*20mm] at (current page.north east) 
      {\begin{tikzpicture}[remember picture, overlay]
        \draw[fill=darkred,darkred] (0,0) rectangle(62mm,20mm);   
        \node[anchor=north west,yshift=-6.1cm-\value{myparts}*20mm,xshift=-60.5mm,minimum height=30mm,inner sep=0mm] at (current page.north east)
        {\parbox[top][30mm][t]{55mm}{\raggedright\color{white} \textsc{\fontfamily{ppl}\selectfont{Part}} \partlabel $\phantom{\textrm{l}}$}};  
        \node[anchor=north east,yshift=-6.1cm-\value{myparts}*20mm,xshift=-63.5mm,text width=\textwidth,minimum height=30mm,inner sep=0mm] at (current page.north east)
              {\parbox[top][30mm][t]{\textwidth}{\raggedleft \color{darkred}{\textsc{\fontfamily{ppl}\selectfont{#1}}}}};
       \end{tikzpicture}
      };
   \end{tikzpicture}
   \gdef\partlabel{}
  } 

\usepackage{pgfornament}

\usepackage{latexsym}
\usepackage{epsfig,amssymb,euscript,slashed}
\usepackage{amsmath}
\usepackage{mathrsfs}
\usepackage{dsfont}
\usepackage{pifont} 
\usepackage{braket}
\usepackage{array,calc,epsfig}
\usepackage{pstricks}
\usepackage{hyperref}
\usepackage{bbm}
\usepackage{enumitem}
\usepackage{fancybox}
\usepackage{multirow}
\usepackage[makeroom]{cancel}
\def\be{\begin{equation}}
\def\ee{\end{equation}}
\def\bseq{\begin{subequations}}
\def\eseq{\end{subequations}}

\def\bea{\begin{eqnarray}}
\def\eea{\end{eqnarray}}
\newcommand\bbone{\ensuremath{\mathbbm{1}}}
\newcommand{\ul}{\underline}
\def\bseq{\begin{subequations}}
\def\eseq{\end{subequations}}

\def\d {{\rm{d}}}
\def\dwb {{\bf{d}}}
\def\cala         {{\cal A}}

\def\calb         {{\cal B}}
\def\calc         {{\cal C}}
\def\cald         {{\cal D}}
\def\cale         {{\cal E}}
\def\calf         {{\cal F}}
\def\calg         {{\cal G}}

\def\calh         {{\cal H}}
\def\cali         {{\cal I}}
\def\calj         {{\cal J}}
\def\calk         {{\cal K}}
\def\call         {{\cal L}}
\def\calm         {{\cal M}}
\def\caln         {{\cal N}}
\def\calo         {{\cal O}}
\def\calp         {{\cal P}}
\def\calq         {{\cal Q}}
\def\calr         {{\cal R}}
\def\cals         {{\cal S}}
\def\calt         {{\cal T}}

\def\calv         {{\cal V}}
\def\calw         {{\cal W}}

\def\calz         {{\cal Z}}

\def\scrl         {{\mathscr L}}
\def\scrm         {{\mathscr M}}

\def\del          {\partial}

\def\ii           {{\rm i}}

\def\Re           {{\rm Re\hskip0.1em}}
\def\Im           {{\rm Im\hskip0.1em}}

\def\Mp        {M_{\rm P}}

\def\half{{\frac12}}

\def\sqr#1#2{{\vcenter{\vbox{\hrule height.#2pt
 \hbox{\vrule width.#2pt height#1pt \kern#1pt \vrule width.#2pt}\hrule
 height.#2pt}}}}

\def\d{\text{d}}

\def\iund{\underline{i}}

\def\slashchar#1{\setbox0=\hbox{$#1$}           
\dimen0=\wd0                                 
\setbox1=\hbox{/} \dimen1=\wd1               
\ifdim\dimen0>\dimen1                        
\rlap{\hbox to \dimen0{\hfil/\hfil}}      
#1                                        
\else                                        
\rlap{\hbox to \dimen1{\hfil$#1$\hfil}}   
/                                         
\fi}

\makeatletter
\renewcommand*\env@matrix[1][*\c@MaxMatrixCols c]{%
	\hskip -\arraycolsep
	\let\@ifnextchar\new@ifnextchar
	\array{#1}}
\makeatother
 
 \definecolor{Color1}{RGB}{96,96,96}
 \definecolor{Color2}{RGB}{192,192,192}
 \definecolor{Color3}{RGB}{245,221,67}
 \definecolor{Color4}{RGB}{250,197,89}
 \definecolor{Color5}{RGB}{17,125,21} 
 \definecolor{Color6}{RGB}{72,197,76} 

\RequirePackage[framemethod=default]{mdframed} 
\newmdenv[skipabove=3pt,
skipbelow=2pt,
rightline=false,
leftline=true,
topline=false,
bottomline=false,
linecolor=darkred,
backgroundcolor=darkred!10,
innerleftmargin=5pt,
innerrightmargin=2pt,
innertopmargin=0pt,
leftmargin=0cm,
rightmargin=0cm,
linewidth=4pt,
innerbottommargin=9pt]{sBox}

\newenvironment{summary}{\begin{sBox}\vspace{3pt}
	}{\end{sBox}}

  \RequirePackage[framemethod=default]{mdframed} 
 \newmdenv[skipabove=3pt,
 skipbelow=2pt,
 rightline=false,
 leftline=true,
 topline=false,
 bottomline=false,
 linecolor=Color4,
 backgroundcolor=Color3!20,
 innerleftmargin=5pt,
 innerrightmargin=2pt,
 innertopmargin=0pt,
 leftmargin=0cm,
 rightmargin=0cm,
 linewidth=4pt,
 innerbottommargin=9pt]{iBox}

 \newenvironment{important}{\begin{iBox}\vspace{3pt}
 	}{\end{iBox}}

   \RequirePackage[framemethod=default]{mdframed} 
 \newmdenv[skipabove=10pt,
 skipbelow=7pt,
 rightline=true,
 leftline=false,
 topline=false,
 bottomline=false,
 linecolor=Color5,
 backgroundcolor=Color6!20,
 innerleftmargin=5pt,
 innerrightmargin=5pt,
 innertopmargin=0pt,
 leftmargin=0cm,
 rightmargin=0cm,
 linewidth=4pt,
 innerbottommargin=0pt]{nBox}

 \RequirePackage[framemethod=default]{mdframed} 
 \newmdenv[skipabove=10pt,
 skipbelow=7pt,
 rightline=false,
 leftline=false,
 topline=false,
 bottomline=false,
 linecolor=Color1,
 backgroundcolor=Color2!20,
 innerleftmargin=5pt,
 innerrightmargin=5pt,
 innertopmargin=0pt,
 leftmargin=0cm,
 rightmargin=0cm,
 linewidth=4pt,
 innerbottommargin=0pt]{lBox}

 \newenvironment{centerbox}{\begin{lBox}\vspace{1 mm}\begin{center}\large\sffamily
	} {\end{center}\vspace{1 mm}\end{lBox}}

\definecolor{purple}{rgb}{0.5,0,1}

\newcommand\Tstrut{\rule{0pt}{2.9ex}}       
\newcommand\Bstrut{\rule[-1.3ex]{0pt}{0pt}} 
\newcommand\TBstrut{\Tstrut\Bstrut}         

\setlist[description]{%
	topsep=30pt,               
	itemsep=5pt,               
	font={\bfseries\sffamily}, 
}

\usepackage{hyperref}
\hypersetup{pdfborder={0 0 0},
	colorlinks=true,
	linkcolor=darkred,
	citecolor=darkred,
	anchorcolor =blue,
	urlcolor=black}

\def\d {{\rm{d}}}

\begin{document}

\frontmatter

\begin{titlepage}
	\begin{center}

		\vspace{1.2cm}
		
		\fontfamily{ppl}\selectfont\Large\textsc{Sede amministrativa}
		
		\fontfamily{ppl}\selectfont\LARGE\textsc{Universit\`a degli Studi di Padova}
		
		\vspace{0.0cm}
		

		\vspace{0.3cm}
		
		\fontfamily{ppl}\selectfont\Large\textsc{Dipartimento di Fisica e Astronomia ``G. Galilei''}
		
		\vspace{1cm}
		
		\includegraphics[width=4cm]{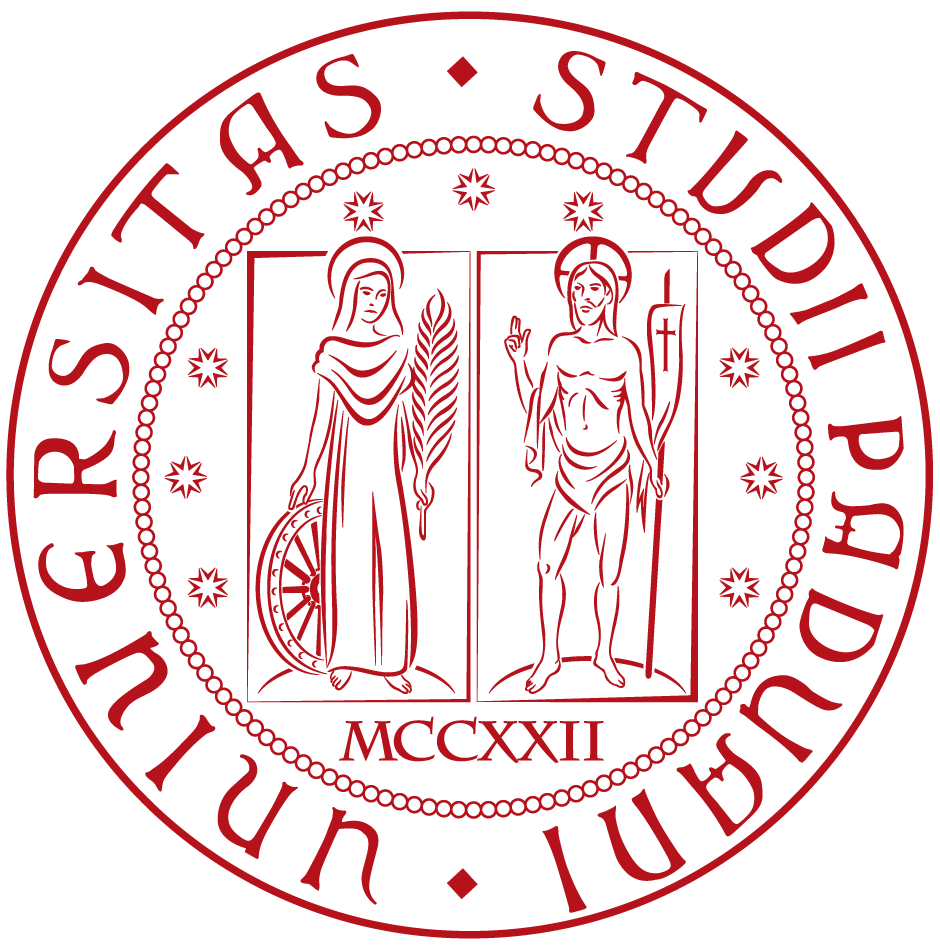}
		
		\vspace{0.3cm}
		
		\fontfamily{ppl}\selectfont\LARGE\textsc{Corso di Dottorato di Ricerca in Fisica}
		
		\vspace{0cm}
		
		\fontfamily{ppl}\selectfont\Large\textsc{xxxii ciclo}

	\end{center}

	\begin{center}
		\noindent\pgfornament[width=10cm,scale=0.7,color=darkred]{88}
	\end{center}

	\vspace{1cm}
	
	\begin{center}
		\fontfamily{ppl}\selectfont\Huge\bfseries{Exploring the Landscape
			\\
			of effective field theories}
		
		\vspace{0.3cm}
		
		
	\end{center}
	
	\vspace{1.7cm}
	
	\noindent{\fontfamily{ppl}\selectfont\large\textsc{Coordinatore}}
	
	\vspace{0.1cm}
	
	\noindent{\Large\fontfamily{ppl}\selectfont{Ch.mo prof. Franco Simonetto}}
	
	\vspace{0.5cm}
	
	\noindent{\fontfamily{ppl}\selectfont\large\textsc{Supervisore}}
	
	\vspace{0.1cm}
	
	\noindent{\Large\fontfamily{ppl}\selectfont{Ch.mo prof. Luca Martucci}}
	
	\vspace{1cm}
	
	\noindent\hfill{\fontfamily{ppl}\selectfont\large\textsc{Dottorando}}
	
	\vspace{0.1cm}
	
	\noindent\hfill{\Large\fontfamily{ppl}\selectfont{Stefano Lanza}}
	
	\vspace{0.7cm}

	\begin{center}
		
		\fontfamily{ppl}\selectfont\Large\textsc{September 2019}
		
	\end{center}
	
\end{titlepage}

\setcounter{page}{0}

\cleardoublepage

\cleardoublepage

\chapter*{Abstract}

String theory is the most promising candidate for a complete theory of quantum gravity. However, in order to obtain relevant four-dimensional phenomenological models, it is necessary to reduce the higher-dimensional string effective descriptions down to four dimensions. The four-dimensional effective field theories so obtained are characterized by a potential which is determined by background fluxes and sources. The set of vacua of the potential constitutes the \emph{Landscape} of string theory, each of the vacua being endowed with its own value of the cosmological constant. A good understanding of the potential which can arise from string theory models is then the premise to properly understand which vacua are admissible within an effective description and inquire their stability.

In this thesis we provide new tools to determine and explore the Landscape of four-dimensional effective field theories originating from string and M-theory. The main aim is to introduce, within four-dimensional effective descriptions, elements that are predicted from string theory. 

To this end, a hierarchy of forms is introduced within the four-dimensional $\mathcal{N}=1$ supergravity theories. The inclusion of gauge three-forms provides a dynamical way to generate flux--induced superpotentials. Instead, gauge two-forms, dual descriptions of axions, may be eventually gauged by the three-forms to generate a superpotential coupling between the different chiral multiplet sectors of the theory. The mutual constraints among the background fluxes, such as tadpole cancellations, are imposed by gauge four-forms.

A hierarchy of objects, to which the gauge forms couple, is then introduced: four-dimensional BPS-strings, membranes and 3-branes enlarge the Landscape, allowing the background fluxes to consistently change transversing different spacetime regions. The Freed-Witten anomaly cancellations and the changing of tadpole cancellation conditions due to background sources are neatly expressed by BPS--junctions of membranes ending on strings and 3-branes ending on membranes. Membrane-mediated domain wall transitions are studied, which determine how the scalar fields flow connecting a vacuum to another of the Landscape. According to the perturbative regime that is scanned only some transitions are allowed, with a dramatic influence on the spectrum of objects that can be consistently incorporated in the four-dimensional description.

This thesis is divided into three parts. In the first part we show how to include a hierarchy of forms and objects in four-dimensional $\mathcal{N}=1$ globally supersymmetric theories. In the second part, such findings are generalized to generic $\mathcal{N}=1$ supergravity theories. In the last part we reformulate the effective field theories arising from compactifications of string and M-theory, providing the consistency conditions that determine which forms and objects can be appropriately included within the effective description.

\chapter*{Acknowledgements}

First and foremost, I would like to thank my Ph.D. supervisor, \emph{Luca Martucci}. The work presented in this thesis would definitely not exist without his guidance and I am simply honored to have been his Ph.D. student. I am and will be always grateful to him for all the interesting discussions that we had, which have greatly contributed to build my knowledge, and for support and continuous encouragements.

\vspace{0.3em}

I am profoundly indebted to \emph{Dmitri Sorokin}, who has also guided me throughout the Ph.D., as mentor and collaborator. I am grateful to him for his constant support, wise advises, endless patience and kindness.

\vspace{0.3em}

I am grateful to the Instituto de F\'isica Te\'orica UAM-CSIC, for the warm hospitality received during my visiting period there. I am deeply thankful to \emph{Fernando Marchesano}, for his kindness, the collaboration and all the delightful discussions that have vastly improved my knowledge.

\vspace{0.3em}

I would like to thank \emph{Igor Bandos}, for the various collaborations and the immense patience.

\vspace{0.3em}

I am deeply grateful to \emph{Irene Valenzuela}, for all the discussions, invaluable helps and precious advices. 

\vspace{0.3em}



This work certainly reflects the influence of all the Physicists with whom I had the honor to discuss, enriching my knowledge. Among them, I would especially like to thank \emph{Stefano Cremonesi, Gianguido Dall'Agata, Miguel Montero, Gabriele Tartaglino-Mazzucchelli, Antoine Van Proeyen} and \emph{Thomas Van Riet}.

\vspace{0.5em}

I would like to thank all the friends and colleagues that I have met throughout my Ph.D.. In this regard, I would like to mention especially \emph{Alessandro Bombini}, \emph{Lorenzo Papini} and \emph{Federico Pobbe}, for the amazing time spent together and support in difficult moments. I also thank, in particular, \emph{Alvaro Herr\'aez} and \emph{Gabriel Larios} for hospitality and nice discussions during my visit at IFT, Madrid. 

\vspace{0.5em}



Last but not the least, I would like to thank my family for having constantly supported me.

\tableofcontents
\cleardoublepage

\mainmatter

\chapter*{Preface}
\label{preface}
\addcontentsline{toc}{chapter}{\nameref{preface}}

The Standard Model is the most accurate theory that describes the interactions among particles. Several experimental checks have proved its validity at the energy scales that are currently probed by the particle detectors, which have culminated with the recent finding of the Higgs boson \cite{Aad:2012tfa}. The Standard Model stands out for its remarkable simplicity and consistency as a quantum theory of interactions. It is unitary and renormalizable, which allow the Standard Model to be treated as a well defined ultraviolet theory, whose validity goes well beyond the energy scales actually scanned.

However, the Standard Model leaves apart the gravitational interactions. Promoting gravity, in its bare Einstein-Hilbert form, to a quantum theory delivers a strict non-renormalizable theory. Although in the early `70s it was recognized that quantum gravity is finite at one-loop \cite{tHooft:1974toh}, the hopes of canceling all the divergences at subsequent loops soon faded away \cite{Goroff:1985sz}. It was then clear that the simple inclusion of gravity into the Standard model could not be provide a well defined quantum theory. Additionally, moving towards the Planck scale new Physics, not predicted by the sole Standard Model, is expected to emerge, from the unification of forces to the black hole Physics. Nevertheless, the Standard Model coupled to gravity can be regarded as an \emph{effective theory}, which is valid for the energy scales well below the Planck mass and could not provide, per se, a complete ultraviolet theory of quantum gravity.

Having at hand a complete quantum theory of gravity could be crucial to better address fundamental problems of modern physics: for instance, the explanations of dark matter and dark energy, as well as a good understanding of inflationary models beg for ultraviolet complete models. Among them, there is the cosmological constant problem, one of the greatest conundrums in modern Physics. The measured value of the cosmological constant is dramatically small, $\Lambda \sim 10^{-122}$ in reduced Planck units \cite{Aghanim:2018eyx}. The problem of its smallness is quantum in nature: the cosmological constant is the vacuum energy, resulting from the sum of all the possible interactions among the particles of the theory and one should have expected it to be of the order of the cut-off of the theory, which could be estimated by the Planck mass. A near-zero value of the cosmological constant would then require quite miraculous cancellations among the different contributions to the vacuum energy.

String theory, along with its parents M- and F-theory, is the most promising candidate for a quantum theory of gravity and could lead important physical insights on phenomena such as the gauge coupling unification and on the cosmological constant, possibly justifying its microscopic origin. However, in order to extract phenomenological, particle models from string theory, one has to pass to an effective description. At low energy, the ten-dimensional string theory, or the eleven-dimensional M-theory, can be approximated with effective field theories, governing the low-energy dynamics of gravity and the matter fields. In order ot get four-dimensional models, one has to properly \emph{compactify} the higher-dimensional theory down to four dimensions and the structure of the compactification determines the spectrum of the four-dimensional fields. 

Such a compactification procedure comes at a high price: plenty of scalar fields are introduced in the four-dimensional description, the \emph{moduli} of the theory, which couple to the matter fields. Their origin relies on the huge arbitrariness that one has in order to pass from a higher-dimensional to a four-dimensional description. They determine a \emph{moduli space}, which is parametrized by their vacuum expectation values (vevs). Moving across the moduli space the effective description may however dramatically change, for the spectrum of objects differs for different regions of the moduli space. For instance, it is known that moving towards certain regions of the moduli space, an infinite tower of particles become massless, signaling the breaking of the effective description \cite{Ooguri:2006in}. Therefore, sticking with a certain EFT implies that the moduli space is only \emph{partially} explorable. In fact, the moduli vevs determine all the physical relevant quantities of the EFT, such as masses and coupling constants. Certain points in the moduli space could well be associated to strongly coupled regimes for which the EFT breaks down.

Studying the moduli space then intertwines two relevant problems: the determination of the region of validity of the effective field theory and the stabilization of the moduli. 
The latter can be achieved with the introduction of a potential for the moduli. However, it is the perturbative regime, for which the putative EFT is supposed to be valid, that determines how the potential is built and whether the vacua are stable. 
Once the moduli are stabilized, in principle, all the four-dimensional Physics is determined. For instance, the on-shell value potential, evaluated at the point where moduli are stabilized, is the effective cosmological constant. The problem of the cosmological constant is then rephrased in questioning which potentials are allowed in a given string theory effective theory and which \emph{Landscape of vacua} they give rise to. 

The arbitrariness in compactifying higher dimensional theories down to four dimensions does not seem to allow for a systematic study of all the possible EFTs. Rather, one may start with a four-dimensional theory and inquire whether such a description is coherent with its quantum gravity ultraviolet completion. Consistency criteria may be formulated, straightly within the four-dimensional description, which dictates whether the EFT can have a string theory origin. This approach has a distinctive \emph{bottom-up} nature, and goes by the name of the \emph{Swampland program}.

In this work we proceed along these lines, pursuing a bottom-up approach. Our main aim is to introduce, within the four-dimensional effective description, elements that are predicted from string theory. We will include generic gauge $p$-forms in four-dimensional theories, along with the objects that they couple to. The former find their origin in the higher-dimensional gauge potentials, while the latter come from higher-dimensional branes. In reshaping the four-dimensional theories, supersymmetry will be our guiding beacon, furnishing an organizing principle for the fields and objects populating the effective theory.

\newpage

\section*{Synopsis}

This work is organized as follows:

	{\textcolor{darkred}{\sffamily{Chapter 1}}} In the introductory Chapter~\ref{chapter:Intro} we illustrate the key ideas that will accompany us throughout the rest of the work by means of simple bosonic models. We show how gauge three-forms dynamically generate a potential for the scalar fields and how membranes, to which gauge three-forms couple, may allow for jumps of the potential between their sides. The inclusion of strings may further allow the axions to transverse periods once they are encircled. Spacetime filling 3-branes also have their own role, constraining the admissible jumps of the potential. Strings, membranes and 3-branes, along with the gauge forms to which they couple, are the basic tools required to `sail' across the Landscape of vacua of an effective theory. At the same time, we show how the spectrum of gauge forms and objects is related to the consistency conditions of a theory and may allow for detecting when a theory is inconsistent, living in the Swampland.
	
	{\textcolor{darkred}{\sffamily{Chapter 2}}} The second chapter is devoted to the implementation of $p$--forms into $\caln=1$ globally supersymmetric multiplets. The spectrum of $p$--forms influences the potential for the scalar fields of the theory. Our attention mostly focuses on the previously least studied cases of gauge three- and four-forms. The former provide a way to dynamically generate either F- or D-term potentials, quadratically and linearly depending on some constants, while the latter constrain the parameters that appear in such dynamically generated potentials. We also revisit the inclusion of gauge two-forms, which we understand as dual to axions, into supersymmetric theories. We show that F-term couplings between axions and another chiral sector may be elegantly reabsorbed into a gauging of the linear multiplets. At the end of the chapter we will be able to couple all the possible gauge $p$--forms to any given $\caln=1$ supersymmetric theory, while still preserving supersymmetry.
	
	{\textcolor{darkred}{\sffamily{Chapter 3}}} In Chapter~\ref{chapter:ExtObj} we introduce BPS--strings, membranes and 3-branes into the $\caln=1$ globally supersymmetric theories given in Chapter~\ref{chapter:GSS}. The link which allows us to couple these extended objects to $\caln=1$ theories is provided by the introduction of \emph{super $p$--forms}, namely superspace defined $p$--forms built out of the supersymmetric multiplets introduced in Chapter~\ref{chapter:GSS}. The guiding principle that we shall follow to write down their supersymmetric actions is the $\kappa$-symmetry: being a local worldvolume fermionic symmetry, $\kappa$-symmetry is the way in which supersymmetry can be (partially) linearly realized over the objects' worldvolumes. The physical consequences of the introductions of these objects in supersymmetric theories are also explored.
	
 	{\textcolor{darkred}{\sffamily{Chapters 4 and 5}}} The fourth and the fifth chapters deal with the extension of the results of Chapters~\ref{chapter:GSS} and~\ref{chapter:ExtObj} to $\caln=1$ supergravity. Our globally supersymmetric findings are easily generalized to the local theories by means of the super-Weyl invariant approach, at the price of introducing an unphysical compensator. In Chapter~\ref{chapter:Sugra} we formulate generic $\caln=1$ Supergravity theories containing any number of gauge two-, three- and four-forms, which are then coupled to BPS--string, membranes and 3-branes in Chapter~\ref{chapter:Sugra_ExtObj}. All the results are given in both the super-Weyl invariant formalism and, after properly gauge-fixing the compensator, in the traditional supergravity formulation.

	{\textcolor{darkred}{\sffamily{Chapter 6}}} Concrete examples of supergravity theories with a hierarchy of forms are given in Chapter~\ref{chapter:EFT}. We reformulate four-dimensional effective field theories originating from compactification of Type II superstring theories and M-theory. The scalar potential, generated by background fluxes, is understood as stemming from integrating out gauge three-forms. The tadpole cancellation conditions, ubiquitous in compactification scenarios, here appear as a gauging of the three-forms by gauge four-forms. The axions are then dualized to gauge two-forms, eventually gauged by three-forms reabsorbing F-terms couplings between different moduli sectors. The proposed reformulation is general and, as we show, can include moduli from both closed and open string sectors, as well as involve nongeometric fluxes.
	
	{\textcolor{darkred}{\sffamily{Chapter 7}}} The seventh chapter is devoted to the consistency conditions that the newly reformulated theories of Chapter~\ref{chapter:EFT} need to satisfy. We show that, according to the perturbative regime, only some extended objects can be included in the effective theory. This has profound physical consequences, since it dictates which portions of the string Landscape are explorable within the same effective theory. 
	
	{\textcolor{darkred}{\sffamily{Appendices}}} The Appendices~\ref{app:Diff_Conv},~\ref{app:Superspace_Conv} and~\ref{app:Sugra_Conv} collect conventions on differential forms and superspace that we use throughout this work and the explicit computations of components of the superfields. The Appendix~\ref{app:SW} delivers an introduction to the super-Weyl invariant formalism in $\caln=1$ supergravity. The interplay between Freed-Witten anomalies and axion monodromies is briefly summarized in Appendix~\ref{app:FW_AM}. 

\newpage

\section*{Papers}

This thesis is elaborated from the following papers:
\begin{itemize}
	\item \fullcite{Farakos:2017jme}
	\item \fullcite{Bandos:2018gjp}
	\item \fullcite{Cribiori:2018jjh}
	\item \fullcite{Bandos:2019qok}
	\item \fullcite{Lanza:2019xxg}
\end{itemize}
and the following proceedings:
\begin{itemize}
	\item \fullcite{Farakos:2017ocw}
	\item \fullcite{Bandos:2019wgy}
	\item \fullcite{Bandos:2019khd}
\end{itemize}

\noindent During the Ph.D. course I have also contributed to:
\begin{itemize}
	\item \fullcite{Holo}
\end{itemize}

\noindent Part of the work hereby presented is original and never published before.


\chapter{Introduction: a toolkit to sail across the Landscape}
\label{chapter:Intro}

A better understanding of the microscopic structure of the effective potential is one of the main scopes of this work. The approach that we follow is, in a certain sense, \emph{bottom-up}, aiming at including, genuinely from a four-dimensional perspective, all the basic elements that string theory predicts, equipped with the proper consistency conditions.

In this chapter we briefly introduce the basic ideas that will accompany us throughout this work. Tightening the links between four-dimensional effective field theories and their higher-dimensional origin will require the introduction of gauge gauge two-, three- and four-forms. Such $p$-forms will provide an alternative, possibly more complete, description for axions, background fluxes and tadpole cancellation conditions.

The introduction of a \emph{hierarchy} of $p$-forms will allow us to introduce in the EFT the extended objects to which they couple, namely strings, membranes and 3-branes. The complete picture draws a spacetime populated by various extended objects, which allow for connecting the different vacua of the Landscape.
\section{String theory, EFTs and the cosmological constant}

In order to construct four-dimensional effective theories originating from string theory, the typical procedure is \emph{top-down} in nature: one starts with a $D$-dimensional effective field theory, formulated in $\scrm_{D}$ and compactify it over an \emph{internal} $(D-4)$-dimensional manifold $X$:
\be
\scrm_{D} = \scrm_{4} \times X\,,
\ee
where $\scrm_{4}$ is the \emph{external} four-dimensional manifold. In this work, we will always consider the internal manifold $X$ to be compact and chosen in such a way that in $\scrm_{4}$ the minimal amount of supersymmetry is preserved. The four-dimensional effective theory is then obtained by reducing the higher-dimensional theory\footnote{We stress that throughout this work we always consider effective theories with \emph{at most} two derivatives acting on the various fields, in both higher- and lower-dimensions.} over $X$ with the spectrum of fields determined by dimensionally reducing the higher-dimensional fields. 

However, the compactification procedure come at a price: many scalar fields appear in the four-dimensional effective theories, which are associated to the admissible deformations of the internal manifold. These fields do actually constitute the \emph{moduli} $\varphi$ of the theory: they are not subjected to any potential and, as such, do not acquire mass. Therefore, the effective theory would be characterized by an infinite family of equivalent vacua, parametrized by the different values of the moduli $\varphi$. Furthermore,  the abundance of such massless particles would be in clear conflict with the observations. In fact, moduli couple to the other fields of the effective theory, in turn determining long--range forces which have been not observed in our Universe. 

The aforementioned, severe issue of compactifications goes by the name of \emph{problem of moduli stabilization}. A viable solution was found in the late `90 \cite{Polchinski:1995sm,Michelson:1996pn,Taylor:1999ii} and further developed in the first years of 2000 \cite{Grana:2000jj,Grana:2001xn,Frey:2002hf,Kachru:2002ns,Kachru:2002he,Gubser:2000vg} (see also  \cite{Grana:2006is,Douglas:2006es,Grimm:2005fa,Becker:2007zj} for reviews): introducing nonvanishing background values for the $p$--form fluxes, which thread the internal manifold $X$, induces a potential for the moduli which may allow for giving them mass, stabilizing them at a scale $m_\varphi$ fixed by the fluxes.  At the level of the four-dimensional effective field theory, this can be understood with the appearance of a potential that acquires the very general form 
\be
\label{Intro_cc_V}
V (\varphi) = \frac12 T^{AB}(\phi)(N_A +f_A(\phi))(N_B +f_B(\phi))+ \hat V(\varphi) 
\ee
where $N_A$ constitute a set of quantized constants and $f_A(\varphi)$ and $\hat V(\varphi)$ are arbitrary functions of the scalar fields. The constants $N_A$ are related to the $p$--form background fluxes, while $f_A(\varphi)$ and $\hat V(\varphi)$ may include both perturbative or non-perturbative contributions. Furthermore, typically the constants $N_A$ cannot be arbitrarily chosen, for they have to obey some mutual constraints, for example, the \emph{tadpole cancellation conditions}. The `effective' cosmological constant is then obtained by simply setting the scalar fields in \eqref{Intro_cc_V} at the vevs $\varphi_*$ at which they are stabilized:
\be
\label{Intro_cc_Vcc}
\Lambda_{\text{effective}} = V (\varphi_*)\;.
\ee

The potential \eqref{Intro_cc_V} has a higher-dimensional origin, but let us now consider a \emph{bottom-up} perspective. Namely, let us assume that we are working with a four-dimensional theory, characterized by some scalar fields $\varphi$, eventually regarded as moduli, that we would like to stabilize: what kind of potential could we choose and how many parameters $N_A$ could we tune to stabilize the moduli? If one wishes the four-dimensional theory to have a proper \emph{quantum} origin, namely from string theory, then one has to stick with the potential \eqref{Intro_cc_V}, the one that string theory predicts. In other words, we do not have \emph{full} freedom within our four-dimensional EFT. 

However, a potential \eqref{Intro_cc_V}, once imported into the four-dimensional theory, is completely \emph{blind} to any constraint that string theory imposes on the quantized constants, as well as the origin of the $N_A$ fluxes gets completely forgotten, being merely treated as constants. A new perspective is required to re-examine the potential and understand the deeper origin of each term of \eqref{Intro_cc_V} so as to tighten the link between the four-dimensional description and its higher-dimensional origin.

To this aim, the first step is the inclusion of the full set of all the possible gauge $p$--forms within the four-dimensional description, which are indeed predicted from a higher dimensional string theory description. In fact, in higher dimensions, some field strengths $F_{q+1} = \d A_q$ appear. The four-dimensional fields which such field strengths give rise are obtained by decomposing $F_{q+1}$ into an external part, living in $\scrm_{4}$ and into an internal part on $X$, schematically as follows:
\be
\label{Intro_Fpdec}
F_{q+1} = \underbrace{\d A_{p}}_{\text{external}} \wedge \underbrace{\omega_{q-p}}_{\text{internal}}\,,
\ee
where $\omega_{q-p}$ is a $(q-p)$--form in $X$. It is then clear that the existence of some gauge $p$--forms $A_p$ in four-dimensions is related to the homology structure of the internal space which allows for decompositions as \eqref{Intro_Fpdec}. However, generically, one should expect that \emph{all} the gauge $p$--forms, for $p=0,\ldots,4$, are present in four dimensions and a `faithful' four-dimensional theory ought to include all of them.

In this regard, we will introduce in the effective four-dimensional theories are gauge three-forms $A_3^A$. These were typically disregarded in the past since they do no carry physical degrees of freedom in four dimensions. We will indeed show in Section~\ref{sec:Intro_Pot} that gauge three-forms can be interpreted as `dual' to the quantized constants $N_A$ so that, once the three-forms are integrated out, the constants $N_A$ appear in the potential. Stated differently, gauge three-forms may dynamically generate entire portions of the potential. Their role does not stop here: if also non-dynamical gauge four-forms $C_4^I$ are introduced, these may gauge the three-forms as $\d A_3^A + Q^A_I C_4^I$ which results in forcing the constant $N_A$ to satisfy some (linear) constraint, mimicking \emph{directly} in four dimensions the tadpole conditions. 

In four-dimensional EFTs, symmetries sometimes will suggest us alternative representations of fields. In some perturbative regimes, the action may develop an approximate `axionic' symmetry for a field $a$, namely $a \to a +c$, with constant $c$. Then, one can alternatively represent the \emph{axion} $a$, regarded as a zero--form, with a gauge two-form $\calb_2$. Via a St\"uckelberg trick, its field strength can be gauged by the three-forms as $\d \calb_2 + c_A A_3^A$ whose interpretation, for the potential \eqref{Intro_cc_V}, is the appearance of a mass term for the axion.  

Furthermore, whenever gauge one-form $A_1^\sigma$ are present in the effective description, these may be gauged by the two-forms as $\d A_1^\sigma + k_\Lambda^\sigma \calb_2^\Lambda$. Such a gauging, does also influence the structure of the potential, albeit in this chapter we will not explore such a possibility.

A \emph{hierarchy} of gauge $p$--forms is then established within the EFT, allowing us to couple the extended objects which are electrically charged under them: strings, membranes and 3-branes. These will be our basic tools to explore the Landscape of vacua. In fact, membranes will make the constants $N_A$ of the potential change across the different regions separated by the membranes. The moduli therefore feel \emph{different} potentials for each region delimited by the membranes and are consequently (classically) stabilized at different vacua on the two sides, with different masses and inducing diverse cosmological constants \eqref{Intro_cc_Vcc}. Membranes then become the objects which allow one to interpolate among different vacua of the Landscape and their very existence is a criterion which determines what vacua one can hope to reach within a single EFT.
Four-dimensional strings will instead tell when an axion is subjected to  a monodromy transformation, while 3-branes make the flux constraint change, according to membrane-induced transitions.

In this chapter, we show how to re-interpret the potential \eqref{Intro_cc_V} in terms of a hierarchy of forms by means of simple bosonic models. Also the role of the extended objects is examined in simple setups, which will allow us to highlight their main physical properties. These models are later imported in globally supersymmetric theories in the Chapters~\ref{chapter:GSS} and~\ref{chapter:ExtObj} and in supergravity in the Chapters~\ref{chapter:Sugra} and~\ref{chapter:Sugra_ExtObj}. 

\section{Generating the potential of effective field theories}
\label{sec:Intro_Pot}

Gauge three-forms are typically neglected in four-dimensional theories, owing to the fact that the gauge redundancy is so vast that it does not allow them to have propagating degrees of freedom \cite{Brown:1987dd}. Nevertheless, their inclusion may influence the very structure of the potential. In fact, the procedure of integrating gauge three-forms out delivers new constants, which were not present before, in the effective description. This rather peculiar phenomenon was firstly explored in supergravity contexts \cite{Aurilia:1980xj}, where gauge three-forms naturally appear after reducing higher--dimensional theories, and later found fertile ground in cosmological \cite{Hawking:1984hk,Brown:1987dd,Brown:1988kg,Duff:1989ah} and inflationary models \cite{Kaloper:2008fb,Kaloper:2011jz,Marchesano:2014mla,Dudas:2014pva,Bielleman:2015ina,Valenzuela:2016yny}, as well as in connections with the strong $CP$-problem \cite{Dvali:2005an,Dvali:2004tma,Dvali:2005zk,Dvali:2013cpa,Dvali:2016uhn,Dvali:2016eay}.

In this section, we develop on the on-shell effect of the gauge three-forms and in which sense they can `generate' nontrivial contributions to the potential.

\subsection{An alternative understanding of the cosmological constant}
\label{sec:Intro_Pot_cc}

One of the first phenomenological applications of gauge three-forms in physics was in understanding the generation of the cosmological constant. In the `80s \cite{Aurilia:1980xj,Hawking:1984hk,Brown:1987dd,Brown:1988kg,Duff:1989ah}, it was recognized that the coupling of three-forms to the simple Einstein-Hilbert action could nontrivially contribute to the value of the cosmological constant. We now briefly recall this model, which provides possibly the simplest example of phenomenological models including three-forms.

The action describing the interaction of a single, real gauge three-form $A_3 = \frac1{3!} A_{mnp} {\rm d} x^m\wedge \d x^n \wedge \d x^p$ with gravity is\footnote{We refer to Appendix~\ref{app:Diff_Conv} for the conventions and useful identities of the differential forms used throughout this work.}

\be
\label{Intro_cc_S3f}
\begin{aligned}
	S &=  \int_{\Sigma} \left( \frac12 \Mp^2 R *\!1  - \Lambda_0 *\!1 -  \frac12 F_4 *\!F_4 \right)+S_{\rm bd}\;.
\end{aligned}
\ee
The integration is here performed over the full spacetime, denoted with $\Sigma$, whose boundary, eventually located at infinity, we will denote with $\del\Sigma$. The first term in \eqref{Intro_cc_S3f} is the usual Einstein-Hilbert term, encoding the dynamics of the graviton $g_{mn}$, and we have also added a `bare' cosmological constant term $\Lambda_0$.\footnote{In writing down the `pure' gravitational part of the action, we have disregarded a possible contribution from the Gibbons-Hawking-York term \cite{York:1972sj,Gibbons:1976ue}, which is present whenever the manifold $\Sigma$ has a nontrivial boundary, in order to ensure the correct variational problem for the components of the metric $g_{mn}$ normal to $\del\Sigma$. Such a boundary term will be reinstated in Section~\ref{sec:LandSwamp_ExtObj_DW} -- see for example \eqref{SL_DW_Sbos} -- where, once the metric is set on-shell, it  nontrivially contributes to the curvature term.} The last term in the parentheses in \eqref{Intro_cc_S3f} expresses just the kinetic terms for the gauge three-form $A_3$, which are written in terms of its four-form field strength $F_4 = \d A_3$ and are invariant under the gauge transformation $A_3 \to A_3 + \d \Lambda_2$, with $\Lambda_2$ a generic two-form.\footnote{The kinetic terms can indeed be recast as
\be
- \frac12 \int_\Sigma  F_4 *\!F_4 = - \frac1{2 \cdot 4!} \int_\Sigma F_{mnpq} F^{mnpq} *\! 1\,,
\ee
where $F_{mnpq} = 4 \del_{[m} A_{npq]}$.}

The last term in the action \eqref{Intro_cc_S3f} deserves particular attention: it collects the boundary contributions which include the gauge three-form $A_3$. These have to be added in order to guarantee that the variational problem for the gauge three-form is well-posed, which can be understood as follows. Let us consider the variation of the action \eqref{Intro_cc_S3f} with respect to $A_3$
\be
\label{Intro_cc_dS3f}
\begin{aligned}
	\delta_{A_3} S &=  - \int_{\Sigma} {\rm d} \delta A_3 *\!F_4 + \delta_{A_3} S_{\rm bd}
	\\
	&= - \int_{\Sigma} \delta A_3 \d *\!F_4  - \int_{\del\Sigma} \delta_{A_3} *\!F_4    + \delta_{A_3}  S_{\rm bd}\;,
\end{aligned}
\ee
where in the second line we have performed an integration by parts and used the Stokes' theorem in the convention \eqref{AppDiff_Stokes}. If one wishes to obtain the equations of motion for the gauge three-form $A_3$ from the variation \eqref{Intro_cc_dS3f}, proper boundary conditions have to be imposed for the variations of $A_3$ over $\del \Sigma$. The first, na\"ive option could be to impose $\delta A_3|_{\del\Sigma} = 0$. However, such a condition is \emph{not} gauge invariant and, as such, has to be discarded. The correct, gauge invariant boundary conditions are
\be
\label{Intro_cc_F4bd}
\delta F_4 |_{\del\Sigma} = \delta \d A_3 |_{\del\Sigma}  = 0\;.
\ee
This implies that $\delta A_3|_{\del\Sigma}$ is not generically zero and whenever a boundary contribution contains a `naked' $\delta A_3$, namely without derivatives acting on it, that contribution has to be canceled. In the present case, this is achieved by selecting the boundary terms in \eqref{Intro_cc_S3f} to be
\be
\label{Intro_cc_Sbd}
S_{\rm bd} = \int_{\del\Sigma} A_3 *F_4 = \frac{1}{3!} \int_{\Sigma} \del_m \left(A_{npq} F^{mnpq} \right) *1 \,,
\ee
whose variation is such that the boundary contributions in \eqref{Intro_cc_dS3f} exactly cancel between one another. 

Having defined the proper boundary conditions, we are now ready to compute the equations of motion for the gauge three-form $A_3$. These can be immediately read from \eqref{Intro_cc_dS3f} and acquire the simple form
\be
\label{Intro_cc_dF4}
\d *\! F_4 = 0\,.
\ee
This equation can be immediately integrated out, obtaining
\be
\label{Intro_cc_dF4sol}
* F_4 = \lambda\,,
\ee
where $\lambda$ is just a real constant, fixed by the boundary conditions $*\!F_4 |_{\del \Sigma} = \lambda$, compatibly with \eqref{Intro_cc_F4bd}. Although fixed at the boundary, one can indeed choose arbitrary values for $\lambda \in \mathbb{R}$. We can then plug the solution \eqref{Intro_cc_dF4sol} into the action \eqref{Intro_cc_S3f} where we started from, obtaining
\be
\label{Intro_cc_S3fos}
\begin{aligned}
	S &=  \int_{\Sigma} \left[ \frac12 \Mp^2 R *\!1  - \left(\Lambda_0 + \frac12 \lambda^2 \right) *\!1  \right]\;.
\end{aligned}
\ee
The \emph{effective} cosmological constant is then the sum of its bare value $\Lambda_0$ and the three-form contribution:
\be
\label{Intro_cc_Leff}
\Lambda_{\rm eff} = \Lambda_0 + \frac12 \lambda^2 \,.
\ee

Some crucial comments are in order. First, we stress that the boundary terms \eqref{Intro_cc_Sbd} have \emph{also} be taken into account in computing \eqref{Intro_cc_S3fos}, since they contribute to the cosmological constant term with
\be
\label{Intro_cc_Sbdos}
S_{\rm bd} = \lambda \int_{\Sigma} F_4 = - \lambda^2 \int_{\Sigma} *1\,.
\ee
By comparing the `three-form action' \eqref{Intro_cc_S3f} with \eqref{Intro_cc_S3fos}, we realize that, in \eqref{Intro_cc_S3f}  we can indeed identify an `off-shell' cosmological constant term as
\be
\int_\Sigma \Lambda_{\text{eff,`off-shell'}} *1 =  \int_{\Sigma} \left(\Lambda_0 *\!1 + \frac12 F_4 *\!F_4 \right)+\int_{\del\Sigma} A_3 *F_4 \,,
\ee
which then acquires the proper meaning of a cosmological constant once the gauge three-form is integrated out via \eqref{Intro_cc_dF4sol}. In this regard, it is crucial the presence of the boundary terms \eqref{Intro_cc_Sbd}. Assume, for example, that $\Lambda_0 = 0$, so that the cosmological constant purely stems from  the gauge three-form $A_3$: if the boundary contributions \eqref{Intro_cc_Sbdos} are neglected, we would get a cosmological $\Lambda_{\rm eff} =  -\frac12 \lambda^2$ with the opposite sign!\footnote{In order to understand why the boundary terms do contribute to the cosmological constant, one can start by adding the topological term $\lambda \int_\Sigma F_4$, written first equality of \eqref{Intro_cc_Sbdos}. Such a topological term, although a divergence, contributes to the cosmological constant since it introduces three-form--graviton vertices of coupling constant $\lambda$.}  The inclusion of the boundary terms in this `three-form formalism' was source of confusion in the past \cite{Hawking:1984hk,Aurilia:1980xj,Duncan:1989ug,Duff:1989ah}. They were indeed first recognized to be a `topologica term' as  in the first equality of  \eqref{Intro_cc_Sbdos} \cite{Aurilia:1980xj,Duff:1989ah}: the constant $\lambda$ was there considered fixed, directly inserted from the start in the action. It was then recognized that such a topological term was indeed a manifestation of a boundary term required by the variational problem \cite{Duncan:1989ug,Brown:1987dd,Brown:1988kg}.

This very simple example should make clear that gauge three-forms, although apparently harmless, can instead be crucial elements for any effective field theory, owing to the fact that they contribute to the potential. In the following section, we will see how such a very simple model can be modified to generate a more general kind of potentials.

\subsection{Potential and gauge three-forms}
\label{sec:Intro_Pot_gen}

With the sole ingredients of gravity and gauge three-forms, one cannot achieve more than obtaining  a constant potential, once the gauge three-forms are set on-shell. However, we can generalize the previous simple model to more general cases pretty straightforwardly. Let us take into account also the presence of a set of $n$ real scalar fields $\phi^a$, $a=1,\ldots, n$. The action describing the interaction of gravity, gauge three-forms and scalar fields takes the general form
\be
\label{Intro_Pot_S3f}
\begin{aligned}
	S &=  \Mp^2 \int_{\Sigma} \left( \frac12 R *1- \frac12G_{ab}(\phi) \d \phi^a \wedge *\d \phi^{b} \right)
	\\
	&\quad\, -\int_\Sigma \Big[\frac12 T_{AB} (\phi)F^A_4*\!F^B_4+ f_A (\phi) F^A_4+
	\hat V (\phi)*\!1\Big] +S_{\rm bd}\;.
\end{aligned}
\ee
Here $ G_{ab}(\phi)$ is a generic field-dependent metric over the field space parametrized by the scalar fields $\phi^a$ and, as such, is assumed to be positive definite at any given point $\phi^a_0$ of the field space. The second line of \eqref{Intro_Pot_S3f} comprises the three-form kinetic terms and the interaction of the gauge three-forms with the scalar fields. More specifically, the kinetic matrix $T_{AB} (\phi)$ is generically allowed to depend on the scalar fields $\phi^a$; we further assume $T_{AB}$ to be symmetric $T_{AB} = T_{BA}$, but it is \emph{not} required to be positive definite.\footnote{Since, as stressed above, gauge three-forms do not carry any physical degrees of freedom, negative--definite kinetic terms are not sources of \emph{bad} ghosts which would imply a lack of unitarity.} Moreover, $f_A (\phi)$ is a further field-dependent quantity, providing the couplings of the fields $\phi^a$ to the the gauge three-forms and $\hat V (\phi)$ collects all the other possible scalar field interactions. 

The last term in the second line of \eqref{Intro_Pot_S3f} indeed collects all the boundary terms involving the gauge three-forms. As explained above, these are necessary to ensure the correct formulation of the three-form variational problem. The proper, gauge invariant boundary conditions that have to be imposed over the gauge three-forms are, in fact
\be
\delta F_4^A |_{\rm bd} \stackrel{!}{=} 0\,,
\ee
rather than $\delta A_3^A  |_{\rm bd} =0 $. It is therefore necessary that, once we take the variation of \eqref{Intro_Pot_S3f} with respect to the gauge three-forms $A_3^A$, \emph{all} the contributions which involve the variations $\delta A_3^A$ over the spacetime boundary $\del \Sigma$ strictly need to vanish. Explicitly, the boundary contributions to the action $S_{\rm bd}$ have to be chosen such that the $A_3^A$--variation of the action \eqref{Intro_Pot_S3f} gives
\be
\label{Intro_Pot_deltaS3f}
\begin{aligned}
	\delta_{A_3^A} S  &= - \int_\Sigma \left[ \d \delta A_3^A  \left( T_{AB}(\phi) *\!F^B_4 + f_A (\phi)\right)\right]+ \delta S_{\rm bd}\,,
	\\
	&= - \int_{\Sigma} \delta A^B_3 \d \left[T_{AB}(\phi) *\!F^A_4+f_A (\phi)\right] \;,
\end{aligned}
\ee
which is achieved by setting
\be
\label{Intro_Pot_Sbd}
S_{\rm bd} = \int_{\del\Sigma}(T_{AB}*\!F^A_4+f_B)A^B_3\,.
\ee
We also note that the action \eqref{Intro_Pot_S3f} is invariant under the gauge transformation  
\be
\label{Intro_Con_A3gauge}
A^A_3\rightarrow A^A_3 + \d \Lambda_2^A\, ,
\ee
for any two-form $\Lambda_2^A$. 

We now regard the second and the third line of \eqref{Intro_Pot_S3f}, given 
\eqref{Intro_Pot_Sbd}, as an `off-shell' potential for the scalar fields
\be
\label{Intro_Pot_Voff}
\begin{aligned}
	-\int_\Sigma V_{\text{`off-shell'}} *1  &=   -\int_\Sigma \Big[\frac12 T_{AB} (\phi)F^A_4*\!F^B_4+ f_A (\phi) F^A_4+
	\hat V (\phi)*\!1\Big] 
	\\
	&\quad\,+ \int_{\del\Sigma}(T_{AB}* F^A_4+f_B)A^B_3\,,
\end{aligned}
\ee
with `off-shell' referring to the presence of the gauge three-forms, which are not yet integrated out. Clearly, as it stands,  \eqref{Intro_Pot_Voff} is also not, strictly speaking, a potential for the scalar fields $\phi^a$, owing to the fact that the gauge three-forms are still there. Let us now integrate out the gauge three-forms. Their equations of motion, immediately read from the variations \eqref{Intro_Pot_deltaS3f}, are
\be
\label{Intro_Pot_dF4eom}
\d \left[T_{AB}(\phi)*\!{F}^B_4+f_A (\phi) \right]=0 \;.
\ee
We may then integrate them, producing 
\be
\label{Intro_Pot_F4eom}
T_{AB} (\phi) *\!{F}^B_4+f_A (\phi)= - N_A \;,
\ee
where $N_A$ are integration constants. Remarkably, for each gauge three-form $A_3^A$ a constant $N_A$ is generated. Classically, $N_A$ are generic real constants; however, in a quantum theory, as it is explored in details in the later Section~\ref{sec:Sugra_Quant}, the constants $N_A$ are quantized, basically because the left hand sides of \eqref{Intro_Pot_F4eom} are the momenta conjugated to the three-forms $A_3^A$. Moreover, the boundary conditions \eqref{Intro_Pot_F4eom} allow also the boundary contributions \eqref{Intro_Pot_Sbd} to be invariant under \eqref{Intro_Con_A3gauge}.

Once \eqref{Intro_Pot_F4eom} are plugged into \eqref{Intro_Pot_Voff} setting the three-forms on-shell, makes it become a proper potential for the scalar fields
\be
\label{Intro_Pot_Von}
\begin{aligned}
	V (\phi)  &=  \frac12 T^{AB}(\phi)(N_A +f_A(\phi))(N_B +f_B(\phi)) +
	\hat V (\phi)
\end{aligned}
\ee
where we have assumed that $T_{AB}$ is invertible, with inverse $T^{AB} = (T_{AB})^{-1}$. Therefore, the action \eqref{Intro_Pot_S3f}, upon using \eqref{Intro_Pot_F4eom} reduces to an action which depends \emph{only} on the scalar fields $\phi^a$ interacting with gravity
\be
\label{Intro_Pot_S3fb}
\begin{aligned}
	S &=  \Mp^2 \int_{\Sigma} \left( \frac12 R *1-\frac12 G_{ab}(\phi) \d \phi^a \wedge *\d \phi^{b} - V (\phi)*\!1\right) 
\end{aligned}
\ee
with the scalar potential $V$ given in \eqref{Intro_Pot_Von}.

The actions \eqref{Intro_Pot_S3f} and \eqref{Intro_Pot_S3fb} are therefore strictly tightened one another, with the latter, purely scalar, being obtained from the former after integrating out the gauge three-forms: in some sense, the actions \eqref{Intro_Pot_S3f} and \eqref{Intro_Pot_S3fb} are \emph{dual} of one another. Indeed, the two formulations may be related by an electro-magnetic like duality. The equations of motion \eqref{Intro_Pot_F4eom} suggests that the (field strengths of the) gauge three-forms $A_3^A$ are `dual' to the constants $N_A$ seen as `zero-form field strength', obeying the trivial Bianchi identities $\d N_A =0$. This is a quite degenerate electro-magnetic duality and not particularly useful per se. One can however translate the duality at the level of the actions as follows. We promote the constants $N_A$ in \eqref{Intro_Pot_S3fb} to real scalar fields $y_A$ and, after that, impose their Bianchi identities -- namely, the constancy of $y_A$ -- via a Lagrange multiplier, which we choose to be a three-form $A_3^A$: 
\be
\label{Intro_Pot_S3fc}
\begin{aligned}
	S &=  \Mp^2 \int_{\Sigma} \left( \frac12 R *1- \frac12G_{ab}(\phi) \d \phi^a \wedge *\d \phi^{b} \right)
	\\
	&\quad\, -\int_\Sigma \Big[\frac12 T^{AB} (\phi)(y_A+ f_A (\phi)) (y_B + f_B(\phi))+
	\hat V (\phi)\Big] *\!1
	\\
	&\quad\,- \int_{\Sigma} \d y_A A_3^A\;.
\end{aligned}
\ee
The action offers two possibilities. On the one hand, one may immediately integrate out the gauge three-forms $A_3^A$. As anticipated, this sets the Bianchi identities for the zero-form field strengths $y_A$
\be
\label{Intro_Pot_BIy}
\d y_A = 0 \quad \Rightarrow \quad y_A = N_A \;\; \text{(constants)}\,,
\ee
reducing the action \eqref{Intro_Pot_S3fc} to \eqref{Intro_Pot_S3fb}.

On the other hand, one can integrate out the scalar fields $y_A$ as
\be
\label{Intro_Pot_yA}
y_A = - T_{AB} (\phi) *\! F_4^B  - f_A (\phi)
\ee
which, plugged into \eqref{Intro_Pot_S3fc}, leads back to the action \eqref{Intro_Pot_S3f} which contains gauge three-forms explicitly, also delivering the proper boundary terms.

The action \eqref{Intro_Pot_S3fc}, to which we refer as \emph{master action}, conveys the joining link between the three-form action \eqref{Intro_Pot_S3f} and an `ordinary', scalar field action \eqref{Intro_Pot_S3fc}, making it explicit the duality between them.

\section{Jumps in the Landscape}
\label{sec:Intro_Jumps}

\begin{wrapfigure}{r}{0.4\textwidth}
	\includegraphics[width=0.4\textwidth]{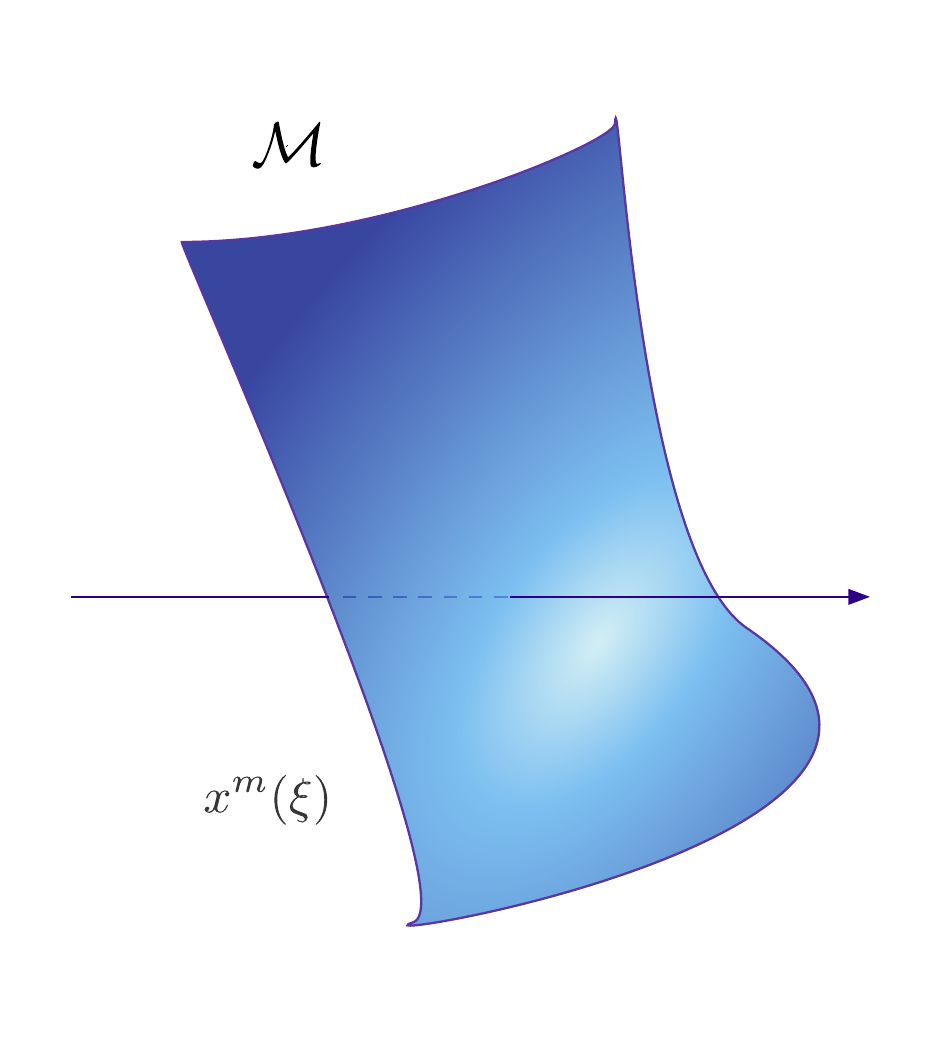}
	\caption{\footnotesize{A membrane as a hypersurface of codimension one described by the embedding \eqref{Intro_Jumps_Memb_Embed}.}}\label{fig:MembG}
\end{wrapfigure}

We have shown that gauge three-forms actively contribute to the determination of the potential for the scalar fields. They indeed generate, by integrating them out, arbitrary constants which, entering the potential, influence the stabilization of the scalar fields as well as the value of the cosmological constant. However, one may wonder that, after all, the three-form perspective does not help much in understanding the physics that lies behind a potential: at the end of the day, the potential dynamically generated by gauge three-forms, as for example \eqref{Intro_Pot_Von}, can be equivalently described in terms of a scalar theory, without gauge three-forms. Their inclusion is also further complicated by the presence of boundary terms, which are necessary for the consistency of the variational problem. The natural question that begs an answer is why gauge three-forms are useful and worth considering within a theory.

Exactly as gauge one-form potentials couple to point particles, with the usual, minimal interaction $q \int A_1$ where the gauge one-form is integrated over the particle worldline, gauge three-forms have their own extended objects to couple with, which are three-dimensional \emph{membranes}. In fact, string theory is not just a theory of particles, for it can include extended objects as well, which manifest themselves also in four dimensions. Let us consider a $(p+1)$--extended objects in higher dimensions. If it `wraps' an internal cycle of dimension $q$ in $X$, then it extends for the residual $(p-q+1)$--dimensions in the external space. In order to get a membrane in four dimensions, then the wrapped cycle has to be of dimension $(p+2)$, as depicted in Fig.~\ref{fig:MembG}. A good four-dimensional theory has to take into account the presence of such objects, properly describing their dynamics.

\begin{figure}[t]
	\centering
	\includegraphics[width=10cm]{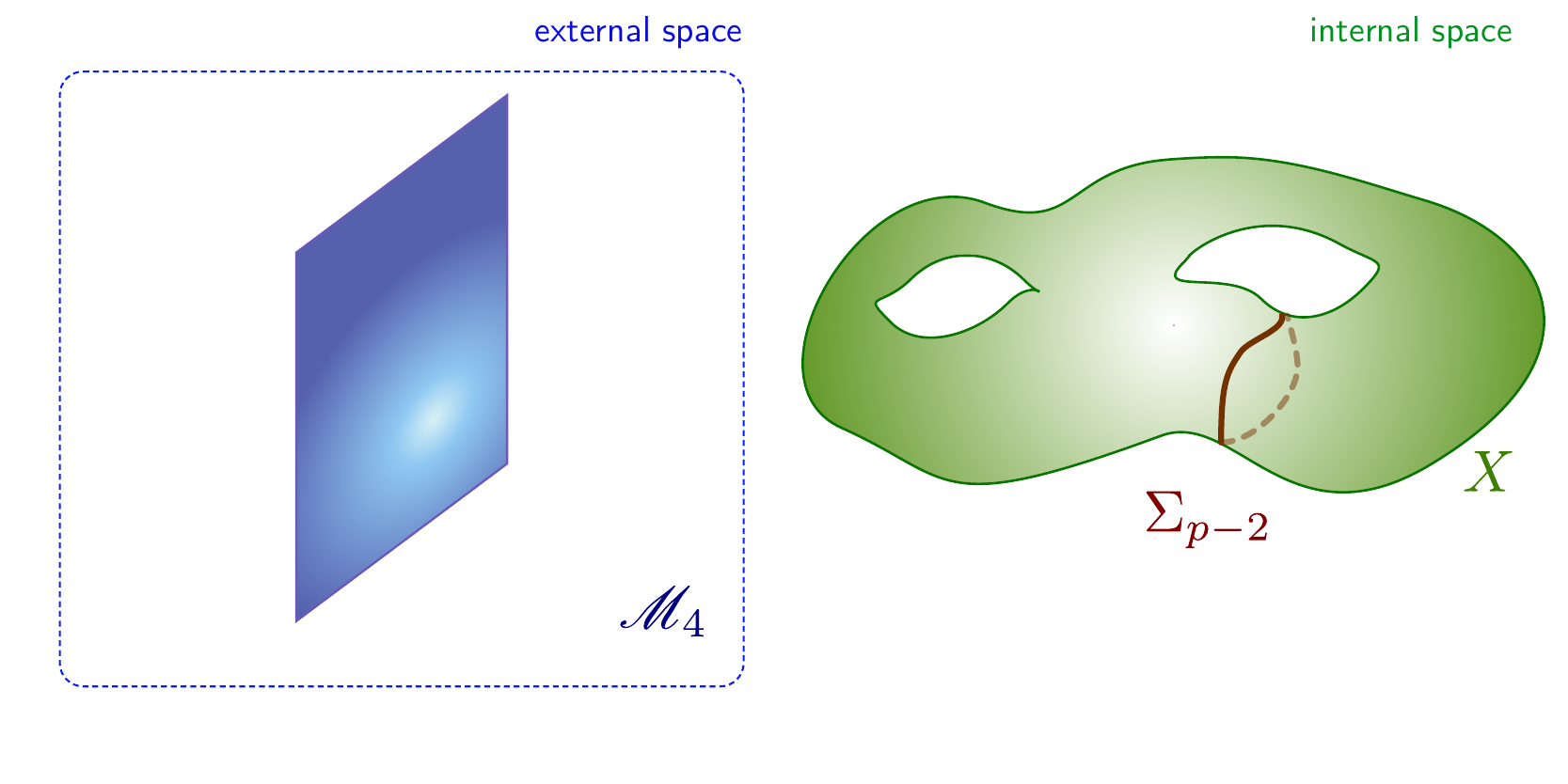}
	\caption{\footnotesize{A $(p+1)$--dimensional object can give birth to a membrane in the four-dimensional, external space if it wraps a $(p-2)$--cycle in the internal space.}}\label{fig:MembHD}
\end{figure}

First, the worldvolume $\calm$ of a membrane is parametrized by three coordinates $\xi^i$, $i=0,1,2$. It is then described as a hypersurface of codimension one in the \emph{target} four-dimensional spacetime via the embedding
\be
\label{Intro_Jumps_Memb_Embed}
\xi^i \quad\mapsto\quad \calm:\; x^m \equiv x^m (\xi)\,,
\ee
as in Fig~\ref{fig:MembG}. With such an embedding, one can define the induced metric over the membrane worldvolume
\be
{\bf h}_{ij} = \frac{\del x^m}{\del \xi^i} \frac{\del x^n}{\del \xi^j} \eta_{mn}\,,
\ee
for a Minkowskian background. The physics of a single membrane, minimally coupled to a set of gauge three-forms $A_3^A$, is governed by the action \cite{Brown:1987dd,Brown:1988kg}
\be
\label{Intro_Jumps_Memb}
S_{\rm memb} = - \int_{\calm} \sqrt{-\det {\bf h}} \, \calt_{\rm memb}(\phi) + q_A \int_\calm A_3^A\;.
\ee
The first term is in the usual Nambu-Goto form and encodes the kinetic terms of the membrane; by construction, it is invariant under worldvolume reparamerizations $\xi^i \to {\xi'}^i(\xi)$. The quantity $ \calt_{\rm memb}(\phi)$ is the \emph{tension} of the membrane, namely its energy per unit area. We have generically assumed that it depends also on the scalar fields of the theory, understood to be evaluated over the membrane worldvolume $\calm$. This general assumption is indeed not uncommon in four-dimensional theories originating from compactification of string or M-theory: there, membranes originate from compactification of higher dimensional objects, naturally producing a four-dimensional tension which depends on the moduli of the compactification. The second term in \eqref{Intro_Jumps_Memb} expresses the minimal U(1) coupling of the membrane to the gauge three-forms $A_3^A$ with charges $q_A$. In a quantum theory, we may consider the membrane charges as quantized, $q_A \in \mathbb{Z}$.

The membrane action contains both the scalar fields $\phi^a$ and the gauge three-forms $A_3^A$ and a full action which describes at once the dynamics of membranes interacting with gravity, three-forms and scalar fields has to be supported by the kinetic terms of all these fields. In other words, the action \eqref{Intro_Jumps_Memb} has to be coupled to the \emph{bulk} action \eqref{Intro_Pot_S3f}, here denoted as $S_{\rm bulk}$:
\be
\label{Intro_Jumps_S3f}
\begin{aligned}
	S &= S_{\rm bulk} - \int_{\calm} \sqrt{-\det {\bf h}} \, \calt_{\rm memb}(\phi) + q_A \int_\calm A_3^A\;\,.
\end{aligned}
\ee

\begin{figure}[h]
	\centering
	\includegraphics[width=10cm]{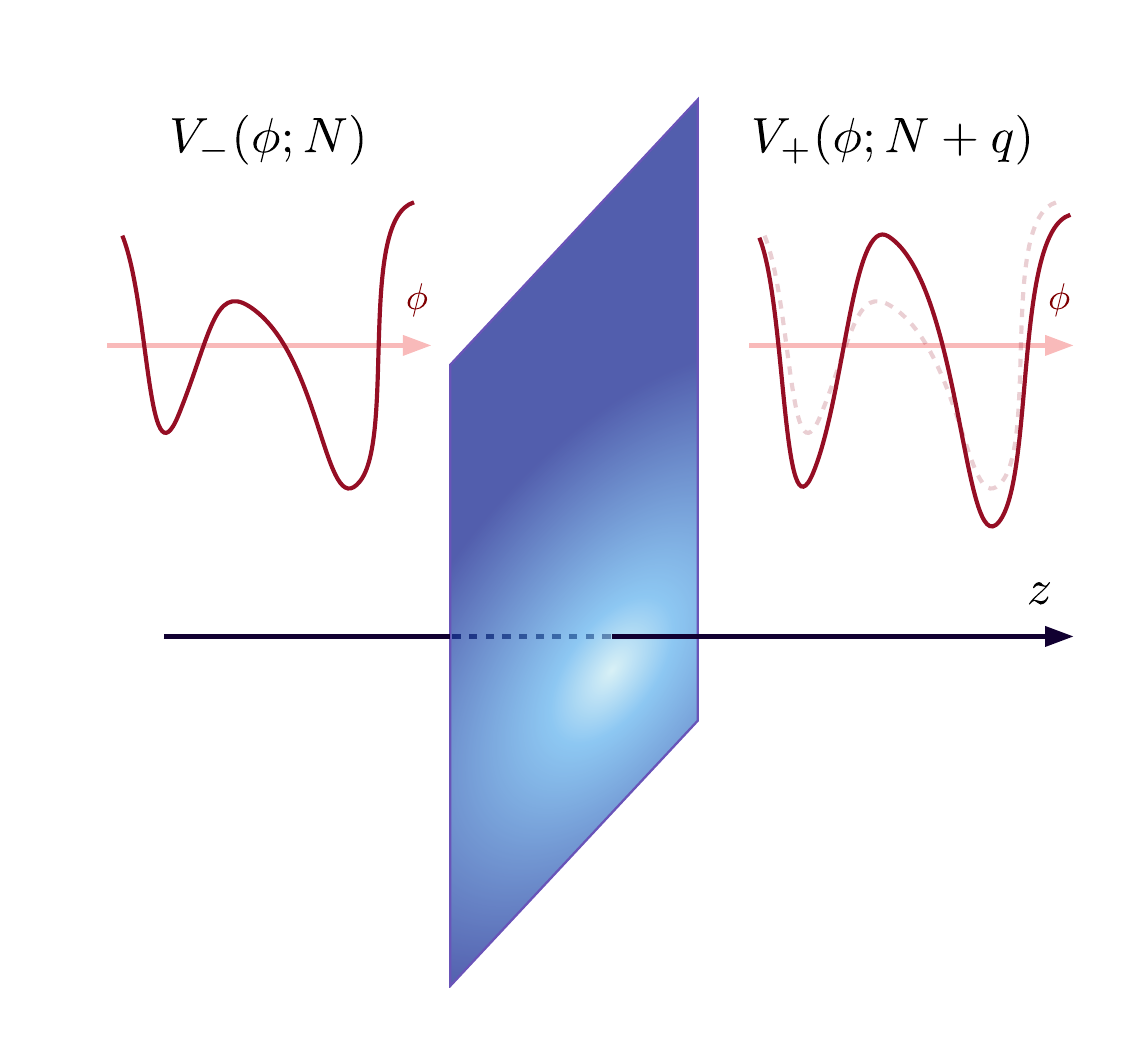}
	\caption{\footnotesize{The membrane separate the spacetime into two regions distinguished by different potentials. As a result, on the two sides the vacua and the associated energies may be different.}}\label{fig:MembVb}
\end{figure}

The action \eqref{Intro_Jumps_S3f} is fit to investigate the influence of the membrane on the description of the bulk theory. For simplicity, let us assume the membrane to span the codimension one hypersurface $z=0$. The equations of motion of the gauge three-forms $A_3^A$ leads to
\be
\label{Intro_Jumps_F4eom}
T_{AB} (\phi) *\!{F}^B_4+f_A (\phi)= - N_A - q_A \Theta (z) \;,
\ee
where $N_A$ are quantized constants. The equation \eqref{Intro_Jumps_F4eom} tells that the constants $N_A$ are shifted by the charges $q_A$ of the membrane, once this is crossed. Once gauge three-forms are set on-shell, the action \eqref{Intro_Jumps_S3f} reduces to an action in the very same form as \eqref{Intro_Pot_S3fb}, adding the membrane Nambu-Goto contribution of \eqref{Intro_Jumps_Memb} and where the potential is defined \emph{differently} on the two sides: 
\be
\label{Intro_Jumps_Von}
\begin{aligned}
 V(\phi)  &=   V_- (\phi;N) \Theta (-z) + V_+(\phi;N+q)  \Theta(z) \,.
\end{aligned}
\ee
The potential is then as in \eqref{Intro_Pot_Von} which is specified by the constants $N_A$ on the left side of the membrane and $N_A+q_A$ on its right:
\be
\label{Intro_Jumps_Vone}
\begin{split}
	V_- (\phi;N) &= \frac12 T^{AB}(\phi)(N_A +f_A(\phi))(N_B +f_B(\phi))  +
	\hat V (\phi)\;,
	\\
	V_+ (\phi;N+q) &= \frac12 T^{AB}(\phi)(N_A + q_A +f_A(\phi))(N_B + q_B +f_B(\phi))  +
	\hat V (\phi) \;,
\end{split}
\ee
as depicted in Fig.~\ref{fig:MembVb}.

\subsection{`Weighting' the flux jumps: how to choose what is dynamical and what not}
\label{sec:Intro_JumpsW}

Let us reconsider the form of the potential \eqref{Intro_Pot_Von} generated by gauge three-forms. There, \emph{all} the constants $N_A$ are dualized to gauge three-forms and, in light of \eqref{Intro_Jumps_F4eom} all of them can be subjected to \emph{jumps} induced by membranes. However, already from \eqref{Intro_Pot_Von}, it is clear that \emph{other} constants, different from $N_A$ may appear in $f_A(\phi)$ or $\hat V(\phi)$. A particular case is when $f_A(\phi) = M_A + \ldots$, where $M_A$ are quantized constants and the dots stand for contributions, possibly field-dependent, which we are presently not interested into. A question that deserves to be addressed is whether these other quantized constants could be generated by the introduction of another set of gauge three-forms $\tilde A_3^A$ along with $A_3^A$. This is a crucial issue: if and only if the constants can be dualized to gauge three-forms, they can undertake jumps, allowing us to explore different vacua of the landscape.

In order to address this issue, let us start with a potential of the form \eqref{Intro_Pot_Von}
\be
\label{Intro_Jumps_Vgen}
V (\phi) = \frac12 T^{\cala \calb}(\phi)(\caln_\cala +f_\cala(\phi))(\caln_\calb +f_\calb(\phi))  + \hat V_0 (\phi)\,.
\ee
Here, $\caln_{\cala}$ are quantized constants spanning a lattice
\be
\label{Intro_Jumps_Gamma}
\Gamma = \{ \caln_{\cala}\, |\, \caln_\cala \in \mathbb{Z}\}\,, \qquad {\rm with}\quad \cala=1,\ldots, \caln
\ee
The constants $N_A$ which can be truly interpreted as stemming from a set of gauge three-forms $A_3^A$ is a subset of $\caln_{\cala}$ and they determine a sublattice $\Gamma_{\rm EFT}$ of $\Gamma$. In order to identify such a sublattice, one can then split the constants $\caln_{\cala}$ as
\be
\label{Intro_Jumps_NA}
\caln_\cala = v^A_\cala N_A + \caln_\cala^{\rm bg}\;,
\ee
where $v^A_\cala$ are $\caln \times N$ matrices with integral entries and $\caln_\cala^{\rm bg}$ collects all the fluxes of $\caln_\cala$ which are \emph{not} interpreted as originated from gauge three-forms. Then, the dynamical sublattice is characterized by the sole $N_A$
\be
\label{Intro_Jumps_GammaDyn}
\Gamma_{\rm EFT} = \{  N_A\; |\; N_A \in \mathbb{Z} \; {\rm and}\; \caln_\cala = v^A_\cala N_A + \caln_\cala^{\rm bg} \} \subset \Gamma\,, 
\ee
with $A=1,\ldots, N \le \caln$. By further introducing
\be
\label{Intro_Jumps_split}
\begin{aligned}
	T^{AB}(\phi) &\equiv v^A_\cala v^B_\calb T^{\cala\calb}(\phi)\,,
	\\
	v^A_\cala f_A (\phi) &\equiv  \caln_\cala^{\rm bg} + f_{\cala}(\phi)\,,
	\\
	\hat V (\phi) &\equiv T^{\cala\calb} (\caln_\cala^{\rm bg} + f_{\cala}(\phi)) (\caln_\calb^{\rm bg} + f_{\calb}(\phi)) + \hat V_0(\phi)\,,
\end{aligned}
\ee
the potential \eqref{Intro_Jumps_Vgen} reduces to \eqref{Intro_Pot_Von}.

The splitting \eqref{Intro_Jumps_NA} is the keystone relation that identifies what flux sublattice we can explore within a single effective theories. Membranes can only induce changes in the dynamical sector $N_A$, while the others $\caln_\cala^{\rm bg}$ are \emph{fixed}; this determines the explorable sublattice 
\be
\label{Intro_Jumps_GammaEFT}
\Gamma_{\rm exp} = \Gamma_{\rm EFT} + \caln_\cala^{\rm bg} \subset \Gamma\;.
\ee
The `background' constants $\caln^{\rm bg}$ then determine equivalence classes
\be
[\caln^{\rm bg}] \in \Gamma_{\rm exp} / \Gamma_{\rm EFT}\,,
\ee
which then label the possible choices of the explorable sublattice.

The picture that we have just illustrated can be alternatively interpreted in terms of membranes that one can include within the effective field theory. Let us assume that in the potential \eqref{Intro_Jumps_Vgen} all the constants $\caln_{\cala}$ can be dualized to gauge three-forms, say $A_3^A$ for $N_A$ and $\tilde A_3^{\tilde A}$ for all the rest of fluxes, above regarded as `background fluxes' $\caln_\cala^{\rm bg}$. Then, membranes exist which couple to gauge three-forms as in \eqref{Intro_Jumps_Memb}, the former with a tension $\calt_{\rm memb}(\phi)$ and the latter with tension $\tilde\calt_{\rm memb}(\phi)$. If the transitions for $\caln_\cala^{\rm bg}$ ought to be forbidden, then the corresponding membranes have to be excluded from the effective theory. In other words, the tensions of such membranes have to be much greater that the tension of the membranes which shift $N_A$
\be
\tilde\calt_{\rm memb}(\phi) \gg \calt_{\rm memb}(\phi)\,.
\ee
For the moment, we do not comment any further on this issue, giving a more complete picture in the later Section~\ref{sec:Intro_Swamp}.

\section{Exploring a constrained Landscape}
\label{sec:Intro_AC}

In typical effective field theories, consistency conditions do \emph{not} allow for choosing arbitrary values of fluxes in $\Gamma$. Indeed, it is common to encounter constraints for the full lattice \eqref{Intro_Jumps_Gamma} of internal fluxes $\caln_\cala$. These may be given by a set of constraints of the form
\be
\label{Intro_Con_calq}
\calq_I\equiv \frac12\cali^{\cala\calb}_I\caln_\cala \caln_\calb+ \tilde Q_I^\cala \caln_\cala + \tilde\calq_I^{\rm bg}=0\,,
\ee
where the index $I$ labels the different conditions, $\cali^{\cala\calb}_I=\cali^{\calb\cala}_I$ defines a symmetric pairing between the fluxes $\caln_A$, ${\tilde Q_I^\cala}\caln_\cala$ stands for a possible linear contribution and $\tilde\calq_I^{\rm bg}$ denotes some additional constant contribution. In string theory the last contribution is typically generated by background sources, such as orientifold planes, branes or curvature corrections. 

As we have explained in the previous section, only the flux sublattice $\Gamma_{\rm EFT}$, labeled by $N_A$, can be understood as stemming dynamically from gauge three-form. Using the splitting \eqref{Intro_Jumps_split}, we can define
\begin{subequations}
	\label{Intro_Con_calqQ}
	\begin{align}
	\cali^{AB}_I&\equiv v^A_\cala v^B_\calb \cali^{\cala\calb}_I\,,\label{Intro_Con_calqQa}
	\\
	\calq^{\rm bg}_I&\equiv \tilde\calq_I^{\rm bg} + \frac12\cali_I^{\cala\calb}\caln_\cala^{\rm bg} \caln_\calb^{\rm bg}\,,\label{Intro_Con_calqQb}
	\\
	Q_I^A&\equiv v^A_\cala \tilde Q^\cala_I+ \cali^{\cala\calb}_I v^A_\cala \caln^{\rm bg}_\calb\,,\label{Intro_Con_calqQc}
	\end{align}
\end{subequations}
and then rewrite the condition \eqref{Intro_Con_calq} as
\be
\label{Intro_Con_calqb}
\calq_I\equiv \frac12\cali^{AB}_IN_AN_B+ Q_I^A N_A + \calq_I^{\rm bg}=0\,,
\ee
manifestly in terms of the dynamical sublattice \eqref{Intro_Jumps_GammaEFT}.

The process of integrating out gauge three-forms, as we explained in Section~\ref{sec:Intro_Pot_gen}, is blind to any constraint, owing to the fact that three-forms generate arbitrary constants spanning the full \eqref{Intro_Jumps_GammaEFT}. It is indeed possible to slightly modify the dualization procedure given in Section~\ref{sec:Intro_Pot_gen} so that, once we integrate out the gauge three-forms, the dynamically generated constants satisfy the constraint \eqref{Intro_Con_calqb}. 

As a first step, it is convenient to impose \eqref{Intro_Con_calqb} at the level of equations of motion, and we will do that by adding, to the action \eqref{Intro_Pot_S3fb}, the following coupling to a set of four-form potentials $C^I_4$
\be
\label{Intro_Con_minimalC4}
\calq_I\int C^I_4\, .
\ee
This is invariant under the gauge transformations
\be
\label{Intro_Con_C4gauge}
C_4^I\rightarrow C_4^I+\d\Lambda_3^I\, ,
\ee
and does not introduce new, unwanted degrees of freedom within the theory. The gauge four-forms here act as Lagrange multipliers, with the constraints \eqref{Intro_Con_calqb} that are set after varying \eqref{Intro_Con_minimalC4} with respect to $C_4^I$.

It is somehow pointless to straightly impose \eqref{Intro_Con_calqb} at the level of the equations of motion in \eqref{Intro_Pot_S3fb}. In fact, in \eqref{Intro_Pot_S3fb}, the fluxes $N_A$ have been already fixed and, rather, the constraints \eqref{Intro_Con_calqb} are to be satisfied \emph{a priori} by the choice of $N_A$ included in \eqref{Intro_Pot_S3fb}. In the dual, three-form perspective the constants are promoted to dynamical variables and there we might require that the set $N_A$, once generated, has to satisfy \eqref{Intro_Con_calqb}. We can try to proceed along the very same lines as in Section~\ref{sec:Intro_Pot_gen}. That is, we promote the constants $N_A$ to real scalar fields $y_A$, with the coupling \eqref{Intro_Con_C4gauge} consistently replaced by
\be
\label{Intro_Con_minC4new}
\int \Big(\frac12\cali_I^{AB}y_A y_B+Q^A_I y_A + \calq_I^{\rm bg}\Big)C_4^I\,.
\ee
In this form, the coupling \eqref{Intro_Con_minC4new} is fit to be added to the master action \eqref{Intro_Pot_S3fc}, arriving at the action
\be
\label{Intro_Pot_S3f4f}
\begin{aligned}
	S &=  \Mp^2 \int_{\Sigma} \left( \frac12 R *1- \frac12G_{ab}(\phi) \d \phi^a \wedge *\d \phi^{b} \right)
	\\
	&\quad\, -\int_\Sigma \Big[\frac12 T^{AB} (\phi)(y_A+ f_A (\phi)) (y_B + f_B(\phi))+
	\hat V (\phi)\Big] *\!1
	\\
	&\quad\,- \int_{\Sigma} \d y_A A_3^A  + \int_\Sigma \Big(\frac12\cali_I^{AB}y_A y_B+Q^A_I y_A + \calq_I^{\rm bg}\Big)C_4^I \;.
\end{aligned}
\ee
Before proceeding, some comments are in order. As it stands, \eqref{Intro_Con_minC4new} breaks the gauge invariance under \eqref{Intro_Con_C4gauge} in the complete action \eqref{Intro_Pot_S3f4f}, which should be re-installed. A general way to restore a symmetry is the `St\"uckelberg trick', which in the considered case amounts to introducing the following St\"uckelberg gauge transformations of the three-form potentials
\be
\label{Intro_Con_A3gauge2}
A^A_3\rightarrow A^A_3-(\cali_{I}^{AB}y_B+Q^A_I)\Lambda_3^I\, .
\ee
Indeed, it is easy to see that in this way the variations under \eqref{Intro_Con_C4gauge}-\eqref{Intro_Con_A3gauge2} of the gauge three-forms $A_3^A$ and four-forms $C_4^I$ of the last line in \eqref{Intro_Pot_S3f4f}  precisely cancel each other. 

This is however not the end of the story, because the contribution $\cali_I^{AB}y_B$ to the charge that defines the three-form gauging \eqref{Intro_Con_A3gauge2} is not a constant. This not only introduces, in a quantum theory, consistency issues regarding the compactness of the three-form gauge symmetry, but actually results in an obstruction to the dualization procedure, basically because the $y_A$ should eventually be expressed in terms of gauge invariant field-strengths of $A^A_3$, which should in turn define their own charges under \eqref{Intro_Con_C4gauge}.       

The only apparent way to get out of this impasse is to  choose the lattice $\Gamma_{\rm EFT}$ of fluxes to  be dualized so that the induced pairings defined in \eqref{Intro_Con_calqQa} vanish
\be
\label{Intro_Con_Iso}
\cali^{AB}_I=0\,.
\ee
In other words, $\Gamma_{\rm EFT}$ must be {\em isotropic} with respect to all the pairings  $\cali_I^{\cala\calb}$  entering the constraints \eqref{Intro_Con_calq}. With such a choice, the constraints \eqref{Intro_Con_calqb} become linear 
\be
\label{Intro_Con_lineartad}
\calq_I=Q^A_IN_A + \calq_I^{\rm bg} =0\, , 
\ee
and one no longer encounters any obstruction  to dualize  the constants $N_A$. Indeed, the term \eqref{Intro_Con_minC4new} reduces to
\be
\label{Intro_Con_minC4newb}
\int \Big(\calq^{\rm bg}_I+Q^A_I y_A\Big)C_4^I\, ,
\ee
and \eqref{Intro_Con_A3gauge2} becomes a well-defined gauge symmetry
\be
\label{Intro_Con_lin3gauge}
A^A_3\rightarrow A^A_3-Q^A_I \Lambda_3^I\,.
\ee
Upon integrating out $y_A$ from the resulting parent Lagrangian one gets the dual action. This can be obtained from the three-form action \eqref{Intro_Pot_S3f} by  adding the term
\be
\label{Intro_Con_C4cc}
\calq^{\rm bg}_I\int C^I_4\, ,
\ee
and replacing $F_4^A=\d A^A_3$ with the field-strengths 
\be
\label{Intro_Con_tildeF4}
\hat F^A_4\equiv F^A_4+Q^A_I C^I_4
\ee
which are gauge invariant under \eqref{Intro_Con_lin3gauge}. As a check, one can rederive the constraints in the dual formulation. In fact, still standing that integrating out the gauge three-forms reproduces \eqref{Intro_Pot_F4eom}, by additionally integrating out the gauge four-forms from \eqref{Intro_Pot_S3f4f}, we would get
\be
\calq^{\rm bg}_I- Q^A_I(T_{AB} (\phi) *\hat F^B_4+f_A (\phi))=0\,,
\ee
which indeed reduce to  \eqref{Intro_Con_lineartad} upon employing \eqref{Intro_Pot_F4eom}.

\subsection{3-branes and `jumping' constraints}
\label{sec:Intro_3branes}

The inclusion of gauge four-form into four-dimensional theories comes with the introduction of the extended objects that they couple to. These are 3-branes, whose four-dimensional worldvolume fills the whole spacetime or just portions thereof. Again, these objects are predicted from higher-dimensional theories: they may be $(p+1)$--brane, wrapping internal $(p-3)$--cycles. Spacetime--filling branes are in fact fundamental in string theory--related constructions, since they actively contribute to the tadpole cancellation conditions, along with O3-planes, which we will encounter in the later Chapter~\ref{chapter:EFT}. 

Consider the above term \eqref{Intro_Con_C4cc}, that we introduced as a particular type of Lagrange multiplier term. One can interpret \eqref{Intro_Con_C4cc} as a Chern-Simons like--term, expressing the coupling of $C_4^I$ to a \emph{spacetime filling 3-brane}, with charges $\calq^{\rm bg}_I$. Indeed, in \eqref{Intro_Con_C4cc}, the four-dimensional 3-brane worldvolume is identified with the spacetime. However, as described by the sole \eqref{Intro_Con_C4cc}, such a `putative' spacetime-filling brane is quite peculiar: we did not include any dynamics fo such a four-dimensional objects, as well as we completely neglected the matter living on its worldvolume.
\begin{wrapfigure}{r}{0.5\textwidth}
	\includegraphics[width=0.47\textwidth]{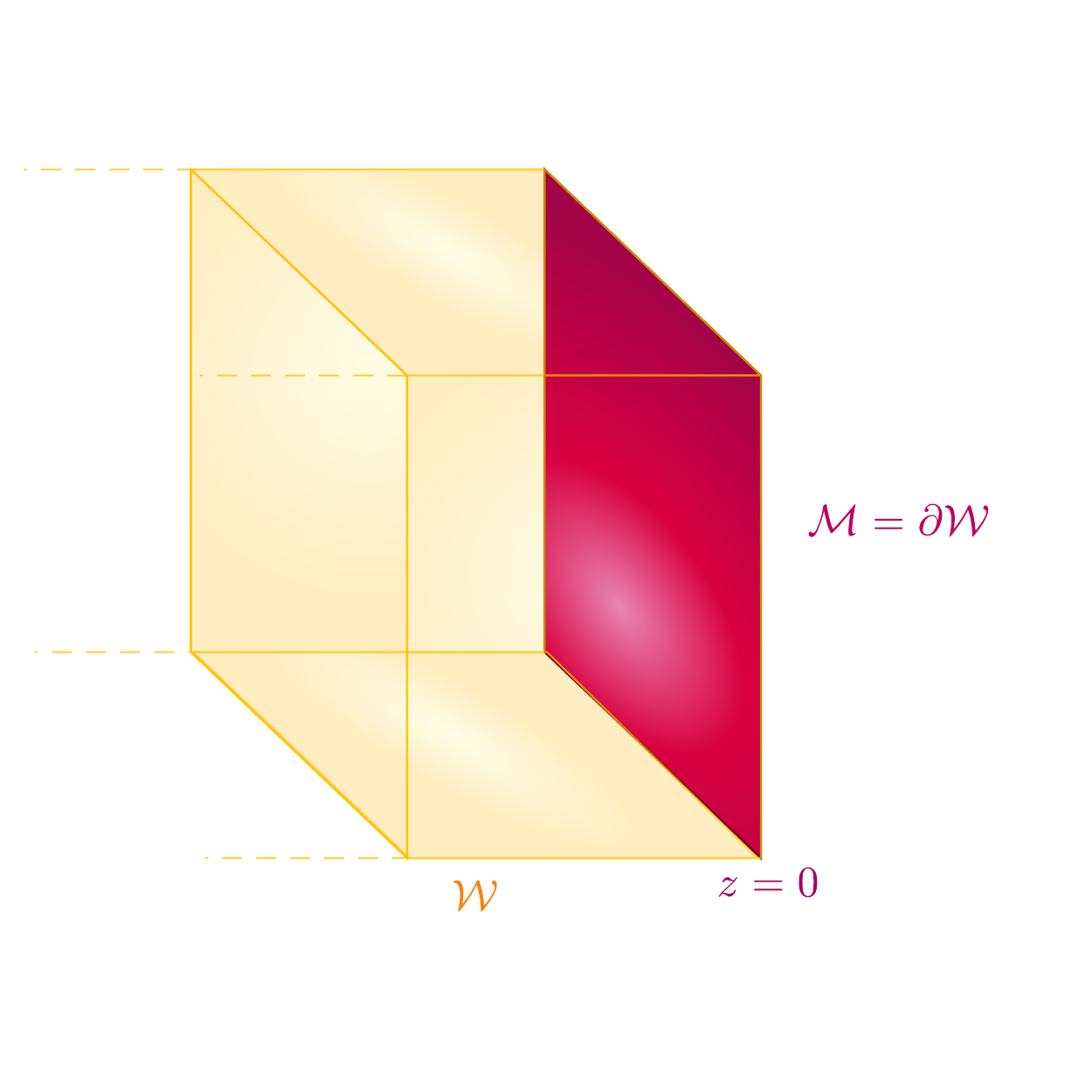}
	\caption{\footnotesize{The 3-brane fills the region $\calw = \{z<0\}$ and a membrane, serving as its boundary, is located at $z=0$.}}\label{fig:MembD3In}
\end{wrapfigure} For the moment, we do not further comment on such nterpretation and, as of now, we content ourselves just thinking of the coupling \eqref{Intro_Con_C4cc} as being a `topological' term for the action \eqref{Intro_Pot_S3f4f}. A more appropriate justification of the term \eqref{Intro_Con_C4cc} as originated from a concrete action of a spacetime-filling brane will be given in later section, pointing out the assumption that leads to single out the term \eqref{Intro_Con_C4cc}.

A special role is played by 3-branes which fill portions of spacetime, which are bounded by membranes, as depicted in Fig.~\ref{fig:MembD3In}. These are quite helpful in understanding how the constraint \eqref{Intro_Con_lineartad} changing after membranes are crossed. Let us reconsider the action \eqref{Intro_Pot_S3f4f} which, as we saw above, generate a Landscape `constrained' by \eqref{Intro_Con_lineartad}; let us further couple a single membrane to \eqref{Intro_Pot_S3f4f} with its action \eqref{Intro_Jumps_Memb}. Then, on the two sides of the membrane, the constants $N_A$ differ by the charges of the membranes $\Delta N_A = q_A$. Clearly, we should expect that, if \eqref{Intro_Con_lineartad} holds on one side of the membrane, it cannot hold on the other side, simply because the fluxes change, making  \eqref{Intro_Jumps_Memb} change as
\be
\label{Intro_Con_NAsh}
Q^A_IN_A + \calq_I^{\rm bg} = 0 \rightarrow Q^A_I (N_A + q_A) + \calq_I^{\rm bg} = 0\;. 
\ee

It is pretty awkward that the constraint \eqref{Intro_Con_lineartad} changes across a membrane, a fact that renders \eqref{Intro_Con_lineartad} only `locally' defined in every spacetime region delimited by the membrane. We can however try to find out a `global' constraint, still of the kind \eqref{Intro_Con_lineartad}. The starting point is the left hand side of the first relation in \eqref{Intro_Con_NAsh}, which introduces the additional $Q^A_I q_A$. This can be alternatively interpreted as a shift on the background value $ \calq_I^{\rm bg} \to  \calq_I^{\rm bg} + q_A Q^A_I$. We can then add the very term to the left hand side of \eqref{Intro_Con_NAsh}, so that the tadpole acquires the same form on \emph{both} the sides of the membrane. And it is a 3-brane that does this job: we may introduce a 3-brane which fills the whole spacetime region $\calw = \{z<0\}$, , which has the membrane as boundary, $\calm = \del \calw$ and whose charges are $\mu_I = Q^A_I q_A$.

To be concrete, the configuration just described is governed by the action
\be
\label{Intro_Con_S3f4f}
\begin{aligned}
	S &=  \Mp^2 \int_{\Sigma} \left( \frac12 R *1- \frac12G_{ab}(\phi) \d \phi^a \wedge *\d \phi^{b} \right)
	\\
	&\quad\, -\int_\Sigma \Big[\frac12 T_{AB} (\phi)\hat F^A_4*\! \hat F^B_4+ f_A (\phi) \hat F^A_4+
	\hat V (\phi)*\!1\Big] 
	\\ 
	&\quad\,
	+ \int_{\del\Sigma}(T_{AB}*\!\hat F^A_4+f_B)A^B_3 + \calq^{\rm bg}_I\int_\Sigma C^I_4
	\\
	&\quad\, - \int_{\calm} \sqrt{-\det {\bf h}} \, \calt_{\rm memb}(\phi) + q_A \int_\calm A_3^A + Q_I^A q_A \int_\calw  C_4^I\,.
\end{aligned}
\ee
It is then clear that the equations of motion of the gauge three-forms $A_3^A$ and those of the gauge four-forms $C_4^I$, once crossing the membrane, lead to the shifts
\be
\label{Intro_Con_shift}
N_A \rightarrow N_A + q_A \,, \qquad \calq_I + q_A Q_I^A\rightarrow \calq_I \,,
\ee
with $\calq_I$ as in  \eqref{Intro_Con_lineartad}. However, the constraint has still the same form
\be
\label{Intro_Con_tadmod}
\calq_I + q_A Q^A_I=Q^A_I (N_A+q_A) + \calq_I^{\rm bg} =0\
\ee
on both the sides of the membrane. One can then regard the shift \eqref{Intro_Con_shift} as changing the background values on the left of the membrane $\calq_I^{\rm bg} + q_A Q_I^A$ to the sole $\calq_I^{\rm bg}$ on the right. Loosely speaking, a 3-brane, which fills the region $z<0$, `dissolves' into fluxes after crossing the membrane. 

We also stress that such a spacetime--region, constituted by a 3-brane delimited by a membrane, could be straightly predicted by the very form of the gauged field strength \eqref{Intro_Con_tildeF4}. In fact, from the last line of \eqref{Intro_Con_S3f4f}, which is invariant under the combined \eqref{Intro_Con_C4gauge} and \eqref{Intro_Con_A3gauge2},  we can infer \eqref{Intro_Con_tildeF4} provide the proper gauge invariant forms living over $\calw$.

\section{Axions and monodromies in the Landscape}
\label{sec:Intro_AM}

Typical effective field theories originating from string theory are populated by \emph{axions}. The axions are regarded as gauge zero-forms, subjected to the gauge transformation
\be
\label{Intro_AM_agauge}
a_\Lambda \quad \rightarrow \quad a_\Lambda + c_\Lambda
\ee
for any given constant $c_\Lambda$. The reason why such zero-forms are common in four-dimensional EFTs can be traced back to their higher-dimensional origin, where gauge $p$-forms $C_p$ appear via their field strengths $F_{p+1} = \d C_p$. Whenever a set of internal, closed $p$-form $\omega^\Lambda_p|_X$ exist, we may decompose $F_{p+1} = \d a_\Lambda (x) \wedge \omega^\Lambda_p|_X$ , with $a_\Lambda (x)$ a set of zero-forms. After integrating over the internal manifold, we are just left with $\d a_\Lambda$ in the four-dimensional theory. The transformations \eqref{Intro_AM_agauge} are then just `remnants' of the higher-dimensional gauge transformation $C_p \to C_p + \d \Lambda_{p-1}$.

At the level of four-dimensional theories, enforcing the gauge symmetry \eqref{Intro_AM_agauge} severely restricts the possible interactions among fields, since the fields $a_\Lambda$ can only appear through their derivatives $\d a_\Lambda$. To be concrete, let us assume that the field content of the effective theory, beside the graviton, is constituted by two sets of real fields: the axions $a_\Lambda$ and real fields $\ell_\Lambda$, in the same number as the axions. We will refer to the latter as \emph{saxions}. The action describing their dynamics may acquire, for example, the form\footnote{We consider this particular form of the action guided by the supersymmetric models -- see, for example, the following \eqref{GSS_AL_Dual1Lagbos} in Section~\ref{sec:GSS_Axions} in global supersymmetry. There, the axions and saxions are naturally paired, sharing the same multiplet as in \eqref{GSS_HF_0Phi}.}
\be
\label{Intro_AM_Slin}
\begin{aligned}
	S &=  \Mp^2 \int_{\Sigma} \left( \frac12 R *1+ \frac12 G^{\Lambda\Gamma}(\ell) \d \ell_\Lambda \wedge *\d \ell_\Gamma + \frac12 G^{\Lambda\Gamma}(\ell) \d a_\Lambda \wedge *\d a_\Sigma \right) \;,
\end{aligned}
\ee
where we have neglected possible axion/saxions mixing terms as well as possible terms with higher number of derivatives. The field metric $G^{\Lambda\Gamma}$, which we here assume to be the same for the two sectors of fields, can only depend on the saxionic sector $\ell_\Lambda$. 

As real scalar fields, the axions $a_\Lambda$ are endowed with one physical degree of freedom each. In four dimensions, gauge two-forms $\calb_2^\Lambda$ also carry a single degree of freedom. Indeed, we can reformulate an axionic theory in terms of another, physically equivalent theory where the axions are replaced by gauge two-forms. The duality of the axions $a_\Lambda$ to gauge two-forms $\calb_2^\Lambda$ can be established directly at level of the action. One can then relax the assumptions that the one-forms $\d a_\Lambda$ are exact, but rather regard them as generic one-forms $\theta_\Lambda$ and add a `dualizing term' to the action \eqref{Intro_AM_Slin} as follows
\be
\label{Intro_AM_Sduala}
\begin{aligned}
	S &=  \Mp^2 \int_{\Sigma} \left( \frac12 R *1+ \frac12 G^{\Lambda\Gamma}(\ell) \d \ell_\Lambda \wedge *\d \ell_\Gamma + \frac12 G^{\Lambda\Gamma}(\ell) \theta_\Lambda \wedge *\theta_\Gamma \right) -   \int_{\Sigma} \theta_\Lambda \calh_3^\Lambda
\end{aligned}
\ee
where, in the last term, we have introduced $\calh_3^\Lambda \equiv \d \calb_2^\Lambda$.  

If one integrates out $\calb_2^\Lambda$ from  \eqref{Intro_AM_Sduala}, the relation $\d \theta_\Lambda = 0$ is obtained. This can be solved by requiring $\theta_\Lambda = \d a_\Lambda$ for a generic zero-form $a_\Lambda$, reducing the action \eqref{Intro_AM_Sduala} to \eqref{Intro_AM_Slin}. 

On the other hand, we can integrate out $\theta_\Lambda$ from \eqref{Intro_AM_Sduala}, leading to
\be
\label{Intro_AM_duality}
\delta \theta_\Lambda :\qquad \Mp^2  *\theta_\Lambda  = G_{\Lambda\Gamma}(\ell) \calh_3^\Gamma \;.
\ee
where $G_{\Lambda\Gamma}$ defined by $G_{\Lambda\Gamma}G^{\Gamma\Pi} = \delta_\Lambda^\Pi$. The relation \eqref{Intro_AM_duality} provides the duality we were looking for, since it properly exchanges the one-forms $\d a_\Lambda$ with their Hodge-dual counterparts $\calh_3^\Lambda$. By plugging \eqref{Intro_AM_duality} into \eqref{Intro_AM_Sduala}, we arrive at an action which now depends on the saxionic sector and the gauge two-forms $\calb_2^\Lambda$ as
\be
\label{Intro_AM_Sdual}
\begin{aligned}
	S &=  \Mp^2 \int_{\Sigma} \left( \frac12 R *1 + \frac12 G^{\Lambda\Gamma}(\ell) \d \ell_\Lambda \wedge *\d \ell_\Gamma \right)  + \frac{1}{2\Mp^2} \int_{\Sigma} G_{\Lambda\Sigma}(\ell) \calh_3^\Lambda \wedge * \calh_3^\Sigma\,.
\end{aligned}
\ee
This action is invariant under the gauge transformations $\calb_2^\Lambda \to \calb_2^\Lambda + \d \alpha_1^\Lambda$, with $\alpha_1^\Lambda$ generic one-forms. Additionally, \eqref{Intro_AM_Sduala} is also invariant under the global shifts
\be
\label{Intro_AM_agaugeB}
\calb_2^\Lambda \quad \rightarrow \quad \calb_2^\Lambda + C_2^\Lambda\,,
\ee
with $C_2^\Lambda$ are constant two-forms, providing the counterparts of the axionic transformations \eqref{Intro_AM_agauge} in two-forms language.

\subsection{Strings and monodronomies}
\label{Intro_AM_Mon}

Gauge two-forms electrically couple to strings. A string spans a two-dimensional hypersurface $\cals$ in the target four-dimensional space $\Sigma$, which is determined by the embedding
\be
\label{Intro_AMS_Embed}
\zeta^i \quad\mapsto\quad \cals:\; x^m \equiv x^m (\zeta)\,,
\ee
where $\zeta^i$, $i=0,1$, are two spacetime coordinates parametrizing the string worldsheet. The embedding \eqref{Intro_AMS_Embed} in turn defines the induced metric over the string worldsheet
\be
{\bm\gamma}_{ij} = \frac{\del x^m}{\del \zeta^i} \frac{\del x^n}{\del \zeta^j} \eta_{mn}\,,
\ee
for a Minkowskian background.

 \begin{wrapfigure}{r}{0.4\textwidth}
	\includegraphics[width=0.4\textwidth]{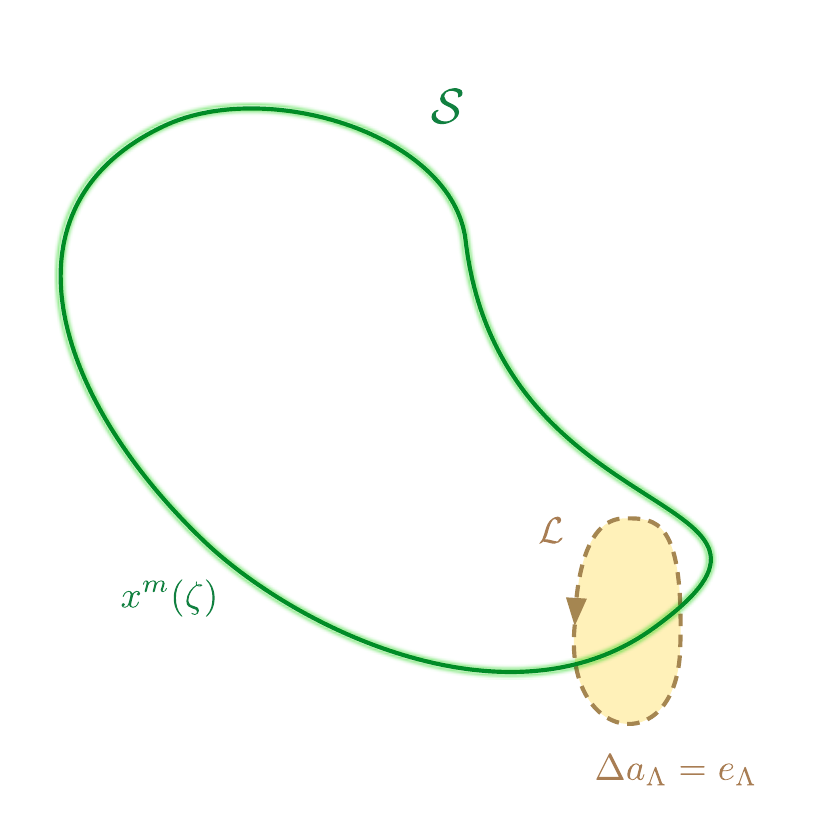}
	\caption{\footnotesize{A membrane as a hypersurface of codimension one described by the embedding \eqref{Intro_AMS_Embed}. It is also depicted the integration path $\call$ used in \eqref{Intro_AM_B2eomint}.}}\label{fig:StringF}
\end{wrapfigure}
A string which is minimally coupled to a set of gauge two-forms $\calb_2^\lambda$ is described by the action
\be
\label{Intro_AMS_S}
S_{\rm string} = - \int_{\cals} \sqrt{-\det {\bm \gamma}} \, \calt_{\rm string}(\ell) + e_\Lambda \int_\cals \calb_2^\Lambda\;.
\ee
Again, the first, Nambu-Goto term describe the motion of the string, encoding the kinetic terms of the string degrees of freedom, and it is invariant under reparamerizations of the worldsheet $\zeta^i \to {\zeta'}^i(\zeta)$. It further depends on the string tension $  \calt_{\rm string}(\ell)$, its mass per unit area, and may generically depends on the saxions $\ell_\Lambda$. The second, Wess-Zumino term expresses instead the minimal coupling of the string to the gauge two-forms $\calb_2^\Lambda$, under which the string has charges $e_\Lambda$.

The full action which describes the interaction of a string with the bulk fields is then obtained by combining \eqref{Intro_AM_Sdual} with \eqref{Intro_AMS_S}:
\be
\label{Intro_AM_SdualS}
\begin{aligned}
	S &=  \Mp^2 \int_{\Sigma} \left( \frac12 R *1+ \frac12 G^{\Lambda\Gamma}(\ell) \d \ell_\Lambda \wedge *\d \ell_\Gamma \right)  +\frac{1}{2\Mp^2} \int_{\Sigma} G_{\Lambda\Sigma}(\ell) \calh_3^\Lambda \wedge * \calh_3^\Sigma
	\\
	&\quad\, - \int_{\cals} \sqrt{-\det {\bm \gamma}} \, \calt_{\rm string}(\ell) + e_\Lambda \int_\cals \calb_2^\Lambda\;.
\end{aligned}
\ee

It is important to stress that it does not make sense to couple a string to the bulk action \eqref{Intro_AM_Slin}, since no kinetic terms for the gauge two-forms appear. However, one can first dualize the axions $a_\Lambda$ to gauge two-forms $\calb_2^\Lambda$  as we did above in \eqref{Intro_AM_duality}, obtaining the action \eqref{Intro_AM_Sdual} and later couple the dual action to a string. The gauge two-forms $\calb_2^\Lambda$, electrically coupled to the string, do indeed provide an alternative representation for the axions $a_\Lambda$. In this sense, we will say that the axions $a_\Lambda$ are `magnetically' coupled to the string \eqref{Intro_AMS_S} and call such a string an `axionic string'.

The coupling of the gauge two-forms to a string has in fact interesting consequences in the dual axion picture. In order to see this, let us compute the equations of motion for the gauge two-forms which originate from the action \eqref{Intro_AM_SdualS}:
\be
\label{Intro_AM_B2eom}
\frac{1}{\Mp^2} \d \left(G_{\Lambda\Gamma} *\! \calh_3^\Gamma\right) =  e_\Lambda \delta_2 (\cals)\;.
\ee
We can now integrate this equation over a disk $\cald$, whose boundary $\call= \del \cald$ is a circle enclosing the string, as depicted in Fig.~\ref{fig:StringF}, obtaining 
\be
\label{Intro_AM_B2eomint}
\frac{1}{\Mp^2} \int_\call \left(G_{\Lambda\Gamma} *\! \calh_3^\Gamma\right) =  e_\Lambda \;.
\ee
Going back to the axionic picture by means of the duality relation \eqref{Intro_AM_duality}, the equation \eqref{Intro_AM_B2eomint} reads
\be
\label{Intro_AM_Da}
\Delta a_\Lambda = e_\Lambda\,,
\ee
which tells that, encircling a string, an axion $a_\Lambda$ is subjected to a \emph{monodromy} transformation that shifts the axion $a_\Lambda$  by the charge $e_\Lambda$ of the string under the dual two-form $\calb_2^\Lambda$.

\subsection{Two-form gaugings and anomalous strings}
\label{sec:Intro_AnomStr}

As stated above, the requirement of invariance under the axionic symmetry \eqref{Intro_AM_agauge} restricts the admissible actions and, in particular, the dependence on the fields of a potential that one could introduce in \eqref{Intro_AM_agauge}. Let us assume that the axions shift by integral values, namely in \eqref{Intro_AM_agauge} $c_\Lambda \in \mathbb{Z}$. Indeed, this is what happens in a quantum theory transversing a string, being the charges $e_\Lambda$ in \eqref{Intro_AM_Da} quantized. An axionic dependent potential could then be built out of, for example, $\cos (2 \pi a_\Lambda)$ or $\sin (2 \pi a_\Lambda)$ but no finite polynomials or a rational function thereof would be allowed as a potential, which are ubiquitous in compactification scenarios. Therefore, another way to realize the symmetry \eqref{Intro_AM_agauge} within a potential has to be found.

The idea relies on a `St\"uckelberg trick', introduced in \cite{Dvali:2005an,Dvali:2004tma,Dvali:2005zk,Dvali:2013cpa,Dvali:2016uhn,Dvali:2016eay}. Let us consider, for simplicity, just a single axion $a$, and dualize it to a gauge two-form $\calb_2$, for which the axionic shift symmetry is realized as \eqref{Intro_AM_agaugeB}. Let us now \emph{gauge} \eqref{Intro_AM_agaugeB}, by promoting $C_2$ to a generic, point-dependent two-form. Clearly, the usual field strength $\calh_3 = \d \calb_2$ would not be gauge-invariant, for it transforms as $\calh_3 \to \calh_3 + \d C_2$. In order to define a proper gauge invariant field strength we then introduced the `gauged' three-forms
\be
\label{Intro_AM_GH3}
\hat \calh_3 = \d \calb_2 + A_3
\ee
and require that, under the gauged \eqref{Intro_AM_agaugeB}, the gauge three-form $A_3$ transforms as
\be
A_3 \quad \rightarrow \quad A_3 - \d C_2
\ee
which is just a gauge transformation for the gauge three-form, which although combines with $\calb_2 \to \calb_2 + C_2$ making \eqref{Intro_AM_GH3} gauge-invariant.

An action for the gauge invariant field strength \eqref{Intro_AM_GH3} is most readily built as
\be
\label{Intro_AM_SGH3}
S = -\frac1{2\Mp^2} \int_\Sigma   \hat\calh_3 \wedge * \hat \calh_3 -  \frac1{2\Mp^4} \int_\Sigma F_4 *\!F_4 + \frac1{\Mp^4} \int_{\del\Sigma} A_3 *F_4
\ee
where the first terms are the gauge-invariant kinetic terms built out of \eqref{Intro_AM_GH3} and the last two terms include the kinetic terms for the gauge three-form $A_3$ \eqref{Intro_cc_S3f}, equipped with the boundary terms \eqref{Intro_cc_Sbd} as discussed in Section~\ref{sec:Intro_Pot_cc}. 

In order to understand the effect of such a gauging for the axionic physics, let us perform the duality of \eqref{Intro_AM_Sduala} in reverse order. Namely, we here promote $\d \calb_2$ to an arbitrary three-form $\Theta_3$ and add a dualizing term as
\be
\label{Intro_AM_SGH3d}
S = -\frac1{2\Mp^2} \int_\Sigma   (\Theta_3 + A_3) \wedge * (\Theta_3 + A_3) -  \frac1{2\Mp^4}  \int_\Sigma F_4 *\!F_4 + \frac1{\Mp^4}  \int_{\del\Sigma} A_3 *F_4  +    \int_{\Sigma} \Theta_3 \wedge \d a\,.
\ee
Integrating out $\Theta_3$ gives
\be
\label{Intro_AM_SGH3io}
\delta \Theta_3: \qquad  \Mp^2 * \d a = \Theta_3 + A_3\,,
\ee
which reduces \eqref{Intro_AM_SGH3d} to
\be
\label{Intro_AM_SGH3db}
S = -\frac1{2} \Mp^2 \int_\Sigma   \d a \wedge * \d a-    \int_{\Sigma} a F_4  -\frac1{2\Mp^4}  \int_\Sigma F_4 *\!F_4 + \frac1{\Mp^4}  \int_{\del\Sigma} A_3 (*F_4 + a)  \,.
\ee
This action is invariant under the shift symmetry \eqref{Intro_AM_agauge}, here interpreted as a remnant of the gauge symmetry of $\calb_2$.

Further integrating out the gauge three-forms gives
\be
\delta A_3: \qquad \d \left( \frac1{\Mp^4} * F_4 + a\right) = 0 \quad \Rightarrow \quad * F_4 = \Mp^4 (n - a)
\ee
where $n$ is an arbitrary, but quantized constant. The action \eqref{Intro_AM_SGH3db} then reduces into an action for the sole axion $a$ as
\be
\label{Intro_AM_SGH3daxion}
S = -\frac1{2} \Mp^2 \int_\Sigma   \d a \wedge * \d a -\frac{\Mp^4}2 \int_\Sigma (a-n)^2 *\!1 \,,
\ee
where the last term provides a \emph{mass} for the axion $a$. Indeed, the model just described was proposed in \cite{Dvali:2005an,Dvali:2004tma,Dvali:2005zk,Dvali:2013cpa,Dvali:2016uhn,Dvali:2016eay} in order to make the QCD axion massive while still preserving the axionic symmetry.

In literature, the model just presented found also applications in inflationary scenarios \cite{Kaloper:2008fb,Kaloper:2011jz}, elaborating on the \emph{natural inflation} models \cite{Freese:1990rb}. In an effective field theory, as our \eqref{Intro_AM_SGH3db}, the shift symmetry \eqref{Intro_AM_agauge} protects the mass term against perturbative corrections. In fact, all the couplings of $a$ to other fields necessarily involve a derivative acting of $a$ and no corrections to the mass term for $a$ can be originated. In gauge theories, however the shift symmetry \eqref{Intro_AM_agauge} may be broken down to a discrete subgroup, due to instanton corrections, which could contribute to the mass term for the axion, but these are typically supressed with respect to the mass term \cite{Kaloper:2011jz}.  These observations make the axion $a$ in \eqref{Intro_AM_SGH3db} a good candidate for the inflaton, with the slow-roll inflation driven by the potential in \eqref{Intro_AM_SGH3daxion}.  The ubiquitous presence of axions in string theory -- along with possibility of realizing the models like \eqref{Intro_AM_SGH3db} in supersymmetric scenarios, as we will see in the following chapters -- allowed for importing the natural inflation models also in EFTs originating from string theory \cite{Marchesano:2014mla,Dudas:2014pva,Dudas:2015lga,Dudas:2015mvk,Buchmuller:2015oma,Bielleman:2015ina,Bielleman:2016olv,Valenzuela:2016yny,Landete:2016cix,Landete:2017amp}, providing their UV completion.

\begin{figure}[h]
	\centering
	\includegraphics[width=6cm]{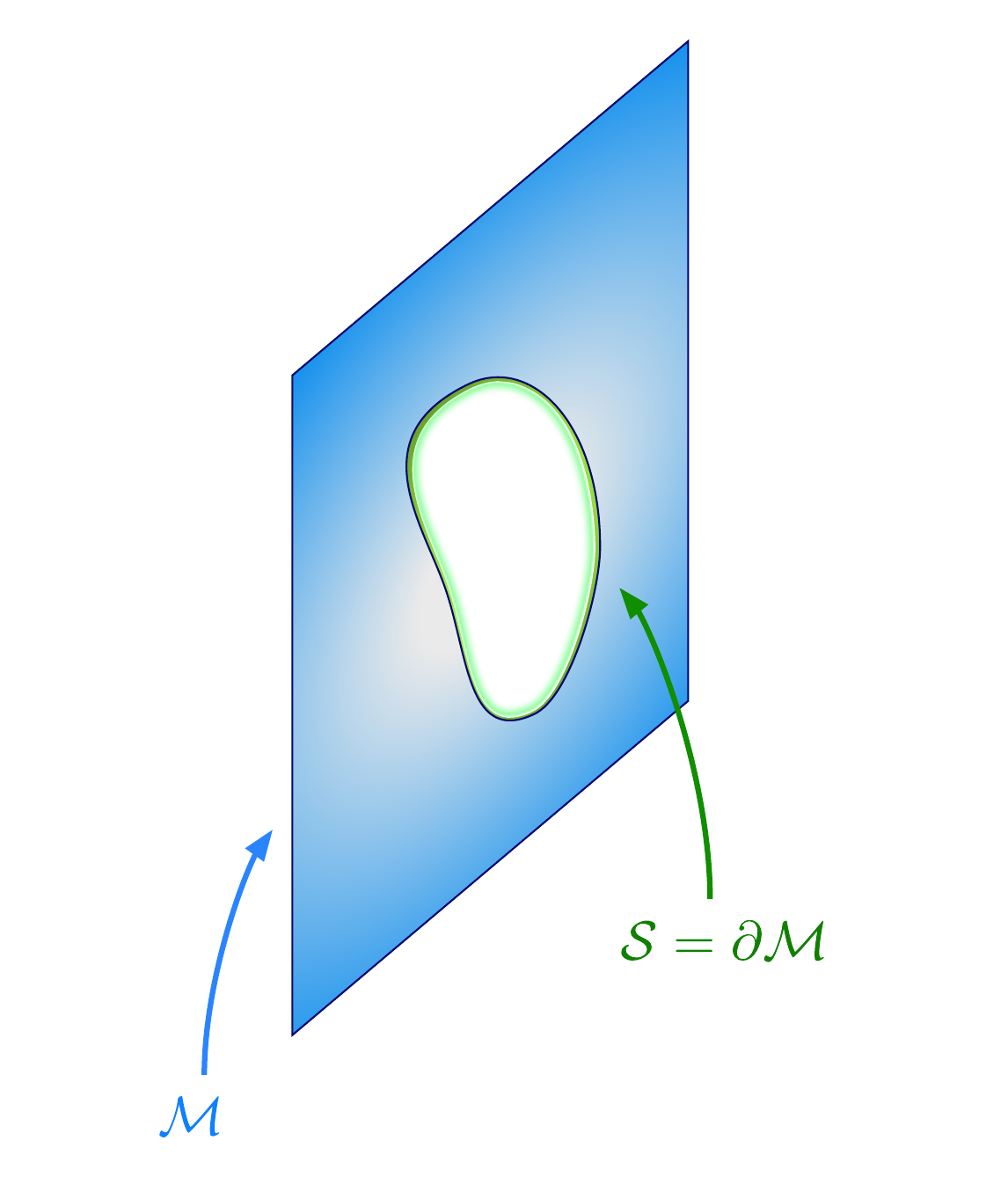}
	\caption{\footnotesize{A membrane with a string as boundary.}}\label{fig:StringHoleIn}
\end{figure}

We now pause to discuss the interpretation of the action \eqref{Intro_AM_SGH3daxion} in terms of the two- and three-form gaugings. In the `on-shell' action \eqref{Intro_AM_SGH3daxion} the axionic shift symmetry  \eqref{Intro_AM_agauge} is broken due to the mass term. This breaking may be regarded as spontaneous, due to the choice of the vacuum $a = n$. In order to ensure the symmetry \eqref{Intro_AM_agauge} over \eqref{Intro_AM_SGH3daxion} one must also require that $n$ may shift, so that \eqref{Intro_AM_SGH3daxion} is invariant under the combined shift
\be
\label{Intro_AM_MSg}
a \rightarrow a + q\,, \qquad n \rightarrow n + q\;.
\ee
This relation finds a cleaner interpretation in terms of extended objects. In fact, the gauged three-form \eqref{Intro_AM_GH3} couples to membranes which have a string as a boundary via an action whose Wess-Zumino term is
\be
\label{Intro_AM_MS}
q\left(\int_\calm A_3 + \int_{\cals} \calb_2\right)\,,
\ee
such that $\del\calm = \cals$ (see Fig.~\ref{fig:StringHoleIn}). Then, the relation \eqref{Intro_AM_MSg} reads as follows: transversing the object defined by the coupling \eqref{Intro_AM_MS}, the axion undertakes a monodromy transformation, being shifted by the charge $q$ of the string, and at the same time the membrane makes the quantized flux shift by the charge $q$ of the membrane so that the effects compensate between each other. In string theory, indeed, a string develops the so-called Freed-Witten anomalies \cite{Freed:1999vc,Maldacena:2001xj}, which are cured by attaching a proper number of branes. Then, configurations of the kind of that depicted in Fig.~\ref{fig:StringHoleIn} are present and it is expected to find symmetries of the kind of \eqref{Intro_AM_MSg} for the potential.

\section{Consistent EFTs and the Swampland}
\label{sec:Intro_Swamp}

Across the previous sections, we have shown how some constants which appear in the potential can be dualized to gauge three-forms and how axions can be alternatively described as gauge two-forms. It is by using the latter representations that one may couple membranes and strings to the effective theory. A special role is deserved to 3-branes, which fill the full or a portion of spacetime and are coupled to gauge four-forms. Using such 3-branes, in Section~\ref{sec:Intro_AC}, we were able to impose some constraints over the flux lattice, with an important caveat: the constraint has to be \emph{linear} with respect to the fluxes $N_A$ which are generated by gauge three-forms; if the constraint is not linear but, for example, quadratic as in \eqref{Intro_Con_calq}, one has to strictly reduce it to a linear constraint. This is achieved by requiring that some fluxes are \emph{not} dynamical and thus not dualized to gauge three-forms. Imposing a constraint as \eqref{Intro_Con_calq} requires to make a choice, that amounts in declaring which set of fluxes is dynamical and which other is not, performing a splitting as \eqref{Intro_Jumps_NA}. In turn, \eqref{Intro_Con_calq} tells the amount of membranes that we can consistently insert in the EFT, forbidding those which would make the fluxes $\caln_{\cala}^{\rm bg}$ shift. In other words, properly choosing the prescription \eqref{Intro_Jumps_NA} tells which are the `good' representations of fields that we may use and the extended objects that does make sense to consistently include in the EFT. Our task, for the moment, is to give very general physical arguments that can justify the splitting \eqref{Intro_Jumps_NA}, leaving a more detailed and precise description to Chapter~\ref{chapter:LandSwamp}.

Any effective field theory is defined by a cut-off scale, which we denote as $\Lambda_{\rm UV}$, that tells the scale above which the effective description is no more valid and has to abandoned eventually in favor of a more complete one. For instance, in string theory EFTs, $\Lambda_{\rm UV}$ could be chosen below the Kaluza-Klein scale, preventing the inclusion of the Kaluza-Klein excitations in the EFT. Other mass scales which can be identified in the EFT are the very masses of the scalar fields $\phi^a$ and we will assume them to be of order $m_\phi$. Given the potential \eqref{Intro_Pot_Von}, we can generically expect that the masses of the scalar fields do depend on the fluxes and that a \emph{hierarchy} of masses is established among the various combinations of the scalar fields. We would like to keep the scalar fields $\phi^a$ within the effective field theory, albeit they are massive. It is then sufficient to require that the cut-off $\Lambda_{\rm UV}$ lies just above the mass scale set by $m_\phi$:
\be
\label{Intro_Sw_mphi}
m_{\phi} \lesssim \Lambda_{\rm UV} \,.
\ee

Now, let us assume that we wish to insert a membrane which causes some fluxes to jump. The potential is differently defined on the two sides of the membrane, with different vacua and generically determining a different mass hierarchy on the two sides. The energy involved in a transition from a vacuum on the left to another on the right may be roughly estimated by $\Delta V \sim \calt_{\rm memb}^2/\Mp^2$. Compatibility with the cut-off $\Lambda_{\rm UV}$ requires $\Delta V \lesssim \Lambda^2_{\rm UV} \Mp^2$, that is
\be
\label{Intro_Sw_memb}
\frac{\calt_{\rm memb}(\phi)}{\Mp^2} \lesssim \Lambda_{\rm UV}\;.
\ee
Furthermore, such a relation has to satisfied by a parametrically large fraction of membranes, so as subsequent membrane--transitions can be undertaken by the fluxes.
%
%

An analogous requirement can be formulated for strings. We then require the tension of the string to be small with respect to the Planck mass as
\be
\label{Intro_Sw_SLstr}
\frac{\calt_{\rm string}(\ell)}{M^2_{\rm P}} \lesssim 1 \,.
\ee

Some crucial notes are in order for these relations. The requirements \eqref{Intro_Sw_memb} and \eqref{Intro_Sw_SLstr} may be read as constraints over the explorable field space. In fact, both the string and membrane tensions are field dependent and then, given certain membranes and strings, specified by their tensions and charges,  \eqref{Intro_Sw_memb}  and \eqref{Intro_Sw_SLstr} tell which is the admissible field region of the field space which can be explored within the EFT. One can alternatively reverse such a reasoning: we can declare which field region we wish to explore and, afterwards, \eqref{Intro_Sw_memb}  and \eqref{Intro_Sw_SLstr}  determine which membranes and strings are admissible in our effective description. This last picture will be particularly useful in string and M-theory EFTs and also connects with the discussion of Section~\ref{sec:Intro_JumpsW}. There, the vevs of the fields can be interpreted as coupling constants and setting a field region identifies the perturbative regime which is described by the EFT.

\begin{figure}[t]
	\centering
	\includegraphics[width=12cm]{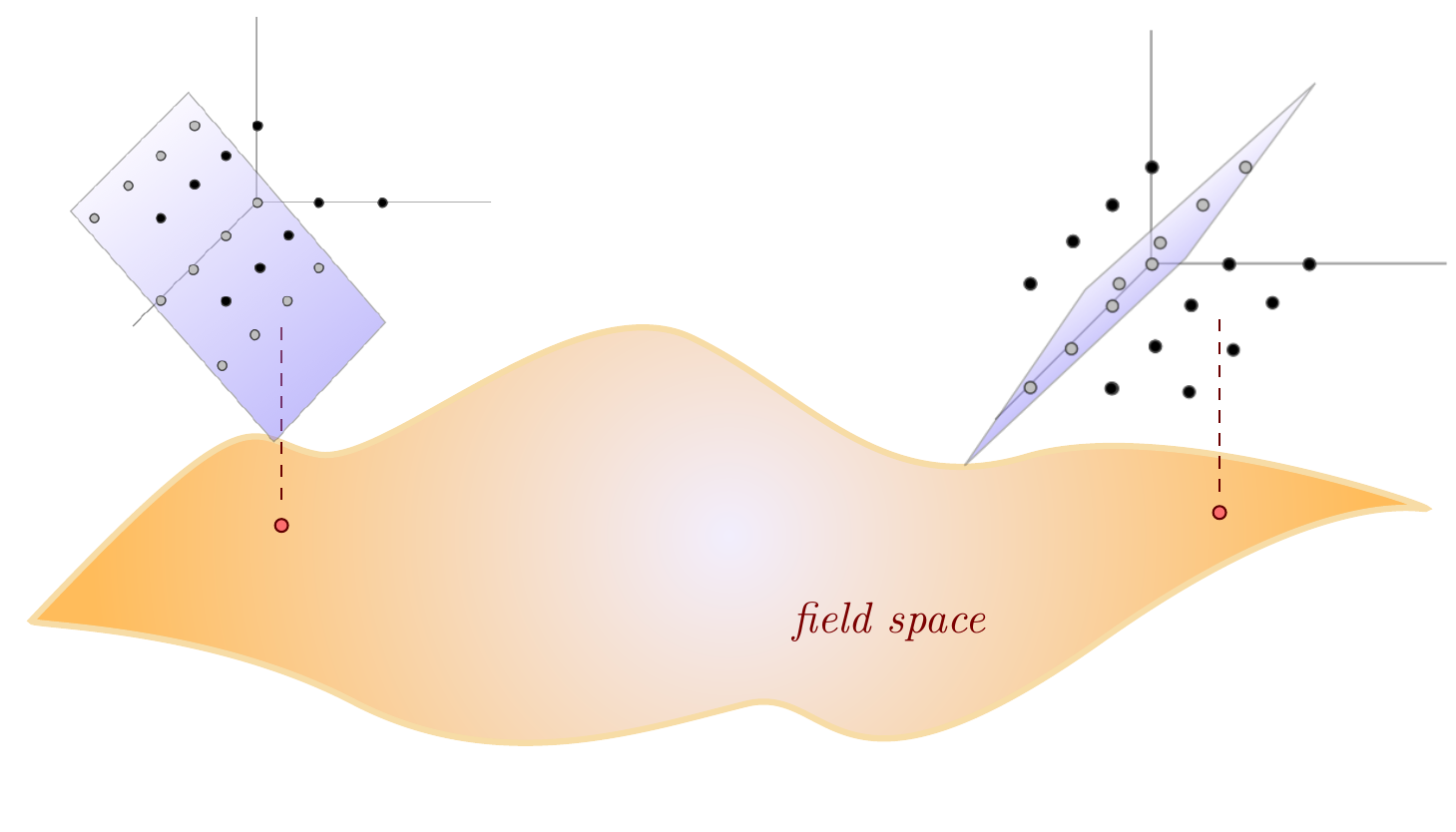}
	\caption{\footnotesize{The dynamical sublattices at different regions in the field space of an effective field theory.} }\label{fig:Intro_EFTLatt}
\end{figure}

Therefore, once we choose the perturbative regime,  \eqref{Intro_Sw_memb}  and \eqref{Intro_Sw_SLstr}  tell the membrane and strings which does make sense to include in the EFTs. The membranes which are excluded are associated to the set of background fluxes $\caln_{\cala}^{\rm bg}$ which are left invariant under any admissible transition in the EFT. In turn, this also determines the charges of the 3-branes which have to be included in the EFT in order to impose the constraints \eqref{Intro_Con_calq}. The membranes which can be included, because they satisfy \eqref{Intro_Sw_memb}, tell which fluxes $N_A$ are dualized to gauge three-forms and constitute the dynamical sublattice \eqref{Intro_Jumps_GammaDyn}. The choice of the dynamical sublattice \eqref{Intro_Jumps_GammaDyn} is then not universal and varies as we move along the field space, as depicted in Fig~\ref{fig:Intro_EFTLatt}. A wrong choice of the membrane and string spectrum, not compatible with the requirements \eqref{Intro_Sw_memb}-\eqref{Intro_Sw_SLstr}  would instead lead the theory to inconsistencies, namely the \emph{Swampland} of the EFTs.

\section{A supersymmetry appetizer}
\label{sec:Intro_Susy}

In this chapter we have considered bosonic models, rarely invoking supersymmetry. However, in order to set the ground for the discussion of the following chapters, we briefly anticipate what to expect from theories which enjoy supersymmetry, focusing on the rigid $\caln=1$ case.

In $\caln=1$ supersymmetry, we understand the scalar fields entering \eqref{Intro_Pot_S3f} as components of \emph{chiral superfields}. In chiral superfields, the scalar fields $\phi^a$ are actually paired so as to build up complex fields $\varphi^i$. The F-term potential is computed starting from a \emph{superpotential} $W(\varphi^i)$, which is holomorphic in $\varphi^i$. We may take the superpotential to acquire the form
\be
\label{Intro_SS_W}
W(\varphi) = N_A \calv^A(\varphi) + \hat W(\varphi)\,,
\ee
where $N_A$ are some constants, $\calv^A(\varphi)$ are generic holomorphic functions of $\varphi^i$, which we will call \emph{periods}, and $\hat W(\varphi)$ denotes other possible contributions. The potential is then computed from the superpotential as
\be
\label{Intro_SS_V}
V(\varphi,\bar\varphi) = K^{\bar\jmath i} W_i \bar W_{\bar \jmath}\,,
\ee
where we have denoted $W_i \equiv \frac{\del W}{\del \varphi^i}$, $\bar W_{\bar\imath} \equiv \frac{\del \bar W}{\del \bar\varphi^{\bar\imath}}$ and $K^{\bar\jmath i}$ is the inverse of the K\"ahler metric $K_{i\bar\jmath}$. The potential \eqref{Intro_SS_V} has then the same form as \eqref{Intro_Pot_S3f}. 

Let us now assume that we wish to include gauge three-forms. The dynamical generation of the scalar potential explained in Section~\ref{sec:Intro_Pot_gen} is translated, in a supersymmetric theory, into the dynamical generation of the part of the superpotential \eqref{Intro_SS_W} which is linear in $N_A$. However, the introduction of gauge three-forms in supersymmetric theories will take place by including \emph{a real or a complex} gauge three-form for each chiral superfield of the theory. Then, if the number of complex scalar fields is $n$, we can generate \emph{at most} $2n$ constants in \eqref{Intro_SS_W}, the others, if present, are necessarily regarded as non-dynamical. 

Furthermore, as already anticipated, the saxions $\ell_\Lambda$ and the gauge two-forms $\calb_2^\Lambda$ also share the same multiplet and hence are always paired, if supersymmetry ought to be preserved.

But the role of supersymmetry is not limited in telling how the numbers of fields are related among one another. The preservation of supersymmetry also strictly fixes the form of the membrane and string actions \eqref{Intro_Jumps_Memb} and \eqref{Intro_AMS_S}. Once we choose a ground state for membranes or strings, the actions \eqref{Intro_Jumps_Memb} and \eqref{Intro_AMS_S} generically break \emph{all} the supersymmetry of the underlying `bulk' theory. The only way to ensure the supersymmetry to be preserved is to require that, over their worldvolumes and worldsheets, membranes and strings enjoy a peculiar fermionic symmetry, which is ultimately identified with the worldvolume preserved supersymmetry. It turns out that this requirements strictly fixes the membrane tension to be
\be
\calt_{\rm memb} =  2 |q_A \calv^A(\varphi)|\,,
\ee
where the periods are the same as those in \eqref{Intro_SS_W}, and the string tension as
\be
\calt_{\rm string} =   |e_\Lambda \ell^\Lambda|\,.
\ee
In short, the membrane and string actions \eqref{Intro_Jumps_Memb} and \eqref{Intro_AMS_S} are fully determined by their charges. Also 3-branes may be imported into supersymmetric settings. However, the absence of a Nambu-Goto term for them greatly simplifies their analysis: they will always preserve supersymmetry.

In what follows, we will take a journey that aims at fully writing the actions presented in this chapter in $\caln=1$ supersymmetric theories, first in global and then in local supersymmetry. This will provide the basic framework that will allow us to re-interpret the effective field theories originating from compactification of string and M-theory and formulate consistency condition thereof according to the supersymmetric objects allowed.

\part{Hierarchy of forms and extended objects in Global Supersymmetry}


\chapter{Hierarchy of forms in global supersymmetry}
\label{chapter:GSS}

In theories enjoying supersymmetry, fields are appropriately collected into \emph{multiplets}. Then, if we wish to import the models studied in the previous chapter into any supersymmetric theory, the first step is to properly accommodate the gauge $p$--forms into such multiplets. Possibly, the most common example of inclusion of gauge forms into a supersymmetric multiplet is that of a one-form $A_1 = A_m \d x^m$. The two-form field strength $F_2 = \d A_1$ neatly appears among the components of a `super-field strength' multiplet $W^\alpha$, along with the gaugino $\lambda^\alpha$ and the non-propagating real scalar field $d$, as required by supersymmetry:
\be
\label{GSS_Intro_Wb}
W^\alpha = \{ F_{mn} = \del_{[m} A_{n]} , \lambda^\alpha , d \}\;.
\ee
In superfield language, this multiplet is represented by a (spinorial) chiral superfield $W_\alpha$ constructed by means of a real multiplet $U$ as 
\be
\label{GSS_Intro_Wa}
W_\alpha = -\frac14 \bar D^2 D_\alpha U\,,
\ee
with $U$ a real multiplet including, in its components, a one-form $A_1$. In the construction \eqref{GSS_Intro_Wa} $U$ plays the role of a gauge potential, conveniently promoted to a superfield. As we shall see below, real linear multiplets are somehow similar to the multiplets $W^\alpha$: they include, in their components, the field strengths of gauge two-forms, the linear multiplets being a super-field strength completion thereof.

The inclusion of gauge three- and four-forms in supersymmetric theories has attracted much less attention in literature. The construction of supersymmetric super-field strength for gauge three-forms was first made in \cite{Gates:1980ay}, with later appearances in \cite{Gates:1983nr,Burgess:1995kp,Binetruy:1996xw,Ovrut:1997ur,Cerdeno:2003us,Groh:2012tf,Dudas:2014pva,Becker:2016xgv}. Possibly, the most prominent application of gauge three-forms into supersymmetric theories has been in examining super-Yang-Mills theories, with the three-form being related to three-dimensional Chern-Simons terms \cite{Farrar:1997fn,Cerdeno:2003us}. The possibility of using gauge three-forms as a tool to dynamically generate a potential was applied, in supersymmetric contexts, only timidly for the cases of field-independent contributions. The generalization to more generic potentials, which include dependences on scalars as in \eqref{Intro_Pot_Von}, requires the introduction of more general multiplets which were not introduced in literature. On the other hand, gauge four-forms, although included in supersymmetric multiplets in \cite{Gates:1980ay}, due to their triviality, were generically excluded from supersymmetric theories.

This chapter has a three-fold aim. First, we generalize the construction of multiplets containing gauge three-forms present in literature. Secondly, we will show how to build generic globally supersymmetric Lagrangians containing gauge two-, three- and four-forms and what is the role of such forms in the Lagrangians. Finally, we show how the different multiplets intertwine among one another, namely how a multiplet which contains a gauge $p$--form may gauge the super-field strength of a $(p-1)$--form. At the end, we will be able to construct, in global supersymmetry, all the \emph{bulk} theories which were presented in the previous chapter. The results of this chapter will be the starting point for the next chapter, where BPS--objects, namely strings, membranes and 3-branes, will be coupled to the bulk theories here introduced.

We use the conventions of \cite{Wess:1992cp} and we refer to Appendix \ref{app:Superspace_Conv} for the complete list of the components of the multiplets that we use throughout this chapter. 
\section{Including gauge three-forms in globally supersymmetric theories}
\label{sec:GSS_TIntro}

To begin with, we show how an ordinary supersymmetric theory, formulated in terms of `standard' chiral multiplets, may be seen as stemming from a dual three-form theory. The \emph{parent} theory which includes gauge three-forms is also required to be supersymmetric. Therefore, the first question to address is: how can we include gauge three-forms into an $\caln=1$ supersymmetric theory? In order to settle the method and for sake of clarity, we present the procedure first working at the more familiar component level and then with the superspace formalism. The basic idea, hereby presented in a simplified setup, will accompany us throughout the rest of this chapter and of this work.

\subsection{The basic idea, in components}
\label{sec:GSS_TIntro_Comp}

Consider the simple Lagrangian \cite{Polonyi:1977pj,Farakos:2017ocw}
\be
\label{GSS_In_CompA}
\call= - \del_m \varphi \del^m \bar \varphi - \ii \psi \sigma^m \del_m \bar{\psi}+ f\bar{f}+ b f+\bar{b} \bar{f}\,,
\ee
where $\varphi$ is a (propagating) complex scalar field, $\psi^\alpha$ a two-component Weyl-fermion and $f$ a complex scalar auxiliary field; moreover, $b$ is just an arbitrary complex constant. The Lagrangian \eqref{GSS_In_CompA} is supersymmetric owing to its invariance under the transformations 
\be
\label{GSS_In_STr}
\begin{aligned}
	\delta \varphi &= \sqrt{2} \epsilon \psi \,,
	\\
	\delta \psi_\alpha &= \ii \sqrt2 \sigma^{m}_{ \alpha \dot\alpha} \bar\epsilon^{\dot\alpha} \del_m \varphi + \sqrt{2} \epsilon^\alpha f\,,
	\\
	\delta f &=  \ii \sqrt{2} \bar{\epsilon} \bar\sigma^m \del_m \psi\,,	
\end{aligned}
\ee
where $\epsilon^\alpha$ is a constant, fermionic parameter. Integrating out $f$ by using its equation of motion, $f = - \bar b$, we arrive at the following Lagrangian
\be
\label{GSS_In_CompB}
\call\big|_{\text{on-shell}} = - \del_m \varphi \del^m \bar \varphi - \ii \psi \sigma^m \del_m \bar{\psi} - |b|^2\,,
\ee
with the last term being just a constant, positive contribution to the vacuum energy $V = |b|^2$.\footnote{In a non-gravitational theory, this contribution could be disregarded. Presently, we will however keep the constant Lagrangian contribution $|b|^2$, for we want to relate it with an analogous contributions from the gauge three-forms -- see below. As stressed in the previous chapter, in a theory coupled to gravity, the constant term $|b|^2$ may be interpreted as a contribution to the cosmological constant. } 

More interestingly, in the previous chapter, we showed that a constant contribution to the Lagrangian may be dynamically generated by setting on-shell some gauge three-forms. We may then inquire how to dynamically generate the constant $b$ which appears in \eqref{GSS_In_CompA}. In the present context, however, we would like to modify the Lagrangian \eqref{GSS_In_CompA} by inserting gauge three-forms in such a way that the Lagrangian is still supersymmetric. It is therefore important that the number of degrees of freedom of \eqref{GSS_In_CompA} and \eqref{GSS_In_CompB} has to match with their three-form Lagrangian counterparts. Then, na\"ively, one should expect that, in the three-form picture, the auxiliary field $f$ is simply replaced by another scalar, built out of a \emph{complex} gauge three-form $C_3$; in addition, gauge invariance forces such a three-form to appear in the Lagrangian only via its field strength $F_4$. In other words, an educated guess is to trade
\be
\label{GSS_In_NaiveTr}
 f \qquad \leftrightarrow \qquad * F_4 = - \frac1{3!} \varepsilon^{mnpq} \del_{[m} C_{npq]} =  - \del^m C_m\,.
\ee
where $C_m$ are the components of the Hodge-dual of the three-form $C_3$
\be
\label{GSS_In_CL_HD}
C^m = - \frac{1}{3!} \varepsilon^{npqm} C_{npq}\,.
\ee
Specifically, we shall set
\be
\label{GSS_In_Trad}
f \equiv - \frac\ii2 *\!\bar F_4
\ee
whose form and factor, here arbitrary, are chosen just for later convenience and will be justified shortly. The Lagrangian \eqref{GSS_In_CompA} is replaced by
\be\label{GSS_In_Comp_DualA}
\begin{split}
	\tilde\call &= - \del_m \bar{\varphi} \del^m \varphi - \ii \psi \sigma^m \del_m \bar{\psi}+ \frac14 |{}^*{F_4}|^2 + \tilde\call_{\rm bd}
	\\
	&= - \del_m \bar{\varphi} \del^m \varphi - \ii \psi \sigma^m \del_m \bar{\psi} - \frac{1}{4 \cdot 4!} F_{mnpq} \bar F^{mnpq} + \tilde\call_{\rm bd}\,, 
\end{split}
\ee
with the boundary terms
\be\label{GSS_In_Comp_DualAbd}
\begin{split}
	 \tilde\call_{\rm bd} &= \frac14 \del_m (C^m\, {}^*{\bar F_4})+\frac14 \del_m(\bar{C}^m \, {}^*{F_4})\,
	 \\&= \frac{1}{4 \cdot 3!} \del_m (C_{npq} \bar F^{mnpq})+\frac{1}{4 \cdot 3!} \del_m ( \bar C_{npq} F^{mnpq})
\end{split}
\ee
With respect to \eqref{GSS_In_CompA}, there are two important differences: first, in \eqref{GSS_In_Comp_DualA} \emph{no parameter appears}; secondly, in \eqref{GSS_In_Comp_DualA} we needed to add the necessary boundary terms $ \tilde\call_{\rm bd}$ to ensure the correct variation of the gauge three-form as stressed in Section~\ref{sec:Intro_Pot}. This Lagrangian may be also rendered supersymmetric invariant, by paying attention that the supersymmetry transformations \eqref{GSS_In_STr} can be rephrased in terms of the new, independent elementary fields. The last line of \eqref{GSS_In_STr} then translates into
\be
\label{GSS_In_STrb}
\begin{aligned}
	\delta C^m &=  -2 \sqrt{2} {\epsilon} \sigma^m \bar \psi\,,	
\end{aligned}
\ee
which prescribes how the gauge three-form components transform under supersymmetry.

The Lagrangians \eqref{GSS_In_CompA} and \eqref{GSS_In_Comp_DualA} are on-shell equivalent. In fact, varying \eqref{GSS_In_Comp_DualA} with respect to the gauge three-form $C_3$ and imposing the gauge invariant boundary conditions $\delta F_4|_\textrm{bd} = 0$, we obtain
\be
\label{GSS_In_Comp_DualF4}
\del_m *\!{F_4} =0 \quad \Rightarrow \quad * {F_4}= 2 c \, , 
\ee
where $c$ is a complex constant determined by the boundary condition for the gauge three-form $C_{3}$. By plugging \eqref{GSS_In_Comp_DualF4} in \eqref{GSS_In_Comp_DualA}, by \emph{also} taking into account the contribution of the boundary terms, we get
\be
\label{GSS_In_Comp_DualB}
\tilde\call\big|_{C_3\, \text{on-shell}} = - \del_m \varphi \del^m \bar \varphi - \ii \psi \sigma^m \del_m \bar{\psi} - |c|^2\,.
\ee
Although identical in form, the on-shell Lagrangians \eqref{GSS_In_CompB} and \eqref{GSS_In_Comp_DualB} encode an important conceptual difference: in contrast with $b$ in \eqref{GSS_In_CompA}, the parameter $c$ that determines the constant Lagrangian contribution in \eqref{GSS_In_Comp_DualB} does not appear in the Lagrangian \eqref{GSS_In_Comp_DualA} from which we started. Rather, the constant $c$ appears only \emph{after} the gauge three-form $C_3$ is integrated out and -- in absence of membranes (see the later Sections~\ref{sec:ExtObj_MembSusy} and~\ref{sec:ExtObj_Summary}) -- is determined by the boundary conditions. By tuning the parameter $c$, \eqref{GSS_In_Comp_DualA} gives birth to all the on-shell Lagrangians of the class \eqref{GSS_In_CompA}. In this sense, the three-form Lagrangian \eqref{GSS_In_Comp_DualA} is a \emph{parent} Lagrangian for \eqref{GSS_In_CompA}.

We have demonstrated that the Lagrangian \eqref{GSS_In_CompA} and \eqref{GSS_In_Comp_DualA}, albeit different, lead to the same class of Lagrangians delivering the same constant contribution to the vacuum energy. One then might wonder whether there is a common starting point which, bifurcating, may lead to either \eqref{GSS_In_CompA} or \eqref{GSS_In_Comp_DualA}. This is possible introducing a complex Lagrange multiplier $x$ and starting from the \emph{master} Lagrangian 
\be
\label{GSS_In_CompM}
\call_{\rm master}= - \del_m \varphi \del^m \bar \varphi - \ii \psi \sigma^m \del_m \bar{\psi}+ f\bar{f}+ \left[x f + \frac\ii{2\cdot 3!} \varepsilon^{mnpq} \bar C_{npq} \del_m x + {\rm c.c.} \right]\,.
\ee
We may follow two distinct paths:
\begin{enumerate}
	\item integrating out the gauge three-form $C_3$ in \eqref{GSS_In_CompM} leads to
	\be
	\label{GSS_In_M_x}
	\delta C_3: \qquad \del_m x = 0 \qquad \Rightarrow \qquad x= b
	\ee
	with $b$ an arbitrary complex constant. Inserting \eqref{GSS_In_M_x} into \eqref{GSS_In_CompM}, the master Lagrangian reproduces \eqref{GSS_In_CompA}. It is worthwhile to notice that the integration of the three-form $C_3$ from \eqref{GSS_In_CompM} may be easily performed since in the master Lagrangian the three-form does \emph{not} appear via its field strength, which implies that no integration by parts involving $C_3$ is needed;
	\item alternatively, integrating out \emph{both} $x$ and $f$ leads to
	\be\label{GSS_In_M_f}
	\begin{aligned}
		\delta x: \qquad f &= \frac\ii2 \del_m \bar C^m \,,
		\\
		\delta f: \qquad x &= - \bar f = \frac\ii2 \del_m C^m\,,
	\end{aligned}
	\ee
	which, substituted in \eqref{GSS_In_CompM}, reproduces \eqref{GSS_In_Comp_DualA} with the proper boundary contributions.
\end{enumerate}

\subsection{The basic idea, in superspace}

The guiding principle for including gauge three-forms into a standard chiral multiplet theory in the previous section was the preservation of the off-shell supersymmetry. We now rephrase the previous discussion in superspace, by appropriately collecting the fields into superfields, the irreducible representations of the supersymmetry algebra. The superfield language provides the most powerful tool to write down manifestly supersymmetric invariant Lagrangians. 

The fields $\varphi$, $\psi^\alpha$ and $f$ contained in the Lagrangian \eqref{GSS_In_CompA} may be collected inside a single supersymmetric object, a \emph{chiral superfield} $\Phi$. It is a special kind of constrained superfield, obeying the condition
\be
\bar D_{\dot \alpha} \Phi = 0\,.
\ee
The superfield $\Phi$ has, in chiral superspace coordinates (see \cite{Wess:1992cp}), the following expansion
\be 
\label{GSS_In_ChiralExp}
\Phi = \varphi + \sqrt{2} \theta \psi + \theta^2 f
\ee
so that we may retrieve the fields $\varphi$, $\psi^\alpha$ and $f$ as the projections
\begin{equation}\label{GSS_In_ChiralComp}
\begin{aligned}
\Phi | &= \varphi \,, \\
D_\alpha \Phi| &= \sqrt 2 \psi_\alpha \,,\\
-\frac14 D^2 \Phi| &= f \,.
\end{aligned}
\end{equation}

The component Lagrangian \eqref{GSS_In_CompA} may be neatly rewritten as the superspace Lagrangian
\be
\label{GSS_In_SS}
\call = \int \d^4\theta\, \Phi \bar \Phi + \left( \int \d^2\theta\, b\, \Phi + {\rm c.c.}\right)\,.
\ee
The replacement \eqref{GSS_In_NaiveTr} can also be performed at the superspace level, by trading the whole multiplet with a new one whose auxiliary field is replaced by (the Hodge dual of) the field strength of a gauge three-form. To this end, we need a superfield which, among its components, may accommodate a complex three-form. A \emph{complex linear multiplet} is a viable option: it is a constrained superfield obeying
\be
\label{GSS_In_CL_constr}
\bar D^2 \Sigma = 0\,, 
\ee
whose components are
\be
\label{GSS_In_CL_comp}
\Sigma = \{s,\sigma,C^m,\kappa,\chi, \vartheta \}\;,
\ee
where $s$ and $\sigma$ are complex scalar fields,  $\kappa$, $\chi$, $\vartheta$ are Weyl fermions and $C^m$ is a \emph{complex} vector. In particular, the latter appears in the $\theta\bar \theta$--component as
\be
\label{GSS_In_CL_compC3}
-\frac12 \bar\sigma^{m\dot\alpha\alpha} [D_\alpha,\bar D_{\dot\alpha}] \Sigma | = C^m\,.
\ee
A three-form may be inserted by simply replacing the complex vector with its Hodge-dual as in \eqref{GSS_In_CL_HD}.\footnote{It is important to notice that such a trading via Hodge-duality simply exchanges, here, (the components of) a one-form with (those of) a three-form. The number of independent components is therefore unaltered by Hodge-duality.} However, such a superfield $\Sigma$ endows definitely too many degrees of freedom with respect to a chiral superfield! We cannot simply exchange $\Phi$ with $\Sigma$ in \eqref{GSS_In_SS}; rather we would expect that the chiral superfield $\Phi$  is traded with another chiral superfield. Indeed, a chiral superfield may be built out of $\Sigma$ by simply taking the projection
\begin{important}
	\be
	\label{GSS_DBT_L}
	S \equiv - \frac14 \bar D^2 \bar \Sigma\,, \qquad\qquad\qquad{\color{darkred}\text{\sf{Double three-form multiplet}}}
	\ee
\end{important}
This is a chiral superfield by construction which has the following expansion in chiral coordinates
\begin{equation}
S = \varphi^S + \sqrt 2 \, \theta \psi^S + \theta^2 f^S\;,
\end{equation}
whose components may be re-expressed in terms of those of $\Sigma$ as
\begin{align}
\label{GSS_In_Scomp}
\varphi^S &= s,\\
\label{compSintermsofSigma2}
\psi^S_\alpha &= \zeta_\alpha \,, \\
\label{compSintermsofSigma3}
f^S &=-\frac\ii2 *\! \bar F_4 = \frac\ii2 \partial_m \bar{C}^m\,.
\end{align}
In its highest, $\theta^2$-component, the superfield \eqref{GSS_DBT_L} endows a complex gauge three-forms, namely two real ones: for this reason, we will call it \emph{double three-form multiplet}. Moreover, the three-form $C_3$ appropriately appears in a gauge invariant way via $* F_4$. This can be traced back to the very definition of the double three-form multiplets \eqref{GSS_DBT_L}. In fact, such multiplets can be seen as \emph{super-field strengths}, which are defined by a complex linear multiplet potential and gauge invariant under the superfield transformation
\be
\label{GSS_In_SGinv}
\Sigma \rightarrow \Sigma + L_1 + \ii L_2\,,
\ee
with $L_1$ and $L_2$ being arbitrary linear multiplets. In fact, the $\theta\bar \theta$--component, of \eqref{GSS_In_SGinv} transforms as the usual three-form gauge field
\be
\label{GSS_In_SGinvComp}
C_{mnp} \rightarrow C_{mnp} + 3 \del_{[m} (\Lambda_1 + \ii \Lambda_2)_{np]}\,.
\ee

With the use of the double three-form superfield \eqref{GSS_DBT_L}, the component Lagrangian \eqref{GSS_In_Comp_DualA} may be rewritten as
\be\label{GSS_In_Dual_SS}
\begin{split}
\tilde\call=& \int \d^4\theta\, S \bar{S} +\tilde \call_{\rm{bd}}
\end{split}
\end{equation}
with
\be\label{GSS_In_Dual_Bd}
\begin{split}
	\tilde\call_{\rm{bd}} &= -\frac14 \left(\int \d^2\theta \bar D^2 - \int \d^2\bar\theta D^2\right) \left[  \left(\frac{1}{4}  \bar D^2 \bar S \right) \bar \Sigma\right]
	+\text{c.c.}
	\\
	&= \frac1{16} [D^2,\bar D^2] \left[  \left(\frac{1}{4}  \bar D^2 \bar S \right) \bar \Sigma\right]+\text{c.c.}\,.
\end{split}
\ee
It can be easily shown that the Lagrangian \eqref{GSS_In_Dual_SS} gives the same component Lagrangian as \eqref{GSS_In_Comp_DualA} with the appropriate boundary terms. It is worthwhile to stress that, in this simple Lagrangian, \emph{no} superpotential term (hence, no arbitrary constant) appears in \eqref{GSS_In_Dual_SS}.

As we did in components, the superspace Lagrangians \eqref{GSS_In_SS} and \eqref{GSS_In_Dual_SS} may be unified into the single superspace \emph{master Lagrangian}
\begin{equation}\label{GSS_In_Mast}
\begin{split}
\call_{\rm master}=& \int \d^4\theta\, \Phi \bar{\Phi} + \left(\int \d^2\theta X\Phi + {\rm c.c.} \right) - \left[ \int \d^2 \theta \left(-\frac14 \bar{D}^2 \right) \left(\bar{X} \Sigma \right) +{\rm c.c.} \right].
\end{split}
\end{equation}
Here $\Phi$ si the same chiral multiplet as the one which appears in \eqref{GSS_In_SS}, $\Sigma$ is the complex linear multiplet \eqref{GSS_In_CL_comp} and $X$ is a chiral multiplet which we treat as a Lagrange multiplier and has the expansion
\be
\label{GSS_In_X}
X = x + \sqrt 2 \, \theta \psi^X + \theta^2 f^{X}\;,
\ee
The particular combination of Grassmannian integrations and chiral projectors that we chose is due to the fact that we strictly want to reproduce the component master Lagrangian \eqref{GSS_In_CompM}, with \emph{no} derivatives acting on the three-form. The two-fold path that we followed in components to retrieve the Lagrangians \eqref{GSS_In_CompA} and \eqref{GSS_In_Comp_DualA} may be performed directly in superspace to relate the Lagrangians \eqref{GSS_In_SS} and \eqref{GSS_In_Dual_SS} as follows:
\begin{enumerate}
	\item in order to recover the Lagrangian \eqref{GSS_In_SS}, we should integrate the complex linear multiplet $\Sigma$ from \eqref{GSS_In_Mast}. Such an integration can be performed at the superspace level with, however, a caveat: the complex linear multiplet is constrained by \eqref{GSS_In_CL_constr} and its variation cannot be directly taken. To integrate it out, we first solve the constraint \eqref{GSS_In_CL_constr} by
	\be
	\label{GSS_In_SigmaPsi}
	\Sigma = \bar D_{\dot \alpha} \bar \Psi^{\dot \alpha}\,,\qquad \bar \Sigma = D^\alpha \Psi_\alpha\,,
	\ee
	where $\Psi_\alpha$ is an \emph{unconstrained} spinorial superfield. Then, we vary \eqref{GSS_In_Mast} with respect to $\Psi_\alpha$ and $\bar \Psi^{\dot \alpha}$, giving
	\be
	\label{GSS_In_Psi}
	\delta \Psi, \, \delta \bar \Psi:\qquad D_{ \alpha} X = 0\quad {\rm and} \quad  \bar D_{\dot \alpha} \bar X = 0\,.
	\ee
	The chirality of $X$ simply sets
	\be
	\label{GSS_In_Psi_X}
	X =  b\,,
	\ee
	with $b$ being an arbitrary complex constant. Hence, \eqref{GSS_In_Mast} reproduces \eqref{GSS_In_SS}. It is important to stress that here, as happened in components, no integration by parts over $\Sigma$ is required to get \eqref{GSS_In_Psi};
	\item in order to get the superspace three-form Lagrangian \eqref{GSS_In_Dual_SS}, we integrate out, from \eqref{GSS_In_Mast}, the Lagrange multiplier $X$ and the ordinary chiral superfield $\Phi$:
	\be
	\begin{split}
	\label{GSS_In_XPhi}
	&\delta X:\qquad \Phi = -\frac{1}{4} \bar{D}^2 \bar{\Sigma} = S\,,
	\\
	&\delta \Phi:\qquad X = \frac{1}{4} \bar{D}^2 \bar{S} \,.
	\end{split}
	\ee
	The first relation expresses, in superfield language, the relation between the components \eqref{GSS_In_ChiralComp} and \eqref{GSS_In_Scomp}. Plugging the solutions \eqref{GSS_In_XPhi} back in the master Lagrangian \eqref{GSS_In_Mast} leads to \eqref{GSS_In_Dual_SS}, with the proper superspace boundary terms \eqref{GSS_In_Dual_Bd}.
\end{enumerate}

The basic idea hereby illustrated is the foundation of the following Sections~\ref{sec:GSS_STF}, \ref{sec:GSS_DTF} and \ref{sec:GSS_MTF}, where it is applied with some variations. We have explained it from the point of view of the components, as well as from that of the superspace. The two points of views have their own characteristics and virtues: on the one hand, the component perspective is more direct, with the drawback that, in order to preserve supersymmetry, we need to rephrase the variations of the fields that we start with in terms of the new, elementary ones as in \eqref{GSS_In_STrb} and ensure that the new multiplet we build is truly an irreducible representation of the supersymmetry algebra. On the other hand, from the superspace perspective, although the component structure is not immediate and has to be computed, it is not necessary to prove the invariance under supersymmetry before and after the relation \eqref{GSS_In_NaiveTr}: the Lagrangian obtained in components is supersymmetric by construction and \eqref{GSS_In_NaiveTr} gets translated into a full superfield relation. For this reason, in what follows, we always start from superspace in order to build up the Lagrangians, while still taking care of making the component structure always explicit, by a direct computation from superspace.

In the next Section~\ref{sec:GSS_DTF} we generalize the construction presented above to more general double three-form multiplets, while in Section~\ref{sec:GSS_STF} the \emph{single} three-form multiplets are considered. In Section~\ref{sec:GSS_MTF} a more general procedure will be given which allow for considering any number of single and double three-form multiplets in a globally supersymmetric theory, so that the dualizations performed in Sections~\ref{sec:GSS_DTF} and~\ref{sec:GSS_STF} will be understood as subcases thereof. A reader who is interested just in such a general dualization may skip directly to Section~\ref{sec:GSS_MTF}.

\section{Double three-form multiplets}
\label{sec:GSS_DTF}

In the previous section, we have shown how, using one double three-form multiplet containing a complex gauge three-form, a positive contribution to the vacuum energy can be generated by integrating out that three-form. In the dual, ordinary chiral multiplet picture, this potential was generated by a linear superpotential $W= b\, \Phi$. In this section, we generalize the previous discussion to an arbitrary number of chiral superfields $\Phi^a$, with $a = 1,\ldots, n$, which will be replaced, as above, by appropriate three-form multiplets. We show how to relate an ordinary chiral theory with the superpotential
\be
\label{GSS_DTF_Wgen}
W_{\text{gen}} = e_a\Phi^a+ m^a \mathcal{G}_{ab}(\Phi) \Phi^b
\ee
with a dual three-form Lagrangian, where the superpotential \eqref{GSS_DTF_Wgen} is not included, but its contribution still enters the potential by integrating out the gauge three-forms \cite{Farakos:2017jme, Farakos:2017ocw}. In \eqref{GSS_DTF_Wgen}, we assume the parameters $e_a$ and $m^a$ to be real, $\calg_{ab}(\Phi)$ to be holomorphic in $\Phi^a$ and $\calg_{ab} (\Phi) = \calg_{ba} (\Phi)$. As we will see in more details in Chapter~\ref{chapter:EFT}, superpotential of the class \eqref{GSS_DTF_Wgen} are very common in compactification scenarios \cite{Gukov:1999ya,Taylor:1999ii}, where $e_a$ and $m^a$ are associated to the background fluxes.

For the sake of generality, we also consider an arbitrary K\"ahler potential for the chiral fields $\Phi$, $K(\Phi,\bar \Phi)$, and include a \emph{spectator} superpotential, $\hat W (\Phi)$. Namely, we start from the ordinary chiral multiplet theory
\be
\label{GSS_DTF_Lchir}
{\mathcal L}=\int \d^4\theta\, K(\Phi,\bar\Phi)+\left[ \int \d^2\theta \big(e_a\Phi^a+ m^a \mathcal{G}_{ab}(\Phi) \Phi^b + \hat W(\Phi)\big)+ {\rm c.c.}\right]\,,
\ee
whose bosonic components are
\be
\label{GSS_DTF_LchirComp}
\begin{split}
	\call |_{\rm bos} &= - K_{a\bar b} \del_m \varphi^a \del^m \bar \varphi^{\bar b} + K_{a\bar b} f^a \bar f^{\bar b} + 
	\\
	&\quad\,+\left[f^a \left(e_a + \calg_{ab}(\varphi) m^b+ \calg_{abc}(\varphi) m^b \varphi^c  + \hat W_a (\varphi)\right) + {\rm c.c.}\right]\,.
\end{split}
\ee
Integrating out the auxiliary fields $f^a$, we arrive at the following on-shell Lagrangian
\be
\label{GSS_DTF_LchirCompOS}
\begin{split}
	\call |_{\rm bos} &= - K_{a\bar b} \del_m \varphi^a \del^m \bar \varphi^{\bar b} - V\,,
\end{split}
\ee
where the potential is
\be
\label{GSS_DTF_VOS}
\begin{split}
	V &= K^{\bar b a} \left(e_a + \calg_{ac}(\varphi) m^c+ \calg_{acd}(\varphi) m^c \varphi^d  + \hat W_a(\varphi)\right)
	\\ 
	&\quad\quad\quad\, \times \left(e_b + \bar \calg_{\bar b \bar e}(\bar \varphi) m^e+ \bar \calg_{\bar b \bar e \bar f}(\bar\varphi) m^e \bar \varphi^{\bar f}  + \bar{\hat W}_{\bar b} (\bar \varphi)\right)\,.
\end{split}
\ee

\subsection{Generating a linear superpotential}

First, let us consider the case in which the superpotential that we are going to generate is linear. Then, in \eqref{GSS_DTF_Wgen} we take the matrix $\calg_{ab}(\Phi)$ to be constant, set
\be
\label{GSS_DTF_ca}
c_a\equiv e_a+\calg_{ab}m^b \, , 
\ee
and rewrite the Lagrangian in the form
\be
\label{GSS_DTF_LchirL}
{\mathcal L}=\int \d^4\theta\, K(\Phi,\bar\Phi)+\left[ \int \d^2\theta \big(c_a\Phi^a + \hat W(\Phi)\big)+ {\rm c.c.}\right]\,.
\ee

As in the previous section, we can promote the constants to
chiral superfield Lagrange multipliers $X_a$ and add a dualizing term containing the complex linear multiplets $\Sigma^{\bar a}$. We then construct the master Lagrangian
\be\label{GSS_DTF_LML}
\begin{split}
	\call_{\rm master}&=\int \d^4\theta\, K(\Phi,\bar\Phi)
	\\
	&\quad\,+\left[\int \d^2\theta \Big(X_a\Phi^a +\frac14 \bar D^2\left(\bar X_{\bar a}\Sigma^{\bar a}\right)\Big)+\int\d^2\theta\,\hat W(\Phi)+\text{c.c.}\right]\,.
\end{split}
\ee
We can now follow the two-fold path of the previous section as follows.

\begin{description}
	\item[Ordinary formulation] First, let us check that the master Lagrangian \eqref{GSS_DTF_LML} reproduces \eqref{GSS_DTF_LchirL}. Re-expressing the complex linear multiplets as $\Sigma^{\bar a} = \bar D \bar \Psi^{\bar a}$ as in \eqref{GSS_In_SigmaPsi} and integrating out $\Psi^{a}$ and $\bar \Psi^{\bar a}$, we get
	\be
	\label{GSS_DTF_Psi_XL}
	\delta \Psi^a:\qquad X_a =  c_a\,,
	\ee
	with $c^a$ being arbitrary complex constants. Substituting \eqref{GSS_DTF_Psi_XL} into \eqref{GSS_DTF_LML}, we re-obtain \eqref{GSS_DTF_LchirL}.
	\item[Three-form formulation] In order to get the superspace three-form Lagrangian \eqref{GSS_In_Dual_SS}, we integrate out, from \eqref{GSS_In_Mast}, the Lagrange multiplier $X$ and the ordinary chiral superfield $\Phi$:
	\be
	\begin{split}
		\label{GSS_DTFL_XPhi}
		&\delta X^a:\qquad \Phi^a = -\frac{1}{4} \bar{D}^2 \bar{\Sigma}^a\,,
		\\
		&\delta \Phi^a:\qquad X_a = \frac{1}{4} \bar{D}^2 K_a - \hat W_a \,.
	\end{split}
	\ee
	The first relation expresses, in superfield language, the trading between the components \eqref{GSS_In_ChiralComp} and \eqref{GSS_In_Scomp}
	\be
	\begin{split}
		\label{GSS_DTF_L}
		&S^a \equiv -\frac{1}{4} \bar{D}^2 \bar{\Sigma}^a\,.
	\end{split}
	\ee
	Plugging the solutions \eqref{GSS_DTFL_XPhi} back into the master Lagrangian \eqref{GSS_DTF_LML} leads to:
	\be\label{GSS_DBTL_Dual_SS}
	\begin{split}
		\tilde\call=& \int \d^4\theta\, K(S, \bar{S}) + \left(\int \d^2\theta\, \hat W(S) + {\rm c.c.}\right) +\tilde \call_{\rm{bd}}
	\end{split}
	\end{equation}
	with the boundary terms
	\be\label{GSS_DBTL_Dual_Bd}
	\begin{split}
	\tilde\call_{\rm{bd}} &= - \frac14 \left(\int \d^2\theta \bar D^2 - \int \d^2\bar\theta D^2\right) \left[ \left( \frac{1}{4} \bar{D}^2 K_a - \hat W_a \right)\bar \Sigma^{a}\right]
	+\text{c.c.}
	\end{split}
	\ee
	Its bosonic components are
	\be
	\label{GSS_DTF_L3fComp}
	\begin{split}
		\call |_{\rm bos} &= - K_{a\bar b} \del_m \varphi^a \del^m \bar \varphi^{\bar b} - \frac{1}{4\cdot4!} K_{a\bar b} F_{mnpq}^a \bar F^{\bar b\, mnpq }
		\\
		&\quad\, +\left[ \frac{\ii}{2\cdot 3!} \varepsilon^{mnpq} \del_m \bar C^a_{npq}\, \hat W_a (\varphi) + {\rm c.c.}\right]\,,
		\\
		&{\rm with}\qquad   \tilde\call_{\rm bd} = \frac{1}{3!} \partial_m \left[\frac\ii2 \bar {C}^{ a}_{npq} \left(-\frac\ii2 K_{a \bar b} F^{\bar b\, mnpq} - \varepsilon^{mnpq} {\hat{W}}_{a}(s) \right)\right] +{\rm c.c.}\,.
	\end{split}
	\ee
	It is simple to prove that this Lagrangian indeed generates a whole family of Lagrangians \eqref{GSS_DTF_LchirCompOS}, provided the identification \eqref{GSS_DTF_ca}. In fact, by integrating out the gauge three-forms via
	\begin{equation}
	\label{GSS_DTF_L3f_io}
	\partial_m \left[-\frac\ii2 K_{a \bar b} F^{\bar b\, mnpq} - \varepsilon^{mnpq}\hat{W}_a(s)\right] = 0\,,
	\end{equation}
	which implies
	\begin{equation}
	\label{GSS_DTF_L3f_iob}
	-\frac\ii2 F^{\bar b}_{mnpq} = K^{\bar b a} \left( c_a +\hat{W}_a(s) \right) \varepsilon_{mnpq}\,,
	\end{equation}
	we arrive at a Lagrangian of the form \eqref{GSS_DTF_LchirCompOS}, with
	\be
	\label{GSS_DTFL_VOS}
	\begin{split}
		V &= K^{\bar b a} \left(c_a   + \hat W_a(\varphi)\right) \left(c_b  + \bar{\hat W}_{\bar b} (\bar \varphi)\right)\,.
	\end{split}
	\ee
\end{description}

\subsection{Generating a nonlinear superpotential}

We now pass to examine the more involved case where $\calg_{ab}(\Phi)$ in \eqref{GSS_DTF_Wgen} is a general symmetric holomorphic matrix. For convenience, we define
\be\label{GSS_DTF_splitGab}
\caln_{ab}=\Re\calg_{ab}\,,\quad\calm_{ab}=\Im\calg_{ab}\,.
\ee
In the following, we assume that $\calm_{ab}$ is invertible, $\det(\calm_{ab})\neq 0$, and let us call $\calm^{ab}$ its inverse. Furthermore we assume, for simplicity, that $\calg_{abc}(\varphi) \varphi^c = 0$ (that is, $\calg_{ab}(\varphi)$ is homogeneous of degree two). This simplifying condition will be relaxed in Section \ref{sec:GSS_MTF}.

In order to get, at once, the ordinary chiral multiplet theory as well as the three-form theory, we start with the master Lagrangian, that is a slight generalization of \eqref{GSS_DTF_LML}
\be\label{GSS_DTF_LM}
\begin{aligned}
	\call_{\rm master} =&\int\d^4\theta\, K(\Phi,\bar\Phi)+\left[\int\d^2\theta\left(X_a\Phi^a+\hat W(\Phi)\right)+\text{c.c.}  \right]\\
	&-\frac14\left[\int\d^2\theta\, \bar D^2\left(\Sigma_a\,\calm^{ab}(X_b-\bar X_{\bar b})\right)+\text{c.c.}\right]\,.
\end{aligned}
\ee

The two-fold path that we followed above is here a little more subtle. Nevertheless, the logic remains the same.
\begin{description}
	\item[Ordinary formulation] Integrating out the complex linear multiplets $\Sigma_{ a} = \bar D \bar \Psi_{a}$ gives
	\be
	\label{GSS_DTF_Psi_X}
	\delta \bar \Psi^{a}:\qquad D_\alpha(\calm^{ab}\,\Im X_b)=0\,,
	\ee
	and we get an analogous relation for the complex conjugate. This is solved by
	\be
	\calm^{ab} \,\Im X_b = m^a\,
	\ee
	for real $m^a$, but, in order to find the most general solution, we set
	\be
	X_b = \Re X_b + \ii \calm_{bc} m^c = \Re (X_b - \calg_{bc} m^c) + \calg_{bc} m^c\,.
	\ee
	The chirality of $X_b$ implies that $\Re (X_b - \calg_{bc} m^c)$ has to be just a (real) constant $e_b$, whence 
	 \be
	 X_b = e_b + \calg_{bc} m^c
	 \ee
	which, plugged into \eqref{GSS_DTF_LM}, allows for recovering \eqref{GSS_DTF_Lchir}.
	\item[Three-form formulation] More interestingly, let us see how, from \eqref{GSS_In_Mast}, we may get a three-form Lagrangian, by integrating out both the Lagrange multipliers $X_a$ and the ordinary chiral superfields $\Phi^a$:
	\be
	\begin{split}
		\label{GSS_DTF_XPhi}
		&\delta X_a:\qquad \Phi^a = \frac14 \bar D^2\left[\calm^{ab}(\Sigma_b-\bar\Sigma_b)\right]\,,
		\\
		&\delta \Phi^a:\qquad X_a = \frac{1}{4} \bar{D}^2 K_a - \hat W_a \,.
	\end{split}
	\ee
	The first relation tells how the \emph{old} chiral superfields are traded, that is $\Phi^a$ are exchanged with the new chiral superfields
	\begin{important}
		\be
		\label{GSS_DTF_NL}
		S^a\equiv\frac 14 \bar D^2\left[\calm^{ab}(\Sigma_b-\bar\Sigma_b)\right] \qquad{\color{darkred}\text{\sf{Nonlinear double three-form multiplet}}}
		\ee
	\end{important}
	which we dub \emph{nonlinear double three-form multiplets}. 
	
	Before computing its components explicitly, let us notice that the complex vector component of $\Sigma_a$ now have to be split as
	\be
	\label{GSS_DTF_NLcompA3}
	-\frac12 \bar\sigma^{m\dot\alpha\alpha} [D_\alpha,\bar D_{\dot\alpha}] \Sigma_a | = \tilde A_a^m - \calg_{ab} A^{b\,m}\,,
	\ee
	and
	\be
	\label{GSS_DTF_NLcompA3HD}
	\tilde A_a^m = - \frac{1}{3!} \varepsilon^{mnpq} \tilde A_{a\,mnp}\,,\qquad A_a^m = - \frac{1}{3!} \varepsilon^{mnpq} A_{a\,mnp}\;.
	\ee
	The split in \eqref{GSS_DTF_NLcompA3} can be traced back to the structure of the nonlinear multiplet \eqref{GSS_DTF_NL}, which shows that the gauge transformations are
	\be
	\label{GSS_DTF_NLGInv}
	\Sigma_a \rightarrow \Sigma_a + L_{1a} - \calg_{ab} L^b_2
	\ee
	or, in components,
	\be
	\label{GSS_DTF_NLSGinvComp}
	C_{a\,mnp} \rightarrow C_{a\,mnp} + 3 \del_{[m} (\Lambda_{1a} -  \bar\calg_{ab}\Lambda_2^b)_{np]}\,,
	\ee
	confirming that the split \eqref{GSS_DTF_NLcompA3} is the correct one that delivers the proper gauge transformations for the both real $A_{3}^a$ and $\tilde A_{3a}$. Accordingly, we define the \emph{complex} four-forms
	\be
	\calf_{4a} \equiv \tilde F_{4a} - \bar \calg_{ab} F^b_4\;,
	\ee
	%
	%
	%
	which, clearly, are not closed (and are not proper field strengths!). 
	
	With respect to the simpler double three-form multiplets \eqref{GSS_DBT_L}, \eqref{GSS_DTF_NL} defines the new multiplets $S^a$ recursively, since also $\calm^{ab}$ depends on them. Nevertheless, we may still set
	\begin{subequations}
	\label{GSS_DTF_NLComp}
	\begin{align}
		S^a | &= \calm^{ab}(\varphi) \left(-\frac14 \bar D^2 \bar \Sigma_b \big|\right) \equiv s^a\,, \label{GSS_DTF_NLComps}
		\\
		-\frac14 D^2 S^a | &\equiv F_{(S)}^a = -\frac\ii2 \calm^{ab} *\! \calf_{4b} + \ii \calm^{ab} \Re (\calg_{bcd} F^c_{(S)} \varphi^d)\,,
		\label{GSS_DTF_NLCompFS}
	\end{align}
	\end{subequations}

	We are now ready to explicitly compute the dual Lagrangian for the nonlinear double three-form multiplets. Inserting \eqref{GSS_DTF_XPhi} into the master Lagrangian \eqref{GSS_DTF_LM}, we get
	\be\label{GSS_DTF_Dual_SS}
	\begin{split}
		\tilde\call=& \int \d^4\theta\, K(S, \bar{S}) + \left(\int \d^2\theta\, \hat W(S) + {\rm c.c.}\right) +\tilde \call_{\rm{bd}}
	\end{split}
	\ee
with the boundary terms
\be\label{GSS_DTF_Dual_Bd}
\begin{split}
	\tilde\call_{\rm{bd}} &= \frac14 \left(\int\d^2\bar\theta D^2-\int\d^2\theta \bar D^2\right)\left(X_a\calm^{ab}\bar\Sigma_b\right)
	+\text{c.c.} \, , 
\end{split}
\ee
where $X_a$ are given in \eqref{GSS_DTF_XPhi}. Its bosonic components, which we write via the Hodge-duals of the gauge three-forms and four-form field strengths for brevity, are
\be
\label{GSS_DTF_NL3fComp}
\begin{split}
	\call |_{\rm bos} &= - K_{a\bar b} \del_m \varphi^a \del^m \bar \varphi^{\bar b} + \frac{1}{4} K_{a\bar b} \calm^{ac} \calm^{bd} *\!\calf_{4c} *\! \bar \calf_{4d}
	\\
	&\quad\, +\left[ -\frac\ii2 \calm^{ab} \hat W_a *\!\calf_{4b} +{\rm c.c.}\right]+\tilde\call_{\rm bd}\,,
\end{split}
\ee	
with
\be
\begin{split}
	\tilde\call_{\rm bd} = \del_m\left\{ \left[\frac14 K_{a\bar b} \calm^{bd} \!*\bar\calf_{4d} \calm^{ac} -\frac\ii2 \calm^{ac} \hat W_a\right] (\tilde A_c^m - \bar \calg_{ce} A^{e\,m})\right\} +{\rm c.c.}
\end{split}
\ee
Here we have preferred to keep, as dynamical scalar fields, the ones of the \emph{old} multiplets $\varphi^a$, rather than \eqref{GSS_DTF_NLComps} due to their simpler kinetic terms.

As a further consistency check, integrating out the gauge three-forms $A_3^A$ and $\tilde A_{3A}$ \emph{separately}, we obtain
\begin{equation}
\begin{split}
	\label{GSS_DTF_NL3f_io}
	& 2 \Re \left[\frac14 K_{a\bar b} \calm^{bd} \!*\bar\calf_{4d} \calm^{ac} -\frac\ii2 \calm^{ac} \hat W_a\right] = - m^c\,,
	\\
	&2\Re \left[\frac14 K_{a\bar b} \calm^{bd} \!*\bar\calf_{4d} \calm^{ac} \bar \calg_{\bar c \bar f} -\frac\ii2 \calm^{ac} \hat W_a \bar\calg_{\bar c \bar f} \right] = - e_f\,,
\end{split}
\end{equation}
where $e_a$ and $m^a$ are real constants. These are simultaneously solved by setting
\begin{equation}
\begin{split}
	\label{GSS_DTF_NL3f_em}
	& \calm^{ab} *\! \bar\calf_{4b} = 2 \ii K^{\bar a c} (e_c + \calg_{cd} m^d + \hat W_c)\;.
\end{split}
\end{equation}
Reinserting \eqref{GSS_DTF_NL3f_em} into \eqref{GSS_DTF_NL3fComp}, the full potential \eqref{GSS_DTF_VOS} is obtained, in the assumption that $\calg_{  a  b c}(\varphi) \varphi^c = 0$.
\end{description}

\section{Single three-form multiplets}
\label{sec:GSS_STF}

There is another viable option to build chiral multiplets whose components contain a gauge three-form, that is by means of a real multiplet as a prepotential. As it is clear from its superspace expansion \eqref{Conv_V}, a real multiplet $U$ contains the components of a single, \emph{real} one-form, $A_1 = A_m \d x^m$, in its $\theta\bar\theta$-part as
\be
-\frac 14 \bar{\sigma}^{\dot\alpha \alpha}_m [D_\alpha, \bar{D}_{\dot\alpha}] U| = A_m\,.
\ee
As in \eqref{GSS_In_CL_HD},  we Hodge-dualize $A_1$ to $A_3 = \frac{1}{3!} A_{npq} \d x^n \d x^p \d x^q$, i.e.
\be
\label{GSS_STF_HD}
A^m = - \frac{1}{3!} \varepsilon^{npqm} A_{npq}\,.
\ee
As we did for the complex linear multiplet $\Sigma$ in \eqref{GSS_DBT_L}, we may define a chiral multiplet out of the real multiplet $U$ by taking its chiral projection
\begin{important}
\be
\label{GSS_STF}
Y \equiv -\frac \ii 4 \bar D^2 U \qquad{\color{darkred}\text{\sf{Single Three-form Multiplet}}}
\ee
\end{important}
The superfield  $Y$ is chiral by construction and therefore enjoys the expansion
\be
\label{GSS_STF_Ex}
Y=y+\sqrt{2}\theta\chi+\theta^2 f_Y\,,
\ee
whose components, in terms of those of $U$, may be computed from the projections\footnote{We refer to \eqref{Conv_Vprojections} for the projections of a generic real multiplet.}
\begin{subequations}
\label{GSS_STF_Comp}
\begin{alignat}{3}
	Y | &=  \left(-\frac \ii 4 \bar D^2 U \right)\Big| &\equiv &\, y \,,
	\\
	D^\alpha Y | &= D^\alpha  \left(-\frac \ii 4 \bar D^2 U \right) \Big|&\equiv &\, \chi^\alpha \,, 
	\\
	-\frac14 D^2 Y | &=	-\frac14 D^2 \left(-\frac \ii 4 \bar D^2 U \right)\Big| &=&\,  \frac12 \left( *F_4+  \ii d  \right) \equiv f_Y \,,  \label{GSS_STF_CompfY}
\end{alignat}
\end{subequations}
As \eqref{GSS_DBT_L} and \eqref{GSS_DTF_NL}, $Y$ differs from an ordinary chiral superfield \eqref{GSS_In_ChiralComp} only in its highest, $\theta^2$-component. However, in contrast with \eqref{GSS_DBT_L} and \eqref{GSS_DTF_NL}, the highest components are not fully replaced by gauge three-forms, since the real, scalar auxiliary field $d$ still remains in \eqref{GSS_STF_Comp}. For this reason, we will refer to the chiral multiplets of the kind \eqref{GSS_STF} as \emph{single three-form multiplets}.
 
Unlike the double three-form multiplets, these have attracted much more attention in the past. They were first constructed in \cite{Gates:1980ay} and later reconsidered, for example, in \cite{Groh:2012tf,Dudas:2014pva,Becker:2016xgv}. Above all, it was recognized that they were useful for the study of super Yang-Mills theory, where three-forms were related to three-dimensional Chern-Simons terms (see \cite{Burgess:1995kp,Farrar:1997fn,Cerdeno:2003us} and the recent work \cite{Bandos:2019qok}).

Since in a single three-form multiplet one \emph{real} gauge three-form appears, in \eqref{GSS_STF_CompfY}, only a real constant may be dynamically generated for each single three-form multiplet in the theory. Therefore, given $n$ single three-form multiplets $Y^a$  \eqref{GSS_STF}, the most general superpotential that we may dynamically generate in the dual chiral picture is
\be
\label{GSS_STF_Wgen}
W_{\text{gen}} = r_a\Phi^a\;,
\ee
with $r_a$ being $n$ \emph{real} constants. In other words, the most general, up to two derivatives, chiral multiplet theory to which we may relate a `dual' single three-form theory is
\be\label{GSS_STF_LCh}
{\mathcal L}=\int \d^4 \theta K(\Phi,\bar\Phi)+\left[\int \d^2\theta\big( r_a\Phi^a+\hat W(\Phi)\big)+\text{c.c.}\right]\,,
\ee
whose (on-shell) components may be easily deduced from \eqref{GSS_DTF_LchirCompOS}
\be
\label{GSS_STF_LchirCompOS}
\begin{split}
	\call |_{\rm bos} &= - K_{a\bar b} \del_m \varphi^a \del^m \bar \varphi^{\bar b} - V\,,
\end{split}
\ee
where the potential is
\be
\label{GSS_STF_VOS}
\begin{split}
	V &= K^{\bar b a} \left(r_a + \hat W_a(\varphi)\right) \left(r_b  + \bar{\hat W}_{\bar b} (\bar \varphi)\right)\,.
\end{split}
\ee

In order to get a three-form Lagrangian, we define the master Lagrangian 
\be\label{GSS_STF_LM}
\begin{aligned}
	{\mathcal L}_{\rm master}=&\int \d^4\theta\,  K(\Phi,\bar\Phi)+\left(\int \d^2\theta\, \hat W(\Phi)+\text{c.c.}\right)\,
	\\
	&+\left[\int \d^2\theta\, X_a\Phi^a+\frac\ii8\int \d^2\theta \bar D^2 \left[ (X_a-\bar X_a)U^a\right] +\text{c.c.}\right]\,,
\end{aligned}
\ee
where we have introduced a set of chiral Lagrangian multiplier $X_a$. The second line in \eqref{GSS_STF_LM} is singled out by the requirement that the three-forms do not appear via their field strengths. In fact, the bosonic components of \eqref{GSS_STF_LM} are
\be\label{GSS_STF_LMComp}
\begin{aligned}
	{\mathcal L}_{\rm master}|_{\rm bos}&= -K_{a\bar b}  \del_m \varphi^a \del^m \bar \varphi^{\bar b}+ K_{a\bar b}  f^a \bar f^b + \left( \hat W_a f^a + {\rm c.c.}\right)
	\\
	&\quad\, + \left[ x_a f^a - \frac{1}{2 \cdot 3!} \varepsilon^{mnpq} A^a_{npq} \del_m x_a  - \frac{\ii}{2} x_a d^a + {\rm c.c. }\right]
	\\
	&\quad\, +\left[ f_{Xa} (\varphi^a - y^a) + {\rm c.c. }\right]\,.
\end{aligned}
\ee

We may now follow two paths.
\begin{description}
\item[Ordinary formulation] The equations of motion of $U^a$, which are unconstrained superfields, can be immediately computed and they set
\be\label{GSS_STF_X}
X_a-\bar X_a=0 \quad\Rightarrow\quad X_a = r_a\,,
\ee
with $r_a$ being real constants. Plugging this solution back into the master Lagrangian \eqref{GSS_STF_LMComp}, we get the ordinary Lagrangian \eqref{GSS_STF_LCh}. The same can be achieved, in components, by integrating out at once, from \eqref{GSS_STF_LMComp} $y^a$, $A^a_3$ and $d^a$.

\item[Three-form formulation] To find the dual three-form formulation, the master Lagrangian has to be fully re-expressed in terms of the real potentials $U^a$ only. To this aim, we integrate out the Lagrange multipliers $X_a$ and the ordinary chiral multiples $\Phi^a$ as
\begin{subequations}\label{GSS_STF_XPhi}
	\begin{align}
	\label{GSS_STF_XX}
	&\delta X_a:\qquad \Phi^a = -\frac\ii4 \bar D^2 U^a \equiv Y^a\,,
	\\
	\label{GSS_STF_Phi}
	&\delta \Phi^a:\qquad X_a = \frac{1}{4} \bar{D}^2 K_a - \hat W_a \,.
	\end{align}
\end{subequations}
The first of these relations provides the sought trading: the old chiral multiplets are exchanged with the single three-form multiplets \eqref{GSS_STF}. Plugging the solutions \eqref{GSS_STF_XPhi} back into \eqref{GSS_STF_LM}, we arrive at
\be\label{GSS_STF_L3f}
\tilde{\mathcal L}=\int \d^4 \theta\, K(Y,\bar Y)+\left[\int\d^2\theta\, \hat W(Y)+\text{c.c.}\right]\,+\tilde {\mathcal L}_{\rm bd},
\ee
with the superspace--formulated boundary terms
	\be\label{GSS_STF_L3fbd}
	\tilde{\mathcal L}_{\rm bd}= 
	-\frac\ii8 \left(\int\d^2\theta \bar D^2-\int\d^2\bar\theta D^2\right)\left[\left(\frac14 \bar D^2 K_A - \hat{W}_A \right)U^A\right]
	+\text{c.c.}
	\, .
	\ee
In components, \eqref{GSS_STF_L3f} reads
\be
\label{GSS_STF_L3fCompA}
\begin{split}
	\call |_{\rm bos} &= - K_{a\bar b} \del_m \varphi^a \del^m \bar \varphi^{\bar b} + \frac14 K_{a\bar b} ( *\! F^a_4+\ii d^a ) ( *\! F^b_4-\ii d^b)
	\\
	&\quad\, +\left[ \frac12 \hat W^a \left( *\! F^a_4+\ii d^a\right) + {\rm c.c.}\right]\,,
\end{split}
\ee
with
\be
\label{GSS_STF_L3fCompAbd}
\begin{split}
	&\tilde\call_{\rm bd} = \frac{1}{3!} \partial_m \left\{ \left[ \varepsilon^{mnpq} \left(\Re \hat W_a + \frac12 \Im K_{a\bar b} d^b +\frac12 \Re K_{a \bar b} *\! F_4^b \right)\right] A_{npq}^a\right\} \,.
\end{split}
\ee
However, with respect to \eqref{GSS_DTF_L3fComp}, still some residual auxiliary fields, namely the $d^a$s, are present. In order to get a Lagrangian in a form similar to \eqref{Intro_Pot_S3f}, we have to integrate them out. By using their equations of motion, we get
\be
\label{GSS_STF_L3fda}
\begin{split}
	d^a &= - \calr^{ab} \left(  \cali_{bc} *\! F_4^c- 2 \Im \hat W_b\right)
\end{split}
\ee
where we have defined $\calr_{ab} \equiv \Re K_{a\bar b}$ and $\cali_{ab} \equiv \Im K_{a\bar b}$, whence \eqref{GSS_STF_L3fCompA} becomes
\be
\label{GSS_STF_L3fComp}
\begin{split}
	\tilde \call |_{\rm bos} &= - K_{a\bar b} \del_m \varphi^a \del^m \bar \varphi^{\bar b} + \frac14 \hat\calr_{ab} *\! F_4^a *\! F_4^b
	\\
	&\quad\, +  \left( \Re \hat W_a + \Im \hat W_c \calr^{cb} \cali_{ba}  \right)*\! F_4^a - \calr^{ab} \Im \hat W_a \Im \hat W_b + \tilde\call_{\rm bd} \,,
\end{split}
\ee
with
\be
\label{GSS_STF_L3fCompbd}
\begin{split}
	&\tilde\call_{\rm bd} = \frac{1}{3!} \partial_m \left\{ \left[ \varepsilon^{mnpq} \left(\Re \hat W_a + \cali_{ab} \calr^{bc} \Im \hat W_c +\frac12 \hat \calr_{ab} *\! F_4^b \right)\right] A_{npq}^a\right\} \,,
\end{split}
\ee
where now $\hat\calr^{ab} = \Re K^{\bar a b}$ and its inverse $\hat\calr_{ab}$.

As a consistency check, after integrating out the gauge three-forms via
\be
\label{GSS_STF_L3fA3os}
\begin{split}
	*\!F_4^a = - 2 \hat \calr^{ab} \left(r_b + \Re \hat W_b + \cali_{bc} \calr^{cd} \Im \hat W_d\right)
\end{split}
\ee
with $r_b$ real constants, it can be shown that the potential \eqref{GSS_STF_VOS} is recovered.

\end{description}
\section{Generic Lagrangians with three-form multiplets}
\label{sec:GSS_MTF}

In the previous sections we have shown how, in a generic supersymmetric theory with $n$ chiral multiplets $\Phi^a$ ($a=1,\ldots,n$), to trade \emph{all} $\Phi^a$ for double three-form multiplets, either \eqref{GSS_DTF_L} or \eqref{GSS_DTF_NL}, or with single three-form multiplets \eqref{GSS_STF}. As we shall see in Chapter~\ref{chapter:EFT}, effective theories arising from string/M-theory typically come with many chiral multiplets, containing different kinds of moduli, but only some of them can be regarded as single or double three-form multiplets. 

The dualization procedures explained in Section \ref{sec:GSS_DTF} and \ref{sec:GSS_STF} can be easily extended to Lagrangians which contain a \emph{spectator} sector of chiral superfields $T^q$ ($q = 1,\ldots,M$), namely chiral fields which are not traded with their three-form counterparts. More explicitly, we may easily promote
\be
K(\Phi, \bar \Phi) \rightarrow K(\Phi, \bar \Phi; T, \bar T)\,, \qquad \hat W(\Phi) \rightarrow \hat W(\Phi, T)\,
\ee
in the formulae of the previous sections, and the dualization procedure would proceed along the very same lines. There is however a complication at the level of the components: the spectators $T^q$, although not participating in the dualization, come with their auxiliary fields $f_T^q$. In order to get a three-form Lagrangian in the form \eqref{Intro_Pot_S3f}, they have to be integrated out and such an integration is, in general, not simple to be performed. We have already encountered this issue in the previous section where, to arrive at \eqref{GSS_STF_L3fComp}, the real auxiliary fields $d^a$ had to be integrated out. Therefore, it is tantalizing to try to formulate a more general, possibly more flexible dualization procedure which, given some chiral superfields, may allow for relating them in any way we prefer to three-form multiplets of any kind.

Consider $n$ chiral superfields $\Phi^a$
\be
\Phi^a = \varphi^a + \sqrt{2} \theta \psi^a + \theta^2 f^a
\ee
and assume that we want to dynamically generate a superpotential of the form
\be
\label{GSS_MTF_Wgen}
W_{\rm gen} = N_A \calv^A(\Phi)
\ee
where $\calv^A(\Phi)$, which we will refer to as \emph{periods}, already introduced in \eqref{Intro_SS_W}, are holomorphic functions of $\Phi^a$. The previous sections have instructed us that the core of the dualization procedure is finding out a master Lagrangian, as in \eqref{GSS_DTF_LML}, \eqref{GSS_DTF_LM}, \eqref{GSS_STF_LM}. Such Lagrangians are `special', in the sense that they are uniquely determined by requiring that the variational problem involving the gauge three-forms is well posed.  We have also seen that the single three-form multiplets \eqref{GSS_STF} contain \emph{minimally} the gauge three-forms: only a real gauge three-form is contained for each multiplet. We may consider them as the basic bricks to build up a general dualization procedure. In the master Lagrangian, however, the real multiplets $P^A$, through which they are defined, appear. Each of them contains a real gauge three-form, which serves to dynamically generate a real constant. Hence, let us introduces $N$ of real potentials $P^A$, whose relevant bosonic components are defined by
\be
\label{GSS_MTF_Pot}
\begin{aligned}
	-\frac 14 \bar{\sigma}^{\dot\alpha \alpha}_m [D_\alpha, \bar{D}_{\dot\alpha}] P^A| &= A_m^A\,,\\ 
	-\frac{\ii}{4} \bar D^2 P^A| &= s^A\,,
	\\
	\frac{\ii}{16} D^2 \bar{D}^2 P^A| &= \frac12\left( \ii d^A - \partial^m A^A_m \right) = \frac{1}{2} \left( \ii d^A + *F_4^A \right) \, .
\end{aligned}
\ee
Along with them, we need to introduce $N$ chiral Lagrangian multiplers $X_A$, with the expansion \eqref{GSS_In_X}. Equipped with all these ingredients, we generalize the master Lagrangian \eqref{GSS_STF_LM} as
\be
\label{GSS_MTF_LM}
\begin{aligned}
	\call_{\rm master} &=\int \d^4\theta\,  K(\Phi,\bar\Phi)+\left(\int \d^2\theta\, \hat W(\Phi)+\text{c.c.}\right)\,
	\\
	&\quad\,+ \left(\int \d^2\theta\, X_A\calv^A(\Phi)+\frac\ii8\int \d^2\theta \bar D^2  (X_A-\bar X_A)P^A  +\text{c.c.}\right)\,.
\end{aligned}
\ee
whose bosonic components have the following form
\be\label{GSS_MTF_LMComp}
\begin{split}
	 \call_{\rm master}|_{\text{bos}} &=-K_{a\bar b}  \del_m \varphi^a \del^m \bar \varphi^{\bar b}+ K_{a\bar b}  f^a \bar f^b + \left( \hat W_a f^a + {\rm c.c.}\right)
	\\
	&\quad\, + \left[ x_A\calv^A_a f^a - \frac{1}{2 \cdot 3!} \varepsilon^{mnpq} A^A_{npq} \del_m x_A  - \frac{\ii}{2} d^A x_A+ {\rm c.c. }\right]
	\\
	&\quad\, +\left[ f_{XA} (\calv^A - s^A) + {\rm c.c. }\right]\,.
\end{split}
\ee
This Lagrangian establishes a link between an ordinary formulation where only chiral multiplets are present and a three-form formulation.

\begin{description}
	\item[Ordinary formulation] From the master Lagrangian \eqref{GSS_MTF_LM}, we may retrieve the formulation in terms of ordinary chiral multiplet by integrating out the degrees of freedom of the real multiplets $P^A$. Integrating $P^A$ from the superspace Lagrangian immediately leads to
	\be\label{GSS_MTF_deltaPAa}
	\delta P^A:\qquad \Im X_A = 0 
	\ee
	which implies that
	\be\label{GSS_MTF_deltaPAb}
	X_A = N_A
	\ee
	with $N_A$ real constants. It is immediate to see that the superspace conditions \eqref{GSS_MTF_deltaPAa} and \eqref{GSS_MTF_deltaPAa} may be also read off from the component Lagrangian \eqref{GSS_MTF_LMComp} upon integrating out, respectively, the auxiliary fields $d^A$ and the three-forms $C_3^A$. Plugging the result \eqref{GSS_MTF_deltaPAb} into \eqref{GSS_MTF_LM}, we get the superspace Lagrangian
	\be\label{GSS_MTF_LCh}
	\begin{aligned}
		\call_{\rm chiral} &=\int \d^4\theta\,  K(\Phi,\bar\Phi)+\left(\int \d^2\theta\, W(\Phi)+\text{c.c.}\right)\,
	\end{aligned}
	\ee
	with the superpotential
	\be\label{GSS_MTF_LChW}
	W (\Phi) \equiv N_A \calv^A(\Phi) + \hat W(\Phi)
	\ee
	After setting the auxiliary fields on-shell, the bosonic components of the Lagrangian \eqref{GSS_MTF_LCh} are 
	\be
	\label{GSS_MTF_LChCompOS}
	\begin{split}
		\call |_{\rm bos} &= - K_{a\bar b} \del_m \varphi^a \del^m \bar \varphi^{\bar b} - V\,,
	\end{split}
	\ee
	where the potential is
	\be
	\label{GSS_MTF_VOS}
	\begin{split}
		V &= K^{\bar b a} \left(N_A \calv^A_a(\varphi)+ \hat W_a(\varphi)\right)
 	\left(N_B \bar \calv^B_{\bar b}(\bar \varphi)+ \bar{\hat W}_{\bar b}(\bar \varphi)\right)\,.
	\end{split}
	\ee
	\item[Three-form formulation] Integrating out the Lagrange multipliers $X_A$, we arrive at the following identification between the holomorphic functions $\calv^A(\Phi)$ and real potentials $P^A$
	\begin{important}%
		\be
		\label{GSS_MTF}
		\calv^A (\Phi) = -\frac \ii 4 \bar D^2 P^A \equiv S^A \qquad{\color{darkred}\text{\sf{Master three-form multiplet}}}
		\ee
	\end{important}
	whose components are
	\begin{subequations}
	\label{GSS_MTF_Comp}
	\begin{alignat}{4}\label{GSS_MTF_Compa}
		\calv^A | &= \calv^A(\varphi) &=& \left(-\frac \ii 4 \bar D^2 P^A \right) \Big| &\equiv &\, s^A \,, 
		\\\label{GSS_MTF_Compb}
		-\frac14 D^2 \calv^A | &=  \calv^A_a(\varphi) f^a &=&	-\frac14 D^2 \left(-\frac \ii 4 \bar D^2 P^A \right)\Big|  &=&\,  \frac12 \left( *F_4^A+  \ii d^A  \right) \equiv f^A \,.
	\end{alignat}
	\end{subequations}
	In contrast with the single or double three-form multiplets \eqref{GSS_STF}, \eqref{GSS_DTF_L} and \eqref{GSS_DTF_NL}, the trading \eqref{GSS_MTF} is less explicit: we are not trading a single chiral superfield $\Phi^a$ with a variant three-form version; rather,  \eqref{GSS_MTF} expresses a recipe to relate, via $\calv^A(\Phi)$ and the relations \eqref{GSS_MTF_Comp}, the components of the old, ordinary chiral superfields $\Phi^a$ with the components of the real potentials $P^A$. For this reason, we will refer to \eqref{GSS_MTF} as \emph{master three-form multiplets}.
	
	We may then proceed as in the previous sections. Further integrating out the ordinary chiral multiplets $\Phi^a$ we get
	\be\label{GSS_MTF_XA}
	\calv^A_a X_A = \frac14 \bar D^2 K_a - \hat W_a\,,
	\ee
	which, plugged back into the master Lagrangian \eqref{GSS_MTF_LM}, gives
	\be\label{GSS_MTF_L3f}
	\tilde{\mathcal L}=\int \d^4 \theta\, K(\Phi (P),\bar \Phi (P))+\left[\int\d^2\theta\, \hat W(\Phi(P))+\text{c.c.}\right]\,+\tilde {\mathcal L}_{\rm bd},
	\ee
	with
	\be\label{GSS_MTF_L3fbd}
	\tilde{\mathcal L}_{\rm bd}= 
	-\frac\ii8 \left(\int\d^2\theta \bar D^2-\int\d^2\bar\theta D^2\right)\left[\left(\frac14 \bar D^2 K_A - \hat{W}_A \right)P^A\right]
	+\text{c.c.} \,.
	\ee

	This Lagrangian, as it stands, has two awkward features: the superfields $\Phi^a$ appearing in \eqref{GSS_MTF_L3f} are not all independent, but have to be re-expressed in terms of the new superfields $S^A$ (hence, in terms of the potentials $P^A$) by inverting \eqref{GSS_MTF}; moreover, it still contains the real, auxiliary fields $d^A$ preventing us to immediately write a three-form Lagrangian in the form \eqref{Intro_Pot_S3f}. A convenient way out is to work directly with the component master Lagrangian \eqref{GSS_MTF_LMComp}, rather than with \eqref{GSS_MTF_LM}. Let us proceed as follows. First, let us integrate out $f_{XA}$ by using their equations of  motion, which set
	\be
	\label{GSS_MTF_IOCompa}	
	\delta\, f_{XA}: \qquad \calv^A(\varphi) = s^A\,.
	\ee
	Then, the integration of $d^A$ can be performed easily, immediately giving
	\be
	\label{GSS_MTF_IOCompb}	
	\delta\, d^A: \qquad \Im x_A = 0\,.
	\ee
	The master Lagrangian then becomes
	\be\label{GSS_MTF_LMCompc}
	\begin{split}
		\call_{\rm master}|_{\text{bos}} &=-K_{a\bar b}  \del_m \varphi^a \del^m \bar \varphi^{\bar b}+ K_{a\bar b}  f^a \bar f^b + \left( \hat W_a f^a + {\rm c.c.}\right)
		\\
		&\quad\, + \left[ x_A\calv^A_a f^a - \frac{1}{2 \cdot 3!} \varepsilon^{mnpq} A^A_{npq} \del_m x_A + {\rm c.c. }\right]
	\end{split}
	\ee
	Further integrating out the auxiliary fields $f^a$ of the oridinary chiral multiplets give
	\be
	\label{GSS_MTF_IOCompc}	
	\delta\, f^a: \qquad \bar f^{\bar b} = - K^{\bar b a} (x_A \calv^A_a+\hat W_a)\,,
	\ee
	which in turn leads to
	\be\label{GSS_MTF_LMCompd}
	\begin{split}
		\call_{\rm master}|_{\text{bos}} &=-K_{a\bar b}  \del_m \varphi^a \del^m \bar \varphi^{\bar b} - K^{\bar b a}\left( \hat W_a + x_A \calv^A_a\right)\left( \bar{\hat W}_{\bar b} + x_B \bar \calv^B_{\bar b}\right)
		\\
		&\quad\, + \left[ - \frac{1}{2 \cdot 3!} \varepsilon^{mnpq} A^A_{npq} \del_m x_A + {\rm c.c. }\right]\;.
	\end{split}
	\ee
	A three-form Lagrangian is obtained with a final integration of $x_A$
	\be
	\label{GSS_MTF_IOCompd}
	\delta\, x_A:\qquad x_A = - T_{AB} (*F_4^B + \Upsilon^B)
	\ee
	where we have introduced
	\begin{subequations}\label{GSS_MTF_TU}
		\begin{align}
		\label{GSS_MTF_TAB}
		T^{AB} &= 2\, \Re \left(K^{\bar b a} \calv^A_a \bar \calv^{\bar B}_{\bar b}\right)\,,
		\\
		\Upsilon^A &\equiv 2\, \Re \left(K^{\bar b a} \calv^A_a \bar{\hat W}_{\bar b}\right) \label{GSS_MTF_UA} \,,
		\end{align}
	\end{subequations}
	and $T_{AB}$ is the inverse of $T^{AB}$. We may then rewrite \eqref{GSS_MTF_LMCompc} as
	\begin{summary}	
	\be
	\label{GSS_MTF_L3fComp}	
	\begin{split}
	\call_{\text{3-form}}|_{\rm bos}&=-K_{a\bar b}  \del_m \varphi^a \del^m \bar \varphi^{\bar b} + \frac{1}{2} T_{AB} *\!F_4^A *\!F_4^B + T_{AB} \Upsilon^A *\!F_4^B  
	\\
	&\quad\,- \left(\hat V - \frac12 T_{AB} \Upsilon^A \Upsilon^B \right)+ \call_{\rm bd}
	\\
	&{\rm with}\qquad   \tilde\call_{\rm bd} = \frac{1}{3!} \partial_m \left[ \varepsilon^{mnpq} {A}^A_{npq} T_{AB} (*\!F_4^B + \Upsilon^B)\right] \,.
	\end{split}
	\ee
	\end{summary}
	and
	%
	\begin{align}
		\hat V &\equiv K^{\bar b a} \hat W_a \bar{\hat W}_{\bar b} \,. \label{GSS_MTF_hatV}
	\end{align}
	%
	It can be easily shown that, after setting the gauge three-forms on-shell
	\be
	\label{GSS_MTF_L3fNA}	
	\delta A_3^A: \qquad T_{AB} (*\! \hat F_4^B + \Upsilon^B) = - N_A\,,
	\ee
	a potential of the same form as \eqref{GSS_MTF_VOS} is obtained.
\end{description}

Before concluding this section, it is quite instructive to see how the cases examined in Sections \ref{sec:GSS_DTF} and \ref{sec:GSS_STF} are recovered from the general approach just introduced.

\subsubsection{Linear superpotential}

Assume that the superpotential \eqref{GSS_MTF_Wgen} which we are going to generate is
\be
\label{GSS_MTF_STF_W}
W_{\rm gen} = r_a \Phi^a\,,
\ee
with $r_a$ being $n$ real constants. From \eqref{GSS_MTF_Wgen}, the periods acquire the simple form
\be
\label{GSS_MTF_STF_Va}
\calv^a(\Phi) \equiv \Phi^a\,.
\ee
According to the recipe given above, the dynamical generation of the constants $r_a$ requires the presence of $n$ real three-forms $A_3^a$. In turn, at the superspace level, this translate into introducing, in the master Lagrangian \eqref{GSS_MTF_LM} $n$ real potentials $P^a$.

The superspace description of the three-form Lagrangian requires to solve the constraint \eqref{GSS_MTF}, which here simply produces \eqref{GSS_STF_Phi}. Namely, all the chiral superfields are replaced with single three-form multiplets $Y^a$. The superspace theory is then \eqref{GSS_STF_L3f}. In components, the three-form Lagrangian is \eqref{GSS_MTF_L3fComp}, with the identification
%
	\begin{align}
	\label{GSS_MTF_STF_TUV}
	T^{ab} = 2\, \Re K^{\bar b a} \,,\qquad 
	\Upsilon^a \equiv 2\, \Re \left(K^{\bar b a}  \bar{\hat W}_{\bar b}\right) \,,
	\qquad 
	\hat V \equiv K^{\bar b a} \hat W_a \bar{\hat W}_{\bar b} \,.
	\end{align}
%

\subsubsection{Maximally nonlinear case}

Let us assume that we want to generate a superpotential of the class
\be
\label{GSS_MTF_DTF_Wgen}
W_{\text{gen}} = e_a\Phi^a+ m^a \mathcal{G}_{ab}(\Phi) \Phi^b
\ee
with $e_a$ and $m^a$ a set of $2n$ real constants, as in Section \eqref{sec:GSS_DTF}. The periods which, as defined from \eqref{GSS_MTF_DTF_Wgen}, are
\be
\calv^A(\Phi)\equiv\left(\begin{array}{c} 
	\Phi^a \\ \calg_{bc}(\Phi) \Phi^c \end{array}\right)\,.
\ee
We need to introduce $2n$ real periods, which we split as
\be
P^A\equiv\left(\begin{array}{r} 
	\calp^a \\ \tilde\calp_b\end{array}\right)\;.
\ee
Now, the constraint \eqref{GSS_MTF} sets
\begin{subequations}\label{GSS_MTS_DTF_c}
	\begin{align}
	\Phi^a&=-\frac{\ii}{4}\bar D^2\calp^a \equiv S^a\, ,\label{GSS_MTS_DTF_cPhi}\\
	\calg_{ab}(\Phi) \Phi^b&=-\frac{\ii}{4}\bar D^2\tilde\calp_a \equiv \tilde S_a\, .
	\label{GSS_MTS_DTF_cG}
	\end{align}
\end{subequations}
In contrast with the previous example, here the form of the multiplets that we expect to describe our three-form theory, the nonlinear three-form multiplets \eqref{GSS_DTF_NL}, do not directly appear. We need to fully solve the constraints \eqref{GSS_MTS_DTF_c} in order to recover the same form of the theory as in Section \ref{sec:GSS_DTF}. Indeed, substituting \eqref{GSS_MTS_DTF_cPhi} into \eqref{GSS_MTS_DTF_cG}, we get the constraint
\be
\bar D^2\big[\calg_{ab}(\calz)\calp^b-\tilde\calp_a\big]=0\,.
\ee
This is solved by setting
\be\label{GSS_MTS_DTF_PhiSigma}
\calg_{ab}(\calz)\calp^b-\tilde\calp_a\equiv -2\Sigma_a\, ,
\ee
where $\Sigma_a$ are generic complex linear multiplets, which allows us to fully re-express the real multiplets in terms of the complex linear multiplets as
\be\label{GSS_MTS_DTF_PPsol}
\calp^a=-2\calm^{ab}\Im\Sigma_b\,, \quad \quad \tilde\calp_a=-2\Im(\bar\calg_{ab}\calm^{bc}\Sigma_c)\, .
\ee
Therefore, plugging these solutions into \eqref{GSS_MTS_DTF_cPhi}, we most readily see that the multiplets $S^a$ do actually coincide with the nonlinear multiplets \eqref{GSS_DTF_NL} of Section~\ref{sec:GSS_DTF}.  

The superspace Lagrangian is then \eqref{GSS_DTF_Dual_SS} with the boundary terms as in \eqref{GSS_DTF_Dual_Bd}. In components, the three-form Lagrangian is \eqref{GSS_MTF_L3fComp}, with
%
	\begin{align}\label{GSS_MTF_DTF_TUV}
	T^{AB} &= 2\Re \begin{pmatrix}
	K^{\bar b a} &  K^{\bar c a} \left(\bar \calg_{ \bar b \bar c  \bar d} \bar \Phi^{\bar d} +\calg_{ \bar b \bar c} \right) \\
	K^{\bar b e} \left( \calg_{  a  e  f}  \Phi^{ f} +\calg_{  a e} \right) &  K^{\bar c e}\left( \calg_{  a  e  f}  \Phi^{ f} +\calg_{  a e} \right) \left(\bar \calg_{ \bar b \bar c  \bar d} \bar \Phi^{\bar d} +\calg_{ \bar b \bar c} \right) \\
	\end{pmatrix} ,
	\\
	\Upsilon^A &\equiv 2\, \Re  \begin{pmatrix}
	K^{\bar b a} \bar{\hat W}_{\bar b}\\
	K^{\bar b c} \left( \calg_{  a  c  d}  \Phi^{ d} +\calg_{  a c} \right) \bar{\hat W}_{\bar b}\\
	\end{pmatrix} \,,
	\\
	\hat V &\equiv K^{\bar b a} \hat W_a \bar{\hat W}_{\bar b} \,. 
	\end{align}
%

\section{Axions and linear multiplets}
\label{sec:GSS_Axions}

As briefly recalled in the introduction of this chapter, gauge two-forms appear directly among the components of linear multiplets. Linear multiplets do contain their three-form field strengths, making them super-field strengths for gauge two-forms. In this section we review how to construct manifestly globally supersymmetric Lagrangians which include linear and chiral multiplets. This will serve as an introduction to build up Lagrangians where both gauge two- and three-forms appear, with the latter gauging the former.

A real linear superfield $L$ obeys the constraints 
\be
\label{GSS_AL_LinConstr}
D^2 L = 0\,,\qquad \bar D^2 L =0\,.
\ee
These reduce the degrees of freedom of $L$ in such a way that a real linear multiplet contains the same amount of propagating scalar fields as a chiral multiplet, namely, focusing on the bosonic sector, a real scalar (1 d.o.f) and a gauge two-form (1 d.o.f.) for the former and a complex scalar for the latter (2 d.o.f.). However, no bosonic non-propagating degrees of freedom are accommodated in linear multiplets. Still, the na\"ive comparison of the degrees of freedom already suggests that there may exist a `duality' which relates chiral and linear multiplets. Such a duality was introduced in \cite{Lindstrom:1983rt} and, for the sake of completeness and to set up the notation, we will summarize it here.

Consider a supersymmetric theory with chiral superfields $\Phi^a$, $a=1, \ldots, n$, which we will here treat as spectators, and some other chiral multiplets $T_\Lambda$, $\Lambda=1,\ldots, M$. Let us assume that the K\"ahler potential depends on $T_\Lambda$ only via their imaginary parts:
\be
\label{GSS_AL_Khyp}
K (\Phi, \bar \Phi; T) \equiv K (\Phi, \bar \Phi; \Im T)\,,
\ee
keeping generic the dependence on $\Phi^a$. The K\"ahler potential is then invariant under the \emph{axionic shifts}
\be
\label{GSS_AL_ShiftS}
T_\Lambda \quad \rightarrow  \quad T_\Lambda + c_\Lambda
\ee
with real constant $c_\Lambda$ and we identify the lowest components $\Re T_\Lambda| = \Re t_\Lambda \equiv a_\Lambda$ as \emph{axions}, the only components influenced by the shift \eqref{GSS_AL_ShiftS}
\be
\label{GSS_AL_Shifta}
a_\Lambda \quad  \rightarrow \quad  a_\Lambda + c_\Lambda\;.
\ee
For this reason, we will sometimes refer to the multiplets $T_\Lambda$ as \emph{axionic multiplets}.

We might therefore trade the axions $a_\Lambda$ with gauge two-forms $\calb_2^\Lambda$ via a usual `electro-magnetic duality', that is
\be
\label{GSS_AL_bosdual}
\d a_\Lambda \quad \leftrightarrow \quad * \d \calb_2^\Lambda\,,
\ee
the number of degrees of freedom does not change and the shift \eqref{GSS_AL_Shifta} simply translates into a shift of the gauge two form $\calb_2 \rightarrow \calb_2 + \Lambda_2$ for constant two-form $\Lambda_2$.

In the same spirit as the one adopted in Section \ref{sec:GSS_TIntro}, we may promote the relation \eqref{GSS_AL_bosdual} at the superspace level passing through a first order formalism. Therefore, we first relax the assumption that in \eqref{GSS_AL_Khyp} $\Im T_\Lambda$ are the imaginary parts of chiral superfields $T_\Lambda$ and rather consider them real, unconstrained superfields $U_\Lambda$.  Then, we start with the Lagrangian
\be
\label{GSS_AL_Dual1}
\call_{\rm dual} = \int \d^4 \theta K(\Phi, \bar \Phi; U) + 2 \int \d^4 \theta\, L^\Lambda U_\Lambda 
\ee
where $L^\Lambda$ are linear multiplets. Two paths may be followed, one leading to the ordinary formulation, solely depending on chiral multiplets with the K\"ahler potential \eqref{GSS_AL_Khyp}, the other to a dual formulation where the linear multiplets $L^\Lambda$ replace the axionic multiplets $T_\Lambda$. In details:

\begin{description}
	\item[Ordinary chiral formulation] In order to retrieve a formulation in terms of chiral multiplets only, we need to integrate out the linear multiplets $L^\Lambda$ from the Lagrangian \eqref{GSS_AL_Dual1}. It is clearly not possible to integrate them out directly, since they are constrained by \eqref{GSS_AL_LinConstr}. However, the constraint \eqref{GSS_AL_LinConstr} can be solved as
	\be
	\label{GSS_AL_Dual1L}
	L^\Lambda = D^\alpha \bar D^2 \Psi_\alpha^\Lambda + \bar D_{\dot \alpha} D^2 \bar \Psi^{\Lambda\dot\alpha}
	\ee
	where $\Psi^\Lambda_\alpha$, $\bar \Psi^{\Lambda \dot\alpha}$ are \emph{unconstrained} spinorial superfields. The -- now legitimate -- variations with respect to $\Psi^\Lambda_\alpha$, $\bar \Psi^{\Lambda \dot\alpha}$ of the Lagrangian \eqref{GSS_AL_Dual1} sets the following constraints on $U_\Lambda$:
	\be
	\label{GSS_AL_Dual1Psi}
	\delta \Psi^\Lambda_\alpha\,, \delta \bar \Psi^{\Lambda\dot\alpha}\,: \qquad  \bar D^2 D^\alpha U_\Lambda=0\,,\quad  D^2 \bar D_{\dot\alpha} U_{\Lambda} = 0\,.
	\ee
	These are solved by
	\be
	\label{GSS_AL_Dual1U}	
	U_\Lambda = \frac{1}{2\ii} (T_\Lambda - \bar T_\Lambda ) = \Im T_\Lambda 
	\ee
	for generic chiral multiplets $T_\Lambda$. Substituting \eqref{GSS_AL_Dual1U} into \eqref{GSS_AL_Dual1}, we get a Lagrangian fully depending on the chiral multiplets $\Phi$ and $T$
	\be
	\label{GSS_AL_Dual1c}
	\call_{\rm chiral} = \int \d^4 \theta K(\Phi, \bar \Phi; \Im T)\,. 
	\ee

	\item[Formulation with linear multiplets] The alternative formulation where the axionic multiplets $T_\Lambda$ are replaced by linear multiplets is obtained by integrating out the real multiplets $U_\Lambda$ from \eqref{GSS_AL_Dual1}, setting
	\be
	\label{GSS_AL_Dual1dU}
	\delta U_\Lambda: \qquad  L^\Lambda = -\frac12 \frac{\del K}{\del U_\Lambda}\;.
	\ee
	Once this is plugged inside \eqref{GSS_AL_Dual1}, we get
	\be
	\label{GSS_AL_Dual1Lag}
	\call_{\rm linear} = \int \d^4 \theta\, \left(K(\Phi, \bar \Phi; U) - \frac{\del K}{\del U_\Lambda}U_\Lambda \right) \equiv \int \d^4 \theta\, F(\Phi, \bar \Phi; L)
	\ee
	where $F(\Phi, \bar \Phi; L)$ is the Legendre transform of the K\"ahler potential $K(\Phi, \bar \Phi; \Im T)$. Indeed, \eqref{GSS_AL_Dual1c} provides a relation as the sought one \eqref{GSS_AL_bosdual}. In fact, assuming for simplicity that the K\"ahler potential depends only on the axionic multiplets $T_\Lambda$, the $[D,\bar D]$--component of \eqref{GSS_AL_Dual1dU} relates
	\be
	\d \Re t_\Lambda = - \frac14 F_{\Lambda \Sigma} *\! \d \calb_2^\Sigma\;.
	\ee
	For completeness, we also give the bosonic components of the Lagrangian \eqref{GSS_AL_Dual1Lag}, which are
	\be
	\label{GSS_AL_Dual1Lagbos}
	\begin{split}
		\call_{\rm linear} |_{\rm bos} &=  -F_{a \bar b} \del \varphi^a \del \bar \varphi^{\bar b}  + \frac14 F_{\Lambda\Sigma} \del_m l^\Lambda \del^m l^\Sigma  + \frac1{4 \cdot 3!} F_{\Lambda\Sigma}  \calh_{mnp}^\Lambda \calh^{\Sigma\,mnp}
		\\
		&\quad\,+ \left[ F_{\bar a \Lambda} \del_m \bar \varphi^{\bar a} \left( - \frac\ii{2 \cdot 3!} \varepsilon^{mnpq} \calh_{mnp}^\Lambda  \right) + {\rm c.c.}\right] + F_{a \bar b} f^a \bar f^{\bar b}\;.
	\end{split}
	\ee
\end{description}

The procedure outlined above can be followed also in the reverse direction: we may start with the Legendre transform \eqref{GSS_AL_Dual1Lag} and, by using a first order formalism, come back to the ordinary, chiral multiplets-only formulation. Since below and in the later Sections \ref{sec:Sugra_LM} this is the procedure that will be preferred, for the sake of completeness, we briefly show how one can obtain \eqref{GSS_AL_Dual1c} from \eqref{GSS_AL_Dual1Lag}. We start with the first order Lagrangian
\be
\label{GSS_AL_Dual2}
\call_{\rm dual} = \int \d^4 \theta F(\Phi, \bar \Phi; L) - 2 \int \d^4 \theta\, L^\Lambda \Im T_\Lambda\,.
\ee
The first piece is just \eqref{GSS_AL_Dual1Lag} where, however, $L^\Lambda$ are understood to be real multiplets rather than linear multiplets; the second piece contains the (imaginary parts of the) chiral multiplets $T_\Lambda$ and serves for the duality. According to the formulation that we prefer, two different roads are ahead.
\begin{description}
	\item[Formulation with linear multiplets] In order to integrate out the chiral multiplets $T_\Lambda$, we first need to solve the chirality constraints $D_\alpha T_\Lambda = 0$, $\bar D^{\dot \alpha} T_\Lambda =0$ and re-express them as\footnote{It is worthwhile to mention that it is here necessary to solve the chiral constraints because in \eqref{GSS_AL_Dual2} the superspace integration is performed over the full set of Grassmannian variables $\theta$, $\bar\theta$. In the master Lagrangians of the previous sections, such as \eqref{GSS_In_Mast}, this was not necessary because the integrations were performed in the chiral superspace.}
	\be
	\label{GSS_AL_Dual2T}
	\Im T_\Lambda = \frac{1}{2\ii} \left(\bar D^2 \bar \Xi_\Lambda - \bar D^2 \Xi_\Lambda\right)
	\ee
	with $\Xi_\Lambda$ unconstrained (complex) superfields. Varying \eqref{GSS_AL_Dual2} with repect to $\Xi_\Lambda$, we get the constraints
	\be
	\label{GSS_AL_Dual2U}
	\delta \Xi_\Lambda\,, \delta \bar \Xi_\Lambda\,: \qquad  D^2 L^\Lambda = 0\,,\quad \bar D^2 L^\Lambda = 0
	\ee
	which tell that $L^\Lambda$ are linear multiplets, retrieving \eqref{GSS_AL_Dual1Lag}.
	
	\item[Ordinary formulation] By integrating out the real multiplets $L^\Lambda$, we get	
	\be
	\label{GSS_AL_Dual2L}
	\delta L^\Lambda\,: \qquad  \Im T_\Lambda = \frac12 \frac{\del F}{\del L^\Lambda}\,,
	\ee
	which, plugged in \eqref{GSS_AL_Dual2}, allows us to re-obtain the ordinary chiral formulation
	\be
	\label{GSS_AL_Dual2Lag}
	\call_{\rm dual} = \int \d^4 \theta\, \left(F(\Phi, \bar \Phi; L) - \frac{\del F}{\del L^\Lambda}L^\Lambda \right) = \int \d^4 \theta\, K(\Phi, \bar \Phi; \Im T) = \call_{\rm chiral}\,.
	\ee
\end{description}

\subsection{Gauged linear multiplets and F-term couplings}
 \label{sec:GSS_GL}

Above we have assumed that no superpotential for the superfields $T_\Lambda$ is present: they only appear in the K\"ahler potential as in \eqref{GSS_AL_Khyp}, thus manifestly preserving the axionic symmetry. But let us now assume that the superfields $T_\Lambda$ do enter in a superpotential and, in particular, they appear via the coupling to the chiral superfields $\Phi^a$ as
\be
\label{GSS_GL_Wc}
W_{\rm coupling} = - c_A^\Lambda T_\Lambda \calv^A(\Phi)\;,
\ee
where the periods $\calv^A(\Phi)$ depend holomorphically on $\Phi^a$. The coupling \eqref{GSS_GL_Wc} is quite peculiar: although it clearly breaks the axionic shift symmetry, a dualization of $T_\Lambda$ with linear multiplets is still possible. As we shall see shortly, this possibility comes with a price, that is the \emph{gauging} of the linear multiplets with \emph{real} multiplets. Such a gauging procedure, here extended and generalized, was already performed in inflationary models \cite{Dudas:2014pva} as well as in \cite{Cerdeno:2003us,Bandos:2019qok} in super-Yang-Mills theory to give mass to \emph{glueballs}.

What we are seeking is a theory which simultaneously contains gauge two-forms sitting inside linear multiplets and gauge three-forms included into multiplets of the sort of \eqref{GSS_MTF}. Therefore, we preliminary start with a theory in the form of \eqref{GSS_AL_Dual2}. We emphasize that, in order to arrive at \eqref{GSS_AL_Dual2}, we have assumed that there is a set of superfields, $\Phi^a$, which is \emph{not} dualized to linear multiplets and is a spectator sector with respect to the axion/two-form duality presented above. We may therefore apply the very procedure of trading chiral multiplets with three-form multiplets of Section \ref{sec:GSS_MTF} to the set of superfields $\Phi^a$. We therefore introduce the following master Lagrangian:
\be
\begin{aligned}
\label{GSS_GL_LM}
\call_{\rm master} &= \int \d^4 \theta F(\Phi, \bar \Phi; \hat L) - 2 \int \d^4 \theta\, \hat L^\Lambda \Im T_\Lambda
\\
&\quad\,+\left(\int \d^2\theta\, \hat W(\Phi)+\text{c.c.}\right)\,
\\
&\quad\,+ \left(\int \d^2\theta\, X_A\calv^A(\Phi)+\frac\ii8\int \d^2\theta \bar D^2  (X_A-\bar X_A)P^A  +\text{c.c.}\right)
\\
&\quad\,+ \left(\frac\ii8\int \d^2\theta \bar D^2  c_A^\Lambda(T_\Lambda-\bar T_{\Lambda})P^A  +\text{c.c.}\right)\,.
\end{aligned}
\ee
The first line is just the Lagrangian \eqref{GSS_AL_Dual2} (where now $\hat L$ are unconstrained real multiplets), upon which we would like to apply the duality procedure of Section \ref{sec:GSS_MTF}. The latter duality is achieved with the third line of \eqref{GSS_GL_LM}. The novel last line of \eqref{GSS_GL_LM}, as we shall see in a moment, is what allows for a dynamical generation of a superpotential in the form \eqref{GSS_GL_Wc}. To this end, let us now show how, from \eqref{GSS_GL_LM}, one can obtain an ordinary chiral multiplet formulation and then a dual formulation with linear and three-form multiplets. 

\begin{description}
	\item[Ordinary chiral formulation] The formulation depending purely on chiral multiplets is obtained by integrating out the real superfields $\hat L^\Lambda$ and $P^A$, which lead, respectively, to
	\be
	\label{GSS_GL_Dual2La}
	\delta L^\Lambda\,: \qquad  \Im T_\Lambda = \frac12 \frac{\del F}{\del L^\Lambda}\;,
	\ee
	and
	\be
	\label{GSS_GL_Dual2Lb}
	\delta P^A\,: \qquad  \Im X_A + c_A^\Lambda \Im T_\Lambda = 0\;,
	\ee
	which implies, due to the chirality of both $X_A$ and $T_\Lambda$, that 
	\be
	\label{GSS_GL_Dual2Lc}
	X_A + c_A^\Lambda T_\Lambda = N_A\,,
	\ee
	with $N_A$ real constants. Plugging \eqref{GSS_GL_Dual2La} and \eqref{GSS_GL_Dual2Lc} into \eqref{GSS_GL_LM}, upon using the relation \eqref{GSS_AL_Dual2Lag}, we get 
	\be
	\label{GSS_GL_Dual2Lag}
	\call_{\rm chiral} = \int \d^4 \theta\, K(\Phi, \bar \Phi; \Im T) + \left(\int\d^2\theta\, W(\Phi;T)+ {\rm c.c.}\right)\;,
	\ee
	with the superpotential
	\be
	\label{GSS_GL_Dual2W}
	W(\Phi;T) = N_A \calv^A(\Phi) -c^A_\Lambda T^\Lambda \calv^A(\Phi)+ \hat W (\Phi) \;.
	\ee

	\item[Formulation with gauged linear multiplets] Obtaining the dual Lagrangian expressed in terms of linear and three-form multiplets is definitely much more involved. To begin with, we integrate out of the master Lagrangian \eqref{GSS_GL_LM} the chiral superfields $T_\Lambda$. Expressing them as in \eqref{GSS_AL_Dual2T} and varying with respect to the unconstrained $\Xi_\Lambda$, we get
	\be
	\label{GSS_GL_Dual2U}
	\delta \Xi_\Lambda\,, \delta \bar \Xi_\Lambda\,: \qquad  D^2 (\hat L^\Lambda - c_A^\Lambda P^A) = 0\,,\quad \bar D^2 (\hat L^\Lambda - c_A^\Lambda P^A)= 0
	\ee
	which are generically solved by setting
	\begin{important}
		\be
		\label{GSS_GL}
		\hat L^\Lambda = L^\Lambda + c_A^\Lambda P^A \qquad{\color{darkred}\text{\sf{Gauged linear multiplets}}}
		\ee
	\end{important}
	which defines the \emph{gauged linear multiplets}. They go by the name `gauged' because in their $\theta \bar \theta$--components there appear
	\be
	\label{GSS_GL_H3g}
	\hat \calh_3^\Lambda = \d \calb_2^\Lambda + c_A^\Lambda  A_3^A
	\ee
	which are the three-form field strengths for $\calb_2^\Lambda$ gauged by the three-forms $A_3^A$ in a St\"uckelberg--like manner.
	
	With \eqref{GSS_GL}, the master Lagrangian \eqref{GSS_GL_LM} becomes
	\be
	\begin{aligned}
		\label{GSS_GL_LMb}
		\call_{\rm master} &= \int \d^4 \theta F(\Phi, \bar \Phi; \hat L) +\left(\int \d^2\theta\, \hat W(\Phi)+\text{c.c.}\right)\,
		\\
		&\quad\,+ \left(\int \d^2\theta\, X_A\calv^A(\Phi)+\frac\ii8\int \d^2\theta \bar D^2  (X_A-\bar X_A)P^A  +\text{c.c.}\right)\;.
	\end{aligned}
	\ee
	It is also useful to write down its bosonic components 
	\be
	\begin{aligned}
	\label{GSS_GL_LMbcomp}
	\call_{\rm master}|_{\rm bos} &= -F_{a \bar b} \del \varphi^a \del \bar \varphi^{\bar b}  + \frac14 F_{\Lambda\Sigma} \del_m l^\Lambda \del^m l^\Sigma  + \frac1{4 \cdot 3!} F_{\Lambda\Sigma} \hat \calh_{mnp}^\Lambda \hat \calh^{\Sigma\,mnp}
	\\
	&\quad\,+ \left[ F_{\bar a \Lambda} \del_m \bar \varphi^{\bar a} \left( - \frac\ii{2 \cdot 3!} \varepsilon^{mnpq} \hat \calh_{mnp}^\Lambda  \right) + {\rm c.c.}\right]
	\\
	&\quad\,+ F_{a \bar b} f^a \bar f^{\bar b}+F_{\Lambda\Sigma} c_A^\Lambda c_B^\Sigma \calv^A(\varphi) \bar \calv^B(\bar\varphi) +
	\\
	&\quad\,+ \left[ \left(\hat W_a(\varphi)- \ii F_{ a \Lambda} c_A^\Lambda \calv^A(\varphi) -\frac\ii2 F_\Lambda c_A^\Lambda \calv^A_a (\varphi) \right) f^a +  {\rm c.c.} \right]
	\\
	&\quad\, + \left[ x_A\calv^A_a f^a - \frac{1}{2 \cdot 3!} \varepsilon^{mnpq} A^A_{npq} \del_m x_A  - \frac{\ii}{2} d^A x_A+ {\rm c.c. }\right]
	\\
	&\quad\, +\left[ f_{XA} (\calv^A - s^A) + {\rm c.c. }\right]\,.
	\end{aligned}
	\ee
	Now, beside the replacement of the K\"ahler potential with its Legendre transform, the master Lagrangian \eqref{GSS_GL_LMb} is formally equivalent to \eqref{GSS_MTF_LM}. We can then proceed along the same lines that we drew in Section \ref{sec:GSS_MTF}.
	
	At the superspace level, integrating out $X_A$ leads to the same identification as \eqref{GSS_MTF}, providing a trading between the ordinary chiral multiplets $\Phi^a$ with three-form multiplets. We then arrive at the superspace Lagrangian
	\be\label{GSS_GL_L3f}
	\tilde{\mathcal L}=\int \d^4 \theta\, F(\Phi (P),\bar \Phi (P);\hat L)+\left[\int\d^2\theta\, \hat W(\Phi(P))+\text{c.c.}\right]\,+\tilde {\mathcal L}_{\rm bd},
	\ee
	with the boundary terms
	\be\label{GSS_GL_L3fbd}
	\tilde{\mathcal L}_{\rm bd}= 
	-\frac\ii8 \left(\int\d^2\theta \bar D^2-\int\d^2\bar\theta D^2\right)\left[\left(\frac14 \bar D^2 F_A - \hat{W}_A \right)P^A\right]
	+\text{c.c.} \,.
	\ee
	As stressed in Section \ref{sec:GSS_MTF}, this Lagrangian is not particularly useful for obtaining its bosonic components. We can however work with the bosonic Lagrangian \eqref{GSS_GL_LMbcomp}. Following the very same steps as in Section \ref{sec:GSS_MTF} to pass from \eqref{GSS_MTF_LM} to \eqref{GSS_MTF_L3fComp}, we arrive at the Lagrangian
	\begin{summary}
	\be
	\label{GSS_GL_L3fComp}	
	\begin{split}
		\tilde \call |_{\rm bos}&=-F_{a \bar b} \del \varphi^a \del \bar \varphi^{\bar b}  + \frac14 F_{\Lambda\Sigma} \del_m l^\Lambda \del^m l^\Sigma  + \frac1{4 \cdot 3!} F_{\Lambda\Sigma} \hat \calh_{mnp}^\Lambda \hat \calh^{\Sigma\,mnp}
		\\
		&\quad\,+ \left[ F_{\bar a \Lambda} \del_m \bar \varphi^{\bar a} \left( - \frac\ii{2 \cdot 3!} \varepsilon^{mnpq} \hat \calh_{mnp}^\Lambda  \right) + {\rm c.c.}\right]
		\\
		&\quad\,+ \frac{1}{2} T_{AB} *\!F_4^A *\!F_4^B + T_{AB} \Upsilon^A *\!F_4^B  
		\\
		&\quad\,- \left(\hat V - \frac12 T_{AB} \Upsilon^A \Upsilon^B \right)+ \call_{\rm bd}
		\\
		&{\rm with}\qquad   \tilde\call_{\rm bd} = \frac{1}{3!} \partial_m \left[ \varepsilon^{mnpq} {A}^A_{npq} T_{AB} (*\!F_4^B + \Upsilon^B)\right] \,.
	\end{split}
	\ee
	\end{summary}
	where we have defined
	\begin{subequations}\label{GSS_GL_TUV}
		\begin{align}
		\label{GSS_GL_TAB}
		T^{AB} &= 2\, \Re \left(F^{\bar b a} \calv^A_a \bar \calv^{\bar B}_{\bar b}\right)\,,
		\\
		\Upsilon^A &\equiv 2\, \Re \left[F^{\bar b a} \calv^A_a \left(\bar{\hat W}_{\bar b} + \ii F_{\bar b \Lambda} c_B^\Lambda \bar \calv^B(\bar \varphi) + \frac\ii2 F_\Lambda c_B^\Lambda \bar \calv^B_{\bar b} (\bar \varphi) \right)\right]\label{GSS_GL_UA} \,,
		\\
		\hat V &\equiv F^{\bar b a} \left( {\hat W}_{a} - \ii F_{a \Lambda} c_A^\Lambda \calv^A( \varphi) - \frac\ii2 F_\Lambda c_A^\Lambda \bar \calv^A_{a} ( \varphi) \right) 
		\\
		&\quad\quad\quad\, \times \left(\bar{\hat W}_{\bar b} + \ii F_{\bar b \Lambda} c_B^\Lambda \bar \calv^B(\bar \varphi) + \frac\ii2 F_\Lambda c_B^\Lambda \bar \calv^B_{\bar b} (\bar \varphi) \right)
		\\
		&\quad\, -F_{\Lambda\Sigma} c_A^\Lambda c_B^\Sigma \calv^A(\varphi) \bar \calv^B(\bar\varphi) \,. \label{GSS_GL_hatV}
		\end{align}
	\end{subequations}
It is important to stress that the Lagrangian \eqref{GSS_GL_L3fComp} is \emph{not} an immediate generalization of the Lagrangian \eqref{GSS_AL_Dual1Lagbos}, to which we add a three-form contribution. In fact, the gauging of the three-form multiplets makes $\hat L$ not linear anymore and significantly changes the component Lagrangian \eqref{GSS_GL_L3fComp}.	
\end{description}

\section{Four-form multiplets}
\label{sec:GSS_4FM}

Finally, to complete the hierarchy, let us explore the possibility of including a gauge four-form into a supersymmetric multiplet. At first, this might appear a useless effort: in four dimensions, a gauge four-form does not carry \emph{any} dynamics and, in some circumstances, they can be set everywhere to zero by a gauge transformation! However, we shall see that, for the bulk theory, they can gauge three-forms, constraining the constants that they can generate and, in the later Section \ref{sec:ExtObj_3-branes}, we will need them in order to build the action for space-time filling 3-branes.

We may follow the same logic as the one we applied in the previous sections to include gauge three-forms into superfields. First, we need a supersymmetric `gauge potential' that can accommodate a four-form. The simplest solution is to take a chiral multiplet $\Gamma$
\be
\label{GSS_4f_Gamma}
\Gamma =  \gamma + \sqrt{2} \theta \psi_\Gamma + \theta^2 f_\Gamma\,,\qquad \bar D_{\dot\alpha} \Gamma = 0 \;,
\ee
serving as a potential, where - say - the imaginary part of its auxiliary field $f_\Gamma$ is replaced by the Hodge-dual of a four-form $C_4 = \frac{1}{4!} C_{mnpq} \d x^m \wedge \d x^n \wedge  \d x^p \wedge \d x^q$
\be
\label{GSS_4f_D}
\Im f_\Gamma = \frac{1}{2\ii} \left(- \frac14 D^2 \Gamma + \frac14 \bar D^2 \bar \Gamma\right) \Big| \equiv -\frac{1}{2 \cdot 4!} \varepsilon^{mnpq} C_{mnpq} = \frac12 *\! C_4
\ee
The gauge transformation of the multiplet \eqref{GSS_4f_Gamma} is immediately given by
\be
\label{GSS_4f_GT}
\Gamma \rightarrow \Gamma + \Xi 
\ee
where $\Xi$ is a single three-form multiplet of the kind \eqref{GSS_STF}
\be
\label{GSS_4f_Xi}
\Xi = \frac14 \bar D^2 U
\ee
whose components are
\be
\label{GSS_4f_GT_Comp}
\begin{aligned}
	\Xi | &=  \left(\frac 14 \bar D^2 U \right)&\equiv &\, \xi \,,
	\\
	D^\alpha \Xi | &= D^\alpha  \left(\frac 14 \bar D^2 U \right)&\equiv &\, \chi^\alpha \,,
	\\
	-\frac14 D^2 \Xi | &=	-\frac14 D^2 \left( \frac 14 \bar D^2 U \right) &=&\,   \frac12 \left(\ii *\!\d \Lambda_3-d\right)  \equiv f_Y \,.
\end{aligned}
\ee
In this way, \eqref{GSS_4f_GT} reproduces, in the imaginary part of its  highest component, the usual gauge transformation 
\be
C_4 \rightarrow C_4 + \d \Lambda_3\,.
\ee
Notably, \eqref{GSS_4f_GT} can be used to gauge away \emph{all} the components of $\Gamma$ but the gauge four-form $C_4$!

Eventually, we may also implement a \emph{complex} gauge four-form into a chiral superfield
\be
\label{GSS_4f_GammaC}
\tilde \Gamma =  \tilde \gamma + \sqrt{2} \theta \tilde \psi_\Gamma + \theta^2 \tilde f_\Gamma\,,\qquad \bar D_{\dot\alpha} \tilde \Gamma = 0 
\ee
by simply regarding its whole complex auxiliary field as the Hodge-dual of a complex four-form $\tilde D_4$ as
\be
\label{GSS_4f_HdC}
f_\Gamma = - \frac14 D^2 \tilde \Gamma \Big| \equiv -\frac{1}{2 \cdot 4!} \varepsilon^{mnpq} \tilde D_{mnpq} = \frac12 *\! \tilde D_4\,.
\ee
The multiplets $\tilde \Gamma$ can be regarded as potential only if they are equipped with the gauge transformation 
\be
\label{GSS_4f_GTC}
\tilde\Gamma \rightarrow \tilde\Gamma + \tilde\Xi \,,
\ee
where $\tilde\Xi$ is a double three-form multiplet of the kind \eqref{GSS_DTF_L}
\be
\label{GSS_4f_XiC}
\tilde\Xi = - \frac14 \bar D^2 \bar\Sigma\,,
\ee
whose components are
\be
\label{GSS_4f_GT_CompC}
\begin{aligned}
	\tilde \Xi | &=  \left(-\frac 14 \bar D^2 \bar \Sigma \right)&\equiv &\, \tilde \xi \,,
	\\
	D^\alpha \tilde \Xi | &= D^\alpha  \left(-\frac 14 \bar D^2 \bar \Sigma \right)&\equiv &\, \tilde \chi^\alpha \,,
	\\
	-\frac14 D^2 \tilde \Xi | &=	-\frac14 D^2 \left( -\frac 14 \bar D^2 \bar \Sigma \right) &=&\,   \frac12*\!\d \Lambda_3 \,.
\end{aligned}
\ee
with $\Lambda_3$ being a complex three-form. It is then immediate to see that the highest component of \eqref{GSS_4f_GTC} correctly reproduces the expected gauge transformation $\tilde D_4 \to \tilde D_4 + \d \Lambda_3$.

\subsection{Implementing constraints via three-form gaugings}
\label{sec:GSS_4FMG}

Let us immediately show a context where the four-form multiplets introduced above are helpful. Assume that the constants $N_A$ which appear in the superpotential \eqref{GSS_MTF_LChW} are not all independent, but satisfy a \emph{linear} constraint of the form
\be
\label{GSS_4f_Constr_Q}
Q_I^A N_A + \calq_I^{\rm bg} = 0
\ee
where $\calq_I^{\rm bg}$, $I=1,\ldots,K$ is a set of fixed `background' real constants -- in analogy with those in \eqref{Intro_Con_calq} --  and $Q_I^A$ is a $K\times N$ matrix. In order to set the constraint \eqref{GSS_4f_Constr_Q} at a Lagrangian level, we need a Lagrange multiplier. We may then use a set of four-form multiplets $\Gamma^I$ and build the term
\be
\label{GSS_4f_Constr_GammaSS}
\call_{\rm constraint} = \ii \int \d^2\theta\, \left( Q_I^AN_A + \calq_I^{\rm bg} \right) \Gamma^I + {\rm c.c.}\,,
\ee
whose components are simply given by
\be
\label{GSS_4f_Constr_GammaComp}
\call_{\rm constraint} = - \left( Q_I^AN_A+ \calq_I^{\rm bg}\right) *\! C_4^I\,.
\ee
It is then immediately clear that the variation of \eqref{GSS_4f_Constr_GammaSS} with respect to $\Gamma^I$ (or, in components, the variation of \eqref{GSS_4f_Constr_GammaComp} with respect to $C_4$) reproduces the desired constraint \eqref{GSS_4f_Constr_Q}.

However, in the three-form picture, the constants $N_A$ are dynamically generated with arbitrary values. Then, it would be tantalizing to see whether the constraint  \eqref{GSS_4f_Constr_Q} could be implemented in the three-form Lagrangians: to be more explicit, we would like a mechanism such that, once we generate the constants $N_A$, these automatically come `dressed' with the constraint \eqref{GSS_4f_Constr_Q}. 

In the spirit of the discussion of the previous sections and recalling the discussion of Section~\ref{sec:Intro_AC}, we may promote the constants $N_A$ which appear in the constraint \eqref{GSS_4f_Constr_Q} to Lagrange multipliers $X_A$. We may then consider the term
\be
\label{GSS_4f_Constr_X}
\call_{\rm constraint} = \ii \int \d^2\theta\, \left( Q_I^AX_A + \calq_I^{\rm bg} \right) \Gamma^I + {\rm c.c.}\,,
\ee
and add it to the master Lagrangian \eqref{GSS_MTF_LM}. The new master Lagrangian is
\be
\label{GSS_GMTF_LM}
\begin{aligned}
	\call_{\rm master} &=\int \d^4\theta\,  K(\Phi,\bar\Phi)+\left(\int \d^2\theta\, \hat W(\Phi)+\text{c.c.}\right)\,
	\\
	&\quad\,+ \left(\int \d^2\theta\, X_A\calv^A(\Phi)+\frac\ii8\int \d^2\theta \bar D^2  (X_A-\bar X_A)P^A  +\text{c.c.}\right)
	\\
	&\quad\,+ \left(\ii \int \d^2\theta\, \left( Q_I^AX_A + \calq_I^{\rm bg}\right) \Gamma^I + {\rm c.c.}\right) \,,
\end{aligned}
\ee
whose bosonic components are
\be
\label{GSS_GMTF_LMComp}
\begin{split}
	\call_{\rm master}|_{\text{bos}} &=-K_{a\bar b}  \del_m \varphi^a \del^m \bar \varphi^{\bar b}+ K_{a\bar b}  f^a \bar f^b + \left( \hat W_a f^a + {\rm c.c.}\right)
	\\
	&\quad\, + \left[ x_A\calv^A_a f^a - \frac{1}{2 \cdot 3!} \varepsilon^{mnpq} A^A_{npq} \del_m x_A  - \frac{\ii}{2} d^A x_A+ {\rm c.c. }\right]
	\\
	&\quad\, +\left[ f_{XA} (\calv^A - s^A) + {\rm c.c. }\right]-  (Q_I^A \,\Re x_A + \calq_I^{\rm bg}) *\!C_4^I  \,.
\end{split}
\ee

\begin{description}
	\item[Ordinary formulation] First, we show how to recover, from \eqref{GSS_GMTF_LM}, the ordinary formulation in terms of the chiral multiplets $\Phi$. The superfields that we need to integrate are $P^A$ and $\Gamma^I$. The former integration, as in \eqref{GSS_MTF_deltaPAa}, sets $X_A$ to be real constants $N_A$
	\be
	\label{GSS_GMTF_deltaPAa}
	\delta P^A:\qquad X_A = N_A\,,
	\ee
	while the latter sets the constraint \eqref{GSS_4f_Constr_Q} over the constants $N_A$. We then obtain the Lagrangian
	\be\label{GSS_GMTF_LCh}
	\begin{aligned}
		\call_{\rm master} &=\int \d^4\theta\,  K(\Phi,\bar\Phi)+\left(\int \d^2\theta\, W(\Phi)+\text{c.c.}\right)\,
	\end{aligned}
	\ee
	with the superpotential
	\be
	\label{GSS_GMTF_LChW}
	\begin{split}
		&W (\Phi) \equiv N_A \calv^A(\Phi) + \hat W(\Phi) \qquad {\rm and}\qquad Q_I^A N_A + \calq_I^{\rm bg} = 0
	\end{split}
	\ee
	\item[Gauged three-form formulation] More interestingly, let us obtain the three-form formulation. As shown in Section \ref{sec:GSS_MTF}, the first step is the integration of the Lagrange multipliers $X_A$. And now a novelty comes: the functions $\calv^A(\Phi)$ are traded with
	\begin{important}%
		\be
		\begin{split}
		\label{GSS_GMTF}
		&\calv^A (\Phi) = -\frac \ii 4 \bar D^2 P^A - \ii Q^A_I \Gamma^I \equiv \hat S^A \;,\qquad  \parbox{10em}{\color{darkred}\sf{Gauged master \\ three-form multiplet}} 
		\end{split}
		\ee
	\end{important}
	where, with respect to \eqref{GSS_MTF}, a new term appears. Due to the simple component structure of the multiplets $\Gamma^I$, the sole, but important effect of the new term is to modify the highest components of the multiplet $S^A$ in \eqref{GSS_MTF_Comp} to 
	\be
	\label{GSS_GMTF_Comp}
	\begin{aligned}
		\calv^A | &= \calv^A(\varphi) &=&\, \hat S^A \big| &\equiv &\, s^A \,, 
		\\
		-\frac14 D^2 \calv^A | &=  \calv^A_a(\varphi) f^a &=&	-\frac14 D^2 \hat S^A \Big|  &=&\,  \frac12 \left( *\hat F_4^A+  \ii d^A  \right) \equiv f^A \,, 
	\end{aligned}
	\ee
	where we have defined the \emph{gauged four-forms}
	\be
	\label{GSS_GMTF_F4g}
	\hat F_4^A \equiv \d A_3^A + Q_I^A C_4^I\,. 		
	\ee

	In order to get the complete superspace description in terms of -- now gauged -- three-form multiplets, a further integration of the old, chiral multiplets $\Phi^a$ is needed. It gives
	\be\label{GSS_GMTFXAio}
	\calv^A_a X_A = \frac14 \bar D^2 K_a - \hat W_a\,,
	\ee
	which, once inserted into \eqref{GSS_MTF_LM}, gives
	\be\label{GSS_GMTF_L3f}
	\begin{split}
	\tilde{\mathcal L} &=\int \d^4 \theta\, K(\Phi (P),\bar \Phi (P))
	\\
	&\quad\,+\left[\int\d^2\theta\, \hat W(\Phi(P))- \ii \int \d^2\theta\,  \calq_I  \Gamma^I+\text{c.c.}\right] +\tilde {\mathcal L}_{\rm bd}\;,
	\end{split}
	\ee
	with the boundary terms
	\be\label{GSS_GMTF_L3fbd}
	\tilde{\mathcal L}_{\rm bd}= 
	-\frac\ii8 \left(\int\d^2\theta \bar D^2-\int\d^2\bar\theta D^2\right)\left[\left(\frac14 \bar D^2 K_A - \hat{W}_A \right)P^A\right]
	+\text{c.c.} \,.
	\ee

	As stressed in Section \ref{sec:GSS_MTF}, the bosonic three-form Lagrangian is more easily obtained directly working with the component form of the master Lagrangian \eqref{GSS_GMTF_LM}. The steps that we need to follow are the same as those from \eqref{GSS_MTF_IOCompa} to \eqref{GSS_MTF_LMCompd} and will not be repeated here. The only difference appears when the final integration of the real Lagrange multipliers $x_A$ is performed, which now gives
	\be
	\label{GSS_GMTF_IOCompd}
	\delta\, x_A:\qquad x_A = - T_{AB} (*\hat F_4^B + \Upsilon^B)
	\ee
	with the very same matrices \eqref{GSS_MTF_TAB} and \eqref{GSS_MTF_UA} and the gauged four-forms \eqref{GSS_GMTF_F4g}. The bosonic components of the (gauged) three-form Lagrangian are then
	\begin{summary}
	\be
	\label{GSS_GMTF_L3fComp}	
	\begin{split}
		\call_{\text{3-form}}|_{\rm bos}&=-K_{a\bar b}  \del_m \varphi^a \del^m \bar \varphi^{\bar b} + \frac{1}{2} T_{AB} *\!\hat F_4^A *\!\hat F_4^B + T_{AB} \Upsilon^A *\!\hat F_4^B  
		\\
		&\quad\,- \left(\hat V - \frac12 T_{AB} \Upsilon^A \Upsilon^B \right)- \calq_I^{\rm bg} *\!C_4^I+ \call_{\rm bd}
		\\
		&{\rm with}\qquad   \tilde\call_{\rm bd} = \frac{1}{3!} \partial_m \left[ \varepsilon^{mnpq} {A}^A_{npq} T_{AB} (*\!\hat F_4^B + \Upsilon^B)\right] \,.
	\end{split}
	\ee
	\end{summary}
	with $\hat V$ as defined in \eqref{GSS_MTF_hatV}.

	It is quite instructive to directly see how the constants $N_A$, appropriately dressed with the constraint \eqref{GSS_4f_Constr_Q}, originate. We then integrate out the gauge three-forms $A_3^A$ 
	\be
	\label{GSS_GMTF_L3fNA}	
	\delta A_3^A: \qquad T_{AB} (*\! \hat F_4^B + \Upsilon^B) = - N_A\,,
	\ee
	as well as the full set of gauge four-forms $C_4^I$
	\be
	\label{GSS_GMTF_L3fC}	
	\delta C_4^I: \qquad T_{AB} (*\! \hat F_4^B + \Upsilon^B) Q_I^A - \calq_I^{\rm bg}  = - N_AQ_I^A - \calq_I^{\rm bg}= 0\,,
	\ee
	where we have employed the on-shell relation \eqref{GSS_GMTF_L3fNA}. Therefore, the Lagrangian that we finally obtain has the same potential as \eqref{GSS_MTF_VOS} where, importantly, the constants $N_A$ are not allowed to take any possible value, due to the restriction \eqref{GSS_4f_Constr_Q}.
	
\end{description}

\section{Three-form vector multiplet}

So far, we have shown how setting on-shell a set of gauge three-forms provides the same effect as an F-term superpotential. It is worthwhile to mention that in global supersymmetry gauge three-forms may also serve to generate a D-term. This requires the promotion of the super-field strength \eqref{GSS_Intro_Wa} to a variant three-form version which accommodates a real gauge three-form in replacement of the real auxiliary field $d$ in \eqref{GSS_Intro_Wb}. This multiplet already appeared, for example, in \cite{Gates:1983nr, Antoniadis:2017jsk} and was later revisited, by properly considering the Lagrangian boundary contribution thereof, in \cite{Cribiori:2018jjh}.

Given a vector multiplet $V$, the minimal Lagrangian which can be built is
\begin{equation}
\label{GSS_VLagS}
\mathcal{L} =\left(\frac14 \int \d^2\theta\, W^\alpha W_\alpha + \text{c.c.}\right)+ \xi \int \d^4\theta\, V
\end{equation}
where $\xi$, a real parameter, is the so called Fayet--Iliopoulos parameter. Focusing only on the bosonic sector for simplicity, we have
\begin{equation}
\label{GSS_VLagOffSa}
\mathcal{L} |_{\text{bos}}= - \frac 14  F^{mn}F_{mn}+\frac12 d^2  +\frac{\xi}{2}\, d\,,
\end{equation}
where we have neglected the total derivative which involves the gauge field. Integrating out the auxiliary field $d$ via its equation of motion $d=-\frac{\xi}{2}$,
we get the on/shell Lagrangian
\begin{equation}
\label{GSS_VLagOnSa}
\mathcal{L}|_{\text{bos, on-shell}} =  - \frac 14  F^{mn}F_{mn}  -\frac{\xi^2}{8}
\end{equation}
with a constant, semi-positive contribution to the vacuum energy $V = \frac{\xi^2}{8}$.

As we did for the dualization of chiral multiplets, instead of starting with \eqref{GSS_VLagS}, we can rather consider the master Lagrangian
\begin{equation}
\begin{split}
\label{GSS_VLagMaster}
\mathcal{L} &=\frac14 \left(\int d^2\theta\, W^\alpha W_\alpha + \text{c.c.}\right) -\frac18 \int d^2\theta\,\bar D^2 (\Lambda V) -\frac18 \int d^2\bar\theta\, D^2 (\Lambda V) 
\\
&\quad\,+\frac18 \left[ \int d^2\theta \bar D^2 (\Lambda \Sigma)+\int d^2\bar\theta D^2 (\Lambda \bar \Sigma)\right]\,,
\end{split}
\end{equation}
which is obtained by promoting the Fayet--Iliopoulos parameter $\xi$ to a real Lagrangian multiplier $\Lambda$ and adding new terms which contain the complex linear multiplet $\Sigma$ encoding the three-form.

\begin{description}
\item[Ordinary vector multiplet formulation] Let us see how, from the Lagrangian \eqref{GSS_VLagMaster}, we get  the usual Lagrangian \eqref{GSS_VLagS}. We then need to eliminate the dependence on the gauge three-form from the master Lagrangian \eqref{GSS_VLagMaster}. Re-expressing $\Sigma$ as in \eqref{GSS_In_SigmaPsi} in terms of unconstrained spinorial superfields $\Psi^\alpha$ and $\bar\Psi^{\dot \alpha}$ and integrating them out, we get
\begin{equation}
D_\alpha \Lambda = 0, \quad \bar D_{\dot\alpha} \Lambda = 0\,,
\end{equation}
which imply $\Lambda$ is just a real constant $\xi$. Once inserted into \eqref{GSS_VLagMaster}, the ordinary vector multiplet Lagrangian \eqref{GSS_VLagS} is recovered. 

\item[Three-form formulation] Now let us follow a second path in order to rephrase the master Lagrangian solely in terms of a new, variant real three-form multiplet. The variation of the Lagrangian \eqref{GSS_VLagMaster} with respect to the Lagrangian multiplier $\Lambda$ gives
\begin{important}
	\begin{equation}
	\label{GSS_VsolV}
	V = \frac{\Sigma + \bar\Sigma}{2} \equiv U, \qquad\qquad\qquad{\color{darkred}\text{\sf{Three-form real multiplet}}}
	\end{equation}
\end{important}
which is a real multiplet containing a gauge three-form as auxiliary degree of freedom \cite{Gates:1983nr, Antoniadis:2017jsk} and for this reason we dub it \emph{three-form real multiplet}. This can be understood as follows. We recall that $\Sigma$ contains in its expansion a complex vector $\mathcal{B}^m$
\begin{equation}
\frac14 \bar{\sigma}^{m\,\dot\alpha \alpha}
\left[D_\alpha,\bar{D}_{\dot{\alpha}}\right] \Sigma| = \mathcal{B}^m \equiv B^m + \ii C^m\,,
\end{equation}
where $B^m \equiv {\rm Re}\, \mathcal{B}^m$ and $C^m \equiv {\rm Im}\, \mathcal{B}^m$. With respect to \eqref{GSS_In_CL_HD}, we here regard only $C^m$ as the Hodge dual of a real three-form $C_{mnp}$
\begin{equation}
C^m = \frac{1}{3!}\varepsilon^{mnpq} C_{npq}\,,
\end{equation}
whose four-form field strength will be denoted $G_4 = \d C_3$. The relevant projection of the real three-form multiplets \eqref{GSS_VsolV} are
\begin{equation}
\label{GSS_Vvara}
\begin{split}
\frac14 \bar{\sigma}^{m\,\dot\alpha \alpha}\left[D_\alpha,\bar{D}_{\dot{\alpha}}\right] U| &=  B^m\\
\frac{1}{16}  D^2 \bar D^2 U| &= \frac12 *\! G_4 - \frac{\ii}{2} \partial_m B^m
\end{split}
\end{equation}
where we have introduced $* G_4$, which is the Hodge-dual of the field strength $G_4$. Comparing \eqref{GSS_Vvara} with the usual projection of an ordinary real multiplet $V$
\begin{equation}
\label{GSS_Vvarb}
\begin{split}
\frac14 \bar{\sigma}^{m\,\dot\alpha \alpha}\left[D_\alpha,\bar{D}_{\dot{\alpha}}\right] V| &=  v^m\\
\frac{1}{16}  D^2 \bar D^2 V| &= \frac{d}{2}-\frac{\ii}{2} \partial_m A^m
\end{split}
\end{equation}
we immediately recognize that, in the dual formulation \eqref{GSS_Vvara}, the auxiliary field $d$ of the usual vector multiplet is here replaced with the Hodge-dual of the field strength of the three-form $C_{mnp}$, namely
\begin{equation}
d \quad \rightarrow \quad  * G_4\,.
\end{equation}

The usual gauge invariance $U \to U + \Phi + \bar \Phi$, which the real multiplets are endowed with, may also be recovered in the dual three-form picture. In fact, the Lagrangian \eqref{GSS_VLagMaster} is invariant under the shift
\begin{equation}
\label{GSS_Ugauge}
\Sigma \rightarrow \Sigma + \Phi + \ii L
\end{equation}
with $\Phi$ and $L$ being, respectively, an arbitrary chiral and real linear superfield. This reflects onto the real three-form multiplet \eqref{GSS_VsolV} as the transformation
\begin{equation}
U \to U + \Phi + \bar \Phi.
\end{equation}
and encodes, in components, the gauge transformation for the three-form $C_{mnp}$ as
\begin{equation}
C_{mnp} \rightarrow C_{mnp} + 3\partial_{[m} \Lambda_{np]}
\end{equation}
where $\Lambda_{mn}$ is an arbitrary real gauge two-form. This enforces the interpretation of $C_{mnp}$ as components of a gauge three-form.

To conclude the construction of the three-form Lagrangian, we compute the variation of the Lagrangian \eqref{GSS_VLagMaster} with respect to the vector multiplet $V$, obtaining
\begin{equation}
\label{GSS_VsolLam}
\Lambda = \frac12\left(D^\alpha W_\alpha+\bar D_{\dot\alpha} \bar W^{\dot \alpha}\right) = D^\alpha W_\alpha\,.
\end{equation}
Substituting \eqref{GSS_VsolV} and \eqref{GSS_VsolLam} in \eqref{GSS_VLagMaster} we get
\begin{equation}
\label{GSS_VLag3f}
\mathcal{L} = \left(\frac14 \int d^2\theta\, W^\alpha W_\alpha + \text{c.c.} \right)+ \mathcal{L}_{\rm bd}
\end{equation}
where the boundary terms are given by
\begin{equation}
\begin{split}
\label{GSS_VLagBd}
\mathcal{L}_{\rm bd} =&-\frac18 \int d^2\theta\,\bar D^2 (\Lambda V) -\frac18 \int d^2\bar\theta\, D^2 (\Lambda V) \\
&+\frac18 \left[ \int d^2\theta\, \bar D^2 (\Lambda \Sigma)+\int d^2\bar\theta\, D^2 (\Lambda \bar \Sigma)\right]\\
=&\, \frac{1}{64} [D^2,\bar D^2] \left(\Lambda (\bar \Sigma -  \Sigma)\right)|\,.
\end{split}
\end{equation}
In components, this Lagrangian is
\begin{equation}
\label{GSS_VLagOffSb}
\mathcal{L} =- \frac 14  F^{mn}F_{mn}  -\frac{1}{2\cdot 4!} G^{mnpq} G_{mnpq}  + \mathcal{L}_{\rm bd}\,,
\end{equation}
with
\begin{equation}
\label{GSS_VLagbdc}
\mathcal{L}_{\rm bd} = \frac{1}{3!}\partial_m \left(G^{mnpq} C_{npq}\right)\,.
\end{equation}
Setting the gauge three-form on-shell we immediately get
\begin{equation}	
G_{mnpq} = \frac{\xi}{2} \varepsilon_{mnpq}
\end{equation}
with $\xi$ a real constant. Plugging this solution back into \eqref{GSS_VLagOffSb}, we obtain the on-shell Lagrangian \eqref{GSS_VLagOnSa}. Hence, the role of the gauge three-form in \eqref{GSS_VLagOffSb} is to dynamically generate the Fayet--Iliopoulos parameter $\xi$ as expectation value of $* G_4$. 

\end{description}

In \cite{Cribiori:2018jjh}, the three-form multiplets \eqref{GSS_VsolV} were introduced alongside with the double three-form multiplets \eqref{GSS_DBT_L} in order to build novel chiral multiplets in $\caln=2$ rigid supersymmetry. The introduction of gauge three-forms allowed us to give a novel, dynamical interpretation of the partial braking of supersymmetry, from $\caln=2$ to $\caln=1$. The supersymmetry breaking parameters are in one-to-one correspondence with the gauge three-forms. Then, by properly choosing the vevs for the gauge three-forms, vacua with different amount of supersymmetry may be generated. A generalization of the procedure introduced in \cite{Cribiori:2018jjh} both in the context of $\caln=1$ supersymmetry, in order to consider more complicated Lagrangians and D-terms, as well as in different scenarios with extended supersymmetry is left for future work. 
\section{A hierachy of forms}

Throughout the sections of this chapter we have introduced a full, complete hierarchy of gauge $p$--forms: each of them is included inside a supersymmetric multiplet, serving as a potential, by means of which we can build a super-field strength, endowed with a proper gauge invariance. We here report the full four-dimensional hierarchy so built, schematically further summarized in Table~\ref{tab:GSS_p-forms}.

\begin{description}
	\item[0-forms] To begin with, we consider real gauge 0-forms, namely \emph{axions} $a(x)$, which transform as
	\be
	\label{GSS_HF_0gt}
	a(x) \quad \rightarrow \quad a(x) +c\;,
	\ee
	where $c$ is just a constant. At the superspace level, axions can be easily included among the components of a chiral multiplet. That is, we split the lowest component of $\Phi$ into real and imaginary parts
	\be
	\label{GSS_HF_0Phi}
	\Phi | = a (x)+ \ii t(x)\;,
	\ee
	and regard the real part $a(x)$ as an axion. An appropriate `field strength', invariant under the transformation \eqref{GSS_HF_0gt}, is simply given by the imaginary part of the superfield $\Phi$
	\be
	\label{GSS_HF_0ImT}
	\Im \Phi = \frac{\Phi - \bar \Phi}{2\ii}\;.
	\ee
	In fact, the superfield $\Im \Phi$ has, as bosonic components
	\be
	\label{GSS_HF_0ImTc}
	\begin{aligned}
		\Im \Phi | &= t\;,
		\\
		\frac 14 \bar{\sigma}^{\dot\alpha \alpha}_m [D_\alpha, \bar{D}_{\dot\alpha}] \Im \Phi |&= \del_m a\;. 
	\end{aligned}
	\ee
	where the axion $a(x)$ only appears with a derivative acting on it.
	\item[1-forms] The inclusion of gauge one-form in superfields has been recalled at the very beginning of this chapter. The components of a gauge one-form appear among those of real superfields $V$ (see \eqref{Conv_V} and \eqref{Conv_Vprojections}) as
	\be 
	\label{GSS_HF_1Am}
	\frac 14 \bar{\sigma}^{\dot\alpha \alpha}_m [D_\alpha, \bar{D}_{\dot\alpha}] V| = A_m\,.
	\ee
	The gauge transformation $A \rightarrow A + \d \Lambda$, with $\Lambda$ a zero-form, is realized via the $[D,\bar D]$--component of the superspace transformation
	\be
	\label{GSS_HF_1gt}
	V \quad \rightarrow \quad V + \frac{1}{2\ii} (\Phi - \bar \Phi)
	\ee
	as it can be easily checked by comparing \eqref{GSS_HF_0ImTc} and \eqref{GSS_HF_1Am}.
	
	A super-field strength can be defined in superspace as
	\be
	\label{GSS_HF_1W}
	W^\alpha = - \frac14 \bar D^2 D^\alpha V\;,
	\ee
	which is clearly invariant under \eqref{GSS_HF_1gt}. The spinorial superfield $W^\alpha$ is chiral and its components are
	\be
	\label{GSS_HF_1Wc}
	W^\alpha = \{ \lambda, F_2, d \}
	\ee
	where $\lambda^\alpha$, the gaugino, is a Weyl spinor, $d$ a real (auxiliary) scalar field and  the two-form field strength $F_2$.
	
	\item[2-forms] Linear superfields contain, among their components, the field strengths $H_3$ of the gauge two-forms $B_2$:
	\be
	L = \{l, \psi, H_3 = \d B_2\}
	\ee
	as can be seen from \eqref{Conv_RealLc}. Therefore, we may regard the linear multiplets as the proper `super-field strengths' which contain the gauge two-forms.
	
	As shown in Section \ref{sec:GSS_Axions}, linear superfields are constrained by \eqref{GSS_AL_LinConstr}, which is solved by
	\be
	\label{GSS_HF_2L}
	L^\Lambda = D^\alpha \bar D^2 \Psi_\alpha^\Lambda + \bar D_{\dot \alpha} D^2 \bar \Psi^{\Lambda\dot\alpha}
	\ee
	where $\Psi^\alpha$ is an unconstrained spinorial superfield, which may be regarded as a prepotential containing the gauge two-forms.
	
	The gauge transformation $B_2 \rightarrow B_2 + \d \Lambda_1$, with $\Lambda_1$ an arbitrary one-form, is realized at the superspace level as
	\be
	\label{GSS_HF_2gt}
	\Psi^\alpha \quad \rightarrow \quad \Psi^\alpha + \Lambda^\alpha \;, \qquad \bar\Psi_{\dot\alpha} \quad \rightarrow \quad \bar\Psi_{\dot\alpha} + \bar \Lambda_{\dot\alpha}\;,
	\ee
	where $W^\alpha$ is the spinorial chiral superfield . Clearly, the linear multiplets \eqref{GSS_HF_2L}, as expected from a field strength, are not affected by \eqref{GSS_HF_2gt}.
	\item[3-forms] In this chapter we have devoted the Sections from \ref{sec:GSS_TIntro} to \ref{sec:GSS_MTF} to the implementation of gauge three-forms into superfields and we will be here very brief. We have seen that the simplest way to implement gauge three-forms is to start with a real superfield $P$. We then replace, by Hodge-duality, the one-form components with three-form components, that is
	\be 
	\label{GSS_HF_3Am}
	-\frac 14 \bar{\sigma}^{\dot\alpha \alpha}_m [D_\alpha, \bar{D}_{\dot\alpha}] P| = \frac{1}{3!} \varepsilon^{mnpq} A_{npq}\,.
	\ee
	We then construct the three-form multiplets
	\be
	\label{GSS_3f}
	Y \equiv -\frac \ii 4 \bar D^2 P\;,
	\ee
	whose components are
	\be
	Y =\{y,\psi_\alpha, d, *F_4\}\;.
	\ee
	The role of such multiplets is to to dynamically generate the constants which appear in the F-term: for each three-form, a single real constant may be generated.
	\item[4-forms] The four-form multiplets were built in Section~\ref{sec:GSS_4FM}. These are chiral multiplets $\Gamma$, required to transform as in \eqref{GSS_4f_GT} under gauge transformations. The gauge transformations allow us to gauge away all the components of $\Gamma$, leaving just the four-form as the only off-shell degree of freedom
	\be
	\Gamma = \{* C_4\}\,.
	\ee
	The usefulness of these multiplets comes when also the three-form multiplets \eqref{GSS_3f} are present in the theory: the four-forms can gauge the three-form multiplets, constraining the constants that can be possibly generated.
\end{description}

\begin{table}[!h]
	\begin{center}
		\begin{tabular}{ |c c c c c | }
			\hline
			\rowcolor{ochre!30}{\bf $\bm{p}$-form} & {\bf in...} & {\bf Super-field strength} & {\bf Gauge tr.} & {\bf Data} 
			\\
			\hline
			\hline
				\cellcolor{darkblue!20} $0$-form & $\Phi$ & $\frac{\ii}{2} (\Phi -\bar \Phi)$ & $-$  &  axions \TBstrut
			\\
			\hline
				\cellcolor{darkblue!20}$1$-form & $V$ & $W^\alpha = - \frac14 \bar D^2 D^\alpha V$ & $\frac{\ii}{2} (\Phi -\bar \Phi)$  & local $U(1)$ \TBstrut
			\\
			\hline
				\cellcolor{darkblue!20}$2$-form & $\Psi^\alpha$ & $L =  D^\alpha \bar D^2 \Psi_\alpha + \bar D_{\dot \alpha} D^2 \bar \Psi^{\dot\alpha}$ & $\Lambda^\alpha$  & (dual) axions  \TBstrut
			\\
			\hline
				\cellcolor{darkblue!20}$3$-form & $U$ & $Y = - \frac{\ii}4 \bar D^2  U$ & $L$  & $W=r \Phi$  \TBstrut
			\\
				\cellcolor{darkblue!20} & $\Sigma$ & $S = - \frac{1}4 \bar D^2  \bar \Sigma$ & $L_1 + \ii L_2$  & $W =c \Phi$  \TBstrut
			\\
				\cellcolor{darkblue!20}& $\Sigma^a$ & $S^a = - \frac{1}4 \bar D^2 \left[ \calm^{ab}(\Sigma_b-\bar\Sigma_b)\right] $ & $L_{1a} - \calg_{ab}L_2^b$  &  $W=e_a\Phi^a + m^a \calg_{ab} \Phi^b $ \TBstrut
			\\
			\hline
				\cellcolor{darkblue!20}$4$-form & $\Gamma$ & $-$ & $-\ii Y$  & $N_A$--constraint \TBstrut
			\\
			\cellcolor{darkblue!20} & $\tilde\Gamma$ & $-$ & $-\ii S$  & $N_A$--constraint \TBstrut
			\\
			\hline
		\end{tabular}
	\end{center}
	\caption{The four-dimensional $\caln=1$ hierarchy of forms. A bosonic $p$--form appears among the components of the superfields in the second column, by means of which a super-field strength may be built; the super-field strength is invariant under the super-gauge transformation enlisted in the fourth column. This table is an extension of the one which appears in \cite{Gates:1980ay}.}
	\label{tab:GSS_p-forms}
\end{table}


\chapter{Extended objects in global supersymmetry}
\label{chapter:ExtObj}

The hierarchy of forms introduced in the previous chapter is here translated into a hierarchy of objects to which the gauge $p$--forms couple. We will build manifestly $\caln=1$ supersymmetric actions for BPS--strings, membranes and 3-branes, which couple, respectively, to gauge two-, three- and four-forms.

However, as will become soon clear, the linear realization of supersymmetry over the worldvolume of an object is definitely nontrivial. If and only if over the worldvolume there exists a certain fermionic gauge symmetry, the so--called $\kappa$-symmetry \cite{deAzcarraga:1982dhu,Siegel:1983hh}, then the bulk supersymmetry may be preserved, being it identified with such a fermionic symmetry. In order to corroborate this interpretation, in the \emph{super-embedding approach} \cite{Bandos:1995zw} (see \cite{Sorokin:1999jx} for a review), it is shown that $\kappa$-symmetry is a genuine supersymmetry realized over the worldvolume of extended objects, as was first established for superparticles in  \cite{Sorokin:1988nj,Sorokin:1989zi,Volkov:1989ky}. As a consequence, in general it will not be possible to preserve all the bulk, ambient space $\caln=1$ supersymmetry, but only a fraction thereof, the other being spontaneously broken.\footnote{In this sense the objects that we will build are $1/n$--BPS, with $1/n$ denoting the fraction of the bulk supersymmetry generators which can be preserved over the worldvolume of the object.}

Two crucial elements will be our guidance for writing down BPS--actions for extended objects: the just mentioned $\kappa$--symmetry and the supersymmetric multiplets introduced in the previous chapter. The latter in fact include, among their components, gauge $p$--forms and bulk scalar fields and they will provide the basic bricks which allow for coupling extended objects at once to the bulk fields.

The gauged versions of the three-form multiplets and linear multiplets introduced in the previous chapter will also find their `composite' objects to couple with: gauged linear multiplets are the proper gauge--invariant objects living over junctions of strings and membranes, while gauged three-form multiplets define the gauge invariant multiplets living on 3-branes with membranes as boundaries.

\section{The free membrane in supersymmetry}
\label{sec:ExtObj_IntroMemb}

Our exposition of the supersymmetric extended objects starts with the simple case of a single, free membrane. With \emph{free membrane} we understand a membrane whose tension is just a constant, independent of the bulk fields, and which does not couple to any gauge three-form; in short, its action is consistent \emph{per se} and is decoupled from the bulk. Nevertheless, as we will see shortly, such a simplified setup allows us, on the one hand, to explain how supersymmetric actions for extended objects can be built and, on the other, to highlight the physics that such objects enjoy.

In usual four dimensional spacetime $\mathbb{R}^{1,3}$, membranes are hypersurfaces of codimension one. The worldvolume of a membrane is parametrized by three coordinates $\xi^i$, $i=1,2,3$ and, as objects living in four dimensional spacetime, they are described via the embedding $\xi^i \mapsto x^m(\xi)$, the $\mathbb{R}^{1,3}$ coordinates $x^m(\xi)$ capturing the motion of the membrane in the target space which is here just the ordinary spacetime. Analogously, we can regard membranes as objects living in the full superspace $\mathbb{R}^{1,3|4}$, with coordinates $\frak{z}^M = (x^m, \theta^\alpha, \bar\theta_{\dot\alpha})$. Still standing that the membrane worldvolume is described by three coordinates $\xi^i$, membranes can be described as superspace objects by the \emph{superspace embedding}
\be\label{ExtObj_FrM_SEmbed}
\xi^\mu\quad\mapsto\quad \calm:\;{\frak z}^M(\xi)=\left(x^m(\xi),\theta^\alpha(\xi),\bar\theta_{\dot\alpha}(\xi)\right)\,. 
\ee
The superspace coordinates ${\frak z}^M(\xi)$ draw an hypersurface, along which the membrane is stretched, in the full ambient space $\mathbb{R}^{1,3|4}$.

With the superspace embedding in the hands, we can now write down the action of a membrane using the very same superspace language that we introduced in Chapter~\ref{chapter:GSS}. A first attempt is to generalize the Nambu--Goto part of the membrane action \eqref{Intro_Jumps_Memb} straightly to superspace as:
\be
\label{ExtObj_FrM_SmembrNG} 
S_{\text{memb,NG}}= -2 \sigma \int_{\calm} \d^3\xi \sqrt{- {\bf h}} \; .
\ee
Here $\sigma$, a real constant, is the membrane tension and ${\bf h}=\det h_{ij}$ is the determinant of the induced metric
\be\label{hmunu}
h_{ij}(\xi)\equiv \eta_{ab}E^a_i(\xi)E^b_j(\xi)\,.
\ee
This is constructed starting from the pull--back of the target superspace supervielbein
\be
E^a_i(\xi) \equiv  \del_i {\frak z}^M(\xi) E^a_M({\frak z}(\xi)) \, .
\ee
However, once the membrane sits in its ground state, the sole Nambu--Goto action \eqref{ExtObj_FrM_SmembrNG} cannot preserve even part of the full $\caln=1$ supersymmetry. This can be most readily understood with a simple counting of the degrees of freedom. Let us consider the membrane ground state for which the membrane spans the hypersurface $z=0$. Over the membrane worldvolume, only a single, physical bosonic degrees of freedom describes the dynamics of the membrane, which is associated to the membrane displacement along $z$ -- below we will be more precise about this statement. However, the fermions $\theta^\alpha$, $\bar\theta_{\dot\alpha}$ would still be there: no multiplet can be built out of $z$ and $\theta$, $\bar \theta$, because there is no pairing between fermionic and bosonic degrees of freedoms. In other words, the membrane worldvolume would break spontaneously \emph{all} the four supersymmetry generators of the $\caln=1$ bulk supersymmetry, making it impossible to linearly realize even a part of it in the whole spacetime.

In order to preserve part of the $\caln=1$ supersymmetry, the Wess-Zumino term
\be
\label{ExtObj_FrM_SmembrWZ} 
S_{\text{memb,WZ}}=  \sigma \int_{\calm} \mathbf{A}_3 
\ee
has to be added to the Nambu--Goto action \eqref{ExtObj_FrM_SmembrNG}, so that the full free membrane action reads
\be
\label{ExtObj_FrM_Smembr} 
S_{\text{memb}}= S_{\text{memb,NG}} + S_{\text{memb,WZ}} =-2 \sigma \int_{\calm} \d^3\xi \sqrt{- {\bf h}} + \sigma \int_{\calm} \mathbf{A}_3\; .
\ee
Here $\mathbf{A}_3$ is a \emph{super-three-form}, defined directly in superspace through the supervielbein forms
\be
\label{ExtObj_FrM_EA} 
E^A = (E^a, E^\alpha, E_{\dot\alpha}) = \d {\frak z}^M E_M^A({\frak z})\,,
\ee
which are supersymmetric invariant by construction (we refer to Appendix \ref{app:Superspace_Conv_Diff} for further details). The definition of $\mathbf{A}_3$ passes through the construction of a proper \emph{super-field strength} $\mathbf{F}_4$, from which the super-three-form is \emph{defined} by requiring\footnote{The differential operator $\dwb$ introduced here is the usual one defined in superspace in \cite{Wess:1992cp} and, differently with respect to the bosonic $\d$, acts from the right (see \eqref{SDiff_d} and \eqref{SDiff_dO}).}
\be
\label{ExtObj_FrM_FdA}
\mathbf{F}_4 =: \dwb \mathbf{A}_3\;.
\ee
The super-field strength $\mathbf{F}_4$ is required to be supersymmetric invariant: it is then natural to build it by using the superveilbein basis \eqref{ExtObj_FrM_EA}; moreover, it has to be closed in the full superspace cohomology, $\dwb \mathbf{F}_4 = 0$. These requirements fix the super-field strength to be
\begin{equation}
\label{ExtObj_FrM_F4}
\begin{aligned}
{\bf F}_{4}   =&\,2 {\ii}\,  E^b\wedge E^a \wedge E^\alpha \wedge E_\beta \sigma_{ab\; \alpha}{}^\beta  - 2 {\ii}\,  E^b\wedge E^a \wedge \bar E_{\dot \alpha} \wedge \bar E^{\dot \beta} \bar \sigma_{ab}{}^{\dot\alpha}{}_{\dot\beta}  \,, \qquad
\end{aligned}
\end{equation}
which implies that, from \eqref{ExtObj_FrM_FdA}, the super-three-form $\mathbf{A}_3$ is given by
\be
\label{ExtObj_FrM_A3}
\begin{aligned}
	{\bf A}_3 =&\,  2  E^a \wedge E^\alpha \wedge E^{\dot\alpha}  \sigma_{a\alpha\dot\alpha}  (\theta^2 - \bar \theta^2)
	\\
	&+  2\ii E^b\wedge E^a \wedge  E^\alpha
	\sigma_{ab\; \alpha}{}^{\beta} \theta_\beta  -  2 \ii  E^b\wedge E^a \wedge  E_{\dot\alpha}
	\bar\sigma_{ab}{}^{\dot\alpha}{}_{\dot\beta} \bar \theta^{\dot \beta}\, .
\end{aligned}
\ee
Remarkably, although \eqref{ExtObj_FrM_F4} is constructed as a supersymmetric invariant object, the super-three-form $\mathbf{A}_3$ is not supersymmetric invariant! This is invariant only up to a total derivative, a feature that will be used in Section~\ref{sec:ExtObj_Memb_CC} in order to compute the central charge--modification to the supersymmetry algebra induced by the membrane.

As of now, we have been vague about the sense in which supersymmetry has to be preserved by the action \eqref{ExtObj_FrM_Smembr}. Indeed, the problem of finding out the amount of preserved supersymmetry is strictly tightened to the symmetries that the action \eqref{ExtObj_FrM_Smembr} enjoys. In global supersymmetry, the action \eqref{ExtObj_FrM_Smembr} is endowed with two symmetries:
\begin{description}
	\item[Reparametrization invariance] By construction, \eqref{ExtObj_FrM_Smembr} is invariant under reparametrizations of the membrane worldvolume
	\be
	\label{ExtObj_FrM_xi}
	\xi \rightarrow \xi'(\xi)\;.
	\ee
	This symmetry was expected already from the bosonic action \eqref{Intro_Jumps_Memb} and holds also at the level of superspace embedding. The simplest way to fix this freedom is to set
	\be
	\label{ExtObj_FrM_Static}
	\xi^i \equiv x^i\, \quad i=0,1,2\;,
	\ee
	which is the so--called \emph{static gauge}. This gauge choice leaves out the fourth coordinate $x^3(\xi)\equiv z(\xi)$, which is not fixed. Therefore, we may regard $z(\xi)$ as the only bosonic physical field which describes the dynamics of the membrane, namely the displacement of the membrane from its ground state, fixed position $z=z_0$;
	\item[$\kappa$-symmetry] Additionally, the action \eqref{ExtObj_FrM_Smembr} is invariant under a \emph{local} fermionic symmetry, which acts on the embedding coordinates as 
	\be
	\label{ExtObj_FrM_kz}
	\begin{split}
		\delta {\frak z}^M (\xi)=\kappa^\alpha(\xi) E^M_\alpha({\frak z}(\xi))+\bar\kappa_{\dot\alpha}(\xi) E^{M\dot\alpha}({\frak z}(\xi))\,,
	\end{split} 
	\ee
	which, more explicitly, reads
	\be
	\label{ExtObj_FrM_kzb}
	\begin{split}
		&\delta x^m (\xi)= \ii \theta(\xi) \sigma^m  \bar\kappa - \ii \kappa \sigma^m \bar\theta(\xi)\;,
		\\
		&\delta \theta^\alpha(\xi) = \kappa^\alpha(\xi)\;,\qquad 	\delta \bar \theta_{\dot\alpha} = \bar\kappa_{\dot\alpha}(\xi)\;.
	\end{split} 
	\ee
	However, the parameters $\kappa^\alpha$ and $\bar \kappa_{\dot\alpha}$ are constrained by the conditions
	\be\label{ExtObj_FrM_kappa}
	\begin{split}
		&\kappa_\alpha =-  ({\Gamma}\bar{\kappa})_\alpha\; , \qquad \bar\kappa^{\dot\alpha} =  -  (\bar{\Gamma}{\kappa})^{\dot\alpha}\,,
	\end{split}
	\ee
	where we have defined
	\be
	\label{ExtObj_FrM_kproj}
	\begin{aligned}
		\Gamma_{\alpha\dot\alpha}&\equiv \frac{\ii\epsilon^{ijk}}{3!\sqrt{-\det h}}\epsilon_{abcd} E^b_iE^c_j E^d_k\,\sigma^a_{\alpha\dot\alpha}\,,
		\\
		\bar\Gamma^{\dot\alpha\alpha}&\equiv \frac{\ii\epsilon^{ijk}}{3!\sqrt{-\det h}}\epsilon_{abcd} E^b_iE^c_j E^d_k\,\bar\sigma^{a\dot\alpha\alpha}\,,
	\end{aligned}
	\ee
	which satisfy
	\be
	\label{ExtObj_FrM_kprojid}
	\Gamma_{\alpha\dot\alpha} \bar \Gamma^{\dot\alpha\beta} = \delta_\alpha^\beta\;.
	\ee
	The constraints \eqref{ExtObj_FrM_kappa} are projection conditions which halves, from four to two, the independent parameters of the fermionic $\kappa$-symmetry. Using the four-component formalism, we can collect $\kappa_\alpha$ and $\bar\kappa^{\dot\alpha}$ into a single Majorana spinor $\bm{\kappa}$
	\be
	\label{ExtObj_FrM_4Dk}
	\bm{\kappa} = \begin{pmatrix}
		\kappa_\alpha \\ \bar\kappa^{\dot\alpha}
	\end{pmatrix}
	\ee
	and introduce the projector
	\be
	\label{ExtObj_FrM_4DP}
	\bm{P} \equiv \frac1{\sqrt2} (\bm{1} - \bm{\Gamma}) = \frac1{\sqrt{2}} \begin{pmatrix}
		\delta_\alpha{}^\beta & - \Gamma_{\alpha\dot\beta} \\
		-\bar\Gamma^{\dot\alpha \beta} & \delta^{\dot\alpha}{}_{\dot\beta}
	\end{pmatrix}\;,
	\ee
	which satisfies $\bm{P}^2=\bm{P}$. Then, the constraints \eqref{ExtObj_FrM_kappa} are solved by setting
	\be
	\bm{\kappa} = \bm{P} \bm{\Psi}\,,
	\ee
	with $\bm{\Psi}$ an arbitrary Majorana spinor, making it explicit the projection structure. 
\end{description}

\begin{figure}[t]
	\centering
	\includegraphics[width=7.2cm]{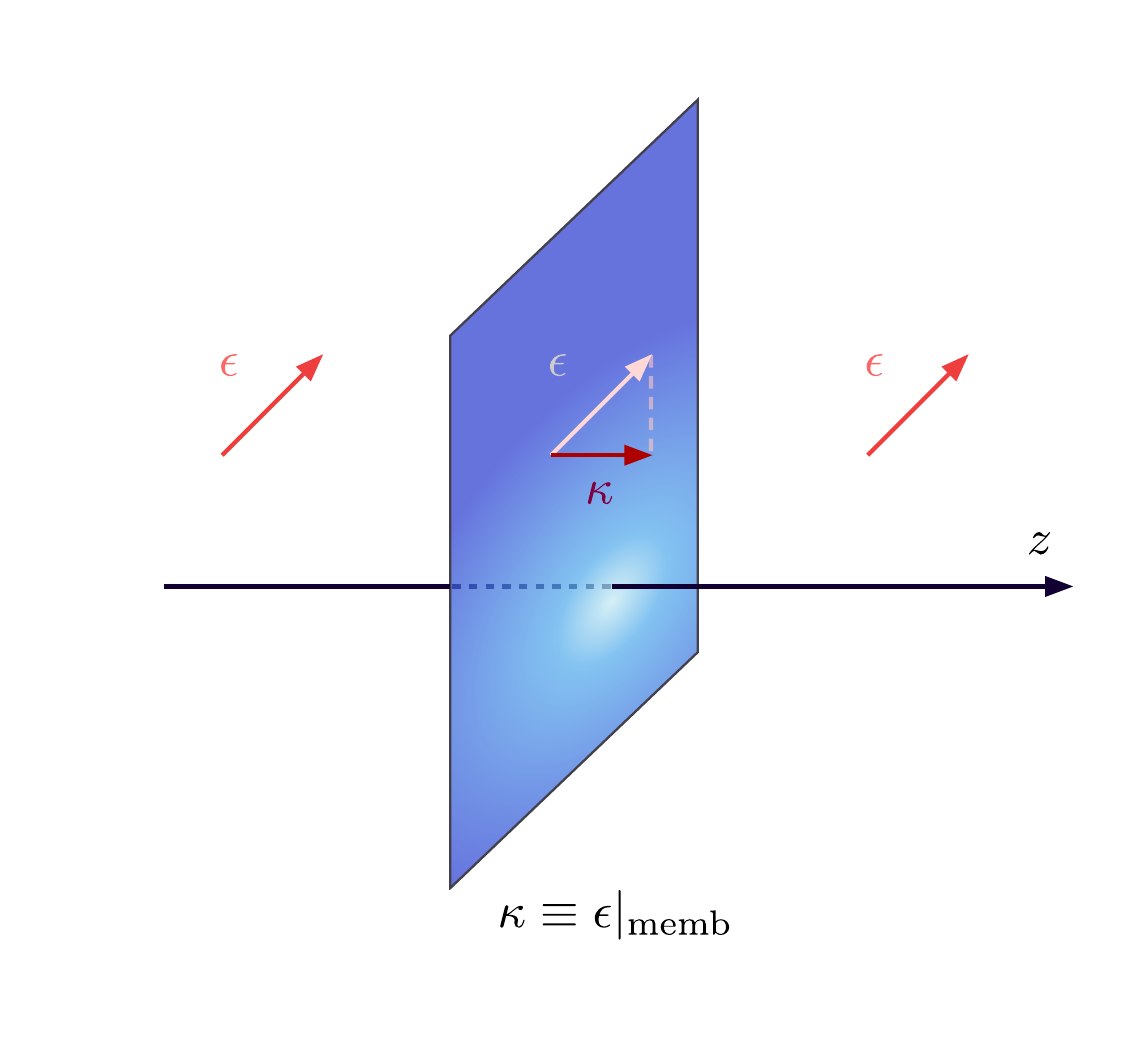} \includegraphics[width=7.2cm]{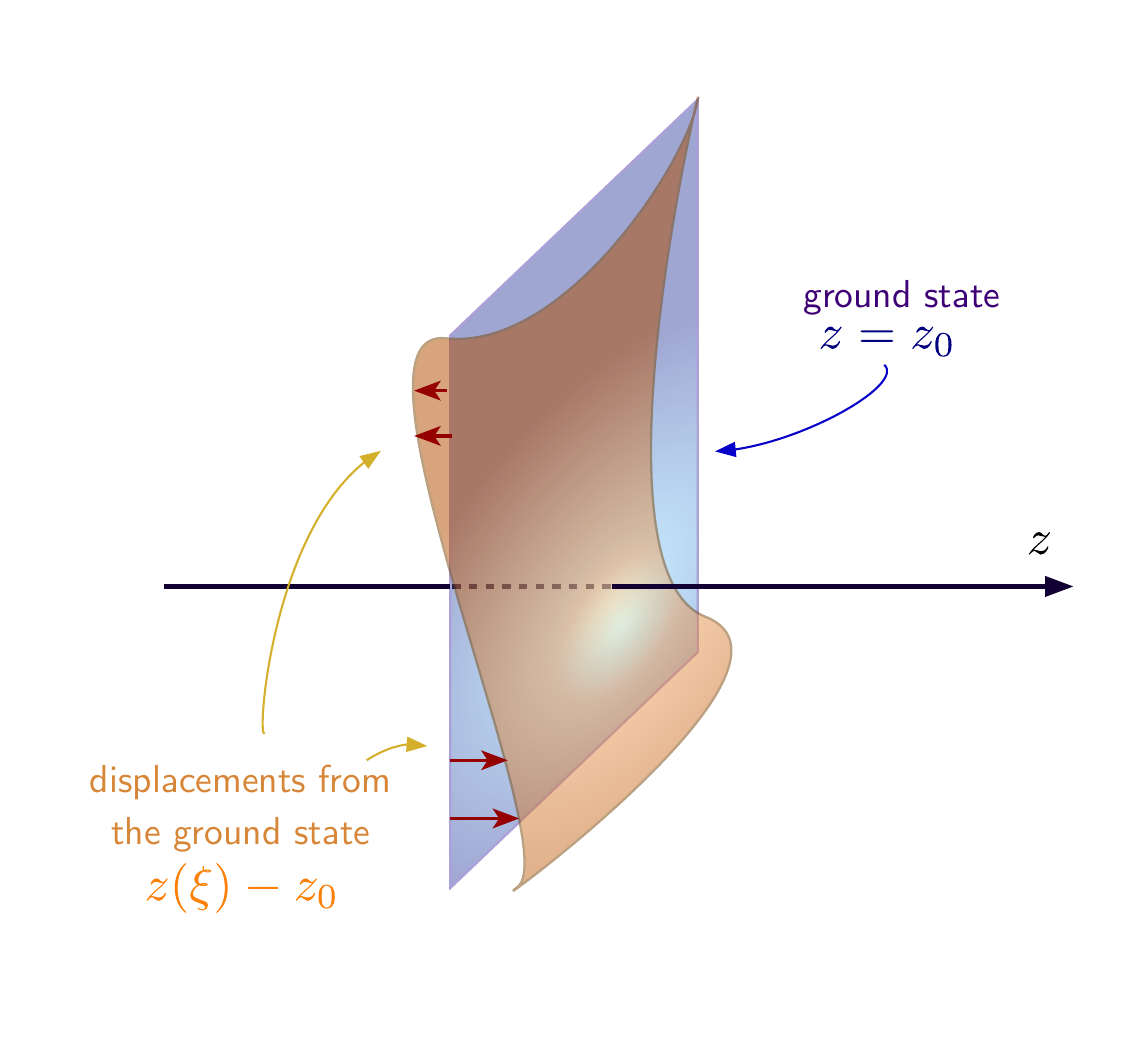}
	\caption{\footnotesize{On the left: the three-dimensional supersymmetry over the membrane is realized if and only if it can be associated with $\kappa$-symmetry. On the right: the translational symmetry along $z$ is spontaneously broken by the membrane ground state; the physical field $z(\xi)$, interpreted as Goldstone boson associated to such breaking, represents the displacement of the membrane with respect to its ground state. }}\label{fig:MembKappaGround}
\end{figure}

It is the $\kappa$-symmetry which allows to linearly realize part of the $\caln=1$ supersymmetry over the membrane worldvolume. Let us recall that the supersymmetry transformations act on the superspace coordinates $\frak{z}^M$ as
\be
\label{ExtObj_FrM_susy}
\begin{split}
	&\delta x^m = \ii \theta \sigma^m  \bar\epsilon - \ii \epsilon \sigma^m \bar\theta \;,\qquad \delta \theta^\alpha = \epsilon^\alpha\,,\qquad 	\delta \bar \theta_{\dot\alpha} = \bar\epsilon_{\dot\alpha}\;,
\end{split} 
\ee
where $\epsilon^\alpha$, $\bar\epsilon_{\dot\alpha}$ are \emph{constant} parameters.  If the entire bulk $\caln=1$ supersymmetry were preserved over the three-dimensional membrane worldvolume, this would imply that, on the membrane, an $\caln=2$ supersymmetry theory would live. However, comparing \eqref{ExtObj_FrM_susy} with \eqref{ExtObj_FrM_kzb} it is evident that, if the membrane has to preserve supersymmetry over its worldvolume, then $\kappa$-symmetry has to be \emph{identified} with supersymmetry over the worldvolume. In other words, in order to preserve supersymmetry, it has to hold that
\be
\epsilon^\alpha |_{\rm memb} = \kappa^\alpha\;,\qquad \bar\epsilon_{\dot\alpha} |_{\rm memb} =  \bar\kappa_{\dot\alpha} \,,
\ee
implying that the supersymmetry parameters obey the very same projection conditions \eqref{ExtObj_FrM_kproj}. In terms of Majorana spinors, this means that supersymmetry is preserved only if we choose the parameter $\bm{\epsilon}_{\rm s}$ such that 
\be
\bm{\epsilon}_{\rm s} = \bm{P} \bm{\epsilon}\,,
\ee
as pictorially depicted in Fig.~\ref{fig:MembKappaGround}. Let us decompose the fermionic fields as
\be
\bm{\theta}(\xi) = \bm{P}  \bm{\theta} (\xi)+ \bm{P}^\bot \bm{\theta}  (\xi)
\ee
where we have introduced the projector $\bm{P}^\bot = \bm{1} + \bm{\Gamma}$, orthogonal to $\bm{P}$, that is $\bm{P} \bm{P}^\bot = 0$. Under the action of supersymmetry transformations, the two components transform as
\begin{subequations}\label{ExtObj_FrM_DirS}
\begin{alignat}{3}
\label{ExtObj_FrM_DirSa}
	&\bm{P}  \bm{\theta} (\xi)\quad & {\text{transform as}} \quad &\bm{P} \delta  \bm{\theta}(\xi) =   \bm{P}  \bm{\epsilon} = \bm{\kappa}(\xi) \,,  &
	\\
\label{ExtObj_FrM_DirSb}	
	&\bm{P}^\bot  \bm{\theta} (\xi)\quad &{\text{transform as}} \quad &\bm{P}^\bot \delta  \bm{\theta}(\xi) =   \bm{P}^\bot  \bm{\epsilon}\,. & 
\end{alignat}
\end{subequations}
The first relation can then be used to gauge away the components $\bm{P}\bm{\theta}$, leaving just $\bm{P}^\bot\bm{\theta}$. Namely, we \emph{gauge fix} the $\kappa$-symmetry so that
\be
\label{ExtObj_FrM_kappagf}
\bm{P}\bm{\theta} = 0\,.
\ee
\vspace{-0.4cm}
\begin{summary}
	Combining this with \eqref{ExtObj_FrM_Static}, we then see that
	\be
	\label{ExtObj_FrM_Phys}
	z(\xi)\,, \qquad \bm{P}^\bot\bm{\theta}(\xi)
	\ee
	are the only physical fields governing the physics of the membrane. The first, $z(\xi)$, describes the displacement of the membrane from its ground state at fixed $z_0$: it is then identified with the Goldstone boson associated to the spontaneous breaking of the bulk translations along $z$ -- see Fig.~\ref{fig:MembKappaGround}; the second, $\bm{P}^\bot\bm{\theta}(\xi)$, is a Goldstino: its nonlinear transformation \eqref{ExtObj_FrM_DirSb} signals the fact that, over the three-dimensional membrane worldvolume, only $\caln=1$ supersymmetry is preserved, rather than the maximal $\caln=2$. In short, ${\frak G} = \{z(\xi), \bm{P}^\bot\bm{\theta}(\xi)\}$ is a Goldstone supermultiplet living over the membrane worldvolume.
\end{summary}
In order to enforce this physical interpretation, it is possible to directly compute the effective three-dimensional Lagrangian for the Goldstone supermultiplet. In fact, for the gauge choices \eqref{ExtObj_FrM_Static}-\eqref{ExtObj_FrM_kappagf}, the veilbein $E_i^a$ are 
\be
E_i^j = \delta_i^j + \ii \bar\theta \bar\sigma^j \del_i \theta + \ii \theta \sigma^j \del_i \bar\theta\;,\qquad E_i^3 =  \del_i z\;,
\ee
and the membrane action \eqref{ExtObj_FrM_Smembr} reduces to
\be
\label{ExtObj_FrM_SmembrEFT} 
S_{\text{memb}}= -2 \sigma \int_{\calm} \d^3x \left(1 +  \del_i z \del^i z +  \ii \bar\theta \bar\sigma^i \del_i \theta + \ii \theta \sigma^i \del_i \bar \theta \right) + \ldots
\ee
where the dots stand for coupling terms which we are presently not interested in. Over the membrane ground state, for fixed $z=z_0$, the action \eqref{ExtObj_FrM_SmembrEFT} clearly reduces to a three-dimensional Volkov-Akulov action \cite{Volkov:1972jx,Volkov:1973ix,Wess:1992cp}, as expected whenever (part of) supersymmetry is nonlinearly realized.


\section{Membranes in supersymmetry}
\label{sec:ExtObj_MembSusy}

The case of free membranes examined in the previous section is definitely too restrictive. Free membranes are completely decoupled from the bulk and they contribute to the total energy of the system just with a constant contribution. However, as already anticipated in Section~\ref{sec:Intro_Jumps} and as will be extensively explored in the following chapters, tensions of membranes that come from reducing higher dimensional objects typically depend on the bulk superfields. Moreover, free membranes do not couple to any gauge field, making the phenomenological models of Section~\ref{sec:Intro_Jumps} not realizable. The question that we want to address is how such dependences may be included by still preserving the supersymmetric properties of the membrane action.

Let us begin with inquiring how to insert a supersymmetric coupling of membranes to gauge three-forms. We would like, in the end, at a bosonic level, to reproduce the same minimal coupling of the membrane to the gauge three-forms $A_3^A$ as in \eqref{Intro_Jumps_Memb}. However, it is clear that, as it is in \eqref{Intro_Jumps_Memb}, such a coupling cannot preserve supersymmetry at all, since the bosonic fields there present need to be paired with an equal number of fermionic fields. If we want to hope to preserve (part of) the bulk supersymmetry, as we did in the previous section, we actually need a proper supersymmetric object which contains, in its lowest bosonic components, a gauge three-form $A_3^A$. More concretely, we seek a \emph{super-three-form} ${\bf A}_3^A$ such that ${\bf A}_3^A| = A_3^A$, which allows for writing a supersymmetric Wess-Zumino term
\be
\label{ExtObj_Memb_SmembrWZ} 
S_{\text{WZ}}= q_A \int_{\calm} {\bf A}_3^A \; .
\ee
In the previous section we have seen that the super-three-form \eqref{ExtObj_FrM_A3} does not preserve supersymmetry and also in the present context we do not expect ${\bf A}_3^A$ to be supersymmetric invariant. In order to build up the super-gauge three-form potentials ${\bf A}_3^A$, it is definitely better to start from the \emph{super-four-form field strengths} ${\bf F}_4^A$. As for \eqref{ExtObj_FrM_F4}, we require them to be closed and such that the lowest bosonic component of ${\bf F}_4^A$ is\footnote{The minus sign which appears in \eqref{ExtObj_Memb_F4A3b} is due to the definition \eqref{ExtObj_Memb_F4d} and the differently defined differentials for bosonic components and superspace forms (see \eqref{AppDiff_dwp} and \eqref{SDiff_d}).} 
\be
\label{ExtObj_Memb_F4A3b}
{\bf F}_4^A | = - F_4^A = - \d A_3^A\,.
\ee
But how to embed gauge three-forms in a super-form? The answer comes from Section \eqref{sec:GSS_MTF}. There we saw that, generically, the field strengths of gauge three-forms can be included in the highest components of the chiral superfields 
\be
\label{ExtObj_Memb_SA}
S^A = - \frac\ii4 \bar D^2 P^A\,,
\ee
which are defined via the potentials $P^A$, whose components contain three-forms as in \eqref{GSS_MTF_Pot}, and are gauge--invariant by construction. The superfields \eqref{ExtObj_Memb_SA} are proper supersymmetric objects, by means of which we can construct the closed super-field strength \cite{Bandos:2010yy,Bandos:2011fw,Bandos:2012gz,Bandos:2018gjp,Bandos:2019wgy,Bandos:2019khd}
\begin{equation}
\label{ExtObj_Memb_F4}
\begin{aligned}
	{\bf F}^A_{4}   &=2 {\ii}  E^b\wedge E^a \wedge E^\alpha \wedge E_\beta \sigma_{ab\; \alpha}{}^\beta\bar S^A  - 2 {\ii}  E^b\wedge E^a \wedge \bar E_{\dot \alpha} \wedge \bar E^{\dot \beta} \bar \sigma_{ab}{}^{\dot\alpha}{}_{\dot\beta} S^A
	\\
	&\quad\, + \frac{\ii}6  E^c\wedge E^b\wedge E^a \wedge E^\alpha \epsilon_{abcd} \sigma^{d}_{\alpha\dot\alpha} \bar{D}^{\dot\alpha} {\bar S}^A
	\\
	&\quad\,+\frac{\ii}6  E^c\wedge E^b\wedge E^a \wedge \bar E_{\dot\alpha} \epsilon_{abcd} \bar\sigma^{d\dot\alpha\alpha} D_{\alpha} {S}^A 
	\\  
	&\quad\, + \frac{1}{96} E^{d} \wedge E^c \wedge E^b \wedge E^a \epsilon_{abcd} \left(D^2 {S}^A+\bar D^2 \bar{S}^A \right) \,. \qquad
	\end{aligned}
\end{equation}
It can be easily checked that, from the components \eqref{GSS_MTF_Pot}, \eqref{ExtObj_Memb_F4A3b} is retrieved. Then, the super-three-forms are \emph{defined} as 
\begin{equation}
\label{ExtObj_Memb_F4d}
{\bf F}^A_{4}   = \dwb {\bf A}^A_3\;.
\end{equation}
Clearly, for any generic super-four-form, we cannot find a three-form potential satisfying \eqref{ExtObj_Memb_F4d}. Nevertheless, by using the definition \eqref{ExtObj_Memb_SA} of the three-form multiplets, \eqref{ExtObj_Memb_F4d} holds for the gauge potential
\be
\label{ExtObj_Memb_A3}
\begin{aligned}
	{\bf A}^A_3 &=  { -}2 {\ii} E^a \wedge E^\alpha \wedge E^{\dot\alpha}  \sigma_{a\alpha\dot\alpha} P^A 
	\\
	&\quad\,+   E^b\wedge E^a \wedge  E^\alpha
	\sigma_{ab\; \alpha}{}^{\beta}{D}_{\beta}P^A +  E^b\wedge E^a \wedge  E_{\dot\alpha}
	\bar\sigma_{ab}{}^{\dot\alpha}{}_{\dot\beta}\bar{D}^{\dot\beta} P^A
	\\&\quad\,+\frac {1} {24}
	E^c \wedge E^b \wedge E^a \epsilon_{abcd} \,\bar{\sigma}{}^{d\dot{\alpha}\alpha}
	[D_\alpha, \bar{D}_{\dot\alpha}]P^A
	\, .
\end{aligned}
\ee
Clearly, it explicitly depends on the real potentials $P^A$. It is indeed the relation \eqref{ExtObj_Memb_SA} which is crucial in obtaining \eqref{ExtObj_Memb_A3}: it is \eqref{ExtObj_Memb_SA} which allows us to trivialize \eqref{ExtObj_Memb_F4} in de Rham-cohomology.

So far, all our efforts have been focused on the Wess-Zumino coupling term, but let us now come to the Nambu-Goto term. The case of the free membrane of the previous section has instructed us that, in order to preserve (part of the) supersymmetry once a membrane ground state is chosen, a close interplay between the Nambu-Goto and Wess-Zumino terms has to take place. Let us write the Nambu-Goto term as
\be
\label{ExtObj_Memb_SmembrNG} 
S_{\text{NG}}= -\int_{\calm} \d^3\xi \sqrt{- h} \; \calt_{\text{memb}} (\Phi) \; ,
\ee
where we have used the same notation as \eqref{ExtObj_FrM_SmembrNG}, albeit here $\calt_{\text{memb}}(\Phi)$ may generically depend on the bulk superfields $\Phi$. It turns out that if and only we choose 
\be
\label{ExtObj_Memb_Tmembr} 
\calt_{\text{memb}}= 2 |q_A \calv^A (\Phi)| = 2 |q_A S^A| \; ,
\ee
where $\calv^A(\Phi)$ are the holomorphic periods introduced in \eqref{GSS_MTF}, then supersymmetry may be preserved over the membrane worldvolume. In fact, in such a case, the combined \eqref{ExtObj_Memb_SmembrNG} and \eqref{ExtObj_Memb_SmembrWZ} terms are invariant under $\kappa$--symmetry transformations of the kind \eqref{ExtObj_FrM_kz}, provided that the parameters $\kappa$ obey the projection conditions
\be
\label{ExtObj_Memb_kappa}
\begin{split}
		&\kappa_\alpha =- \frac {q_A S^A}{|q_A S^A|} ({\Gamma}\bar{\kappa})_\alpha\; , \qquad \bar\kappa^{\dot\alpha} =  - \frac {q_A \bar S^A}{|q_A S^A|} (\bar{\Gamma}{\kappa})^{\dot\alpha}\,,
		\\
		&\kappa^\alpha = \frac {q_A S^A}{|q_A S^A|} (\bar{\kappa}{\bar \Gamma})^\alpha\; , \qquad \quad  \bar\kappa_{\dot\alpha} =   \frac {q_A \bar S^A}{|q_A S^A|} ({\kappa}{\Gamma})_{\dot\alpha}\,,
\end{split}
\ee
where we have introduced the matrices $\Gamma_{\alpha\dot\alpha}$ and $\bar\Gamma^{\dot\alpha\alpha}$ as in \eqref{ExtObj_FrM_kproj}.

\begin{summary}
Therefore, the BPS-action of a membrane minimally coupled to the gauge three-forms $A_3^A$ is
	\be
	\label{ExtObj_Memb_S}
	S_{\rm memb}=-2\int_\calm \d^3\xi\,|q_A S^A|\sqrt{-\det {\bf h}}+q_A\int_\calm{\bf A}^A_3\;.
	\ee
We stress that, given the charges $q_A$, the form of this action is strictly determined by $\kappa$--symmetry.
\end{summary}

\subsection{Proof of $\kappa$-symmetry}
\label{sec:ExtObj_MembSusy_kappa}

In this section we give a detailed proof of the $\kappa$--symmetry of the action \eqref{ExtObj_Memb_S}. Although $\kappa$--symmetry transformations directly act on the embedding coordinates ${\frak z}^M(\xi)$ as in \eqref{ExtObj_FrM_kz}, it is convenient to see how they  act over the super-vielbeins. The $\kappa$--transformations are defined by requiring
\be
\label{ExtObj_Memb_Dk_kappa}
\begin{split}
	&i_\kappa E^a =\delta_\kappa {\frak z}^M E_M^a({\frak z})=0\; , \qquad  \\
	&i_\kappa E^\alpha =\delta_\kappa {\frak z}^M E_M^\alpha ({\frak z})=\kappa^\alpha\; ,\qquad i_\kappa E^{\dot\alpha} =\delta_\kappa {\frak z}^M E_M^{\dot\alpha} ({\frak z})=\bar{\kappa}{}^{\dot\alpha}\; . \qquad 
\end{split}
\ee
In the following we will need the variations of the bosonic supervielbein
\bea
\label{ExtObj_Memb_Dk_kappaE}
&\delta_\kappa E^a = \dwb i_\kappa E^a+ i_\kappa \dwb  E^a = -2\ii E^\alpha 
(\sigma^a\bar{\kappa})_\alpha  + 2\ii  ({\kappa}\sigma^a)_{\dot\alpha} E^{\dot\alpha}  \; , 
\eea
and of generic chiral (anti-chiral) superfields $T$ ($\bar T$)
\bea
\label{ExtObj_Memb_Dk_kappaT}
&\delta_\kappa  T =  \kappa^\alpha {D}_\alpha T\; , 
\qquad \delta_\kappa  \bar T = \bar \kappa_{\dot\alpha} \bar{{D}}^{\dot\alpha} \bar T =  \bar{{D}}_{\dot\alpha} \bar T \bar\kappa^{\dot\alpha} \, .
\eea
The variation of the action \eqref{ExtObj_Memb_S} splits into
\be
\label{ExtObj_Memb_Dk_dS}
\delta_\kappa S_{\rm memb} = \delta_\kappa S_{\rm memb,NG}+ \delta_\kappa S_{\rm memb,WZ}\,,
\ee
and we consider separately the two pieces.

First, let us start with the variation of the Nambu-Goto part \eqref{ExtObj_Memb_SmembrNG}. We rewrite the volume measure as 
\be
\label{ExtObj_Memb_2vol}
\d^3\xi  \sqrt{-\det {\bf h}}= \frac 1 3 {}^{*_3}\!E_a\wedge E^a\; , 
\ee
where  we have introduced the worldvolume Hodge dual of the supervielbein, defined as
\be
\label{ExtObj_Memb_*E}
{}^{*_3}\!E^A \equiv \frac 1 2 \d\xi^j\wedge \d\xi^i \sqrt{-\det {\bf h}} \epsilon_{ijk}h^{kl} E_l^A
\; . \qquad 
\ee
Using \eqref{ExtObj_Memb_*E}, we may easily compute the  variation of the Nambu-Goto part of the action with respect to the embedding coordinates ${\frak z}^M(\xi)$, which reads
\be\label{ExtObj_Memb_vNG}
\begin{split}
	\delta_\kappa S_{\text{memb,NG}} &= - 2\int_{\calm} {}^{*_3}\!E_a\wedge \delta E^a\, |T| -  \int_{\calm} \d^3\xi\sqrt{-\det {\bf h}}
	\frac {T \delta\bar{T} +  \delta T\, \bar{T} } {|T|}
	\\
	&=  4\ii \int_{\calm} {}^{*_3}\!E_a\wedge  E^\alpha 
	(\sigma^a\bar{\kappa})_\alpha |T| + 4\ii \int_{\calm} {}^{*_3}\!E_a\wedge  E_{\dot \alpha} (\bar\sigma^a \kappa)^{\dot\alpha}\, |T|
	\\
	&\quad\,-  \int_{\calm} \d^3\xi\sqrt{-\det {\bf h}}
	\frac {T  \bar D_{\dot\alpha}  \bar{T}  \bar \kappa^{\dot\alpha} } {|T|} - \int_{\calm} \d^3\xi\sqrt{-\det {\bf h}} \frac {\bar{T}  D^\alpha T \kappa_\alpha } {|T|}
	\; ,
\end{split}
\ee
where we have introduced the shorthand notation $T \equiv q_A S^A$.

Let us now come to the variation of the Wess-Zumino term, which can be recast as
\be
\label{ExtObj_Memb_vWZ0} 
\delta S_{\text{memb,WZ}} = q_A \int_{\calm} \delta {\bf A}_3^A = q_A \int_{\calm}(i_{\kappa} \dwb {\bf A}_3^A+ \dwb \, i_{ \kappa} {\bf A}_3^A)\,. 
\ee
For the moment, let us consider just \emph{closed} membranes, namely, membranes without any boundary (this assumption will be relaxed in Section \eqref{sec:ExtObj_Strings_MS}). Then, the second, boundary contribution in \eqref{ExtObj_Memb_vWZ0} simply cancels out
\be
\label{ExtObj_Memb_vWZhyp} 
\del \calm = \varnothing \qquad \Rightarrow \qquad q_A \int_{\calm} \dwb \, i_{ \kappa} {\bf A}_3^A  = q_A \int_{\del \calm} i_\kappa {\bf F}_4^A\stackrel{!}{=} 0\,,
\ee
reducing \eqref{ExtObj_Memb_vWZ0} to
\be
\label{ExtObj_Memb_vWZ} 
\delta S_{\text{memb,WZ}} = q_A \int_{\calm} \delta {\bf A}_3^A = q_A \int_{\calm}(i_{\kappa} \dwb {\bf A}_3^A+ \dwb \, i_{ \kappa} {\bf A}_3^A)=q_A \int_{\calm} i_{\kappa} {\bf A}_3^A\,. 
\ee
We now have
\be\label{ExtObj_Memb_iA3} 
\begin{split}
	i_{\kappa} {\bf F}_4^A= \;  & 4\ii E^b\wedge E^a \wedge E^\alpha  (\sigma_{ab}\kappa)_{\alpha}\bar{S}^A - 4\ii E^b\wedge E^a \wedge  E_{\dot\alpha} (\bar{\sigma}_{ab}\bar{\kappa})^{\dot{\alpha}} S^A \\  
	& + \frac{\ii}6 E^c\wedge E^b\wedge E^a  \epsilon_{abcd}  (\kappa \sigma^d)_{\dot \alpha} \bar{D}^{\dot\alpha}{ \bar S}^A  +\frac{\ii}6 E^c\wedge E^b\wedge E^a \epsilon_{abcd}  (\bar\kappa \bar\sigma^d)^{\alpha}  D_{\alpha} { S^A} 
	\; .
\end{split}
\ee
It is straightforward, from the definition \eqref{ExtObj_FrM_kproj}, to get the following identities
\be
\label{ExtObj_Memb_Use_Id_a} 
\begin{aligned}
	&E^b\wedge E^c\wedge E^\beta \sigma_{bc\, \beta}{}^\alpha = -\, {}^{*_3}\!E^a\wedge E^\beta (\sigma_a\,{\bar\Gamma})_{\beta}{}^\alpha \; ,
	\\
	&E^b\wedge E^c\wedge \bar E_{\dot\beta} \bar \sigma_{bc} {}^{\dot\beta}{}_{\dot\alpha} =  \, {}^{*_3}\!E^a\wedge \bar E_{\dot \beta} (\bar \sigma_a\,{\Gamma})^{\dot\beta}{}_{\dot \alpha} \; ,
\end{aligned}
\ee
by means of which the first line of \eqref{ExtObj_Memb_iA3} may be written as
\be
\label{ExtObj_Memb_vWZ1} 
\begin{split}
	&4\ii{}^{*_3}\!E^a\wedge E^\beta (\sigma_a\,{\bar\Gamma} \kappa)_{\beta}{} \bar{S}^A + 4\ii {}^{*_3}\!E^a\wedge  E_{\dot \beta} ( \bar \sigma_a\,{\Gamma} \bar\kappa)^{\dot\beta} S^A\,,
	\; .
\end{split}
\ee
and using 
\be
\label{ExtObj_Memb_Use_Idb}
\begin{aligned}
	&\d^3\xi \sqrt{-\det {\bf h}}\, {\Gamma} = 
	\frac \ii {3!}  \sigma^a\epsilon_{abcd}
	E^b\wedge E^c\wedge E^d  \, , \qquad \d^3\xi \sqrt{-\det {\bf h}}\, {\bar\Gamma} = 
	\frac \ii {3!}  \bar \sigma^a\epsilon_{abcd}
	E^b\wedge E^c\wedge E^d  \, , 
\end{aligned}
\ee
the second line becomes
\be
\label{ExtObj_Memb_vWZ2} 
\begin{split}
	&    \d^3 \xi\, \sqrt{-\det {\bf h}}\,  (\kappa \Gamma )_{\dot \alpha} \bar{{D}}^{\dot\alpha}\bar{ S}^A + \d^3 \xi\, \sqrt{-\det {\bf h}}\,  (\bar \kappa \bar \Gamma)_{\alpha}  {D}^{\alpha}{ S}^A
	\; .
\end{split}
\ee
Collecting \eqref{ExtObj_Memb_vWZ1} and \eqref{ExtObj_Memb_vWZ2}, the variation of the Wess-Zumino term \eqref{ExtObj_Memb_vWZ} becomes
\be
\label{ExtObj_Memb_vWZb} 
\begin{split}
	\delta S_{\text{memb,WZ}} &=   4\ii \int_{\calm} {}^{*_3}\!E^a\wedge E^\beta (\sigma_a\,{\bar\Gamma} \kappa)_{\beta}{} \bar{ T} +4\ii \int_{\calm} {}^{*_3}\!E^a\wedge  E_{\dot \beta} ( \bar \sigma_a\,{\Gamma} \bar\kappa)^{\dot\beta}T
	\\
	&\quad\, + \int_{\calm} \d^3 \xi\, \sqrt{-\det {\bf h}}\,  \bar{{D}}_{\dot\alpha}\bar{ T}  (\bar \Gamma \kappa )^{\dot \alpha}+\int_{\calm}\d^3 \xi\, \sqrt{-\det {\bf h}}\, {D}^{\alpha}{ T} ( \Gamma \bar \kappa)_{\alpha}  
\end{split}
\ee

It is immediate to check that this variation cancels \eqref{ExtObj_Memb_vNG} if and only if the $\kappa$--parameters obey \eqref{ExtObj_Memb_kappa}.

\subsection{Central charge for a membrane}
\label{sec:ExtObj_Memb_CC}

Although the $\mathcal{N}=1$ does not have a central extension in a strictly group theoretical meaning, the presence of extended objects does modify the supersymmetry algebra in a very similar way that a central extension does. This is a well known issue, solved first in the seminal work by Olive and Witten \cite{Witten:1978mh} for solitons, later generalized to more general extended objects (see, for example, \cite{deAzcarraga:1989mza,Hughes:1986fa,Hughes:1986dn,Bergshoeff:1996tu,Sorokin:1997ps,Bergshoeff:1998vx,Shifman:2009zz,Azcarraga:2011hqa}). 

In a supersymmetric theory, we can define the conserved currents $j^m_\alpha(x)$, $\bar \jmath^{m\dot\alpha}(x)$. In presence of extended objects, the usual anti-commutators of the conserved currents
\be
\label{ExtObj_Memb_MCGaa}
\{j^0_\alpha(x), j^0_\beta(x')\}
=0\,,\qquad \{ {\bar\jmath}^{0\dot\alpha}(x), {\bar\jmath}^{0\dot \beta}(x')\}
=0\,,
\ee
get modified so as to include a central extension
\be
\label{ExtObj_Memb_MCGa}
\{j^0_\alpha(x), j^0_\beta(x')\}
=m_{\alpha\beta}(x,x')\,,\qquad \{ {\bar\jmath}^{0\dot\alpha}(x), {\bar\jmath}^{0\dot \beta}(x')\}
=\bar m^{\dot \alpha \dot \beta}(x,x')\,.
\ee
The integration over spacetime, whenever well defined, allows for defining the 
\emph{central, topological charges} $M_{\alpha\beta}$ and $\bar M^{\dot \alpha \dot \beta}$ as
\be
\label{ExtObj_Memb_MCGb}
\{Q_\alpha, Q_\beta\}
=M_{\alpha\beta}\,,\qquad \{\bar Q^{\dot\alpha}, \bar Q^{\dot \beta}\}
=\bar M^{\dot \alpha \dot \beta}\,.
\ee

Localized extended objects, such as membranes, modifies the supersymmetry algebra by introducing a central charge--like term as in \eqref{ExtObj_Memb_MCGb} \cite{deAzcarraga:1989mza,Sorokin:1997ps}. The central charge is associated with the \emph{quasi--invariance} of the Wess-Zumino term under superymmetry transformations. Namely, under the supersymmetry transformations, the Wess-Zumino term is invariant up to a total derivative
\be
\delta_\epsilon S_{\rm memb, WZ} = \epsilon^\alpha \int  \d \Delta_\alpha + \bar\epsilon_{\dot\alpha} \int  \d \bar\Delta^{\dot\alpha} + \ldots\;,
\ee
defining the current two-forms $\Delta_\alpha (x)$ and $\bar\Delta^{\dot\alpha} (x)$. The conserved currents $j^m_\alpha$ and $\bar \jmath^{m\dot\alpha}$  have to take into account of these `anomalous' pieces and are modified to
\be
\tilde j^m_\alpha = j^m_\alpha - \Delta_\alpha^m\,,\qquad \bar{\tilde\jmath}^{m\dot\alpha} = \bar\jmath^{m\dot\alpha}- \bar\Delta^{m\dot\alpha}
\ee
so that $j^0_\alpha$, $\bar \jmath^{0\dot\alpha}$ now satisfy the algebra \eqref{ExtObj_Memb_MCGa} with
\be
\begin{aligned}
	m_{\alpha\beta}(x,x') &= - \{j^0_\alpha(x),\Delta^0_\beta(x')\} - \{j^0_\beta(x),\Delta^0_\alpha(x')\}
	\\
	\bar m^{\dot\alpha\dot\beta}(x,x') &= - \{{\bar\jmath}^{0\dot\alpha}(x),\bar\Delta^{0\dot\beta}(x')\} - \{{\bar\jmath}^{0\dot\beta}(x),\bar\Delta^{0\dot\alpha}(x')\}
\end{aligned}
\ee
Once integrated, they immediately give
\be
\begin{aligned}
	M_{\alpha\beta} &= - \int \left(\delta_{\epsilon^\alpha} \Delta_\beta+ \delta_{\epsilon^\beta} \Delta_\alpha \right)\;,
	\\
	\bar M^{\dot\alpha\dot\beta} &= - \int \left(\delta_{\bar\epsilon_{\dot\alpha}} \bar\Delta^{\dot\beta}+ \delta_{\bar\epsilon_{\dot\beta}} \bar\Delta^{\dot\alpha} \right)\;.
\end{aligned}
\ee
From the variation of the Wess-Zumino term \eqref{ExtObj_Memb_vWZ}
\be\label{ExtObj_Memb_vWZbd} 
\delta S_{\rm memb, WZ} |_{\rm bd} = q_A \int_{\calm} \delta {\bf A}_3^A|_{\rm bd}  = q_A \int_{\calm} \dwb i_{\epsilon} {\bf A}_3^A
\ee
we may identify the current two-forms
\be
\epsilon^\alpha \Delta_\alpha(x) \equiv q_A i_{\epsilon} {\bf A}_3^A \,,\qquad \bar\epsilon_{\dot\alpha}\bar\Delta^{\dot\alpha}(x) \equiv q_A i_{\bar\epsilon} {\bf A}_3^A\,.
\ee
Considering the membrane in its static ground state, we easily get that the only relevant terms for $\Delta_\alpha (x)$ and $\bar\Delta^{\dot\alpha} (x)$ are
\be
\Delta_\alpha(x) =q_A  E^b \wedge E^a \sigma_{ab\alpha}{}^\gamma D_\gamma P^A \,,\qquad \bar\Delta^{\dot\alpha}(x) =q_A E^b \wedge E^a \bar\sigma_{ab}{}^{\dot\alpha}{}_{\dot\gamma} \bar D^{\dot\gamma} P^A \,.
\ee
whence
\be
\label{ExtObj_Memb_CG_finb}
\begin{aligned}
M_{\alpha\beta} &= -4\ii q_A\int E^b \wedge E^a\, \sigma_{ab\alpha}{}^\gamma \varepsilon_{\beta\gamma}  \bar \calv^A(\bar\Phi) \,, 
\\ 
\bar M^{\dot\alpha\dot\beta} &= 4\ii q_A\int E^b \wedge E^a\, \bar\sigma_{ab}{}^{\dot\alpha}{}_{\dot\gamma}\varepsilon^{\dot\beta\dot\gamma} \calv^A(\Phi) \,.
\end{aligned}
\ee
The central charges then depend on the very same combination $q_A \calv^A (\Phi)$ whose absolute value defines the tension of the membrane in \eqref{ExtObj_Memb_S}. As an anticipation of what will be discussed in Section~\ref{sec:ExtObj_DWM}, we also notice that the combination $q_A \calv^A (\Phi)$ that enters the central charge is the same as the one that appears in the superpotential \eqref{GSS_MTF_Wgen}, which was there generated by the gauge three-forms.
\section{Domain Walls}
\label{sec:ExtObj_DW}

In four-dimensional theories other BPS-objects may appear, namely \emph{domain walls} which, like membranes, extend along three dimensions. However, unlike membranes, these are not generically \emph{structureless}: they are not hypersurfaces of strict codimension one, for they are also extended along the fourth dimension, although they are `mostly' localized within a certain interval thereof. Domain walls may be born whenever a theory exhibits multiple vacua. Here, we will summarize their main features in global supersymmetry which will be relevant for the discussion of the next section, where we will study the interplay between membranes and domain walls. The literature on the subject is vast and, in our treatment, we will mostly follow \cite{Cvetic:1992bf,Cvetic:1992sf,Cvetic:1996vr,Shifman:2009zz}, to which we refer for further references and discussions.

Let us consider a classical theory, with field content $\Psi^a$ and with a potential $V(\Psi, \bar \Psi)$.  In the classical framework, the vacua of the theory can be computed by simply extremizing the potential, setting $\del_{\Psi^a} V = 0$. Such a condition determines the space of vacua $\scrm_{\rm vac}$. Generically, we may assume that $\scrm_{\rm vac}$ is made up by different disconnected components, that is, the zero-th homotopy group is nontrivial
\be
\pi_0 (\scrm_{\rm vac}) \neq \varnothing\;.
\ee
Two stable vacua, belonging to two path-disconnected components to $\scrm_{\rm vac}$, may fill two different regions of spacetime, which are separated by \emph{domain walls}. Domain walls are indeed defined as \emph{solitonic} solutions of the Minkowskian equations of motion which interpolate between two stable vacua. 

The construction of domain wall solutions is generically not easy. Presently we focus just on a particular, but important class of domain walls, namely \emph{BPS--domain walls}. These are solitonic solutions which preserve part of the bulk supersymmetry in a sense that will be shortly discussed. Moreover, the vacua among which a BPS--domain wall interpolate are supersymmetric. Furthermore, as an additional simplifying assumption, we consider the domain walls to be flat and symmetric for $SO(1,2)$ rotations in the $x^i$--directions, with $i=0,1,2$. This implies that the profile of the domain wall is solely determined by the dynamics along the fourth direction $x^3 \equiv z$.

To be concrete, let us consider a supersymmetric theory whose field content is just made up by chiral multiplets $\Phi^a$ (which, for the moment, we assume to be just ordinary chiral multiplets as \eqref{Conv_ChiralB}) and with an F-term potential solely determined by a superpotential $W(\Phi)$. Superymmetric vacua can be easily found by requiring
\be
\label{ExtObj_DW_dW}
\frac{\del}{\del \varphi^a} W (\varphi) \stackrel{!}{=} 0\;.
\ee
Let us assume that this condition leads to (at least) two disconnected set of vacua $\scrm_{\text{vac},-}, \scrm_{\text{vac},+} \subset \scrm_{\text{vac}}$. Let su pick up two vacua belonging to distinct subsets, $\varphi_-^a \in \scrm_{\text{vac},-}$ and $\varphi_+^a \in \scrm_{\text{vac},+}$.  We can imagine the vacuum $\varphi_-^a$  as located at $z \to  -\infty$ and $\varphi_+^a$ as reached when the scalar fields $\varphi^a$ move towards $z \rightarrow + \infty$:
\be
\label{ExtObj_DW_Vacua}
\begin{aligned}
	\varphi_-^a \equiv \varphi^a |_{-\infty} &\qquad \text{vacuum reached for } z\rightarrow-\infty\;,
	\\
	\varphi_+^a \equiv \varphi^a |_{+\infty} &\qquad \text{vacuum reached for } z\rightarrow+\infty\;.
\end{aligned}
\ee
In this setup, a domain wall is the extended solitonic object which is determined by how the field profiles vary, subjected to the boundary condition \eqref{ExtObj_DW_Vacua}.

In order to see how a BPS--domain wall solution can be constructed, let us first make use of the simplifying assumption that the dynamics is fully determined by the $z$-direction, that translates in the fact that all the components of the chiral multiplets $\Phi^a$ nontrivially depend just on $z$:
\be
\label{ExtObj_DW_Fields}
\varphi^a \equiv \varphi^a(z)\,,\qquad \psi^a_\alpha \equiv \psi^a_\alpha(z)\,\qquad f^a \equiv f^a(z)\,.
\ee

In order the domain wall to preserve supersymmetry, it is necessary that its profile, specified by the field configurations \eqref{ExtObj_DW_Fields}, identifies a nontrivial killing spinor which allows the supersymmetry variations to vanish. Since we are just focusing on bosonic components, a killing spinor is straightforwardly identified by requiring the variations of the chiralini $\psi^a_\alpha$ to vanish, that is
\be
\label{ExtObj_DW_susyvar}
\delta \psi_\alpha^a \big|_{\rm DW}  = \sqrt{2} \ii \sigma_{\alpha\dot\alpha}^m \bar \epsilon^{\dot \alpha} \del_m \varphi^a + \sqrt{2} \epsilon_\alpha f^a \big|_{\rm DW} \overset{!}{=} 0 \,,
\ee
where $\epsilon^\alpha$ is the global supersymmetry parameter. Hence, setting the auxiliary fields $f^a$ on-shell
\be
\label{ExtObj_DW_susyvar2}
\ii \sigma^3_{\alpha \dot \alpha} \bar\epsilon^{\dot \alpha} \del_z \varphi^a =  K^{\bar b a} \bar W_{\bar b}\epsilon_\alpha\;.
\ee
Provided that the supersymmetry parameter $\epsilon^\alpha$ satisfies the projection condition
\begin{important}
\be
\label{ExtObj_DW_proj}
\epsilon_\alpha = \mp \ii e^{\ii \vartheta}\sigma^3_{\alpha \dot \alpha} \bar\epsilon^{\dot \alpha} 
\ee
\end{important}
with real $\vartheta$ then, \eqref{ExtObj_DW_susyvar2} is satisfied when
\begin{important}
	\be
	\label{ExtObj_DW_flow}
	\del_z \varphi^a = \mp e^{\ii \vartheta}  K^{\bar b a} \bar W_{ \bar b}\qquad\qquad{\color{darkred}\text{\sf{Flow equation}}}
	\ee
\end{important}

Some comments to these crucial relations are in order. The projection condition \eqref{ExtObj_DW_proj} singles out the combination of original supersymmetric parameters which are still well defined over the domain wall. Analogously to \eqref{ExtObj_FrM_kappa}, \eqref{ExtObj_DW_proj} halves the degrees of freedom of the supersymmetric parameter $\epsilon$: in this sense, the domain walls under considerations are \emph{$\half$--BPS} objects. Furthermore, the relation \eqref{ExtObj_DW_flow} dictates how the profiles of the scalar fields vary along the transverse direction for the given superpotential. 

Above, we have also stated that the domain walls are solitons, which satisfy the classical equations of motions, but so far we have not exploited them. Using the simplifyng assumption \eqref{ExtObj_DW_Fields}, the action, integrating out the auxiliary fields $f^a$ and restricting just to the bosonic components, acquires the simplified form
\be
\label{ExtObj_DW_Sbulk}
S_{\rm bos} = \int \d^4 x\, \call_{\rm bos} = \int \d^4 x\,  \left[ -K_{a \bar b} \del_z \varphi^a \del_z \bar\varphi^{\bar b} - V(\varphi,\bar\varphi)\right]\;,
\ee
where $V(\varphi,\bar\varphi) = K^{\bar b a} W_a \bar W_{\bar b}$. It is immediate to show that the scalar profile \eqref{ExtObj_DW_flow} is a solution of the equations of motion which can be computed from \eqref{ExtObj_DW_Sbulk}
\be
- K_{a \bar b \bar c} \del_m \varphi^a \del^m \bar \varphi^{\bar c} + \del_m \left( K_{a \bar b} \del^m \varphi^{a}\right)- K^{\bar e d} K_{d\bar b\bar c} K^{\bar c a} \bar W_{\bar e} W_a-K^{\bar c a} W_a \bar W_{\bar c \bar b}=0\;,
\ee
satisfying them identically. Therefore, we conclude that solving \eqref{ExtObj_DW_flow} with the boundary conditions \eqref{ExtObj_DW_Vacua} identifies the domain wall as a supersymmetric, solitonic extended object. 

In the following sections we explore two important properties of the domain walls: in Section~\ref{sec:ExtObj_DW_T} we see how to properly define an energy for domain walls and in Section~\ref{sec:ExtObj_DW_CG} we show how the presence of a domain wall modifies the supersymmetry algebra by introducing a central charge. Finally, in Section~\ref{sec:ExtObj_DW_Ex} we give a simple example where we explicitly compute the BPS domain wall solution.

\subsection{Tension of a BPS-domain wall}
\label{sec:ExtObj_DW_T}

Although domain walls are objects (mostly) localized along the $z$-direction, they are infinitely extended along the three spacetime direction $x^i$. Hence it does not make sense, in general, to define an \emph{energy} for the domain wall configuration. Instead, more appropriately, we define an \emph{energy per unit area}, or a tension $\calt_{\rm DW}$, for the domain wall as
\be
\label{ExtObj_DW_T}
S_{\text{on-shell}} \equiv - \int \d^3 x  \;  \calt_{\rm DW}\,,
\ee
where the action is understood as evaluated over the domain wall configuration specified by the flow equation \eqref{ExtObj_DW_flow}. 

As a first step, we rewrite the action \eqref{ExtObj_DW_Sbulk} in the more practical and elegant \emph{BPS--form}
\be
\label{ExtObj_DW_Sb}
\begin{split}
	S_{\rm bos} = \int \d^3 x\,\d z\,  \Bigg[&-K_{a\bar b} \left(\del_z \varphi^a \pm  e^{\ii \beta} K^{\bar c a} \bar W_{\bar c}\right) \left( \del_z \bar\varphi^{\bar b} \pm  e^{-\ii \beta} K^{\bar b d}  W_d \right) 
	\\
	&\pm \left(\del_z \varphi^a W_a e^{-\ii \beta} + \del_z \bar\varphi^{\bar b} \bar W_{\bar b} e^{\ii \beta} \right)  \Bigg]\,,
\end{split}
\ee
where $\beta$ is a \emph{generic} real constant. By rearranging the terms in the second line, the action may be written as 
\be
\label{ExtObj_DW_Sd}
\begin{split}
	S_{\rm bos} &=   \int \d^3 x\,\d z\,\Bigg[-K_{a\bar b} \left(\del_z \varphi^a \pm  e^{\ii \beta} K^{\bar c a} \bar W_{\bar c}\right) \left( \del_z \bar\varphi^{\bar b} \pm e^{-\ii \beta} K^{\bar b d}  W_d \right) \Bigg]
	\\
	&\quad\,\pm 2 \int \d^3 x\,\d z\,  \frac{\d}{\d z} \Re \left(W e^{-\ii \beta}\right)\,,
\end{split}
\ee
which, upon performing an integration of the last line, leads to
\be
\label{ExtObj_DW_Se}
\begin{split}
	S_{\rm bos} &=   \int \d^3 x\,\d z\,\Bigg[-K_{a\bar b} \left(\del_z \varphi^a \pm e^{\ii \beta} K^{\bar c a} \bar W_{\bar c}\right) \left( \del_z \bar\varphi^{\bar b} \pm  e^{-\ii \beta} K^{\bar b d}  W_d \right) \Bigg]
	\\
	&\quad\,\pm 2 \int \d^3 x\,\Re \left[(W_{+\infty}-W_{-\infty} )e^{-\ii \beta}\right]\,.
\end{split}
\ee
The (semi--)positive definiteness of the first line, implies that the energy per unit area of the field configuration, which we here generically denote with $\sigma$, is
\be
\label{ExtObj_DW_BPSine}
\sigma  \geq \mp\, 2\, \Re \left[(W_{+\infty}-W_{-\infty} )e^{-\ii \beta}\right]\,.
\ee
which is in the form of a BPS bound. 

Indeed, BPS domain walls satisfies \eqref{ExtObj_DW_proj} and, choosing $\beta = \alpha$, from \eqref{ExtObj_DW_Se} we most readily get
\be
\label{ExtObj_DW_bounda}
\calt_{\rm DW} = \mp\,2\, \Re \left[(W_{+\infty}-W_{-\infty} )e^{-\ii \alpha}\right]
\ee
It is also possible to show that the scalar profile \eqref{ExtObj_DW_flow} implies the existence of the \emph{integral} of motion
\be 
\label{ExtObj_DW_Int}
\frac{\d}{\d z} \Im (W e^{-\ii\alpha}) = 0 \qquad \Rightarrow \qquad \Im \left[e^{-\ii \alpha}(W_{+\infty}-W_{-\infty}) \right]=0\,.
\ee
Further combing this with the request the tension $\calt_{\rm DW}$ to be non-negative definite strictly correlates $\alpha$ with the choice of sign in \eqref{ExtObj_DW_bounda}. In fact, we need to choose
\begin{subequations}
\label{ExtObj_DW_alphasign}
\begin{alignat}{3}
	\alpha &= \arg (W_{+\infty} - W_{-\infty}) + (2k +1) \pi\, &\quad \text{with}\quad k\in \mathbb{Z} \quad  &\text{for the upper sign}\,,&
	\\
	\alpha &= \arg (W_{+\infty} - W_{-\infty}) + 2k \pi\,&\quad \text{with}\quad k\in \mathbb{Z} \quad  &\text{for the lower sign}\,.&
\end{alignat}
\end{subequations}

Finally, with the choice \eqref{ExtObj_DW_alphasign}, the tension of the BPS--domain wall \eqref{ExtObj_DW_bounda} becomes
\begin{important}
	\be
	\label{ExtObj_DW_bound}
	\calt_{\rm DW} = 2\, |W_{+\infty}-W_{-\infty} | \qquad{\color{darkred}\text{\sf{Tension of a BPS domain wall}}}
	\ee
\end{important}

In view of \eqref{ExtObj_DW_BPSine}, the tension \eqref{ExtObj_DW_bound} is the strongest bound on the energy per unit area. Notably, such a bound is uniquely determined by the value of the superpotential at the asymptotic vacua.

\subsection{Central charge for a BPS--domain wall}
\label{sec:ExtObj_DW_CG}

Domain walls, as extended objects, modify the supersymmetry algebra as in \eqref{ExtObj_Memb_MCGb} \cite{Witten:1978mh}. In this case, it is however more convenient to compute directly the commutators \eqref{ExtObj_Memb_MCGb}. We need to consider tensorial representations of the supersymmetry generators $Q_\alpha$, $\bar Q^{\dot \alpha}$, which can be immediately read from the variations of the fermions. The supersymmetry variations of the fermions of the bulk superfields are 
\be
\label{ExtObj_DW_CG_susyvar1}
\begin{aligned}
	\delta \chi_\alpha^a &=  \{\chi_\alpha^a, Q_\beta\} \zeta^\beta  +  \{\chi_\alpha^a, \bar Q_{\dot \beta}\} \bar\zeta^{\dot \beta}\,,
	\\
	\delta \bar \chi^{\dot\alpha a} &=  \{Q^\beta, \bar \chi^{\dot\alpha a}\} \zeta_\beta +  \{\bar Q^{\dot \beta}, \bar \chi^{\dot \alpha a}\} \bar\zeta_{\dot\beta}\,.
\end{aligned}
\ee
which, on the domain wall background specified by \eqref{ExtObj_DW_Fields} and \eqref{ExtObj_DW_flow}, read
\be
\label{ExtObj_DW_CG_susyvar2}
\begin{aligned}
	\delta \chi_\alpha^a &=    \sqrt{2} \ii \sigma^3_{\alpha \dot \alpha} \bar\zeta^{\dot \alpha} \del_z \varphi^a - \sqrt{2} \zeta_\alpha K^{\bar b a} \bar W_{ b}\,,
	\\
	\delta \bar \chi^{\dot\alpha \bar b} &=  \sqrt{2} \ii \bar\sigma^{3\dot\alpha \alpha} \zeta_{\alpha} \del_z \bar\varphi^{\bar b} - \sqrt{2} \bar\zeta^{\dot\alpha} K^{\bar b a} W_{ a}\,.
\end{aligned}
\ee
By comparing \eqref{ExtObj_DW_CG_susyvar1} with \eqref{ExtObj_DW_CG_susyvar2}, we most readily get
\be
\label{ExtObj_DW_CG_charge}
\begin{aligned}
	Q_\alpha &=  \int \d x^1\d x^2 \d z\,\left[ -\ii\sqrt{2} \sigma^0_{\alpha \dot\beta} \bar\chi^{\dot \beta \bar b} \bar W_{I}+ \sqrt{2} \sigma^3_{\alpha \dot\beta} \bar\sigma^{0\dot\beta \gamma} \chi_{\gamma}^a K_{a\bar b} \del_z \bar\varphi^{\bar b}\right],
	\\
	\bar Q^{\dot\alpha} &=  \int \d x^1\d x^2 \d z\,\left[ \ii\sqrt{2} \bar\sigma^{0\dot\alpha \beta} \chi_{\beta}^a W_{a}+ \sqrt{2} \bar\sigma^{3\dot\alpha \beta} \sigma^0_{\beta \dot\gamma} \bar\chi^{\dot\gamma \bar b} K_{a\bar \bar b} \del_z \varphi^a\right]\,,
\end{aligned}
\ee
where we have employed the equal-time anti-commutation relation 
\be
\label{ExtObj_DW_CG_ac}
\{\chi_\alpha^a(x), \bar \chi_{\dot \beta}^{\bar b}(x')\} = \ii  K^{\bar b a} \delta^{(3)} (x-x') \sigma^0_{\alpha \dot\beta}\,.
\ee
Hence, by explicitly computing the anti-commutators in \eqref{ExtObj_Memb_MCGb}, we get
\be
\label{ExtObj_DW_CG_exp}
\begin{aligned}
	M_{\alpha \beta}  &= -8\int \d x^1\d x^2 \d z\,  \varepsilon_{\beta\gamma} \sigma^{03}{}_\alpha{}^\gamma\,\bar W_{\bar b} \del_z \bar\varphi^{\bar b}\,,
	\\
	\bar M^{\dot\alpha \dot \beta }&= -8\int \d x^1\d x^2 \d z\,  \bar\sigma^{03 \dot\alpha}{}_{\dot\gamma} \varepsilon^{\dot\gamma\dot\beta}\, W_a \del_z \varphi^a\,. 
\end{aligned}
\ee
which, integrated, give the topological charges
\be
\label{ExtObj_DW_CG_fin}
\begin{aligned}
	M_{\alpha \beta}  &= -8\int \d x^1\d x^2\,\varepsilon_{\beta\delta} \sigma^{03}{}_\alpha{}^\delta\, (\bar W_{+\infty}-\bar W_{-\infty}) \,,
	\\
	\bar M^{\dot\alpha \dot \beta }&= -8\int \d x^1\d x^2\,  \bar\sigma^{03 \dot\alpha}{}_{\dot\gamma} \varepsilon^{\dot\gamma\dot\beta} (W_{+\infty}-W_{-\infty} ) \,.
\end{aligned}
\ee

\subsection{A simple example}
\label{sec:ExtObj_DW_Ex}

Before concluding this section, we show a very simple example of explicit computation of a BPS--domain wall profile.

Consider a supersymmetric theory where a single, ordinary chiral superfield $\Phi$ is present and described by the K\"ahler potential and superpotential
\be
\label{ExtObj_DW_Ex_KW}
K (\Phi, \bar \Phi) = \Phi \bar \Phi\,,\qquad \qquad W(\Phi) = \frac{m^2}{\lambda} \Phi -\frac{\lambda}{3} \Phi^3\;,
\ee
with $m>0$ and $\lambda$ real parameters.

This model admits two distinct supersymmetric vacua given by
\be
\label{ExtObj_DW_Ex_dW}
\del_\Phi W \stackrel{!}{=} 0 \qquad \Rightarrow \qquad \Phi_\pm | = \varphi_\pm = \pm \frac{m}{\lambda}\;,
\ee
at which the superpotential takes the value $W_\pm = \pm \frac23 \frac{m^3}{\lambda^2}$. let us assume that the vacuum determined by $\varphi_-$ is reached when $z \rightarrow - \infty$, while the vacuum $\varphi_+$ when $z \rightarrow +\infty$. 

Let us choose $\alpha = 0$, so that \eqref{ExtObj_DW_Int} is trivially satisfied. Then, in order to achieve a positive tension for the domain wall configuration, in \eqref{ExtObj_DW_bounda} the lower sign has to be chosen, leading to
\be
\calt_{\rm DW} = \frac43 \frac{m^3}{\lambda^2}\;.
\ee
Equivalently, we could have chosen $\alpha = \pi$ and the upper sign in \eqref{ExtObj_DW_bounda} obtaining the same tension.

\begin{figure}[t]
	\centering
	\includegraphics[width=6cm]{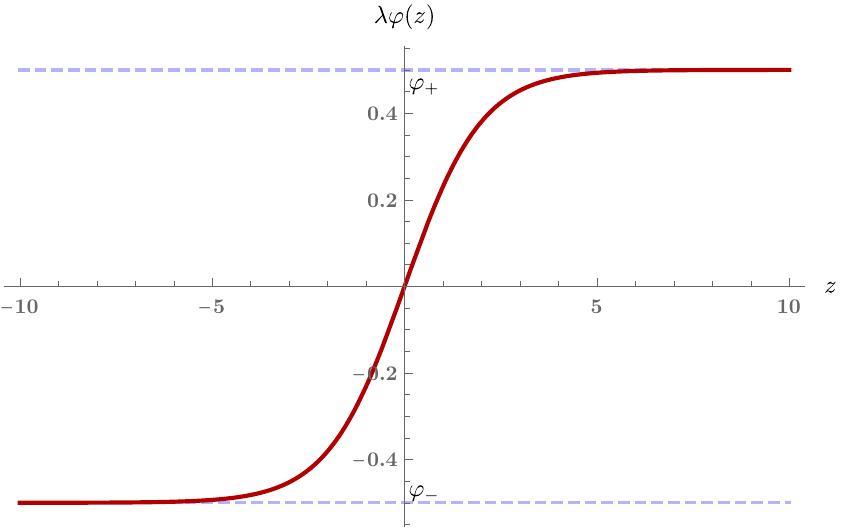} \includegraphics[width=6cm]{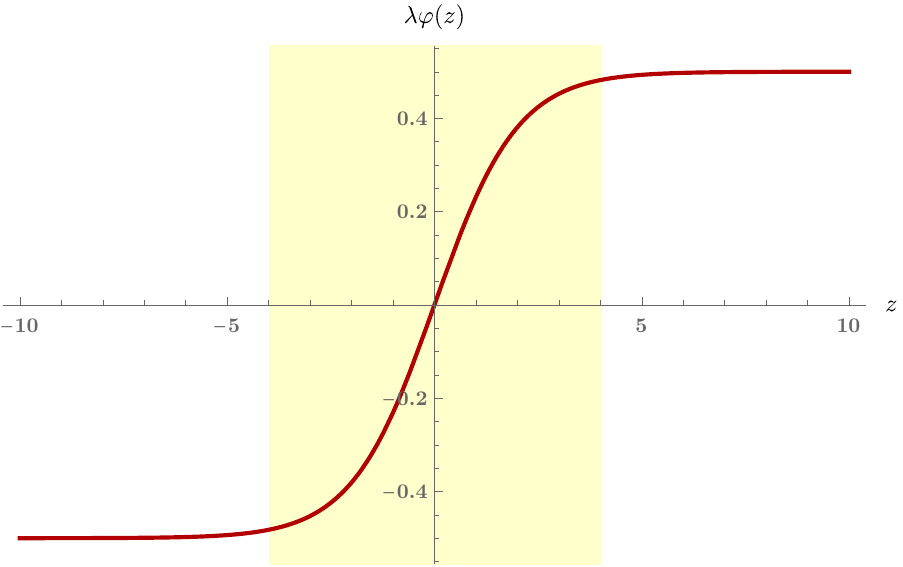}
	\caption{\footnotesize{The domain wall profile as specified by \eqref{ExtObj_DW_Ex_flowsol}, where we chose $m= \frac12$. The wall profile interpolates between the asymptotic vacua $\varphi_-$ and $\varphi_+$. Its width, highlighted in the figure on the right and given by the region where $\varphi$ evolves, is roughly estimated by $\sim 2/m$.}}\label{fig:DW}
\end{figure}

The domain wall profile is determined by the flow equation \eqref{ExtObj_DW_flow}, which, in this simple model, becomes
\be
\label{ExtObj_DW_Ex_flow}
\del_z \varphi = \mp \frac{e^{\ii \alpha}}{\lambda}  \left(m^2 - \lambda^2 \bar \varphi^2 \right) =  \frac1\lambda  \left(m^2 - \lambda^2 \bar \varphi^2 \right)\;.
\ee
Since $\Im \varphi =0$ is a solution to \eqref{ExtObj_DW_Ex_flow}, we can assume the field $\varphi$ to be real. Then, the differential equation \eqref{ExtObj_DW_Ex_flow} is solved by
\be
\label{ExtObj_DW_Ex_flowsol}
\varphi = \frac{m}{\lambda} \tanh (m z)\,.
\ee
It can be easily checked that this solution is such that $\lim\limits_{z \to \pm \infty} \varphi(z)= \varphi_\pm$, as also depicted in Fig.~\ref{fig:DW}. The domain wall profile is nontrivial around $z=0$: its \emph{width} can be roughly estimated from the region where the field $\varphi$ varies, that is $\sim 2/m$ , as highlighted in Fig.~\ref{fig:DW}.

\section{Domain Walls sourced by Membranes}
\label{sec:ExtObj_DWM}

We have introduced membranes and domain walls separately and indeed it seems that they are objects that do not \emph{talk} between one another: membranes are fundamental BPS--objects, which are part of the spectrum of a theory; domain walls, instead, arise only if the theory admits them as solitonic solutions. However, they are somewhat similar in nature: they cover three spacetime directions and they are both BPS, spontaneously breaking half of the bulk generators. 

This section is devoted to explore how new domain wall solution can be generated by membranes. After qualitatively describing how membranes may induce new domain walls, we will pass to examine the mathematical properties that configurations which merge membranes and domain walls enjoy.

\subsection{Membranes creating new domain walls}
\label{sec:ExtObj_DWM_Diff}

The membranes that we introduced in Sections \ref{sec:ExtObj_IntroMemb} and \ref{sec:ExtObj_MembSusy} are \emph{fundamental} objects, governed by actions of the form \eqref{ExtObj_FrM_Smembr} or \eqref{ExtObj_Memb_S}, using which they are directly included into a supersymmetric theory. Furthermore, they are \emph{structureless}. That is, owing to their very definition via the embedding \eqref{ExtObj_FrM_SEmbed} as hypersurfaces in the ambient superspace, in the ordinary spacetime, when they are frozen in their ground state, they are strictly hypersurfaces of codimension one. In the static gauge, for example, a membrane fully covers the $x^i$--directions, but it is just a point, $z=z^0$, along the fourth.



The presence of charged membranes within a theory does not just serve to add more energy to the system, but also to divide the spacetime into two regions with two \emph{different} potentials, as already explained in Section~\ref{sec:Intro_Jumps}. Eventually, the sets of vacua on the two sides of the membrane are different. Membranes, in other words, provide a new theory, with new vacua and, within this theory, one might look for the admissible domain wall solutions which interpolate between the vacua on one side to the others on the other side. In this sense membranes and domain walls are extended objects that can coexist and interestingly create new nonperturbative effects within a theory. Remarkably, an ordinary chiral multiplet theory without membranes, may not admit any domain wall solution (for example, if the theory has a trivial $\pi_0 (\scrm_{\rm vac})$, because the vacuum is just a point); however, once we couple a membrane to the theory, new vacua are born and domain wall solutions can exist. In Section~\ref{sec:ExtObj_DWM_Ex} we will indeed see a simple example when this happens.

It is also important to remark that it is not obvious that domain walls `enclosing' a membrane may exist. A domain wall is in fact a \emph{full} solitonic solution which includes the backreaction of the scalars of the theory, as dictated by the flow equations and these, generically, may not be possible to be solved, preventing the formation of a domain wall. The study of the domain walls with a membrane at its core is the object of the next sections, where we compute the energy and the central charges for such a configuration, whenever a domain wall solution exists.

\subsection{BPS-condition and flow equations}
\label{sec:ExtObj_DWM_Susy}

The analysis of BPS--domain wall solutions performed in \eqref{sec:ExtObj_DW} may be easily extended to the case where membranes are present. Our aim is to explicitly show how the presence of membranes can \emph{create} new domain wall solutions.  In order to ease the exposition, we focus on the case where there is just a single, flat membrane which divides the spacetime into two distinct regions. Moreover, we will work in the static gauge \eqref{ExtObj_FrM_Static}: the membrane stretches along three spacetime directions, say $x^i$ with $i=0,1,2$, and the transverse direction is most readily identified with $z \equiv x^3$. We also assume the membrane to be fixed at $z=0$, frozen in its ground state.

In the general case, some chiral, \emph{bulk} superfields $S^A$ are coupled to the membrane, as in Section \ref{sec:ExtObj_MembSusy}. We assume them to be the three-form superfields \eqref{GSS_MTF}, with chiral expansion
\be
S^A= s^A + \sqrt{2} \theta^\alpha \psi_{S\alpha}^A+\frac12 \left( *F_4^A+  \ii d^A  \right) \theta^2\,,
\ee
where the gauge three-forms $A^A_3$ appear in their highest components through the field strengths $F_4^A$. As in the previous section, we will work with the simplifying assumption that the profile of the fields depends only on the coordinate $z$ transverse to the membrane
\be
s^A \equiv s^A(z)\,,\qquad \psi^A_{S\alpha} \equiv \psi^A_{S\alpha}(z)\,, \qquad A_3^A \equiv A_3^A(z)\,, \qquad d^A \equiv d^A(z)\;.
\ee
However we stress that the scalar fields $s^A(\varphi(z))$ do actually depend on the sets of the physical fields $\varphi^a$ through which they can be expressed (see \eqref{GSS_MTF_Compa}). We refer to Section~\ref{sec:GSS_MTF} for a detailed discussion regarding the correspondence between ordinary chiral fields and three-form multiplets $S^A$. Moreover, the physical fields $s^A$ (and $\varphi^a$) and $\psi^A_{(S)}$ are continuous in the whole spacetime, although their derivatives may be discontinuous at $z=0$, where the membrane is encountered.

The full action describing the bulk and the membrane dynamics is
\be
\label{ExtObj_DWM_Stot}
S = S_{\rm bulk} + S_{\rm memb}\,,
\ee
where the bulk action, defined by the Lagrangian \eqref{GSS_MTF_L3f}, is
\be
\begin{split}
	S_{\rm bulk} &= \int \d^4 x\, \call_{\rm bulk} 
	\\
	&\quad\, =\int \d^4 x\,  \left( \int \d^4 \theta\, K(\Phi (P),\bar \Phi (P))+\left[\int\d^2\theta\, \hat W(\Phi(P))+\text{c.c.}\right]+ \tilde\call_{\rm bd}   \right)
\end{split}
\ee
and $S_{\rm memb}$ is given in \eqref{ExtObj_Memb_S}. For simplicity we shall focus just on bosons and set the fermions to zero. Then, the bulk action is the very same as that given in \eqref{GSS_MTF_L3fComp}. 

In order to get an `ordinary' chiral theory and connect the present discussion with that of the previous section, we have to get rid of the gauge three-forms. We then integrate them out as
\be
\label{ExtObj_DWM_A3eom}
T_{AB}(*\hat{F}^B_4+\Upsilon^B)= -N_A - q_A \Theta (z)\;,
\ee
where the last term is due to the Wess-Zumino term of \eqref{ExtObj_Memb_S}, which makes the constants $N_A$ jump with the membrane charge.

As explained in Section \ref{sec:GSS_MTF}, we then arrive at an on-shell Lagrangian of the form
\be
\label{ExtObj_DWM_Sbulk}
S_{\rm bulk} = \int \d^4 x\, \call_{\rm bulk} = \int \d^4 x\,  \left[ -K_{a\bar b} \del_z \varphi^a \del_z \bar\varphi^{\bar b} - V(\varphi,\bar\varphi)\right]\;.
\ee
Importantly, due to the presence of the membrane, the potential on the two sides of the membrane is \emph{different}, being it specified by different constants, that is 
\be
\label{ExtObj_DWM_Vbulk}
\begin{aligned}
	&V(\varphi,\bar\varphi) =   \Theta(-z) V_- (\varphi,\bar\varphi) + \Theta(z) V_+ (\varphi,\bar\varphi) \;,
\end{aligned}
\ee
with
\be
\begin{aligned}
	V_- (\varphi,\bar\varphi) &=  K^{\bar b a}   (N_A\calv^A_a(\varphi) + \hat W_a(\varphi)) (N_B\bar\calv^B_{\bar b}(\bar\varphi) +\bar{\hat W}_{\bar b}(\bar\varphi) )  \,,
	\\
	V_+ (\varphi,\bar\varphi) &= K^{\bar b a}   [(N_A+q_A)\calv^A_a(\varphi) + \hat W_a(\varphi)] [(N_B+q_B)\bar\calv^B_{\bar b}(\bar\varphi) +\bar{\hat W}_{\bar b}(\bar\varphi) ]   \,.
\end{aligned}
\ee
The potential  \eqref{ExtObj_DWM_Vbulk} can be seen as originating from the superpotential
\be
\label{ExtObj_DWM_W}
W(\varphi) = \Theta(-z) W_- (\varphi) + \Theta(z) W_+ (\varphi)
\ee
with the usual relation $V = K^{\bar b a}W_a \bar W_{\bar b}$, where $W_-$ and $W_+$ are, respectively, the \emph{on--shell superpotential} on the left and on the right of the membrane, which are partly dynamically generated by the gauge three-forms and given by
\be
\begin{aligned}
	W_-  (\varphi)\equiv N_A \calv^A(\varphi) + \hat W (\varphi) \qquad {\rm and} \qquad  W_+  (\varphi)\equiv (N_A+q_A) \calv^A(\varphi) + \hat W (\varphi)\,.
\end{aligned}
\ee
It is also convenient to define
\be
\label{ExtObj_DWM_DW}
\Delta W(\Phi) \equiv W_+(\Phi) - W_- (\Phi)= q_A \calv^A(\Phi)
\ee
so that, when the scalar fields are evaluated at $z=0$, $\calt_{\text{memb}} = 2 |\Delta W|$.

Generically, the classical vacua of the potentials $V_-(\varphi,\bar \varphi)$ and $V_+(\varphi,\bar \varphi)$ are different and we may study the domain wall solutions which interpolate between the set of vacua on the left of the membrane to those on the right. Let us then consider a vacuum $\varphi_-^a$ on the left of the membrane and $\varphi_+^a$ on the right. A domain wall is a solitonic solution of the field equations which interpolates between $\varphi_-^a$, which we may assume to be the asymptotic field configuration at $z\to-\infty$, and $\varphi_+^a$, which is reached at $z\to+\infty$; in between, the fields may nontrivially vary.

We are still interested in BPS--domain walls and the preservation of (part of) the bulk supersymmetry requires the supersymmetry variations of the fields to vanish over the domain wall solution. Setting to zero the variations of the chiralini as in \eqref{ExtObj_DW_susyvar} leads to the flow equation 
\be
\label{ExtObj_DWM_flow}
\del_z \varphi^a = \mp e^{\ii \vartheta}  K^{a\bar b} \bar W_{ \bar b}\;,
\ee
with the superpotential defined in \eqref{ExtObj_DWM_W}, provided that the supersymmetry parameters satisfy the projection condition
\be
\label{ExtObj_DWM_proj}
\epsilon_\alpha = \mp \ii e^{\ii \alpha}\sigma^3_{\alpha \dot \alpha} \bar\epsilon^{\dot \alpha} \;.
\ee
Consistency with the $\kappa$-symmetry of the membrane requires that, at $z=0$, the supersymmetry parameter is \emph{identified} with the $\kappa$-parameter:
\begin{important} 
\be
\label{ExtObj_DWM_kappa}
\epsilon_\alpha |_{z=0} \stackrel{!}{=} \kappa_\alpha  = -\ii \frac{q_A \calv^A(\varphi)}{|q_A \calv^A(\varphi)|} \Big|_{z=0} \sigma^3_{\alpha \dot \alpha} \bar\kappa^{\dot \alpha}  \;.
\ee
\end{important}
Setting $\vartheta \equiv \arg [q_A \calv^A(\varphi)] |_{z=0}$, the previous relation implies that
\be
\label{ExtObj_DWM_kappab}
e^{\ii \alpha} \stackrel{!}{=} \pm e^{\ii \vartheta}
\ee
needs to hold over the membrane worldvolume.
 
In the following sections we first compute, as we did for sourceless domain walls, the tension of the domain wall-plus-membrane system and later the central charge. Finally, we present a very simple example where we explicitly compute the domain wall solution where a membrane is present.

\subsection{Tension of a BPS-domain wall sourced by a single membrane}
\label{sec:ExtObj_DWM_T}

Even though a membrane is present, we may still define the tension of a domain wall enclosing the membrane as in \eqref{ExtObj_DW_T}, where the action to be set on-shell is the full \eqref{ExtObj_DWM_Stot}. Once we set the three-forms on-shell on the two sides of the membrane, in order to compute the tension, we can proceed as in Section \ref{sec:ExtObj_DW_T}. In fact, we may still rewrite the action in the BPS--form
\be
\label{ExtObj_DWM_Sd}
\begin{split}
	S_{\rm bos} &=  \int \d^3 x\,\d z\,\Bigg[-K_{a\bar b} \left(\del_z \varphi^a \pm  e^{\ii \beta} K^{\bar c a} \bar W_{\bar c}\right) \left( \del_z \bar\varphi^{\bar b} \pm e^{-\ii \beta} K^{\bar b d}  W_d \right) \Bigg]
	\\
	&\quad\,\pm 2 \int \d^3 x\,\d z\,  \frac{\d}{\d z} \Re \left(W e^{-\ii \beta}\right)- \int \d^3 x\, \d z\, \delta(z) \calt_{\text{memb}} \,.
\end{split}
\ee
where $\beta$ is an arbitrary real constant. and the superpotential is given in \eqref{ExtObj_DWM_W}. In order to perform the integration in the last line, let us notice that, from \eqref{ExtObj_DWM_W}, we easily get
\be
\frac{\d}{\d z} W = W_a \del_z \varphi^a + \Delta W \delta(z)
\ee
and then \eqref{ExtObj_DWM_Sd} gives
\be
\label{ExtObj_DWM_Se}
\begin{split}
	S_{\rm bos} &=     \int \d^3 x\,\d z\,\Bigg[-K_{a\bar b} \left(\del_z \varphi^a \pm e^{\ii \beta} K^{\bar c a} \bar W_{\bar c}\right) \left( \del_z \bar\varphi^{\bar b} \pm  e^{-\ii \beta} K^{\bar b d}  W_d \right) \Bigg]
	\\
	&\quad\,\pm 2 \int \d^3 x\,\Re \left[(W_{+\infty}-W_{-\infty} )e^{-\ii \beta}\right]
	\\
	&\quad\,- \int \d^3 x\, \d z\, \delta(z) \left[\calt_{\text{memb}} \mp 2 \Re (\Delta W e^{-\ii \beta}) \right]\,.
\end{split}
\ee
Furthermore, although the superpotential is discontinuous, the integral of motion \eqref{ExtObj_DW_Int} is present also here
\be 
\label{ExtObj_DWM_Int}
\frac{\d}{\d z} \Im (W e^{-\ii\alpha}) = 0 \qquad \Rightarrow \qquad \Im \left[e^{-\ii \alpha}(W_{+\infty}-W_{-\infty}) \right]=0\,.
\ee

The third line of \eqref{ExtObj_DWM_Se} includes the contributions localized over the membrane worldvolume. However, upon using the consistency condition \eqref{ExtObj_DWM_kappab} between bulk supersymmetry and $\kappa$--symmetry, they cancel out
\be
\calt_{\text{memb}} \mp 2 \Re (\Delta W e^{-\ii \beta}) |_{z=0} = \calt_{\text{memb}} - 2 \Re (\Delta W e^{-\ii \vartheta}) |_{z=0} = 0 
\ee

Hence, the on-shell action simply reads
\begin{important}
	\be
	\label{ExtObj_DWM_bound}
	\calt_{\rm DW} = 2\, |W_{+\infty}-W_{-\infty} | \qquad{\color{darkred}\small\text{\sf{Tension of a BPS domain wall sourced by a membrane}}}
	\ee
\end{important}
which is formally equivalent to \eqref{ExtObj_DWM_bound} with, however, the superpotential asymptotic values computed from \eqref{ExtObj_DWM_W}, in turn determined by the presence of the membrane.

\subsection{Central charge of a BPS-domain wall sourced by a membrane}
\label{sec:ExtObj_DWM_CG}

In the case where membranes are present,  the topological charges which modify the supersymmetry algebra have a twofold origin: first, they come from the very existence of the membrane, in analogy with \cite{deAzcarraga:1989mza} and \eqref{ExtObj_Memb_CG_finb}; moreover, another contribution come from the full domain-wall configuration interpolating among vacua, in analogy with \cite{Witten:1978mh,Cvetic:1992sf} and our \eqref{ExtObj_DW_CG_fin}. 

By combining \eqref{ExtObj_Memb_CG_finb} and \eqref{ExtObj_DW_CG_fin}, we immediately get the central charges for the whole configuration of a domain wall supported by a static membrane 
\be
\label{CG_fin_tot}
\begin{aligned}	
	M_{\alpha \beta}  &= -8\int \d x^1\d x^2\,\varepsilon_{\beta\delta} \sigma^{03}{}_\alpha{}^\delta\, (\bar W_{+\infty}-\bar W_{-\infty}) \,,
	\\
	\bar M^{\dot\alpha \dot \beta }&= -8\int \d x^1\d x^2\,  \bar\sigma^{03 \dot\alpha}{}_{\dot\gamma} \varepsilon^{\dot\gamma\dot\beta} (W_{+\infty}-W_{-\infty} ) \,,
\end{aligned}
\ee
with the superpotential defined as in \eqref{ExtObj_DWM_W}.

\subsection{A simple example}
\label{sec:ExtObj_DWM_Ex}

Consider a supersymmetric theory with just a single chiral superfield $\Phi$ and characterized by
\be
\label{ExtObj_DWM_Ex_KWa}
K (\Phi, \bar \Phi) = \Phi \bar \Phi\,,\qquad \qquad W(\Phi) = n \Phi - \frac12 \Phi^2\;,
\ee
with $n$ a real constant. This model exhibits just a \emph{single} supersymmetric vacuum, located at $\varphi = n$. Therefore, according to the discussion of Section \ref{sec:ExtObj_DW}, it is not possible to construct any nontrivial domain wall within this theory. 

However, in Section \ref{sec:GSS_MTF} we saw that the linear part of the superpotential in \eqref{ExtObj_DWM_Ex_KWa} can be seen as generated by a single three-form multiplet. Namely, we consider a dual three-form theore, described by the Lagrangian \eqref{GSS_MTF_L3f}, with just one single three-form multiplet $S$ (as \eqref{GSS_STF}), including a gauge three-form $A_3$ in its components:
\be
\label{ExtObj_DWM_Ex_KWS}
K (S, \bar S) = S \bar S\,,\qquad \qquad \hat W(S) = - \frac12 S^2\;.
\ee
Once the gauge three-form is integrated out, we reget the same theory as the one specified by \eqref{ExtObj_DWM_Ex_KWa}, where $n$ is regarded as an arbitrary integration constant. 

The true advantage of dealing with the theory in the form \eqref{ExtObj_DWM_Ex_KWS} is that now we can couple a membrane, charged with charge $q$ under the gauge three-form $A_3$. Let us assume, for simplicity, that the membrane is located at $z=0$ and frozen in its static ground state. Then, the membrane influences the equation of motion for the gauge three-forms as in \eqref{ExtObj_DWM_A3eom}, making the constant $n$ shift to $n+q$ once pass across $z=0$. In turn, this implies that the superpotential of the dual chiral multiplet theory is
\be
\label{ExtObj_DWM_Ex_W}
W(\Phi) = W_{\rm gen}(\Phi) - \frac12 \Phi^2\;,
\ee
where the dynamically generated part \emph{differs} on the two sides:
\be
W_{\rm gen}(\Phi)  = (n + q \Theta(z)) \Phi \,.
\ee
As a result, two different potentials on the two sides of the membrane originate, each one with its own suspersymmetric vacuum, determined by 
\be
\label{ExtObj_DWM_Ex_dW}
\begin{aligned}
	&\del_\Phi W_{-} \stackrel{!}{=} 0 \qquad \Rightarrow \qquad \Phi_- | = n\;,
	\\
	&\del_\Phi W_{+} \stackrel{!}{=} 0 \qquad \Rightarrow \qquad \Phi_+ | = n + q\;,
\end{aligned}
\ee
with $W_+$ and $W_-$ defined from \eqref{ExtObj_DWM_Ex_W} as
\be
W =: W_- \Theta(-z)+ W_+ \Theta(z)\;.
\ee
At the vacua, the superpotential is
\be
\label{ExtObj_DWM_Ex_Wvac}
W_{-\infty} = \frac{c^2}{2}\;,\qquad W_{+\infty} = \frac{(c+q)^2}{2}\;.
\ee
Now, domain walls can exist which interpolate between $\Phi_-$  and $\Phi_+$ from the left to the right of the membrane. 

\begin{figure}[h]
	\centering
	\includegraphics[width=8cm]{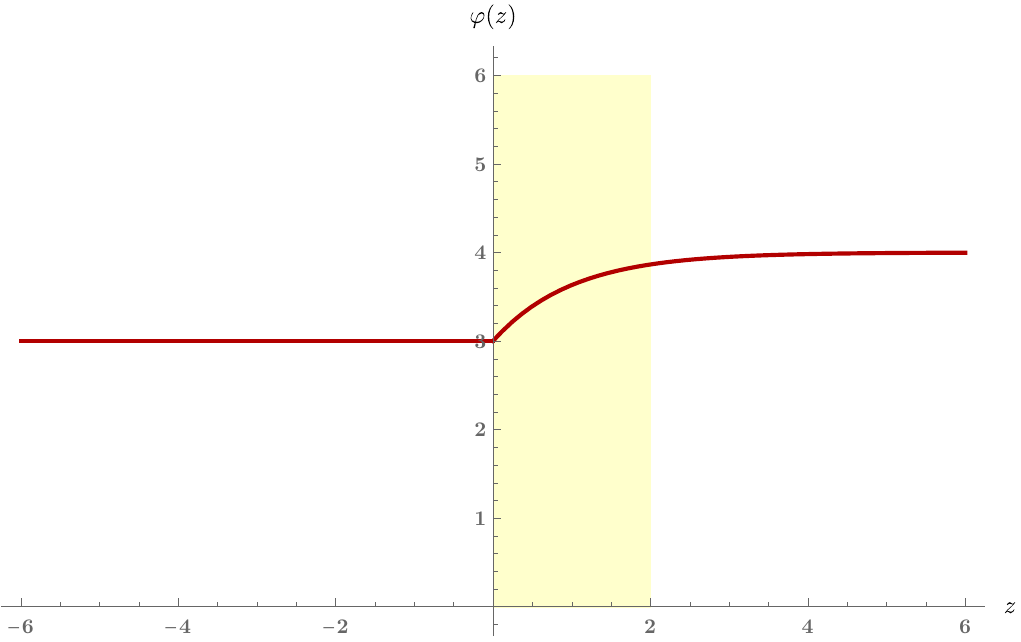}
	\caption{\footnotesize{The domain wall profile described by \eqref{ExtObj_DWM_Ex_varphisol}. The membrane is located at $z=0$: there, although $\varphi$ is continuous, is derivative is not. In yellow is highlighted the region where the scalar field varies.} }\label{fig:DWM}
\end{figure}

Let us now determine $\alpha$. The integral of motion \eqref{ExtObj_DWM_Int} implies that
\be
\Im (e^{-\ii\alpha} W_{+\infty}) \equiv C \equiv \Im (e^{-\ii\alpha} W_{-\infty}) 
\ee
with $C$ a fixed constant. Consistency with \eqref{ExtObj_DWM_Ex_Wvac} and positivity of the domain wall tension \eqref{ExtObj_DWM_bound} requires $C=0$ and $\alpha = 2 \pi k$, $k\in \mathbb{Z}$, for the upper sign choice or $\alpha = (2 k+1) \pi$ for the lower sign in \eqref{ExtObj_DWM_Se}. Then, the flow equations \eqref{ExtObj_DWM_flow} become
\be
\begin{cases}
	\del_z \varphi = - \bar \varphi + c  &{\text{for}}\quad z<0
	\\
	\del_z \varphi = - \bar \varphi + c + k &{\text{for}}\quad z>0
\end{cases}
\ee
Further assuming $\varphi$ to be real for simplicity, these are simply solved by
\if{}
\be
\begin{cases}
	\varphi (z)= n + A_- e^{-z}  &{\text{for}}\quad z<0
	\\
	\varphi  (z)=  n + q + A_+ e^{-z} &{\text{for}}\quad z>0
\end{cases}
\ee
$A_- = 0$ and $A_+ = -q$
\fi{}
\be
\label{ExtObj_DWM_Ex_varphisol}
\begin{cases}
	\varphi (z)= n   &{\text{for}}\quad z<0
	\\
	\varphi  (z)=  n + q (1- e^{-z}) &{\text{for}}\quad z>0
\end{cases}
\ee
where the integration constants are determined requiring $\varphi$ to be continuous at $z=0$. The wall profile is plotted in Fig.~\ref{fig:DWM}.

This example should have made concrete an important, remarkable physical fact that was drawn in the general discussion at the beginning of this section: the very existence of membranes open the possibilities of getting new vacua, eventually connected by domain walls as examined here.

\section{Strings in Supersymmetry}
\label{sec:ExtObj_Strings}

In Section~\ref{sec:GSS_Axions} we have seen how to construct generic supersymmetric Lagrangians containing some linear multiplets $L^\Lambda$, which play the role of super-field strengths for some gauge two-forms $\calb_2^\Lambda$. Within such a framework, we may then consider fundamental, structureless objects which electrically couple to $\calb_2^\Lambda$. These objects are \emph{fundamental strings}, of the kind which we already introduced in Section~\ref{sec:Intro_AM}: they are \emph{structureless} objects, namely of strictly codimension two. Here, however, we would like to take a step forward: we want to write down an action for strings, generically coupled to the $\calb_2^\Lambda$ gauge fields, in a manifestly supersymmetric way. The strategy which we follow is the same as the one we adopted for membranes in the Sections~\ref{sec:ExtObj_IntroMemb} and \ref{sec:ExtObj_MembSusy}. 

A fundamental string, whose worldsheet is parametrized by the coordinates $\zeta^i$ ($i=1,2$), spans a superspace hypersurface $\cals$ which is described by the superspace embedding
\be
\label{ExtObj_Str_Sws}
\zeta^i\quad\mapsto\quad \cals:\;{\frak z}^M(\xi)=\left(x^m(\zeta),\theta^\alpha(\zeta),\bar\theta_{\dot\alpha}(\zeta)\right). 
\ee
Consequently, we define the induced metric on the string as
\be
\label{ExtObj_Str_Sindh}
{\bm \gamma}_{ij} \equiv E_i^a \eta_{ab} E_j^b \,,
\ee
where we have introduced the pull--backs of the target superspace supervielbein
\be
E^a_i (\zeta) \equiv  \del_i {\frak z}^M(\zeta) E^a_M({\frak z}(\zeta)) \, .
\ee
Furthermore, it is necessary to promote the gauge two-forms $\calb_2^\Lambda$ to a super--gauge two-form ${\bf B}_2^\Lambda$ whose lowest bosonic component just coincides with $\calb_2^\Lambda$. As we will see shortly, the super--gauge two-form ${\bf B}_2^\Lambda$ are defined in terms of the linear multiplets $L^\Lambda$, exactly as ${\bf A}_3^A$ in \eqref{ExtObj_Memb_A3} are defined by the potentials $P^A$.

Equipped with these ingredients, we can introduce the supersymmetric action for a fundamental string:
\begin{important}
	\be
	\label{ExtObj_Str_S}
	S_{\rm string} = -\int_\cals\d^2\zeta\, |e_\Lambda  L^\Lambda|\sqrt{-\det {\bm \gamma}}  + e_\Lambda\int_\cals {\bf B}^\Lambda_2\,.
	\ee
\end{important}
The first part of the action \eqref{ExtObj_Str_S} encodes the dynamics of the string and defines the string tension
\be
\label{ExtObj_Str_Ts}
\calt_{\rm string} = |e_\Lambda  l^\Lambda|\,.
\ee
The second part of the action \eqref{ExtObj_Str_S}, that is the Wess-Zumino term, expresses the coupling of the gauge two-forms $\calb_2^\Lambda$ to the string with charges $e_\Lambda$.

As stressed for membranes, the  super-two-form ${\bf B}^\Lambda_2$ is defined in terms of closed super-field strength three-form ${\bf H}^\Lambda_3=\dwb {\bf B}_2^\Lambda$,
\be
\label{ExtObj_Str_H3}
\begin{aligned}
	{\bf H}^\Lambda_3= \dwb {\bf B}_2^\Lambda &= - 2{\ii} E^a \wedge E^\alpha \wedge \bar E^{\dot\alpha}  \sigma_{a\alpha\dot\alpha}{ L}^\Lambda 
	\\ 
	&\quad\, +  E^b\wedge E^a \wedge  E^\alpha
	\sigma_{ab\; \alpha}{}^{\beta} D_{\beta}{ L}^\Lambda +  E^b\wedge E^a \wedge   \bar E^{\dot\alpha}
	\bar\sigma_{ab}{}^{\dot\beta}{}_{\dot\alpha}\bar D_{\dot\beta}{ L}^\Lambda
	\\
	&\quad\,+\frac {1} {24}
	E^c \wedge E^b \wedge E^a \epsilon_{abcd}\, \bar{\sigma}{}^{d\dot{\alpha}\alpha}
	[D_\alpha, \bar D_{\dot\alpha}]{ L}^\Lambda
	\, ,
\end{aligned}
\ee
and it is the unique closed super-three-form which can be constructed from the linear multiplets \cite{Bandos:2003zk}. It can be immediately seen that its lowest bosonic component is 
\be
{\bf H}_3^\Lambda | = \calh_3^\Lambda = \d \calb_2^\Lambda
\ee
so that \eqref{ExtObj_Str_S} reduces to \eqref{Intro_AMS_S} once we restrict to the bosonic components.

As for the membrane action \eqref{ExtObj_FrM_Smembr} and \eqref{ExtObj_Memb_S}, the string action \eqref{ExtObj_Str_S} is invariant under worldsheet reparametrizations by construction. Moreover, it enjoys a local, fermionic $\kappa$-symmetry specified by the parameters $\kappa_\alpha (\zeta)$ and $\bar\kappa^{\dot\alpha}(\zeta)$. As is demonstrated in Section~\ref{sec:ExtObj_Strings_kappa}, requiring
\be
\label{ExtObj_Str_kappa_b}
\delta_\kappa S_{\rm string} \stackrel{!}{=} 0
\ee
implies that the fermionic parameters $\kappa_\alpha (\zeta)$ and $\bar\kappa^{\dot\alpha}(\zeta)$ obey the projection conditions
\be
\label{ExtObj_Str_kappa}
\kappa_\alpha=-\frac{e_\Lambda L^\Lambda}{|e_\Lambda L^\Lambda|}\Gamma_{\alpha}{}^\beta\kappa_\beta\,,\qquad \bar \kappa^{\dot \alpha}=-\frac{e_\Lambda L^\Lambda}{|e_\Lambda L^\Lambda|}\bar\Gamma^{\dot\alpha}{}_{\dot\beta}\bar \kappa^{\dot\beta} \,,
\ee
where we have defined
\be
\label{ExtObj_Str_proj}
\Gamma_{\alpha}{}^\beta\equiv \frac{1}{\sqrt{-\det{\bm \gamma}}}\epsilon^{ij}E^a_iE^b_j(\sigma_{ab})_\alpha{}^\beta\,,\qquad \bar\Gamma^{\dot\alpha}{}_{\dot\beta}\equiv \frac{1}{\sqrt{-\det{\bm \gamma}}}\epsilon^{ij}E^a_iE^b_j(\bar\sigma_{ab})^{\dot\alpha}{}_{\dot\beta}\,.
\ee
satisfying
\be
\label{ExtObj_Str_proj_id}
\Gamma_{\alpha}{}^\gamma \Gamma_{\gamma}{}^\beta = \delta^\beta_\alpha\,,\qquad \bar \Gamma^{\dot \alpha}{}_{\dot\gamma} \bar \Gamma^{\dot\gamma}{}_{\dot \beta} = \delta_{\dot\beta}^{\dot\alpha}\,.
\ee

\subsection{Proof of $\kappa$--symmetry}
\label{sec:ExtObj_Strings_kappa}

In this section we provide a detailed demonstration of the $\kappa$-symmetry of the string action \eqref{ExtObj_Str_S}. We want to show that the variation of the string action  \eqref{ExtObj_Str_S} is zero under local fermionic transformations, specified by the parameters $\kappa_\alpha (\zeta)$ and $\bar\kappa^{\dot\alpha}(\zeta)$, provided that the local parameters satisfy \eqref{ExtObj_Str_kappa}. The action of such transformations on the super-vielbeins is given by \eqref{ExtObj_Memb_Dk_kappa} and \eqref{ExtObj_Memb_Dk_kappaE}. 

We split the variation of \eqref{ExtObj_Str_S} into two pieces, the variation of the Nambu-Goto part and that of the Wess-Zumino part
\be
\label{ExtObj_Str_Dk_dS}
\delta_\kappa S_{\rm string} = \delta_\kappa S_{\rm string,NG}+ \delta_\kappa S_{\rm string,WZ}\,.
\ee

The variation of the Nambu-Goto part $S_{\rm string,NG}$ can be easily computed by first rewriting the world-sheet measure as
\be
\label{ExtObj_Str_2vol}
\d^2\zeta  \sqrt{-{\bm \gamma}}= - \frac 1 2 {}^{*_2}\!E_a\wedge E^a\,, 
\ee
where we have defined the action of worldvolume Hodge duality operation ${}^{*_2}$ as
\be
\label{ExtObj_Str_*E}
{}^{*_2}\!E^M \equiv  \d\xi^i \sqrt{-{\bm \gamma}} \epsilon_{ij} \gamma^{jl} E_l^M
\; . \qquad 
\ee
Then, using \eqref{ExtObj_Memb_Dk_kappaE}, we immediately get
{\small\be\label{ExtObj_Str_vNG}
\begin{split}
	\delta S_{\text{string,NG}} &=  \int_{\cals} {}^{*_2}\!E_a\wedge \delta E^a\, |e_\Lambda L^\Lambda| -  \int_{\cals} \d^2\zeta \sqrt{-{\bm \gamma}}\,
	\frac {e_\Sigma L^\Sigma} {|e_\Sigma L^\Sigma|} e_\Lambda \left(\kappa^\alpha D_\alpha L^\Lambda + \bar \kappa_{\dot \alpha} \bar D^{\dot \alpha} L^\Lambda\right)
	\\
	&=  -2\ii \int_{\cals} {}^{*_2}\!E_a\wedge  E^\alpha \sigma^a_{\alpha\dot\alpha} \bar \kappa^{\dot \alpha}|e_\Lambda L^\Lambda|  + 2\ii \int_{\cals} {}^{*_2}\!E_a \wedge  \kappa^\alpha \sigma^a_{\alpha\dot\alpha} \bar E^{\dot \alpha} |e_\Lambda L^\Lambda|
	\\
	&\quad\,-  \int_{\cals} \d^2\zeta \sqrt{-{\bm \gamma}}\,
	\frac {e_\Sigma L^\Sigma} {|e_\Sigma L^\Sigma|} e_\Lambda \kappa^\alpha D_\alpha L^\Lambda -  \int_{\cals} \d^2\zeta \sqrt{-{\bm \gamma}}\,
	\frac {e_\Sigma L^\Sigma} {|e_\Sigma L^\Sigma|} e_\Lambda  \bar \kappa_{\dot \alpha} \bar D^{\dot \alpha} L^\Lambda
	\; ,
\end{split}
\ee}

The variation of the Wess-Zumino term is
\be\label{ExtObj_Str_vWZ} 
\delta S_{\text{string,WZ}} = e_\Lambda \int_{\cals} \delta_\kappa {\bf B}_2^\Lambda = e_\Lambda \int_{\cals}(i_{\kappa} \dwb {\bf B}_2^\Lambda+ \dwb \, i_{\kappa} {\bf B}_2^\Lambda)=e_\Lambda \int_{\cals} i_{\kappa} {\bf H}_3^\Lambda\,. 
\ee
where, in order to get the second equality, we have assumed that the string has no boundary. Using \eqref{ExtObj_Memb_Dk_kappa}, from \eqref{ExtObj_Str_H3}, we get
\be
\label{ExtObj_Str_iH3} 
\begin{split}
	 i_{\kappa} {\bf H}_3^\Lambda= \;  & -2\ii E^a \wedge E^\alpha  (\sigma_{a}\bar \kappa)_{\alpha} L^\Lambda + 2 \ii E^a \wedge  (\kappa \sigma_a)_{\dot{\alpha}}  \bar E^{\dot\alpha}  L^\Lambda \\  
	& - E^b\wedge E^a  (\kappa \sigma_{ab})^\beta D_\beta L^\Lambda  + E^b\wedge E^a  \bar D_{\dot \beta} L^\Lambda ( \bar \sigma_{ab} \bar \kappa)^{\dot \beta}
	\; .
\end{split}
\ee
By using
\be
\label{ExtObj_Str_Use_Id_a} 
\begin{aligned}
	&E_a\wedge E^{\alpha}\, \sigma^a_{\alpha\dot \alpha } = - {}^{*_2}\!E_a\wedge E^{\beta} \sigma_{\beta \dot\alpha}^a \bar\Gamma^{\dot\alpha}{}_{\dot\alpha}\; ,
	\\
	&E^a\wedge \bar E_{\dot \alpha}\, \bar\sigma_a^{\dot \alpha \alpha} = {}^{*_2}\!E^a\wedge \bar E_{\dot \alpha} \bar\sigma^{\dot\alpha \beta}_a \Gamma_\beta{}^\alpha\; ,
\end{aligned}
\ee
the first line may be written as
\be
\label{ExtObj_Str_vWZ1} 
\begin{split}
	&2\ii{}^{*_2}\!E^a\wedge E^\alpha (\sigma_a\,{\bar\Gamma} \bar \kappa)_{\alpha}{} L^\Lambda -2 \ii {}^{*_2}\!E^a\wedge  \bar E_{\dot \alpha} ( \bar \sigma_a\,{\Gamma} \kappa)^{\dot\alpha} L^\Lambda
	\; ;
\end{split}
\ee
indeed, by means of the identities
\be
\label{ExtObj_Str_Use_Idb}
\begin{aligned}
	&\d^2\zeta \sqrt{-{\bm \gamma}}\, {\Gamma} = 
	E^a \wedge E^b \sigma_{ab} \, , 
	\\
	&\d^2\zeta \sqrt{-{\bm \gamma}}\, {\bar \Gamma} = 
	- E^a \wedge E^b \bar \sigma_{ab} \, .
\end{aligned}
\ee
the second line becomes
\be
\label{ExtObj_Str_vWZ2} 
\begin{split}
	&    -\d^2 \zeta\, \sqrt{-{\bm \gamma}}\,  (\kappa \Gamma)^\beta D_\beta - \d^2 \zeta\, \sqrt{-{\bm \gamma}}\,  (\bar \kappa \Gamma)_{\dot \beta} \bar D^{\dot \beta} L^\Lambda 
	\; .
\end{split}
\ee
Hence
\be
\label{ExtObj_Str_vWZb} 
\begin{split}
	\delta S_{\text{string,WZ}} &=   2\ii e_\Lambda \int_\cals{}^{*_2}\!E^a\wedge E^\alpha (\sigma_a\,{\bar\Gamma} \bar \kappa)_{\alpha}{} L^\Lambda -2 \ii e_\Lambda \int_\cals {}^{*_2}\!E^a\wedge  E_{\dot \alpha} ( \bar \sigma_a\,{\Gamma} \kappa)^{\dot\beta} L^\Lambda
	\\
	&\quad\, -e_\Lambda \int_\cals \d^2 \zeta\, \sqrt{-{\bm \gamma}}\,  (\kappa \Gamma)^\beta D_\beta L^\Lambda -e_\Lambda \int_\cals \d^2 \zeta\, \sqrt{-{\bm \gamma}}\,  (\bar \kappa \bar \Gamma)_{\dot \beta} \bar D^{\dot \beta} L^\Lambda 
\end{split}
\ee
Then, it is immediate to see that \eqref{ExtObj_Str_vWZb} cancels \eqref{ExtObj_Str_vNG} provided that
\be
\label{ExtObj_Str_vkappa}
\kappa_\alpha=-\frac{e_\Lambda L}{|e_\Lambda L|}\Gamma_{\alpha}{}^\beta\kappa_\beta\,,\qquad \bar \kappa^{\dot \alpha}=-\frac{e_\Lambda L}{|e_\Lambda L|}\bar\Gamma^{\dot\alpha}{}_{\dot\beta}\bar \kappa^{\dot\beta} \,.
\ee

\subsection{Membranes ending on Strings}
\label{sec:ExtObj_Strings_MS}

In Section \ref{sec:ExtObj_MembSusy} we just considered membranes without any boundary: these could be, for instance, membranes which are infinitely stretched along three spacetime direction, as hyperplanes, or closed membranes. This assumption was crucial, in Section \ref{sec:ExtObj_MembSusy_kappa}, to demonstrate the $\kappa$-symmetry of the membrane action \eqref{ExtObj_Memb_S} (see, in particular, \eqref{ExtObj_Memb_vWZ} and \eqref{ExtObj_Memb_vWZhyp}, where this hypothesis enters). Let us now re-examine the issue, assuming that the membrane ends on a string, that is
\begin{figure}[t]
	\centering
	\includegraphics[width=5cm]{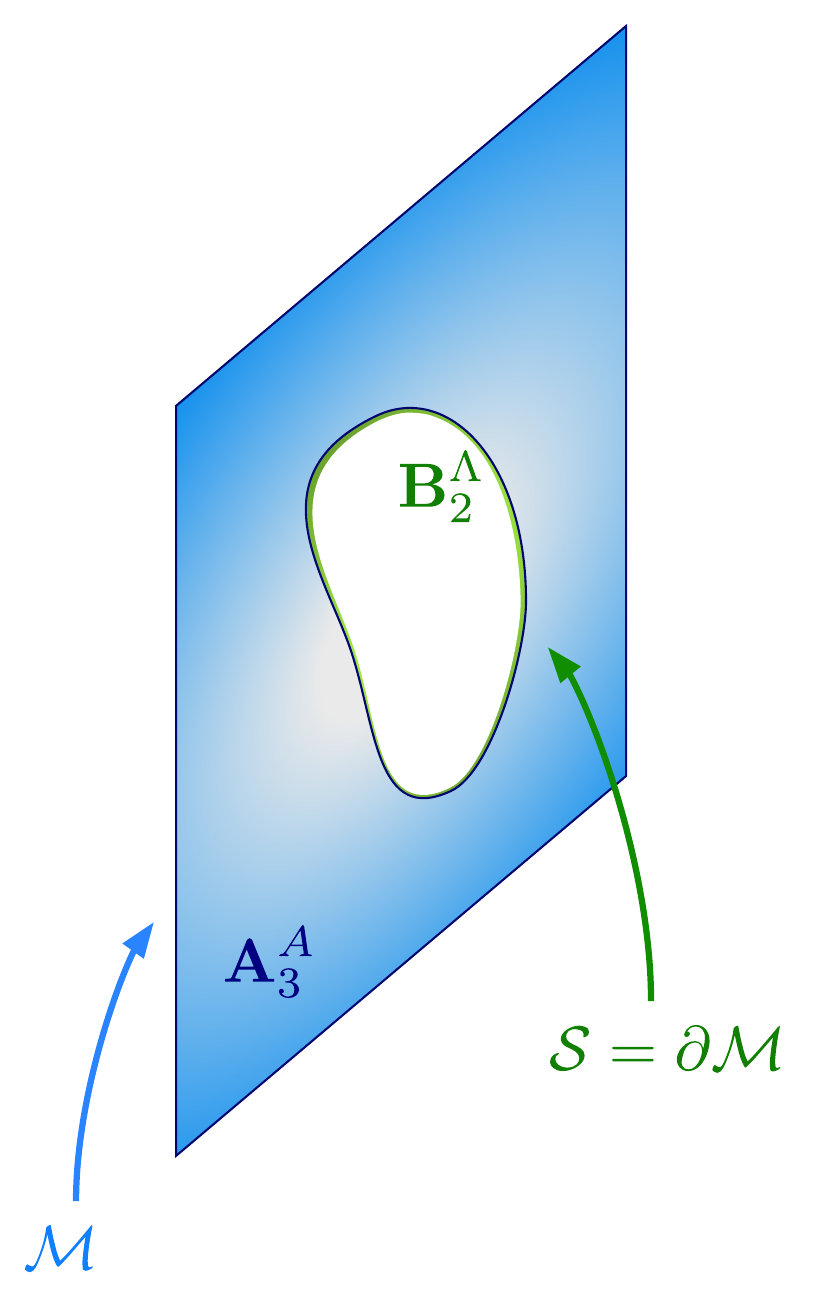}
	\caption{\footnotesize{A membrane, coupled to the gauge three-forms ${\bf A}_3^A$, whose boundary is a BPS--string, coupled to the gauge two-forms ${\bf B}_2^\Lambda$.} }\label{fig:SMemb}
\end{figure}	
\be
\label{ExtObj_SM_dMS}
\del \calm = \cals
\ee
as depicted in Fig.~\ref{fig:SMemb}. Then, even though we assume the $\kappa$-symmetry parameter to satisfy \eqref{ExtObj_Memb_kappa}, the membrane action \eqref{ExtObj_Memb_S} is \emph{not} $\kappa$-symmetric, due to the boundary term 
\be
\label{ExtObj_SM_vmemb}
\delta S_{\text{memb}}|_{\kappa} =  q_A \int_{\del\calm} i_\kappa {\bf A}_3^A\,.
\ee
This lack of variation clearly prevents us to consider such membranes as BPS-objects. However, this is not the end of the story. Assume that the boundary is a BPS--string described by an action of the same form as \eqref{ExtObj_Str_S}. The full variation under $\kappa$--transformations at the junction is then
\be
\label{ExtObj_SM_vtot}
(\delta S_{\text{memb}} + \delta S_{\text{string}})|_{\kappa} = \delta S_{\text{string,NG}}|_{\kappa} + e_\Lambda \int_{\cals} i_{\kappa} {\bf H}_3^\Lambda +  q_A \int_{\cals} i_\kappa {\bf A}_3^A
\ee
with $\kappa$ obeying \eqref{ExtObj_Memb_kappa}. It is therefore reasonable to try to cancel the last membrane contribution by appropriately modifying the string Nambu-Goto term. In order to see what such a modification has to be, rewrite $q_A = e_\Lambda c_A^\Lambda$, so that \eqref{ExtObj_SM_vtot} becomes
\be
\label{ExtObj_SM_vtotb}
(\delta S_{\text{memb}} + \delta S_{\text{string}})|_{\kappa} =  \delta S_{\text{string,NG}}|_{\kappa} + e_\Lambda \int_{\cals} i_{\kappa} \left({\bf H}_3^\Lambda + c_A^\Lambda {\bf A}_3^\Lambda \right)\,.
\ee
With respect to \eqref{ExtObj_Str_vWZ} that we encountered in proving the $\kappa$-symmetry of BPS--strings, here the field strengths ${\bf H}_3^\Lambda$ are \emph{gauged} by some combinations of the three-form potentials coupling to the membrane. A direct inspection of \eqref{ExtObj_SM_vtotb}, recalling the definitions of super-forms \eqref{ExtObj_Memb_A3} and \eqref{ExtObj_Str_H3}, shows that the variation \eqref{ExtObj_SM_vtotb} is zero provided that we modify the Nambu-Goto term of the BPS-string as
\be
S_{\text{string,NG}}= -\int_\cals\d^2\zeta\, |e_\Lambda  \hat L^\Lambda|\sqrt{-\det {\bm \gamma}} \,,
\ee
where the old linear multiplets $L^\Lambda$ are replaced by their gauged versions
\be
\label{ExtObj_Str_Lgauged}
\hat L^\Lambda \equiv L^\Lambda + c_A^\Lambda P^A\,,
\ee
and if the parameters $\kappa$, beside satisfying \eqref{ExtObj_Memb_kappa}, \emph{also} obey the projection conditions \eqref{ExtObj_Str_kappa}. The projections \eqref{ExtObj_Memb_kappa} and \eqref{ExtObj_Str_kappa} are, in general, two independent conditions that the spinor $\kappa$ has to satisfy. These both halves the number of the independent parameters, making the junction a $\frac14$--BPS object.

\begin{summary}
	To summarize, the full supersymmetric action of a membrane, coupled to the bulk scalars $s^A$ and charged under the gauge three-forms $A_3^A$, whose boundary is a string, charged under the gauge two forms $\calb_2^\Lambda$, is
	\be
	\label{ExtObj_SM_junction}
	\begin{split}
		S_{\text{BPS-junction}} &=-2\int_\calm \d^3\xi\,|e_\Lambda c_A^\Lambda S^A|\sqrt{-\det {\bf h}}+e_\Lambda c_A^\Lambda\int_\calm{\bf A}^A_3 +
		\\
		&\quad\,-\int_\cals\d^2\zeta\, |e_\Lambda  \hat L^\Lambda|\sqrt{-\det {\bm \gamma}}  + e_\Lambda\int_\cals {\bf B}^\Lambda_2\;,
	\end{split}
	\ee
	with the superfields $S^A$ and $\hat L^\Lambda$ defined in \eqref{ExtObj_Memb_SA} and \eqref{ExtObj_Str_Lgauged}.
\end{summary}

\section{Spacetime filling 3-branes}
\label{sec:ExtObj_3-branes}

Finally, to complete the hierarchy of four-dimensional BPS-objects, let us consider 3-branes. For the moment, let us assume that these objects fill the whole spacetime with their worldvolume $\calw$ and do not have boundaries. We should expect that 3-branes couple to some gauge four-forms $C_4^I$. As discussed in Section \ref{sec:GSS_4FM}, the variant chiral superfields \eqref{ExtObj_Memb_F4}  are those which properly contain (real) gauge four-forms in their highest components and we may use them to construct the super-four-form potentials
{\small \be
\begin{split}
	\label{ExtObj_3B_C4}   
	{\bf C}^I_{4}   &=   E^b\wedge E^a \wedge \bar E^{\dot\alpha} \wedge \bar  E^{\dot\beta }\bar{\sigma}_{ab\; \dot{\alpha}\dot{\beta}} \Gamma^I + E^b\wedge E^a \wedge E^\alpha \wedge E^\beta \sigma_{ab\; \alpha\beta}\bar{ \Gamma}^I \
	\\ 
	&\quad\, -\frac{1}6  E^c\wedge E^b\wedge E^a \wedge \bar E^{\dot\alpha} \epsilon_{abcd} \sigma^d_{\alpha\dot\alpha} D^{\alpha}{ \Gamma^I}  +\frac{1}6 E^c\wedge E^b\wedge E^a \wedge E^\alpha \epsilon_{abcd} \sigma^d_{\alpha\dot\alpha} \bar{D}^{\dot\alpha}\bar{ \Gamma}^I 
	\\
	&\quad\,  +\frac{\ii}{96} E^{d} \wedge E^c \wedge E^b \wedge E^a \epsilon_{abcd} \left(D^2 \Gamma^I
	-\bar{D}^2
	\bar{ \Gamma}^I \right) 
	\,.\qquad
\end{split}
\ee}
Its lowest bosonic component can be immediately computed from \eqref{GSS_4f_D} and is 
\be
\label{ExtObj_3B_C4c} 
\begin{split}
	{\bf C}^I_{4} | &= \frac1{4!} (2\, \Im f_\Gamma^I) \varepsilon_{mnpq}\d x^m \wedge \d x^n \wedge \d x^p \wedge \d x^q 
	\\
	&= \frac{1}{4!} C^I_{mnpq} \d x^m \wedge \d x^n \wedge \d x^p \wedge \d x^q = C^I_4 
\end{split}  
\ee
This is formally similar to the super-field strength \eqref{ExtObj_Memb_F4} of three-form potentials. As such, also ${\bf C}^I_{4}$ is closed
\be
\dwb {\bf C}^I_{4} = 0\,.
\ee

We can immediately write down a BPS-action for 3-branes as
\begin{important}
	\be
	\label{ExtObj_3B_S}
	S_{\text{3-brane}}=\mu_I\int_\calw {\bf C}^I_4
	\ee
\end{important}
where we have not included \emph{any} Nambu-Goto term. In fact, owing to the closure of ${\bf C}_4^I$ and assuming that the 3-brane has no boundary, the action \eqref{ExtObj_3B_S} is already $\kappa$--symmetric, the proof of $\kappa$--symmetry being trivial 
\be
\label{ExtObj_3B_Skappa}
\delta_\kappa S_{\text{3-brane}} =  \mu_I\int_\calw\left( i_\kappa \dwb {\bf C}^I_4 + \dwb i_\kappa  {\bf C}^I_4\right) = 0\,,
\ee
with no projection condition over the parameters $\kappa$. In other words, we may directly choose $\kappa$ to be the spacetime supersymmetry parameter and the 3-brane worldvolume still enjoys the full $\caln=1$ supersymmetry.

\subsection{3-branes ending on membranes}

\begin{figure}[t]
	\centering
	\includegraphics[width=7cm]{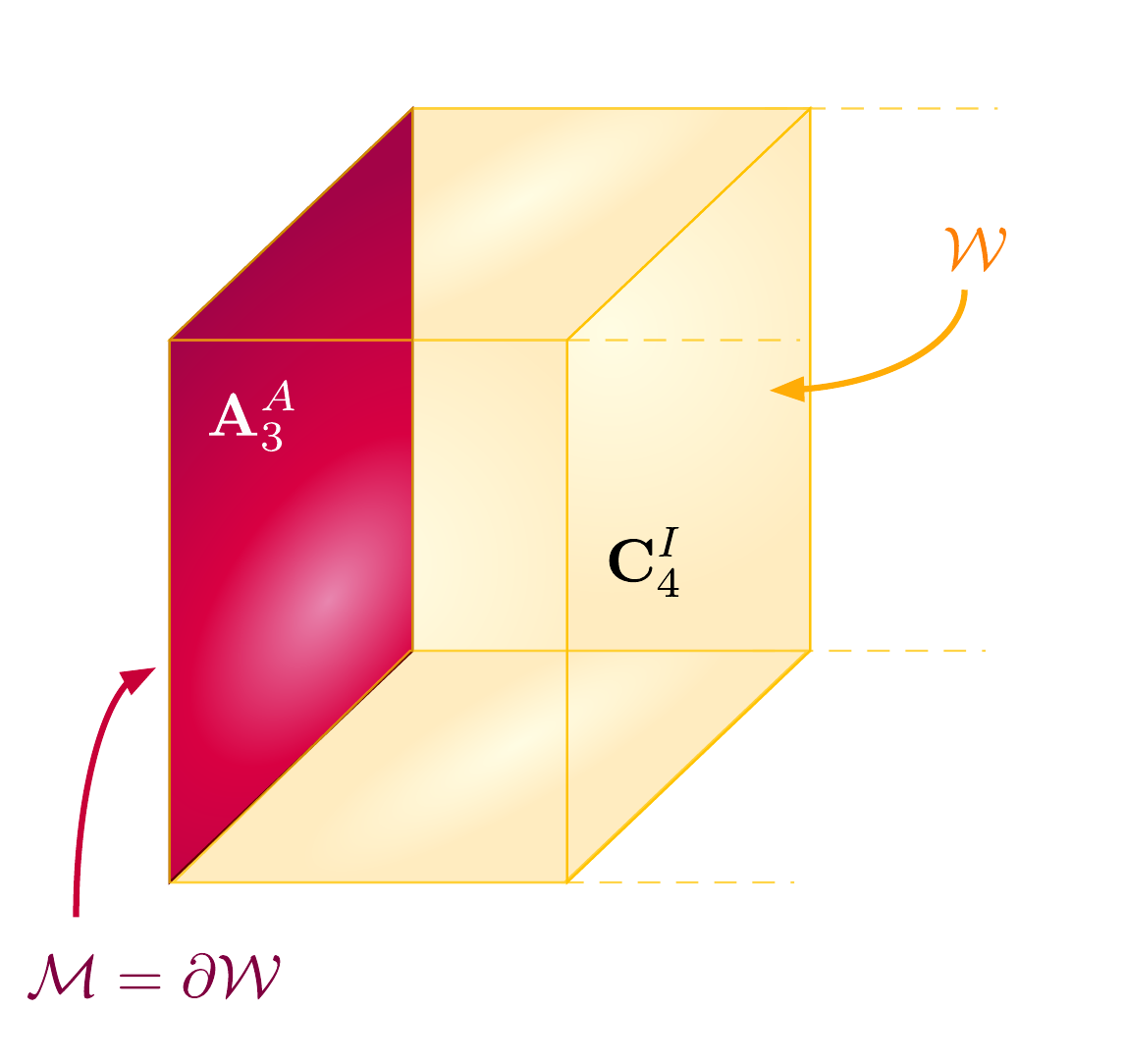}
	\caption{\footnotesize{A 3-brane, coupled to the gauge four-forms ${\bf C}_4^I$, whose boundary is a BPS--membrane, coupled to the gauge three-forms ${\bf A}_3^A$.} }\label{fig:MembD3}
\end{figure}

The (trivial) proof of $\kappa$-symmetry for 3-branes in \eqref{ExtObj_3B_Skappa} relies on the fact that 3-branes do not have any boundary. However, let us now assume that this is not the case and, rather, that the 3-brane ends on a BPS--membrane, that is $\del \calw = \calm$ (see Fig.~\ref{fig:MembD3}). Then, the variation under $\kappa$-transformations of the the 3-brane action \eqref{ExtObj_3B_Skappa} is no more zero, due to the boundary contribution
\be
\label{ExtObj_M3B_vmemb}
\delta_\kappa S_{\text{3-brane}} = - \mu_I \int_{\del\calw} i_\kappa  {\bf C}^I_4 =  - \mu_I \int_{\calm} i_\kappa  {\bf C}^I_4\,.
\ee
Therefore, over the membrane worldvolume, the full variation of the action is
\be
\label{ExtObj_M3B_vtot}
 \delta S_{\text{memb}} + \delta S_{\text{3-brane}}= \delta S_{\text{memb,NG}}|_{\kappa}+ q_A \int_{\calm} i_{\kappa} {\bf F}_4^A -  \mu_I \int_{\calm} i_\kappa  {\bf C}^I_4\,.
\ee
We may hope to cancel the last, 3-brane boundary contribution by conveniently modifying the Nambu-Goto part of the membrane action. In order to achieve this, let us decompose $\mu_I = q_A Q_I^A $, so that the variation reads \eqref{ExtObj_M3B_vtot}
\be
\label{ExtObj_M3B_vtotb}
\delta S_{\text{memb}} + \delta S_{\text{3-brane}} =  \delta S_{\text{memb,NG}}|_{\kappa} - q_A \int_{\calm} i_{\kappa} \left( - {\bf F}_4^A + Q^A_I {\bf C}_4^I \right)
\ee
Remarkably, in the last \emph{modified} Wess-Zumino term appear the gauged four-forms
\be
- {\bf F}_4^A| + Q^A_I {\bf C}_4^I | = \d A_3^A + Q_A^I C_4^I
\ee
Recalling the definitions \eqref{ExtObj_Memb_F4} and \eqref{ExtObj_3B_C4}, from the proof of membrane $\kappa$-symmetry that we gave in Section \ref{sec:ExtObj_MembSusy_kappa}, it is immediate to see that \eqref{ExtObj_M3B_vtotb} vanishes provided that the $\kappa$--parameter does obey \eqref{ExtObj_Memb_kappa} \emph{and} the Nambu-Goto part of the membrane action is modified as
\be
S_{\text{memb,NG}}= -2 \int_\calm\d^3\xi\, |q_A  \hat S^A|\sqrt{-\det {\bf h}} 
\ee
where the chiral three-form multiplets $S^A$ are substituted by their gauged versions
\be
\label{ExtObj_M3B_SAgauged}
\hat S^A  \equiv S^A - \ii Q^A_I \Gamma^I\,.
\ee

\begin{summary}
	To summarize, the full supersymmetric action of a 3-brane whose boundary is a membrane, coupled to the bulk scalars $s^A$ and charged under the gauge three-forms $A_3^A$, is
	\be
	\label{ExtObj_M3B_region}
	\begin{split}
		S_{\text{BPS-region}} &=-2\int_\calm \d^3\xi\,|q_A \hat S^A|\sqrt{-\det {\bf h}}+q_A\int_\calm{\bf A}^A_3 + q_A Q^A_I\int_\calw {\bf C}^I_4
	\end{split}
	\ee
	with the superfields $\hat S^A$ defined in \eqref{ExtObj_M3B_SAgauged}.
\end{summary}

Moreover, the action \eqref{ExtObj_M3B_region} is invariant under the combined gauge transformation:
\begin{subequations}\label{ExtObj_M3B_CombGauge}
	\begin{align}
	{\bf C_4}^I&\quad\rightarrow\quad {\bf C_4}^I+\d {\bf \Lambda}^I_3\,,\\
	{\bf A}^A_3&\quad\rightarrow\quad {\bf A}^A_3-q_AQ^A_I{\bf \Lambda}^I_3\,,
	\end{align}
\end{subequations}
where, in analogy with \eqref{ExtObj_Str_H3}, ${\bf \Lambda}^I_3$ is defined as
\be\label{s3gaugeb}
\begin{aligned}
	{\bf \Lambda}^I_3=&\,  { -}2 {\ii} E^a \wedge E^\alpha \wedge \bar E^{\dot\alpha}  \sigma_{a\alpha\dot\alpha}\Xi^I 
	 \\ &+  E^b\wedge E^a \wedge  E^\alpha
	\sigma_{ab\; \alpha}{}^{\beta}{D}_{\beta}\Xi^I+ E^b\wedge E^a \wedge  \bar E^{\dot\alpha}
	\bar\sigma_{ab}{}^{\dot\beta}{}_{\dot\alpha}\bar{D}_{\dot\beta}\Xi^I
	\\&+\frac {1} {24} 
	E^c \wedge E^b \wedge E^a \epsilon_{abcd} \,\bar{\sigma}{}^{d\dot{\alpha}\alpha}
	[D_\alpha, \bar{D}_{\dot\alpha}]\Xi^I 
	\, . \qquad
\end{aligned}
\ee
Indeed, \eqref{ExtObj_M3B_CombGauge} provides the supersymmetric promotion of the bosonic gauge transformation foreseen in \eqref{Intro_Con_lin3gauge}.
\section{A hierarchy of objects, a hierarchy of gaugings}
\label{sec:ExtObj_Summary}

We conclude this chapter by providing the complete actions coupling the bulk theories examined in Chapter~\ref{chapter:GSS} with the extended objects just introduced. We will outline, for each of the cases examined, the physical consequences, for the bulk theory, of the introduction of the extended objects.

\begin{centerbox}
	Three-forms coupled to membranes
\end{centerbox}

\begin{figure}[h]
	\centering
	\includegraphics[width=8cm]{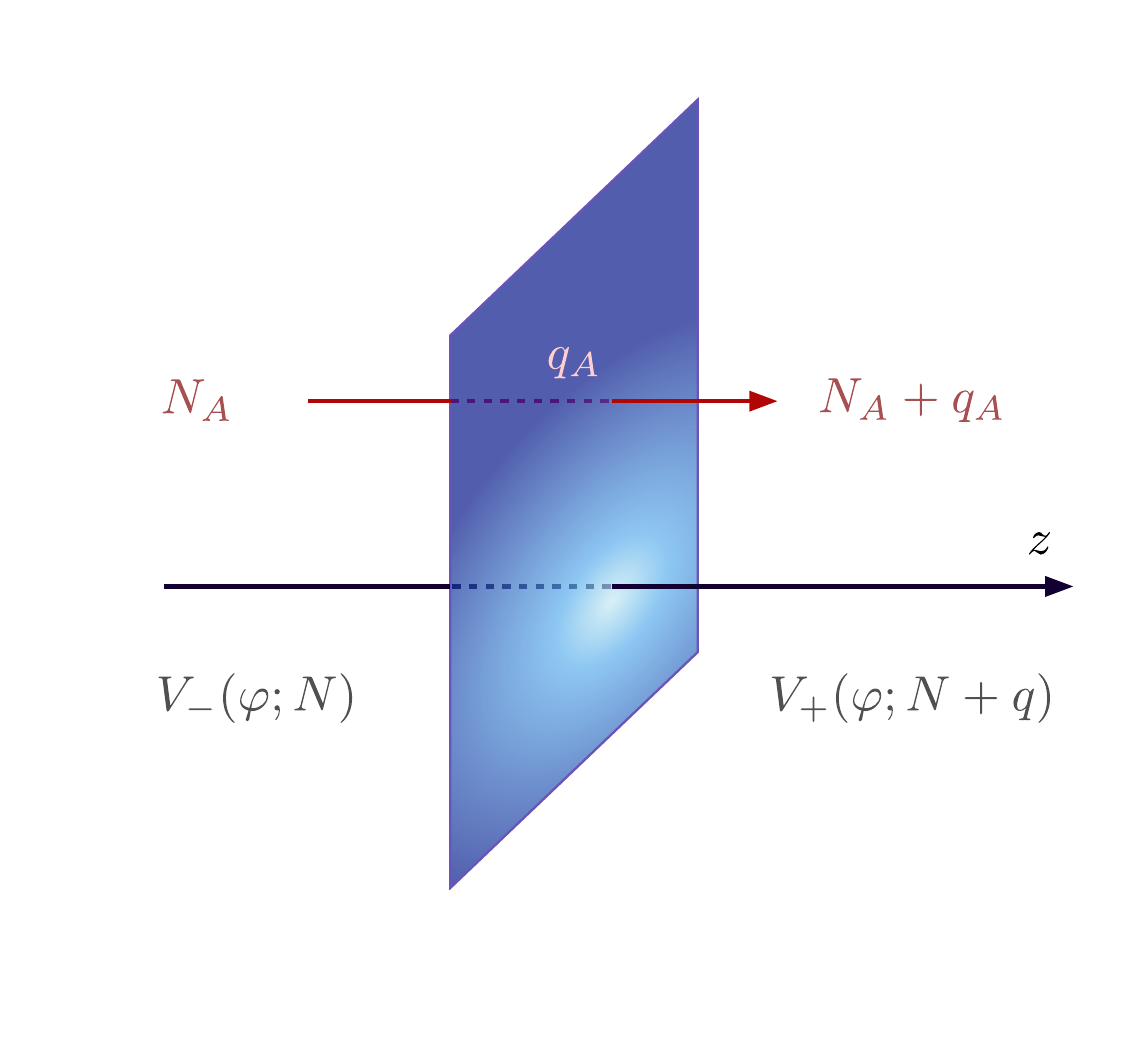}
	\caption{\footnotesize{The presence of a membrane of charges $q_A$ under the gauge three-forms $A_3^A$ influences the potential, for the constant $N_A$ shift making the potential change its shape on the two sides of the membrane.} }\label{fig:MembV}
\end{figure}

The complete action describing the coupling of a single membrane (which we assume to be localized at $z=0$) to gauge three-form is obtained combining \eqref{GSS_MTF_L3fComp} with \eqref{ExtObj_Memb_S}. We get 
\be
\label{ExtObj_Sum_Smem}
\begin{aligned}
S |_{\rm bos} &= - \int_{\Sigma}  K_{a\bar b} \d \varphi^a \wedge *\d \bar\varphi^{\bar b}
\\
&\quad\, -\int_\Sigma \Big[\frac12 T_{AB}F^A_4*\!F^B_4+ T_{AB}\Upsilon^AF^B_4+\Big(
\hat V-\frac12 T_{AB}\Upsilon^A \Upsilon^B\Big)*\!1\Big] 
\\
&\quad\,+\int_{\del\Sigma}T_{AB}(* F^A_4+\Upsilon^A)A^B_3
\\
&\quad\,-2\int_\calm \d^3\xi\,|q_A \calv^A(\varphi)|\sqrt{-\det {\bf h}}+q_A\int_\calm A_3^A\;,
\end{aligned}
\ee
where $\Sigma$ denotes the four-dimensional spacetime. Integrating out the gauge three-forms leads to the equations of motion
\be
\label{ExtObj_Sum_F4eom}
T_{AB}(*\!{F}^B_4+\Upsilon^B)= -N_A - q_A \Theta (z)\;,
\ee
with  $N_A$ real constants. The resulting on-shell theory is the ordinary theory
\be
\label{ExtObj_Sum_Sbulk}
S |_{\rm bos}  =  \int  \left[ -K_{a\bar b} \d \varphi^a \wedge *\d \bar\varphi^{\bar b}  - V(\varphi,\bar\varphi) *1\right]\;.
\ee
where, however, the potential is different on the two sides of the membrane
\be
\label{ExtObj_Sum_Vbulk}
\begin{aligned}
	&V(\varphi,\bar\varphi) =   \Theta(-z) V_- (\varphi,\bar\varphi) + \Theta(z) V_+ (\varphi,\bar\varphi) 
\end{aligned}
\ee
with
\be
\begin{aligned}
	V_- (\varphi,\bar\varphi) &=  K^{\bar b a}   (N_A\calv^A_a(\varphi) + \hat W_a(\varphi)) (N_B\bar\calv^B_{\bar b}(\bar\varphi) +\bar{\hat W}_{\bar b}(\bar\varphi) )  \,,
	\\
	V_+ (\varphi,\bar\varphi) &= K^{\bar b a}   [(N_A+q_A)\calv^A_a(\varphi) + \hat W_a(\varphi)] [(N_B+q_B)\bar\calv^B_{\bar b}(\bar\varphi) +\bar{\hat W}_{\bar b}(\bar\varphi) ]   \,.
\end{aligned}
\ee
Therefore, the role of a membrane is to make the constants $N_A$, which appear in the superpotential \eqref{GSS_MTF_Wgen}, jump of an amount given by the charge of the membrane, as depicted in Fig.~\ref{fig:MembV}.

\begin{centerbox}
	Three- and four-forms coupled to membranes and 3-branes
\end{centerbox}

Combining  \eqref{GSS_GMTF_L3fComp} with \eqref{ExtObj_M3B_region}  we can build up the action which couples, at once, gauge three- and four-forms to membranes and 3-branes, with the latter having the former as boundaries:
\be
\label{ExtObj_Sum_SM3B}
\begin{aligned}
	S |_{\rm bos} &= - \int_{\Sigma}  K_{a\bar b} \d \varphi^a \wedge *\d \bar\varphi^{\bar b}
	\\
	&\quad\, -\int_\Sigma \Big[\frac12 T_{AB} \hat F^A_4*\! \hat F^B_4+ T_{AB}\Upsilon^A\hat F^B_4+\Big(
	\hat V-\frac12 T_{AB}\Upsilon^A \Upsilon^B\Big)*\!1\Big] 
	\\
	&\quad\,+\int_{\del\Sigma}T_{AB}(* \hat F^A_4+\Upsilon^A)A^B_3 + \int_\Sigma  \calq_I^{\rm bg} C^I_4
	\\
	&\quad\,-2\int_\calm \d^3\xi\,|q_A \calv^A(\varphi)|\sqrt{-\det {\bf h}}+q_A\int_\calm A^A_3 +q_A Q^A_I\int_\calw C^I_4\;.
\end{aligned}
\ee
Again the dual, on-shell chiral theory has the very same form as \eqref{ExtObj_Sum_Sbulk} and, owing to the presence of membranes, the potential is generically different on the two sides of the membrane. The equations of motion for the gauge three-forms $A_3^A$ are as in \eqref{ExtObj_Sum_F4eom}; additionally, here, in order to get the ordinary chiral theory, we also need to integrate out the gauge four-forms $C_4^I$, obtaining 
\be
-T_{AB} (*\! \hat F_4^B + \Upsilon^B) Q_I^A + q_A Q_I^A-  \calq_I^{\rm bg}   = 0\,,
\ee
in all the spacetime. Indeed, the action \eqref{ExtObj_Sum_SM3B} provides the globally supersymmetric version of the model introduced in Section~\ref{sec:Intro_3branes}.
The role of 3-branes having membranes as boundaries is to `accompany' the changing of the tadpole cancellation condition due to the change of the constants $N_A$.

\begin{centerbox}
	Two-forms coupled to strings
\end{centerbox}

In the absence of a superpotential for the chiral fields $\Phi^a$, merging \eqref{GSS_AL_Dual1Lagbos} with \eqref{ExtObj_Str_S} we get the full action describing a BPS--string coupled to a bulk theory where both linear and chiral multiplets are present
\be
\label{ExtObj_Sum_SStr}
\begin{aligned}
	S |_{\rm bos} &= \int_{\Sigma} \left(- F_{a\bar b} \d \varphi^a \wedge *\d \bar\varphi^{\bar b} + \frac14 F_{\Lambda \Sigma}  \d l^\Lambda \wedge *\d l^\Sigma + \frac14 F_{\Lambda\Sigma} \calh^\Lambda_3 \wedge *\calh^\Sigma_3 \right)
	\\
	&\quad\, +  \int_\Sigma \left(\frac{\ii}{2} F_{\bar a \Sigma}\, \d \bar \varphi^{\bar a} \wedge \calh^{\Sigma}_3  + {\rm c.c.}\right)
	\\
	&\quad\, -\int_\cals\d^2\zeta\, |e_\Lambda  l^\Lambda|\sqrt{-\det {\bm \gamma}}  + e_\Lambda\int_\cals \calb^\Lambda_2\;.
\end{aligned}
\ee
In the electro-magnetic dual theory, after transversing a string, the axions shift with the charge of the string
\be
\label{ExtObj_Sum_ashift}
a_\Lambda \quad \rightarrow \quad a_\Lambda + e_\Lambda\;. 
\ee

\begin{centerbox}
	Two- and three-forms coupled to a BPS-junction
\end{centerbox}

\begin{figure}[h]
	\centering
	\includegraphics[width=8cm]{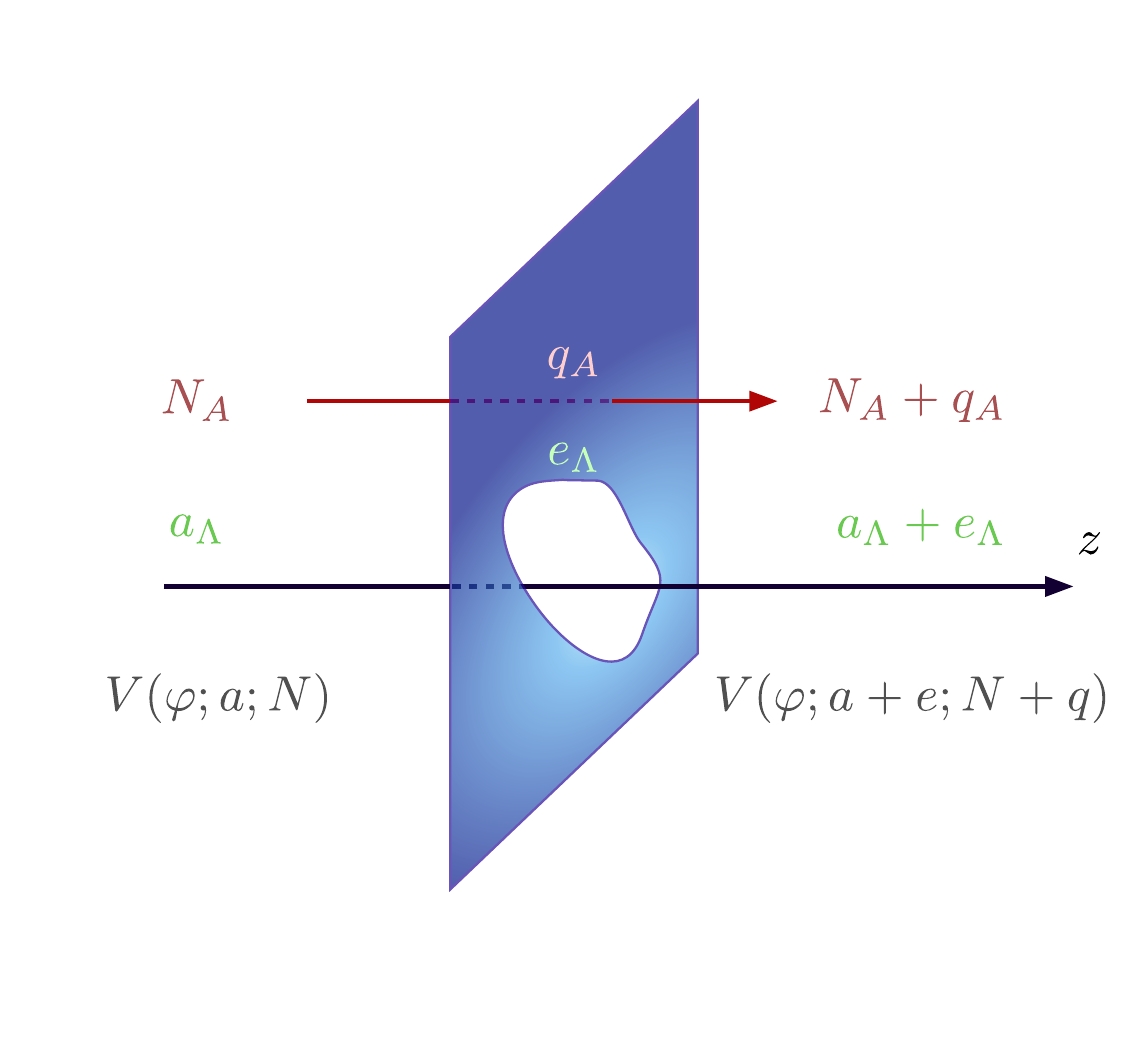}
	\caption{\footnotesize{Crossing simultaneously a string and a membrane ending on the string makes the constants $N_A$ and the axions $a_\Lambda$ to shift so as the potential is left invariant on the two sides.}}\label{fig:MembH}
\end{figure}

The full action that governs the coupling of a BPS--junction to a bulk theory is obtained from  \eqref{GSS_GL_L3fComp} and \eqref{ExtObj_SM_junction}:
	\be
	\begin{aligned}
		S |_{\rm bos} &= \int_{\Sigma} \left(- F_{a\bar b} \d \varphi^a \wedge *\d \bar\varphi^{\bar b} + \frac14 F_{\Lambda \Sigma}  \d l^\Lambda \wedge *\d l^\Sigma + \frac14 F_{\Lambda\Sigma} \hat \calh^\Lambda_3 \wedge *\hat \calh^\Sigma_3 \right)
		\\
		&\quad\,+  \int_\Sigma \left(\frac{\ii}{2} F_{\bar a \Sigma}\, \d \bar \varphi^{\bar a} \wedge \hat \calh^{\Sigma}_3  + {\rm c.c.}\right)
		\\
		&\quad\, -\int_\Sigma \Big[\frac12 T_{AB} F^A_4*\! F^B_4+ T_{AB}\Upsilon^A F^B_4+\Big(
		\hat V-\frac12 T_{AB}\Upsilon^A \Upsilon^B\Big)*\!1\Big] 
		\\
		&\quad\,+\int_{\del\Sigma}T_{AB}(* F^A_4+\Upsilon^A)A^B_3 
		\\
		&\quad\, -\int_\cals\d^2\zeta\, |e_\Lambda  l^\Lambda|\sqrt{-\det {\bm \gamma}}  + e_\Lambda\int_\cals \calb^\Lambda_2
		\\
		&\quad\,-2\int_\calm \d^3\xi\,|e_\Lambda c^\Lambda_A \calv^A(\varphi)|\sqrt{-\det {\bf h}}+e_\Lambda c^\Lambda_A\int_\calm A^A_3 \;.
\end{aligned}
\ee
Transversing \emph{both} the string and the membrane, combining the results \eqref{ExtObj_Sum_F4eom} and \eqref{ExtObj_Sum_ashift}, we get the combined shift
\be
\label{ExtObj_Sum_Nashift}
N_A \quad \rightarrow \quad N_A + e_Ac_A^\Lambda\,, \qquad a_\Lambda \quad \rightarrow \quad a_\Lambda + e_\Lambda\;. 
\ee
In dual, ordinary chiral theory described by the superpotential \eqref{GSS_GL_Dual2W}
\be
\label{ExtObj_Sum_Dual2W}
W(\Phi;T) = N_A \calv^A(\Phi) -c^A_\Lambda T^\Lambda \calv^A(\Phi)+ \hat W (\Phi) 
\ee
the shift \eqref{ExtObj_Sum_Nashift} \emph{has no net effect!} In other words, a BPS--junction so built connects two equivalent theories sharing the same potential. This is depicted in Fig.~\ref{fig:MembH}.

\begin{centerbox}
	Two-, three- and four-forms coupled to a BPS-junction and a BPS-region
\end{centerbox}

As an example of more complete networks that we can build with the objcets presented in this chapter, we consider the simultaneous presence of a BPS--junction and a BPS--region within the same theory. Combining \eqref{GSS_GMTF_L3fComp} and \eqref{GSS_GL_L3fComp} with \eqref{ExtObj_SM_junction} and \eqref{ExtObj_M3B_region}, we get the full action 
\be
\begin{aligned}
	S |_{\rm bos} &= \int_{\Sigma} \left(- F_{a\bar b} \d \varphi^a \wedge *\d \bar\varphi^{\bar b} + \frac14 F_{\Lambda \Sigma}  \d l^\Lambda \wedge *\d l^\Sigma + \frac14 F_{\Lambda\Sigma} \hat \calh^\Lambda_3 \wedge *\hat \calh^\Sigma_3 \right)
	\\
	&\quad\,+  \int_\Sigma \left(\frac{\ii}{2} F_{\bar a \Sigma}\, \d \bar \varphi^{\bar a} \wedge \hat \calh^{\Sigma}_3  + {\rm c.c.}\right)
	\\
	&\quad\, -\int_\Sigma \Big[\frac12 T_{AB} \hat F^A_4*\! \hat F^B_4+ T_{AB}\Upsilon^A \hat F^B_4+\Big(
	\hat V-\frac12 T_{AB}\Upsilon^A \Upsilon^B\Big)*\!1\Big] 
	\\
	&\quad\,+\int_{\del\Sigma}T_{AB}(* \hat F^A_4+\Upsilon^A)A^B_3 + \int_\Sigma \calq_I^{\rm bg} C^I_4
	\\
	&\quad\, -\int_\cals\d^2\zeta\, |e_\Lambda  l^\Lambda|\sqrt{-\det {\bm \gamma}}  + e_\Lambda\int_\cals \calb^\Lambda_2
	\\
	&\quad\,-2\int_\calm \d^3\xi\,|e_\Lambda c^\Lambda_A \calv^A(\varphi)|\sqrt{-\det {\bf h}}+e_\Lambda c^\Lambda_A\int_\calm A^A_3 
	\\
	&\quad\,-2\int_{\calm'} \d^3\xi\,|q_A \calv^A(\varphi)|\sqrt{-\det {\bf h}}+q_A\int_{\calm'} A^A_3 
	\\
	&\quad\, + q_A Q^A_I\int_\calw C^I_4\;.
\end{aligned}
\ee
The worldvolumes are chosen as
\be
\del \calm = \cals\,,\qquad \del \calw = \calm'\;,
\ee
and the membranes described by the worldvolumes $\calm$ and $\calm'$ span different hypersurfaces (for example, one lozalized at $z=0$, the other at $z=z_0 \neq 0$). In fact, a membrane which serves as the boundary of a 3-brane cannot simultaneously end on a string, otherwise the latter would be `the boundary of a boundary'. Consistency then requires the compatibility condition
\be
Q_I^A c_A^\Lambda = 0
\ee
to hold. In turn, this implies that the gauge three-forms that are gauged by four-forms as in \eqref{GSS_GMTF_F4g} do not simultaneously participate to the gauging of three-form field strength as in \eqref{GSS_GL_H3g}.



\part{Hierarchy of forms and extended objects in Supergravity}


\chapter{Hierarchies of forms and gaugings in supergravity}
\label{chapter:Sugra}

As of now, we have extensively explored the implementation of $p$--forms into globally supersymmetric multiplets and how to construct their super-field strengths, whenever a gauge invariance is present. With such multiplets, we were also able to couple extended objects to generic $\caln=1$ supersymmetric theories by preserving totally or partially the supersymmetry of the bulk theory. 

This chapter is devoted to the extension of the results of Chapters \ref{chapter:GSS} and \ref{chapter:ExtObj} to locally supersymmetric theories. Preliminarily, however, we will briefly introduce the super-Weyl invariant approach to supergravity \cite{Kaplunovsky:1994fg,Buchbinder:1995uq} which is further examined in greater details in Appendix \ref{app:SW}. For our purposes, the super-Weyl invariant approach has some advantages: still relying on the superfield formalism, it allows us to write manifestly supersymmetric invariant action, and the components of super-Weyl invariant Lagrangians, up to some proper covariantizations, have a striking resemblance to those of globally supersymmetric Lagrangians. The apparent price to be paid is the introduction of an unphysical compensator which, as we will see, will indeed be a useful tool to achieve a canonical normalization of the Einstein-Hilbert term with a little effort.

Such a formalism will allow us to easily generalize the globally supersymmetric results to locally supersymmetric theories: we will be able to introduce gauge three-forms in any given $\caln=1$ supergravity theory to dynamically generate $F$-term potentials, introduce gauge four-forms to constrain the constants that we may so generate and couple gauge two-forms, included in linear multiplets eventually gauged by three-form multiplets. As in the Chapter \ref{chapter:GSS}, we will end up with a now locally supersymmetric theory where a hierarchy of forms is present.  What is developed in this chapter will serve as the basic framework to study and reformulate effective theories originating from compactification of string/M-theory.

\section{The super-Weyl invariant approach to Supergravity}
\label{sec:Sugra_SW}

Although discussed in details in Appendix \ref{app:SW}, in order to keep this chapter self-contained, we briefly summarize the super-Weyl invariant approach to supergravity. Consider a theory with $n$ chiral multiplets $\Phi^i$, whose components
\be
\label{Sugra_SW_Phi}
\Phi^i = \{\varphi^i,\chi^i_\alpha, F^i_{(\Phi)}\}
\ee
contain propagating complex scalars $\varphi^i$ and Weyl-fermions $\chi^i_\alpha$ and the nonpropagating complex scalar fields $F^i_{(\Phi)}$. The promotion of supersymmetry to a local symmetry of the theory further introduces new degrees of freedom, collected into the \emph{supergravity multiplet}
\be
\label{Sugra_SW_calr}
\calr = \{e_a^m,\psi^m_\alpha, G^a, M\}
\ee
where $e_a^m$ is the veilbein encoding the propagating bosonic degrees of the graviton, $\psi^m_\alpha$ is the gravitino, the fermionic superpartner of the graviton, and the nonpropagating fields $G^a$ and $M$ are a real vector and a complex scalar, required for a proper definition of $\calr$ as an off-shell multiplet. Up to two derivatives, the most general, locally supersymmetric Lagrangian which describes the coupling of the chiral multiplets $\Phi^i$ to the supergravity multiplet $\calr$ is
\be
\label{Sugra_SW_L}
\call = \int \d^4\theta\,E\, \Omega(\Phi, \bar\Phi) + \left( \int \d^2\Theta\,2\cale\, W(\Phi) +{\rm c.c.}\right)\,.
\ee
Here we have denoted with $E$ the Berezinian super-determinant of the super-vielbein $E^A_M$ and $\d^2\Theta\,2\cale$ the chiral superspace measure, which is the covariantization of $\int \d^2\theta$ of global supersymmetry. The Lagrangian is then fully specified by the \emph{kinetic section} $\Omega(\Phi, \bar \Phi)$, which defines the K\"ahler potential $K(\Phi, \bar \Phi)$ as
\be
\label{Sugra_SW_Om}
\Omega(\Phi, \bar \Phi) \equiv -3\, e^{-\frac13 K(\Phi,\bar \Phi)}\,,
\ee
and the holomorphic \emph{superpotential} $W(\Phi)$. 

In $\caln=1$ supergravity, the chiral fields $\Phi^i$ describe a Hodge--K\"ahler manifold, namely a K\"ahler manifold $\scrm_{\rm scalar}$ such that there exists a line bundle $\scrl \rightarrow \scrm_{\rm scalar}$, whose first Chern class equals the cohomology class of the K\"ahler form $J$: $c_1(\scrl) = [J]$. The superpotential $W(\Phi)$, as well as the chiralini $\chi^i_\alpha$, the gravitino $\psi^m_\alpha$ and the auxiliary fields $F^i_{\Phi}$, $M$ are then interpreted as sections of the line bundle $\scrl$. Then, once a K\"ahler transformation is performed
\be
\label{Sugra_SW_KT}
K (\Phi, \bar \Phi) \rightarrow K (\Phi, \bar \Phi)  + h(\Phi) + \bar h(\bar \Phi)\,,
\ee
also the superpotential nontrivially transforms
\be
\label{Sugra_SW_WT}
W (\Phi) \rightarrow e^{-h(\Phi)}W(\Phi) \,,
\ee
along with the fermions and the auxiliary fields. This makes the K\"ahler invariance of the Lagrangian \eqref{Sugra_SW_L} not manifest. In fact, if we want to recover its K\"ahler invariance, we additionally need to exploit \emph{super-Weyl transformations}. These, which are a superspace generalization of the more common Weyl transformations, act on the supervielbeins so that the superspace measures get transformed to
\be
\label{Sugra_SW_E}
\d^4 \theta\, E \rightarrow \d^4 \theta\,  E\, e^{-\frac{\Upsilon + \bar \Upsilon}3}\;,\quad  \d^2 \Theta\, 2 \cale \rightarrow   \d^2 \Theta\, 2 \cale\, e^{\Upsilon}\,,
\ee
with $\Upsilon$ a generic chiral field, compensating \eqref{Sugra_SW_KT} and \eqref{Sugra_SW_WT} by choosing $\Upsilon = h(\Phi)$.

One might wonder whether there exists a smarter way to implement the K\"ahler invariance in supergravity Lagrangians of the kind of \eqref{Sugra_SW_L}, which does not need the pass through super-Weyl transformations. The answer resides in the so-called super-Weyl invariant approach, first introduced in \cite{Kaplunovsky:1994fg}. At the core of the approach is the introduction of an unphysical chiral field $U$, serving the role of a \emph{compensator}, which transforms, under super-Weyl transformations, as
\be
\label{Sugra_SW_U}
U \rightarrow e^{\Upsilon} U\;.
\ee
We then introduce a new set of $n+1$ chiral superfields$Z^a$, $a=1,\ldots,n$. These carry a dependence on both the compensator $U$ and the physical superfields $\Phi^i$, which can however be split as
\be
\label{Sugra_SW_Za}
Z^a (U, \Phi) = U g^a(\Phi)
\ee
where $g^a(\Phi)$ are holomorphic sections, which depend on the physical fields only and are inert under super-Weyl transformations. We rewrite \eqref{Sugra_SW_L} as
\be
\label{Sugra_SW_LSW}
\call = \int \d^4\theta\,E\, \calk(Z, \bar Z) + \left( \int \d^2\Theta\,2\cale\, \calw(Z) +{\rm c.c.}\right)\,,
\ee
Here, $\calk(Z, \bar Z)$ and $\calw(Z)$ satisfy the homogeneity properties
\be
\label{Sugra_SW_KWhom}
\calk(\lambda Z,\bar\lambda\bar { Z})=|\lambda|^{\frac 23}\calk({Z},\bar { Z})\;,\qquad
\calw(\lambda Z)=\lambda\calw(Z)\;.
\ee
and, extracting the compensator $U$ using \eqref{Sugra_SW_Za}, are then related to the usual K\"ahler potential $K(\Phi,\bar \Phi)$ and superpotential $W(\Phi)$ via
\be
\label{Sugra_SW_KWphys}
\calk (Z,\bar Z) \equiv -3 |U|^{\frac23} e^{-\frac{K(\Phi,\bar \Phi)}{3}}\;, \qquad \calw(Z) \equiv \frac{1}{\Mp^3}U W(\Phi)\;,
\ee
where we have set $K(\Phi, \bar \Phi) \equiv K(g(\Phi), \bar g(\bar \Phi))$ and $W(\Phi) \equiv  W(g(\Phi))$.
It is immediate to check that, using \eqref{Sugra_SW_Za} and \eqref{Sugra_SW_KWhom}, the Lagrangian \eqref{Sugra_SW_LSW} is invariant under the sole super-Weyl transformations. Moreover, the K\"ahler invariance is here manifest, in fact emerging as the residual freedom in the parametrization \eqref{Sugra_SW_Za}. That is, \eqref{Sugra_SW_Za} is unaffected by the transformations
\be
U \rightarrow e^{h(\Phi)} U\,, \qquad g(\Phi) \rightarrow e^{-h(\Phi)}  g(\Phi)\,,
\ee
which, applied to \eqref{Sugra_SW_KWphys}, clearly reproduce the transformations \eqref{Sugra_SW_KT}--\eqref{Sugra_SW_WT}.

As anticipated above, the super-Weyl invariant approach has also virtues at the level of the bosonic components. From \eqref{Sugra_SW_LSW} we get the bosonic Lagrangian 
\be
\label{Sugra_SW_LSWcomp}
\begin{split}
	e^{-1} \call_{\rm bos} &= -\frac16\calk\, R - \calk_{a\bar b} D_\mu z^a \bar D^\mu \bar z^b + \calk_{a\bar b}  f^a \bar f^b + \left( \hat\calw_a f^a + {\rm c.c.}\right)\,.
\end{split}
\ee
where we have defined $f^a \equiv  \bar M z^a - F^a_\calz$ and introduced the derivatives
\be
\label{Sugra_SW_Da}
D_\mu z^a = \del_\mu z^a + \ii A_\mu z^a\,\qquad {\rm with}\quad	A_\mu = \frac{3\ii}{2\calk} (\calk_a \del_\mu z^b - \calk_{\bar b} \del_\mu \bar z^b)\,.
\ee
which are covariant with respect to the $U(1)$--bundle associated to $\scrl$. 

The bosonic components \eqref{Sugra_SW_LSWcomp} have an astonishingly simple form: beside the introduction of the covariant derivatives and the curvature term, they are formally equal to those of rigid superymmetry (see, for example, \eqref{GSS_DTF_LchirComp})! However, such a neat appearance is just the result of a hidden redundancy in the Lagrangian \eqref{Sugra_SW_LSWcomp}, that is the dependence on the super-Weyl compensator $U$. The presence of $U$ can indeed be used in our favor in order to immediately pass to the Einstein-frame. Concretely, we gauge fix $U$ in such a way that its lowest bosonic component $u$ is
\be
\label{Sugra_SW_u}
u=\Mp^3\, e^{\frac12 K(\varphi,\bar\varphi)}\, ,
\ee
which sets, for instance, $\calk=-3 M^2_{\rm P}$ and, as explained in Section \ref{app:SW_C}, leads to the usual supergravity Lagrangian
\be
\label{Sugra_SW_LSWcompgf}
\begin{split}
	e^{-1} \call_{\rm bos} &= \frac{1}2 \Mp^2 R - \Mp^2 K_{i\bar \jmath} \del_m \varphi^i \del^m \bar \varphi^{\bar \jmath}- V(\varphi,\bar\varphi) \;,
\end{split}
\ee
where the potential $V (\varphi,\bar\varphi)$  is given by the usual Cremmer et al. formula \cite{Cremmer:1978iv}
\be
\label{Sugra_SW_Cremmeretal}
V (\varphi,\bar\varphi) = \frac{e^{K}}{\Mp^2} \left( K^{\bar\jmath i} D_i W \bar D_{\bar \jmath} \bar W - 3 |W|^2 \right)\;.
\ee
\section{Gauge three-forms and F-term potential}
\label{sec:Sugra_MTF}

In Chapter \ref{chapter:GSS}, across the Sections  \ref{sec:GSS_TIntro} to \ref{sec:GSS_MTF}, we have seen how to implement gauge three-forms into globally supersymmetric theories with arbitrary K\"ahler potential and superpotential. Presently, we shall see how to generalize those constructions to supergravity theories by using the super-Weyl invariant formalism introduced in the previous section. Here, rather than reproducing each of the cases examined in Chapter~\ref{chapter:GSS} separately one by one,  we will directly use the more general recipe given in Section~\ref{sec:GSS_MTF}. Namely, we shall start with a master Lagrangian with a generic number of chiral multiplets and three-forms and relate them via master three-form multiplets. We shall however see how, from this generic viewpoint, the local versions of the models of Sections \ref{sec:GSS_STF} and \ref{sec:GSS_DTF} are retrieved.

Consider a superpotential of the form
\be
\label{Sugra_MTF_calw}
\calw (Z) =  N_A \calv^A(Z) + \hat \calw(Z)\;.
\ee
Here $N_A$, $A=0,\ldots,N$, are \emph{real} constants and consistency with the homogeneity properties \eqref{Sugra_SW_KWhom} require the \emph{periods} $\calv^A(Z)$ and the additional superpotential contribution $\hat \calw(Z)$ to be homogeneous of degree one. Our goal, as in Section~\ref{sec:GSS_MTF}, is to generate, via gauge three-forms, a contribution to the potential which generates the same effect as the superpotential
\be
\label{Sugra_MTF_calwgen}
\calw_{\rm gen}(Z) = N_A \calv^A(Z)\,.
\ee
By extracting the compensator $U$ as
\be
\label{Sugra_MTF_WcalvA}
\calv^A(Z) = U \calv^A(g(\Phi)) \equiv U \Pi^A(\Phi)\,,
\ee
with $\Pi^A$ holomorphic in $\Phi^i$, and gauge fixing the super-Weyl invariance as in \eqref{Sugra_SW_u}, the physical superpotential that we are going to generate via gauge three-forms is
\be
\label{Sugra_MTF_Wphys}
W_{\rm gen}(\Phi) = \Mp^3 N_A \Pi^A (\Phi)\,.
\ee
The recipe of Section \ref{sec:GSS_MTF} can now be applied to the local case by a simple covariantization. We start with a chiral multiplet theory \eqref{Sugra_SW_LSW}, with the superpotential of the class \eqref{Sugra_MTF_calw}. In order to build a master Lagrangian, the term \eqref{Sugra_MTF_calwgen} of the superpotential that we are going to generate is promoted to
\be
\label{Sugra_MTF_calwgenp}
N_A \calv^A(Z) \quad \rightarrow \quad X_A\calv^A(Z)
\ee
where $X_A$ are chiral Lagrange multipliers with components $\{x_A, \psi^{\alpha(X)}_A, F_A^{(X)} \}$. We then introduce the master Lagrangian
{\small \be
\label{Sugra_MTF_LM}
\begin{aligned}
	\call_{\rm master} &=\int \d^4\theta\,E\, \calk(Z, \bar Z) + \left( \int \d^2\Theta\,2\cale\, \hat \calw(Z) +{\rm c.c.}\right)\,
	\\
	&\quad\,+ \left(\int \d^2\Theta\,2\cale\, X_A\calv^A(Z)+\frac\ii8\int \d^2\Theta\,2\cale\, (\bar \cald^2 - 8 \calr)  \left[(X_A-\bar X_A)P^A \right]  +\text{c.c.}\right)\,.
\end{aligned}
\ee}
The last term contains a set real superfields $P^A$. As we will see shortly, as in Section \ref{sec:GSS_MTF}, the role of the last term is either to set $X_A$ to constants or to provide the correct duality between ordinary and three-form chiral multiplets. In the following, we will also need the bosonic components of the master Lagrangian \eqref{Sugra_MTF_LM}, which read 
\be\label{Sugra_MTF_LMComp}
\begin{split}
	e^{-1}\call_{\rm master}|_{\text{bos}} &=-\frac16 \calk R-\calk_{a\bar b}  D_m z^a D^m z^{\bar b}+ \calk_{a\bar b}  f^a \bar f^{\bar b} + \left( \hat\calw_a f^a + {\rm c.c.}\right)
	\\
	&\quad\, + \Bigg[ x_A\calv^A_a f^a - \frac{1}{2 \cdot 3!\, e} \varepsilon^{mnpq} A^A_{npq} \del_m x_A  - \frac{\ii}{2} d^A x_A 
	\\
	&\quad\quad\quad\,+ f_{XA} (\calv^A - s^A) + x_A \left(\bar M \calv^A  -\Re (\bar M s^A)\right) + {\rm c.c. }\Bigg]\,,
\end{split}
\ee
which are formally similar to those of \eqref{GSS_MTF_LMComp}, if we turn off the fields of the supergravity multiplet. Two paths can be followed which can lead either to an ordinary formulation \eqref{Sugra_SW_LSW} or to a three-form formulation.

\begin{description}
	\item[Ordinary formulation] Integrating the real superfields $P^A$ from the Lagrangian \eqref{Sugra_MTF_LM} immediately imposes
	\be\label{Sugra_MTF_deltaPAa}
	\delta P^A:\qquad \Im X_A = 0\;.
	\ee
	The chirality of $X_A$ further constraints them to be real constants $N_A$
	\be
	\label{Sugra_MTF_deltaPAb}
	X_A = N_A\;.
	\ee
	Once \eqref{Sugra_MTF_deltaPAb} is plugged into \eqref{Sugra_MTF_LM}, we immediately obtain the Lagrangian \eqref{Sugra_SW_LSW} with the superpotential \eqref{Sugra_MTF_calw}
	\be
	\label{Sugra_MTF_LSW}
	\call = \int \d^4\theta\,E\, \calk(Z, \bar Z) + \left[ \int \d^2\Theta\,2\cale\, (N_A \calv^A(Z)+ \calw(Z)) +{\rm c.c.}\right]\,.
	\ee
	After gauge-fixing the super-Weyl invariance, the superpotential is
	\be
	\label{Sugra_MTF_Wphysgf}
	W(\Phi) = \Mp^3 N_A \Pi^A (\Phi) + \hat W(\Phi)\,,
	\ee
	which defines the potential as in \eqref{Sugra_SW_LSWcompgf}.
	
	The possibility of recovering the ordinary formulation from the master Lagrangian \eqref{Sugra_MTF_LM} can also be seen at the level of bosonic components. Integrating out the auxiliary fields $d^A$, $f_{XA}$ and the three-forms $C_3^A$ from \eqref{Sugra_MTF_LMComp}, we immediately obtain that $x_A$ are real constants.
	
	This proves the consistency of the master Lagrangian with the ordinary formulation \eqref{Sugra_SW_LSW}.

	\item[Three-form formulation] In order to get a three-form Lagrangian, we need to integrate out the Lagrange multipliers $X_A$ and (part of) the ordinary chiral fields $Z^a$. Integrating out the Lagrange multipliers $X_A$ results in the following relation between the periods and the real superfields $P^A$
	\begin{important}%
		\be
		\label{Sugra_MTF}
		\calv^A (Z) = -\frac \ii 4 (\bar \cald^2 - 8 \calr)P^A \equiv S^A \qquad{\color{darkred}\text{\sf{Master three-form multiplet}}}
		\ee
	\end{important}
	whose components are
	\begin{subequations}
	\label{Sugra_MTF_Comp}
	\begin{align}	
		\label{Sugra_MTF_Compa}
		\calv^A | &= \calv^A(z) &=& \left(-\frac \ii 4  (\bar \cald^2 - 8 \calr) P^A \right) \Big| &\equiv &\, s^A \,, 
		\\
		\label{Sugra_MTF_Compb}
		-\frac14 \cald^2 \calv^A | &=  \calv^A_a(z) F^a_{(Z)} &=&	-\frac14 \cald^2 \left(-\frac \ii 4 (\bar \cald^2 - 8 \calr) P^A \right)\Big|  &\equiv&\,  F^A_{(S)} \,, 
	\end{align}
	\end{subequations}
	with
	\be
	\label{Sugra_MTF_Compc}
	\begin{aligned}
		 F^A_{(S)} = \frac12 \left(*\!F_4^A + \ii d^A\right) + \Re (\bar M s^A)\,.
	\end{aligned}
	\ee
	Loosely speaking, \eqref{Sugra_MTF}, via the component relations \eqref{Sugra_MTF_Comp}, \eqref{Sugra_MTF_Compc} provides the translation between ordinary chiral fields $Z^a$ and three-form multiplets. As we shall see in the example of the next section, the number of $Z^a$ multiplets which can be promoted to three-form multiplets is determined by the periods $\calv^A(Z)$. For this reason, we shall call $S^A$ \emph{master three-form multiplets}.
	
	Integrating out the chiral superfields $\Phi^a$ gives
	\be\label{Sugra_MTF_XA}
	\calv^A_a X_A = \frac14 (\bar \cald^2-8 \calr) \calk_a - \hat \calw_a\,,
	\ee
	which, plugged back into the master Lagrangian \eqref{Sugra_MTF_LM}, gives
	\be\label{Sugra_MTF_L3f}
	\tilde{\mathcal L}=\int \d^4\theta\,E\, \calk(Z(P), \bar Z(P)) + \left[ \int \d^2\Theta\,2\cale\, \hat \calw(Z(P)) +{\rm c.c.}\right] +\tilde {\mathcal L}_{\rm bd},
	\ee
	with the boundary terms
	\be\label{Sugra_MTF_L3fbd}
	\tilde{\mathcal L}_{\rm bd}= 
	-\frac\ii8 \left[\int\d^2\Theta\,2\cale\, (\bar \cald^2 - 8 \calr)-\int\d^2\bar \Theta\,2\bar \cale\, (\cald^2 - 8 \bar\calr)\right]\left(X_A P^A\right)
	+\text{c.c.} \,,
	\ee
	with $X_A$ defined by \eqref{Sugra_MTF_XA}.  The superspace Lagrangian \eqref{Sugra_MTF_L3f} is then specified in terms of the superfields $Z^a$, some of which do depend on $P^A$ through \eqref{Sugra_MTF} and their highest components contain gauge three-forms. At first sight, it is clear that the Lagrangian \eqref{Sugra_MTF_L3f}, due to the nontrivial dependence on the real potentials $P^A$, does not fit well to compute the bosonic components of generic models containing an arbitrary number of gauge three-forms. In order to more easily extract the bosonic components of the three-form Lagrangian, it is better to start from the master Lagrangian \eqref{Sugra_MTF_LM}, with components
	\eqref{Sugra_MTF_LMComp}. Rather then integrating whole superfields at once, as done above, we integrate out the components one by one. First, integrating out the auxiliary fields of the Lagrange multipliers $X_A$ immediately gives 
	\be
	\label{Sugra_MTF_IOCompa}	
	\delta\, f_{XA}: \qquad \calv^A(\varphi) = s^A\,,
	\ee
	compatibly with \eqref{Sugra_MTF_Compa};
	further integrating $d^A$ results in setting $x_A$ to be real Lagrange multipliers
	\be
	\label{Sugra_MTF_IOCompb}	
	\delta\, d^A: \qquad \Im x_A = 0\,.
	\ee
	Up to this point, the master Lagrangian becomes
	\be\label{Sugra_MTF_LMCompc}
	\begin{split}
		e^{-1}\call_{\rm master}|_{\text{bos}} &=-\frac16 \calk R-\calk_{a\bar b}  D_m \varphi^a D^m \varphi^{\bar b}+ \calk_{a\bar b}  f^a \bar f^{\bar b} + \left( \hat\calw_a f^a + {\rm c.c.}\right)
		\\
		&\quad\, + \Bigg[ x_A\calv^A_a f^a - \frac{1}{2 \cdot 3!\, e} \varepsilon^{mnpq} A^A_{npq} \del_m x_A  + {\rm c.c. }\Bigg]\,.
	\end{split}
	\ee
	Let us now proceed by integrating out the fields $f^a$
	\be
	\label{Sugra_MTF_IOCompc}	
	\delta\, f^a: \qquad \bar f^{\bar b} = - \calk^{\bar b a} (x_A \calv^A_a+\hat \calw_a)\,,
	\ee
	which recast the Lagrangian \eqref{Sugra_MTF_LMCompc} in the simple form
	{\small\be\label{Sugra_MTF_LMCompd}
	\begin{split}
		e^{-1}\call_{\rm master}|_{\text{bos}} &=-\frac16 \calk R-\calk_{a\bar b}  D_m \varphi^a D^m \bar \varphi^{\bar b} - \calk^{\bar b a}\left( \hat \calw_a + x_A \calv^A_a\right)\left( \bar{\hat \calw}_{\bar b} + x_B \bar \calv^B_{\bar b}\right)
		\\
		&\quad\, + \left[ - \frac{1}{2 \cdot 3!e} \varepsilon^{mnpq} A^A_{npq} \del_m x_A + {\rm c.c. }\right]\;.
	\end{split}
	\ee}
	From this Lagrangian, it is then easy to get a three-form Lagrangian. It is just sufficient to integrate out the now--real Lagrange multipliers $x_A$
	\be
	\label{Sugra_MTF_IOCompd}
	\delta\, x_A:\qquad x_A = - T_{AB} (*F_4^B + \Upsilon^B)\,,
	\ee
	where we have defined
	\begin{subequations}
		\begin{align}
		\label{Sugra_MTF_TAB}
		T^{AB} &= 2\, \Re \left(K^{\bar b a} \calv^A_a \bar \calv^{\bar B}_{\bar b}\right)\,,
		\\
		\Upsilon^A &\equiv 2\, \Re \left(K^{\bar b a} \calv^A_a \bar{\hat \calw}_{\bar b}\right) \label{Sugra_MTF_UA} \,,
		\end{align}
	\end{subequations}
	and $T_{AB}$ is the inverse of $T^{AB}$. Finally, we arrive to the Lagrangian 
	\be
	\label{Sugra_MTF_LdComp}	
	\begin{split}
		e^{-1}\tilde \call|_{\rm bos}&=-\frac16 \calk R-\calk_{a\bar b}  D_m z^a D^m \bar z^{\bar b} + e^{-1}\call_{\text{3-form}}|_{\rm bos}\,,
	\end{split}
	\ee
	Here we have collected the three-form and potential contributions in $e^{-1}\call_{\text{3-form}}|_{\rm bos}$, which reads
	\be
	\label{Sugra_MTF_L3fComp}	
	\begin{split}
		e^{-1}\call_{\text{3-form}}|_{\rm bos}&=  \frac{1}{2} T_{AB} *\!F_4^A *\!F_4^B+ T_{AB} \Upsilon^A *\!F_4^B 
		\\
		&\quad\, - \left(\hat V - \frac12 T_{AB} \Upsilon^A \Upsilon^B \right)+ e^{-1}\call_{\rm bd}
	\end{split}
	\ee
	with the boundary terms
	\be
	\label{Sugra_MTF_L3fCompbd}	
	\begin{split}
		\call_{\rm bd} = \frac{1}{3!} \partial_m \left[ \varepsilon^{mnpq} {A}^A_{npq} T_{AB} (*\!F_4^B + \Upsilon^B)\right] \,,
	\end{split}
	\ee
	and
	\begin{subequations}
		\begin{align}
		\hat V &\equiv \calk^{\bar b a} \hat \calw_a \bar{\hat \calw}_{\bar b} \,. \label{Sugra_MTF_hatV}
		\end{align}
	\end{subequations}
	Up to now, the steps that we have followed to obtain a generic three-form Lagrangian are the same -- up to covariantization and some field redefinitions -- to the ones of Section \ref{sec:GSS_MTF}. But this was expected, since we have worked within super-Weyl invariant Lagrangians. 
	
	In supergravity, however, if we want to get Lagrangians expressed solely in terms of physical fields, an additional step is required: we need to gauge-fix the super-Weyl invariance so as to eliminate the dependence on the unphysical compensator $U$. As explained in the previous section, the gauge-fixing \eqref{Sugra_SW_u} allows us to get a Lagrangian in the Einstein-frame. Employing \eqref{Sugra_SW_u}, as explained in more details in Appendix \ref{app:SW}, the Lagrangian \eqref{Sugra_MTF_L3fComp} becomes
	\be
	\label{Sugra_MTF_L3fComp_gf}	
	\begin{split}
		e^{-1}\tilde \call|_{\rm bos}&=\frac12 \Mp^2  R- \Mp^2 K_{i\bar \jmath}  \del_m \varphi^i \del^m \bar \varphi^{\bar \jmath} + \frac{1}{2} T_{AB} *\!F_4^A *\!F_4^B
		\\
		&\quad\, + T_{AB} \Upsilon^A *\!F_4^B  
		- \left(\hat V - \frac12 T_{AB} \Upsilon^A \Upsilon^B \right)+ e^{-1}\tilde\call_{\rm bd}
		\\
		&{\rm with}\qquad   \tilde\call_{\rm bd} = \frac{1}{3!} \partial_m \left[ \varepsilon^{mnpq} {A}^A_{npq} T_{AB} (*\!F_4^B + \Upsilon^B)\right] \,,
	\end{split}
	\ee
	where now the three-form part of the Lagrangian is also re-expressed in terms of the physical quantities $K(\Phi,\bar \Phi)$, $\Pi^A(\Phi)$ and $\hat W(\Phi)$ as
	\begin{subequations}\label{Sugra_MTF_TUVgf}
		\begin{align}
		\label{Sugra_MTF_Tgf}
		T^{AB}&=2 M^4_{\rm P}\,e^{K} \Re\left(K^{\bar\jmath i}D_i\Pi^A\bar D_{\bar\jmath}\bar\Pi^B-3\Pi^A\bar\Pi^B\right)\,,
		\\
		\label{Sugra_MTF_Ugf}
		\Upsilon^A&=2 M_{\rm P}\,e^{K}\Re\left(K^{\bar\jmath i}D_i\hat W\bar D_{\bar\jmath}\bar\Pi^A-3\hat W\bar\Pi^A\right)\, ,
		\\
		\label{Sugra_MTF_hatVgf}
		\hat V&=\frac{e^K}{M^2_{\rm P}}\left(K^{\bar\jmath i}D_i \hat W\bar D_{\bar\jmath}\overline{ \hat W}-3|\hat W|^2\right)\, .
		\end{align}
	\end{subequations}
	It is immediate to see, as a consistency check, that integrating out the gauge three-forms as
	\be
	\label{Sugra_MTF_L3fNA}	
	\delta A_3^A: \qquad T_{AB} (*\! \hat F_4^B + \Upsilon^B) = - N_A\,,
	\ee
	reduces \eqref{Sugra_MTF_L3fComp_gf} to \eqref{Sugra_SW_LSWcompgf} with the superpotential \eqref{Sugra_MTF_Wphysgf}.
\end{description}

\subsection{Relevant examples}
\label{sec:Sugra_MTF_Examples}

Before moving on and see how gauge four- and two-forms may be implemented into supergravity theories, let us pause for a moment to examine some relevant examples where the previous discussion finds concrete applications. Some of the models here presented will naturally emerge in EFT originating from string/M-theory compactifications. As a summary, the examples here examined are collected in Table \ref{tab:Sugra_three-form_Wgen}.

\begin{table}[!h]
	\begin{center}
		\begin{tabular}{ |c c c c c | }
			\hline
			\rowcolor{ochre!30}{\bf Scalars} &  {\bf Three-forms} & {\bf $\bm{W_{\rm gen}}$} & {\bf Superfields} & {\bf $\bm{\tilde{\call}}$}  \TBstrut
			\\
			\hline
			\hline
			$1$ & $1$ & $\lambda$ & $S = - \frac\ii4 (\bar \cald^2 - 8\calr)P$  &  \eqref{Sugra_MTF_Ex_minSTF_L3fCompc}  \TBstrut
			\\
			\hline
			$1$ & $2$ & $\omega$ & $S = - \frac14 (\bar \cald^2 - 8\calr) \bar\Sigma$  &  \eqref{Sugra_MTF_Ex_minDTF_L3fCompc}  \TBstrut
			\\
			\hline
			$n+1$ & $n+1$ & $r_0 + r_i \Phi^i$ & $S^a = - \frac\ii4 (\bar \cald^2 - 8\calr)P^a$  &  \eqref{Sugra_MTF_Ex_STF_L3fCompc}  \TBstrut
			\\
			\hline
			$n+1$ & $2(n+1)$ & \small{$e_0 + e_i \Phi^i + m^a \calg_{a}(\Phi)$} & \small{$S^a =  \frac\ii2 (\bar \cald^2 - 8\calr) (\calm^{ab} \Im \Sigma_b)$}  &  \eqref{Sugra_MTF_Ex_DTF_L3fCompc}  \TBstrut
			\\
			\hline
		\end{tabular}
	\end{center}
	\caption{The relevant examples of supergravity models where (part of) the potential is dynamically generated by gauge three-forms. The number of scalars include the super-Weyl compensator.}
	\label{tab:Sugra_three-form_Wgen}
\end{table}

\subsubsection{Minimal single three-form Supergravity}

As stressed in the introductory Chapter \ref{chapter:Intro}, one of the main reasons why gauge three-forms were first considered into any effective theory was to generate a cosmological constant, eventually dynamically neutralized by subsequent membrane nucleation. One of the first appearances of gauge three-forms in supegravity was to find out similar models in contexts where supersymmetry is preserved. For instance, in \cite{Duff:1980qv,Aurilia:1980xj,Hawking:1984hk,Brown:1987dd,Brown:1988kg,Duff:1989ah,Duncan:1989ug,Bousso:2000xa,Feng:2000if,Wu:2007ht}, the cosmological constant is understood as originating from a real gauge three-form;  in the later \cite{Ovrut:1997ur,Farakos:2016hly} is inserted into the supergravity multiplet by preserving supersymmetry. Here, we rephrase that simple model in our super-Weyl invariant approach.

Our aim is to dynamically generate a field--independent cosmological constant, namely we wish to generate a potential $W_{\rm gen} = \Mp^3 \lambda$, with $\lambda$ a real, dimensionless constant. We then assume that no physical fields $\Phi^i$ are present and we set $K(\Phi,\bar \Phi)=0$ and $\hat W(\Phi) =0$. In the super-Weyl invariant approach, such requests translate into setting
\be
\label{Sugra_MTF_Ex_minSTF_KW}	
\calk(U,\bar U) =  - 3 |U|^\frac23\,,\qquad \hat \calw(U) = 0\,,
\ee
and, since we are going to generate just a single constant, there is only one period, which, by homogeneity, can be chosen to be just
\be
\calv(U) = U\,.
\ee

Generating a single real constant $\lambda$ requires only one real three-form and we need to consider only a single real prepotential $P$. Then, the super-Weyl invariant Lagrangian \eqref{Sugra_MTF_L3f} simply becomes
\be
\label{Sugra_MTF_Ex_minSTF_LSS}
\tilde\call = \int \d^4\theta\,E\, |S|^\frac23 + \tilde\call_{\rm bd}
\ee
with $S$ given by
\be
\label{Sugra_MTF_Ex_minSTF}
S = -\frac{\ii}{4} (\bar \cald^2 - 8 \calr) P 
\ee
and the boundary terms are given by \eqref{Sugra_MTF_L3fbd} with $X= - \frac14 (\bar \cald^2 - 8 \calr) U^{-\frac23} \bar U^\frac13$. In components, from \eqref{Sugra_MTF_L3fComp}, we immediately get
\be
\label{Sugra_MTF_Ex_minSTF_L3fCompb}	
\begin{split}
	e^{-1}\tilde\call|_{\rm bos}&=\frac12 {|u|^\frac23} R + \frac{1}{3 |u|^\frac43} \del_m u \del^m \bar u - \frac{1}{12 |u|^\frac43}  (*\!F_4)^2+ e^{-1} \tilde\call_{\rm bd}
	\\
	&{\rm with}\qquad   \tilde\call_{\rm bd} = - \partial_m \left( \frac{1}{6 \cdot 3! |u|^\frac43} \varepsilon^{mnpq} {A}_{npq}  *\!F_4\right)\,.
\end{split}
\ee
Gauge fixing $u = \Mp^3$ as in \eqref{Sugra_SW_u}, we arrive at
\be
\label{Sugra_MTF_Ex_minSTF_L3fCompc}	
\begin{split}
	e^{-1}\tilde\call|_{\rm bos}&=\frac12 {\Mp^2} R  - \frac{1}{12 \Mp^4}  (*\!F_4)^2+ e^{-1} \tilde\call_{\rm bd}
	\\
	&{\rm with}\qquad   \tilde\call_{\rm bd} = - \partial_m \left( \frac{1}{6 \cdot 3! \Mp^4} \varepsilon^{mnpq} {A}_{npq}  *\!F_4\right)\,.
\end{split}
\ee
as can be also directly computed from \eqref{Sugra_MTF_L3fComp_gf}, considering just a single period $\Pi = 1$.  

\begin{summary}
At the superfield level, the action \eqref{Sugra_MTF_Ex_minSTF_L3fCompc} is described by just one superfield, namely the supergravity one \eqref{Sugra_SW_calr}, where, however, the imaginary part of the complex auxiliary field $M$ has been replaced with the real gauge three-form $A_3$:
	\be
	\label{Sugra_SW_calr_STF}
	\calr_{\rm single} = \{e_a^m,\psi^m_\alpha, G^a, \Re M, *F_4\}\,,
	\ee
Its action is
	\be
	\label{Sugra_SW_STF_Sugra}
	\call_{\rm single} = \left(-3 \int \d^2\Theta\,2\cale\, \calr_{\rm single} + {\rm c.c.}\right) + \tilde\call_{\rm bd}
	\ee
In this sense, \eqref{Sugra_MTF_Ex_minSTF_L3fCompc} is a \emph{variant} minimal supergravity, which we call \emph{single three-form supergravity} \cite{Ovrut:1997ur,Farakos:2017jme}.
\end{summary}

By integrating out the gauge three-form $A_3$, we immediately get
\be
* F_4 = 6 \Mp^4 \lambda\,,
\ee
whence, on-shell, \eqref{Sugra_MTF_Ex_minSTF_L3fCompc} reduces to 
\be
\label{Sugra_MTF_Ex_minSTF_L3fCompOS}	
\begin{split}
	e^{-1}\tilde\call|_{\rm bos}&=\frac12 {\Mp^2} R  + 3\Mp^4\lambda^2 \,, \qquad V = -3\Mp^4\lambda^2
\end{split}
\ee
with a constant, negative potential.

\subsubsection{Minimal double three-form Supergravity}

A slight generalization of the previous example consists in generating a cosmological constant term which, although still negative definite, depends on a complex parameter $\omega \equiv \lambda_1 + \ii \lambda_2$ (with $\lambda_1$ and $\lambda_2$ real). Namely, our goal is to generate, with the help of two real gauge three-forms, a superpotential $W_{\rm gen} = \Mp^3 \omega$. In super-Weyl invariant language, this means that in \eqref{Sugra_MTF_calwgen} we need two periods
\be
\label{Sugra_MTF_Ex_minDTF_calw}
\calw(U) = N_A \calv^A(U) \qquad {\rm with} \qquad N_A \equiv \begin{pmatrix}
	\lambda_1 \\ \lambda_2
\end{pmatrix}\,,\qquad \calv^A(U) \equiv U \begin{pmatrix}
1 \\ \ii 
\end{pmatrix}\,.
\ee
We still work with the assumptions that no matter fields are present, so that, in the super-Weyl invariant approach, \eqref{Sugra_MTF_Ex_minSTF_KW} still holds. 

In order to regard $\calw(U)$ as generated by gauge three-forms, we need to introduce two real potentials, $P^A =( \calp ,\tilde \calp)^t$, to generate the real constants $\lambda_1$ and $\lambda_2$, building the complex $\omega$. The trading prescribed by \eqref{Sugra_MTF} then reads
\be
\label{Sugra_MTF_Ex_minDTF_U}
U = - \frac\ii4 (\bar \cald^2 - 8 \calr) \calp\,,\qquad \ii U = - \frac\ii4 (\bar \cald^2 - 8 \calr) \tilde \calp\,.
\ee 
These two conditions may be rearranged as
\be
\label{Sugra_MTF_Ex_minDTF_Uc}
(\bar \cald^2 - 8 \calr) (\calp - \ii \tilde \calp) = 0\,,
\ee
which is solved by setting
\be
\label{Sugra_MTF_Ex_minDTF_Sigma}
\ii \calp + \tilde \calp = -2 \Sigma\,,
\ee
with $\Sigma$ a complex linear multiplet. Then, $U$ is fully determined by the complex linear multiplet as
\be
\label{Sugra_MTF_Ex_minDTF_S}
U =  - \frac\ii4 (\bar \cald^2 - 8 \calr) (-2 \Im \Sigma) = - \frac14 (\bar \cald^2 - 8 \calr) \Sigma \equiv S\,,
\ee
which provides the Supergravity generalization of the double three-form multiplet \eqref{GSS_DBT_L}.

In the super-Weyl invariant approach, the superspace Lagrangian has exactly the same form as \eqref{Sugra_MTF_Ex_minSTF_LSS}, where now $S$ is given by \eqref{Sugra_MTF_Ex_minDTF_S}. After gauge-fixing the super-Weyl invariance as in \eqref{Sugra_SW_u}, we arrive at the Lagrangian
\be
\label{Sugra_MTF_Ex_minDTF_L3fCompc}	
\begin{split}
	e^{-1}\tilde\call|_{\rm bos}&=\frac12 {\Mp^2} R  - \frac{1}{12 \Mp^4}  |*\!G_4|^2+ e^{-1} \tilde\call_{\rm bd}
	\\
	&{\rm with}\qquad   \tilde\call_{\rm bd} = - \partial_m \left( \frac{1}{6 \cdot 3! \Mp^4} \varepsilon^{mnpq} {C}_{npq}  *\!G_4 + {\rm c.c.}\right)\,,
\end{split}
\ee
where $G_4 = \d C_3$ is the field strength of the complex three-form entering the components of $\Sigma$ as in \eqref{ConvSG_Sigma}.
\begin{summary}
	At the superfield level, the action \eqref{Sugra_MTF_Ex_minDTF_L3fCompc} is described by just the supergravity multiplet \eqref{Sugra_SW_calr}, where now the full complex auxiliary field $M$ has been traded with the complex gauge three-form $C_3$:
	\be
	\label{Sugra_SW_calr_DTF}
	\calr_{\rm double} = \{e_a^m,\psi^m_\alpha, G^a, *G_4\}\,,
	\ee
	Its action is
	\be
	\label{Sugra_SW_DTF_Sugra}
	\call_{\rm double} = \left(-3 \int \d^2\Theta\,2\cale\, \calr_{\rm double} + {\rm c.c.}\right) + \tilde\call_{\rm bd}
	\ee
	In this sense, \eqref{Sugra_MTF_Ex_minSTF_L3fCompc} is another \emph{variant} version of the minimal supergravity, which we call \emph{double three-form supergravity} \cite{Farakos:2017jme}.
\end{summary}

Integrating out the complex gauge three-form $C_3$ we immediately get
\be
* G_4 = 6 \Mp^4 \lambda\,,
\ee
with $\lambda$ a complex constant, whence, on-shell, \eqref{Sugra_MTF_Ex_minDTF_L3fCompc} becomes
\be
\label{Sugra_MTF_Ex_minDTF_L3fCompOS}	
\begin{split}
	e^{-1}\tilde\call|_{\rm bos}&=\frac12 {\Mp^2} R  + 3\Mp^4|\omega|^2 \,, \qquad V = -3\Mp^4 (\lambda_1^2+\lambda_2^2)\,.
\end{split}
\ee
with again a constant, negative potential.

\subsubsection{Linear superpotential}

Consider now the more involved case where a generic number $n+1$ of complex scalars $z^a$, $a=0,\ldots, n$, including the super-Weyl compensator, enter the superpotential $\calw(z) = N_A \calv^A(z)$, $A=1,\ldots, N \leq n+1$. We may assume the matrix $\calv^A_a(z)$ to have maximal rank $N$. It is however convenient  choose `adapted' coordinates $z^a=(z^A,\tilde z^{\tilde a})$ such that $\calv^A(z)\equiv  z^A$ and the homogeneous superpotential generated by the gauge three-forms takes the form $\calw_{\rm gen}=N_A z^A$. In this picture, clearly, $\tilde z^{\tilde a}$ constitute a spectator sector, with zero super-Weyl weight.  

In the $z^A$-sector, we may isolate the compensator $u$ by setting $z^0=u$ and $z^i=u\varphi^i$, with $i=1,\ldots,n$, so that $\Pi^0=1$ and $\Pi^i=\varphi^i$.  In the spectator sector, the fields $\tilde z^{\tilde a}$ can be immediately identified with the physical fields $\varphi^{\tilde a}$. We will then collect $\varphi^\alpha = (\varphi^i, \varphi^{\tilde a})$. Then, before dualization, the superpotential takes the form $W=M^3_{\rm P}(N_0+N_i \varphi^i)$. For instance, we will encounter this kind of superpotential when we will discuss M-theory compactifications on $G_2$-holonomy spaces in Section  \ref{sec:EFT_Mth}. 

In the dual three-forms picture, where $N_A$ are promoted to dynamical variables, we then introduce $N$ gauge three-forms $A^A_3$ included in the same number of real multiplets $P^A$. Then, the constraint \eqref{Sugra_MTF} reads
\be
\label{Sugra_MTF_Ex_STF}	
\calv^A (Z) = Z^A = - \frac\ii4 (\bar \cald^2 - 8 \calr) P^A
\ee
The dual Lagrangian is 
\be
\label{Sugra_MTF_Ex_STF_L3fCompc}	
\begin{split}
	e^{-1}\tilde \call|_{\rm bos}&=\frac12 \Mp^2  R- \Mp^2 K_{\alpha \bar \beta}  \del_m \varphi^\alpha \del^m \bar \varphi^{\bar \beta} + \frac{1}{2} T_{AB} *\!F_4^A *\!F_4^B
	\\
	&\quad\, + T_{AB} \Upsilon^A *\!F_4^B  
	- \left(\hat V - \frac12 T_{AB} \Upsilon^A \Upsilon^B \right)+ e^{-1}\tilde\call_{\rm bd}\,,
	\\
	&{\rm with}\qquad   \tilde\call_{\rm bd} = \frac{1}{3!} \partial_m \left[ \varepsilon^{mnpq} {A}^A_{npq} T_{AB} (*\!F_4^B + \Upsilon^B)\right] \,.
\end{split}
\ee
The three-form kinetic matrix $T_{AB}$ is the inverse of $T^{AB}$ as defined in \eqref{Sugra_MTF_Tgf}, This can be most readily computed by observing that $D_i\Pi^0=K_i$ and $D_i\Pi^j=\delta_i^j+\varphi^i K_j$, which gives
\be
\label{Sugra_MTF_Ex_TABsingle}
\begin{split}
	T^{AB}_{\rm single}&=\left(\begin{array}{cc} T^{00} & T^{0 j}\\
		T^{i0} & T^{i j}
	\end{array}\right)
	\\
	&=2M^4_{\rm P} e^K \Re \left(\begin{array}{cc} \alpha & K^{\alpha\bar\jmath}K_\alpha+\alpha \bar \phi^{\bar\jmath}\\
		K^{i\bar \beta}K_{\bar \beta }+\alpha  \phi^{i} & K^{i{\bar\jmath}}+K^{i\bar \beta} K_{\bar \beta} \bar \phi^{\bar \jmath} + K^{\alpha\bar\jmath} K_\alpha \phi^i + \alpha \phi^i \bar \phi^{\bar\jmath}
	\end{array}\right)
\end{split}
\ee
where we have defined the quantity
\be
\alpha= K^{\bar\beta \alpha} K_{\bar\beta} K_\alpha -3\,.
\ee

\subsubsection{Maximally nonlinear case}

We now consider the maximal case, where the number of gauge three-forms $N$ equals the number of \emph{real} scalars $2n+2$ entering the superpotential $\calw = N_A \calv^A(z)$. and we are still assuming $\calv^A_a$ be of maximal rank. For the sake of generality, we keep the spectator sector $\tilde z^{\tilde a}$, with zero super-Weyl weights. Locally in field space, we can make a field-redefinition such that 
\be
\label{Sugra_MTF_Ex_maxrank}
\calv^A(z)\equiv\left(\begin{array}{c} 
	z^a \\
	\calg_a(z)
\end{array}\right)\,.
\ee
and conveniently introduce a set of $2(n+1)$ real prepotential
\be
	P^A\equiv\left(\begin{array}{r} 
	\calp^a \\ \tilde\calp_b\end{array}\right)\, ,
\ee
whose $[\cald,\bar\cald]$--components are the gauge three-forms $A_3^A$ and $\tilde A_{3A}$. The constraint \eqref{Sugra_MTF} then splits as follows: 
\begin{subequations}\label{Sugra_MTF_Ex_ZGa}
	\begin{align}
	Z^a&=S^a\equiv-\frac{\ii}{4} (\bar \cald^2 - 8\calr)\calp^a\, ,\label{Sugra_MTF_Ex_Za}\\
	\calg_{a}(Z)&=\tilde S_a=-\frac{\ii}{4} (\bar \cald^2 - 8\calr)\tilde\calp_a\, \label{Sugra_MTF_Ex_Ga}.
	\end{align}
\end{subequations}
By substituting the first condition into the second, we get
\be
(\bar \cald^2 - 8\calr)\big[\calg_{ab}(\calz)\calp^b-\tilde\calp_a\big]=0\, ,
\ee
which can be explicitly solved by setting
\be
\label{Sugra_MTF_Ex_DTF_phiSigma}
\calg_{ab}(Z)\calp^b-\tilde\calp_a\equiv -2\Sigma_a\, ,
\ee
where $\Sigma_a$ are arbitrary complex linear multiplets, obeying the constraint \eqref{ConvSG_Sigmaconstr}. We can then invert \eqref{Sugra_MTF_Ex_DTF_phiSigma} into 
\be
\label{Sugra_MTF_Ex_DTF_PPsol}
\calp^a=-2\calm^{ab}\Im\Sigma_b\;, \quad \quad \tilde\calp_a=-2\Im(\bar\calg_{ab}\calm^{bc}\Sigma_c)\; .
\ee
Hence, at the superspace level, the full Lagrangian is given by
\be\label{Sugra_DTF_L3f}
\tilde{\mathcal L}=\int \d^4\theta\,E\, \calk(S, \bar S; \tilde Z, \bar {\tilde Z}) + \left[ \int \d^2\Theta\,2\cale\, \hat \calw(Z; T) +{\rm c.c.}\right] +\tilde {\mathcal L}_{\rm bd},
\ee
with the boundary terms properly given by \eqref{Sugra_MTF_L3fbd}.

Gauge-fixing the super-Weyl invariance by first setting $z^0=u$, $z^i = u \varphi^i$ and imposing \eqref{Sugra_SW_u}, we arrive at the dual Lagrangian
\be
\label{Sugra_MTF_Ex_DTF_L3fCompc}	
\begin{split}
	e^{-1}\tilde \call|_{\rm bos}&=\frac12 \Mp^2  R- \Mp^2 K_{\alpha \bar \beta}  \del_m \varphi^\alpha \del^m \bar \varphi^{\bar \beta} + \frac{1}{2} T_{AB} *\!F_4^A *\!F_4^B
	\\
	&\quad\, + T_{AB} \Upsilon^A *\!F_4^B  
	- \left(\hat V - \frac12 T_{AB} \Upsilon^A \Upsilon^B \right)+ e^{-1}\call_{\rm bd}
	\\
	&{\rm with}\qquad   \tilde\call_{\rm bd} = \frac{1}{3!} \partial_m \left[ \varepsilon^{mnpq} {A}^A_{npq} T_{AB} (*\!F_4^B + \Upsilon^B)\right] \,,
\end{split}
\ee
where now the three-form kinetic matrix $T_{AB}$ is the inverse of the matrix
\be
\begin{split}
	T^{AB}&=\left(\begin{array}{c c c}  &  & T^{0}_{ b}
		\\
		\multicolumn{2}{c}{\smash{\raisebox{.5\normalbaselineskip}{$T_{\rm single}^{ab}$}}} & T^{i}_{b}
		\\
		T^{0}_a & T^{j}_a  & T_{a b}
	\end{array}\right)\,,
\end{split}
\ee
with $T_{\rm single}^{ab}$ as given in \eqref{Sugra_MTF_Ex_TABsingle} and the other matrix elements 
\be
\label{Sugra_MTF_Ex_TABdouble}	
\begin{aligned}
	T^{0}_{a} & = 2M^4_{\rm P} e^K \Re \left(K^{\bar \jmath \beta} K_\beta \bar \calg_{\bar a \bar \jmath} + \alpha \bar \calg_{\bar a}\right) \;,
	\\
	T^{i}_{a} & = 2M^4_{\rm P} e^K \Re \left(K^{\bar\jmath i} \bar\calg_{\bar a \bar\jmath} + K^{\bar \beta i} K_{\bar \beta} \bar\calg_{\bar a} \bar\varphi^{\bar\jmath}+ K^{\bar\jmath \beta} K_\beta  \varphi^i \bar\calg_{\bar a \bar\jmath}  + \alpha \varphi^i \bar\calg_{\bar a} \right)\;,
	\\
	T_{ab} & = 2M^4_{\rm P} e^K \Re \left( K^{\bar\jmath i}\calg_{ia} \bar\calg_{\bar\jmath \bar b}+ 2 K^{\bar\jmath \beta}  K_\beta \calg_a \bar\calg_{\bar\jmath \bar b}+\alpha \calg_a \bar \calg_{\bar b}\right)\;.
\end{aligned}
\ee

\section{Gauge four-forms and constraints}
\label{sec:Sugra_4FM}

The four-form multiplets introduced in Section~\ref{sec:GSS_4FM} in global supersymmetry may be straightforwardly introduced in supergravity as well. Since their introduction is a simple generalization of the global case, here we will be briefer, recalling the basic features of such multiplets.

Let us consider a chiral multiplet $\Gamma$, which satisfies the constraint $\bar \cald_{\dot \alpha} \Gamma = 0$. In order to implement a gauge four-form among its components, it is just sufficient to trade -- say -- the imaginary part of the auxiliary field $F_{(\Gamma)}$ of $\Gamma$ with the Hodge-dual of a real four-form $C_4$:
\be
\label{Sugra_4FM_CompF}
\Im F_{(\Gamma)} = \frac12 *\!C_4\;.
\ee
Hence, the components of a four-form multiplet are
\be
\label{Sugra_4FM_Comp}
\begin{aligned}
	\Gamma | &= \gamma \,,
	\\
	\cald^\alpha \Gamma | &= \chi^\alpha_{(\Gamma)} \,,
	\\
	-\frac14 \cald^2 \Gamma | &=   F_{(\Gamma)} \equiv \Re F_{(\Gamma)} + \frac\ii2 *\!C_4  \,.
\end{aligned}
\ee
Interpreting $\Gamma$ as a super-gauge potential, it has to transform via a super-gauge transformation. Such a transformation is of the kind
\be
\label{Sugra_4FM_GT}
\Gamma \rightarrow \Gamma + \Xi
\ee
where $\Xi$ is a single three-form multiplet as \eqref{Sugra_MTF_Ex_minSTF}, that is
\be
\label{Sugra_4FM_Xi}
\Xi = \frac14 (\bar\cald^2-8\calr) P
\ee
whose components are
\be
\label{Sugra_4FM_XiComp}
\begin{aligned}
	\Xi | &=  \left[\frac 14 (\bar \cald^2-8\calr) U \right] \bigg| &\equiv &\, \xi \,,
	\\
	\cald^\alpha \Xi | &= \cald^\alpha  \left[\frac 14 (\bar \cald^2-8\calr) U \right] \bigg|&\equiv &\, \chi^\alpha \,,
	\\
	-\frac14 \cald^2 \Xi | &=	-\frac14 \cald^2 \left[\frac 14 (\bar \cald^2-8\calr) U \right] \bigg| &=&\,   \frac12 \left(\ii *\!\d \Lambda_3 - d\right) + \ii \Im (\bar M \xi) \,.
\end{aligned}
\ee
It is then clear that a transformation of the kind \eqref{Sugra_4FM_GT} may gauge away all the components of $\Gamma$, but the gauge four-form, which transforms, as expected, as
\be
\label{Sugra_4FM_GT_c}
C_4 \rightarrow C_4 + \d \Lambda_3\,.
\ee

\subsection{Implementing constraints via three-form gaugings}
\label{sec:Sugra_G4FM}

Consider a three-form theory as \eqref{Sugra_MTF_L3fComp_gf}. Once gauge three-forms are integrated out as in \eqref{Sugra_MTF_L3fNA}, they dynamically generate any real arbitrary constants $N_A$. However, as we shall see in details in the next chapters, in typical compactification scenarios, the constants $N_A$ are indeed constrained. In particular, assume that they satisfy the \emph{linear} constraint
\be
\label{Sugra_4FM_Constr}
Q_I^A N_A + \calq_I^{\rm bg} = 0\,.
\ee
with $\calq_I^{\rm bg}$, $I=1,\ldots,L$, fixed real constants and $Q_I^A$ real $L \times (N+1)$ matrices. As already pointed out in Section \ref{sec:GSS_4FMG}, such constraints can indeed be implement dynamically into the three-form Lagrangian \eqref{Sugra_MTF_L3fComp_gf} and it is the peculiar structure of the four-form multiplets which, regarding them as Lagrange multipliers, serve to set the constraints \eqref{Sugra_4FM_Constr}. 

Preliminarily, we notice that, also in Supergravity, a term of the form
\be
\label{Sugra_4f_Constr_GammaSS}
\call_{\rm constraint} = \ii \int \d^2\Theta\,2\cale\, \left( Q_I^AN_A+ \calq_I^{\rm bg}\right) \Gamma^I + {\rm c.c.}\,,
\ee
locally supersymmetric generalization of \eqref{GSS_4f_Constr_GammaSS}, where $\Gamma^I$ are a set of $L$ four-form multiplets, serves at the scope of setting the constraint \eqref{Sugra_4FM_Constr} once we integrate out the superfields $\Gamma^I$. This is also clear at the component level, where, due to the fact that we can gauge away all the components of $\Gamma^I$ but the four-forms $C_4^I$ in their highest components, \eqref{Sugra_4f_Constr_GammaSS} simply reduces to
\be
\label{Sugra_4f_Constr_GammaComp}
e^{-1}\call_{\rm constraint}| = - \left( Q_I^AN_A + \calq_I^{\rm bg} \right) *\! C_4^I\,.
\ee
It is then immediate to recognize that integrating out the gauge four-forms $C_4^I$ sets the constraints \eqref{Sugra_4FM_Constr}.

However, the terms \eqref{Sugra_4f_Constr_GammaSS} and \eqref{Sugra_4f_Constr_GammaComp}, as they stand, are not fit to be implemented into the three-form Lagrangians \eqref{Sugra_MTF_L3fComp_gf}. There, the fluxes $N_A$ are dynamically generated, as such they are not present directly in \eqref{Sugra_MTF_L3fComp_gf} and it is pointless to constraint them as in  \eqref{Sugra_4FM_Constr} straightly from the begininng. Rather, we need a mechanism which is able to generate at once the fluxes $N_A$ along with the constraints \eqref{Sugra_4FM_Constr}. We therefore promote \eqref{Sugra_4f_Constr_GammaSS} to
\be
\label{Sugra_4f_Constr_XSS}
\call_{\rm constraint} = \ii \int \d^2\Theta\,2\cale\, \left( Q_I^AX_A+ \calq_I^{\rm bg}\right) \Gamma^I + {\rm c.c.}\,,
\ee
where $X_A$ are Lagrange multipliers. The term \eqref{Sugra_4f_Constr_XSS} can then be straightly added to the master Lagrangian \eqref{Sugra_MTF_LM}, resulting in
\be
\label{Sugra_GMTF_LM}
\begin{aligned}
	\call_{\rm master} &=\int \d^4\theta\,E\, \calk(Z, \bar Z) + \left( \int \d^2\Theta\,2\cale\, \hat \calw(Z) +{\rm c.c.}\right)\,
	\\
	&\quad\,+ \left(\int \d^2\Theta\,2\cale\, X_A\calv^A(Z)+\frac\ii8\int \d^2\Theta\,2\cale\, (\bar \cald^2 - 8 \calr)  \left[(X_A-\bar X_A)P^A \right]  +\text{c.c.}\right)
	\\
	&\quad\,+ \left[\ii \int \d^2\Theta\,2\cale\, \left( Q_I^AX_A + \calq_I^{\rm bg} \right) \Gamma^I + {\rm c.c.} \right]\;,
\end{aligned}
\ee
whose components are
\be\label{Sugra_GMTF_LMComp}
\begin{split}
	e^{-1}\call_{\rm master}|_{\text{bos}} &=-\frac16 \calk R-\calk_{a\bar b}  D_m \varphi^a D^m \varphi^{\bar b}+ \calk_{a\bar b}  f^a \bar f^{\bar b} + \left( \hat\calw_a f^a + {\rm c.c.}\right)
	\\
	&\quad\, + \Bigg[ x_A\calv^A_a f^a - \frac{1}{2 \cdot 3!\, e} \varepsilon^{mnpq} A^A_{npq} \del_m x_A  - \frac{\ii}{2} d^A x_A 
	\\
	&\quad\quad\quad\,+ f_{XA} (\calv^A - s^A) + x_A \left(\bar M \calv^A  -\Re (\bar M s^A)\right) + {\rm c.c. }\Bigg]
	\\
	&\quad\,-  (Q_I^A \,\Re x_A+ \calq_I^{\rm bg}) *\!C_4^I\;.
\end{split}
\ee
Given the master Lagrangian, we may decide to either re-get an ordinary formulation in terms of chiral multiplets or obtain a formulation where (gauged) three-forms are present. The steps to get both the formulations closely follow those of global supsersymmetry of Section \ref{sec:GSS_4FMG} and we will here be very brief.

\begin{description}
	\item[Ordinary formulation] Integrating out the real superfields $P^A$ and the four-form multiplets $\Gamma^I$ from the master Lagrangian \eqref{Sugra_GMTF_LM}, we get $X_A = N_A$, with $N_A$ real constants constrained by \eqref{Sugra_4FM_Constr}. We therefore go back at \eqref{Sugra_MTF_LSW}.
	
	\item[Gauged three-form formulation] More interestingly, we may keep the potentials $P^A$ and $\Gamma^I$ and rather integrate out the Lagrange multipliers $X_A$ and the ordinary chiral superfields $Z^a$. The latter integration reproduces \eqref{Sugra_MTF_XA}, while the former gives
	\begin{important}%
		\be
		\label{Sugra_GMTF}
		\calv^A (Z) = -\frac \ii 4 (\bar \cald^2 - 8 \calr)P^A - \ii Q_I^A \Gamma^I \equiv \hat S^A  \qquad \parbox{10em}{\color{darkred}\sf{Gauged master \\ three-form multiplet}}
		\ee
	\end{important}
	which are the master three-form multiplets introduced in \eqref{Sugra_MTF}, augmented with a gauging of the three-forms. their components differ form \eqref{Sugra_MTF_Comp} only for \eqref{Sugra_MTF_Compc}, which here becomes
	\be
	\label{Sugra_GMTF_Compc}
	\begin{aligned}
		F^A_{(S)} = \frac12 \left(*\!\hat F_4^A + \ii d^A\right) + \Re (\bar M s^A)\,,
	\end{aligned}
	\ee
	where we have defined the gauged four-form field strengths
	\be
	\label{Sugra_GMTF_F4g}
	\hat F_4^A \equiv \d A_3^A + Q_I^A C_4^I\,. 		
	\ee

	In order to recover the full bosonic Lagrangian, it is convenient to work with the bosonic components of the master Lagrangian \eqref{Sugra_GMTF_LMComp}. By closely following the same steps as in Section \ref{sec:GSS_4FMG}, after gauge-fixing the super-Weyl invariance as in \eqref{Sugra_SW_u}, we get
	\be
	\label{Sugra_GMTF_L3fComp_gf}	
	\begin{split}
		e^{-1}\tilde \call|_{\rm bos}&=\frac12 \Mp^2  R- \Mp^2 K_{i\bar \jmath}  \del_m \varphi^i \del^m \bar \varphi^{\bar \jmath} + \frac{1}{2} T_{AB} *\!\hat F_4^A *\! \hat F_4^B
		\\
		&\quad\, + T_{AB} \Upsilon^A *\!\hat F_4^B  
		- \left(\hat V - \frac12 T_{AB} \Upsilon^A \Upsilon^B \right)- \calq_I^{\rm bg}*\!C_4^I + e^{-1}\call_{\rm bd}
		\\
		&{\rm with}\qquad   \tilde\call_{\rm bd} = \frac{1}{3!} \partial_m \left[ \varepsilon^{mnpq} {A}^A_{npq} T_{AB} (*\!\hat F_4^B + \Upsilon^B)\right] \,,
	\end{split}
	\ee
	where we have employed the same definitions as in \eqref{Sugra_MTF_TUVgf}. It can be easily shown, in complete analogy with the global case \eqref{GSS_GMTF_L3fNA}-\eqref{GSS_GMTF_L3fC}, that integrating out the gauge three- and four-forms we indeed get the usual Cremmer et al. potential, specified by the superpotential \eqref{Sugra_MTF_Wphysgf}, where the constants $N_A$ generated by the gauge three-forms are constrained by \eqref{Sugra_4FM_Constr}.
\end{description}
\section{Gauge two-forms and axions}
\label{sec:Sugra_LM}

Consider a theory where, among the matter multiplets, we can single out an axionic sector made up by the chiral multiplets $T_\Lambda$. Explicitly, the kinetic part of the Lagrangian reads
\be
\label{Sugra_AL_Lc}
\call_{\rm chiral} = \int \d^4 \theta\,E\, \calk(Z, \bar Z; \Im T)\;,
\ee
which depends on the sector $T_\Lambda$ only through their imaginary parts, while keeping a generic dependence on the other chiral multiplets $Z^a$, including the super-Weyl compensator. Consistency of super-Weyl transformation with the chirality of $T_\Lambda$ requires the axionic multiplets $T_\Lambda$ to carry zero super-Weyl weight and the homogeneity properties of the kinetic section are
\be
\calk(\lambda Z, \bar \lambda \bar Z; \Im T) = |\lambda|^\frac23 \calk(Z, \bar Z; \Im T)\,.
\ee

As seen in global supersymmetry in Section \ref{sec:GSS_Axions}, there exists a dual description where the axionic multiplets are traded with linear multiplets $L^\Lambda$. In supergravity, linear multiplets satisfy the covariant constraints
\be
(\cald^2 - 8\bar \calr) L^\Lambda = 0\,,\qquad (\bar \cald^2 - 8 \calr) L^\Lambda = 0\,,
\ee
and their bosonic components are a real scalar field $l^\Lambda$ and a real field strength $\calh_3^\Lambda$ of a gauge two form $\calb_2^\Lambda$, which can be regarded as the electro-magnetic dual of the axions $\Re T_\Lambda | =a_\Lambda$. The kinetic Lagrangian \eqref{Sugra_AL_Lc}  gets replaced by
\be
\label{Sugra_AL_L}
\call_{\rm linear} = \int \d^4 \theta\,E\, \calf(Z, \bar Z; L)\,,
\ee
where $\calf(Z, \bar Z; L)$, similarly to $F(\Phi, \bar\Phi;L)$ in \eqref{GSS_AL_Dual1Lag} for the global case, is the Legendre transform of the kinetic function $\calk(Z, \bar Z; \Im T)$. Unlike their chiral counterparts, the linear multiplets $L^\Lambda$ are required to have super-Weyl weights $(\frac13,\frac13)$, whence the homogeneity property of the kinetic function are
\be
\calf(\lambda Z, \bar \lambda \bar Z; |\lambda|^{\frac23} L) = |\lambda|^\frac23 \calf(Z, \bar Z; L)\,.
\ee

In the following, after introducing a first order formalism which relates the Lagrangians  \eqref{Sugra_AL_Lc} and \eqref{Sugra_AL_L}, we shall see how linear multiplets can be gauged by the three-form multiplets introduced in Section~\ref{sec:Sugra_MTF}, along the lines of the global case examined in Section~\ref{sec:GSS_GL}.

\subsection{Axion/two-form duality in supergravity}
\label{sec:Sugra_LMDual}

In order to establish a connection between the ordinary formulation in terms of chiral multiplets with a dual formulation where the axionic multiplets are replaced with linear multiplets, we consider the super-Weyl invariant Lagrangian
\be
\label{Sugra_AL_Dual2}
\call_{\rm dual} = \int \d^4 \theta\,E\, \calf(Z, \bar Z; L) - 2 \int \d^4 \theta\,E\,  L^\Lambda \Im T_\Lambda\,.
\ee
where here $L$ are generic real multiplets. The two pictures are related as follows.

\begin{description}
	\item[Ordinary chiral formulation] The formulation in terms of the multiplets $T_\Lambda$ is obtained by integrating out the real multiplets $L^\Lambda$ from \eqref{Sugra_AL_Dual2} as
	\be
	\label{Sugra_AL_Dual2L}
	\delta L^\Lambda\,: \qquad  \Im T_\Lambda = \frac12 \frac{\del \calf}{\del L^\Lambda}\,.
	\ee
	Substituting this relation in \eqref{Sugra_AL_Dual2} gives
	\be
	\begin{aligned}
		\call_{\rm dual} &= \int \d^4 \theta\, E\,\left(\calf(Z, \bar Z; L) - \frac{\del \calf}{\del L^\Lambda}L^\Lambda \right)
		\\
		&= \int \d^4 \theta\, E\, \calk(Z, \bar Z; \Im T) = \call_{\rm chiral}\,.
	\end{aligned}
	\label{Sugra_AL_Dual2Lag}
	\ee
	The purely chiral multiplet formulation is then obtained requiring that the kinetic functions $\calk(\Phi, \bar \Phi; \Im T)$ and $\calf(\Phi, \bar \Phi; L)$ are related by a Legendre transformation.
	
	\item[Formulation with linear multiplets] In order to integrate out the chiral multiplets $T_\Lambda$, we first need to solve the chirality constraints $\cald_\alpha T_\Lambda = 0$, $\bar \cald^{\dot \alpha} \bar T_\Lambda =0$ and re-express them as
	\be
	\label{Sugra_AL_Dual2T}
	\Im T_\Lambda = \frac{1}{2\ii} \left[(\bar \cald^2 - 8 \calr) \bar \Xi_\Lambda -  (\cald^2-8 \bar\calr)  \Xi_\Lambda\right]
	\ee
	with $\Xi_\Lambda$ unconstrained (complex) superfields. Varying \eqref{Sugra_AL_Dual2} with repect to $\Xi_\Lambda$, we get the constraints
	\be
	\label{Sugra_AL_Dual2U}
	\delta \Xi_\Lambda\,, \delta \bar \Xi_\Lambda\,: \qquad   (\cald^2-8 \bar\calr)  L^\Lambda = 0\,,\quad (\bar \cald^2 - 8 \calr) L^\Lambda = 0
	\ee
	which tell that $L^\Lambda$ are linear multiplets, retrieving \eqref{Sugra_AL_L}.

	In the super-Weyl invariant formalism, the bosonic components of the Lagrangian \eqref{Sugra_AL_L} are
	{\small\be
	\label{Sugra_AL_Lbos}
	\begin{split}
		e^{-1} \call_{\rm bos} &= -\frac16\tilde \calf\,R - \calf_{a\bar b} D z^a \bar D \bar z^b  + \frac1{4} \calf_{\Lambda \Sigma} \del_\mu l^\Lambda \del^\mu l^\Sigma + \frac1{4\cdot 3!} \calf_{\Lambda \Sigma} \calh_{\mu\nu\rho}^\Lambda  \calh^{\Sigma\mu\nu\rho}
		\\
		&\quad\,+ \left(\frac\ii{2\cdot 3 !} \calf_{\bar a \Sigma} \varepsilon^{\mu\nu\rho\sigma} \calh^{\Sigma}_{\nu\rho\sigma} \bar D_\mu \bar z^{\bar a}+ {\rm c.c.}\right)\;.
	\end{split}
	\ee}
	Before gauge fixing the super-Weyl invariance, we introduce the Legendre transform of the K\"ahler potential
	\be\label{Sugra_AL_F}
	F(\varphi,\bar\varphi;\ell)=K+2\ell^\Lambda\Im t_\Lambda \, ,
	\ee
	where we have introduced the variables
	\be
	\label{Sugra_AL_lK}
	\ell^\Lambda\equiv -\frac{1}{2}\frac{\del K}{\del \Im t_\Lambda}\,,
	\ee
	related to the lowest components of the linear multiplets $l^\Lambda$ as
	\be
	\label{Sugra_AL_lgf}
	l^\Lambda =M^2_{\rm P}\, \ell^\Lambda \,.
	\ee
	In \eqref{Sugra_AL_Lbos}, we may then gauge fix the super-Weyl invariance by setting
	\be
	\label{Sugra_AL_u}
	u=\Mp^3\, e^{\frac12 \tilde F (\phi,\bar\phi)}\, , \quad {\rm with} \quad \tilde F = F -\ell^\Lambda F_\Lambda\,,
	\ee
	which is indeed analogous to \eqref{Sugra_SW_u}, because of the identification $\tilde F (\varphi,\bar\varphi; \ell) = K (\varphi,\bar\varphi; \Im t (\ell)) $.  However, we stress that gauge-fixing the super-Weyl invariance in the presence of linear multiplets is a quite involved procedure \cite{Lanza:2019xxg}, which is discussed in details in Appendix~\ref{app:SW}. The bosonic components of the gauge-fixed Lagrangian reassemble into a rather simple form
	\be
	\label{Sugra_AL_Lgf}
	\begin{split}
		e^{-1} \call_{\rm bos} &= \frac{\Mp^2}2 R  - \Mp^2 F_{i\bar \jmath}  \del \phi^i \bar \del \bar \phi^{\bar\jmath} + \frac{\Mp^2 }4 F_{\Lambda\Sigma} \del_\mu \ell^\Lambda \del^\mu \ell^\Sigma \\
		&\quad\,+ \frac{1}{4\cdot 3! \Mp^2} F_{\Lambda\Sigma}\calh_{\mu\nu\rho}^\Lambda \calh^{\Sigma\mu\nu\rho}
		+ \left\{ \frac\ii{2\cdot 3 !} F_{\bar \imath \Sigma} \varepsilon^{\mu\nu\rho\sigma}\calh^{\Sigma}_{\nu\rho\sigma} \del_\mu \bar \phi^{\bar\imath} + {\rm c.c.}\right\} \,.
	\end{split}
	\ee

\end{description}

\subsection{Gauged linear multiplets and F-term couplings}
\label{sec:Sugra_GL}

Linear multiplets do not only provide an alternative description for axionic multiplets, but they can also dynamically generate F-term couplings between the axionic multiplets and another chiral multiplet sector. To be concrete, let us consider a theory which, as above, is made up by two sectors: an axionic one $T_\Lambda$ and an ordinary chiral one $Z^a$, which also includes the super-Weyl compensator. Consider a superpotential of the form
\be
\label{Sugra_GL_W}
\calw (Z; T) =  N_A \calv^A(Z) - c_A^\Lambda T_\Lambda  \calv^A(Z)  + \hat \calw(Z)\,.
\ee
In Section \ref{sec:Sugra_MTF} we saw that $N_A$ can be interpreted as integration constants coming from integrating out a set of gauge three-forms. Hence, the first part of the superpotential is dynamically generated by exchanging, according to the recipe \eqref{Sugra_MTF}, (part of) the multiplets $Z^a$ with three-form multiplets $S^A$. 

We want to show that also the second piece of the superpotential \eqref{Sugra_GL_W} may be indeed dynamically generated in the following sense: if we move to the dual picture where the axionic multiplets $T_\Lambda$ are exchanged with linear multiplets $L^\Lambda$, the coupling term $c_A^\Lambda T_\Lambda  \calv^A(Z)$ is fully re-absorbed into a gauging of the linear multiplets $L^\Lambda$ with the real multiplets $P^A$ \cite{Dudas:2014pva}. Therefore, in a gauge fixed picture, our aim is to show that the full superpotential in a purely, ordinary chiral multiplet theory
\be
\label{Sugra_GL_Wgf}
W (\Phi; T) = \underbrace{ \Mp^3 N_A \Pi^A (\Phi)}_\text{generated by $A_3^A$}  -\underbrace{\Mp^3 c_A^\Lambda T_\Lambda  \Pi^A(\Phi)}_\text{generated by gauging $L^\Lambda$} + \underbrace{\hat W (\Phi)}_{\text{spectator}}\,.
\ee
collapses to just the spectator part $\hat W(\Phi)$ once we pass to a parent theory with three-form multiplets and gauged linear multiplets.

The joining link between a chiral theory and the dual one that we are looking for is provided once again by a master Lagrangian, which here is
{\small\be
\begin{aligned}
\label{Sugra_GL_LM}
\call_{\rm master} &= \int \d^4 \theta\,E\, \calf(Z, \bar Z; \hat L) - 2 \int \d^4 \theta\,E\,  \hat L^\Lambda \Im T_\Lambda
\\
&\quad\, + \left( \int \d^2\Theta\,2\cale\, \hat \calw(Z) +{\rm c.c.}\right)
\\
&\quad\,+ \left(\int \d^2\Theta\,2\cale\, X_A\calv^A(Z)+\frac\ii8\int \d^2\Theta\,2\cale\, (\bar \cald^2 - 8 \calr)  \left[(X_A-\bar X_A)P^A \right]  +\text{c.c.}\right)
\\
&\quad\,+ \left(\frac\ii8\int \d^2\Theta\,2\cale\, (\bar \cald^2 - 8 \calr)  \left[c_A^\Lambda (T_\Lambda-\bar T_\Lambda)P^A \right]  +\text{c.c.}\right)\,.
\end{aligned}
\ee}
The first line is just \eqref{Sugra_AL_Dual2} and the second line contains the spectator super-Weyl superpotential $\hat \calw(Z)$. The third line, directly coming from \eqref{Sugra_MTF_LM}, provides the trading between the multiplets $Z^a$ and $S^A$. The last line is new and it will be responsible for the generation of the F-term coupling. Let us now see how, from \eqref{Sugra_GL_LM}, we may retrieve either an ordinary theory or a dual one.

\begin{description}
	\item[Ordinary chiral formulation] A formulation where only chiral multiplets are present is obtained integrating out the the real superfields $P^A$ and $\hat L^\Lambda$ from the master Lagrangian \eqref{Sugra_GL_LM}. The former integration sets
	\be
	\label{Sugra_GL_dP}
	\delta P^A\,: \qquad  \Im X_A + c_A^\Lambda \Im T_\Lambda = 0\,,
	\ee
	which, employing the chirality of both $X_A$ and $T_\Lambda$, implies
	\be
	\label{Sugra_GL_NA}
	X_A = N_A - c_A^\Lambda T_\Lambda\,,
	\ee
	with $N_A$ arbitrary real constants. The variation of the master Lagrangian \eqref{Sugra_GL_LM} with respect to the real superfields $\hat L^\Lambda$ instead gives
	\be
	\label{Sugra_GL_dL}
	\delta \hat L^\Lambda\,: \qquad  \Im T_\Lambda = \frac12 \frac{\del \calf}{\del \hat L^\Lambda}\,.
	\ee
	Plugging both \eqref{Sugra_GL_NA} and \eqref{Sugra_GL_dL} into \eqref{Sugra_GL_LM} we arrive at the Lagrangian
	\be
	\label{Sugra_GL_Lchir}
	\call_{\rm chiral} = \int \d^4\theta\,E\, \calk(Z, \bar Z; \Im T) + \left( \int \d^2\Theta\,2\cale\, \calw(Z; T) +{\rm c.c.}\right)
	\ee
	with the super-Weyl invariant superpotential
	\be
	\label{Sugra_GL_LchirW}
	\calw (Z; T) = N_A \calv^A(\Phi) - c_A^\Lambda T_\Lambda \calv^A(\Phi) + \hat \calw(Z)\;.
	\ee
	\item[Formulation with linear multiplets] The steps which allow for obtaining a Lagrangian fully given in terms of (gauged) linear multiplets and three-form multiplets is definitely much more involved. At the superspace level, the first step is to integrate out the axionic chiral superfields $T_\Lambda$. Re-expressing them in terms of unconstrained complex superfields $\Xi_\Lambda$ as in \eqref{Sugra_AL_Dual2T}, the variations with respect to the latter give
	\be
	\label{Sugra_GL_Dual2U}
	\delta \Xi_\Lambda\,, \delta \bar \Xi_\Lambda\,: \qquad   (\cald^2-8 \bar\calr) (\hat L^\Lambda - c_A^\Lambda P^A) = 0\,,\quad (\bar \cald^2 - 8 \calr) (\hat L^\Lambda - c_A^\Lambda P^A) = 0\,,
	\ee
	which are solved by
	\begin{important}
	\be
	\label{Sugra_GL}
	\hat L^\Lambda =  L^\Lambda + c_A^\Lambda P^A  \,,
	\ee
	\end{important}
	with $L^\Lambda$ generic linear multiplets. The relation \eqref{Sugra_GL} defines a \emph{gauged} counterpart of the linear multiplets $L^\Lambda$ and their $[\cald,\bar\cald]$--components contain the three-form field strengths $\calh_3^\Lambda$ gauged by the three-forms $A_3^A$:
	\be
	\label{Sugra_GL_hatH3}
	\hat\calh_3^\Lambda = \calh_3^\Lambda + c_A^\Lambda A_3^A\,.
	\ee

	The variation with respect to the Lagrange multipliers $X_A$ indeed gives the same trading as in \eqref{Sugra_MTF} between the \emph{old} chiral superfields $Z^a$ and the three-form multiplets $S^A$. A final integration of the chiral multiplets $Z^a$ leads to
	\be\label{Sugra_GL_XA}
	\calv^A_a X_A = \frac14 (\bar \cald^2-8 \calr) \calf_a - \hat \calw_a\,.
	\ee
	The master Lagrangian \eqref{Sugra_GL_LM} becomes
	\be\label{Sugra_GL_L3f}
	\tilde{\mathcal L}=\int \d^4\theta\,E\, \calf(Z(P), \bar Z(P), \hat L) + \left[ \int \d^2\Theta\,2\cale\, \hat \calw(Z(P)) +{\rm c.c.}\right] +\tilde {\mathcal L}_{\rm bd},
	\ee
	with the boundary terms
	\be\label{Sugra_GL_L3fbd}
	\tilde{\mathcal L}_{\rm bd}= 
	-\frac\ii8 \left[\int\d^2\Theta\,2\cale\, (\bar \cald^2 - 8 \calr)-\int\d^2\bar \Theta\,2\bar \cale\, (\cald^2 - 8 \bar\calr)\right]\left(X_A P^A\right)
	+\text{c.c.} \,,
	\ee
	with $X_A$ defined by \eqref{Sugra_GL_XA}.  
	
	At the level of components, it is however better to work with the master Lagrangian \eqref{Sugra_GL_LM}. Within the super-Weyl invariant formalism, the procedure of integrating out the fields proceeds along the very same lines as in the previous Section \ref{sec:Sugra_LMDual} and we refer to the Appendix \ref{app:SW} for further details regarding how that procedure is generalized to supergravity. The super-Weyl invariant Lagrangian which includes both gauge two- and three-forms is
	\be
	\label{Sugra_CLgaugedb}
	\begin{split}
		e^{-1} \call_{\rm bos} &= -\frac16\tilde \calf\,R - \calf_{a\bar b} D z^a \bar D \bar z^b  + \frac1{4} \calf_{\Lambda \Sigma} \del_\mu l^\Lambda \del^\mu l^\Sigma 
		\\
		&\quad\,+ \frac1{4\cdot 3!} \calf_{\Lambda \Sigma} \hat\calh_{\mu\nu\rho}^\Lambda \hat\calh^{\Sigma\mu\nu\rho}
		+ \left(\frac\ii{2\cdot 3 !} \calf_{\bar a \Sigma}
		 \varepsilon^{\mu\nu\rho\sigma}\hat\calh^{\Sigma}_{\nu\rho\sigma} \bar D_\mu \bar z^{\bar a}+ {\rm c.c.}\right)
		 \\
		 &\quad\,+ e^{-1} \call_{\text{3-forms}}\,,
	\end{split}
	\ee
	where the three-form Lagrangrian has the same form as \eqref{Sugra_MTF_L3fComp} with, however
	\begin{subequations}\label{Sugra_GL_TUV}
		\small{\begin{align}
		T^{AB}(z,\bar z)&\equiv 2\,\Re\left(\calf^{\bar b a}\,\calv_a^A\bar\calv_{\bar b}^B\right)\,,\label{Sugra_GL_T}
		\\
		\Upsilon^A(z,\bar z)&\equiv 2\Re \left[\calf^{\bar b a} \left(\bar{\hat\calw}_{\bar b} + \frac{\ii}{2} \calf_\Lambda c_{B}^\Lambda \bar\calv^B_{\bar b} + \ii \calf_{\Lambda \bar b} c_F^\Lambda \bar \calv^{\bar F} \right) \calv_{a}^A \right]\,,\label{Sugra_GL_U}
		\\
		\begin{split}
		\hat V(z,\bar z)&\equiv \calf^{\bar b a} \left(\hat\calw_a - \frac{\ii}{2} \calf_\Lambda c_A^\Lambda  \calv^A_a - \ii \calf_{\Lambda a} c_E^\Lambda \calv^E \right) \times
		\\
		&\quad \quad\,\times \left(\bar{\hat\calw}_{\bar b} + \frac{\ii}{2} \calf_\Lambda c_{B}^\Lambda \bar\calv^B_{\bar b} + \ii \calf_{\Lambda \bar b} c_F^\Lambda \bar \calv^{\bar F} \right)
		 - \calf_{\Lambda \Sigma} c_A^\Lambda c_B^\Sigma \calv^A \bar \calv^B\,.\label{Sugra_GL_V}
		\end{split}
		\end{align}}
	\end{subequations}

	We can now proceed to gauge-fixing the super-Weyl invariance by following the same procedure described above for ungauged linear multiplets. After having imposed the Einstein frame condition  \eqref{Sugra_AL_u}, redefined the linear multiplets as in \eqref{Sugra_AL_lgf} and introduced the  Legendre transform \eqref{Sugra_AL_F} of the K\"ahler potential, we arrive at
	\be
	\label{Sugra_GL_Lgf}
	\begin{split}
		e^{-1} \call_{\rm bos} &= \frac{\Mp^2}2 R  - \Mp^2 F_{i\bar \jmath}  \del \phi^i \bar \del \bar \phi^{\bar\jmath} + \frac{\Mp^2 }4 F_{\Lambda\Sigma} \del_\mu \ell^\Lambda \del^\mu \ell^\Sigma 
		\\&\quad\,+ \frac{1}{4\cdot 3! \Mp^2} \hat\calh_{\mu\nu\rho}^\Lambda \hat\calh^{\Sigma\mu\nu\rho}
		+ \left\{ \frac\ii{2\cdot 3 !} F_{\bar \imath \Sigma} \varepsilon^{\mu\nu\rho\sigma}\hat\calh^{\Sigma}_{\nu\rho\sigma} \del_\mu \bar \phi^{\bar\imath} + {\rm c.c.}\right\} 
		\\
		&\quad\,+ e^{-1} \call_{\text{three-forms}}\,,
	\end{split}
	\ee
	where the three-form Lagrangian has the same form as \eqref{Sugra_MTF_L3fComp} with now
	\begin{subequations}\label{Sugra_GL_TUVgf}
		\small{\begin{align}
		T^{AB}&\equiv 2M^4_{\rm P}\, e^{\tilde F}\,\Re\left({F}^{\bar \jmath i}\,D_i  \Pi^A \bar D_{\bar \jmath} \bar\Pi^B-(3 - \ell^\Lambda \tilde F_\Lambda) \Pi^A\bar\Pi^B\right)\,,\label{Sugra_GL_Tgf}
		\\
		\begin{split}
		\Upsilon^A&\equiv 2M_{\rm P}\, e^{\tilde F}\Re \Bigg[{F}^{\bar \jmath i}\,\left( D_{\bar \jmath}  \bar{\hat W}  + \frac{\ii}{2} M_{\rm{P}}^3 F_\Lambda c_{B}^\Lambda D_{\bar \jmath} \bar \Pi^B + \ii M_{\rm{P}}^3 F_{\Lambda \bar \jmath} c_B^\Lambda \bar \Pi^{\bar B}\right) D_i\Pi^A 
		\\
		&\quad\quad\, - (3 - \ell^\Lambda \tilde F_\Lambda) \left(\bar{\hat W}+ \frac\ii2 M_{\rm{P}}^3 F_\Lambda c_D^\Lambda \bar\Pi^D\right) \Pi^A - \ii M_{\rm{P}}^3 \tilde F_{\Lambda} c_B^\Lambda \bar \Pi^{\bar B} \Pi^A\Bigg]\,,\label{Sugra_GL_Ugf}
		\end{split}
		\\
		\begin{split}
		\hat V&\equiv \frac{e^{\tilde F}}{M_{\rm P}^2} {F}^{\bar \jmath i} \left( D_{i}  {\hat W} - \frac{\ii}{2} M_{\rm{P}}^3 F_\Lambda c_{A}^\Lambda D_{i} \Pi^A - \ii M_{\rm{P}}^3 F_{\Lambda i} c_A^\Lambda \Pi^{A} \right) \times
		\\
		&\quad\,\qquad\qquad  \times \left(D_{\bar \jmath}  \bar{\hat W} + \frac{\ii}{2} M_{\rm{P}}^3 F_\Lambda c_{B}^\Lambda D_{\bar \jmath} \bar \Pi^B + \ii M_{\rm{P}}^3 F_{\Lambda \bar \jmath} c_B^\Lambda \bar \Pi^{\bar B} \right)
		\\
		&\quad\,- (3 - \ell^\Lambda \tilde F_\Lambda) \frac{e^{\tilde F}}{M_{\rm P}^2} \left|\hat W - \frac\ii2 M_{\rm{P}}^3 F_\Lambda c_A^\Lambda \Pi^A \right|^2 - M_{\rm{P}}^4 e^{\tilde F} F_{\Lambda \Sigma} c_A^\Lambda c_B^\Sigma \Pi^A \bar \Pi^B
		\\
		&\quad\,- M_{\rm{P}} e^{\tilde F}  \left[ - \ii c_B^\Sigma \tilde F_\Sigma \Pi^B \left( \bar{\hat W} + \frac\ii2 M_{\rm{P}}^3 F_\Lambda c_A^\Lambda \bar\Pi^A \right)+{\rm c.c.}\right] .\label{Sugra_GL_Vgf}
		\end{split}
		\end{align}}
	\end{subequations}

\end{description}

\section{Quantization conditions and dualities}
\label{sec:Sugra_DualQuant}

We conclude this chapter discussing two important properties that the actions introduced above enjoy: the compactness of the gauge symmetry, which allows us to infer the quantization conditions for the constants $N_A$, and the symmetries of the actions, which will be related, in the next chapter, to the BPS--extended objects.

\subsection{Three-forms and quantization conditions}
\label{sec:Sugra_Quant}

In the discussion so far, the integration constants $N_A$ introduced in \eqref{Sugra_MTF_L3fNA} in the dual three-form picture, or in the superpotential \eqref{Sugra_MTF_Wphys} in the ordinary chiral multiplet theory, have been treated as arbitrary real parameters. However, this is really so if the gauge three-forms $A_3^A$ are associated to  non-compact gauge symmetries. On the other hand, constructions from string theory, as well as purely four-dimensional arguments (for instance in \cite{Banks:2010zn}), indicate that in consistent quantum gravitational theories all gauge symmetries should be compact. In practice, given any compact three-dimensional submanifold $\cale$, this means that the integrals
\be
\label{Sugra_Quant_q}
{\frak q}^A  \equiv \int_\cale A^A_3\;,
\ee
are periodic. It is then natural to normalize the gauge three-form fields so that \eqref{Sugra_Quant_q} have  $2\pi$--periodicity.

The compactness of the gauge symmetries implies quantization conditions on the corresponding field strengths. As in \cite{Kaloper:2011jz}, a simple way to identify these conditions is to relate our system to a  $1$-dimensional theory by performing the dimensional reduction of the four-dimensional theory $\mathbb{R}\times \cale$ along the compact space-like $\cale$.  Let us focus on the four-form part of the bosonic action \eqref{Sugra_MTF_L3fComp_gf}, namely 
\be
\int_{\mathbb{R}} \d t\, L_{\rm flux}\equiv- \frac12 \int_{\mathbb{R}\times\cale} T_{AB} F_4^A * F_4^B
\ee
where $t$ parametrizes the time direction $\mathbb{R}$.
In the  one-dimensional effective theory we can  compute the momenta conjugated to the angles ${\frak q}^A$, namely 
\be\label{Sugra_Quant_p}
\begin{aligned}
	{\frak p}_A&\equiv \frac{\del L_{\rm flux}}{\del\dot{\frak q}^A}=- T_{AB} * F_4^B\;,
\end{aligned}
\ee
where $\dot {\frak q}^A$ denote the  derivatives of the angles with respect to $t$. Quantum mechanically, the momenta must be integrally quantized, that is ${\frak p}_A \in\mathbb{Z}$, since the angles are $2\pi$-periodic.
On the other hand, by comparing \eqref{Sugra_Quant_p} and \eqref{Sugra_MTF_L3fNA} we see that  ${\frak p}_A$ coincide  with the integration constants $N_A$, leading to the quantization condition
\be\label{bulkquant}
N_A \in\mathbb{Z} \, . 
\ee
This shows how  the compactness of the gauge symmetries  implies the quantization of the constants appearing in the effective superpotential \eqref{Sugra_MTF_Wphys}.
This is in agreement with what is expected from explicit string theory models, in which  $N_A$ usually measure quantized internal fluxes, as we will see in details in Chapter~\ref{chapter:EFT}. However, we stress that  the three-form formulation has allowed for a purely four-dimensional derivation of this fact.

\subsection{Symmetries of the dual action}
\label{sec:Sugra_Dual}

In general, there may exist a duality group $G_{\rm dual}$ of transformations which acts on the quantized constants $N_A$ as well as on the EFT fields.  Hence, in a traditional  EFT depending on fixed non-vanishing fluxes $N_A$, part or all of the duality group is explicitly  broken. By using the dual  formulation in terms of three-forms, the constants $N_A$ are traded for dynamical three-forms $A^A_3$. In such a formulation $G_{\rm dual}$ acts as an actual symmetry group of the action, which is  only {\em spontaneously} broken.

More concretely, given a homogeneous kinetic section $\calk(z,\bar z)$ and a superpotential \eqref{Sugra_MTF_calwgen}, a class of  dualities are given by isometries $z^a\rightarrow z'^a$ which leave $\calk(z,\bar z)$ and $\hat\calw(z)$ invariant and act linearly on the periods 
\be\label{perduality}
\calv^A(z)\rightarrow \calv^A(z')=R^A{}_B\calv^B(z)\, .
\ee 
In the presence of non-vanishing  flux quanta $N_A$, the superpotential   \eqref{Sugra_MTF_calwgen} is clearly not invariant under such a transformation, which may be regarded as a `spurionic' symmetry if $N_A$ transform in an opposite way: 
$N_A\rightarrow N_A'=(R^{-1})^B{}_AN_B$.\footnote{Notice that, by imposing the Weyl-fixing and splitting the homogeneous periods $\calv^A$ as in \eqref{Sugra_MTF_WcalvA}, the periods $\Pi^A$ may transform linearly only up to a pre-factor, which should then be compensated by a K\"ahler tranformation of $K$.} Recalling the procedure followed for constructing the three-form formulation, or more directly from  explicit formulas like \eqref{Sugra_MTF_L3fComp_gf}, with \eqref{Sugra_MTF_TUVgf}, or \eqref{Sugra_CLgaugedb}, with \eqref{Sugra_GL_TUV}, it is clear that the three-form theory is exactly invariant under the duality transformation provided that the three-form potentials transform as follows, 
\be
A^A_3\rightarrow A'^A_3=R^A{}_BA^B_3\, .
\ee
Such a symmetry is  spontaneously broken once a certain vacuum sector specified by  \eqref{Sugra_MTF_L3fNA} is selected. 

Another class of possible duality transformations appear in the models with superpotentials of the forms \eqref{Sugra_GL_W}. These are associated with the integral shifts $t_\Lambda\rightarrow t_\Lambda +n_\Lambda$, with $n_\Lambda\in\mathbb{Z}$, and are explicitly broken, even though the explicit breaking may be compensated by shifting $N_A$  to $N_A'=N_A+c_A^\Lambda n_\Lambda$. 
To analyze this case, let us consider the corresponding Weyl-invariant EFT  with three-forms $A_3^A$  and chiral fields $(z^a,t_\Lambda)$. By extending the formulas \eqref{Sugra_MTF_TAB}--\eqref{Sugra_MTF_UA} in order to include the chiral fields $t_\Lambda$ in an obvious way and to replace $\hat\calw$ with  $\hat\calw'$ as defined in \eqref{Sugra_GL_W}, it is immediate to check that a  shift $t_\Lambda\rightarrow t_\Lambda+n_\Lambda$ induces the shifts
$\Upsilon^A\rightarrow \Upsilon^A-n_\Lambda c^\Lambda_B T^{AB}$ and $\hat V\rightarrow \hat V-n_\Lambda c^\Lambda_B h^B+\frac12 n_\Lambda n_\Sigma   c^\Lambda_Ac^\Sigma_BT^{AB}$. It  follows that the three-form action \eqref{Sugra_MTF_L3fComp_gf} is exactly invariant under such transformations. Hence, also in this case, the duality transformations are \emph{proper} symmetries of the three-form action, which are only spontaneously broken by the choice of a vacuum sector. We also notice that, on the one hand, the same conclusion would hold also in the presence of corrections depending on $e^{2\pi\ii k^\Lambda t_\Lambda}$, with $k^\Lambda\in\mathbb{Z}$, which break the continuous shift symmetries, but preserve the discrete ones.  On the other hand, in absence of such corrections, one may make a further step and dualize the chiral fields $t_\Lambda$ to the linear multiplet bosons  $(l^\Lambda,\calb^\Lambda_2)$. In the resulting formulation the original shift symmetries are  traded for the compact gauge symmetry of the two-form potentials $\calb^\Lambda_2$.

\chapter{BPS--extended objects and domain walls in supergravity}
\label{chapter:Sugra_ExtObj}

Having introduced gauge two-, three- and four-forms into generic supergravity theories, we may couple them to BPS--strings, membranes and 3-branes. As was explained in Chapter~\ref{chapter:ExtObj} for globally supersymmetric theories, the key elements that we need to introduce to properly couple such BPS--objects are super-gauge forms. These allow us to write down the actions coupling the BPS--objects to the bulk sector in a manifestly supersymmetric manner. Once the charges of the objects are specified, the form of their action is strictly determined by $\kappa$--symmetry. Indeed, as we will see, the construction of their BPS actions proceeds along the same lines as for the global case of Chapter \ref{chapter:ExtObj}. Therefore, in this chapter we will be very brief and we will focus on highlighting the differences of the local case with the respect to the globally supersymmetric one.

Particular attention will be devoted to membranes. As already explored in the previous chapters, membranes determine spacetime regions with different potential for the scalar fields. As shown in Section~\ref{sec:ExtObj_DWM} for global supersymmetry, new domain wall solutions, which are supported by membranes, may appear within a theory. Presently, we will generalize those constructions to supergravity, highlighting the modifications induced by gravity.

\section{Extended BPS--objects in supergravity}
\label{sec:Sugra_ExtObj}

In supergravity we may introduce the very same hierarchy of BPS--objects that we included in globally supersymmetric theories in Chapter~\ref{chapter:ExtObj}. We shall introduce the BPS--objects by increasing codimension, starting with 3-branes, then moving to membranes and finally strings. We will also show how to construct BPS string/membrane junctions as well as BPS membrane/3-brane regions.

\subsection{Supersymmetric 3-branes}
\label{sec:Sugra_ExtObj_3b}

We start with supersymmetric spacetime filling 3-branes, with no boundary. They are specified by the super-embedding
\be
\xi^i \quad \mapsto \quad  {\frak z}^M = (x^m(\xi), \theta^\alpha(\xi),\bar\theta_{\dot\alpha}(\xi))\;,
\ee
where $\xi^i$, with $i=0,\ldots,3$, are the 3-brane worlvolume coordinates. The action of a supersymmetric spacetime filling 3-brane is given by
\begin{important}%
\be
\label{Sugra_ExtObj_3b_S}
S_{\text{3-brane}} = \mu_I\int_\calw {\bf C}^I_4\; ,
\ee
\end{important}
where $\calw$ is the 3-brane world-volume and $\mu_I$ denotes the charges of the 3-branes. The super-four-forms ${\bf C}^I_{4}$ are a covariantization of \eqref{ExtObj_3B_C4}
{\small\begin{eqnarray} 
\label{Sugra_ExtObj_3b_C4}   
{\bf C}^I_{4}   &=&   2 E^b\wedge E^a \wedge E^\alpha \wedge E^\beta \sigma_{ab\; \alpha\beta}\bar{ \Gamma}^I  +  2 E^b\wedge E^a \wedge \bar E^{\dot\alpha} \wedge \bar  E^{\dot\beta }\bar{\sigma}_{ab\; \dot{\alpha}\dot{\beta}} \Gamma^I \qquad \nonumber \\  && {-} \frac{1}6  E^c\wedge E^b\wedge E^a \wedge \bar E^{\dot\alpha} \epsilon_{abcd} \sigma^d_{\alpha\dot\alpha} {\cal D}^{\alpha}{ \Gamma^I}  +\frac{1}6 E^c\wedge E^b\wedge E^a \wedge E^\alpha \epsilon_{abcd} \sigma^d_{\alpha\dot\alpha} \bar{{\cal D}}^{\dot\alpha}\bar{ \Gamma}^I \nonumber \\  &&  +\frac{\ii}{96} E^{d} \wedge E^c \wedge E^b \wedge E^a \epsilon_{abcd} \left[({\cal D}^2-24\bar{\calr})
{ \Gamma^I} -(\bar{{\cal D}}^2-24\calr)
\bar{ \Gamma}^I \right]
\,.
\end{eqnarray}}
These are constructed out of the four-form multiplets $\Gamma^I$ introduced in Section \ref{sec:Sugra_4FM}, which contain, among their components \eqref{Sugra_4FM_Comp}, real gauge four-forms $C_4^I$. It is immediate to show that, using \eqref{Sugra_4FM_Comp}, the purely bosonic components of \eqref{Sugra_ExtObj_3b_C4} are
\be
{\bf C}^I_{4}   | = C_4^I = \frac{1}{4!} C^I_{mnpq} \d x^m \wedge \d x^n \wedge \d x^p \wedge \d x^q\;.
\ee
The super-four-forms \eqref{Sugra_ExtObj_3b_C4} are the \emph{unique} closed super-forms which can be constructed with the four-form multiplets $\Gamma^I$. As explained in Section \ref{sec:ExtObj_3-branes}, the closure of ${\bf C}^I_{4}$, combined with the fact that the 3-brane has no boundary, makes the action \eqref{Sugra_ExtObj_3b_S} to be trivially invariant under any superdiffeomorphism and hence, in particular, under any $\kappa$-transformation. Being the $\kappa$-parameters unconstrained, we can trivially choose the $\kappa$-parameters to coincide, over the full 3-brane worldvolume, with the local supersymmetry parameters. Therefore, the 3-brane specified by the action \eqref{Sugra_ExtObj_3b_S} preserves the \emph{complete} $\caln=1$ bulk supersymmetry.

It may seem quite awkward that we have not included any Nambu-Goto term to the action \eqref{Sugra_ExtObj_3b_S}. In Section~\ref{sec:LandSwamp_Hierarchy_IIB_S3b}, for Type IIB EFTs, we will give a top-down justification of why (and under which hypotheses) an action of the sort of \eqref{Sugra_ExtObj_3b_S} may be born from a complete 3-brane action, which also include a Nambu-Goto term. For the moment let us notice that, eventually, we may also add a supersymmetric Nambu-Goto term in the action \eqref{Sugra_ExtObj_3b_S}, but it turns out that such an addition comes with a huge price: supersymmetry is fully spontaneously broken, being it realized only non-linearly. Such a 3-brane can be identified with a \emph{Goldstino brane} of the kind introduced in \cite{Bandos:2015xnf,Bandos:2016xyu}. 

As we shall see in Chapter~\ref{chapter:LandSwamp}, the aforementioned problem of including a Nambu--Goto term intertwines with the definition of the charges $\mu_I$. Within the four-dimensional description that we are considering throughout this chapter, nothing seems to strictly fix the sign of the charges $\mu_I$: whatever sign we choose for $\mu_I$, the action \eqref{Sugra_ExtObj_3b_S} is still invariant under the full $\caln=1$ local supersymmetry. But we do know that, in typical (orientifolded) string compactifications, where both 3-branes and anti-3-branes are present, only one choice of signs for $\mu_I$ is consistent with a linear realization of the $\caln=1$ local supersymmetry, the opposite choice being associated to the Goldstino branes. It would then be interesting to see whether such correlation could be understood also at pure four-dimensional level.

Furthermore, since the 3-branes that we are considering here do fill entire spacetime regions, the worldvolume fields living on them cannot be generically neglected. Nevertheless, in this work, we shall assume that this is the case. We treat the action \eqref{Sugra_ExtObj_3b_S} as a topological term once coupled to a bulk action and then, apart from contributing to the total four-form charge $\calq_I$ by $\mu_I$, it does not contain much more physical content. Due to its peculiar nature of being completely decoupled from the rest, the action \eqref{Sugra_ExtObj_3b_S} can be added to any supergravity action.


\subsection{Supersymmetric membranes}
\label{sec:Sugra_ExtObj_Memb}

We now pass to supersymmetric objects of codimension one, the BPS--membranes. Membranes are described by three worldvolume coordinates $\xi^i$, with $i=0,1,2$ and, as objects living in the ambient superspace, are defined via the super-embedding
\be
\xi^i \quad \mapsto \quad  {\frak z}^M = (x^m(\xi), \theta^\alpha(\xi),\bar\theta_{\dot\alpha}(\xi))\;.
\ee

As in Section~\ref{sec:ExtObj_MembSusy}, we would like to couple membranes minimally to some gauge three-forms $A_3^A$ with charges $q_A$. Therefore, to begin with, we need a proper generalization to superspace Wess-Zumino term which appears in \eqref{Intro_Jumps_Memb}. This promotion of the Wess-Zumino term is obtained, as explained in Section~\ref{sec:ExtObj_MembSusy}, by introducing a super-gauge three-form ${\bf A}^A_3$:
\be
\label{Sugra_ExtObj_M_Swz}
S_{\text{memb,WZ}} = q_A\int_\calm {\bf A}^A_3\;,
\ee
where $\calm$ is the membrane worldvolume. We already saw in Section~\ref{sec:Sugra_MTF} that the real multiplets $P^A$ are the proper suspersymmetric objects that contain a real three-form among their components (see, for example, \eqref{ConvSG_P}). The super-three-form ${\bf A}^A_3$ is then built as a manifestly supersymmetric object with the supervielben $E^A$ and the real multiplets $P^A$ as
\be
\label{Sugra_ExtObj_M_A3}
\begin{aligned}
	{\bf A}^A_{3}=&\,  { -}2 {\ii} E^a \wedge E^\alpha \wedge \bar E^{\dot\alpha}  \sigma_{a\alpha\dot\alpha}P^A +   E^b\wedge E^a \wedge  E^\alpha
	\sigma_{ab\; \alpha}{}^{\beta}{\mathcal D}_{\beta}P^A \\ &+ E^b\wedge E^a \wedge  \bar E^{\dot\alpha}
	\bar\sigma_{ab}{}^{\dot\beta}{}_{\dot\alpha}\bar{\mathcal D}_{\dot\beta}P^A  
	\\&+\frac {1} {24} 
	E^c \wedge E^b \wedge E^a \epsilon_{abcd} \,\left(\bar{\sigma}{}^{d\dot{\alpha}\alpha}
	[{\mathcal D}_\alpha, \bar {\mathcal D}_{\dot\alpha}]P^A+8G^d P^A\right)
	\, . \qquad
\end{aligned}
\ee
Such a super-three-form is the covariantization of \eqref{ExtObj_Memb_A3} and it is immediate to show that, using the components \eqref{ConvSG_P}, if we restrict ourselves to the bosonic components
\be
\label{Sugra_ExtObj_M_A3b}
{\bf A}^A_{3} |_{\text{bos}} = A_3^A = \frac1{3!} A^A_{mnp} \d x^m \wedge \d x^n \wedge \d x^p\,.
\ee
The closed four-form form super-field strength is given by 
\be
\label{Sugra_ExtObj_M_F4}
\begin{aligned}
	{\bf F}^A_{4}   &= \dwb {\bf A}^A_{3}= \;   2 {\ii}  E^b\wedge E^a \wedge E^\alpha \wedge E_\beta \sigma_{ab\; \alpha}{}^\beta\bar S^A  - 2 {\ii}  E^b\wedge E^a \wedge \bar E_{\dot \alpha} \wedge \bar E^{\dot \beta} \bar \sigma_{ab}{}^{\dot\alpha}{}_{\dot\beta} S^A
	\\  
	&\quad\, + \frac{\ii}6  E^c\wedge E^b\wedge E^a \wedge \bar E^{\dot\alpha} \epsilon_{abcd} \sigma^d_{\alpha\dot\alpha} {\cal D}^{\alpha}{S}^A
	\\
	&\quad\, +\frac{\ii}6 E^c\wedge E^b\wedge E^a \wedge E^\alpha \epsilon_{abcd} \sigma^d_{\alpha\dot\alpha} \bar{{\cal D}}^{\dot\alpha}\bar{S}^A 
	\\  
	&\quad\, + \frac{1}{96} E^{d} \wedge E^c \wedge E^b \wedge E^a \epsilon_{abcd} \left[({\cal D}^2-24\bar{\calr})
	{ S}^A+(\bar{{\cal D}}^2-24\calr) 
	\bar{S}^A \right]
		\,, \qquad
\end{aligned}
\ee
which is the covariantization of \eqref{ExtObj_Memb_F4}.\footnote{Comparing this with \eqref{Sugra_ExtObj_3b_C4}, these coincide provided the identification $S^A \leftrightarrow \ii \Gamma^I$. The mismatch of the $\ii$-factor is just due to the convention we are using for the four-form multiplets $\Gamma^I$ in \eqref{Sugra_4FM_Comp}.} Importantly enough, such super-four-forms do depend on the three-form multiplets $S^A$ \eqref{Sugra_MTF}, which are gauge-invariant by construction. Restricting to the bosonic components, we obtain
\be
\label{Sugra_ExtObj_M_F4b}
{\bf F}^A_{4} |_{\text{bos}} = \dwb {\bf A}^A_{3} |_{\text{bos}} = - \d A_3^A = - \frac1{3!} \del_{[m} A^A_{npq]} \d x^m \wedge \d x^n \wedge \d x^p \wedge \d x^q\;.
\ee
The construction of this kind of super-three-forms and their super-field strengths in supergravity was firstly performed in \cite{Bandos:2010yy,Bandos:2011fw,Bandos:2012gz} and later generalized in
\cite{Bandos:2018gjp,Bandos:2019wgy,Bandos:2019khd,Bandos:2019lps}.  It is important to stress here that the very possibility to relate \eqref{Sugra_ExtObj_M_A3} and \eqref{Sugra_ExtObj_M_F4} via the relation ${\bf F}^A_{4}= \dwb {\bf A}^A_{3}$ comes from the definition of the multiplets \eqref{Sugra_MTF} and how those are related to the real prepotentials $P^A$. Taking $S^A$ as general chiral multiplets, quite similarly to what we did in \eqref{Sugra_ExtObj_3b_C4}, may allow for defining a closed four-form, but this, not being exact in the superspace de Rham cohomology, does not define a field strength.

As explained in details in Section \ref{sec:ExtObj_MembSusy}, the sole action \eqref{Sugra_ExtObj_M_Swz} spontaneously breaks all the four generators of the bulk $\caln=1$ supersymmetry. In order to overcome this problem, we need to add another contribution to the Wess-Zumino term such that the resulting action enjoys a fermionic gauge symmetry, the $\kappa$-symmetry, of the kind of \eqref{ExtObj_FrM_kappa}.  Assuming that the membrane has \emph{no} boundary -- a request which will be relaxed in the later Section~\ref{sec:Sugra_ExtObj_MembString} -- it turns out that this can be achieved by adding
\be
\label{Sugra_ExtObj_Memb_Sng}
S_{\rm memb,NG}=-2\int_\Sigma \d^3\zeta\,|q_A S^A|\sqrt{-\det {\bf h}}
\ee
to \eqref{Sugra_ExtObj_M_Swz}, where $q_A$ are the same charges that appear in the Wess-Zumino term \eqref{Sugra_ExtObj_M_Swz} and we have defined ${\bf h}_{ij}\equiv \eta_{ab}E^{a}_iE^{b}_j$, with $E^{a}_i$  the pull-back of the bulk super-vielbein to the worldvolume $\calm$. Therefore, the BPS--action of a membrane minimally coupled to the gauge three-forms $A_3^A$, in the super-Weyl invariant formalism is
\begin{important}
	\be
	\label{Sugra_ExtObj_Memb_S}
	S_{\rm memb}=-2\int_\calm \d^3\xi\,|q_A S^A|\sqrt{-\det {\bf h}}+q_A\int_\calm{\bf A}^A_3\;,
	\ee
\end{important}
which, in turn, defines the membrane tension
\be
\label{Sugra_ExtObj_Memb_Tm}
\calt_{\rm memb}= 2 |q_A s^A| = 2 |q_A \calv^A(\varphi)|\;,
\ee
where the fields are assumed to be evaluated over the membrane worldvolume. The action \eqref{Sugra_ExtObj_Memb_S} naturally couples to the bulk action defined by \eqref{Sugra_MTF_L3fComp}. As its globally supersymmetric counterpart \eqref{ExtObj_Memb_S}, the action \eqref{Sugra_ExtObj_Memb_Tm} enjoys the local fermionic $\kappa$--symmetry \eqref{ExtObj_FrM_kz}, with the parameter $\kappa^\alpha$ satisfying the projection condition
\be
\label{Sugra_ExtObj_Memb_kappa}
\begin{split}
	&\kappa_\alpha =- \frac {q_A S^A}{|q_A S^A|} ({\Gamma}\bar{\kappa})_\alpha\; , \qquad \bar\kappa^{\dot\alpha} =  - \frac {q_A \bar S^A}{|q_A S^A|} (\bar{\Gamma}{\kappa})^{\dot\alpha}\,.
\end{split}
\ee
As a result, the action \eqref{Sugra_ExtObj_Memb_S} spontaneously breaks half of the bulk supersymmetry generators, making it possible to linearly realize only a three-dimensional $\caln=1$ supersymmetry over the membrane worldvolume. Furthermore, as in the global case,  \eqref{Sugra_ExtObj_Memb_S} is also invariant under worldvolume reparametrizations \eqref{ExtObj_FrM_xi} by construction. Additionally, however, the bulk super-diffeomorphism invariance may allow to always choose a frame where the membrane is located at $z=0$ and looks static.

\begin{summary}
Gauge-fixing the super-Weyl invariance as in \eqref{Sugra_SW_u}, the membrane action becomes	
	\be
	\label{Sugra_ExtObj_Memb_Sgf}
	S_{\rm memb}=-2 \Mp^3 \int_\calm \d^3\xi\, e^{\frac{K}2}|q_A \Pi^A(\Phi)|\sqrt{-\det {\bf h}}+q_A\int_\calm{\bf A}^A_3\;,
	\ee
which defines the gauge-fixed, physical membrane tension
\be
\label{Sugra_ExtObj_Memb_Tmgf}
\calt_{\rm memb}= 2  \Mp^3 e^{\frac{K}2}|q_A \Pi^A(\varphi)|\;.
\ee
The action \eqref{Sugra_ExtObj_Memb_Sgf} couples to the bulk contribution \eqref{Sugra_MTF_L3fComp_gf}.
\end{summary}


\subsection{BPS--3-brane/membrane region}
\label{sec:Sugra_ExtObj_3bMemb}

In the previous Section~\ref{sec:Sugra_ExtObj_3b} we have just considered spacetime filling 3-branes, whose simple action \eqref{Sugra_ExtObj_3b_S}, combined with the absence of any boundary, trivially preserves the full $\caln=1$ supersymmetry. But let us now relax this assumption and assume that the 3-brane does not fill the spacetime entirely, but just a \emph{region} thereof delimited by a BPS--membrane of the kind studied in Section \ref{sec:Sugra_ExtObj_Memb}, that is $\del \calw = \calm$ -- see Fig.~\ref{fig:MembD3}. 

The presence of a nontrivial boundary makes the action \eqref{sec:Sugra_ExtObj_3b} no more invariant under bulk superdiffeomorphisms, owing to the contribution
\be
\label{Sugra_ExtObj_M3B_vmemb}
\delta_\kappa S_{\text{3-brane}} = - \mu_I \int_{\del\calw} i_\kappa  {\bf C}^I_4 =  - \mu_I \int_{\calm} i_\kappa  {\bf C}^I_4\;,
\ee
which is localized over the membrane serving as a boundary. However, at the boundary, the variation of the membrane action has also to be taken into account, Under the action of $\kappa$-symmetry transformations, the full variation of the combined 3-brane/membrane action is
\be
\label{Sugra_ExtObj_M3B_vtotb}
\delta_\kappa S_{\text{memb}} + \delta_\kappa S_{\text{3-brane}} =  \delta S_{\text{memb,NG}}|_{\kappa} - q_A \int_{\calm} i_{\kappa} \left( - {\bf F}_4^A + Q^A_I {\bf C}_4^I \right)\;,
\ee
where we have decomposed $\mu_I = Q_I^A q_A$. We can make the variation vanish if and only if we choose the membrane Nambu--Goto term as
\be
\label{Sugra_ExtObj_M3B_Sng}
S_{\text{memb,NG}}= -2 \int_\calm\d^3\xi\, |q_A  \hat S^A|\sqrt{-\det {\bf h}} 
\ee
where, with respect \eqref{Sugra_ExtObj_Memb_S}, the three-form multiplets appear in their \emph{gauged} version \eqref{Sugra_GMTF}.

Therefore, in the super-Weyl invariant formalism, the full supersymmetric action of a 3-brane region delimited by a BPS membrane, coupled to the bulk scalars $s^A$ and charged under the gauge three-forms $A_3^A$, is
\be
\label{Sugra_ExtObj_M3B_region}
\begin{split}
	S_{\text{BPS-region}} &=-2\int_\calm \d^3\xi\,|q_A \hat S^A|\sqrt{-\det {\bf h}}+q_A\int_\calm{\bf A}^A_3 + q_A Q^A_I\int_\calw {\bf C}^I_4\;,
\end{split}
\ee
with the superfields $\hat S^A$ defined in \eqref{Sugra_GMTF}.

We also notice that the action \eqref{Sugra_ExtObj_M3B_region} is invariant under the combined gauge transformation:
\begin{subequations}
	\label{Sugra_ExtObj_M3B_sgforms}
	\begin{align}
	{\bf C_4}^I&\quad\rightarrow\quad {\bf C_4}^I+\d {\bf \Lambda}^I_3\,,\\
	{\bf A}^A_3&\quad\rightarrow\quad {\bf A}^A_3-q_AQ^A_I{\bf \Lambda}^I_3\,,
	\end{align}
\end{subequations}
with
\be
\label{Sugra_ExtObj_M3B_s3gaugeb}
\begin{aligned}
	{\bf \Lambda}^I_3=&\,  { -}2 {\ii} E^a \wedge E^\alpha \wedge \bar E^{\dot\alpha}  \sigma_{a\alpha\dot\alpha}\Xi^I +  E^b\wedge E^a \wedge  E^\alpha
	\sigma_{ab\; \alpha}{}^{\beta}{\cald}_{\beta}\Xi^I \\ &+ E^b\wedge E^a \wedge  \bar E^{\dot\alpha}
	\bar\sigma_{ab}{}^{\dot\beta}{}_{\dot\alpha}\bar{\cald}_{\dot\beta}\Xi^I
	\\&+\frac {1} {24} 
	E^c \wedge E^b \wedge E^a \epsilon_{abcd} \,\left(\bar{\sigma}{}^{d\dot{\alpha}\alpha}
	[\cald_\alpha, \bar{\cald}_{\dot\alpha}]\Xi^I+8 G^d\Xi^I \right)
	\, . \qquad
\end{aligned}
\ee

\begin{summary}
	After gauge-fixing the super-Weyl invariance as in \eqref{Sugra_SW_u}, we get
	\be
	\label{Sugra_ExtObj_M3B_regiongf}
	\begin{split}
		S_{\text{BPS-region}} &=-2\int_\calm \d^3\xi\,e^{\frac{K}2}|q_A \hat \Pi^A(\Phi)|\sqrt{-\det {\bf h}}+q_A\int_\calm{\bf A}^A_3 
		\\
		&\quad\,+ q_A Q^A_I\int_\calw {\bf C}^I_4
	\end{split}
	\ee
	with the superfields $\hat \Pi^A$ defined from \eqref{Sugra_GMTF} and \eqref{Sugra_MTF_WcalvA}.
\end{summary}


\subsection{Supersymmetric strings}
\label{sec:Sugra_ExtObj_Strings}

Finally, let us examine the last fundamental BPS--objects, the supersymmetric strings of codimension two. The construction of a proper supersymmetric action for strings proceeds along the same lines as for membranes \cite{Bandos:2003zk,Bandos:2019qok,Bandos:2019lps}. 

A supersymmetric string, whose worldsheet $\cals$ is parametrized by two coordinates $\zeta^i$, $i=0,1$, is embedded in the ambient superspace via
\be\label{Sugra_ExtObj_St_Emb}
\zeta^i \quad\mapsto\quad \cals:\;{\frak z}^M(\zeta)=\left(x^m(\zeta),\theta^\alpha(\zeta),\bar\theta_{\dot\alpha}(\zeta)\right)\,. 
\ee
We want our supersymmetric strings to be minimally charged under some gauge two-forms $\calb_2^\Lambda$, with charges $e_\Lambda$, as in \eqref{Intro_AMS_S}. The first step to build a putative supersymmetric action for a string is to promote to superspace the Wess-Zumino term in \eqref{Intro_AMS_S}. We then consider 
\be
\label{Sugra_ExtObj_Str_Swz}
S_{\rm string,WZ} = e_\Lambda\int_\cals {\bf B}^\Lambda_2\;,
\ee
where ${\bf B}^\Lambda_2$ is a \emph{super-gauge two-form} whose purely bosonic component is just the ordinary $\calb_2^\Lambda$:
\be
{\bf B}^\Lambda_2 |_{\text{bos}} = \calb_2^\Lambda = \frac12 \calb_{mn}^\Lambda  \d x^m \wedge \d x^n\;.
\ee
As a super-form, ${\bf B}^\Lambda_2$ is defined through its super-three-form field-strength ${\bf H}^\Lambda_3$, by requiring
\be
{\bf H}^\Lambda_3=: \dwb {\bf B}^\Lambda_2\,.
\ee
The super-form ${\bf H}^\Lambda_3$ is built in terms of the supervielbeins $E^A$ and the linear multiplets $L^\Lambda$, the proper supersymmetric and gauge invariant objects which contain, among their components, the field strengths $\calh_3^\Lambda = \d \calb_2^\Lambda$ (see \eqref{ConvSG_L}). Indeed, requiring ${\bf H}^\Lambda_3$ to be a closed super-form singles out a unique super-three-form which can be constructed out of the linear multiplets $L^\Lambda$ \cite{Adamietz:1992dk,Bandos:2003zk}, which is
\be
\label{Sugra_ExtObj_Str_H3}
\begin{aligned}
	{\bf H}^\Lambda_3=&  - 2{\ii} E^a \wedge E^\alpha \wedge \bar E^{\dot\alpha}  \sigma_{a\alpha\dot\alpha}{ L}^\Lambda \\ & +   E^b\wedge E^a \wedge  E^\alpha
	\sigma_{ab\; \alpha}{}^{\beta}{\mathcal D}_{\beta}{ L}^\Lambda + E^b\wedge E^a \wedge   \bar E^{\dot\alpha}
	\bar\sigma_{ab}{}^{\dot\beta}{}_{\dot\alpha}\bar{\mathcal D}_{\dot\beta}{ L}^\Lambda
	\\&+\frac {1} {24}
	E^c \wedge E^b \wedge E^a \epsilon_{abcd} \,(\bar{\sigma}{}^{d\dot{\alpha}\alpha}
	[{\mathcal D}_\alpha, \bar{\mathcal D}_{\dot\alpha}]+8G^d){ L}^\Lambda
	\, .
\end{aligned}
\ee
Using the components \eqref{ConvSG_L}, it can be easily checked that, if we restrict ourselves to the bosonic components only 
\be
{\bf H}^\Lambda_3 |_{\text{bos}} = \dwb {\bf B}^\Lambda_2 |_{\text{bos}} = \d \calb_2^\Lambda = \frac12 \del_{[m} \calb^\Lambda_{np]} \d x^m \wedge \d x^n \wedge \d x^p\;.
\ee

As for the case of membranes, the Wess--Zumino term \eqref{Sugra_ExtObj_Str_Swz} alone breaks spontaneously all the bulk $\caln=1$ supersymmetry. In order to preserve part of the supersymmetry we need to add, to \eqref{Sugra_ExtObj_Str_Swz}, a Nambu--Goto term. The final, supersymmetric action of a BPS-string, in the super-Weyl invariant formalism, is
\begin{important}
	\be
	\label{Sugra_ExtObj_Str_S}
	S_{\rm string} = -\int_\cals\d^2\zeta\, |e_\Lambda  L^\Lambda|\sqrt{-\det {\bm{\gamma}}}  + e_\Lambda\int_\cals {\bf B}^\Lambda_2\,.
	\ee
\end{important}
It can be seen that \eqref{Sugra_ExtObj_Str_S} reduces to \eqref{Intro_AMS_S} once we restrict to the bosonic components and it defines the string tension as
\be
\label{Sugra_ExtObj_TStr}
\calt_{\text{string}} = |e_\Lambda  l^\Lambda|\;,
\ee
with the scalar fields $l^\Lambda$ assumed to be evaluated over the string worldsheet $\cals$. The super-Weyl invariant action can be naturally coupled to the actions defined by \eqref{Sugra_AL_Lbos}, which also contain the dynamics of the linear multiplets.

The form of the action \eqref{Sugra_ExtObj_Str_S} is strictly determined by $\kappa$-symmetry. In fact, the action \eqref{Sugra_ExtObj_Str_S} is invariant under the fermionic  $\kappa$-transformations \eqref{ExtObj_FrM_kappa}, where the parameter $\kappa^\alpha$ satisfies the projection condition
\be
\label{Sugra_ExtObj_Str_k}
\kappa_\alpha=-\frac{e_\Lambda  L^\Lambda}{|e_\Lambda L^\Lambda|}\Gamma_{\alpha}{}^\beta\kappa_\beta\, ,
\ee
with $\Gamma_{\alpha}{}^\beta$ defined in \eqref{ExtObj_Str_proj}. The bulk supersymmetry, with local parameter $\epsilon_\alpha(x)$ can then be preserved, over the string worldsheet, only if the $\kappa$-parameters can be identified $\epsilon_\alpha(x)$, namely
\be
\kappa_\alpha(\zeta) \equiv \epsilon_\alpha(x)|_{\text{string}}\,.
\ee
Hence, the supersymmetric strings described by the action \eqref{Sugra_ExtObj_Str_S} are $\half$--BPS objects, preserving only half of the bulk supersymmetry over their worldsheet, the other half being spontaneously broken.

	\begin{summary}
	Before concluding this section, we also report the Einstein-frame string action
	\be
	\label{Sugra_ExtObj_Str_Sgf}
	S_{\rm string} = -\Mp^2 \int_\cals\d^2\zeta\, |e_\Lambda  L^\Lambda|\sqrt{-\det {\bm{\gamma}}}  + e_\Lambda\int_\cals {\bf B}^\Lambda_2\,.
	\ee
	where we have employed the change of variables \eqref{Sugra_AL_lgf} and which defines the physical string tension
	\be
	\label{Sugra_ExtObj_TStrgf}
	\calt_{\text{string}} = \Mp^2 |e_\Lambda  \ell^\Lambda|\;.
	\ee
	This action couples directly to the bulk action defined by \eqref{Sugra_AL_Lgf}.
\end{summary}

\subsection{BPS--membrane/string junction}
\label{sec:Sugra_ExtObj_MembString}

In Section \ref{sec:Sugra_ExtObj_Memb} we assumed that the membrane had no boundaries: this assumption is crucial in order to build the $\kappa$--symmetric action \eqref{ExtObj_Memb_S}, allowing for mutual the cancellation of the variation of the Wess-Zumino and Nambu-Goto terms as given in \eqref{Sugra_ExtObj_M_Swz} and \eqref{Sugra_ExtObj_Memb_Sng}. However, let us now assume that the membrane possesses a boundary and it is given by a BPS-string of the kind examined in the previous section, that is $\del \calm = \cals$. Then, the variation of the action  \eqref{ExtObj_Memb_S} under the $\kappa$--transformation, with parameters obeying  \eqref{Sugra_ExtObj_Memb_kappa}, is no more zero, due to the appearance of the term
\be
\label{Sugra_ExtObj_SM_vmemb}
\delta_\kappa S_{\text{memb}} =  q_A \int_{\del\calm} i_\kappa  {\bf A}^A_3 =  q_A  \int_{\cals} i_\kappa  {\bf A}^A_3\;,
\ee
which is localized over the string at the boundary of the membrane. Such a variation may be compensated by that of the string located at the boundary. We write the total variation at the boundary as
\be
\label{Sugra_ExtObj_SM_vtotb}
\delta S_{\text{string}} + \delta S_{\text{memb}} =  \delta S_{\text{string,NG}}|_{\kappa} + e_\Lambda \int_{\calm} i_{\kappa} \left(  {\bf H}_3^\Lambda + c^\Lambda_A {\bf A}_3^A \right)\;,
\ee
where, we recall, we are assuming the $\kappa$-parameters to satisfy \eqref{Sugra_ExtObj_Memb_kappa} and we have decomposed $q_A = c_A^\Lambda e_\Lambda$. We can then make the variation vanish by requiring the string Nambu-Goto term to be
\be
\label{Sugra_ExtObj_SM_Sng}
S_{\text{string,NG}}= -\int_\cals\d^2\zeta\, |e_\Lambda  \hat L^\Lambda|\sqrt{-\det {\bf {\bm \gamma}}} 
\ee
where, with respect \eqref{Sugra_ExtObj_Str_S}, the linear multiplets appear in their gauged version \eqref{Sugra_GL}. With the choice \eqref{Sugra_ExtObj_SM_Sng}, the variation \eqref{Sugra_ExtObj_SM_vtotb} vanishes provided that the $\kappa$-parameters satisfy \eqref{Sugra_ExtObj_Str_k} in addition to \eqref{Sugra_ExtObj_Memb_kappa}.

Therefore, in the super-Weyl invariant approach, the full supersymmetric action describing a junction of a membrane, coupled to the bulk scalars $s^A$ and charged under the gauge three-forms $A_3^A$, with a string, charged under the gauge two forms $\calb_2^\Lambda$, is
\begin{important}	
	\be
	\label{Sugra_ExtObj_SM_junction}
	\begin{split}
		S_{\text{BPS-junction}} &=-2\int_\calm \d^3\xi\,|e_\Lambda c_A^\Lambda S^A|\sqrt{-\det {\bf h}}+e_\Lambda c_A^\Lambda\int_\calm{\bf A}^A_3 +
		\\
		&\quad\,-\int_\cals\d^2\zeta\, |e_\Lambda  \hat L^\Lambda|\sqrt{-\det {\bm \gamma}}  + e_\Lambda\int_\cals {\bf B}^\Lambda_2\;,
	\end{split}
	\ee
\end{important}
with the superfields $S^A$ and $\hat L^\Lambda$ defined, respectively, in \eqref{Sugra_MTF} and \eqref{Sugra_GL_hatH3}. Owing to the two projection conditions \eqref{Sugra_ExtObj_Str_k} and \eqref{Sugra_ExtObj_Memb_kappa}, at the junction only a fourth of the original $\caln=1$ supersymmetry is preserved.

\begin{summary}	
	Gauge-fixing the super-Weyl invariance as in \eqref{Sugra_SW_u}, we obtain the following Einstein-frame action for a $\frac14$--BPS junction
	\be
	\label{Sugra_ExtObj_SM_junctiongf}
	\begin{split}
		S_{\text{BPS-junction}} &=-2\Mp^3 \int_\calm \d^3\xi\,e^{\frac K2}|e_\Lambda c_A^\Lambda S^A|\sqrt{-\det {\bf h}}+e_\Lambda c_A^\Lambda\int_\calm{\bf A}^A_3 +
		\\
		&\quad\,- \Mp^2 \int_\cals\d^2\zeta\, |e_\Lambda  \hat {L}^\Lambda|\sqrt{-\det {\bm \gamma}}  + e_\Lambda\int_\cals {\bf B}^\Lambda_2\;,
	\end{split}
	\ee
	which is coupled to the action defined by \eqref{Sugra_GL_Lgf}.
\end{summary}

\section{Hierarchies of forms and BPS-objects in supergravity}
\label{sec:Sugra_HG}

In the section, for the sake of completeness, we provide the complete actions coupling the bulk theories examined in Chapter~\ref{chapter:Sugra} with the extended objects just introduced. We will outline, for each of the cases examined, the physical consequences, for the bulk theory, of the introduction of the extended objects. We will give the action directly in the Einstein frame.

\begin{centerbox}
	Three-forms coupled to membranes
\end{centerbox}

The complete action describing the coupling of a single membrane located at $z=0$ to a set of gauge three-forms $A_3^A$ is obtained combining \eqref{Sugra_MTF_L3fComp_gf} with \eqref{Sugra_ExtObj_Memb_Sgf}:
\be
\label{Sugra_Sum_Smem}
\begin{aligned}
	S |_{\rm bos} &=  \Mp^2 \int_{\Sigma} \left( \frac12 R *1- K_{i\bar \jmath} \d \varphi^i \wedge *\d \bar\varphi^{\bar \jmath} \right)
	\\
	&\quad\, -\int_\Sigma \Big[\frac12 T_{AB}F^A_4*\!F^B_4+ T_{AB}\Upsilon^AF^B_4+\Big(
	\hat V-\frac12 T_{AB}\Upsilon^A \Upsilon^B\Big)*\!1\Big] 
	\\
	&\quad\,+\int_{\del\Sigma}T_{AB}(* F^A_4+\Upsilon^A)A^B_3
	\\
	&\quad\,-2\Mp^3 \int_\calm \d^3\xi\,e^{\frac K2}|q_A \Pi^A(\varphi)|\sqrt{-\det {\bf h}}+q_A\int_\calm A_3^A\;,
\end{aligned}
\ee
where $\Sigma$ denotes the four-dimensional spacetime and the quantities $T^{AB}$, $\Upsilon^A$ and $\hat V$ are given in \eqref{Sugra_MTF_TUVgf}. If we integrate out the gauge three-forms $A_3^A$ as
\be
\label{Sugra_Sum_F4eom}
T_{AB}(*\!{F}^B_4+\Upsilon^B)= -N_A - q_A \Theta (z)\;,
\ee
with  $N_A$ real constants, we obtain a chiral theory in the usual form \eqref{Sugra_SW_LSWcompgf}. However, the standard Cremmer et al. potential is different on the two sides, for the superpotential is different and is given by
\be
\label{Sugra_Sum_Wbulk}
\begin{aligned}
	&W(\Phi) =   \Theta(-z) W_- (\Phi) + \Theta(z) W_+ (\Phi) \;,
\end{aligned}
\ee
with
\be
\label{Sugra_Sum_Wpm}
\begin{aligned}
	&W_-(\Phi) =  \Mp^3 N_A \Pi^A(\Phi) + \hat W(\Phi)\;, 
	\\
	&W_+(\Phi) =  \Mp^3 (N_A+q_A) \Pi^A(\Phi) + \hat W(\Phi)\;.
\end{aligned}
\ee
Therefore a membrane allows the quantized constants $N_A$ to jump of an amount given by the charge of the membrane under the gauge three-forms $A_3^A$ which generate $N_A$ dynamically.

\begin{centerbox}
	Three- and four-forms coupled to membranes and 3-branes
\end{centerbox}

The action which couples gauge three- and four-forms, the latter gauging the former, to a membrane and a 3-brane, filling the spacetime region $\calw = \{z<0\}$, is obtained by combining  \eqref{Sugra_GMTF_L3fComp_gf} with \eqref{Sugra_ExtObj_M3B_regiongf}:
\be
\label{Sugra_Sum_SM3B}
\begin{aligned}
	S |_{\rm bos} &=  \Mp^2 \int_{\Sigma} \left( \frac12 R *1- K_{i\bar \jmath} \d \varphi^i \wedge *\d \bar\varphi^{\bar \jmath} \right)
	\\
	&\quad\, -\int_\Sigma \Big[\frac12 T_{AB} \hat F^A_4*\! \hat F^B_4+ T_{AB}\Upsilon^A\hat F^B_4+\Big(
	\hat V-\frac12 T_{AB}\Upsilon^A \Upsilon^B\Big)*\!1\Big] 
	\\
	&\quad\,+\int_{\del\Sigma}T_{AB}(* \hat F^A_4+\Upsilon^A)A^B_3 + \int_\Sigma \calq_I^{\rm bg} C^I_4
	\\
	&\quad\,-2\Mp^3 \int_\calm \d^3\xi\,e^{\frac K2}|q_A \calv^A(\varphi)|\sqrt{-\det {\bf h}}+q_A\int_\calm A^A_3 + q_A Q^A_I\int_\calw C^I_4\;.
\end{aligned}
\ee
The on-shell chiral multiplet theory has the very same form as \eqref{Sugra_SW_LSWcompgf}  with the a different superpotential on the two spacetime regions separated by the membrane as in \eqref{Sugra_Sum_Wbulk}. The equations of motion for the gauge three-forms $A_3^A$ are the same as \eqref{Sugra_Sum_F4eom}; however, here the constants $N_A$ are not arbitrary but constrained, as can be seen integrating out the gauge four-forms $C_4^I$:
\be
\label{Sugra_Sum_TadNA}
(N_A  + q_A)Q_I^A + \calq_I^{\rm bg} = 0 \;,
\ee
in both the spacetime regions. A 3-brane with the BPS--membrane as boundary has then the role to change the constraint \eqref{Sugra_4FM_Constr} accompanying the flux shift due to the membrane charge; in this way, the constraint \eqref{Sugra_4FM_Constr} can be satisfied in both the spacetime regions dissected by the membrane.

\begin{centerbox}
	Two-forms coupled to strings
\end{centerbox}

In the absence of a superpotential for the chiral fields $\Phi^a$, merging \eqref{Sugra_AL_Lgf} with \eqref{Sugra_ExtObj_Str_Sgf} we get the full action describing a BPS--string coupled to a bulk theory where both linear and chiral multiplets are present
\be
\label{Sugra_Sum_SStr}
\begin{aligned}
	S |_{\rm bos} &= M^2_{\rm P} \int_\Sigma \left( \frac12\, R *1   -F_{i\bar \jmath}\,  \d \varphi^i \wedge *\d \bar \varphi^{\bar\jmath} + \frac14 F_{\Lambda\Sigma}  \d \ell^\Lambda \wedge * \d \ell^\Sigma \right)
	\\
	&\quad\,+\frac{1}{4 M_{\rm P}^2} \int_\Sigma F_{\Lambda\Sigma}\,  \hat\calh_3^\Lambda \wedge * \hat \calh^{\Sigma}_3 +  \int \left(\frac{\ii}{2} F_{\bar \imath \Sigma}\, \d \bar \varphi^{\bar\imath} \wedge \hat\calh^{\Sigma}_3  + {\rm c.c.}\right)
	\\
	&\quad\, -\Mp^2  \int_\cals\d^2\zeta\, |e_\Lambda  \ell^\Lambda|\sqrt{-\det {\bm \gamma}}  + e_\Lambda\int_\cals B^\Lambda_2\;.
\end{aligned}
\ee
In the electro-magnetic dual theory, after encircling a string, the axions shift by the charge of the string
\be
\label{Sugra_Sum_ashift}
a_\Lambda \quad \rightarrow \quad a_\Lambda + e_\Lambda\;. 
\ee

\begin{centerbox}
	Two- and three-forms coupled to a BPS-junction
\end{centerbox}

The full action that governs the coupling of a BPS--junction to a bulk theory is obtained from  \eqref{Sugra_GL_Lgf} and \eqref{Sugra_ExtObj_SM_junctiongf}:
\be
\begin{aligned}
	S |_{\rm bos} &= M^2_{\rm P} \int \left( \frac12\, R *1   -F_{i\bar \jmath}\,  \d \varphi^i \wedge *\d \bar \varphi^{\bar\jmath} + \frac14 F_{\Lambda\Sigma}  \d \ell^\Lambda \wedge * \d \ell^\Sigma \right)
	\\
	&\quad\,+\frac{1}{4 M_{\rm P}^2} \int F_{\Lambda\Sigma}\,  \hat\calh_3^\Lambda \wedge * \hat \calh^{\Sigma}_3 +  \int \left(\frac{\ii}{2} F_{\bar \imath \Sigma}\, \d \bar \varphi^{\bar\imath} \wedge \hat\calh^{\Sigma}_3  + {\rm c.c.}\right)
	\\
	&\quad\, -\int_\Sigma \Big[\frac12 T_{AB} F^A_4*\! F^B_4+ T_{AB}\Upsilon^A F^B_4+\Big(
	\hat V-\frac12 T_{AB}\Upsilon^A \Upsilon^B\Big)*\!1\Big] 
	\\
	&\quad\,+\int_{\del\Sigma}T_{AB}(* F^A_4+\Upsilon^A)A^B_3 
	\\
	&\quad\, -\Mp^2 \int_\cals\d^2\zeta\, |e_\Lambda  \ell^\Lambda|\sqrt{-\det {\bm \gamma}}  + e_\Lambda\int_\cals B^\Lambda_2
	\\
	&\quad\,-2\Mp^3 \int_\calm \d^3\xi\,e^{\frac {\tilde F}2}|e_\Lambda c^\Lambda_A \calv^A(\varphi)|\sqrt{-\det {\bf h}}+e_\Lambda c^\Lambda_A\int_\calm A^A_3 \;.
\end{aligned}
\ee
Transversing \emph{both} the string and the membrane, combining the results \eqref{Sugra_Sum_F4eom} and \eqref{Sugra_Sum_ashift}, we get the combined shift
\be
\label{Sugra_Sum_Nashift}
N_A \quad \rightarrow \quad N_A + e_Ac_A^\Lambda\,, \qquad a_\Lambda \quad \rightarrow \quad a_\Lambda + e_\Lambda\;. 
\ee
In dual, ordinary chiral theory described by the superpotential \eqref{GSS_GL_Dual2W}
\be
\label{Sugra_Sum_Dual2W}
W(\Phi;T) = N_A \calv^A(\Phi) -c_A^\Lambda T_\Lambda \calv^A(\Phi)+ \hat W (\Phi) 
\ee
the shift \eqref{Sugra_Sum_Nashift} \emph{has no net effect!} In other words, a BPS--junction so built connects two equivalent theories sharing the same potential.

Eventually, the above actions can be combined to obtain supersymmetric effective actions for  more complicated networks of 3-branes, membranes and strings, like those considered in e.g. \cite{Evslin:2007ti,BerasaluceGonzalez:2012zn}. A detailed treatment of these more involved configurations is left for future work.

\section{Sailing through the Landscape}
\label{sec:LandSwamp_ExtObj_DW}

Within a four-dimensional theory, membranes play a central role. As stressed for the global case in Section~\ref{sec:ExtObj_DWM}, they make the potential change once they are crossed, consequently modifying the space of vacua $\scrm_{\rm vac}$ in the various spacetime regions which they dissect. This, in turn, leads to a change of the mass spectrum across the different spacetime regions. 

We devote this section to study how vacua can change and are connected across the spacetime. We shall focus on the case of supersymmetric domain walls, which determine \emph{stable} regions, where fields sit at certain vevs. Such a configuration is of course possible if a potential admits a space of vacua $\scrm_{\rm vac}$ with nontrivial zeroth homotopy group, namely it is made up by path-disconnected components. Domain walls are stable, solitonic objects which interpolate between two vacua belonging to different path-disconnected components.  

We will generalize the globally supersymmetric results of Sections~\ref{sec:ExtObj_DW} and ~\ref{sec:ExtObj_DWM} to generic supergravity theories. However, here, we directly consider the case of domain walls supported by membranes, obtaining the case without membranes as a limit thereof.

\subsection{Domain Walls in Supergravity: the flow equations and the BPS--conditions}

Let us consider a generic supergravity theory, whose matter content is described by $n$ physical chiral superfields $\Phi^i$. Our aim is to study domain walls induced by membranes, coupled to some gauge three-forms. We consider the simplest case where just a single membrane is present, covering the spacetime hypersurface $\calm= \{z=0\}$ and we restrict ourselves to the bosonic components only. As stressed in the previous sections, an ordinary supergravity formulation is not fit to fully describe nontrivial couplings of the membrane to the bulk theory and we should rather consider the dual three-form formulation \eqref{Sugra_Sum_Smem}. The action \eqref{Sugra_Sum_Smem} encodes a potential via gauge three-forms; that is, in order to obtain a potential, we need to integrate out the gauge three-forms as we did in \eqref{Sugra_Sum_F4eom}. We then arrive at an `ordinary' supergravity formulation
\be
\label{SL_DW_Sbos}
\begin{aligned}
	S |_{\rm bos} &=   \int_{\Sigma} \left( \frac12  R *1-   K_{i\bar \jmath} \d \varphi^i \wedge *\d \bar\varphi^{\bar \jmath}- V(\varphi,\bar\varphi) * 1 \right) 
	\\
	&\quad\,  - \int_\calm \sqrt{-\det {\bf h}}\, \calt_{\rm memb}(\varphi) + \int_{\del\Sigma} K *1
\end{aligned}
\ee
where we have set $\Mp = 1$ to ease the following exposition. Owing to the presence of the membrane, the potential differs on the two sides of the membrane
\be
\label{SL_DW_V}
V (\varphi (z),\bar\varphi (z))= V_- (\varphi,\bar\varphi) \Theta (-z) +  V_+ (\varphi,\bar\varphi) \Theta (z) \,.
\ee
Working with $\caln=1$ supersymmetric theories, the potential is computed by the usual Cremmer et al. formula \eqref{Sugra_SW_Cremmeretal}, with the superpotential undertaking a step once the membrane is encountered as in \eqref{Sugra_Sum_Wpm}. The last term of \eqref{SL_DW_Sbos} is the Gibbons-Hawking-York boundary terms \cite{York:1972sj,Gibbons:1976ue,Ceresole:2006iq}, containing the extrinsic curvature $K^{\rm ext}= g^{ab} K_{ab}^{\rm ext}$ is the extrinsic curvature, with $K_{ab}^{\rm ext} =\frac12  n^m \frac{\del g_{ab}}{\del x^m}$.

Domain walls are solitonic objects, being solutions of both the Minkowskian scalar and gravitational equations of motion. In general, without further assumptions, it is pretty difficult to extract solutions of the equations of motion from the generic action \eqref{SL_DW_Sbos}. However, we are interested in the simplest family of domain wall solutions, namely \emph{flat} and \emph{static} domain walls, which enjoy an SO(1,2) symmetry along three spacetime directions. It is then useful to split the spacetime coordinates $x^m$ into $(x^i,x^3\equiv z)$, $i=0,1,2$. A flat domain wall fully covers the spacetime directions $x^i$ and has a nontrivial profile only along the fourth direction $z$. With respect to the globally supersymmetric case examined in the Sections~\ref{sec:ExtObj_DW} and~\ref{sec:ExtObj_DWM}, here gravity also plays a role and influences the solution. In fact, as an extended object, a domain wall `backreacts' on the spacetime metric, modifying the geometry. Compatibly with the SO(1,2) symmetry, we choose the following ansatz for the spacetime metric \cite{Vilenkin:1984hy,Cvetic:1992bf,Ceresole:2006iq}
\be
\label{SL_DW_ds2}
\d s^2=e^{2D(z)}\d x^i\d x_i+\d z^2 \, ,
\ee
where the \emph{warping} $D(z)$ solely depends on the fourth coordinate $z$.  As a further simplifying assumption, the scalar fields $\varphi^i$ are allowed to depend only on the transverse coordinate 
\be
\label{SL_DW_varphiz}
\varphi^i=\varphi^i(z)\;.
\ee
We further assume that $ \varphi^i(z)$ are continuous along $z$, while their derivatives may be discontinuous at $z=0$, where the membrane is located.

The profile of a domain wall is determined by solving some \emph{flow equations}, which dictates how the scalar fields evolve across the spacetime. We are interested in domain walls which preserve (part of) the bulk supersymmetry. The analysis that we are going to carry is pretty similar to that of domain walls in standard supergravity (for example in \cite{Cvetic:1992bf, Cvetic:1992st, Cvetic:1992sf, Cvetic:1993xe,Cvetic:1996vr,  Ceresole:2006iq}, to which we refer for further details). In order to preserve supersymmetry, the variations of the bulk fields need to vanish over the background specified by \eqref{SL_DW_ds2}, leading to a set of Killing spinor equations. Presently, since we are focusing on bosonic domain walls, it is just sufficient to require that the variations of the bulk fermions need to vanish.

The supersymmetry variations of the gravitino $\psi_m{}^\alpha$ and the chiralini $\chi_\alpha^i$ are
\be
\label{SL_DW_susy}
\begin{aligned}
	\delta \psi_m{}^\alpha &= -2 \hat{\cald}_m \epsilon^\alpha -\ii  e_m{}^c e^{-\frac{K}{2}} W (\varepsilon \sigma_c \bar{\epsilon})^\alpha\,,
	\\
	\delta \chi^i_\alpha &= \sqrt{2} \epsilon_\alpha e^{\frac{K}{2}} K^{\bar \jmath i} ( \overline{W}_{\bar \jmath}+K_{\bar \jmath}\overline{W}) -\ii \sqrt{2} \sigma_{\alpha\dot{\beta}}{}^a \bar{\epsilon}^{\dot{\beta}} \del_a \varphi^i\, .
\end{aligned}
\ee
Here, $\epsilon^\alpha = \epsilon^\alpha (z)$ is the local supersymmetry parameter and the action of the covariant derivative on the supersymmetry parameter is given by 
\be
\begin{split}
	\hat{\cald}_m \epsilon^\alpha &\equiv \del_m \epsilon^\alpha + \epsilon^\beta \omega_{m\beta}{}^\alpha - \frac{\ii}{2} \cala_m \epsilon^\alpha \, ,
\end{split}
\ee
with the $U(1)$ K\"ahler connection
\be
\cala_m = \frac{\ii}{2} \left(K_i \del_m \varphi^i - K_{\bar \imath} \del_m \bar{\varphi}^{\bar\imath} \right)\,. 
\ee
The variations \eqref{SL_DW_susy} evaluated over the background \eqref{SL_DW_ds2} read
\begin{subequations}
\label{SL_DW_susydw}
\begin{align}
	\delta \psi_z{}^\alpha = & -2 \dot\epsilon^\alpha +\ii \cala_z  \epsilon^\alpha-\ii e^{\frac{K}{2}} W (\varepsilon \sigma_{\underline{z}} \bar{\epsilon})^\alpha\,,\label{SL_DW_susydwa}
	\\
	\delta \psi_i{}^\alpha = & e^D \left[ \dot{D}\ (\epsilon \sigma_{\underline z}\bar{\sigma}_{\iund})^\alpha - \ii e^{\frac{K}{2}} W (\varepsilon \sigma_{\iund} \bar{\epsilon})^\alpha\right]\,,\label{SL_DW_susydwb}
	\\
	\delta \chi^i_\alpha = &\sqrt{2} \epsilon_\alpha e^{\frac{K}{2}} K^{\bar \jmath i} ( \overline{W}_{\bar \jmath}+K_{\bar \jmath}\overline{W})  -\ii \sqrt{2} \sigma_{\alpha\dot{\beta}}{}^{\underline z} \bar{\epsilon}^{\dot{\beta}} \dot \varphi^i\, ,\label{SL_DW_susydwc}
\end{align}
\end{subequations}
where the dot corresponds to the derivative with respect to $z$ and the underlined indices are flat indices. Preserving supersymmetry requires such variations to vanish
\be
\label{SL_DW_susydw0}
	\delta \psi_m{}^\alpha \stackrel{!}{=} 0\,,\qquad \delta \chi^i_\alpha \stackrel{!}{=} 0\;.
\ee
These conditions are solved by first imposing that the local supersymmetry parameter $\epsilon_\alpha(z)$ obeys the projection conditions
\begin{important}
\be
\label{SL_DW_eproj}
\epsilon_\alpha (z)= \mp \ii e^{\ii\alpha}(\sigma_{\ul 3})_{\alpha\dot\alpha}\bar\epsilon^{\dot\alpha} (z)\,,
\ee
\end{important}
which, plugged into \eqref{SL_DW_susydwb}-\eqref{SL_DW_susydwc}  and setting \eqref{SL_DW_susydw0}, lead to the \emph{flow equations}
\begin{important}
\begin{subequations}\label{SL_DW_Flow}
	\begin{align}
	\dot \varphi^i &= \mp e^{\frac 12 K(\varphi,\bar\varphi)+\ii\vartheta(y)}K^{\bar\jmath i}(\overline W_{\bar\jmath}+K_{\bar\jmath} \overline W)\;,\label{SL_DW_Flowa}
	\\
	\dot D &= \pm e^{\frac 12 K(\varphi,\bar\varphi)}|W| \;, \label{SL_DW_Flowb}
	\end{align}
\end{subequations}
\end{important}
where  $\dot D \equiv \frac{\d }{\d z}D$ and   $\vartheta(z)= \vartheta(\varphi(z),\bar{\varphi}(z))$ is the phase of the superpotential $W$ 
namely 
\be
\label{SL_DW_vartheta} 
W=e^{\ii\vartheta}|W|\;.
\ee 
In addition to \eqref{SL_DW_Flow}, it can be shown that \eqref{SL_DW_susydwa} and \eqref{SL_DW_eproj} imply that the phase $\vartheta$ satisfies
\be
\label{SL_DW_thetaflow}
\dot\vartheta=- \Im\left(\dot\varphi^i K_i\right) 
\ee
across the flow. In the following, we shall assume that $\vartheta$ is \emph{continuous} along the flow, also when the membrane is encountered. 

Some comments to the crucial relations that we have found are in order. The first of these relations, \eqref{SL_DW_Flowa}, in analogy with \eqref{ExtObj_DW_flow}, tells how the scalar fields evolve, passing from the vacuum $\varphi^i_{-\infty}$ at the asymptotic left of the membrane to the vacuum $\varphi^i_{+\infty}$, located at its far right. Then, \eqref{SL_DW_Flowa} is the relation determining the scalar profile of the domain wall solution. The second relation \eqref{SL_DW_Flowb}, which does not appear in the globally supersymmetric treatment of Section~\ref{sec:ExtObj_DWM}, is connected to the geometry, expressing the backreaction of the domain wall, including the membrane, over the spacetime metric \eqref{SL_DW_ds2}.

We are now also in the position to describe in which sense supersymmetry is preserved. The discussion proceeds along the same lines of Section~\ref{sec:ExtObj_DWM} and here we will be briefer. The projection condition \eqref{SL_DW_eproj} tells that only half the bulk supersymmetry is preserved by the full domain wall solution. Over the membrane worldvolume, for the membrane static ground state, supersymmetry is realized via the fermionic $\kappa$--symmetry. In order to preserve half of the supersymmetry generators \emph{also} on the membrane worldvolume, we need the supersymmetry parameter \eqref{SL_DW_eproj} to be identified with the $\kappa$--symmetry parameter of \eqref{Sugra_ExtObj_Memb_kappa}. We must then require that, at $z=0$ where the membrane sits in its ground state
\begin{important} 
	\be
	\label{SL_DWM_kappa}
	\epsilon_\alpha |_{z=0} \stackrel{!}{=} \kappa_\alpha  = -\ii \frac{q_A \calv^A(\varphi)}{|q_A \calv^A(\varphi)|} \sigma^3_{\alpha \dot \alpha} \bar\kappa^{\dot \alpha} \Big|_{z=0} \;.
	\ee
\end{important}
%
%
This is compatible with \eqref{SL_DW_eproj}. In fact, due to the continuity of the phase, over the membrane we can identify $\vartheta\equiv \arg [q_A \calv^A(\varphi)] |_{z=0}$.

In order to rewrite the flow equations in a more compact form, it is convenient to introduce a \emph{flowing covariantly holomorphic superpotential} \cite{Ceresole:2006iq} \footnote{Note that this quantity matches with the super-Weyl superpotential of \eqref{Sugra_SW_KWhom}, after having gauge fixed the super-Weyl compensator as in \eqref{Sugra_SW_u}.}
\be
\label{SL_DW_covsup}
\calz(\varphi,\bar\varphi)\equiv e^{\frac12 K(\varphi,\bar\varphi)}W=e^{\frac12 K(\varphi,\bar\varphi)}\left[\Theta(z)W_+(\varphi)+\Theta(-z)W_-(\varphi)\right]\,.
\ee
As in \eqref{Sugra_Sum_Wpm} and \eqref{SL_DW_V}, the dependence of $\calz$ on $z$ is both explicit, through the step functions, and implicit, through the scalar fields $\varphi^i = \varphi^i(z)$. The jump induced by the membrane over the quantity $\calz$ can be computed by taking the limits of the difference of the right and left covariantly holomorphic superpotential \eqref{SL_DW_covsup} as
\be
\label{SL_DW_calzjump}
\Delta\calz \equiv \lim_{\varepsilon\rightarrow  0}\left(\calz|_{z=\varepsilon}-\calz|_{z=-\varepsilon}\right)= e^{\frac12 K (\varphi,\bar\varphi)}\left(q_A \Pi^A (\varphi)\right)|_{z=0}\,.
\ee
The absolute value of $\Delta \calz$ is related to the (gauge-fixed) membrane tension \eqref{Sugra_ExtObj_Memb_Tmgf} as
\be
\label{SL_DW_effTM}
\calt_{\rm memb} \equiv 2\,e^{\frac12 K(\varphi,\bar\varphi)}\left|q_A \Pi^A(\varphi)\right|_{z=0}=2|\Delta\calz|
\ee
and at $z=0$ its phase $e^{\ii\vartheta(z)}$ enters the $\kappa$-symmetry projector \eqref{Sugra_ExtObj_Memb_kappa}. 

The previous formulas may be written in a more compact way in terms of $\calz$, rather than of the superpotential $W$; the translation between the formulas makes use of the holomorphicity of $W$, which gives
\be
\label{SL_DW_partialmod}
\partial_{\bar{\jmath}}|W|+\ii\partial_{\bar{\jmath}}\vartheta \; |W|=0,
\ee
consequently relating the derivatives of $\calz$ with the K\"ahler covariant derivatives of the superpotential as
\be
\label{SL_DW_dZ}
\del_{\bar\jmath}|\calz|\equiv \frac12 e^{\ii\vartheta}e^{\frac12 K}\bar D_{\bar\jmath}\bar W\,.
\ee
Hence, in terms of $\calz$, the usual Cremmer et al. potential \eqref{Sugra_SW_Cremmeretal} becomes 
\be
\label{SL_DW_VZ}
V(\varphi,\bar\varphi) =K^{\bar\jmath i} D_i \calz \bar D_{\bar\jmath} \bar \calz - 3 |\calz|^2 = 4 K^{\bar\jmath i} \del_i |\calz| \del_{\bar \jmath} |\calz| - 3 |\calz|^2
\ee
and the flow equations \eqref{SL_DW_Flow} take the simpler form
\cite{Cvetic:1992bf, Cvetic:1992st, Cvetic:1992sf, Cvetic:1993xe,Cvetic:1996vr, Ceresole:2006iq}
\begin{important}
\begin{subequations}
	\label{SL_DW_Flow2}
	\begin{align}
	\dot \varphi^i &= \mp 2K^{\bar\jmath i}\,\del_{\bar\jmath}|\calz|\;,\label{SL_DW_Flow2a}
	\\
	\dot D &= \pm |\calz|\;. \label{SL_DW_Flow2b}
	\end{align}
\end{subequations}
\end{important}

The equations \eqref{SL_DW_Flow2} govern the domain wall profile. The points where the flow starts and ends are the fixed points for the \eqref{SL_DW_Flow2}. Owing to \eqref{SL_DW_dZ}, the flow equation \eqref{SL_DW_Flow2a} has fixed-point solutions provided by the supersymmetric vevs $\varphi^i_*$ (such that $D_{\bar\jmath}W|_{\varphi_*}=0$).  Then the solution of \eqref{SL_DW_Flow2b}  is $D=-|\calz_*|z+\text{constant}$, which corresponds to an AdS space of radius $1/|\calz_*|$ for $\calz_*\neq 0$ and to flat space if $\calz_*=0$. Hence, a  regular  BPS domain wall  interpolates between two different supersymmetric vacua and its geometry is asymptotically AdS or flat for $z\rightarrow\pm\infty$.      

As in AdS/CFT contexts  \cite{Girardello:1998pd,Freedman:1999gp}, one may then define a monotonic c-function which tells the direction of the flow. To this aim, let us preliminarily recall that the phase $\vartheta(z)$ is required to be a continuous function. In other words, the phase of $\calz$ does not to change in passing through the membrane, so that we have $\Delta \calz=e^{\ii\vartheta(0)}|\Delta \calz|$. This requirement, together with the covariant holomorphicity of $\calz$ and \eqref{SL_DW_Flow2a}, implies that
\be\label{SL_DW_dcalz}
\frac{\d |\calz|}{\d z}=2\Re\left(\dot\varphi^i\del_i|\calz|\right) +\frac12 \calt_{\rm memb}\,\delta(z)\,,
\ee
which follows from \eqref{SL_DW_covsup}. As in \cite{Ceresole:2006iq}, we can combine \eqref{SL_DW_Flow2} and \eqref{SL_DW_dcalz} to get the following relation
\be
\label{SL_DW_Ctheorem}
\dot C=4K^{i\bar\jmath}\del_i|\calz|\del_{\bar\jmath}|\calz|+\frac12 \calt_{\rm memb}\delta(z)\geq 0 , 
\ee
where we have introduced $C(z)\equiv -\dot D(z)$.  Equation \eqref{SL_DW_Ctheorem} shows the contribution of the membrane to the monotonic flow of $C(z)$, which `jumps up' by $\frac12 \calt_{\rm memb}$ at $z=0$.\footnote{A similar equation was also derived in \cite{Haack:2009jg} by dimensionally reducing ten-dimensional flow equations in the presence of effective membranes corresponding to D-branes wrapped along internal cycles.}    

Let us assume that $\calz$ is nowhere vanishing and focus on the lower sign choice in \eqref{SL_DW_Flow2}. Then,  $|\calz|_{z=+\infty}>|\calz|_{z=-\infty}$, equation \eqref{SL_DW_Flow2a} tells us that $|\calz|=C$. Hence \eqref{SL_DW_Ctheorem} also implies that $|\calz|$ monotonically increases as we move from $z=-\infty$ to $z=+\infty$. Clearly, in the case $|\calz|_{z=+\infty}<|\calz|_{z=-\infty}$, the upper choice of sign in the flow equations \eqref{SL_DW_Flow2} implies that $|\calz|$ is monotonically decreasing, while \eqref{SL_DW_Ctheorem} still holds, since in that case the sign-reversed \eqref{SL_DW_Flow2a} becomes $|\calz|=\dot D\equiv -C$. Due to this `symmetry', in the following we will just consider the \emph{lower} choice of signs in \eqref{SL_DW_eproj} and \eqref{SL_DW_Flow2}.

As a final note, let us assume to remove our assumption that $\calz$ is nowhere vanishing, supposing that $\calz|_{z_0}=0$ at some transversal coordinate $y_0$. Then, $|\calz|$ must necessarily be monotonically increasing for $z>z_0$ and decreasing for $z<z_0$, so that $\calz$ can vanish only at $z_0$. This implies that for $z>z_0$ the flow equations \eqref{SL_DW_Flow2} with the upper sign hold, while for $y<y_0$ one must use the lower signs.


\subsection{BPS action and domain wall tension}

We now compute the tension of the domain wall configuration enclosing the membrane. The starting point is the action \eqref{Sugra_Sum_Smem}, with a membrane coupled to a bulk supergravity action. Once gauge three-forms are set on-shell, we arrive at the action \eqref{SL_DW_Sbos}, with potential defined as in \eqref{SL_DW_Sbos}. Using the covariantly holomorphic superpotential \eqref{SL_DW_covsup}, we may rewrite \eqref{SL_DW_Sbos} as
\be
\label{SL_DWT_Sbos}
\begin{aligned}
	S |_{\rm bos} &=   \int_{\Sigma} \left( \frac12 R *1-  K_{i\bar \jmath} \d \varphi^i \wedge *\d \bar\varphi^{\bar \jmath} \right) +\int_{\del\Sigma} K *1
	\\
	&\quad\,- \int_\Sigma (K^{\bar\jmath i} D_i \calz \bar D_{\bar\jmath} \bar \calz - 3 |\calz|^2)*1  - \int_\calm \sqrt{-\det {\bf h}}\, \calt_{\rm memb}(\varphi,\bar\varphi) \;.
\end{aligned}
\ee
Now, employing the metric ansantz \eqref{SL_DW_ds2}, we get the on-shell value for the curvature
\be
R = - 6 \left[2 (\dot D)^2 + \ddot  D \right] \,,
\ee
and
\be
\label{SL_DWT_Sbosgrav}
\begin{aligned}
	   \int_{\del\Sigma} K *1 = \frac{\d}{\d z}e^{3D} \Big|_{+\infty} - \frac{\d}{\d z} e^{3D} \Big|_{-\infty} 
\end{aligned}
\ee
and we recognize that the `pure gravitational' part of \eqref{SL_DWT_Sbos} simply reduces to
\be
\frac12   \int_{\Sigma} R *1+  \int_{\del\Sigma} K *1 = 3   \int_\Sigma   (\dot D)^2 *1\,.
\ee

Following \cite{Ceresole:2006iq}, we may now rewrite the action in the elegant \emph{BPS--form}
{\small\be
\label{SL_DWT_SbosBPS}
\begin{aligned}
	S |_{\rm bos} &=   \int\d^3 x\int\d z\, e^{3D}\left[3\big(\dot D+|\calz|\big)^2-K_{i\bar\jmath}\big(\dot\varphi^i -2K^{i\bar k}\del_{\bar k}|\calz|\big)\big(\dot{\bar\varphi}^{\bar\jmath} -2K^{l\bar \jmath}\del_l|\calz|\big)\right]
	\\
	&\quad\,-2  \int\d^3 x\int\d z\, e^{3D}\left[3\dot D|\calz|+2\Re\big(\dot\varphi^i\del_i|\calz|\big)\right] - \int \d^3x \int \d z e^{3D} \delta(z)\, \calt_{\rm memb} (\varphi,\bar\varphi)\;.
\end{aligned}
\ee}
On the other hand, using \eqref{SL_DW_covsup} we can write the second line of \eqref{SL_DWT_SbosBPS} in the form
\be
\begin{aligned}
&-2\int\d^3 x\int\d z\,\left[ \frac{\d}{\d z}\big(e^{3D}|\calz|\big) -\frac12 \delta(z)\calt_{\rm memb} e^{3D}\right] - \int \d^3x \int \d z e^{3D} \delta(z)\, \calt_{\rm memb} 
\\
&= -2\int\d^3 x\left[\big( e^{3D}|\calz|)|_{z=+\infty}-\big(e^{3D}|\calz|\big)|_{z=-\infty}\right]\;,
\end{aligned}
\ee
where the terms localized over the membrane perfectly cancel. Then, \eqref{SL_DWT_SbosBPS} reduces to the
\be
\label{SL_DWM_redS}
\begin{aligned}
	S_{\rm red}=&\int\d^3 x\int\d z\, e^{3D}\left[3\big(\dot D+|\calz|\big)^2-K_{i\bar\jmath}\big(\dot\varphi^i -2K^{i\bar k}\del_{\bar k}|\calz|\big)\big(\dot{\bar\varphi}^{\bar\jmath} -2K^{l\bar \jmath}\del_l|\calz|\big)\right]
	\\
	&-2\int\d^3 x\left[\big( e^{3D}|\calz|)|_{z=+\infty}-\big(e^{3D}|\calz|\big)|_{z=-\infty}\right].
\end{aligned}
\ee
This reduced action is identical in form to the one obtained in \cite{Ceresole:2006iq} in the absence of membranes, basically because of the observed reciprocal cancellations of various terms localised on the membrane. 

Hence, as in the absence of membranes, the extremization of the BPS action \eqref{SL_DWM_redS} precisely reproduces the bulk flow equations \eqref{SL_DW_Flow2}. Indeed, on any solution of the flow equations, we get
\be
\label{SL_DWM_redSos}
S_{\rm red}|_{\text{on-shell}}=-2\int\d^3 x\left[\big( e^{3D}|\calz|)|_{z=+\infty}-\big(e^{3D}|\calz|\big)|_{z=-\infty}\right]=-\int\d^3 \tilde x \,\calt_{\rm DW} , 
\ee
where on the slices of constant $y$ we have introduced coordinates $\tilde x^i=e^{D(y)}x^i$, so that $\d^3 \tilde x$ denotes the physical volume, and the domain wall tension is
\begin{important}
	\be
	\label{SL_DWMT}
	\calt_{\rm DW}=2\big(|\calz|_{z=+\infty}-|\calz|_{z=-\infty}\big)\;.
	\ee
\end{important}

It is however important to stress that \eqref{SL_DWMT} is only formally identical to the formula obtained in the absence of membranes \cite{Cvetic:1992bf, Cvetic:1992st, Cvetic:1992sf, Cvetic:1993xe,Cvetic:1996vr, Ceresole:2006iq}. However one should keep in mind that it includes the contribution of the membrane. 
This can be seen by splitting the overall change of $|\calz|$ in the bulk and membrane contributions\footnote{In \cite{Haack:2009jg} the same conclusion was reached starting from a  ten-dimensional description of similar domain wall solutions. }
\be
\label{SL_DW_tension2}
\calt_{\rm DW}=2\big(|\calz|_{z=+\infty}-\lim_{\varepsilon\rightarrow 0}|\calz|_{z=\varepsilon}\big)+2\big(\lim_{\varepsilon\rightarrow 0}|\calz|_{z=-\varepsilon}-|\calz|_{z=-\infty}\big)+ \calt_{\rm memb}\,.
\ee

From \eqref{SL_DWMT} we see that our working assumption $|\calz|_{z=+\infty}> |\calz|_{z=-\infty}$ guarantees that $\calt_{\rm DW}>0$. The case  $|\calz|_{z=+\infty}< |\calz|_{z=-\infty}$ (with still nowhere vanishing $\calz$)  can be obtained by changing $z\rightarrow -z$ in the above steps, so that the sign-reversed flow equations \eqref{SL_DW_Flow2} extremize the corresponding BPS reduced action and the tension is given by $\calt_{\rm DW}=2\big(|\calz|_{z=-\infty}-|\calz|_{z=+\infty}\big)$. 

Furthermore, as mentioned above, at the end of the previous section, the case in which there is a vanishing point of $z_0$ of $\calz$ can be obtained by gluing two regions along which $|\calz|$ flows in opposite directions, first decreasing from $|\calz|_{z=-\infty}$ to $0$ and then increasing to $|\calz|_{z=+\infty}$. The above arguments can be easily adapted to this case as well and give  $\calt_{\rm DW}=2\big(|\calz|_{z=-\infty}+|\calz|_{z=+\infty}\big)$, again as in the absence of membranes \cite{Cvetic:1992bf, Cvetic:1992st, Cvetic:1992sf, Cvetic:1993xe,Cvetic:1996vr, Ceresole:2006iq}. However,  in this case the membrane sitting at $z_0$ would have vanishing localized tension, $\calt_{\rm DW}=0$. This would signal breaking of the validity of the effective action.

\section{A mini-landscape example}

To exemplify the discussion of the previous section, let us now consider a concrete simple model for which domain wall solutions can be computed explicitly. In the super-Weyl, chiral multiplet formulation the theory is described by two superfields $z^a=(z^0,z^1)$, associated with the prepotential 
\be\label{SL_DWEx_expre}
\calg=-\ii z^0 z^1.
\ee
The periods are then defined as in \eqref{Sugra_MTF_Ex_maxrank}
\be\label{SL_calvA}
\calv^A = \begin{pmatrix}
	z^0 \\ z^1 \\ -\ii z^1 \\ -\ii z^0
\end{pmatrix}\;.
\ee

The super-Weyl invariance can then be gauge-fixed by choosing, for example
\begin{equation}
\label{SL_DWEx_exfI}
	z^0 = 1\,, \qquad z^1 = - \ii \phi\,,
\end{equation}
where $\phi$ is the only physical field. The gauge-fixed potential acquires the form \eqref{Sugra_MTF_Wphys}, with $N_A = (e_0,e_1,-m^0,-m^1)$
\be\label{SL_DWEx_exSup}
W(\Phi)=(e_0+\ii m^1)-\ii (e_1 +\ii m^0)\Phi \, .
\ee

The dual three-form picture is easily identified from the discussion of Section~\ref{sec:Sugra_MTF}. We would like to dynamically generate all the constants $N_A$. We will then need, as for the maximal nonlinear case of Section~\ref{sec:Sugra_MTF_Examples}, four gauge three-forms, properly included into a couple of double three-form multiplets. Since $\calg$ is quadratic, the $2\times 2$ matrix
\be
\calg_{ab}=\ii\calm_{ab}=-\ii\left(\begin{array}{cc} 0 & 1\\
	1 & 0\end{array}\right)
\ee
is constant, and the constraint \eqref{Sugra_MTF}, which defines the two double three-form multiplets $S^a$ in terms of the complex linear superfields $\Sigma_a=(\Sigma_0,\Sigma_1)$,  becomes linear
\be
S^0= - \frac\ii2(\bar\cald^2-8\calr)\Im\Sigma_1\,,\quad S^1=-\frac\ii2(\bar\cald^2-8\calr)\Im\Sigma_0 \, . 
\ee
At the component level, they include the four three-forms
\be\label{SL_DWEx_exgauge3}
\left(\begin{array}{l} A^0_{(3)} \\
	A^1_{(3)}\\
	\tilde A_{(3)0}\\
	\tilde A_{(3)1}
\end{array}\right)
\ee
where, for clarity, we have introduced the change of notation $A^A_{3}\rightarrow A^A_{(3)}$, with field strengths $F^0_{(4)},F^1_{(4)},\tilde F_{(4)0}, \tilde F_{(4)1}$. The three-form Lagrangian is then \eqref{Sugra_MTF_L3f}, where, in the following, we will put to zero superpotential $\hat \calw$  for simplicity. 

As of now we have not assumed any particular form for the K\"ahler potential. We will regard $z^a$ as coordinates for a one-dimensional \emph{special K\"ahler manifold}, whose K\"ahler potential is of the same form as \eqref{EFT_IIA_Kk}. Then, after gauge-fixing the super-Weyl invariance, we arrive at
\be\label{SL_DWEx_exKP}
K(\phi,\bar\phi)=-\log\left(4\Im\phi\right)+ \hat K_0\;.
\ee
where $\hat K_0$ is an additional, constant contribution. In particular, we see that in order to have a well defined K\"ahler potential we should require
\be\label{SL_DWEx_phipos}
\Im\phi>0\,.
\ee

\subsection{Bosonic action, three-forms  and $SL(2,\mathbb{Z})$ dualities}

The three-form Lagrangian may more conveniently re-written by introducing the following $SL(2,\mathbb{Z})$--covariant basis \cite{Bandos:2018gjp}
\be\label{SL_DWEx_excalf}
\calf_{(4)0}=\tilde F_{(4)0}-\ii F_{(4)}^1\,,\quad \calf_{(4)1}=\tilde F_{(4)1}-\ii F_{(4)}^0 \, . 
\ee
and we will collect $\calf_{(4)a} =(\calf_{(4)0}, \calf_{(4)1})$. It is then possible to show that the gauge three-form action \eqref{Sugra_MTF_L3fComp_gf} reduces to
\be\label{SL_DWEx_exbaction}
S= \int_\Sigma\left[\frac12 R *\!1 -  \frac{\d \phi \wedge *\!\d \bar\phi}{4(\Im\phi)^2} - T^{ab}(\phi) \bar\calf_{4a} *\!\calf_{4b}\right]+S_{\rm bd}
\ee
with the boundary contribution
\be
\label{SL_DWEx_exbactionbd}
S_{\rm bd}= \int_{\del \Sigma} (\tilde{A}_{b(3)}-\calg_{ac} A^c_{(3)}) T^{ab}  *\!\calf_{4b}+ {\rm c.c.}
\ee
and 
\be\label{SL_DWEx_excalt}
T^{ab} (\phi)= \frac{e^{-\hat K_0}}{6\,\Im\phi}\left(\begin{array}{rc} 1 {\ \ \ \ \ \ } & \ii\phi -  \Im\phi\\
	-\ii\bar\phi -  \Im\phi & |\phi|^2\end{array}\right)
\, , 
\ee
where $\hat K_0$ is the constant appearing in the complete K\"ahler potential \eqref{SL_DWEx_exKP}.
It is known that the group of symmetries associated with the special K\"ahler structure defined by the prepotential \eqref{SL_DWEx_expre} is $SL(2,\mathbb{R})$ (see for instance \cite{Freedman:2012zz}). More precisely, an element
\be
\left(\begin{array}{rr} a & b\\
	c & d\end{array}\right)\in SL(2,\mathbb{R}) 
\ee
acts on $\phi$ as follows
\be\label{SL_DWEx_phiSL2R}
\phi\rightarrow\frac{a\phi+b}{c\phi+d}\,.
\ee
In the assumption of the quantization conditions of Section~\ref{sec:Sugra_Quant}, this reduces to $SL(2,\mathbb{Z})$ (with $a,b,c,d\in\mathbb{Z}$), which we will interpret as the duality group of the model. This duality symmetry $SL(2,\mathbb{Z})$ is embedded into the group of symplectic transformations $Sp(4,\mathbb{Z})$, see e.g. \cite{Freedman:2012zz}. Here we focus on the $SL(2,\mathbb{Z})$ generators 
\be
\mathfrak{t}=\left(\begin{array}{rr} 1 & 1\\
	0 & 1\end{array}\right),\quad\mathfrak{s}=\left(\begin{array}{rr}\ 0 &\ 1\\
	-1 & 0\end{array}\right) 
\ee
which correspond to the following $Sp(4,\mathbb{Z})$ transformations
\be\label{SL_DWEx_SL2ZS}
\cals(\mathfrak{t})=\left(\begin{array}{rrrr} 1 &\ 0 &\ 0 &\ 0\\
	0 & 1 & 0 & 1\\
	-1& 0 & 1 & 0\\
	0 & 0 & 0 & 1
\end{array}\right),
\quad\cals(\mathfrak{s})=
\left(\begin{array}{rrrr} 
	0 &\ 0 &\ 1 &\ 0\\
	0 & 0 & 0 & 1\\
	-1& 0 & 0 & 0\\
	0 &-1 & 0 & 0
\end{array}\right)\,.
\ee
One can check that,  applying $\cals(\mathfrak{t})$ and $\cals(\mathfrak{s})$ to the (bosonic component of)  the symplectic vector \eqref{SL_calvA}, one gets the $\mathfrak{t}$- and $\mathfrak{s}$-actions on $\phi$ as in \eqref{SL_DWEx_phiSL2R}.\footnote{Under $\mathfrak{s}$ one needs to make also the change  $Y\rightarrow \phi Y$, which can be reabsorbed by a K\"ahler transformation $K\rightarrow K-\log\phi -\log\bar\phi$.} Furthermore, one can verify that $[\cals(\mathfrak{s})\cals(\mathfrak{t})]^3=\bbone$, which together with $\cals(\mathfrak{s})^2=-\bbone$  implies that \eqref{SL_DWEx_SL2ZS}  generate a four-dimensional representation of $SL(2,\mathbb{Z})$. 

By applying \eqref{SL_DWEx_SL2ZS} to \eqref{SL_DWEx_exgauge3}, one can then get the transformation properties of the gauge three-forms under the $SL(2,\mathbb{Z})$ duality group. Consider  the complex field-strengths \eqref{SL_DWEx_excalf} and organize them into a 2-components vector
\be\label{SL_DWEx_calfvec}
\vec\calf_{(4)}\equiv \left(\begin{array}{r}
	\calf_{(4)0}\\
	\calf_{(4)1}
\end{array}\right) \, . 
\ee
Under  $SL(2,\mathbb{Z})$ we have  $\vec\calf_{(4)}\rightarrow U\vec\calf_{(4)} $, with
\be\label{SL_DWEx_USL(2,Z)}
U(\mathfrak{t})= \left(\begin{array}{rr} 
	1 & -\ii\\
	0 & 1
\end{array}\right) 
\,,\quad 
U(\mathfrak{s})=
\left(\begin{array}{rr}
	0 & -\ii\\
	-\ii & 0\end{array}\right)\,.
\ee
Notice that these matrices satisfy $U^\dagger\sigma_1 U=\sigma_1$ and $\det U=1$, i.e.\  they are elements of $SU(1,1)$ (defined with respect to the $\mathbb{C}^2$ metric $\sigma_1$), which is known to be isomorphic to $SL(2,\mathbb{R})$. In other words,  $U(\mathfrak{t})$ and $U(\mathfrak{s})$ generate the $SU(1,1)$ representation of $SL(2,\mathbb{Z})$.

Using \eqref{SL_DWEx_excalt}, one can also check that the $2\times 2$ matrix $T^{ab}(\phi)$ transforms as follows 
\be
\begin{aligned}
	T(\phi+1)&=  U(\mathfrak{t})^{\dagger -1 } T(\phi) U(\mathfrak{t})^{-1} \, , \\
	T\Big(-\frac1\phi\Big)&=U(\mathfrak{s})^{\dagger -1 } T(\phi) U(\mathfrak{s})^{-1} \, . 
\end{aligned}
\ee
Observing that the combination $\tilde{A}_{(3)I}-\bar\calg_{IK} A^K_{(3)}$ appearing in the boundary term \eqref{SL_DWEx_exbactionbd} transforms as $\calf_{(4)I}$, one can readily check that the bulk and boundary terms in the action \eqref{SL_DWEx_exbaction} are separately invariant under the $SL(2,\mathbb{Z})$ duality group. 
This shows that the $SL(2,\mathbb{Z})$ duality group is indeed a symmetry of the action \eqref{SL_DWEx_exbaction}.


\subsection{The mini-landscape of vacua}
\label{SL_DWEx_sec:landscape}

Let us study the vacua of the action \eqref{SL_DWEx_exbaction} (in the absence of membranes). As discussed in Section \ref{sec:Sugra_MTF}, one can first integrate out the gauge three-forms by  picking up a particular symplectic constant vector 
\be\label{SL_DWEx_exconst}
\left(\begin{array}{c}
	m^0 \\ m^1 \\ e_0 \\ e_1\end{array}\right)\in\mathbb{Z}^4\,,
\ee
and rewriting \eqref{SL_DWEx_exbaction} in the form
\be\label{SL_DWEx_exbaction2}
S=\int_\Sigma\left[\frac12 R *\!1 -\frac{\d \phi \wedge * \d \bar\phi}{4(\Im\phi)^2} -V(\phi,\bar\phi)\right],
\ee
where $V$ is a conventional $\caln=1$ potential 
\be\label{SL_DWEx_expot}
\begin{aligned}
	V(\phi,\bar\phi)&= e^{K+\hat K_0}\Big(K^{\phi\bar\phi}|D_\phi W|^2-3|W|^2\Big)\\
	&=-\frac{e^{\hat K_0}}{2\Im\phi} \Big[(m^1)^2+(e_0)^2+4(m^0m^1+e_0e_1)\Im\phi 
	\\
	&\quad~~~~~~~~~~~~~~ +2(m^0e_0-m^1e_1)\Re\phi+((m^0)^2+(e_1)^2)|\phi|^2\Big]\,, 
\end{aligned}
\ee
with $W$  and $K$ are as in  \eqref{SL_DWEx_exSup} and \eqref{SL_DWEx_exKP}, respectively. Generically, the effective action \eqref{SL_DWEx_exbaction2} is not invariant under $SL(2,\mathbb{Z})$ transformations of $\phi$ (unless we  appropriately transform also the integration constants $m^a,e_a$). However, given the `microscopic' formulation with three-forms we started from, we can regard this breaking as {\em spontaneous} rather than explicit. 

Let us first consider the simplest  possibility:  $m^0=m^1=e_0=e_1=0$. In this case $W\equiv 0$ and hence $V\equiv 0$. So, we have a one-dimensional moduli space of vacua parametrized by an arbitrary expectation value of $\phi$. As standard in similar situations, one should identify two vacua related by an  $SL(2,\mathbb{Z})$  duality transformation. In view of the restriction \eqref{SL_DWEx_phipos}, the  moduli space of the inequivalent vacua can be identified with the familiar fundamental domain
\be\label{SL_DWEx_fundom}
\Big\{-\frac12\leq\Re\phi\leq \frac12\Big\}\cap \Big\{|\phi|\geq 1\Big\}\,.
\ee

On the other hand, this moduli space is drastically modified by any non-trivial set of constants \eqref{SL_DWEx_exconst}. It is useful to introduce the complex numbers
\be
\alpha_0\equiv e_0-\ii m^1\,,\quad \alpha_1\equiv e_1-\ii m^0
\ee
taking values in $\mathbb{Z}+\ii\mathbb{Z}$. Notice that the vector
\be
\vec\alpha\equiv  \left(\begin{array}{r}
	\alpha_0\\
	\alpha_1
\end{array}\right)
\ee
transforms as \eqref{SL_DWEx_calfvec} under the $SL(2,\mathbb{Z})$ duality tranformations, that is, in the fundamental $SU(1,1)$ representation generated by \eqref{SL_DWEx_USL(2,Z)}. The bosonic component of the superpotential \eqref{SL_DWEx_exSup} takes the form $W=\bar\alpha_0-\ii\bar\alpha_1\phi$ and the corresponding supersymmetric vacuum expectation value of $\phi$ (such that $D_\phi W|_{\phi_*}=0$) is 
\be\label{SL_DWEx_exvacua}
\phi_*=\ii\frac{\alpha_0}{\alpha_1}\,.
\ee
Taking into account \eqref{SL_DWEx_phipos}, we require
\be\label{SL_DWEx_exposcond}
\Im\phi_*= \frac{1}{|\alpha_1|^2}\Re(\alpha_0\bar\alpha_1) 
\ee
to be finite and positive.
In particular, this implies that the cases
$\alpha_1=0$ and $\Re(\alpha_0\bar\alpha_1)=0$ must be discarded.  
Then, given a certain $\vec\alpha$, \eqref{SL_DWEx_exvacua} is the only extremum of the potential \eqref{SL_DWEx_expot}. 

The condition $\Im\phi_*>0$ is equivalent to requiring that
\be\label{SL_DWEx_expos}
\Re(\alpha_0\bar\alpha_1)\equiv \frac12 \vec\alpha{}^\dagger \sigma_1\vec\alpha>0 . 
\ee
At the supersymmetric vacua \eqref{SL_DWEx_exvacua} the covariantly holomorphic superpotential $\calz$ takes the value
\be\label{SL_DWEx_calz*}
\calz_*=e^{\frac12\hat K_0}\frac{\bar\alpha_1}{|\alpha_1|}\sqrt{\Re(\alpha_0\bar\alpha_1)}
\ee
and the potential reduces to
\be
V_*=-3|\calz_*|^2=-3\,e^{\hat K_0}\Re(\alpha_0\bar\alpha_1)\,, 
\ee
which is strictly negative in view of \eqref{SL_DWEx_expos}, and thus determines the constant curvature of the AdS vacuum. 
The AdS radius (where, we recall, natural units $M_{\rm P}=1$) is identified with the inverse of
\be\label{SL_DWEx_exmodZ}
|\calz_*|= e^{\frac12\hat K_0}\sqrt{\Re({\alpha_0\bar\alpha_1})}\, . 
\ee
Since $\alpha_0$ and $\alpha_1$ are integrally quantized, we should assume that $e^{\frac12\hat K_0}\ll 1$ in order  to be within the regime of reliability of our effective supergravity, which is equivalent to $|\calz_*|\ll 1$, therefore the AdS radius is much larger than the Planck length.

Now the $SL(2,\mathbb{Z})$ duality group  of the theory relates a vacuum \eqref{SL_DWEx_exvacua} associated with a certain set  of constants \eqref{SL_DWEx_exconst} (and a corresponding effective superpotential \eqref{SL_DWEx_exSup}) to another vacuum associated with a different set of constants and effective superpotential. 
It follows that, chosen a certain (generic) set of constants \eqref{SL_DWEx_exconst},  the domain of  $\phi$ is the entire upper half-plane \eqref{SL_DWEx_phipos}  (up to some possible residual and non-generic identification),  and not the fundamental domain  \eqref{SL_DWEx_fundom}. This is a purely four-dimensional realization of the flux-induced monodromy effects observed in string compactifications, see for instance \cite{Marchesano:2014mla} for a recent discussion.

\begin{figure}[h]
	\centering\includegraphics[width=7cm]{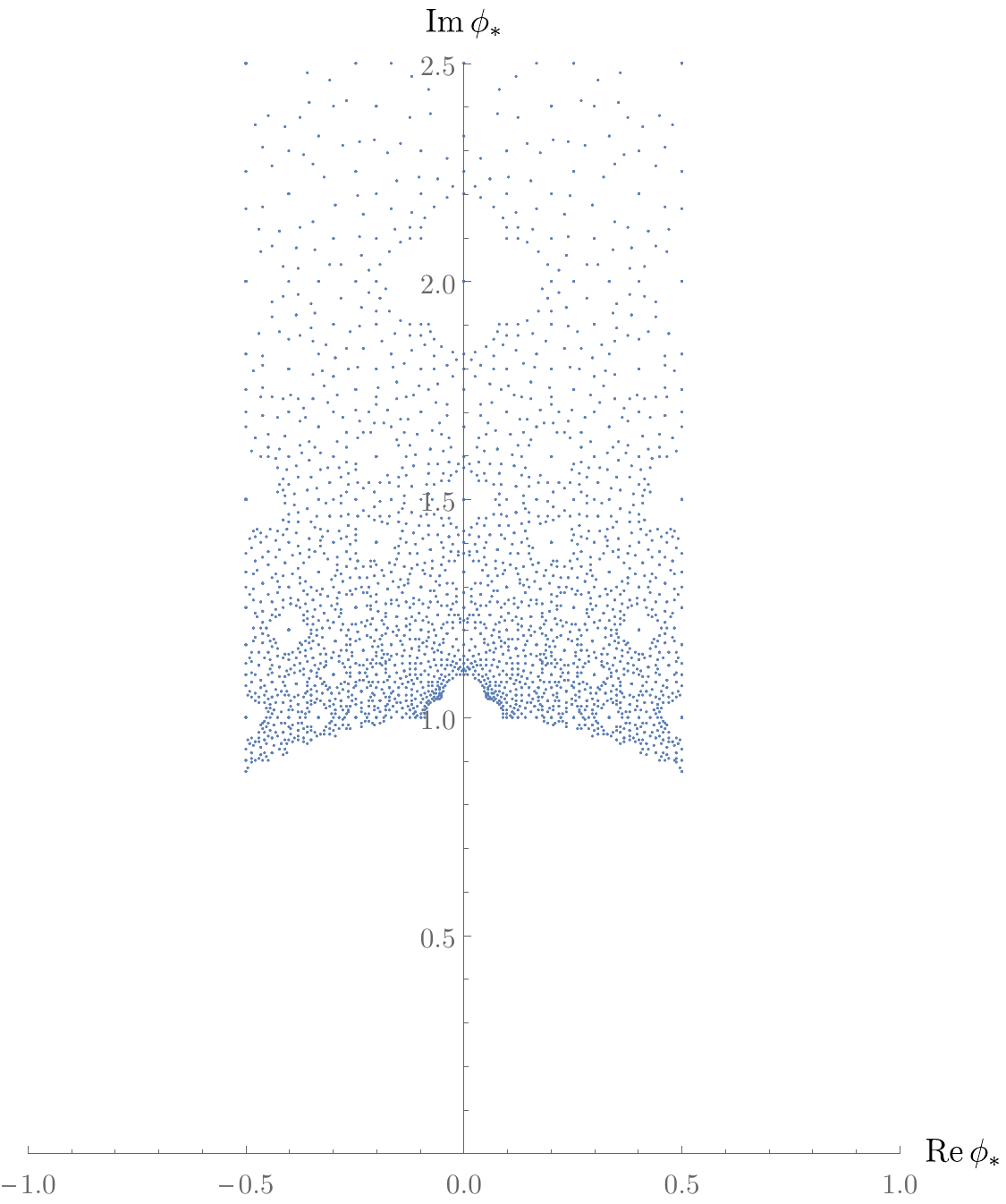}
	\caption{\footnotesize A sampling of vacua \eqref{SL_DWEx_exvacua} filling the fundamental domain \eqref{SL_DWEx_fundom}, for the values of the constants $e_I,m^J \in [-11,11]$.}
	\label{fig:SL_vacua}
\end{figure}

On the other hand, in order to identify the {\em inequivalent} vacua, corresponding to inequivalent choices of the constants  \eqref{SL_DWEx_exconst} and of the corresponding effective potentials,  we can restrict ourselves to  the vacua \eqref{SL_DWEx_exvacua} which sit in the fundamental domain of \eqref{SL_DWEx_fundom}.
The set of such vacua is plotted in Fig.~\ref{fig:SL_vacua}. A similar  set of vacua appears in the simplest models of type IIB flux compactifications on a rigid Calabi--Yau  \cite{Denef:2004ze}, in which $\phi$ can be identified with the  axion-dilaton. In the type IIB models one needs to impose the tadpole cancellation condition, which adds a constraint on the set of allowed vacua. In our formulation with gauge three-forms, the tadpole cancellation condition can be implemented as outlined at the end of Section \ref{sec:Sugra_4FM}.


\subsection{Domain walls  between aligned vacua}
\label{SL_DWEx_sec:exDWs}

In this section we explicitly construct a class of  domain walls of the kind discussed in the previous section, encapsulating a membrane. The membrane will be considered static, frozen at its position $y=0$.\footnote{Here, with respect to the previous section, we have called the fourth spacetime direction $y$ instead of $z$ in order not to make confusion with the chiral fields $z^a$.} These relate pairs of vacua $\phi|_{-\infty}=\phi_*$ and $\phi|_{+\infty}=\phi_*'$ of the form \eqref{SL_DWEx_exvacua}, corresponding to two sets of constants $\alpha_a$ and $\alpha_a'$ respectively. We make the simplifying assumption that the phases of $\calz_*$ and $\calz'_*$ are aligned  and that the phase of  $\calz(y)$ remains constant along the flow.
Of course, in order to have a (non-trivial) domain wall  $|\calz_*|$ and $|\calz'_*|$ should be different and then the corresponding vacua cannot be related by a $SL(2,\mathbb{Z})$ duality.
From \eqref{SL_DW_thetaflow} we see that we should impose $\Im(\dot\phi\,\del_\phi K)=0$ with $K$ as in \eqref{SL_DWEx_exKP}.  This is possible only if $\Re\phi$ is constant and equals to
\be\label{SL_DWEx_exaxion}
\Re\phi_*= -\frac{\Im(\alpha_0\bar\alpha_1)}{|\alpha_1|^2} . 
\ee
Clearly, we should also require $\Re\phi_*'=\Re\phi_*$\,.
Hence, 
\be
v(y) \equiv \Im \phi(y) 
\ee
is the only dynamical real field along the flow.  Again, we will assume that $|\calz|$ is always increasing along the flow, which drives the field $v$ from $v|_{-\infty}=v_*$ towards $v|_{+\infty}=v_*'$ and, at $y=0$, it crosses a membrane of charges $p^a,q_b$ such that
\be
q_0-\ii p^1=\alpha_0-\alpha_0'\,,\quad q_1-\ii p^0=\alpha_1-\alpha'_1\,.
\ee

The equations of the flow \eqref{SL_DW_Flow2} are governed by the growth of $|\calz|$. In particular,   equation \eqref{SL_DW_Flow2a} reduces to
\be\label{SL_DWEx_flowv}
\dot{v} = 4v^2\, \frac{\d}{\d v} |\calz|\,.
\ee
For $y<0$, $|\calz|$ takes the following form
\be\label{SL_DWEx_Zv}
|\calz(v)|=\frac{e^{\frac12\hat K_0}}{2\sqrt{v}}\left[|\alpha_1|v+\frac{\Re(\alpha_0\bar\alpha_1)}{|\alpha_1|}\right]\,.
\ee
For $y>0$ the from of $|\calz|$ is obtained by replacing $\alpha_a$ with $\alpha_a'$ in \eqref{SL_DWEx_Zv}.

On the left of the membrane, $v_*$ is a global minimum of $|\calz|$. Hence, it is a repulsive fixed point of \eqref{SL_DWEx_flowv}, a flow is triggered and $v$ is driven away from $v_*$, letting the value of $|\calz|$ increase. When the membrane is reached at $y=0$, $v$ and consequently $|\calz|$ have evolved to certain values $v(0)$ and $|\calz|_{y=0}$. Here, the solution of the flow equations on the left should be glued to the one on the right. We are then led to impose the continuity of $v$ across $y=0$ while still keeping a growing $|\calz|$. However, since on the right of the membrane $v_*'$ is also a global minimum of $|\calz|$, $v_*'$ is a repulsive (rather than attractive) fixed point of \eqref{SL_DWEx_flowv}. Hence the solution to the flow equations is such that  $v$ reaches the value $v_*'$ at $y=0$ and then  remains constant
\be
v(y) = v_*' \quad \text{for}\;\,\,y\geq 0\,.
\ee
Correspondingly,  $|\calz|$ starts from $|\calz_*|$ at $y=-\infty$ and smoothly grows  until it reaches the membrane. At this point it jumps up to $|\calz'_*|$ and then remains constant (see Fig. \ref{fig:SL_flowZ} for an example). Hence, on the right of the membrane, the background is just the AdS vacuum solution.

Recalling 
\eqref{SL_DW_tension2}, we  see that the bound
\be\label{SL_DWEx_DZbound}
\calt_{\rm DW} \geq   \calt_{\rm memb}\,
\ee
is saturated if and only if on the left-hand side of the membrane $|\calz|$ is also constant. In the following we will first examine  the case in which the bound \eqref{SL_DWEx_DZbound} is saturated, leading to trivial flow equations on both sides of the membrane, and then we will consider an example for which the inequality \eqref{SL_DWEx_DZbound} strictly holds. 

As a warm up, 
let us assume that on the left of the membrane $\vec\alpha = 0$, so that the potential \eqref{SL_DWEx_expot} and $\calz$ are identically zero. Then, for $y<0$, the flow equations \eqref{SL_DW_Flow2} are trivial and are immediately solved by taking $v$ and $D$ to be arbitrary constants. In particular, with no loss of generality, we can choose $D(y)\equiv0$ for $y<0$. 
Therefore, on the left of the membrane, the bulk is always at a fixed Minkowski vacuum. As discussed above, on the right of the membrane, the bulk is at its supersymmetric AdS vacuum, in which $v$ takes the constant value $v_*'$. Hence, by continuity, we should impose that $v(y)\equiv v_*'$ also for $y<0$.  Furthermore, by imposing also the continuity of the warp factor, we must set $D(y)=-|\calz'_*|y$ for $y>0$. 

It is worthwhile to mention that this particular case of trivial flow for both $y < 0$ and $y > 0$ can be realized only  when on the left-hand side the vacuum is Minkowski, owing to the freedom in choosing any constant value of $v$ for $y<0$.

\begin{figure}[h]
	\centering\includegraphics[width=6cm]{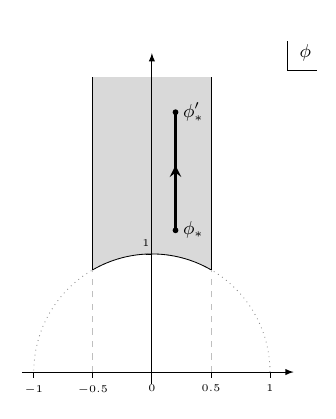}
	\caption{\footnotesize The fundamental domain  of $\phi$. Along the domain wall, $\phi$ flows up along the vertical line specified by $\Re \phi_*$.}
	\label{fig:SL_FDf}
\end{figure}

Let us now consider a more involved example, for which the flow on the left side of the membrane is nontrivial. For any choice of initial constants $\alpha_0,\alpha_1$ and any $k\in\mathbb{Q}$ such that
\be
k\alpha_1\in\mathbb{Z}+\ii\mathbb{Z}
\ee
we can choose a jump to new constants
\be\label{SL_DWEx_constchoice}
\alpha_0'=\alpha_0+k\alpha_1\,,\quad \alpha_1'=\alpha_1 \, , 
\ee
which clearly satisfies $\Re\phi_*'=\Re\phi_*$. Notice that
\be
\Im\phi'_*=\Im\phi_*+k\,.
\ee
The flow moves along the vertical direction of the upper-half-plane parametrized by $\phi$ (see Fig. \ref{fig:SL_FDf}). With no loss of generality, we take $k>0$, so that $\Im\phi'_*>\Im\phi_*$.   From \eqref{SL_DWEx_calz*}, one can also see that $|\calz_*'|>|\calz_*|$ which is the default  assumption in the previous section. 

Under these restrictions, the initial and final values of $v(y)$ are
\be
v_*=\frac{\Re(\alpha_0\bar\alpha_1)}{|\alpha_1|^2}\,,\quad~~~~ v_*'=\frac{\Re(\alpha_0\bar\alpha_1)}{|\alpha_1|^2}+k 
\ee
and the membranes charges are
\be\label{SL_DWEx_exMcharges}
p^0=0\,,\quad p^1=k\,\Im\alpha_1\,,\quad q_0=-k\,\Re\alpha_1\,,\quad q_1=0\,.
\ee

\begin{figure}[ht!]
	\centering\includegraphics[width=9cm]{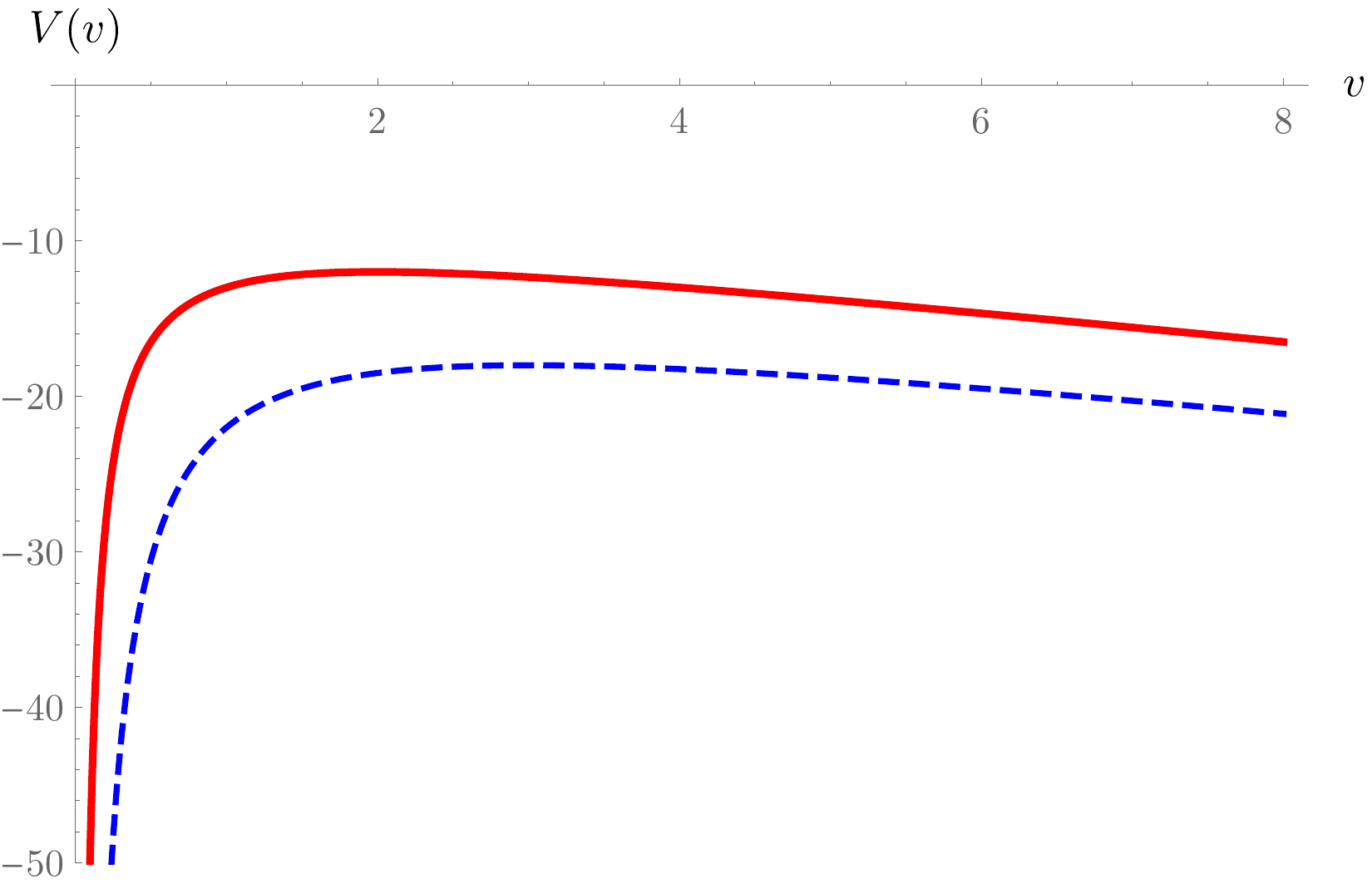}
	\caption{\footnotesize The potential \eqref{SL_DWEx_expot} for the choice of the constants $e_1 = m^0 =1$, $e_0 = m^1= 2$, $k=1$ and keeping $\Re\phi = 0$. The solid red line refers to the potential on the left of the membrane, while the dashed blue line to that on the right. This potential exhibits, on the left of the membrane, a supersymmetric AdS critical point located at $v_*=2$ and, on the right, a supersymmetric AdS critical point at $v_*'=3$.}
	\label{fig:SL_Example_Pot}
\end{figure}

\begin{figure}[ht!]
	\centering\includegraphics[width=8cm]{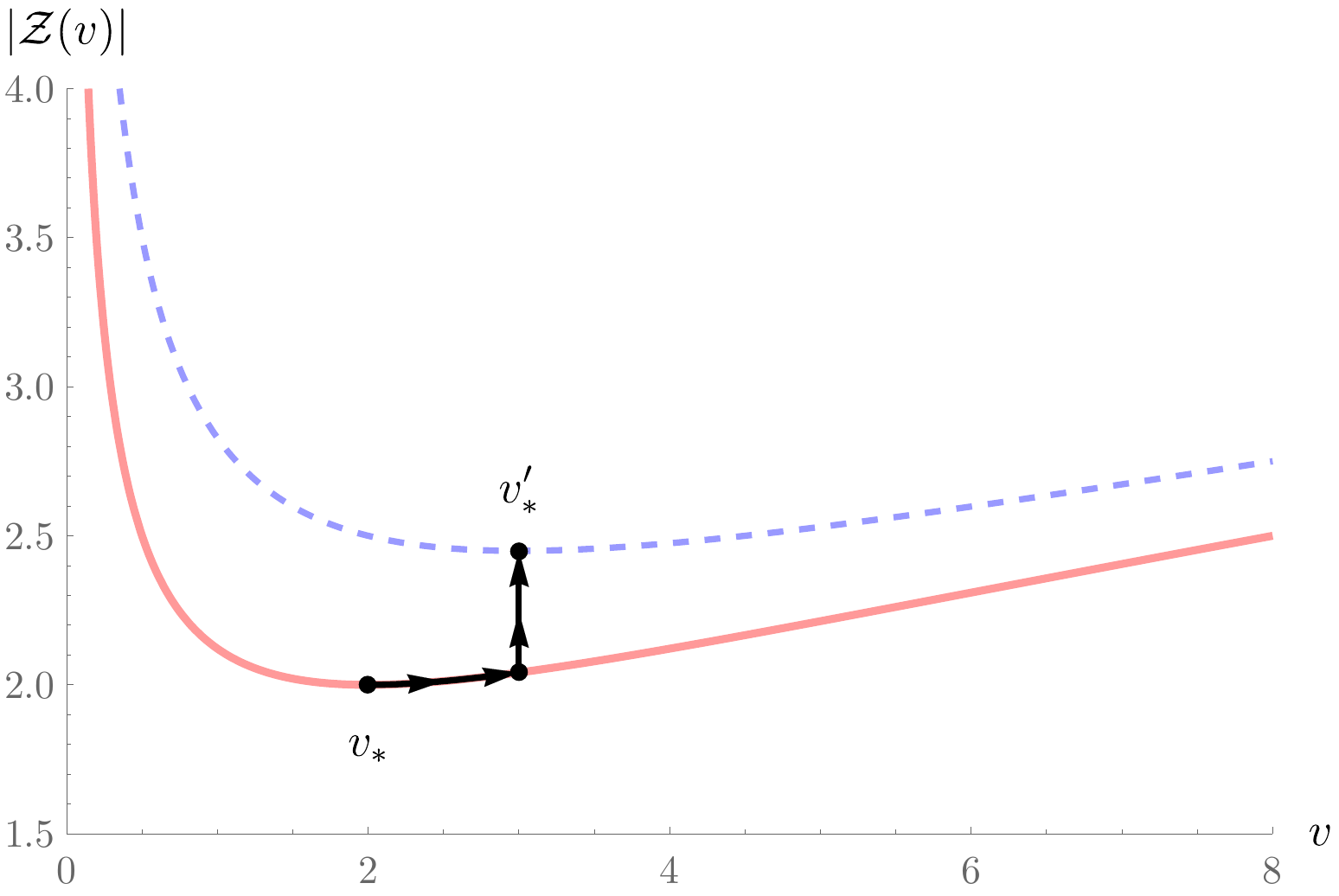}
	\caption{\footnotesize The flow of $|\calz|$ for the same set of parameters as in Fig. \ref{fig:SL_Example_Pot}. The solid red line refers to $|\calz|$ on the left of the membrane, while the dashed blue line to that on the right. The flow drives $v_*$ towards the value $v_*'$, at which $|\calz|$ jumps so that $v$ is located at new supersymmetric vacuum on the right.}
	\label{fig:SL_flowZ}
\end{figure}

We can now compute the function  $\calz(v,y)$ corresponding to our setting
\be
\calz(v,y)=\frac{\bar\alpha_1 e^{\frac12\hat K_0}}{2|\alpha_1|\sqrt{v}}\left[|\alpha_1| v+\frac{\Re(\alpha_0\bar\alpha_1)}{|\alpha_1|}+k|\alpha_1|\Theta(y)\right].
\ee
In agreement with \eqref{SL_DW_dcalz}, $\calz(v,y)$ is discontinuous at $y=0$ and the width of the discontinuity is set by the tension of the membrane with the charges \eqref{SL_DWEx_exMcharges}
\be
\lim_{\varepsilon\rightarrow 0 }\Big|\calz(y_{\rm M}+\varepsilon)-\calz(y_{\rm M}-\varepsilon)\Big|=\frac{k|\alpha_1|e^{\frac12\hat K_0}}{2\sqrt{v|_{y=0}}}\equiv \frac12 \calt_{\rm memb}.
\ee
An example for the flow of $|\calz|$ is depicted in Fig. \ref{fig:SL_flowZ}.

Consider now the flow equation \eqref{SL_DWEx_flowv}. For the examples under consideration, it takes the explicit form
\be
\dot v=e^{\frac12\hat K_0}\sqrt{v}\left[|\alpha_1|v-\frac{\Re(\alpha_0\bar\alpha_1)}{|\alpha_1|}-k|\alpha_1|\Theta(y)\right]\label{SL_DWEx_exflow1}\, , 
\ee
which is solved by
\be
v(y)=\left\{\begin{array}{l}
	v_*\coth^2\left[\frac12|\calz_*|(y+c)\right]\quad~~\text{for $y\leq 0$} \, ,
	\\
	v_*' \quad~~~~~~~~~~~~~~~~~~~~~~~~~~~~~\text{for $y\geq 0$} \, . 
\end{array}\right.\,
\ee
The integration constant $c$ must be negative, $c<0$, and  is fixed by the continuity at $y=0$, which imposes
$v_*\coth^2\left[\frac12c\,|\calz_*|\right]=v_*'
$
and   always admits a solution.

\begin{figure}[ht]
	\centering
	\includegraphics[width=7cm]{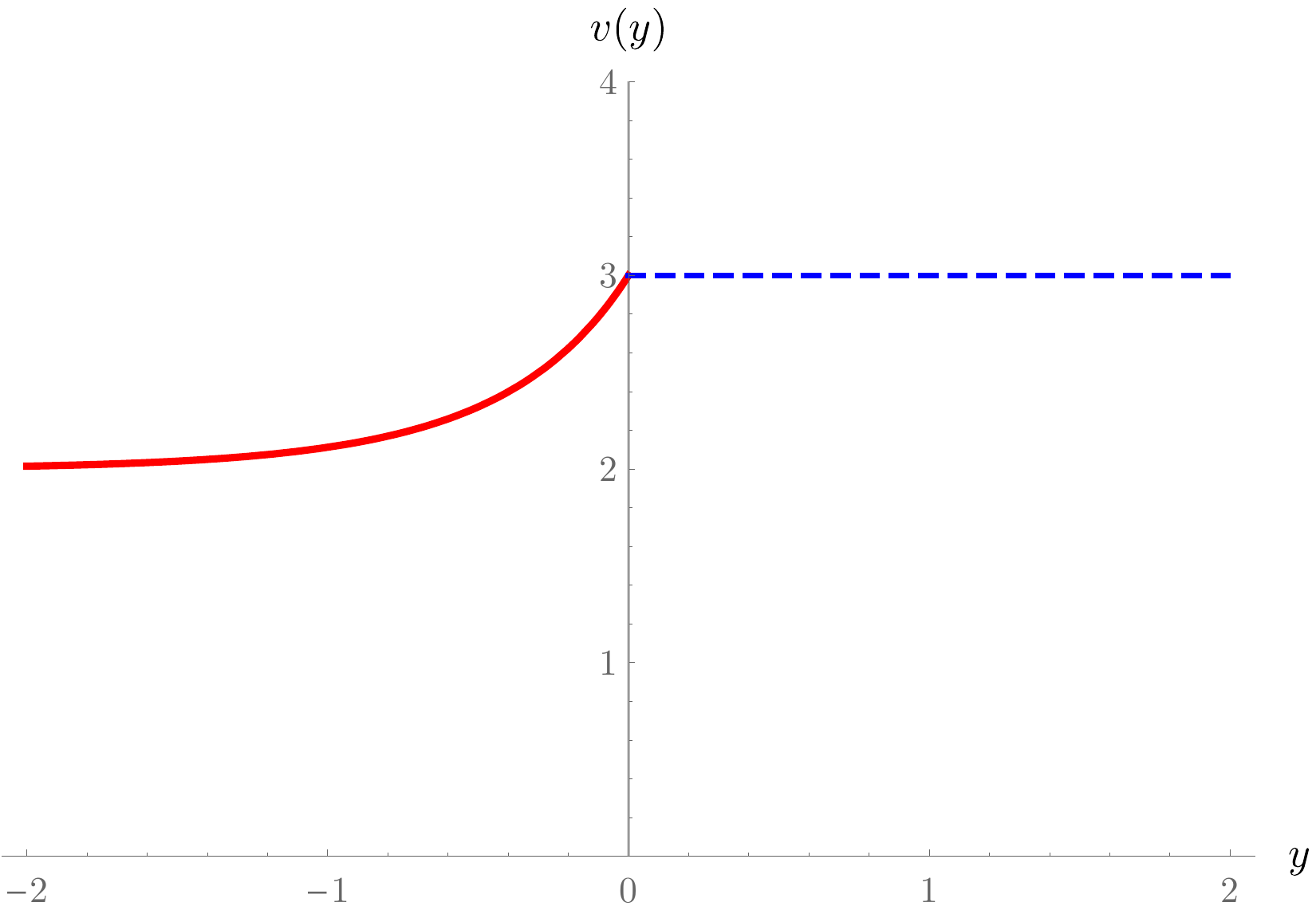} \includegraphics[width=7cm]{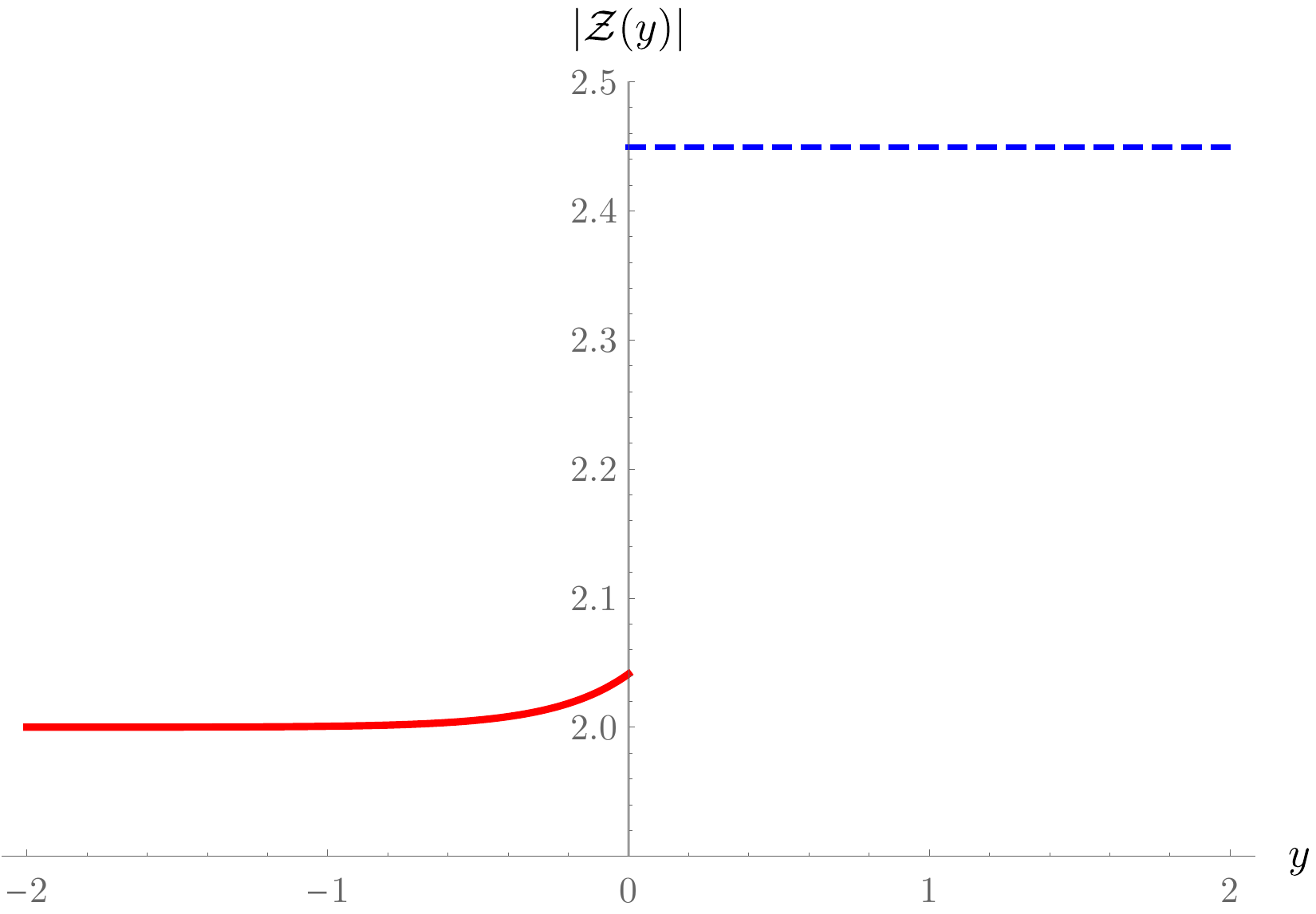}
	
	\centering
	\includegraphics[width=7cm]{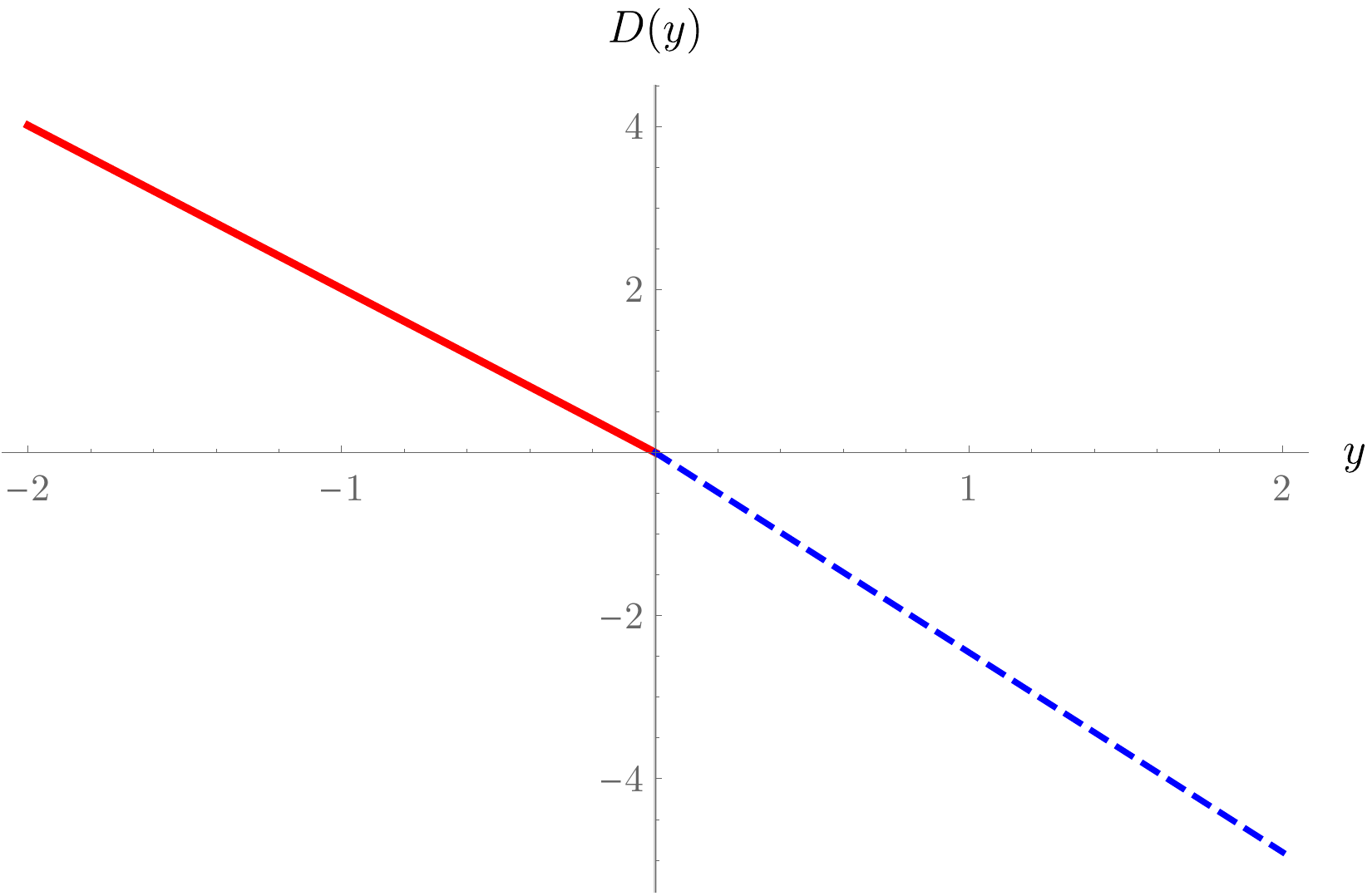} \includegraphics[width=7cm]{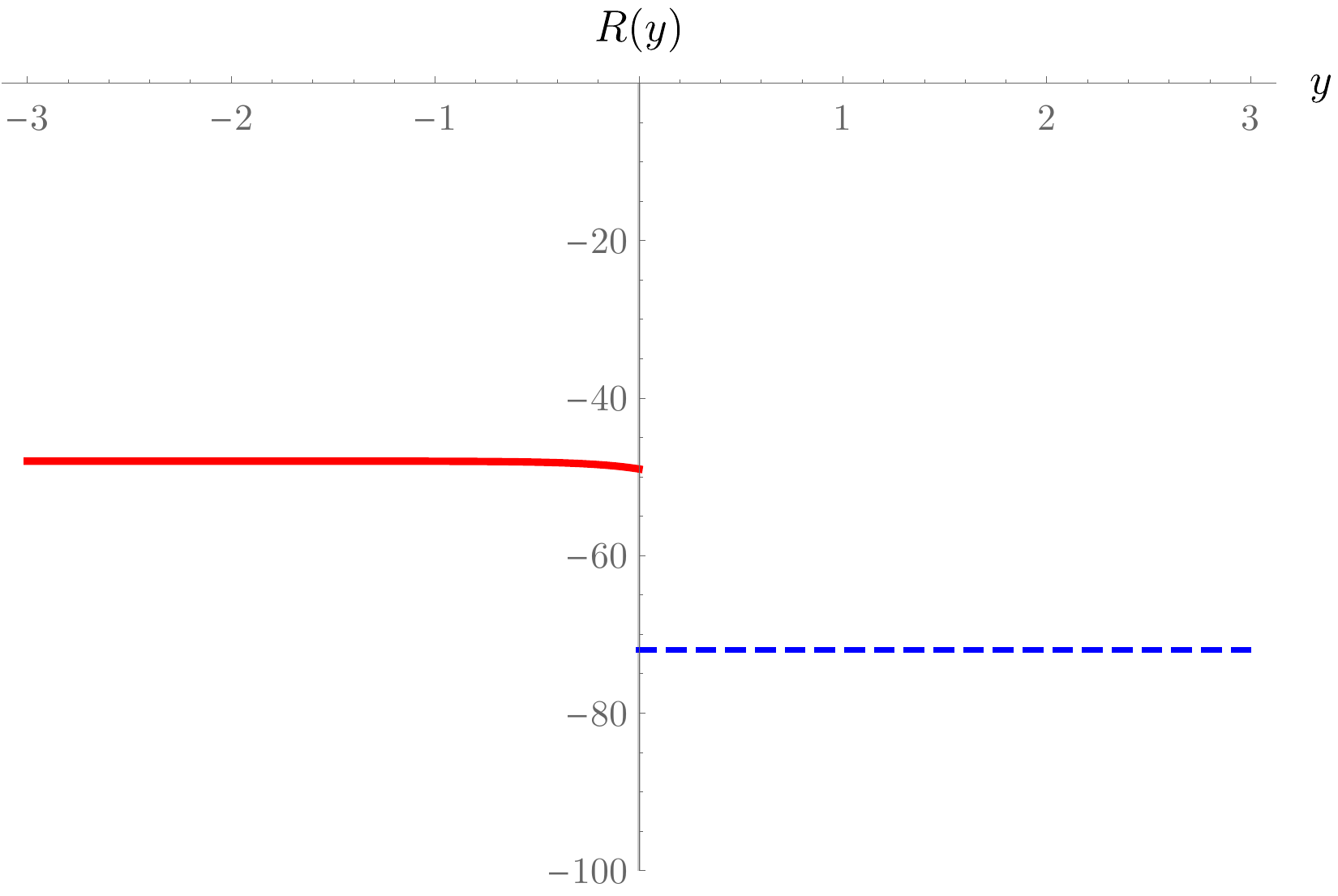}
	
	\caption{\footnotesize Above are depicted all the solutions to the flow equations \eqref{SL_DWEx_exflow1}, \eqref{SL_DWEx_exflow2} for the same set of parameters as in Fig.\,\ref{fig:SL_Example_Pot}. The solid red lines refer to the quantities in the region on the left of the membrane, while the dashed blue lines to those on the right. On the top left there is the evolution of the scalar field: starting from the critical point on the left of the membrane, the field $v$ is driven towards the critical point on the right of the membrane. On  the  top  right  there  is  the modulus of the covariantly  holomorphic superpotential $|\calz|$, which is always increasing. On the bottom left there is the warping $D(y)$, which is always decreasing, using which the curvature, on the bottom right, can be obtained. As expected from the AdS vacua, the curvature is at a fixed positive value when the field $v$ reaches the vacua and, even though not explicitly shown here, is singular at the point $y=0$. In the figures $y$ is $|\calz_*|^{-1}$ units.}
	\label{fig:SL_Memb_Simple_Example_ClassI}
\end{figure}
Since the bulk exhibits a non-trivial flow only on the left-hand side of the membrane, the membrane tension is given by the vacuum expectation value of $\phi_*'$ to the right of the membrane 
\be
\calt_{\rm memb}=\frac{k|\alpha_1|^2e^{\hat K_0}}{|\calz'_*|}.
\ee

We still have to solve the equation for the warping  \eqref{SL_DW_Flow2b}, which in the present case reads
\be
\dot D=-\frac{e^{\frac12\hat K_0}}{2\sqrt{v}}\left[|\alpha_1|v+\frac{\Re(\alpha_0\bar\alpha_1)}{|\alpha_1|}+k|\alpha_1|\Theta(y)\right]\label{SL_DWEx_exflow2}.
\ee
It also admits an analytic solution given by 
\be
D(y)=\left\{\begin{array}{l}
	d+{e^{-\frac12\hat K_0}\left(\log(-\sinh u)+\log\cosh u\right)}\quad~\text{for $y\leq 0$}\,,\\
	- |\calz_*'| y\quad~~~~~~~~~~~~~~~~~~~~~~~~~~~~~~~~~~~~~~~~~\text{for $y\geq 0$}\,, 
\end{array}\right.
\ee
where we have set to zero an arbitrary additive constant  and   $u(y)\equiv\frac12 |\calz_*|(y+c)$. The integration constant $d$  is fixed by imposing the continuity of $D(y)$ at $y=0$.

A couple of final comments. Notice that on the left-hand side of the membrane  the deviation of the complete solution from the  AdS vacuum   is concentrated within a length of the  same order of the  AdS radius $|\calz_*|^{-1}$. Hence, in this sense, the domain wall may be considered as `thick'. Furthermore, clearly, we can make a coordinate redefinition $y\rightarrow y-y_{\rm M}$ to get a solution with the membrane localised  at any point $y_{\rm M}$.

\part{Exploring the Landscape of effective field theories}


\chapter{Effective field theories from compactifications of string and M-theory}
\label{chapter:EFT}

In the previous chapters we gave a general recipe to introduce gauge three-forms into generic $\caln=1$ supergravity theories while still preserving supersymmetry. Their net effect is to generate, once they are integrated out, a potential which includes a quadratic dependence on some  quantized constants, corresponding to a superpotential which linearly depends on those constants. Such constants can be eventually constrained, an effect that in the dual three-form picture rephrases into a gauging of the three-forms. We have also illustrated how to couple linear multiplets to the dual theory, whose role is to provide an alternative description of the axions and also incorporate, via gauging, an F-term coupling between the axionic and other chiral sectors of the theory. The conclusions drawn in the previous chapter are general and hold for any four-dimensional $\caln=1$ supergravity theory. 

Concrete, important applications of the new formulation that we propose are effective field theories emerging from compactifications of string and M-theory. Although our formulation applies to more general cases, we will be mainly interested in superstring Type II EFTs compactified over orientifolded Calabi-Yau manifolds and M-theory compactified over $G_2$--manifolds. In this chapter we will be focusing on the \emph{bulk} theories, neglecting the possible presence of membranes or strings in the four-dimensional theories, which will be included in the next chapter. The material presented here is partly derived from \cite{Lanza:2019xxg}, using which we can re-obtain known results in the literature  \cite{Groh:2012tf,Bielleman:2015ina,Carta:2016ynn,Farakos:2017jme,Bandos:2018gjp,Herraez:2018vae,Escobar:2018tiu,Escobar:2018rna} but also significantly extend them.
\section{Towards a reformulation of EFTs from compactifications}
\label{sec:EFT_intro}

Before directly considering specific examples of effective theories, let us qualitatively explain the general ideas and the method that will be applied in the following sections. The main aim of this chapter is to provide a new formulations of the EFTs directly from a four-dimensional perspective, catching however all the elements that are predicted from string theory.

The realm within which we move are four-dimensional $\caln=1$ EFTs obtained after compactifying higher dimensional string and M-theory over proper compact manifolds. We shall consider effective theories with at most two-derivatives. Our first step will be to reformulate the theory in the super-Weyl invariant approach introduced in Section~\ref{sec:Sugra_SW}. The kinetic terms of the fields entering the EFT are then fully encoded into a real kinetic section $\calk$.\footnote{In this regard, we will also be neglecting higher derivatives terms on the graviton, such as those considered in \cite{Becker:2001pm,Liu:2013dna,Grimm:2013bha,Grimm:2014efa,Grimm:2015mua,Grimm:2017pid,Grimm:2017okk}, which may be related to the backreaction of the fluxes over the metric and $\alpha'$--corrections.} As we shall see, the superfield content splits into two sectors: an axionic sector $T_\Lambda$ of chiral superfields, distinguished by the fact that $\calk$ depends explicitly only on $\Im T_\Lambda$ and with zero super-Weyl weight, and the rest of the chiral superfields $Z^a$, with super-Weyl weight $(1,0)$, which depends on both the compensator $U$ and the physical chiral superfields $\Phi^i$. In the EFT, a potential is introduced via a superpotential $\calw$, which, as we shall see in the various examples, acquires the typical form
\be
\label{EFT_Intro_W}
\calw (Z; T) =  \caln_{\cala} \calv^\cala (Z) - c_\cala^\Lambda T_\Lambda  \calv^\cala(Z)  + \hat \calw_0(Z)\,.
\ee
The quantized constants $\caln_A$, with $\cala = 1,\ldots, \caln$ are originated from background fluxes threading the manifold upon which we are compactifying and determine the \emph{maximal lattice} $\Gamma$; the parameters $c_\cala^\Lambda$, which provide couplings between the axionic sector $T_\Lambda$ and the other chiral superfields $Z^a$, also depends on the quantized background fluxes. All the other contributions, which may include, for example, perturbative and nonperturbative corrections, are generically collected into $\hat \calw_0(Z)$.

%
%

A reformulation of the ordinary EFTs so formulated passes through regarding $\caln_{\cala}$ as generated by gauge three-forms and, dualizing $T_\Lambda$ to linear multiplets $L^\Lambda$, further reabsorb the coupling term $c_\cala^\Lambda T_\Lambda  \calv^\cala(Z) $ of the superpotential in a gauging of the three-forms. The relevant issue that has to be addressed in the EFTs that we will encounter is to what extent the procedure explained in Chapter~\ref{chapter:Sugra}. Namely, we need to answer the question: \emph{how many fluxes can truly be promoted to dynamical variables?} and \emph{how many axions can be regarded as dual to gauge two-forms?} 

In answering the first question, two main restrictions are encountered:
\begin{itemize}
	\item[-] \underline{supersymmetry} requires the fields of the compactifications to be properly collected into supermultiplets. The moduli of the compactifications need to be paired with gauge three-forms. Then, if the maximal amount of scalar moduli is $n+1$ (including the compensator), we \emph{cannot} promote to gauge three-forms a number $N>2(n+1)$ of fluxes;
	\item[-] the background fluxes are constrained by the \underline{tadpole cancellation conditions}. As we shall see in the specific examples in Sections~\ref{sec:EFT_IIB_Tad} and~\ref{sec:EFT_IIA_Tad}, these acquire the general form
	\be
	\label{EFT_Intro_Tada}
	\calq_I\equiv \frac12\cali^{\cala\calb}_I\caln_\cala \caln_\calb+ \tilde Q_I^\cala \caln_\cala+\tilde\calq_I^{\rm bg}=0\,,
	\ee
	where the index $I$ labels the different tadpole conditions, $\cali^{\cala\calb}_I=\cali^{\calb\cala}_I$ defines a symmetric pairing between the fluxes $\caln_A$, ${\tilde Q_I^\cala}\caln_\cala$ stands for a possible linear contribution of fluxes to the tadpoles, and $\tilde\calq_I^{\rm bg}$ denotes some background `charge' that needs to be canceled by the flux contribution. In string theory the last contribution is typically generated by orientifolds or curvature corrections (see the following \eqref{EFT_IIB_Tadb} and \eqref{EFT_IIA_Tadb} for examples). As was anticipated in the introductory Section~\ref{sec:Intro_AC} and as concretely explained in supergravity models in Section~\ref{sec:Sugra_G4FM}, a consistent implementation of the constraint \eqref{EFT_Intro_Tada} requires it to be \emph{linear} in the fluxes generated by the gauge three-forms. Then, we need to identify a subset $N_A$ of $\caln_{\cala}$ as in \eqref{Intro_Jumps_NA}, satisfying $\cali^{\cala\calb} = 0$, leaving untouched some others $\caln_{\cala}^{\rm bg}$. This reduces \eqref{EFT_Intro_Tada} to
	\be
	\label{EFT_Intro_Tadb}
	\calq_I\equiv \calq_I^{\rm bg}+\frac12\cali^{AB}_IN_AN_B+ Q_I^A N_A=0\,.
	\ee
\end{itemize}

The above issues apparently leave apart the axionic sector made up by the multiplets $T_\Lambda$. As explained in Section~\ref{sec:Sugra_GL}, we can reabsorb the F-term couplings $c_A^\Lambda  T_\Lambda \calv^A(Z)$ by introducing gauged linear multiplets \cite{Dudas:2014pva}. The introduction of such objects requires the presence in the EFT of gauge two-forms and paired `saxionic' fields. We again stress that in dualizing $T_\Lambda$ to $L^\Lambda$ we are assuming that \emph{no gauge three-form is paired to the lowest components of}  $T_\Lambda$ (in other words,  $T_\Lambda$ cannot be simultaneously regarded as three-form multiplets). Furthermore, the gauging of the gauge two-forms by gauge three-forms strictly requires the gauging parameters $c_A^\Lambda$ to depend on $\caln_\cala^{\rm bg}$ only, strictly correlating with how we choose $N_A$ in \eqref{EFT_Intro_Tadb}. 

Finally, we will come up with an alternative four dimensional theory, which is fully determined by the Legendre transform $\calf(Z,\bar Z; \hat L)$ of the kinetic section $\calk(Z, \bar Z; T)$ and with a superpotential $\hat \calw(Z)$, which depends on the fixed background fluxes $\caln_\cala^{\rm bg}$.

%


\section{Type IIB effective field theories}
\label{sec:EFT_IIB}

Effective four-dimensional $\caln=1$ theories originating from the compactification of ten-dimensional Type IIB string theory provide some of the simplest examples where the above general discussion can find concrete applications. After having recalled the basic features of orientifolded Calabi-Yau compactifications with background fluxes, we will show how the four-dimensional EFT can be rephrased within the language of the previous chapter, introducing gauge three-forms to generate background fluxes and gauge two-forms in replacement of axions.

\subsection{From 10D to 4D}
\label{sec:EFT_IIB_10d}

In ten dimensions, the bosonic massless spectrum of Type IIB superstring theory consists of the graviton $\tilde g_{MN}$, the NS-NS gauge two-form $\tilde B_2$, the dilaton $\tilde \phi$, which stem from the NS-NS sector, and gauge zero-, two- and four-forms $\tilde  C_0$, $\tilde C_2$ and $\tilde C_4$, from the R-R sector.\footnote{Here and in the following sections, the tilde denotes the ten-dimensional quantities.} In the Einstein frame, the low energy effective action governing their dynamics is \cite{Giddings:2001yu,Grimm:2004uq,Ibanez:2012zz}
\be
\label{EFT_IIB_S10d}
\begin{aligned}
	S_{\rm IIB}^{(10)} &= \frac{1}{2 \kappa_{10}^2} \int \left(  \tilde R * 1 - \frac12 \d \tilde \phi \wedge *\d \tilde \phi - \frac12e^{-\tilde \phi}  \tilde H_3 \wedge * \tilde H_3\right)
	\\
	&\quad\,-\frac{1}{2 \kappa_{10}^2} \int \left(\frac12 e^{2 \tilde \phi} \tilde F_1 \wedge * \tilde F_1 + \frac12 e^{\tilde\phi} \tilde F_3 \wedge * \tilde F_3 + \frac14 \tilde F_5 \wedge  *\tilde F_5\right)
	\\
	&\quad -\frac{1}{4 \kappa_{10}^2} \int \tilde C_4 \wedge \tilde H_3  \wedge \tilde F_3 + S_{\rm loc}\;,
\end{aligned}
\ee
with $2\kappa_{10}^2=\ell_s^8/(2\pi)$ and where we have introduced the gauge invariant field strengths
\be
\label{EFT_IIB_Fs}
\begin{aligned}
	\tilde F_1 &= \d \tilde C_0\,, \qquad \qquad  \tilde H_3 = \d \tilde B_2\,,
	\\
	\tilde F_3 &= \d \tilde C_2 - \tilde C_0 \d \tilde B_2\,,
	\\
	\tilde F_5 &= \d \tilde C_4 -\frac12 \d \tilde B_2 \wedge \tilde C_2 +\frac12 \tilde B_2 \wedge \d \tilde C_2\,.
\end{aligned}
\ee
In order to obtain a correct counting of the degrees of freedom, the self-duality condition $\tilde F_5 = * \tilde F_5$ has to be further imposed, either at the level of the equations of motion or slightly modifying the above action by following \cite{Pasti:1995tn,Pasti:1996vs,Pasti:1996va,Maznytsia:1998xw}. Furthermore, $S_{\rm loc}$ generically includes all the contribution from localized sources, such as D-branes and O-planes. 

A four-dimensional $\caln=1$ supergravity effective theory is obtained by compactifying the ten-dimensional theory \eqref{EFT_IIB_S10d} on an orientifolded Calabi-Yau background $X={\rm CY}_3/\calr$. The ten-dimensional background is chosen to be in the block diagonal form
\be
\label{EFT_IIB_ds2}
	\d s^2 = e^{2 A(y)} g_{mn} \d x^m \d x^n + e^{-2 A(y)} g_{\mu\bar\nu} \d y^{\mu} \d \bar y^{\bar\nu}\;,
\ee
where $x^m$, with $m=0,\ldots,3$, denote the \emph{external} four-dimensional coordinates and $y^\mu$, with $\mu=1,2,3$, are the \emph{internal} complex coordinates parametrizing the Calabi-Yau three-fold ${\rm CY}_3$, with $g_{\mu\bar\nu}$ an hermitean metric. The warp factor $e^{2A(y)}$, which depends on the internal coordinates, is due to the presence of sources and background fluxes. We here choose the orientifold involution $\calr$ such that its action over the holomorphic Calabi-Yau three-form is
\be
\label{EFT_IIB_calr}
\calr \Omega = - \Omega \,.
\ee
The fixed loci in $X$ are then either points or hypersurfaces of complex codimension one. This implies that the allowed spacetime filling objects for such a background are D3-branes and O3-planes, which are just points in the internal manifold or D7-branes and O7-planes wrapping internal four-cycles. 

The four-dimensional fields building up the EFT are identified with the zero modes on $X$ of the ten-dimensional fields. These are in one-to-one correspondence with the harmonic forms of $X$. Presently, we focus on identifying only the scalar sector of the four-dimensional EFT. To begin with, we introduce the \emph{periods} as
\be
\label{EFT_IIB_PiA}
\Mp^3 \Pi^A(\varphi)=\frac{\pi}{\ell^3_{\rm s}}\int_{\Sigma_A} \Omega = \frac{\pi}{\ell^3_{\rm s}}\int_X \Omega\wedge \varpi^A\,,
\ee
where $\varpi^A$, $A = 0,\ldots, b_3^-$ a basis of $H_-^3(X;\mathbb{Z})$ and $\Sigma_A$ a basis of dual cycles. The periods $\Pi^A$ do depend on the complex structure parameters $\varphi^i$, $i=1,\ldots,h_-^{2,1}$. Then, $\Omega$ decomposes as
\be
\Omega = \frac{\ell^3_{\rm s}}{\pi} \Mp^3 \Pi^A(\phi) \varpi_A\,,
\ee
with $\varpi_A$ that integrate, over a cycle $\Sigma_B$, as $\int_{\Sigma_B} \varpi_A = \delta_A^B$.
To be  more explicit, we may choose a symplectic basis $\{\alpha_a, \beta^b\}$, $a=0,1,\ldots,h^{(2,1)}_-(X)$ of $H_-^3(X,\mathbb{Z})$ as\footnote{We recall that a symplectic basis is such that 
\be
	\int_X \alpha_a \wedge \beta^b = - \int_X   \beta^b \wedge \alpha_a = \delta^a_b\,,\qquad 	\int_X \alpha_a \wedge \alpha_b =	\int_X \beta^a \wedge \beta^b=0\,.
\ee
By Poincar\'e duality, we can relate this three-form basis with a dual basis of symplectic cycles $\{\Sigma^a, \tilde\Sigma_b\}$. Given the relations $\alpha_a = {\rm PD}[\Sigma^a]$, $\beta^a = {\rm PD}[\tilde\Sigma_a]$, a symplectic basis of cycles is defined as
\be
\begin{split}
	\Sigma^a \cap \tilde \Sigma_b &:= \int_{\tilde \Sigma_b} \alpha_a = - \int_{\Sigma^a} \beta^b = \delta_a^b\,,
	\\
	\Sigma^a \cap \Sigma^b &:=  \int_{\Sigma^a} \alpha_b = 0\,, \qquad \qquad \tilde \Sigma_a \cap \tilde \Sigma_b &:=  \int_{\tilde\Sigma_a} \beta^b = 0\,,
\end{split}
\ee
where the symbol $\cap$ denotes the intersection number between two cycles (see, for example \cite{Candelas:1987is,Candelas:1990rm,Denef:2008wq}).
}
\be
\Omega = f^a \alpha_a  - \calg_a\beta^a\,.
\ee
The periods $(f^a(\varphi), \calg_a (\varphi))$, in terms of a symplectic basis of cycles $\{\Sigma^a, \tilde\Sigma_b\}$, are given by
\be
f^a(\varphi) = \int_{\tilde\Sigma_a} \Omega \;,\qquad \quad \calg_a (\varphi) = \int_{\Sigma^a} \Omega \;.
\ee
By construction, they depend on the complex structure, but they are not most readily \emph{good} coordinates for the complex structure moduli space, owing to their redundancy. In fact, it can be shown \cite{Candelas:1990rm} that the periods $\calg_a$ can be fully re-expressed in terms of $f^a$. Indeed, as function of $f^a$, $\calg_a(f)$ are homogeneous of degree one and can be regarded as derivatives $\calg_a = \frac{\del \calg}{\del f^a}$ of a \emph{prepotential} $\calg$, homogeneous of degree two \cite{Craps:1997gp}. Moreover, due to the very definition of $\Omega$ as a section of a line bundle, $f^a$ provide projective coordinates, one of which is unphysical and can be gauge-fixed. We then conclude that the complex structure moduli space $\scrm_{\rm cs}$ is a special K\"ahler manifold, parametrized by the projective coordinates $f^a$ and described by the K\"ahler potential
\be
\label{EFT_IIB_Kcs}
K_{\rm cs} = - \log \left[ \ii \left( f^a \bar \calg_a - \bar f^a \calg_a \right) \right]\,.
\ee
In the super-Weyl invariant formulation, the complex structure moduli are included into $n+1$ chiral multiplets $z^a$ with super-Weyl weights (1,0). Then, we introduce a super-Weyl compensator $u$, lowest component of the chiral superfield $U$, and define
\be
z^a = u f^a (\phi)\,.
\ee
The kinetic section $\calk_{\rm cs}$ associated to the complex structure sector is then
\be
\calk_{\rm cs} (z,\bar z) = - 3 |U|^{\frac23} e^{-\frac13 K_{\rm cs}(\varphi,\bar\varphi)}\;,
\ee
where we have isolated the compensator using the homogeneity properties of the K\"ahler potential.

The four-dimensional dilaton $\phi$ and the axion $C_0$ assemble into the holomorphic \emph{axio-dilaton} as
\be
\tau = C_0 + \ii e^{-\phi}\,, 
\ee
and contributes to the K\"ahler potential with
\be
\label{EFT_IIB_Ktau}
K_\tau = - \log \left[-\ii (\tau - \bar \tau)\right]\,. 
\ee

In order to identify the remaining moduli, it is necessary to expand the K\"ahler form $J$ and both the NS-NS and R-R potentials into a basis of harmonic forms of $H_+^2(X,\mathbb{Z})$ in accordance with their parity under $\calr$. The K\"ahler form $J$ decomposes as
\be
J = v^\alpha (x)\omega_\alpha \,, \qquad \qquad \alpha= 1,\ldots, h^{(1,1)}_+ (X)\,,
\ee
where  $\omega_\alpha$ is an harmonic basis for $H^{(1,1)}_+ (X,\mathbb{Z})$. The NS-NS field strength $\tilde H_3$ and the R-R field strengths $\tilde F_3$ and $\tilde F_5$ split as
\be
\label{EFT_IIB_Fred}
\begin{aligned}
\tilde H_3 &= \d b^k (x) \wedge \omega_k  + \overline{H}_3 \,, 
\\
\tilde F_3 &= \d c^k (x) \wedge \omega_k  + \overline{F}_3 \,, \qquad \qquad k = 1,\ldots, h^{(1,1)}_- (X)\,,
\\
\tilde F_5 &= \d C_ \alpha  (x) \wedge \tilde\omega^\alpha  \,, \; \qquad  \qquad \qquad \alpha = 1,\ldots, h^{(2,2)}_+ (X)\,,
\end{aligned}
\ee
where $\omega_k$ and $\tilde\omega^\alpha$ are, respectively, an harmonic basis for $H^{(1,1)}_- (X,\mathbb{Z})$ and $H^{(2,2)}_+ (X,\mathbb{Z})$. We also included the \emph{background} three-form fluxes $\overline{H}_3$, $\overline{F}_3$ which are expanded as\footnote{We here adopt the convention such that a gauge $p$--form $C_p$ has dimensions $[C_p]=-p$ in units of energy and its field strength $F_{p+1} = \d C_p$ dimensions $[F_{p+1}]=-p+1$. }
\be
\label{EFT_IIB_Fbgred}
\begin{aligned}
	\overline{H}_3 &= \ell_s^2  h_A \varpi^A \equiv \ell_s^2 m^a_H \alpha_a  - \ell_s^2 e_a^H \beta^a \,, 
	\\
	\overline{F}_3 &=  \ell_s^2  m_A \varpi^A  \equiv \ell_s^2 m^a_F \alpha_a  - \ell_s^2 e_a^F \beta^a \,, 
\end{aligned}
\ee
where $m_A = (e_a^F, m^a_F)$ and $h_A = (e_a^H, m^a_H)$  are quantized constants.

We can then collect the scalars $c^k$, $b^k$ into $h_-^{(1,1)}(X)$ chiral coordinates $G^k = c^k -\tau b^k$. The K\"ahler deformations $v^\alpha$ and the 4D scalars $C_\alpha$ do not however find an immediate identification as K\"ahler coordinates. The proper chiral fields $T^\alpha$ serving as coordinates are rather convoluted functions of $v^\alpha$, $C_\alpha$, $G^k$ and the axio-dilaton $\tau$. We will not need their expressions here and we refer to \cite{Grimm:2004uq,Carta:2016ynn} for further details. Together, $T^\alpha$, $G^k$ and $\tau$ do provide $h^{(1,1)}(X)+1$ coordinates for a K\"ahler manifold described by the K\"ahler potential
\be
\label{EFT_IIB_Kk}
K_{\rm k} = - \log \left[-\ii (\tau - \bar \tau)\right] - 2 \log \left[\frac16 \calk(\tau, T, G)\right] \equiv - \log \left[-\ii (\tau - \bar \tau)\right] - 2 \log V_{\rm E}\,,
\ee
where $V_{\rm E}$ is the internal volume of $X$, measured in the Einstein frame.
The set of moduli which the four-dimensional EFT includes is further summarized in Table~\ref{tab:EFT_IIB_mod}. In the super-Weyl invariant language, we assume that both the dilaton and the superfields $G^k$ and $T^\alpha$ do \emph{not} carry any super-Weyl weight.

The full kinetic section for the moduli is then
\be
\calk(Z,\bar Z; \Psi,\bar \Psi) = \calk_{\rm cs}(Z,\bar Z)\calk_{\rm k}(\Psi,\bar \Psi)  = - 3 |U|^\frac23 e^{-\frac{K_{\rm cs} (\Phi,\bar \Phi)}{3}} e^{-\frac{K_{\rm k} (\Psi,\bar \Psi)}{3}}\;,
\ee
with the superfields $\Phi^i$ and $\Psi^{\hat{a}}$ whose lowest components are, respectively, the complex structure moduli $\phi^i$ and the K\"ahler coordinates $\psi^{\hat{a}}$. While the superfields $Z^a = U g^a(\Phi)$ carry super-Weyl weights $(1,0)$, the superfields $\Psi^{\hat{a}}$ do not carry any super-Weyl weight. It is worthwhile to mention that the different weights of $Z^a$ and $\Psi^{\hat{a}}$ make the homogeneity properties \eqref{Sugra_SW_KWhom} realized only by  $\calk_{\rm cs}(Z,\bar Z)$, regarding $\Psi^{\hat{a}}$ as a `spectator' sector.

After gauge-fixing the super-Weyl invariance as in \eqref{Sugra_SW_u}, we retrieve K\"ahler potential
\be
\label{EFT_IIB_K}
K = K_{\rm cs} + K_{\rm k} = -  \log \left[ \ii \left( f^a \bar \calg_a - \bar f^a \calg_a \right) \right]- \log \left[-\ii (\tau - \bar \tau)\right] - 2 \log V_{\rm E}
\ee
which reflects the factorized structure of the moduli space
\be
\label{EFT_IIB_Mmod}
\scrm_{\rm moduli} = \scrm_{\rm cs}^{h_-^{(2,1)}} \times \scrm_{\rm k}^{h^{(1,1)}+1}
\ee
with the former, $\scrm_{\rm cs}$, being a special K\"ahler manifold parametrized by the projective coordinates $f^a$ and the latter, $\scrm_{\rm k}$, a K\"ahler manifold parametrized by $\psi^{\hat{a}}=(\tau, G^a, T^\alpha)$.  
 We also stress that the factorized structure \eqref{EFT_IIB_Mmod} and the K\"ahler potential \eqref{EFT_IIB_K} hold for the large volume and constant warping approximations. We do not expect them to remain unaltered beyond such approximations and we refer to \cite{Martucci:2014ska,Martucci:2016pzt} for the definition of K\"ahler potentials taking into account a nontrivial warping. Still, one can explicitly check that, in the large $\Im\tau$ and large $V_{\rm E}$ limit, the warped K\"ahler potentials of \cite{Martucci:2014ska,Martucci:2016pzt} are well approximated by \eqref{EFT_IIB_K}.

\begin{table}[!h]
\begin{center}
\begin{tabular}{ |c c c c | }
\hline
\rowcolor{ochre!30}{\bf Fields} & {\bf Number} & {\bf Data} & {\bf Manifold}
\\
\hline
\hline
$g^a(\varphi)$ & $h_-^{(2,1)}(X)$ & complex structure moduli & $\scrm_{\rm cs}$\TBstrut
\\
\hline
$(C_0, \phi)$ & 1 & axio-dilaton & \multirow{3}{*}{$\scrm_{\rm k}$} \TBstrut
\\
\multirow{1}{*}{$(b^k,c^k)$} & \multirow{1}{*}{$h_-^{(1,1)}(X)$}\TBstrut & \multirow{2}{*}{K\"ahler moduli} &  
\\
\multirow{1}{*}{$(v^\alpha,C_\alpha)$} \TBstrut & \multirow{1}{*}{$h_+^{(1,1)}(X)$} & &
\\
\hline
\end{tabular}
\end{center}
\caption{How the 4D fields originating form compactification of Type IIB superstring theory reassemble as lowest components of $\caln=1$ chiral superfields.}
\label{tab:EFT_IIB_mod}
\end{table}

The presence of the three-form background fluxes $\overline{H}_3$, $\overline{F}_3$ introduces a potential for the complex structure moduli and the axio-dilaton. The potential is determined by the GVW--superpotential \cite{Gukov:1999ya,Giddings:2005ff} which, in super-Weyl invariant language is
\be
\label{EFT_IIB_calw}
\calw(z;\tau)= U \frac{\pi}{\ell^5_{\rm s}}\int_X \Omega\wedge (\overline{F}_3-\tau \overline{H}_3) = m_A \calv^A(z) - \tau h_A \calv^A(z)\, .
\ee
Here we have included the super-Weyl compensator as required by the homogeneity properties \eqref{Sugra_SW_KWhom}\footnote{Here, with a little abuse of notation, we are denoting with $\tau$ both the axio-dilaton as well as the superfield which contains it as lowest component $\tau (x,\theta,\bar \theta)| = \tau(x)$.} and introduced the periods
\be
\label{EFT_IIB_calvA}
\calv^A (z) = u \Pi^A (\varphi)\;,
\ee
which are homogeneous of degree one as in \eqref{Sugra_MTF_WcalvA}. Once gauge-fixed, the periods \eqref{EFT_IIB_calvA} reduce to \eqref{EFT_IIB_PiA} and the superpotential becomes
\be
\label{EFT_IIB_calwgf}
W(\Phi;\tau)= \Mp^3 \left(m_A \Pi^A(\Phi) - \tau h_A \Pi^A(\Phi) \right)\, .
\ee

\subsection{The tadpole cancellation condition}
\label{sec:EFT_IIB_Tad}

The background three-forms $\overline{H}_3$, $\overline{F}_3$ introduce, in the four-dimensional EFT, a set of $2 b^3_-(X)$ quantized fluxes $(m_A, h_A)$. However, $(m_A, h_A)$ cannot be arbitrarily chosen: due to the \emph{tadpole cancellation condition}, they are strictly tightened together and also related to the numbers of D3-branes and O3-planes which are present in the theory.

The tadpole cancellation condition originates from the ten-dimensional Bianchi identities involving the four-form $\tilde C_4$, which here, due to the self-duality of $\tilde F_5$, coincide with their equations of motion:
\be
\label{EFT_IIB_Tada}
\d \tilde F_5 = \tilde H_3 \wedge \tilde F_3 + 2 \kappa^2_{10} \mu_3 \rho_3^{\rm local}\,,
\ee
with $\mu_3 = 2\pi \ell_s^{-4}$. We can now integrate \eqref{EFT_IIB_Tada} over the full internal, compact manifold $X$: standing that the term on the left hand side of \eqref{EFT_IIB_Tada} integrates to zero, upon using the decompositions \eqref{EFT_IIB_Fred} and \eqref{EFT_IIB_Fbgred}, we get
\be
\label{EFT_IIB_Tadb}
0 =I^{AB} h_A m_B  + N_{\rm D3} - \frac12 N_{\rm O3}\,.
\ee
%
%
%
Here we have defined the anti-symmetric intersection number
\be
\label{EFT_IIB_Int}
I^{AB}\equiv \int_X \varpi^A\wedge \varpi^B\,, \qquad I^{AB} = - I^{BA}\,,
\ee
while $N_{\rm D3}$ denotes the number of D3-branes in the ${\rm CY}_3$ double cover  (namely, counting \emph{separately} D3-branes and their orientifold images, rather than counting the D3-branes in the quotient space ${\rm CY}_3/\calr$, where we would count just half as many branes).\footnote{We recall that in our conventions, the same as \cite{Ibanez:2012zz}, the charge of an O$p$-plane, with respect to that of a D$p$-brane charged under the same R-R form, is
\be
\label{EFT_IIB_QOD}
\calq_{\text{O}p} = -2^{p-4} \calq_{\text{D}p}\,.
\ee}

\subsection{Reformulating the 4D EFT}
\label{sec:EFT_IIB_4d}

We are now ready to reformulate the four-dimensional EFT by introducing a hierarchy of gauge forms. The forthcoming discussion will directly involve the complex structure moduli and the axio-dilaton, while the remaining K\"ahler moduli will be `spectators'. Hence, for the sake of clarity, we will set $h_-^{(1,1)}(X) = 0$, so that the internal volume $V_{\rm E}$ in \eqref{EFT_IIB_Kk} does not depend on the axio-dilaton $\tau$ \cite{Grimm:2004ua}. 

The superpotential \eqref{EFT_IIB_calw} depends on $2 b^3_-(X)$ background fluxes $(m_A,h_A)$, which determine the maximal lattice $\Gamma$ of fluxes 
\be
\label{EFT_IIB_4dGamma}
\Gamma  =  \left\{(m_A, h_A)\; |\; m_A, h_A \in \mathbb{Z}  \right\}=  \{ \caln_\cala \}  \,  .
\ee
However, due to the constraint \eqref{EFT_IIB_Tadb}, it is clear that they are not \emph{all} independent. Only one sublattice, constituted by either $m_A$ or $h_A$ can be regarded as originated from three-form potentials. For simplicity, let us choose the former R-R fluxes as the dynamical sublattice
\be
\label{EFT_IIB_4dGammadyn}
\Gamma_{\rm EFT}   =   \left\{(m_A, 0)\; |\; m_A \in \mathbb{Z} \right\} \, .
\ee
This is indeed an isotropic lattice in the sense of \eqref{Intro_Con_Iso}. In fact, comparing the tadpole cancellation condition \eqref{EFT_IIB_Tada} with the general \eqref{EFT_Intro_Tada}, we recognize that
\be\label{EFT_IIB_cali}
\cali \equiv \left(\begin{array}{cc}
	0 & -I \\
	I & 0
\end{array}
\right)\, ,
\ee
and \eqref{EFT_IIB_4dGammadyn} clearly constitute null directions with respect to $\cali^{\cala\calb}$.

Then, given an initial flux background $\caln_\cala^{\rm bg}=(m_A^{\rm bg},h_A^{\rm bg})$, the set of fluxes that is accessible within the 4D EFT description is 
\be
\label{EFT_IIB_4dGammaexp}
\Gamma_{\rm exp}  \, \equiv \,  \Gamma_{\rm EFT}+\caln^{\rm bg}  \, = \, \left\{(m_A, h_A^{\rm bg}) \;|\;  m_A \in \mathbb{Z},\  h_A^{\rm bg}\, {\rm fixed}  \right\} \, ,
\ee
which is an affine sublattice of the flux lattice $\Gamma$. These sublattices are parametrized by quotient elements $[\caln^{\rm bg}] \in \Gamma/\Gamma_{\rm EFT}$, which can be identified with the set of NS-NS fluxes $h_A$. In other words, as of now, the superpotential $\calw(z;\tau)$ in the super-Weyl invariant formalism will be so built: 
\be
\label{EFT_IIB_calwgen}
\calw(z;\tau)= \underbrace{m_A \calv^A(z)}_\text{generated by three-forms}  \quad \underbrace{- \tau h_A \calv^A(z)}_\text{`spectator' sector $\to \hat \calw(z;\tau)$}\, ,
\ee
where we have neglected, for simplicity, additional corrections.
In this setup, the fluxes $h_A$ are then considered frozen, fixed constants and, consequently, we may rewrite the constraint \eqref{EFT_IIB_Tadb} in the linearized form \eqref{Sugra_4FM_Constr} as
\be
\label{EFT_IIB_4dTad}
Q^A m_A + \calq^{\rm bg} = 0\,, \qquad {\rm with} \quad Q^A \equiv - I^{AB} h_B\,, \quad \calq^{\rm bg} =  N_{\rm D3} - \frac12 N_{\rm O3}\,.
\ee
The dynamical generation of the sublattice \eqref{EFT_IIB_4dGammadyn} requires the inclusion, in the four-dimensional theory, of $b_-^3(X)$ gauge three-form potential, the same amount as $m^A$ in \eqref{EFT_IIB_4dGammadyn}. 

Gauge three-forms may be included into a generic EFT following the steps traced in Section~\ref{sec:Sugra_MTF}. Here we would also like to have a ten-dimensional understanding of why gauge three-forms have to be present in the four-dimensional theory, whether they are predicted and what is their origin. Clearly, it is not possible to obtain, by dimensional reduction, any four-form field strength from  the usual \eqref{EFT_IIB_Fs}. However, as was noticed in \cite{Bielleman:2015ina}, we may pass to a `democratic' formulation where, beside the electric potentials $C_0$, $C_2$, $B_2$, their magnetic duals $C_{8} $, $C_6$, $B_6$ also appear. Reducing their field strengths, 4D four-form field strengths naturally appear
\be
\label{EFT_IIB_4dDualF}
\begin{aligned}
	\tilde F_7 &=  \d A_3^A(x) \wedge \varpi_A +\ldots \,, \qquad
	\tilde H_7 &=  \d \check{A}_3^A(x) \wedge \varpi_A  +\ldots \,.
\end{aligned}
\ee
The maximal amount of gauge three-forms $(A_3^A,\check{A}_3^A)$ which are predicted by dimensional reduction is then $2 \times 2 b^3_-(X)$, as the dimensionality of $\Gamma$. In order to dynamically generate the sublattice \eqref{EFT_IIB_4dGammadyn} it is just sufficient to consider the subset $A_3^A$, dual to the R-R potentials $F_3$.

We have now at hands all the basic elements to reshape the 4D EFT presented in Section~\ref{sec:EFT_IIB_10d}. Notably, the number of gauge three-forms $A_3^A$ is twice as the number of complex scalars $z^a$: this is the maximally nonlinear case of Section~\ref{sec:Sugra_MTF_Examples}, described by nonlinear double three-form multiplets of the kind of \eqref{Sugra_MTF_Ex_ZGa}. We can then introduce the multiplets
\be
S^A \equiv \begin{pmatrix}
	S^a \\ \calg_a (S)
\end{pmatrix}
\ee
with
\be
\label{EFT_IIB_4dS}
S^a \equiv-\frac{\ii}{4}(\bar \cald^2 - 8 \calr) \calp^a\,,\qquad \tilde S_a = \calg_a (S) \equiv-\frac{\ii}{4}(\bar \cald^2 - 8 \calr)  \calp_a\,,
\ee
which neatly collect, in their components, the complex structure moduli $z^a$ (with the super-Weyl compensator), as well as the gauge three-forms $A_3^A = (A_3^a, \tilde A_{3a})$, with the latter embedded into $P^A = (\calp^a , \tilde\calp_a)$ as in \eqref{ConvSG_P}. Furthermore, the dynamical implementation of the tadpole cancellation condition \eqref{EFT_IIB_4dTad} requires the inclusion in the EFT of one four-form multiplet $\Gamma$ of the kind of \eqref{Sugra_4FM_Comp}, whose effect is the appearance of a four-form $C_4$ and the gauging of $S^A$ as
\be
\label{EFT_IIB_4dSGaug}
S^A  \; \rightarrow \hat S^A \equiv -\frac \ii 4 (\bar \cald^2 - 8 \calr)P^A - \ii Q^A \Gamma \,,
\ee
resulting into a gauging of the three-forms as
\be
\label{EFT_IIB_F4gaug}
\hat F_4^A = \d A_3^A + Q^A C_4 = \d A_3^A - I^{AB} h_B C_4\,.
\ee
Therefore, in the super-Weyl invariant language we arrive at the 4D EFT Lagrangian
\be
\label{EFT_IIB_4dL}
\begin{aligned}
	\tilde\call &=  \int \d^4\theta\,E\,\calk_{\rm cs}(\hat S,\bar {\hat S}) e^{-\frac{K_{\rm k}(\tau,\bar \tau; T, \bar T)}{3}}   + \left( \int \d^2\Theta\,2\cale\, \hat\calw(\hat S;\tau ) +{\rm c.c.}\right)\,,
	\\
	&\quad\, + \left[\ii \int \d^2\Theta\,2\cale\, \calq^{\rm bg}  \Gamma + {\rm c.c.} \right] + \tilde \call_{\rm bd}\;.
\end{aligned}
\ee
After fixing the super-Weyl invariance we arrive at the four-dimensional action determined by \eqref{Sugra_GMTF_L3fComp_gf} as
\be
\label{EFT_IIB_4Sa}
\begin{aligned}
S |_{\rm bos} &=  \Mp^2 \int_{\Sigma} \left( \frac12 R *1- K_{M\bar N}\,  \d \varphi^M \wedge *\d \bar \varphi^{\bar N} - \frac{1}{4(\Im \tau)^2} {\rm d} \tau \wedge *{\rm d} \bar\tau \right)
\\
&\quad\, -\int_\Sigma \Big[\frac12 T_{AB} \hat F^A_4*\! \hat F^B_4+ T_{AB}\Upsilon^A\hat F^B_4+\Big(
\hat V-\frac12 T_{AB}\Upsilon^A \Upsilon^B\Big)*\!1\Big] 
\\
&\quad\,+\int_{\del\Sigma}T_{AB}(* \hat F^A_4+\Upsilon^A)A^B_3 + \int_\Sigma \calq^{\rm bg} C_4\;.
\end{aligned}
\ee
Here we have collectively denoted the physical complex structure moduli and the (complexified) K\"ahler moduli by $\varphi^M = (\varphi^i, t^{\hat\alpha})$, with $i=1,\ldots,h^{2,1}_-$ and $\hat\alpha = 1,\ldots, h_+^{1,1}$. The three-form kinetic terms are defined as in \eqref{Sugra_MTF_TUVgf} with 
%
\begin{align}
\label{EFT_IIB_hatWgf}	
\hat W(\varphi;\tau) &=  -\Mp^3 \tau h_A \Pi^A(\Phi) 
\end{align}
%
and the periods as those entering \eqref{EFT_IIB_calwgf}.

The action \eqref{EFT_IIB_4Sa} already provides a three-form origin for the R-R flux potential which, owing to the gauging \eqref{EFT_IIB_F4gaug}, are generated \emph{dressed} with the constraint \eqref{EFT_IIB_4dTad}. Still, we do not content ourselves with the formulation \eqref{EFT_IIB_4Sa} and make a step forward. The NS-NS originated superpotential $\hat W(\varphi;\tau)$, which in \eqref{EFT_IIB_hatWgf} we regarded as a spectator, may be reabsorbed into a gauging of the two-form dual to the axion $C_0$. In order to see this, first, we dualize the axio-dilaton $\tau$ to the scalar component $\ell$ of a linear multiplet by using \eqref{Sugra_AL_lK}
\be\label{EFT_IIB_ltau}
\ell = \frac{1}{2 \Im \tau}\,.
\ee
The field metric is now determined by the Legendre transform \eqref{Sugra_AL_F} of the K\"ahler potential \eqref{EFT_IIB_K}, that is
\be\label{EFT_IIB_F}
F =  \log \ell+ 1 -2\log V_{\rm E}+K_{\rm cs}\,.
\ee
Owing to the block  diagonal structure of the metric, the Lagrangian \eqref{Sugra_GL_Lgf} reduces to 
\be
\label{EFT_IIB_SLeg}
\begin{split}
	S|_{\rm bos} &= M^2_{\rm P} \int_\Sigma \left( \frac12\, R *1   -F_{M\bar N}\,  \d \varphi^M \wedge *\d \bar \varphi^{\bar N} - \frac1{4\ell^2}   \d \ell \wedge * \d \ell \right)
	\\
	&\quad - \frac{1}{M_{\rm P}^2} \int_\Sigma \left( \frac1{4\ell^2} \hat\calh_3 \wedge * \hat \calh_3 \right)  + \calq^{\rm bg} \int_\Sigma C_4
	\\
	&\quad\, -\int_\Sigma \Big[\frac12 T_{AB} \hat F^A_4*\! \hat F^B_4+ T_{AB}\Upsilon^A \hat F^B_4+\Big(
	\hat V-\frac12 T_{AB}\Upsilon^A \Upsilon^B\Big)*\!1\Big] 
	\\
	&\quad\,+\int_{\del\Sigma}T_{AB}(* \hat F^A_4+\Upsilon^A)A^B_3 \;.
\end{split}
\ee
The field strength $\hat\calh_3$ is gauged as in \eqref{Sugra_GL_hatH3}, with the charges $c_A^\Lambda \to h_A$, that is 
\be
\label{EFT_IIB_tildeH3}
\hat \calh_3\equiv \calh_3 + h_A A^A_3\, .
\ee
The potential is fully encoded in the three-form part of the Lagrangian, whose quantities are determined by
\be
\begin{aligned}
	T^{AB}& = 2M_{\rm P}^4 e^{\tilde F}\,\Re\left(K^{\bar \jmath i}_{\rm cs}\,D_i  \Pi^A \bar D_{\bar \jmath} \bar\Pi^B+   \Pi^A\bar\Pi^B\right)\,,
	\\
	\Upsilon^A& = -\frac{M_{\rm P}^4 e^{\tilde F}}{\ell}\Im \left(  h_B K^{\bar \jmath i}_{\rm cs} \bar D_{\bar \jmath} \bar \Pi^B  D_i \Pi^A -  h_B \bar \Pi^{\bar B} \Pi^A\right)\,,
	\\
	\hat V &=  \frac{M_{\rm P}^4 e^{\tilde F}}{4  \ell^2} \left( h_A h_B K^{\bar \jmath i}_{\rm cs} D_i \Pi^A \bar D_{\bar \jmath} \bar \Pi^B +    h_A h_B \Pi^A \bar \Pi^B \right)\,,
\end{aligned}
\ee
as it can be easily computed from the general formulas \eqref{Sugra_GL_TUVgf}.

\begin{table}[!h]
	\begin{center}
		\begin{tabular}{ |c c c c | }
			\hline
			\rowcolor{ochre!30}{\bf Superfields} & {\bf Bosonic fields} & {\bf Number} & {\bf Data} 
			\\
			\hline
			\hline
			\cellcolor{darkblue!20} & \multirow{2}{*}{$z^a(\varphi)$}& \multirow{3}{*}{$h_-^{(2,1)}(X)+1$} &   complex structure deformations 
			\\
			\cellcolor{darkblue!20} & & & R-R axions 
			\\
			\multirow{-3}{*}{\cellcolor{darkblue!20}$\hat S^a$} & $A_3^A$, $\tilde A_{3A}$ & & gauge three-forms 
			\\
			\hline
			\cellcolor{darkblue!20} $T^{\hat\alpha}$ &$t^{\hat\alpha}$ & $h_+^{(1,1)}(X)$ & K\"ahler deformations
			\\
			\hline
			\cellcolor{darkblue!20}  & $\ell$& \multirow{2}{*}{$1$} &   (dual of) dilaton
			\\
			\multirow{-2}{*}{\cellcolor{darkblue!20}$\hat L$} & $\calb_2$ & & $C_0$--axion 
			\\
			\hline
			\cellcolor{darkblue!20} $\Gamma$ & $C_4$ & $1$ & gauge four-form \TBstrut
			\\
			\hline
		\end{tabular}
	\end{center}
	\caption{The 4D superfields leading to a new reformulation of $\caln=1$ Type IIB EFTs.}
	\label{tab:EFT_IIB_sf}
\end{table}

The spectrum of the newly reformulated Type IIB EFTs is further summarized in Table~\ref{tab:EFT_IIB_sf}. We stress, again, that in this reformulation the R-R fluxes are treated as dynamical variables, while the NS-NS are \emph{frozen}. Hence, as will be performed in Section~\ref{sec:LandSwamp_Hierarchy_IIB}, only the R-R fluxes can undertake membrane--mediated transitions. Additionally, considering the NS-NS fixed allowed us to both implement the tapdole cancellation condition and define the gaugings of the linear multiplet, dual to the axio-dilaton. Of course, a different choice of dynamical lattice \eqref{EFT_IIB_4dGammadyn} would have led to a different theory, with a different hierarchy of forms. In Section~\ref{sec:LandSwamp_Hierarchy_IIB} we will however show that the choice \eqref{EFT_IIB_4dGammadyn} is the most natural to be taken.

\section{Type IIA effective field theories}
\label{sec:EFT_IIA}

Effective fields theories originating from Type IIA superstring theory provide other examples of theories which can be rephrased in the dual formulation of Chapter~\ref{chapter:Sugra}. With respect to the case of Type IIB EFTs examined in the previous section, background fluxes generate a superpotential which involve both the entire sets of K\"ahler and complex structure moduli \cite{DeWolfe:2002nn,Grimm:2004uq}. After briefly recalling the basic elements entering the 4D EFTs originating from ten dimensions, we show how the closed string superpotential can be interpreted as dynamically generated by a set of gauge three-forms. Furthermore, we will show how the dual three-form picture can take into account the presence of nongeometric background fluxes \cite{Aldazabal:2006up,Wecht:2007wu,Ibanez:2012zz,Gao:2017gxk} and open string moduli \cite{Marchesano:2014iea,luca2}.

\subsection{From 10D to 4D}
\label{sec:EFT_IIA_10d}
 
The bosonic massless, closed string spectrum of  Type IIA superstring theory comprehends the graviton $\tilde g_{MN}$, the gauge two-form $\tilde B_2$ and the dilaton $\tilde \phi$ originating from the NS-NS sector and a gauge one-form $\tilde A_1$ and a gauge three-form $\tilde C_3$ from the R-R sector. The ten-dimensional, $\caln=2$ supersymmetric action describing their dynamics, taking into account a possible nonvanishing Romans mass $m^0$, is \cite{Grimm:2004uq,Ibanez:2012zz,Blumenhagen:2013fgp}
\be
\label{EFT_IIA_S10d}
\begin{aligned}
	S_{\rm IIA}^{(10)} &= \frac{1}{2 \kappa_{10}^2} \int \left( \tilde R * 1 - \frac12 \d \tilde \phi \wedge *\d \tilde \phi - \frac12 e^{-\tilde \phi} \tilde H_3 \wedge * \tilde H_3\right)
	\\
	&\quad\,-\frac{1}{4 \kappa_{10}^2} \int \left(e^{\frac52 \tilde\phi} (m^0)^2  *1 + e^{\frac32 \tilde \phi} \tilde G_2 \wedge * \tilde G_2 + e^{\frac12\tilde\phi} \tilde G_4 \wedge * \tilde G_4 \right)
	\\
	&\quad\, - \frac{1}{4 \kappa_{10}^2}\int   \Bigg[ \tilde B_2 \wedge \d \tilde C_3 \wedge \d \tilde C_3 + \frac{m^0}{3} (\tilde B_2)^3 \wedge \d \tilde C_3 + \frac{(m^0)^2}{20} (\tilde B_2)^5\Bigg]\
	\\
	&\quad\,+ S_{\rm loc}\;,
\end{aligned}
\ee
where we have defined the gauge invariant field strengths
\be
\label{EFT_IIA_Fs}
\begin{aligned}
	\tilde G_2 &= \d \tilde A_1 + m^0 B_2\,, \qquad \qquad  \tilde H_3 = \d \tilde B_2\,,
	\\
	\tilde G_4 &= \d \tilde C_3 + \d \tilde A_1 \wedge  \tilde B_2 - \frac12 m^0 \tilde B_2 \wedge \tilde B_2\,,
\end{aligned}
\ee
and $S_{\rm loc}$ generically collects all the contributions from localized sources. 

The compactification of the ten-dimensional theory over an orientifolded Calabi-Yau manifold $X={\rm CY}_3/\calr$ produces a four-dimensional $\caln=1$ supergravity EFT. The compactification metric ansatz is in the form \eqref{EFT_IIB_ds2}. However, we here choose the action of the orientifold involution $\calr$ on the Calabi-Yau holomorphic three-form $\Omega$ as
\be
\label{EFT_IIA_calr}
\calr \Omega =  e^{2\ii \theta} \bar\Omega \,,
\ee
still acting trivially on the external coordinates $x^m$. The fixed loci under \eqref{EFT_IIA_calr} are spacetime filling O6-planes, wrapping special Lagrangian three-cycles in $X$. 

The identification of the fields building up the four-dimensional theory proceeds as in Section~\ref{sec:EFT_IIB_10d}. We expand the ten-dimensional fields into a basis of harmonic forms on $X$, the zero-modes, which are ultimately identified with the four-dimensional fields, being in one-to-one correspondence with the harmonic forms. Presently, we are just interested in the scalar sector of fields and how the moduli assemble into $\caln=1$ superfields.

To begin with, let us notice that the K\"ahler form $J$ and the NS-NS two-form $\tilde B_2$ are both odd under the orientifold involution \eqref{EFT_IIA_calr}. We can then expand them into a basis $\{\omega_i\}$, $i=1,\ldots, h^{(1,1)}_-(X)$ of harmonic two-forms of $H^{(1,1)}(X;\mathbb{Z})$ as
\be
J = t^i (x)\omega_i \,, \qquad B_2 =  b^i (x)\omega_i\,, \qquad \qquad i= 1,\ldots, h^{(1,1)}_- (X)\,.
\ee
We can combine them into the complexified K\"ahler form $J_{\rm c} = B_2 + \ii J$, which straightforwardly identifies the complex scalar fields
\be
\label{EFT_IIA_phia}
\varphi^i(x) \equiv \int_{C_{i}} J_{\rm c} = b^i (x) + \ii t^i(x)\,,
\ee
where $\{C_i\}$ is a basis of orientifold-odd two-cycles, Poicar\'e dual to the basis $\{\omega_i\}$. It can be shown that the moduli space $\scrm_{\rm k}$, described by the complexified K\"ahler deformations \eqref{EFT_IIA_phia}, is a special K\"ahler manifold. By introducing the homogeneous coordinates $f^a(\phi)$, with $a=0,1,\ldots, h^{(1,1)}_-(X)$, the geometry of $\scrm_{\rm k}$ is fully encoded into a prepotential $\calg(f)$, which is homogenous of degree two in $f^a$ and determines the K\"ahler potential
\be
\label{EFT_IIA_Kk}
K_{\rm k} = - \log \left[ \ii \left( \bar f^a  \calg_a - f^a \bar \calg_a \right) \right]\,.
\ee
In the large-volume limit we can set
\be
\label{EFT_IIA_calg}
\calg(z)=-\frac{1}{6f^0}\,\kappa_{ijk}f^if^kf^k+\frac12 a_{ab}\,f^af^b+\calo(e^{2\pi\ii \varphi})\, ,
\ee
where $\kappa_{ijk}$ are triple-intersection numbers and the terms $a_{ab}z^az^b$ and $\calo(e^{2\pi\ii \varphi})$ encode the possible perturbative and non-perturbative $\alpha'$-corrections, respectively. However, away from the large volume limit, $\calg(z)$ can eventually have a more general form.  

In order to pass to a super-Weyl invariant formulation, we introduce $h^{(1,1)}_-(X)+1$ chiral multiplets $z^a$ with super-Weyl weights (1,0), related to $f^a$ as
\be
z^a = u f^a (\phi)\,,
\ee
where we have introduced the compensator $U$, whose lowest component is $u$.
The kinetic section $\calk_{\rm k}$ expressing the kinetic terms of the K\"ahler moduli is
\be
\calk_{\rm k} (z,\bar z) = - 3 |U|^{\frac23} e^{-\frac13 K_{\rm k}(\varphi,\bar\varphi)}\;,
\ee
where we have isolated the compensator using the homogeneity properties \eqref{Sugra_SW_KWhom}. Gauge-fixing the super-Weyl invariance amounts in setting, for example, $z^0 = u$ and $z^i = u \varphi^i$, with $u$ as in \eqref{Sugra_SW_u}.

The identification of the complex structure moduli is slightly more involved. The complex structure deformations are encoded into the Calabi-Yau holomorphic three-form $\Omega$, which is defined only up to a real rescaling $\Omega \to \Omega e^{-\Re h(z)}$, with $h(z)$ a generic holomorphic function\footnote{Generically, $\Omega$ is a section of a line bundle $\scrl \to X$ and, as such, transforms as $\Omega \to \Omega e^{- h(z)}$ for a generic holomorphic function $h(z)$. This reduces to just a real rescaling by  \eqref{EFT_IIA_calr}.}. A gauge invariant quantity can then be defined as $\calc \Omega$ by introducing a compensator $\calc= e^{-\tilde\phi - \ii \theta} e^{\frac12(K_{\rm cs} - K_{\rm k})}$ transforming as $\calc \to \calc e^{\Re h(z)}$. One may then decompose $\calc\Omega$ into a symplectic basis $(\alpha_\lambda, \beta^\kappa) \in H^3(X;\mathbb{Z})$ as
\be
\label{EFT_IIA_Omegaa}
\calc\Omega = \alpha_\lambda (\calc g^\lambda)- \beta^\kappa (\calc\calf_\kappa)
\ee
with the periods $(g^\kappa, \calf_\lambda(z))$ defined as
\be
\label{EFT_IIA_csper}
g^\kappa = \int_X \Omega \wedge \beta^\kappa\,,\qquad \calf_\lambda (g) = \int_X  \Omega \wedge \alpha_\lambda\,.
\ee
In order to exhibit the action of the orientifold projection on $\Omega$, $H^3(X;\mathbb{Z})$ can be decomposed into its even and odd-parts under the involution $\calr$, $H^3_+(X;\mathbb{Z}) \oplus H^3_-(X;\mathbb{Z})$ and consequently define the symplectic bases $(\alpha_K,\beta^Q) \in H^3_+(X;\mathbb{Z})$ and  $(\beta^K,\alpha_Q) \in H^3_-(X;\mathbb{Z})$. Then, \eqref{EFT_IIA_calr} implies that
\be
\Im (\calc g^K) = \Re (\calc \calf_K) = 0\,,\qquad \Re (\calc g^Q) = \Im (\calc \calf_Q) = 0\,,
\ee
halving the number of independent periods, reducing the expansion \eqref{EFT_IIA_Omegaa} to
\be
\label{EFT_IIA_Omega}
\calc \Omega = \Re (\calc g^K) \alpha_K + \ii \Im (\calc g^Q) \alpha_Q - \Re (\calc \calf_Q) \beta^Q - \ii \Im (\calc \calf_K) \beta^K\,,
\ee
which encode $h^{(2,1)}(X) + 1$ independent parameters describing the complex structure deformations. Indeed, the R-R three-form $C_3$, even under $\calr$, enjoys the expansion
\be
\label{EFT_IIA_C3}
C_3 = \xi^K(x) \alpha_K - \xi_Q (x)\beta^Q \;,
\ee
which identifies the four-dimensional axions $(\xi^K,\xi_Q )$.

Combining the four-dimensional fields originating from $\Omega$ and $C_3$, we can identify the complexified three-forms
\be
\Omega_{\rm c} = C_3 + \ii \Re (\calc \Omega)\,,
\ee
and we identify the complex coordinates for the complex structure space $\scrm_{\rm cs}$ as
\be
\label{EFT_IIA_Omegacexp}
\begin{split}
	N^K &= \int \Omega_{\rm c} \wedge \beta^K \equiv \xi^K +\ii n^K\,,
	\qquad
	U_Q &= \int \Omega_{\rm c} \wedge \alpha_\Lambda \equiv \xi_Q +\ii u_Q\,.
\end{split}
\ee
These provide $h^{(2,1)}(X)+1$ K\"ahler coordinates for the complex structure moduli space $\scrm_{\rm cs}$,  which is described by the K\"ahler potential
\be
\label{EFT_IIA_Kcs}
\begin{split}
	K_{\rm cs} &= -2 \log \left[ \frac\ii2 \int_X  \calc \Omega \wedge \overline{\calc \Omega} \right]\\
	&= -2 \log \left[ \Im (\calc g^Q) \Re (\calc \calf_Q) - \Re (\calc g^K) \Im (\calc \calf_K)\right] =: - \log e^{-4 D}
\end{split}
\ee
where we have defined the four-dimensional dilaton $D$.

The scalar fields entering the Type IIA EFTs are schematically summarized in Table~\ref{tab:EFT_IIA_mod}. The moduli space has then a factorized structure
\be
\label{EFT_IIA_Mmod}
\scrm_{\rm moduli} = \scrm_{\rm k}^{h_-^{(1,1)}} \times \scrm_{\rm cs}^{h^{(2,1)}+1}
\ee
described by the K\"ahler potential
\be
\label{EFT_IIA_K}
K= K_{\rm k} + K_{\rm cs}= - \log \left[ \ii \left( \bar f^a  \calg_a - f^a \bar \calg_a \right) \right]- \log e^{-4 D}\,.
\ee
In super-Weyl invariant language, we identify $z^a$ as the lowest components of the superfields $Z^a$, which carry super-Weyl weight $(1,0)$. Furthermore, we identify $N^K$ and $U_Q$ with the lowest components of chiral superfields which, with an abuse of notation, we call $N^K$ and $U_Q$ as well; we assume that the superfields $N^K$ and $U_Q$ do not carry any super-Weyl weight. It will also be convenient to collect them into the superfields $T_\Lambda = (N^K, U_Q)$. The kinetic section $\calk$ which determines the kinetic terms is then
\be
\calk(Z,\bar Z; T, \bar T) = \calk_{\rm k}(Z,\bar Z) e^{-\frac{K_{\rm cs} (T ,\bar T)}{3}}  = - 3 |U|^\frac23 e^{-\frac{K_{\rm k} (\Phi,\bar \Phi)}{3}} e^{-\frac{K_{\rm cs} (T ,\bar T)}{3}}\;.
\ee

\begin{table}[!h]
	\begin{center}
		\begin{tabular}{ |c c c c | }
			\hline
			\rowcolor{ochre!30}{\bf Fields} & {\bf Number} & {\bf Data} & {\bf Manifold}
			\\
			\hline
			\hline
			$\varphi^i$ & $h_-^{(1,1)}(X)$ & K\"ahler moduli & $\scrm_{\rm k}$\TBstrut
			\\
			\hline
			$T_\Lambda = (N^K, U_\Lambda)$ & $h^{(2,1)}(X)+1$ & complex structure moduli & $\scrm_{\rm cs}$\TBstrut
			\\
			\hline
		\end{tabular}
	\end{center}
	\caption{How the 4D fields originating form compactification of Type IIA superstring theory reassemble as lowest components of $\caln=1$ chiral superfields.}
	\label{tab:EFT_IIA_mod}
\end{table}

A superpotential for the moduli fields can be introduced by threading the Calabi-Yau manifold $X$ with background fluxes. In the democratic formulation, along with the R-R gauge potentials $C_1$, $C_3$, we also introduce $C_5$, $C_7$ and $C_9$, with the gauge invariant field strengths $F_0 \equiv m^0$, $G_2$, $G_4$ accompanied by their duals $G_{10} = * G_0$, $G_8 = * G_2$, $G_6 = * G_4$ \cite{Bergshoeff:2001pv,Carta:2016ynn,Bielleman:2015ina}. We collectively arrange the gauge potentials $C_{2p+1}$ into the polyform
\be
{\bf C} = C_1 + C_3 + C_5 +C_7 +C_9
\ee
which identifies the so-called \emph{C-basis} \cite{Bergshoeff:2001pv}. This is distinguished from the \emph{A-basis} 
\be
{\bf A} = {\bf C} \wedge e^{-B_2} \,,
\ee
in terms of which we introduce the gauge invariant field strengths
\be
{\bf G} = (\d {\bf A} + \overline{{\bf G}}) \wedge e^{B_2} = \d {\bf C} - H_3 \wedge {\bf C} + \overline{{\bf G}} \wedge e^{B_2} \,.
\ee
Here we have introduced the background R-R fluxes as the formal sum $\overline{{\bf G}}$ of harmonic forms on $X$. The A-basis is particularly useful to convey the quantization condition for the fluxes. That is, given a $(p+1)$--cycle $\Sigma_{p+1}$ in $X$ not intersecting a localized source, the charge quantization reads
\be
\frac{1}{\ell_s^p}\int_{\Sigma_{p+1}} \left(\d A_p + \overline{G}_{p+1} \right)  \in \mathbb{Z}\;.
\ee
Therefore, the integral background fluxes are defined as
\be
\label{EFT_IIA_em}
m^0 = \ell_s \overline{G}_0\,,\qquad m^i = -\frac1{\ell_s} \int_{\Sigma_{i}} \overline{G}_2\,, 
\qquad e_i = \frac1{\ell_s^3} \int_{\tilde\Sigma^{i}} \overline{G}_4 \,,\qquad e_0 = \frac1{\ell_s^5} \int_{X} \overline{G}_6\,,  \ee
where $\Sigma_{i}$ and $\tilde\Sigma^{i}$ constitute bases of cycles of $H_2^-(X;\mathbb{Z})$ and $H_4^+(X;\mathbb{Z})$ respectively. We are now in the position to write down the superpotential generated by the background fluxes as
\be
\label{EFT_IIA_calwRR}
\calw_{\text{R-R}} =  U \int_X{\bf{G}} \wedge e^{\ii J}  = U \int_X \overline{\bf{G}} \wedge e^{J_{\rm c}} \,,
\ee
where we have introduced an overall compensator in order to guarantee the super-Weyl weight $(1,0)$. If we define the quantities
\be
\label{EFT_IIA_calwRRmV}
\calv^A(Z) = \begin{pmatrix}
	Z^a \\ \calg_a(Z)
\end{pmatrix}\,, \qquad e_A =  \begin{pmatrix}
e_a \\ m^a
\end{pmatrix}\,,
\ee
then, the superpotential \eqref{EFT_IIA_calwRR} may be rewritten in the more compact form
\be
\label{EFT_IIA_calwRRsw}
\calw_{\text{R-R}} = e_A \calv^A( Z) \,.
\ee
A superpotential involving the complex structure moduli is instead introduced assuming that the Calabi-Yau is threaded with background NS-NS fluxes $\overline{H}_3$
\be
\label{EFT_IIA_hbg}
\overline{H}_3 = h^Q \alpha_Q-h_K \beta^K 
\ee
with $h_K$ and $h^Q$ the quantized fluxes defined by
\be
\frac{1}{\ell_s^2} \int_{\Sigma^Q} \overline{H}_3 =  h_Q\,,\qquad \frac{1}{\ell_s^2} \int_{\Sigma_K} \overline{H}_3 =  h^K\,,
\ee
where $\{\Sigma_K, \Sigma^Q\}$ is a basis of cycles in $H_3^-(X;\mathbb{Z})$, Poincar\'e dual of $\{\beta^K,\alpha_Q\}$. In the super-Weyl invariant language, a superpotential for the complex structure moduli is given by
\be
\label{EFT_IIA_calwNS}
\calw_{\text{NS-NS}}(U; N) = U \int_X \Omega_{\rm c} \wedge \overline{H}_3  =: -U \left(h_K N^K + h^L N_L\right) \equiv -U h^\Lambda T_\Lambda \,,
\ee
where the super-Weyl compensator $U$ is introduced to get the correct super-Weyl weight. Therefore, collecting \eqref{EFT_IIA_calwRRsw} and \eqref{EFT_IIA_calwNS}, we get the superpotential
\be
\label{EFT_IIA_calw}
\calw(Z;T) = e_A \calv^A(Z) - U h^\Lambda T_\Lambda  + \hat \calw(Z;N) 
\ee
where $ \hat \calw(Z;N) $ generically collects all the other contributions beyond those presently considered. After gauge-fixing the super-Weyl invariance, the superpotential \eqref{EFT_IIA_calw} is
\be
\label{EFT_IIA_calwgf}
W (\Phi ;T) = \Mp^3 \left( e_A \Pi^A (\Phi)- h^\Lambda T_\Lambda \right) + \hat W(\Phi) \,,
\ee
where the periods may be chosen as
\be
\label{EFT_IIA_PiA}
\Pi^A  = \begin{pmatrix}
	1 \\   \varphi^i   \\ \calg_a(\varphi)
\end{pmatrix} \,.
\ee

\subsection{The tadpole cancellation condition}
\label{sec:EFT_IIA_Tad}

The set of tadpole cancellation conditions that we will be interested in are those originating from the seven-form $\tilde C_7$, which couples to D6-branes and O6-planes. The tadpole cancellation can be more easily understood in a `democratic' formulation where also $C_5$, $C_7$ and $C_9$ are considered \cite{Bergshoeff:2001pv,Ibanez:2012zz} -- we will briefly comment on such a formulation in the next section. It  emerges as equations of motion for $\tilde C_7$, which coincide with the Bianchi identities for the field strengths $\tilde G_2$. The relevant contribution of $\tilde C_7$ to the ten-dimensional action are
\be
\begin{split}
\label{EFT_IIA_TadSC7}
S |_{\tilde C_7} &\supset -\frac{1}{4\kappa_{10}^2} \int_{\calm_4 \times X}  e^{\frac32 \tilde \phi}\d  \tilde C_7 \wedge * \tilde G_8 + \mu_6 \sum\limits_{\alpha} q_\alpha \int_{\calm_4 \times \Pi^\alpha} \tilde C_7
\\
&=  -\frac{1}{4 \kappa_{10}^2} \int_{\calm_4 \times X}  e^{\frac32 \tilde \phi} \d \tilde C_7 \wedge * \tilde G_8 + \mu_6 \sum\limits_{\Lambda} q_\Lambda \int_{\calm_4 \times X} \tilde C_7 \wedge \delta_3(\Pi^\alpha)\,,
\end{split}
\ee
where the sum over $\alpha$ involves separately the D6-branes, wrapping some internal three-cycle $\Pi^\alpha$, their orientifold images, wrapping the image cycles $\calr \Pi^\alpha$ and the O6-planes.\footnote{We recall that, due to the chosen convention \eqref{EFT_IIB_QOD}, given the charge $q^{\rm D6}_\alpha =1$ under $\tilde C_7$ of a D6-brane and their orientifold image, counted separately, the charge of an O6-plane is $q_\alpha^{\rm O6}=-4$.} From \eqref{EFT_IIA_TadSC7} we immediately read the Bianchi identity for $\tilde G_2$
\be
\label{EFT_IIA_Tadeom}
\d * \tilde G_8 = \d \tilde G_2 = 2 \kappa_{10}^2  \mu_6 \sum\limits_{\alpha} q_\alpha \delta_3(\Pi^\alpha)\,.
\ee
Integrating it over the internal, compact manifold $X$ provides the cohomological relation
\be
\label{EFT_IIA_Tad}
\sum\limits[\Pi^\alpha] +  [\calr \Pi^\alpha]  -4 [\Pi_{\text{O6}}] = m^0 [\Pi_H]\,,
\ee
which we will re-express as
\be
\label{EFT_IIA_Tadb}
\calq^\Lambda_{\rm bg} - m^0 h^\Lambda =0\,.
\ee
where $\calq^\Lambda_{\rm bg}$ generically takes into account the contributions of the local sources.
In contrast with the Type IIB case, where the Bianchi identity \eqref{EFT_IIB_Tada} provides just a single constraint, here the tadpole conditions are counted by the dimension of the odd integral cohomology $H^3_-(X;\mathbb{Z})$.

\subsection{Reformulating the 4D EFT}
\label{sec:EFT_IIA_4d}

The four-dimensional Type IIA EFTs are equipped with a set of $2h_-^{(1,1)} + h^{(2,1)}+3$ fluxes $(e_A, h^\Lambda)$, spanning the lattice 
\be
\label{EFT_IIA_4dGamma}
\Gamma  =  \left\{(e_A, h^\Lambda)\; |\; e_A, h^\Lambda \in \mathbb{Z}  \right\}=  \{ \caln_\cala \}  \,  .
\ee
The admissible flux configurations are however constrained due to the tadpole cancellation condition \eqref{EFT_IIA_Tad}. The tadpole condition \eqref{EFT_IIA_Tad} further signals that it is not possible to regard all the fluxes as originated from integrating out three-forms, but only a sublattice of them. The simplest choice seems to assume the NS-NS background fluxes to be fixed at given background values $h^\Lambda_{\rm bg}$ and the R-R fluxes to be dynamically generated. Namely, the flux sublattice that we can span with such a choice is
\be
\label{EFT_IIA_4dGammadyn}
\Gamma_{\rm exp}  \, \equiv \,  \Gamma_{\rm EFT}+\caln^{\rm bg}  \, = \, \left\{(e_A, h^{\Lambda}_ {\rm bg}) \;|\;  e_A \in \mathbb{Z},\  h^{\Lambda}_{\rm bg}\, {\rm fixed}  \right\} \, .
\ee
This reflects, upon the super-Weyl superpotential \eqref{EFT_IIA_calw}, in the fact that we will regard the R-R contributions \eqref{EFT_IIA_calwRRsw} as dynamically generated by a set of gauge three-forms, while the other contributions are, for the moment, considered as spectators
\be
\label{EFT_IIA_calwgen}
\calw(z;\tau)= \underbrace{e_A \calv^A(z)}_\text{generated by three-forms}  \quad \underbrace{- U  h^\Lambda T_\Lambda + \hat \calw(Z;T)}_\text{`spectator' sector $\to \calw'(z;\tau)$}\, .
\ee
Having set the NS-NS fluxes as fixed, the tadpole cancellation condition \eqref{EFT_IIA_Tad} becomes linear with respect to the dynamically generated fluxes, which here is just the Romans mass:
\be
\label{EFT_IIA_4dTad}
m^0 Q^\Lambda  + \calq_{\rm bg}^\Lambda=0\,, \qquad {\rm with} \quad Q^\Lambda \equiv  - h^\Lambda \,.
\ee
It is then immediate to see that, since $\cali^{\cala\calb}_\Lambda \to \cali^{\Sigma}_{0\Lambda} = \delta_\Sigma^\Lambda \delta_{0}^\cala$, the subset $e_A$ constitute an isotropic sublattice in the sense of \eqref{Intro_Con_Iso}. 

The dynamical generation of the sublattice \eqref{EFT_IIA_4dGammadyn} requires the introduction  of  a set of $2 h_-^{(1,1)}(X)+2$ gauge three-form potentials, the same amount as $e_A$ in \eqref{EFT_IIB_4dGammadyn}. Their origin from a ten-dimensional theory may be better understood switching to a democratic formulation \cite{Bergshoeff:2001pv,Carta:2016ynn}. Along with the Romans mass $m^0$ and $\tilde G_2$, $\tilde G_4$, we also consider $\tilde G_{10}$, $\tilde G_8$ and $\tilde G_6$. They obey $*\tilde G_{10} = m^0$, $*\tilde G_{8} = \tilde G_2$, $*\tilde G_6 =G_4$, which can be imposed at the level of equations of motion. The presence of the higher-form potentials $\tilde C_5$, $\tilde C_7$, $\tilde C_9$ allows one to get, by dimensional reduction, the following sets of gauge three-forms in 4D EFT:
\begin{alignat}{4}
\label{EFT_IIA_4d4f}
	&\tilde G_4 &&= \d A_3^0 + \ldots\,, &&\qquad  \tilde G_6 &&= \d A_3^i \wedge \omega_i + \ldots\,,
	\\
	&\tilde G_8 &&= \d \tilde A_{3i} \wedge \omega^i + \ldots \,, &&\qquad  \tilde G_{10} &&= \d \tilde A_{30} \wedge {\rm vol}_X+ \ldots \,,
\end{alignat}
where the dots stand for other contributions where we are presently not interested into.
Together, the gauge three-forms $\{A_3^a,\tilde A_{3a}\}$ are a set of $2h_-^{(1,1)}+2$ gauge three-forms. This has to be compared with the number of physical complex scalars $\varphi^i$, $h_-^{(1,1)}$, which \emph{actively} participate in the dynamically generated part of the superpotential. This case falls into the \emph{maximally nonlinear} class of examples of Section~\ref{sec:Sugra_MTF_Examples}.  The supersymmetric EFT which appropriately include the gauge three-forms is described in terms nonlinear double three-form multiplets of the kind of \eqref{Sugra_MTF_Ex_Za}
\be
\label{EFT_IIA_4dS}
S^a \equiv-\frac{\ii}{4}(\bar \cald^2 - 8 \calr) \calp^a\,, \qquad S_a = \calg_a (S) \equiv-\frac{\ii}{4}(\bar \cald^2 - 8 \calr) \tilde \calp_a\,,
\ee
which collect at once the K\"ahler moduli (plus the super-Weyl compensator) $z^a$ and the gauge three-forms $A_3^A$ among their components. 

The gauge three-forms so included would arbitrarily generate any possible constant $e_A \in \mathbb{Z}$. The tadpole constraint \eqref{EFT_IIA_Tad} is enforced upon the dynamically generated $e_A$ by inserting, in the EFT, a set of gauge four-forms $C_4^\Lambda$ gauging the three-forms $\{A_3^a, \tilde A_{3a}\}$. Following the recipe of Section~\ref{sec:Sugra_4FM}, we introduce $b_3^-(X)$ gauge four-forms, one for each of the constraints imposed by \eqref{EFT_IIA_4dTad}. Applying \eqref{Sugra_GMTF}, with the definitions \eqref{EFT_IIA_4dTad}, it turns out that the only multiplets that is subjected to the gauging is
\be
\label{EFT_IIA_4dSGaug}
S_0 = \calg_0(S^a)  \; \rightarrow \hat S_0 \equiv -\frac \ii 4 (\bar \cald^2 - 8 \calr) \tilde \calp_{0}+ \ii h^\Lambda \Gamma_\Lambda\,.  
\ee
This results also into the fact that in the four-dimensional theory, rather than $\tilde F_{40} $, its gauged counterpart appears
\be
\hat {\tilde F}_{40} = \d \tilde A_{30}  - h^\Lambda C_{4\Lambda}\;.
\ee

To recap, up to to this point we have arrived at the superspace Lagrangian
\be
\label{EFT_IIA_4dLA}
\begin{aligned}
	\tilde\call &= \int \d^4\theta\,E\,\calk_{\rm k}(\hat S,\bar {\hat S}) e^{-\frac{K_{\rm cs}(T,\bar T)}{3}}   + \left( \int \d^2\Theta\,2\cale\, \calw'(\hat S;\tau ) +{\rm c.c.}\right)\,,
	\\
	&\quad\, + \left[\ii \int \d^2\Theta\,2\cale\, \calq_{\rm bg}^\Lambda  \Gamma_\Lambda + {\rm c.c.} \right] + \tilde \call_{\rm bd}
\end{aligned}
\ee
with the remnant superpotential
\be
\label{EFT_IIA_calwB}
\calw'(S;N) =  -S^0 h^\Lambda T_\Lambda  + \hat \calw(S) \;.
\ee
The components of \eqref{EFT_IIA_4dLA}, after gauge-fixing the super-Weyl invariance, are the same as \eqref{Sugra_GMTF_L3fComp_gf}, with the three-form kinetic parts given by \eqref{Sugra_MTF_TUVgf} and \eqref{Sugra_MTF_Ex_TABdouble} (see also, in comparison, \cite{Carta:2016ynn,Herraez:2018vae,Escobar:2018tiu,Escobar:2018rna,Marchesano:2019hfb}).

This not the end of the story. The complex structure multiplets $T_\Lambda$ may be regarded as axionic multiplets for the Lagrangian \eqref{EFT_IIA_4dLA} and can then be dualized to linear multiplets $L^\Lambda$ following the procedure of Section~\ref{sec:Sugra_LMDual}. In this picture, the R-R axions $\xi_\Lambda$ are then replaced with gauge two-forms $\calb_2^\Lambda$, while the saxions $n_\Lambda$ are replaced by $\ell^\Lambda$, the lowest components of the linear multiplets. Furthermore, comparing the superpotential \eqref{EFT_IIA_calwB} with \eqref{Sugra_GL_W} considered in Section~\ref{sec:Sugra_GL}, we immediately recognize that the first contribution in \eqref{EFT_IIA_calwB}, which couples the compensator $S^0$ to the complex structure superfields $T_\Lambda$, can be reabsorbed in a gauging of the newly added linear multiplets. Hence, we trade $L^\Lambda$ with their gauged variants
\be
\label{EFT_IIA_La}
L^\Lambda \rightarrow \hat L^\Lambda = L^\Lambda + h^\Lambda \calp^0\;,
\ee
which, at the level of components, reflects into the gauging
\be
\label{EFT_IIA_H3G}
\calh_3^\Lambda = \d \calb_2^\Lambda \rightarrow \hat\calh_3^\Lambda = \d \calb_2^\Lambda  + h^\Lambda A_3^0 \;.
\ee

The full spectrum of superfields which build the theory, along with the bosonic components that they include, is summarized in Table~\ref{tab:EFT_IIA_sf}. At the superspace level, the Lagrangian \eqref{EFT_IIA_4dLA} becomes
\be
\label{EFT_IIA_4dL}
\begin{aligned}
	\tilde\call &= -3 \int \d^4\theta\,E\,\calf(\hat S,\bar {\hat S}; \hat L)  + \left( \int \d^2\Theta\,2\cale\, \hat\calw(\hat S) +{\rm c.c.}\right)\,,
	\\
	&\quad\, + \left[\ii \int \d^2\Theta\,2\cale\, \calq^{\rm bg}_\Lambda  \Gamma^\Lambda + {\rm c.c.} \right] + \tilde \call_{\rm bd}
\end{aligned}
\ee
where the only remnant piece of the superpotential is $\hat \calw (S)$. 

\begin{table}[!h]
	\begin{center}
		\begin{tabular}{ |c c c c | }
			\hline
			\rowcolor{ochre!30}{\bf Superfields} & {\bf Bosonic fields} & {\bf Number} & {\bf Data} 
			\\
			\hline
			\hline
			\cellcolor{darkblue!20} & \multirow{2}{*}{$z^a(\varphi)$}& \multirow{3}{*}{$h_-^{(1,1)}(X)+1$} &   K\"ahler deformations 
			\\
			\cellcolor{darkblue!20} & & & R-R axions 
			\\
			\multirow{-3}{*}{\cellcolor{darkblue!20}$\hat S^a$} & $A_3^A$, $\tilde A_{3A}$ & & gauge three-forms 
			\\
			\hline
			\cellcolor{darkblue!20}  & $\ell^\Lambda(t)$& \multirow{2}{*}{$h^{(2,1)}(X)+1$} &   complex structure deformations
			\\s
			 \multirow{-2}{*}{\cellcolor{darkblue!20}$\hat L^\Lambda$} & $B_2^\Lambda$ & & NS-NS axions 
			\\
			\hline
			\cellcolor{darkblue!20} $\Gamma^\Lambda$ & $C_4^\Lambda$ & $h^{(2,1)} + 1$ & gauge four-forms \TBstrut
			\\
			\hline
		\end{tabular}
	\end{center}
	\caption{How the 4D fields originating form compactification of Type IIA superstring theory reassemble as lowest components of $\caln=1$ chiral superfields.}
	\label{tab:EFT_IIA_sf}
\end{table}

In components, the Lagrangian \eqref{EFT_IIA_4dL} reads 
\be
\label{EFT_IIA_SLeg}
\begin{split}
	S|_{\rm bos} &= M^2_{\rm P} \int \left( \frac12\, R *1   -K_{\text{k}\,i\bar \jmath}\,  \d \varphi^i \wedge *\d \bar \varphi^{\bar\jmath} + \frac14 F_{\Lambda\Sigma}  \d \ell^\Lambda \wedge * \d \ell^\Sigma \right)
	\\
	&\quad\,+\frac{1}{4 M_{\rm P}^2} \int F_{\Lambda\Sigma}\,  \hat\calh_3^\Lambda \wedge * \hat \calh^{\Sigma}_3 +  
	\\
	&\quad\, -\int_\Sigma \Big[\frac12 T_{AB} \hat F^A_4*\! \hat F^B_4+ T_{AB}\Upsilon^A \hat F^B_4+\Big(
	\hat V-\frac12 T_{AB}\Upsilon^A \Upsilon^B\Big)*\!1\Big] 
	\\
	&\quad\,+\int_{\del\Sigma}T_{AB}(* \hat F^A_4+\Upsilon^A)A^B_3 
	\\
	&\quad + \calq_{\rm bg}^\Lambda\int_\Sigma C_{4\Lambda}
\end{split}
\ee
with the three-form part specified by \eqref{Sugra_GL_TUVgf} with $\Pi^A$ as in \eqref{EFT_IIA_PiA}. Moreover, the Legendre transform of the K\"ahler potential \eqref{EFT_IIA_K} is here given by
\be\label{SL_IIA_F}
F(\varphi,\bar\varphi;\ell)=K+2\ell^\Lambda\Im t_\Lambda \, ,
\ee
where we have introduced the variables
\be
\label{SL_IIA_ell}
\ell^\Lambda = -\frac{1}{2}\frac{\del K_{\rm cs}}{\del \Im t_\Lambda}\,,
\ee
and we recall that $\tilde F(\varphi,\bar\varphi;\ell) = K(\varphi,\bar\varphi; \Im t(\varphi, \bar\varphi;\ell))$.

\subsection{Including non-geometric fluxes}

Non-geometric fluxes can be included by extending the NS-NS sector to
\be\label{EFT_IIA_NGF}
c^\Lambda_A= (c^\Lambda_0,c^\Lambda_i, c^{\Lambda j},c^{\Lambda 0})\equiv  (h^\Lambda, \omega^\Lambda_i, Q^{\Lambda j}, R^\Lambda)\, ,
\ee
where 
$h^\Lambda$ counts the  $H_3$ flux quanta, while $\omega^\Lambda_i$, $R^\Lambda$, $Q^{\Lambda j}$ represent the geometric and non-geometric fluxes \cite{Aldazabal:2006up,Wecht:2007wu,Ibanez:2012zz,Gao:2017gxk}.
A superpotential which takes into account the contribution of the non-geometric sector is a slight extension of \eqref{EFT_IIA_calwNS}
\be
\label{EFT_IIA_NGWNS}
\calw_{\text{NS-NS}}(Z;T) = U \int \Omega_{\rm c}\wedge \frak{D} (e^{J_c})\,.
\ee
Here $\frak{D}$ is the twisted differential operator
\be
\frak{D} = \d + \overline{H}_3 \wedge\; -\, \omega\, \lhd\; +\, Q \rhd\; - R\, \bullet\;\,,
\ee
where the actions of the operators over the cohomology classes is such that \cite{Gao:2017gxk,Ihl:2007ah}
\be
\begin{aligned}
	\omega\, \lhd\; &:\quad H^p(X,\mathbb{Z}) \rightarrow H^{p+1}(X,\mathbb{Z})
	\\
	Q\, \rhd\; &:\quad H^p(X,\mathbb{Z}) \rightarrow H^{p-1}(X,\mathbb{Z})
	\\
	R\, \bullet\;&:\quad H^p(X,\mathbb{Z}) \rightarrow H^{p-3}(X,\mathbb{Z})
\end{aligned}
\ee
Expanding the integrand in \eqref{EFT_IIA_NGWNS} into an harmonic basis over the Calabi-Yau, gives the superpotential
\be
\calw_{\text{NS-NS}}(Z;T) = -U T_\Lambda \left(h^\Lambda  + \omega_{i}^\Lambda \Phi^i +\frac12 \kappa_{ijk} Q^{\Lambda i} \Phi^j \Phi^k - \frac16 \kappa_{ijk} \Phi^i \Phi^j \Phi^k R^\Lambda\right)
\ee
which can be recast in the more compact form 
\be
\calw_{\text{NS-NS}}(\calz; T) = -c^\Lambda_A T_\Lambda \calv^A (Z;T)  \;.
\ee
where we have used the quantities have been defined so as
\be
{\frak D} e^{J_c} =: -\left(h^\Lambda  + \omega_{i}^\Lambda \Phi^i +\frac12 \kappa_{ijk} Q^{\Lambda i} \Phi^j \Phi^k - \frac16 \kappa_{ijk} \Phi^i \Phi^j \Phi^k R^\Lambda\right) \alpha_\Lambda\;.
\ee

The four-dimensional tadpole cancellation condition requires that the contribution$ \int C_7 \wedge \frak{D}\, \overline{\mathbf{G}}$, once we expand $C_7$, is canceled by the contributions of background charges; namely, the tadpole cancellation condition \eqref{EFT_IIA_4dTad} now becomes
\be
\label{EFT_IIA_NGtad}
\calq_{\rm bg}^\Lambda - h^\Lambda m^0 + \omega_{i}^\Lambda m^i  - e_i Q^{\Lambda i} - e_0 R^\Lambda = 0\,.
\ee
These can be written in the form \eqref{EFT_Intro_Tadb} by making the index change $(\ldots)_I\rightarrow (\ldots)^\Lambda$, grouping the fluxes into  $\caln_\cala=(N_A,c^\Lambda_B)\in\Gamma$ 
and correspondingly decomposing 
\be
\label{IIANGcali}
(\cali^\Lambda)^{\cala\calb} = \begin{pmatrix} 0 & \delta_\Theta^\Lambda I^{AD}  \\
	-  \delta_\Sigma^\Lambda I^{CB}& 0 \end{pmatrix}
\ee
with
\be
I^{AB}= \begin{pmatrix}
	0 & 0 & 0 & -1 \\
	0 & 0 & -\delta^i_j & 0 \\
	0 &  \delta_i^j &0 & 0 \\
	-1 & 0 & 0 & 0 
\end{pmatrix}\,,
\ee
which identifies the charges
\be
\label{EFT_IIA_NGQ}
Q^\Lambda \equiv - ( R^\Lambda, Q^{\Lambda i}, - \omega^\Lambda_i, h^\Lambda)\,.
\ee

The path that we followed in the previous section to get the dual formulation can be applied to the present case as well, taking care of the change of the tadpole condition, from \eqref{EFT_IIA_4dTad} to the more complicated \eqref{EFT_IIA_NGtad}. The gauging of the three-form multiplets of \eqref{Sugra_GMTF}, due to \eqref{EFT_IIA_NGQ}, now involve all the multiplets
\be
\begin{aligned}
	\hat S^0 &= - \frac\ii4(\bar\cald^2-8\calr)\calp^0+ \ii  R^\Lambda \Gamma_\Lambda\,,
	\\
	\hat S^i &= - \frac\ii4(\bar\cald^2-8\calr)\calp^i +\ii  Q^{\Lambda i} \Gamma_\Lambda\,,
	\\
	\hat{ S}_i &=-\frac\ii4(\bar\cald^2-8\calr)\tilde\calp_i -\ii \omega_{i}^\Lambda \Gamma_\Lambda\,,
	\\
	\hat{ S}_0 &=-\frac\ii4(\bar\cald^2-8\calr)\tilde\calp_0 + \ii  h^\Lambda \Gamma_\Lambda\,,
\end{aligned}
\ee
resulting into a gauging of \emph{all} the gauge three-form entering the four dimensional theory as
\be
\label{EFT_IIA_NGF4}
\begin{aligned}
	\hat F_4^0 &= \d A_3^0 - R^\Lambda C_{4\Lambda}\,,
	\\
	\hat F_4^i &= \d A_3^i - Q^{\Lambda i} C_{4\Lambda}\,,
	\\
	\hat{F}_{4i} &=\d A_{3i} + \omega_i^\Lambda C_{4\Lambda}\,,
	\\
	\hat{F}_{40} &=\d A_{30} - h^\Lambda C_{4\Lambda}\,.
\end{aligned}
\ee
Of course, this reflects also in a change of the gauging of the linear multiplets \eqref{EFT_IIA_La} which now, in order to incorporate all the non-geometric fluxes, gets modified to
\be
\hat L^\Lambda = L^\Lambda + c_A^\Lambda \calp^A
\ee
which, in components, implies the gauging of the two-forms as
\be
\label{EFT_IIA_NGH3}
\begin{aligned}
	\hat \calh_3^\Lambda &= \d \calb_2^\Lambda + h^\Lambda A_3^0 + \omega_i^\Lambda A_3^i + Q^{\Lambda i} \tilde A_{3i} +  R^\Lambda  \tilde A_{30}\;.
\end{aligned}
\ee
The four-dimensional bosonic action is the very same as \eqref{EFT_IIA_SLeg}, provided however the redefinition of the gauged two- and three-forms according to \eqref{EFT_IIA_NGH3} and \eqref{EFT_IIA_NGF4}.

\subsection{Including open string moduli}

The \emph{open string} moduli sector can also be introduced without many difficulties.  In this case, we will just focus on moduli originating from spacetime filling D6 branes \cite{Carta:2016ynn}. Let us for simplicity turn off the geometric and non-geometric fluxes, restricting $\Gamma$ so that
\be\label{EFT_IIA_CYcond}
\omega^\Lambda_i= R^\Lambda= Q^{\Lambda j}=0\,,
\ee 
in order to deal with proper Calabi-Yau orientifold compactifications. The chiral moduli $\chi_\alpha\simeq \chi_\alpha+1$ of a D6-brane wrapping a special Lagranian 3-cycle $S$ (including D6 and image D6) combine Wilson lines and geometric moduli, both labelled by $\alpha=1,\ldots, b^-_1(S)$.  These couple to the bulk moduli  through a contribution to the superpotential \cite{Marchesano:2014iea,luca2}.   In  super-Weyl invariant notation, this contribution takes the form\footnote{In the presence of metric fluxes D6-branes also develop superpotentials quadratic in $\chi_\alpha$, see e.g. \cite{Marchesano:2006ns}.}
\be
\label{EFT_IIA_D6coupling}
\calw_{\rm D6}=-n^\alpha_ a\,\chi_\alpha z^a\,.
\ee
Here $n^\alpha_ 0\in\mathbb{Z}$ correspond to the D6-brane world-volume flux quanta, which take values into $H^2_-(S;\mathbb{Z})$ and are then also labelled by $\alpha=1,\ldots, b^-_1(S)$. The remaining quantized coupling constants are given by the intersection numbers 
\be
n^\alpha_i=\int_S\omega_i\wedge \eta^\alpha\,,
\ee
where $\omega_i$ is the basis of $H^2_-(X;\mathbb{Z})$ used to identify  the complexified K\"ahler moduli $\varphi^i$, while $\eta^\alpha$ is the basis of $H^1_-(S)$ used to identify the world-volume moduli $\chi_\alpha$ \cite{Marchesano:2014iea}. Comparing \eqref{EFT_IIA_D6coupling} with \eqref{Sugra_GL_W}, it is clear that also the coupling \eqref{EFT_IIA_D6coupling} can be interpreted as produced from a two-form gauging of the linear multiplets dual to the world-volume chiral fields  $\chi_\alpha$. To be more precise, let us denote by $\hat\calb_2^\alpha$ the corresponding two-form potentials and split $A^A_3$ into $(A^a_3,\tilde A_{3 a})$. Then, \eqref{EFT_IIA_D6coupling} corresponds to a gauging 
\be
\label{EFT_IIA_B2D6a}
\hat\calb_2^\alpha\rightarrow \hat\calb_2^\alpha+n^\alpha_a\Lambda^a_2\,,
\ee 
under the transformation $A^a_3\rightarrow A^a_3+\d\Lambda^a_2$, implying the introduction of the gauge invariant three-forms
\be
\label{EFT_IIA_B2D6}
\hat \calh_3^\alpha \rightarrow  \hat \calh_3^\alpha + n^\alpha_a A_3^a\,.
\ee
In doing so, we are assuming the fluxes $n^\alpha_a$ must be treated as non-dynamical in the EFT. This also fits well with the observation made in \cite{Herraez:2018vae} that the matrix $T^{AB}$ becomes non-invertible in the presence of such fluxes.


Again, the four-dimensional bosonic action acquires the same form as \eqref{EFT_IIA_SLeg}, provided however the redefinition of the gauged three-forms \eqref{EFT_IIA_NGF4} and the gauged two-forms given by \eqref{EFT_IIA_NGH3}. However, the Legendre transform $F$ will also depend on the (duals of) the moduli $\chi_\alpha$ and the associated two-forms appear via the gauge-invariant field strengths \eqref{EFT_IIA_B2D6}.

\section{M-theory effective field theories}
\label{sec:EFT_Mth}

At very strong coupling, when $g_s \gg 1$, the ten-dimensional Type IIA effective theory is no longer valid and we should rather pass to the eleven-dimensional M-theory description. Indeed, the eleven-dimensional M-theory spectrum is naturally endowed with a gauge three-form, which  will reflect, as we will see shortly, in the presence of multiple gauge three-forms in the four-dimensional spectrum, and thus seems naturally fit to apply the dualization procedure developed in Chapter~\ref{chapter:Sugra}. 

It is known that, once compactified on a circle, M-theory reproduces the weakly coupled Type IIA supergravity, but M-theory and Type IIA are definitely not quite the same. For example, no Romans mass is allowed in the M-theory description \cite{Aharony:2010af} and no tadpole cancellation condition constrains the background fluxes. These peculiarities forces us to treat the M-theory EFTs separately from the Type IIA ones. In the following we will first consider the case where the seven-dimensional manifold upon which we compactify the eleven-dimensional M-theory enjoys a $G_2$--holonomy. This request will be later relaxed and we will also provide connections with the Type IIA case examined in the previous section.

\subsection{From 11D to 4D}
\label{sec:EFT_Mth_11D}

The eleven-dimensional massless, bosonic spectrum of M-theory is made up just by the graviton $g_{MN}$ and a gauge three-form $\tilde C_3$. The low energy supergravity EFT is  \cite{Grimm:2004ua,Becker:2007zj,Ibanez:2012zz}
\be
\label{EFT_Mth_S11d}
\begin{aligned}
	S_{\text{M-theory}}^{(11)} &= \frac{1}{ 2\kappa_{11}^2} \int \left( \tilde R * 1 - \frac12  \tilde G_4 \wedge * \tilde G_4 -\frac1{6} \tilde C_3 \wedge \tilde G_4 \wedge \tilde G_4\right)\;,
\end{aligned}
\ee
where we have introduced the four-form field strength $\tilde G_4 = \d \tilde C_3$.

A four-dimensional $\caln=1$ supergravity theory can be obtained by compactifying M-theory over a proper seven manifold $Y$. Presently, we shall assume that the manifold $Y$ is equipped with a $G_2$--holonomy.  The manifold $Y$, similarly to a Calabi-Yau manifold, is embedded with a real, covariantly constant and harmonic three-form $\Phi_3$, satisfying $\d_Y \Phi_3 = 0$. 

The identification of the four-dimensional fields entering the effective theory proceeds along the same lines as for Type II superstring theories. The scalar fields have a two-fold origin. Introducing a basis of harmonic forms $\{\check\rho_{\check i}\}$ of $H^3(X)$, with $\check{\imath}=1,\ldots, b^3(X)$, we may decompose the eleven dimensional three-form $\tilde C_3$ as
\be
\label{EFT_Mth_Cex}
\tilde C_3 = c^{\check{\imath}} \check\rho_{\check \imath}\,,
\ee
as well as the covariantly constant three-form $\Phi_3$ as
\be
\label{EFT_Mth_Phiex}
\Phi_3 = \phi^{\check{\imath}} \check\rho_{\check{\imath}}\,.
\ee
The zero modes $c^{\check{\imath}}$ and $\phi^{\check{\imath}}$ identify the complex coordinates
\be
\label{EFT_Mth_Si}
\check{S}^{\check{\imath}} \equiv \int_{\Sigma^{\check{\imath}}} (\tilde C_3 + \ii \Phi_3) = c^{\check{\imath}} + \ii \phi^{\check{\imath}}\,.
\ee
where we have introduced a basis $\Sigma^{\check{\imath}}$ of three-cycles in $H^3(X;\mathbb{Z})$, Poicar\'e dual to $\check\rho_{\check \imath}$. The coordinates $\check{S}^{\check{\imath}}$ parametrize a K\"ahler manifold with K\"ahler potential 
\be
\label{EFT_Mth_K}
K_{G_2} = -3 \log \left(\frac{1}{7\kappa_{11}^2} \int_X \Phi_3 \wedge * \Phi_3 \right)\,.
\ee
At the superfield level, we may regard $\check{S}^{\check{\imath}}$ as the lowest components of a set of $b^3(X)$ superfields which, with a slight abuse of notation, we will call $\check{S}^{\check{\imath}}$ as well.

A superpotential for the chiral superfields $\check{S}^{\check{\imath}}$ is introduced by turning on some background fluxes $\overline{G}_4$ and its dual $\overline{G}_7$. Expanding $\overline{G}_4$ in an harmonic basis $\{\check\zeta^{\check\jmath} \}$ of $H^4(X;\mathbb{Z})$, with $\check\jmath = 1,\ldots, b^3(X)$, dual to $\{\check\rho_{\check\imath}\}$, as
\be
\label{EFT_Mth_Gex}
\overline{G}_4 = n_{\check\jmath} \check\zeta^{\check\jmath} 
\ee
and $\overline{G}_7$ as
\be
\overline{G}_4 = n_0 \d {\rm vol}_Y
\ee
we recognize that the available number of fluxes is $b^3(X)+1$. Then, a superpotential for the chiral fields $S^{\check{\imath}}$ acquires the form \cite{Beasley:2002db,Grimm:2004uq,House:2004pm}
\be
\label{EFT_Mth_W}
\calw_{G_2} = U \left[\int_X G_7 + \int_X \left(\tilde  C_3 + \ii \Phi_3\right) \wedge \tilde G_4 \right]
\ee
where $G_4 \equiv \d \tilde C_3 +  \overline{G}_4$ includes also the contribution of the background four-form fluxes. The contributions from $G_7$ is actually tied to the one of $G_4$ via \cite{Beasley:2002db}
\be
\label{Mth_Wcon}
\int_X \left(G_7 + \frac12 C_3 \wedge G_4 \right) = 0  
\ee
so that \eqref{EFT_Mth_W} can be also written as
\be
\label{EFT_Mth_Wa}
\calw_{G_2} =U  \int_X \left(\frac12 C_3 + \ii \Phi\right) \wedge G_4
\ee
As stressed in \cite{Beasley:2002db}, the factor of $\frac12$ is crucial to make the superpotential holomorphic in $S^{\check\imath}$. In fact, now expanding $C_3$, $\Phi$ and $\overline{G}_4$ as in \eqref{EFT_Mth_Cex}, \eqref{EFT_Mth_Phiex} and \eqref{EFT_Mth_Gex}, \eqref{EFT_Mth_W} gives
\be
\label{EFT_Mth_Wb}
\calw_{G_2}  =U \left(n_0+n_{\check\imath}   \check{S}^{\check\imath}\right)\,.
\ee
The introduction of $U$, interpreted here as a compensator, is necessary in the super-Weyl formulation, in order to ensure the superpotential \eqref{EFT_Mth_Wb} to properly acquire super-Weyl weight $(1,0)$. In the following, it will also convenient to collect the fluxes as $n_\cali = (n_0, n_{\check\imath})$, with $\cali = 0,1,\ldots, b^3(Y)$. Accordingly, we also define the \emph{periods} $\calv^\cali(S) = (U, U \check{S}^{\check\imath})$, the superpotential being rewritten as
\be
\label{EFT_Mth_Wp}
\calw_{G_2}  =n_{\cali} \calv^\cali (U,S)\,.
\ee

\subsection{Reformulating the 4D EFT}
\label{sec:EFT_Mth_4D}

The potential of the four-dimensional theory is characterized by $b^3(Y)+1$ background fluxes $(n_0, n_{\check\imath})$, spanning a lattice
\be
\label{EFT_Mth_4dGamma}
\Gamma  =  \left\{(n_0,n_{\check\imath})\; |\; n_0,n_{\check\imath} \in \mathbb{Z}  \right\}=  \{ n_\cali \}  \,  .
\ee
In contrast with the Type II EFTs, no tadpole cancellation condition constrains the admissible values $n_{\cali}$. We may then regard \emph{all} the lattice $\Gamma$ as generated by a set of gauge three-forms $A_3^\cali$. Comparing the number of complex scalars upon which the superpotential \eqref{EFT_Mth_Wb} depends, $b^3(Y)$, with the dimensionality of the lattice \eqref{EFT_Mth_4dGamma}, we immediately understand that the case at hand falls into the linear superpotential family examined in Section~\ref{sec:Sugra_MTF_Examples}. We then trade all the periods $\calv^I$ with single three-form multiplets as in \eqref{Sugra_MTF_Ex_STF}
\be
\label{EFT_Mth_STF}	
\calv^\cali (z) = - \frac\ii4 (\bar \cald^2 - 8 \calr) P^\cali \equiv S^\cali \,,
\ee
each of them including, among its bosonic components, the moduli $s^{\check\imath}$, along with the compensator $u$, via $z^\cali(u,s)$ as well as the gauge three-forms $A_{3}^\cali$
\be
\label{EFT_Mth_STFc}	
S^\cali = \{z^\cali, A_3^\cali\} \,.
\ee
The dual three-form theory is then immediately read off from \eqref{Sugra_MTF_Ex_STF_L3fCompc} and is given by
\be
\label{EFT_Mth_4Sa}
\begin{aligned} 
	S |_{\rm bos} &=  \Mp^2 \int_{\Sigma} \left( \frac12 R *1- K_{\check{\imath} \bar{\check{\jmath}}} \d s^{\check{\imath}} \wedge *\d \bar s^{\bar {\check\jmath}} \right)
	\\
	&\quad\, - \frac12  \int_\Sigma T_{\cali\calj} F^\cali_4*\!  F^\calj_4 +\int_{\del\Sigma}T_{\cali\calj} * F^\cali_4 A_3^\calj
\end{aligned}
\ee
with  $T_{\cali \calj}$ defined from \eqref{Sugra_MTF_Ex_TABsingle}, with $\alpha =4$. The spectrum of the dual theory is summarized in Table~\ref{tab:EFT_IMth_sf}.

\begin{table}[!h]
	\begin{center}
		\begin{tabular}{ |c c c c | }
			\hline
			\rowcolor{ochre!30}{\bf Superfields} & {\bf Bosonic fields} & {\bf Number} & {\bf Data} 
			\\
			\hline
			\hline
			\cellcolor{darkblue!20}  &  &  &  $\Phi$--deformations
			\\
			\cellcolor{darkblue!20} & \multirow{-2}{*}{$z^{\cali}$} &  & $C_3$--axions 
			\\
			\multirow{-3}{*}{\cellcolor{darkblue!20}$S^{\cali}$} & $A_3^\cali$ & \multirow{-3}{*}{$b^{3}(Y)$} & gauge three-forms
			\\
			\hline
		\end{tabular}
	\end{center}
	\caption{How the 4D fields originating form compactification of M-theory theory reassemble as lowest components of $\caln=1$ chiral superfields in the dual picture.}
	\label{tab:EFT_IMth_sf}
\end{table}

\subsection{Connection to Type IIA compactifications}
\label{sec:EFT_Mth_IIA}

In order to connect this picture with the Type IIA orientifold compactifications of Section \eqref{sec:EFT_IIA}, let us assume that the internal manifold $Y$ can be decomposed as \cite{Grimm:2004uq}
\be
\label{EFT_Mth_Xd}
Y = \frac{X \times S^1}{\hat\calr}
\ee
where $X$ is a Calabi-Yau three-fold, $S^1$ a circle with radius $R$ and $\hat\calr$ acts on $X$ as an anti-holomorphic involution and on $S^1$ by inverting the coordinate $y \to -y$ (in other words, the only forms along $S^1$ which we should consider belong to the class $H^1_-(S^1)$).

To begin with, we split the eleven-dimensional forms $\Phi_3$ and $C_3$ in components along the Calabi-Yau $X$ and the circle $S^1$. The choice \eqref{EFT_Mth_Xd} makes the third cohomology class of $Y$ split as
\be
H^3(Y) = H^3_+(X) \oplus [H^2_-(X) \wedge H^1_-(S^1)]
\ee
and we can then expand $C_3$ and $\Phi_3$ as 
\be
\label{EFT_Mth_expb}
\begin{aligned}
	C_3 &= \hat B_2 \wedge \d y + \sqrt{2} \hat C_3 
	\\
	\Phi_3 &= \hat J \wedge \d y + \sqrt{8} \Re (\calc \Omega)
\end{aligned}
\ee
where the hatted-quantities have only internal legs.

In order to compute the superpotential \eqref{EFT_Mth_W}, we will explicitly need the four-form field strength $G_4$ with only internal legs
\be
\label{EFT_Mth_bg}
\begin{aligned}
	G_4|_Y  = \overline{H}_3 \wedge \d y + \sqrt{2}\overline{G}_4
\end{aligned}
\ee
where $\overline{H}_3 $ and $\overline{F}_4 $ are the background fluxes, as well as $G_7|_Y$ decomposed as
\be
\label{EFT_Mth_bg2}
\begin{aligned}
	G_7|_Y = \overline{G}_6 \wedge \d y
\end{aligned}
\ee
where $\overline{F}_6$ is an internal six-form flux. As in Type IIA, we now define the complexified forms
\be
J_c = \hat B_2 + \ii J\,,\qquad \Omega_c = \hat C_3 + 2\ii \Re (\calc\Omega)\,.
\ee
Therefore, we immediately recognize that the chiral coordinates \eqref{EFT_Mth_Si} emerging from M-theory
\be
\check{S}^{\check\imath} \check\rho_{\check\imath} = J_c \wedge \d y + \sqrt{2} \Omega_c 
\ee
do actually include at once both the K\"ahler and the complex structure moduli, which constituted the scalar spectrum considered in Section \ref{sec:EFT_IIA_10d} (see Table~\ref{tab:EFT_IIA_mod}).

Then, the superpotential becomes
\be
\label{EFT_Mth_WIIA}
\calw_{G_2} = U \int_X (\overline{G}_6+ J_c \wedge \overline{G}_4 + \Omega_c \wedge \overline{H}_3)\,.
\ee
Expanding now over an internal basis of $X$, we get
\be
\calw_{G_2} = U(e_0+e_i \Phi^i + h^\Lambda T_\Lambda)
\ee
where the normalization of the background fluxes $e_i$ and $h^\Lambda$ is chosen so as to match with \eqref{EFT_IIA_calwRRsw}-\eqref{EFT_IIA_calwNS}.

We get a four-dimensional theory, which is consistent with those obtained from Type IIA compactifications. This is indeed a subcase of the superpotentials appearing in Type IIA \eqref{EFT_IIA_calwRRsw}-\eqref{EFT_IIA_calwNS}. By consistency with the strongly coupled regime, the Romans mass $m^0$ has necessarily  to be zero \cite{Aharony:2010af}. Here also $m^i = 0$, but we will see in the next section how to reintroduce them. Hence, the full superpotential which come from compactifying on $G_2$-holonomy seven-manifolds, translated in terms of moduli for $X$,  is the same as \eqref{EFT_IIA_calw}, provided that in the latter we set $\hat \calw =0$ and $m^a = 0$. The three-form theory \eqref{EFT_Mth_4Sa} indeed can be retrieved from the Type IIA case \eqref{EFT_IIA_SLeg} employing these assumptions.

\subsection{Relaxing the $G_2$--holonomy assumption}
\label{sec:EFT_Mth_G2}

The assumption of $G_2$-holonomy of the internal seven-manifold allows only for the expansion \eqref{EFT_Mth_Cex}-\eqref{EFT_Mth_Phiex} of the three-forms $C_3$ and $\Phi_3$. This, in turn, influences the structure of the superpotential \eqref{EFT_Mth_W} generated by the background fluxes. From the Type IIA perpspective, the superpotential \eqref{EFT_Mth_Wb} cannot contain two-form background fluxes, as the $m^i$ appearing in \eqref{EFT_IIA_calwgf}. However, we may relax the $G_2$--holonomy request, so that the three-form $\Phi_3$ is not closed anymore, that is $\d_Y \Phi_3 \neq 0$. Then, the expansions \eqref{EFT_Mth_Cex} and \eqref{EFT_Mth_Phiex} are performed over a basis which is not harmonic, that is, in particular, $\d_X \check\rho^{\check\imath} \neq 0$.

The generalization of the superpotential \eqref{EFT_Mth_Wa} is 
\cite{DallAgata:2003txk,House:2004pm,Micu:2006ey,Derendinger:2014wwa,Andriolo:2018yrz}
\be
\label{Mth_Waltd}
\begin{split}
	\calw_{G_2} &=U \int_X  \frac12 (C_3 + \ii \Phi_3) \wedge \d (C_3 + \ii \Phi_3) 
	\\
	&= U \int_X \left[\overline{G}_7+ (\tilde C_3 + \ii \Phi_3) \wedge \overline{G}_4 + \frac12 (\tilde C_3 + \ii \Phi_3) \wedge \d (\tilde C_3 + \ii \Phi_3) \right]
	\\
	&=U   \left(n_0 + n_{\check\imath} \check{S}^{\check\imath} +  \frac12 m_{\check\imath\check\jmath} \check{S}^{\check\imath} \check{S}^{\check\jmath}\right)\,,
\end{split}
\ee
where we have defined the symmetric matrix
\be
m_{\check\imath\check\jmath} \equiv \int_X \check\rho_{\check\imath} \wedge \d \check\rho_{\check\jmath}\;,
\ee
whose entries play the role of quantized fluxes. The Type IIA superpotential \eqref{EFT_IIA_calwgf} can now be recovered from the compactification over $G_2$-structure manifolds. The expansions \eqref{EFT_Mth_expb} of $C_3$ and $\Phi$ read
\be
\label{Mth_expIIAc}
\begin{aligned}
	C_3 &= \hat B_2 \wedge \sigma + \sqrt{2} \hat C_3 + [\overline{B}_2 \wedge \d y + \sqrt{2} \overline{C}_3]
	\\
	\Phi_3 &= \hat J \wedge \sigma + \sqrt{8} \Re (C \Omega)
\end{aligned}
\ee
where $\sigma = \d y + \overline{A}_1$ and $\d \sigma = \overline{F}_2 = m^i \omega_i$ represents the failure of the closure of $\Phi$. Hence,
\be
G_4|_Y = \d_Y C_3 = \hat B_2 \wedge \overline{F}_2 + \overline{H}_3 \wedge \d y + \sqrt{2} \overline{F}_4
\ee
Substituted in the first line of \eqref{Mth_Waltd} and integrating over the $S^1$ gives
\be
\label{Mth_WIIAc}
\calw_{G_2} = U \int_X \left(\overline{F}_6+ J_c \wedge \overline{F}_4 + \frac12 J_c \wedge J_c \wedge \overline{F}_2 + \Omega_c \wedge \overline{H}_3\right)\,.
\ee
A final expansion over a over an internal basis of $X$ gives
\be
\calw_{G_2} =U \left(e_0-e_i \Phi^i + \frac12 \kappa_{ijk} m^i \Phi^j \Phi^k + h^\Lambda T_\Lambda \right)
\ee
The choice of which fluxes to generate dynamically then proceeds as in Section \eqref{sec:EFT_IIA_4d}, with the simplification $m^0=0$ and no tadpole cancellation condition. The bosonic theory is the same as the one obtained in \eqref{EFT_IIA_SLeg} for vanishing Romans mass: this, in particular, implies that in \eqref{EFT_IIA_SLeg} the gauge three-forms are \emph{not} gauged, with no gauge four-form appearing because no tadpole condition has to be enforced.

The four-dimensional bosonic action in then again as \eqref{EFT_IIA_SLeg}, but now no gauge three-form associated to $m^0$ appears: namely, in \eqref{EFT_IIA_SLeg} we should set $\tilde A_{30} =0$.

\section{The F-theory uplift}
\label{sec:EFT_Fth}

Before concluding this chapter, let us briefly comment on the possible uplift of the previous constructions to the F-theory case. In Section~\ref{sec:EFT_IIB} we have seen how, in  a weak-coupling regime with only O3-planes and D3-branes, we may choose an isotropic sublattice $\Gamma_{\rm EFT}\subset \Gamma$. In more general flux compactifications, the choice of $\Gamma_{\rm EFT}$ is less obvious. This is indeed what happens if D7-branes wrapping holomorphic surfaces are present \cite{Gomis:2005wc,luca1} and we include their world-volume fluxes into $\Gamma$, or we go to a strongly coupled F-theory regime. In order to illustrate this point let us consider an F-theory compactification on a smooth elliptically fibred  Calabi-Yau four-fold $Y$ -- see \cite{Denef:2008wq,Weigand:2010wm,Weigand:2018rez} for reviews. We can then identify $\Gamma$ with the lattice of `transversal' fluxes $G_4 \in H^4(Y;\mathbb{Z})_{\rm T}$ \cite{Dasgupta:1999ss},\footnote{For smooth elliptically fibered Calabi-Yau four-folds there is no half-integral correction to the flux quantization \cite{Collinucci:2010gz}.} that is, whose Poincar\'e dual 4-cycle has vanishing intersection number with any pair of divisors of $Y$. By introducing an appropriate basis of transversal cocycles  $\alpha^\cala \in H^4(Y;\mathbb{Z})_{\rm T}$, we can expand 
\be
G_4=\caln_\cala\,\alpha^\cala\, ,
\ee
and introduce the symmetric pairing
\be\label{Fpairing}
\cali^{\cala\calb}\equiv \int_Y\alpha^\cala\wedge\alpha^\calb\, .
\ee
The D3-charge tadpole condition then takes the form \eqref{EFT_IIB_4dTad}, with the additional contribution $-\frac1{24}\chi(Y)$ to $\calq_{\rm bg}$, where $\chi(Y)$ is the Euler characteristic of $Y$. Furthermore, the flux-induced superpotential is given by
\be\label{Ftsup}
\calw(z)=\caln_\cala \calv^\cala(z) \quad~~~~\text{with}\quad~~~~ \calv^\cala(z) \equiv \frac{\pi}{\ell^6_{\rm M}}\int_Y \Omega_4\wedge \alpha^\cala\, ,
\ee
where  $\Omega_4$ is the (dimensionless) holomorphic (4,0)-form on $Y$ and $\ell^6_{\rm M}$ is the M-theory Planck length. The chiral fields $z^a$ ($a=0,\ldots,h^{3,1}$) parametrize $Y$-complex structure moduli and include the super-Weyl compensator.  

Let us now pick an isotropic sublattice $\Gamma_{\rm EFT}\subset \Gamma$. By splitting $\caln_\cala$ as in \eqref{Intro_Jumps_NA}, and defining $\calv^A(z)\equiv v^A_\cala\calv^\cala(z)$, we can write the superpotential \eqref{Ftsup} in the form
\be\label{Ftsupb}
\calw(z)=N_A\,\calv^A(z)+\hat\calw(z)\quad~~~~\text{with}\quad~~~~ \hat\calw(z)\equiv \caln^{\rm bg}_\cala\,\calv^\cala(z)\,.
\ee
By making an Hodge decomposition of $H^4(Y;\mathbb{C})_{\rm T}$, one realises that the pairing \eqref{Fpairing} has signature $(2+h^{2,2}_{\rm T}, 2h^{3,1})$. Then the dimension of any maximal isotropic sublattice $\Gamma_{\rm EFT} \subset \Gamma$  is {\em at most} $2h^{3,1}$. This means there are {\em always} enough complex scalars $z^a$ to allow for a supersymmetric dualization of the flux quanta $N_A$ spanning $\Gamma_{\rm EFT}$ to three-form potentials $A^A_3$. 
\footnote{For a general Calabi-Yau four-fold, the pairing \eqref{Fpairing} has signature $(b^+_4, b_4^-)$, with $b^+_4-b^-_4=chi/3 + 32$,  If $h^{2,1}=0$, from the }

As a concrete example, leaving a more detailed discussion of other settings to the future, we can consider the model spelled out in  \cite{Collinucci:2008pf},  for which $\dim \Gamma=23320$, $h^{3,1}=3878$, $h^{2,2}_{\rm T}=h^{2,2}-2=15562$. In  this case, following our general prescription, one would obtain an EFT with $\dim \Gamma_{\rm EFT}=7756$ three-form potentials $A^A_3$ and $\dim (\Gamma/\Gamma_{\rm EFT})=15564$ non-dynamical fluxes. The $7756$ three-form potentials may be accommodated, together with the Weyl compensator and $3877=h^{3,1}-1$ of the $Y$-complex structure moduli,  into  $3878$ double three-form multiplets. Notice that there does not appear any `natural' choice of the sublattice $\Gamma_{\rm EFT}$ and of the  $Y$-complex structure modulus which is excluded from these double three-form multiplets. Analogously, differently from what we observed in the previous subsection, there is no obvious  hierarchy between the energy scales of the potential and the membrane tensions.

It is instructive to observe how such `democracy' is removed by going to weak coupling.  In this limit, the total amount of fluxes  $23320$ splits into $2\times 300$ bulk R-R plus NS-NS three-form fluxes and $22720$ D7-brane fluxes, while the moduli $z^a$ include the overall Weyl compensator, the axio-dilaton $\tau$, $149$ bulk complex-structure moduli, and $3728$ D7-brane geometric moduli. The $149$ complex-structure moduli and the Weyl compensator can combine with 300 R-R bulk fluxes into double three-form multiplets, as we did in the previous subsections. The axio-dilaton $\tau$ can again be dualized to a linear multiplet which is gauged, with charge defined by the $300$ NS-NS fluxes. The  $7756-300=7456$ D7-brane fluxes in $\Gamma_{\rm EFT}$, can  then be  accommodated, together with the   D7-brane moduli,  into other 3728 double three-form multiplets. The remaining   15264 D7-brane fluxes stand non-dynamical and contribute to $\hat\calw$ in  \eqref{Ftsupb}.

Leaving a more detailed study of the three-form formulation of the EFT for general F-theory compactifications -- as well as their weak-coupling limits --  to the future, we close this section by briefly discussing the microscopic origin of the three-form gauging 
\be\label{F3gauge}
A^A_3\rightarrow A^A_3-Q^A\Lambda_3\quad~~~~\text{with}\quad~~~~ Q^A=\cali^{\cala\calb}v_{\cala}^A\caln_\calb^{\rm bg}\, .
\ee
This can be understood from the perspective of the dual  M-theory compactified to three dimensions on an elliptically fibred Calabi-Yau four-fold $Y$ and the derivation is similar to the weakly-coupled IIB case discussed in Section~\ref{sec:EFT_IIB_4d}. One can start from the eleven-dimensional Bianchi identity
\be
\d F_7-\frac12 F_4\wedge F_4=(\text{M2-brane charge})\, .
\ee
By appropriately reducing  $F_7$, $F_4$ and the corresponding potentials to three dimensions one gets a set of gauge-invariant three-forms which are dual to the F-theory field-strengths $\hat F^A_4=\d A^A_3+Q^A C_4$ .  These are gauge invariant under $C_4\rightarrow C_4+\d\Lambda_3$ provided that $A^A$ transform as in \eqref{F3gauge}. In the weak-coupling limit, this generalizes the gauging of three-forms discussed in the previous subsections by including a sector supported on the D7-branes.


\chapter{Consistency conditions of EFTs and the {spectrum} of BPS--objects }
\label{chapter:LandSwamp}

The previous chapter led us to a full reformulation of the four-dimensional effective theories originating from string and M-theory. The introduction of the gauge three-forms into the EFTs allowed us to give a dynamical interpretation of the appearance of the background fluxes. However, not all of the background fluxes that enter the four-dimensional EFT can be generically dualized to gauge three-forms. The tadpole cancellation conditions and the Freed-Witten anomalies force us to treat some of the fluxes as frozen: these fluxes in fact identify the gauging parameters for the gauge two- and three-forms, which cannot be rendered dynamical. Accordingly, in the previous chapter, in both Type IIA and Type IIB cases, we \emph{arbitrarily} chose which fluxes can be dualized and those which may not.

We will now fill the gap of the previous chapter and provide some physical arguments to enforce the choices of dynamical sublattices of Sections~\ref{sec:EFT_IIB} and~\ref{sec:EFT_IIA}. These arguments rely on the amount and type of extended objects, either strings, membranes or 3-branes, which can be included into the effective field theory. The spectrum of objects is indeed strictly tightened to cut-off of the EFT as well as on the masses of the scalar fields and how these change across the spacetime: namely, crossing membranes and encircling strings induce changes in the background fluxes and axions, leading to a change in the mass spectrum. Further developing the ideas settled in the introductory Section~\ref{sec:Intro_Swamp}, we shall see how forbidding some transitions, basically by requiring the full consistency of the effective description, naturally leads to the exclusion of some objects in the EFT. 

The picture that will emerge at the end of the chapter convey an EFT whose spectrum of objects is strictly fixed by the perturbative regimes that the EFT is supposed to scan. The inclusion of other objects in the theory, as well as an improper dualization of fluxes to gauge three-forms or of axions to two-forms, may lead the EFT to the \emph{Swampland} of inconsistent EFTs.
\section{Extended objects from higher dimensions}
\label{sec:LandSwamp_ExtObj_HD}

In Section~\ref{sec:Sugra_ExtObj} we have introduced supersymmetric 3-branes, membranes and strings. Writing down a supersymmetric action for them required to enforce $\kappa$-symmetry, which strictly tightens together  the tensions for these objects to their charges. From a different viewpoint, in ten-dimensional string theory $p$-branes can be coupled, understood as hypersurfaces spanning a $(p+1)$--dimensional volume. Once we reduce the ten-dimensional theory down to 4D, a plethora of objects stem from such $p$-branes and we end up with a four-dimensional theory which naturally include a hierarchy of BPS--objects, as those of Section~\ref{sec:Sugra_ExtObj}. 

We will briefly recall how the reduction can be performed, showing that the tension of the reduced objects naturally depends on the moduli of the compactification. Our aim is to get a grasp of how the BPS--objects of Section~\ref{sec:Sugra_ExtObj} may be generated from string theory and we will be very schematic, referring to \cite{Ibanez:2012zz,Font:2019cxq} and references therein for a more detailed discussion.

In ten dimensions and in the string frame\footnote{We recall that, in the string frame, the Einstein-Hilbert term of the Type II effective actions is normalized as \cite{Ibanez:2012zz}
\be
\label{SL_EO_EH10D}
S_{\rm EH, II}^{(10)} = \frac{1}{2\kappa^2_{10}} \int e^{-2\phi} R *1 
\ee
with $\phi$ the ten-dimensional dilaton such that $g_s = e^\phi$ and $\kappa^2_{10} = \ell_{\rm s}^8/ (4\pi) $, where $\ell_{\rm s} = 2\pi \sqrt{\alpha'}$ (with respect to \cite{Ibanez:2012zz}, our $\ell_{\rm s}^{\rm here} = 2\pi l_s^{\rm there}$). We also recall that reducing \eqref{SL_EO_EH10D} to four dimensions, we get the following identification for the four-dimensional Planck mass
\be
\label{SL_EO_MP}
M_{\rm P} = \frac{\sqrt{4 \pi  V_6}}{\ell_{\rm s} g_s}\,,
\ee
where $V_6$ is the volume of $X$ as computed in the string frame -- see \eqref{SL_EO_Vp-2}.},the action of a D$p$-brane stretching along the cycle $(p+1)$-dimensional cycle $\Sigma_{p+1}$ is given by
\be
\label{SL_EO_Sp10D}
S_{\text{D}p}  = S_{\text{D}p, \rm DBI}  + S_{\text{D}p, \rm CS} \,.
\ee
The first, Dirac-Born-Infeld term is
\be
\label{SL_EO_SpNG10D}
S_{\text{D}p, \rm DBI}  = -\mu_p \int_{\Sigma_{p+1}} \d^{p+1}\sigma e^{-\phi} \sqrt{-\det\left(P[g_{MN}+B_{MN}]\right) }\,,
\ee
where $g_{MN}$ and $B_{MN}$ are, respectively, the ten-dimensional metric and the components of the NS-NS two-form, with $P$ denoting their pullback to the brane worldvolume $\Sigma_{p+1}$. Additionally, here and below we have neglected possible contributions from the worldvolume gauge field. This term encodes the kinetic terms of the $p$-brane and $\mu_p$ denotes its tension which, in terms of the string length $\ell_{\rm s}$, is given by
\be
\mu_p = \frac{2\pi}{\ell_{\rm s}^{p+1}}\,.
\ee
The second, Chern-Simons term in \eqref{SL_EO_Sp10D} expresses the minimal coupling of the $p$-brane to a gauge $p$-form $C_{p+1}$ 
\be
\label{SL_EO_SpCS10D}
S_{\text{D}p, \rm CS}  = \mu_p \int_{\Sigma_{p+1}} P[C_{p+1}e^{-B_2}]\,,
\ee
where, again, $P$ denotes the pull-back to the brane worldvolume. The sign and charge $\mu_p$ in \eqref{SL_EO_SpCS10D} are fixed by requiring that the full action \eqref{SL_EO_Sp10D} preserves supersymmetry. Namely, it has to be $\kappa$-symmetric so that, when \eqref{SL_EO_Sp10D}  is evaluated at its ground state, only half of the supersymmetry generators are spontaneously broken, making the D$p$-branes $1/2$--BPS objects, in analogy with our four-dimensional membrane counterparts \eqref{Sugra_ExtObj_Memb_Sgf}.  We also recall that the action of the NS5-branes is the same as \eqref{SL_EO_Sp10D}, provided however that the DBI and CS-terms in \eqref{SL_EO_Sp10D} get multiplied by $1/g_{\rm s}$.

Instead, anti-D$p$-branes, denoted with $\overline{\text{D}p}$, spontaneously break all the supersymmetry: their action is the very same as \eqref{SL_EO_Sp10D}, with, however, an opposite Chern-Simons term.
Such a difference of relative sign does not allow the $\kappa$-symmetry to be realized, making it impossible to preserve even a portion of the bulk supersymmetry over the worldvolume.\footnote{We stress that the choices of sign depend on the background, namely from the solutions of the Killing spinor equations that dictate the amount of preserved supersymmetry (quite in analogy with what we did in Section~\ref{sec:LandSwamp_ExtObj_DW} for four-dimensional domain walls).}

Other objects, already introduced in the previous chapter, are O-planes: they preserve supersymmetry, albeit their DBI-term has an opposite sign with respect to \eqref{SL_EO_Sp10D}
\be
\label{SL_EO_SOp10D}
S_{{\text{O}p}}  = \mu_p \int_{\Sigma_{p+1}} \d^{p+1}\sigma e^{-\phi} \sqrt{-\det\left(P[g_{MN}]\right) } -\mu_p \int_{\Sigma_{p+1}} P[C_{p+1}]\,.
\ee
The sign of the Chern-Simons term has been also changed with respect to  \eqref{SL_EO_Sp10D} in order to ensure these objects to preserve supersymmetry.

In the next sections, we will be mostly interested in retrieving membranes and strings in four-dimensional EFTs. For simplicity, let us focus just on D$p$-branes. In order to get four-dimensional membranes, we can start with a  D$p$-brane and assume that it wraps an internal $(p-2)$-cycle $\Sigma_{p-2}$, so that it spans a three-dimensional hypersurface $\calm$ in the external space. As a simplifying assumption, we shall take the cycle $\Sigma_{p+1}$ as the product of $\Sigma_{p-2}$ and the external membrane worldvolume. The 4D tension can be easily computed by dimensionally reducing the DBI--term \eqref{SL_EO_SpNG10D} as:
\be
\label{SL_EO_SNG4D}
S_{\rm memb, NG} = - \int_{\calm}\d^3\sigma\, \calt_{\bf q}(\phi)\sqrt{-{\bf h}}\,,
\ee
with the tension
\be
\label{SL_EO_Tred}
\calt_{\bf q}(\phi) = \frac{1}{4\sqrt{\pi}} g_s^2 M_{\rm P}^3 \frac{V_{p-2} }{V_6^{3/2}} 
\ee
and we have defined the volumes in the string frame as
\be
\label{SL_EO_Vp-2}
V_{p-2} \equiv \ell_{\rm s}^{2-p} {\rm vol}\, (\Sigma_{p-2})\;.
\ee
It is however important to stress that the volumes of the cycles $ {\rm vol}\, (\Sigma_{p-2})$ are \emph{moduli dependent} and therefore we expect also the tension \eqref{SL_EO_Tred} to generically depend on the moduli of the compactification. Furthermore, if we wish the membrane to be a supersymmetric object in four dimensions, in the sense of Section~\ref{sec:Sugra_ExtObj_Memb}, one needs to properly choose the internal cycles $\Sigma_{p-2}$ (for example, in Type IIB, the three-cycles wrapped by a D5-brane need to be \emph{special Lagrangian}): this is a rather important requirement, but its discussion goes beyond the scope of this work and we refer, for example, to \cite{Martucci:2005ht,Martucci:2006ij} for detailed discussions. The relation presented above is quite general and does not rely on the bulk sector. We will however show in the examples of the next sections that the tension \eqref{SL_EO_Tred} reduces to precisely to the expression \eqref{Sugra_ExtObj_Memb_Sgf} that we obtained, in four dimensions, from arguments that were genuinely based on $\kappa$--symmetry.

The same reasoning can be applied to get BPS--strings in the four-dimensional EFT. A supersymmeytric string is obtain by wrapping a D$p$-brane over a $\Sigma_{p-1}$ internal cycle, spanning a two-dimensional spacetime surface $\cals$ in the external space. A direct reduction of \eqref{SL_EO_SpNG10D}, in the absence of worldvolume fluxes, leads to the Nambu-Goto term
\be
\label{SL_EO_SNG4Ds}
S_{\rm string, NG} = - \int_{\cals}\d^2\sigma\, \calt_{\bf q}(\phi)\sqrt{-{\bm \gamma}}\,,
\ee
with the tension
\be
\label{SL_EO_Treds}
\calt_{\bf q}(\phi) = \frac{1}{2} g_s M_{\rm P}^2 \frac{V_{p-1} }{V_6}\,.
\ee

Supersymmetric 3-branes may, in principle, be treated equivalently as membranes and strings. However, the 3-branes introduced in Section~\ref{sec:Sugra_ExtObj_3b} are quite peculiar, for the Nambu-Goto term is absent. Still, as we will see explicitly in the Type IIB case in Section~\ref{sec:LandSwamp_Hierarchy_IIB_S3b}, those 3-branes constitute a `degenerate case' of the extended objects treated here, because a mutual cancellation between the Nambu-Goto and Chern-Simons part of the reduced \eqref{SL_EO_Sp10D} will lead to an action which is the sole \eqref{Sugra_ExtObj_3b_S}.

\section{Perturbative regimes and spectrum of objects}
\label{sec:LandSwamp_Hierarchy_Gen}

A membrane allows for describing potential with different shapes in two regions of spacetime, but there are some caveats. Within the same effective theory, it is possible to pass from a potential $V(\varphi;N_A)$, depending on some quantized constants $N_A$, to the potential $V(\varphi;N_A + q_A)$, where the constants get shifted by $q_A$, if and only if the constants $N_A$ are dual to a set of gauge three-forms $A_3^A$ and there exists a membrane, with charges $q_A$ under $A_3^A$, which allows for the flux transition $N_A \to N_A+q_A$. Although generically a domain wall solution which interpolates between the vacua on the two sides may not exist, the presence of a membrane is still a necessary condition to connect the vacua of $V(\varphi;N_A)$ with those of $V(\varphi;N_A + q_A)$.

As explained in Section~\ref{sec:EFT_intro}, generically the potential depends on the fluxes $\caln_\cala$, but only a portion thereof can be properly understood as dynamical. In fact, whenever there are some constraints on the flux lattice such as \eqref{EFT_Intro_Tada}, we are forced to consider some fluxes \emph{fixed} at a background value. The fluxes which are `frozen' \emph{cannot} be subjected to transitions: in fact, they are not dualized to gauge three-forms and no membrane coupling is present in the effective theory. In this section, we elaborate more on the distinction of which fluxes are dynamical and which are not, further developing the general discussion of Section~\ref{sec:Intro_Swamp}. We will give physical arguments that corroborate the discussion of Section~\ref{sec:EFT_intro}, which rely on the spectrum of membranes that is allowed in the effective theory.

Any effective field theory is characterized by a cut-off scale $\Lambda_{\rm UV}$, below which the theory is valid. We will assume that such a cut-off is field independent. If we wish to disregard the Kaluza-Klein excitations, we need to choose 
\be
\label{SL_HG_mKK}
\Lambda_{\rm UV} \lesssim  m_{\rm KK}\,,
\ee
where $m_{\rm KK}$ is the lowest mass of the Kaluza-Klein excitations. We will always work in the assumption that \eqref{SL_HG_mKK} is satisfied, consequently identifying $m_{\rm KK}$ as the upper bound for the cut-off scale $\Lambda_{\rm UV}$. In the effective field theory, plenty of moduli are present, which we generically denote with $\chi$ and the presence of a potential may render them massive. Their mass, of generic order $m_\chi$, clearly depends on the fluxes $\caln_{\cala}$. Since we want to keep the moduli within the EFT description, we assume that they are stabilized at masses $m_\chi$ lower than the cut-off $\Lambda_{\rm UV}$:
\be
\label{SL_HG_mchi}
m_{\chi} \lesssim \Lambda_{\rm UV}   \,.
\ee
Both the Kaluza-Klein scale $m_{\rm KK}$ as well the moduli masses $m_\chi$ are moduli dependent. Therefore, the combined conditions \eqref{SL_HG_mKK}  and \eqref{SL_HG_mchi} identifies a region within the moduli space in which the effective field theory is valid, for a given choice of the cutoff $\Lambda_{\rm UV}$.

The cut-off $\Lambda_{\rm UV}$ determines, in turn, which objects we are legitimate to include within the effective theory. Let us first consider membranes. We would like to treat the membrane semi-classically, neglecting their quantum fluctuations. Indeed, the quantum spectrum of a supersymmetric membrane is continuous, with no mass gap \cite{deWit:1988xki}: including the quantum excitations of membranes would then introduce an infinite tower of fields. In order to avoid such a tower, it is necessary to assume that the tension of the membrane is greater that the cut-off $\Lambda_{\rm UV}$ in the following sense:
\begin{important}
\be
\label{SL_HG_SCmemb}
\frac{\calt_{\rm memb}}{\Lambda^3_{\rm UV}}\gtrsim 1\qquad{\color{darkred}\text{\sf{Semi-classical membranes requirement}}}
\ee
\end{important}

Let us now assume that on the two sides of the membrane we can select two different vacua, with the moduli acquiring that may acquire masses of order $m_\chi$ depending on the background fluxes. We can now split the fluxes as in \eqref{Intro_Jumps_NA}, namely into a set $N_A$ which is dynamical and into another $\caln^{\rm bg}_\cala$ regarded as frozen. A hierarchy of masses is originated, which is fixed by both the sectors. A membrane can indeed induce transitions between a vacuum on the left to another on the right and let us assume that a domain wall solution, interpolating between them, exists. In short, we are assuming that it is legitimate to consider transitions between the vacua. Across the flow which leads to a change of the vevs of the fields, the masses $m_\chi$, which do depend on the particular point of the moduli space and thus on the background fluxes, generically change. We now require that \emph{the allowed transitions are those which do not significantly change the hierarchy of masses}. In other words, we assume that the hierarchy is basically fixed by the frozen sector $\caln^{\rm bg}_\cala$ and that any change in the dynamical sector $N_A$ has not enough energy to completely re-shuffle the masses $m_\chi$.

Furthermore, the energy involved in transitions across the membrane can roughly be estimated from the difference of the potential at the vacua as
\be
|\Delta V|^2 \simeq  \Mp^2 \left||\calw_{+\infty}|^2  - |\calw_{-\infty}|^2 \right| \simeq  \calt^2_{\rm memb} \Mp^{-4} \,.
\ee
%
%
%
Therefore, if a transition is allowed, the energy involved has to be lower than the cut-off of the theory
\begin{important}
\be
\label{SL_HG_mLcond}
\frac{\calt_{\rm memb}}{M^2_{\rm P}} \lesssim \Lambda_{\rm UV} \qquad{\color{darkred}\text{\sf{Compatibility with UV cut-off}}}
\ee
\end{important}
This relation becomes indeed the criterion that has to be respected by the split \eqref{Intro_Jumps_NA}. Namely, the membranes that we can include in the four-dimensional EFT must obey \eqref{SL_HG_mLcond} and the allowed three-forms are those which couple to such membranes. This indeed determines which fluxes \emph{cannot} have a three-form origin and have to be regarded as $\caln_{\cala}^{\rm bg}$, which determine the gaugings induced by the tadpole cancellation conditions. 

It is also important to stress another important feature of the relation \eqref{SL_HG_mLcond}. All the quantities that appear in \eqref{SL_HG_mLcond} are moduli dependent, making \eqref{SL_HG_mLcond} be related to the particular region of the moduli space that we are exploring. The relation \eqref{SL_HG_mLcond} is, in particular, related to the coupling constants of the theory, leading to the following interpretation: given a certain perturbative regime that we want to scan within the EFT, \eqref{SL_HG_mLcond}  tells what membranes can be included into the effective theory. As an example, in the EFTs originating from string theory, \eqref{SL_HG_mLcond} depends on the string coupling constant $g_{\rm s}$ and, as we shall see in the next two sections, choosing whether we want to explore the regions $g_{\rm s} \lesssim  1$, $g_{\rm s} \sim 1$ or $g_{\rm s} \gtrsim  1$ different kind of membranes may be included.

\begin{figure}[t]
	\centering
	\includegraphics[width=13cm]{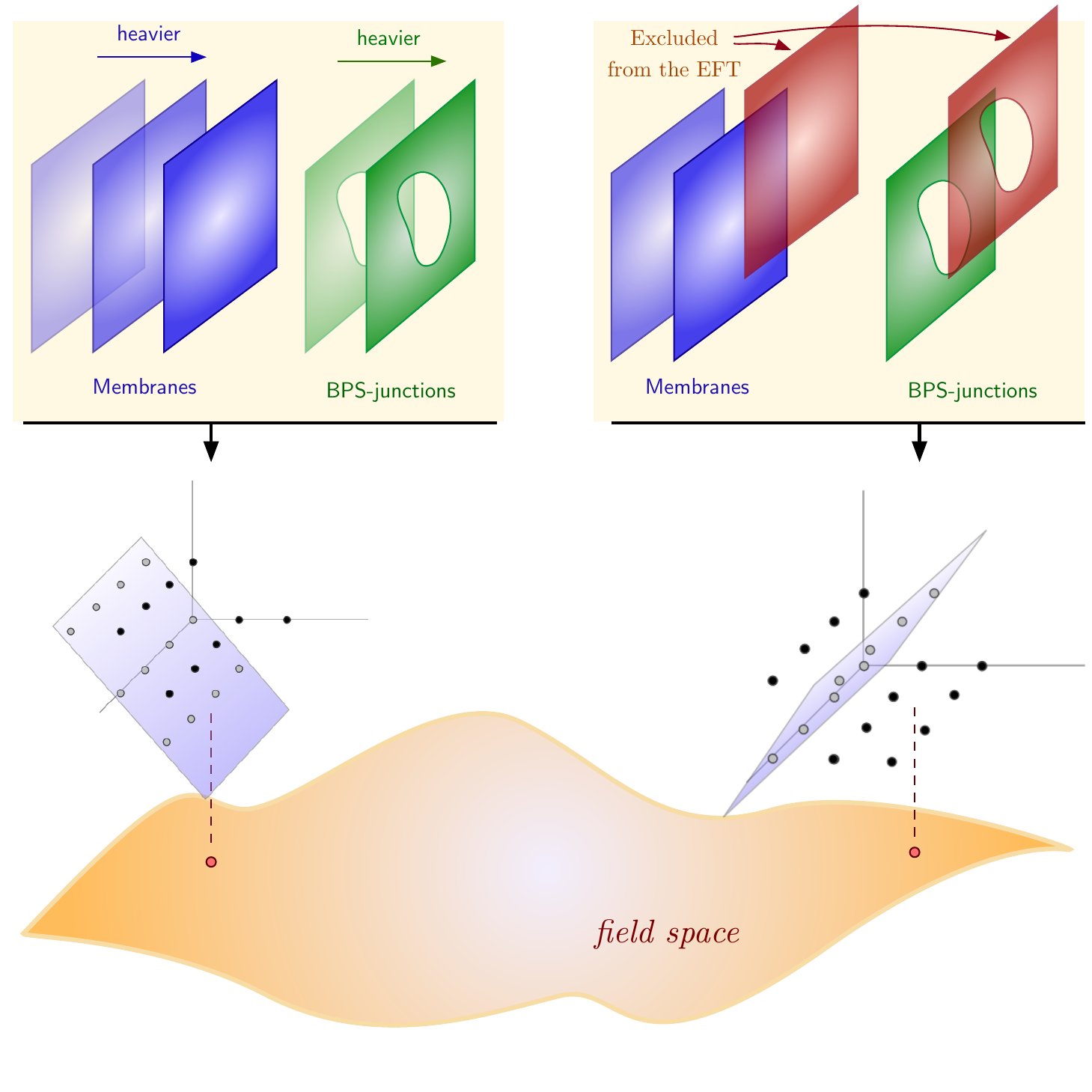}
	\caption{\footnotesize{At different points in the field space, the hierarchy of objects change: some membranes and strings may become lighter than others. Accordingly, the dynamical sublattice $\Gamma_{\rm EFT}$ changes. An EFT is then characterized differently according to the perturbative regime that we want to scan.} }\label{fig:Intro_EFTLattb}
\end{figure}   

As of now, our discussion has been focused on membranes only, but can be also extended to BPS--strings, with some modifications and additional notes. As we noticed for membranes, if we wish to treat strings as semi-classical objects, their tensions has to satisfy
\begin{important}
\be
\label{SL_HG_SCstr}
\frac{\calt_{\rm string}}{\Lambda^2_{\rm UV}}\gtrsim 1  \qquad{\color{darkred}\text{\sf{Semi-classical strings requirement}}}
\ee
\end{important}
In an EFT where also semiclassical membranes are present, this criterion combines with \eqref{SL_HG_SCmemb}. However, strings develop the so--called Freed-Witten anomalies as briefly summarized in Appendix~\ref{app:FW_AM}. These are cured by attaching, in the effective description, a certain amount of membranes to the string, creating a BPS--junction as that examined in Section~\ref{sec:Sugra_ExtObj_MembString}. An anomaly--free theory has then to consistently include the whole BPS--junction. In terms of the tension of the string, in analogy with \eqref{SL_HG_mLcond}, we require
\begin{important}
	\be
	\label{SL_HG_SLstr}
	\frac{\calt_{\rm string}}{M^2_{\rm P}} \lesssim 1 \, \qquad{\color{darkred}\text{\sf{Compatibility with Freed-Witten anomaly cancellations}}}
	\ee
\end{important}
However, in contrast with \eqref{SL_HG_mLcond}, no $\Lambda_{\rm UV}$ appears. This can be traced back to the fact that the backreaction of the strings grow only logarithmically with the distance and is negligible with respect to that of membranes, which grows linearly. In Type IIB case, we will deliver an example of cancellation of Freed-Witten anomalies with BPS--objects, relating the criteria \eqref{SL_HG_SCstr}, \eqref{SL_HG_SLstr}  for strings, \eqref{SL_HG_SCmemb}, \eqref{SL_HG_mLcond} for membranes with the consistency of the EFT.


To summarize, the region of the moduli space that we would like to explore selects the perturbative regime of the EFT. Within such region, the compatibility of the EFT dictates which elements, namely field representations and extended objects, may be consistently included. As a result, depicted in Fig.~\ref{fig:Intro_EFTLattb}, the effective descriptions change moving across the field space.

\section{A hierarchy of objects in Type IIB EFTs}
\label{sec:LandSwamp_Hierarchy_IIB}

In Section~\ref{sec:EFT_IIB_4d} we arbitrarily regarded the R-R fluxes $m_A$ as dynamical, by properly generating them by using a set of $b_3^-(X)$ gauge three-forms, and the NS-NS fluxes $h_A$ as frozen, entering the gauging of both gauge two- and three-forms. There, such splitting seemed rather arbitrary, but we will now justify it in light of the general discussion of Section~\ref{sec:LandSwamp_Hierarchy_Gen}. The first step is the introduction of a proper cut-off $\Lambda_{\rm UV}$, which dictates what kinds of membranes and string may be included in the EFT.

\subsection{Identification of the EFT scales}
\label{sec:LandSwamp_Hierarchy_IIB_scales}

Preliminarily, we may characterize the EFT by introducing a moduli-independent UV cut-off scale $\Lambda_{\rm UV}$, which has however to be related to other mass scales entering the EFT. There are two natural scales that may enter the effective theory: the Kaluza-Klein mass scale $m_{\rm KK}$ and the masses of the moduli $m_\chi$, which are induced by the background fluxes.  On the one hand, the Kaluza-Klein scale $m_{\rm KK}$ sets an upper bound on the cut-off, so that we are legitimate to discard all the Kaluza-Klein excitation modes from the effective theory; on the other, we may assume that the induced masses $m_\chi$ for the moduli lie below the cut-off $\Lambda_{\rm UV}$. In summary, we choose the cut-off in such a way that
\be 
\label{SL_IIB_mphiKK}
m_\chi \lesssim \Lambda_{\rm UV}\lesssim m_{\rm KK}\,.
\ee

\begin{figure}[!th]
	\centering
	\includegraphics[width=7cm]{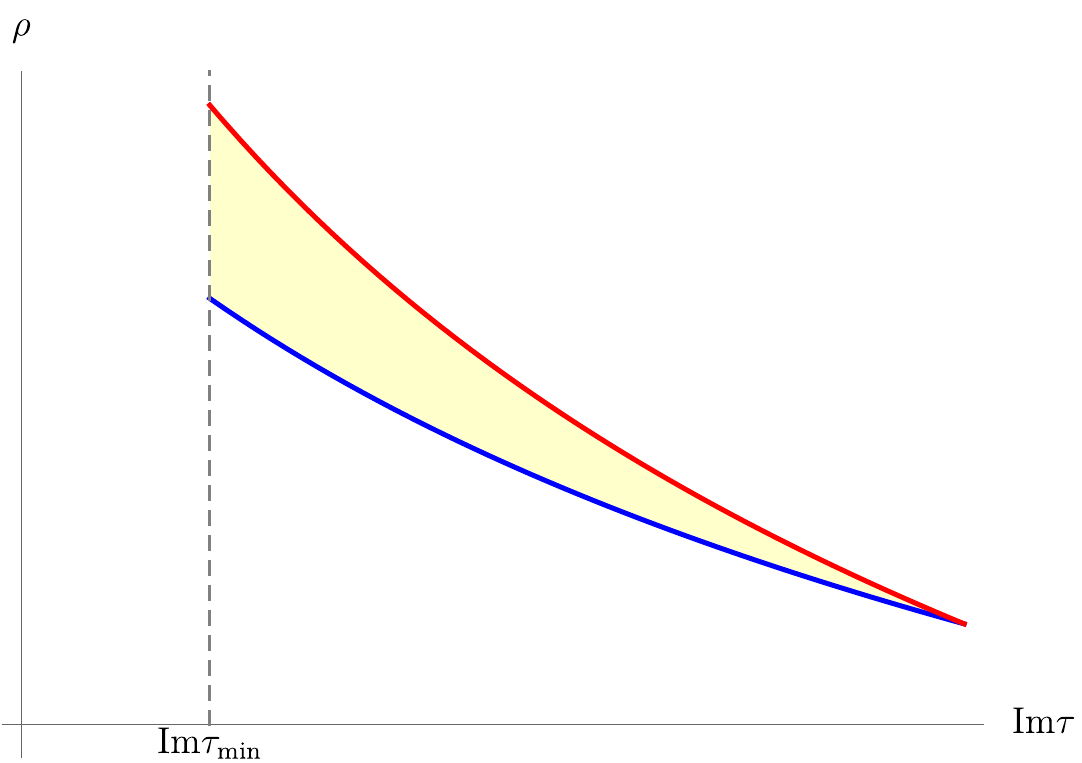}
	\caption{\footnotesize{ Region of validity of the EFT specified by \eqref{SL_IIB_taurange}.} }\label{fig:SL_RegVt}
\end{figure}

The relation \eqref{SL_IIB_mphiKK} indeed specifies a region inside the moduli space as follows. The Kaluza-Klein mass $m_{\rm KK}$ can be easily computed from \eqref{EFT_IIB_S10d}, obtaining
\be\label{SL_IIB_LKK}
m_{\rm KK}\simeq \frac{1}{\rho^2\,\Im \tau }\,M_{\rm P}\, ,
\ee 
where, for the sake of clarity, we have dropped all the $2\pi$s and $\calo(1)$ factors. We have also introduced the volume modulus
\be\label{SL_IIB_rho}
\rho\equiv V^{\frac13}_{\rm s}\, ,
\ee 
with $V_{\rm s}$ the string-frame volume of the internal space, measured in string units. In order to get an estimate over the moduli mass scale $m_\chi$, we focus on the induced mass $m_\phi$ of the four-dimensional dilaton $\phi$. Then, from the usual Cremmer et al. potential \eqref{Sugra_SW_Cremmeretal}, with the superpotential \eqref{EFT_IIB_calwgf}, one can obtain the following estimation for the dilaton mass:
\be
\label{SL_IIB_Lflux}
m_\phi \simeq \frac{|h|\|\Pi(\phi)\|}{\rho^3\,\Im\tau}\,M_{\rm P}\,,
\ee 
where $|h|$ represents the typical $h_A$ flux quanta and  $\left\| \Pi(\phi)\right\|^2$ schematically  denotes  a contribution of the form $e^{K_{\rm cs}}K_{\rm cs}^{i\bar\jmath}\Pi^A_i\Pi^B_j$, $e^{K_{\rm cs}}\Pi^A\Pi^B$, \ldots We  assume that these terms are all of the same finite order. We also employed the relations
\be
\label{SL_IIB_eK}
e^{K_{\rm k}} =  \frac{e^{4\phi}}{V_6^2} \,,\qquad e^{K_{\rm cs}} = \frac{1}{8 V_6} \,,
\ee
as can be deduced from \eqref{EFT_IIB_Kcs} and \eqref{EFT_IIB_Kk}. Here we have introduced  the internal volume $V_6$ as measured in the string units, that is $V_6 \equiv \frac{1}{3!} \int_X J^3$.Therefore the requirement \eqref{SL_IIB_mphiKK} translates into the following EFT condition on $\Im\tau$ and $\rho$: 
\be
\label{SL_IIB_taurange}
\rho^2\,\Im \tau\ \lesssim\  \frac{M_{\rm P}}{\Lambda_{\rm UV}}\  \lesssim \ \frac{ \rho^3\,\Im \tau}{|h|\|\Pi\|}    \, ,
\ee
which specifies the region depicted in Fig.~\ref{fig:SL_RegVt}. 

From these conditions we obtain the following minimal and maximal values of $\rho$ and $\Im\tau$ respectively, allowed by the EFT bounds \eqref{SL_IIB_LKK}
\be
\rho_{\rm min}\simeq |h|\|\Pi\|\ ,\ \quad \Im\tau_{\rm max}\simeq \frac{1}{|h|\|\Pi\|}\left(\frac{M_{\rm P}}{\Lambda_{\rm UV}}\right)\, .
\ee
Hence, if the combination $|h|\|\Pi\|$ entering the estimate of $m_\phi$ in \eqref{SL_IIB_LKK} is moderately large, the conditions \eqref{SL_IIB_taurange} guarantee the geometric regime. Furthermore, by taking $\Lambda_{\rm UV}\ll M_{\rm P}$,  $\Im\tau$ can reach very large values. On the other hand, we will take 
\be
\label{SL_IIB_taurange2}
\Im\tau\geq \Im\tau_{\rm min}\,,
\ee
for any $\Im\tau_{\rm min}$ which should be large but much smaller than $M_{\rm P}/\Lambda_{\rm UV}$, so that
\be
\rho^2_{\rm max}=\frac{1}{\Im\tau_{\rm min}} \left(\frac{M_{\rm P}}{\Lambda_{\rm UV}}\right)
\ee
is large too.

\subsection{A hierarchy of membranes}
\label{sec:LandSwamp_Hierarchy_IIB_M}

Let us now pass to the dual formulation in terms of three-forms. For the moment, let us assume that we can dualize both the R-R and NS-NS fluxes to gauge three-forms. Then, accordingly, we should couple all the possible membranes which make both the kinds of fluxes jump. 

Indeed, in Type IIB EFTs, four-dimensional membranes may have only two origins: they can come from either D5-branes or NS5-branes wrapped over internal special Lagrangian three-cycles $\Sigma_A$; we will call the membranes so obtained as `R-R membranes' and `NS-NS membranes', respectively.  The former, carrying charges $q^{\text{R-R}}_A$ under the gauge three-forms, provide the jumps of the R-R fluxes as $\Delta m_A =q^{\text{R-R}}_A$; the latter do instead make the NS-NS fluxes jumps as $\Delta h_A =q^{\text{NS-NS}}_A$, with $q^{\text{NS-NS}}_A$ denoting their charges. Their tension can be obtained from the general reduction formula \eqref{SL_EO_Tred}, with the volume of the internal three-cycle $\Sigma^\Lambda$ given, schematically, by
\be
{\rm vol} (\Sigma^\Lambda) \sim  q_A \left|\int_{\Sigma^\Lambda} \Omega \right| \sim  |q_A \calv^A(\varphi)|
\ee
where $q_A$ are some quantized charges, either $q^{\text{R-R}}_A$ or $q^{\text{NS-NS}}_A$. Then, the tensions of R-R and NS-NS membranes are
\be
\calt_{\rm memb}^{\text{R-R}}\simeq  \Mp^3 e^{\frac K2} |m_A \Pi^A|\,, \quad \quad \calt_{\rm memb}^{\text{NS-NS}}\simeq  \Mp^3 \frac{e^{\frac K2}}{g_{\rm s}} |h_A \Pi^A|\,,
\ee
coherently with the four-dimensional counterpart \eqref{Sugra_ExtObj_Memb_Tmgf} predicted form $\kappa$-symmetry. We get the following estimations of the tensions of the R-R and NS-NS membranes wrapped over the very same cycle $\Sigma_A$
\be
\label{SL_IIB_caltcond0}
\frac{\calt_{\rm memb}^{\text{R-R}}}{M^2_{\rm P}}\simeq  \frac{|q^{\text{R-R}}|\|\Pi\|}{\rho^3(\Im\tau)^2}\, M_{\rm P}\,, \quad \quad \frac{\calt_{\rm memb}^\text{NS-NS}}{M^2_{\rm P}}\simeq  \frac{|q^{\text{NS-NS}}|\|\Pi\|}{\rho^3\Im\tau}\, M_{\rm P}\,.
\ee
In Table~\ref{tab:IIB_charges} we have collected all the relevant membrane and flux information. 

\begin{table}[!h]
	\begin{center}
		\begin{tabular}{| c c c c c c |}
			\hline
			\rowcolor{ochre!30} Dp-brane & wrapped over... & Charge & Flux & $\calt_{\rm memb} /M^3_{\rm P}$ & Dynamical
			\\\hline\hline
			D5 &  three-cycles $\Gamma^A$ & $q^{\text{R-R}}_A$ &	$m_A = \int_{\Sigma^A}  \overline{F}_3$ & $g_s^2 \rho^{-3}$ & \cmark \TBstrut
			\\
			\hline
			\rowcolor{darkred!20}  NS5 & three-cycles $\Gamma^A$ &$q_A^{\text{NS-NS}}$ &	$h_A = \int_{\Sigma^A}  \overline{H}_3$ & $g_s \rho^{-3}$ & \xmark \TBstrut
			\\
			\hline
		\end{tabular}
	\end{center}
	\caption{The branes originating the membranes which give the jumps over the RR fluxes $m_A$ and the NS-NS fluxes $h_A$.}
	\label{tab:IIB_charges}
\end{table}

Now we should question whether it makes sense to include both the kinds of membranes within the cut-off $\Lambda_{\rm UV}$ determined from \eqref{SL_IIB_mphiKK}. First, we can easily relate the tensions to the Kaluza-Klein scale \eqref{SL_IIB_LKK}, obtaining
\be
\label{SL_IIB_caltvsKK}
\frac{\calt_{\rm memb}^{\text{R-R}}}{M^2_{\rm P}}\simeq  \frac{|q^{\text{R-R}}|}{|h| \Im\tau} \frac{\rho_{\rm min}}{\rho}\, m_{\rm KK}\,, \quad \quad \frac{\calt_{\rm memb}^\text{NS-NS}}{M^2_{\rm P}}\simeq  \frac{|q^\text{NS-NS}|}{|h|}\frac{\rho_{\rm min}}{\rho}\, m_{\rm KK}\,.
\ee
Hence, both R-R and NS-NS membranes do satisfy the condition \eqref{SL_HG_mLcond} for $\Lambda_{\rm UV} \simeq m_{\rm KK}$ in the large volume, weak coupling region in which we are working.

However, it is also clear that, in the weak string coupling regime $g_{\rm s} = (\Im \tau)^{-1} \ll 1$, the NS-NS membranes are much heavier that the R-R membranes wrapped over the same internal three-cycle, since their tensions are enhanced with a factor $1/g_{\rm s}$:
\be
\frac{\calt_{\rm memb}^{\text{R-R}}}{\calt_{\rm memb}^{\rm NS-NS}} \simeq \frac{1}{\Im \tau} = g_{\rm s}\;.
\ee
In other words, at low energy, the only membranes which we can insert into the EFT are the R-R membranes; the NS-NS membranes are excluded because the jumps that they induce over the NS-NS fluxes would involve greater energies with respect to the cut-off $\Lambda_{\rm UV}$. We are then led to the dynamical flux sublattice \eqref{EFT_IIB_4dGammadyn} or, equivalently, to the explorable sublattice \eqref{EFT_IIB_4dGammaexp}. 

We can also be more precise, trying to quantify the cutoff $\Lambda_{\rm UV}$  that realizes such a situation. First, combining \eqref{SL_IIB_caltcond0} with \eqref{SL_IIB_Lflux}, we can relate the membrane tensions to the dilaton mass as
\be
\label{SL_IIB_vsmphi}
\frac{\calt_{\rm memb}^{\text{R-R}}}{M^2_{\rm P}}\simeq  \frac{|q^{\text{R-R}}|}{|h| \Im\tau} \, m_\phi\,, \quad \quad \frac{\calt_{\rm memb}^\text{NS-NS}}{M^2_{\rm P}}\simeq  \frac{|q^{\text{NS-NS}}|}{|h|}\,  m_\phi\,.
\ee
Therefore, the energetic condition including R-R membranes and leaving out the NS-NS membranes is
\be
\label{SL_IIB_RRcond}
\frac{\calt_{\rm memb}}{M^2_{\rm P}} \lesssim m_\phi\,.
\ee
Namely, it is just sufficient to choose the low energy cut-off to be just slightly above $m_\phi$, while still being well below the Kaluza-Klein mass scale:
\be
m_\phi \lesssim \Lambda_{\rm UV}\lesssim m_{\rm KK}\,,
\ee
as foreseen in \eqref{SL_IIB_mphiKK}. In this case, the condition \eqref{SL_IIB_RRcond} is satisfied by a \emph{parametrically large} number of R-R membranes, but only `marginally' by NS-NS membranes. 

On the other hand, one may read \eqref{SL_IIB_RRcond} by following a reverse reasoning. That is, \eqref{SL_IIB_RRcond} selects the region of validity of the EFT description within $\Gamma_{\rm EFT}$, which include R-R membranes but exclude the NS-NS ones, and restricts
\be
\label{SL_IIB_RRcond2}
\frac{|q^\text{R-R}|}{|h| \Im \tau}  \lesssim 1\, .
\ee
Such region will have a minimal radius set by $|h|\,\Im\tau_{\rm min}$ and for large values of this quantity one may effectively work with a lattice of fluxes $\Gamma_{\rm EFT}$.

As a final remark, we also stress that the semi-classical requirement \eqref{SL_HG_SCmemb} is indeed satisfied for both R-R and NS-NS membranes, legitimating us to treat them semi-classically. In fact, from \eqref{SL_IIB_caltcond0}, we obtain the estimates
\be
\label{SL_IIB_SCmemb}
\frac{\calt_{\rm memb}}{\Lambda_{\rm UV}^3} \simeq \frac{|q|\|\Pi\|}{\rho^3(\Im\tau)^2}\left(\frac{M_{\rm P}}{\Lambda_{\rm UV}}\right)^3\,,
\ee
so that, in the above range of $\Im\tau$ and $\rho$, \eqref{SL_IIB_SCmemb} is always satisfied.

\subsection{Including strings and 3-branes}
\label{sec:LandSwamp_Hierarchy_IIB_S3b}

Membranes are crucial to identify the dynamical fluxes, but they are not the only extended objects that one can include in four-dimensional Type IIB EFTs. After having chosen the flux sublattice \eqref{EFT_IIB_4dGammadyn}, in Section~\ref{sec:EFT_IIB_4d}, we dualized the axio-dilaton to a linear multiplet -- see \eqref{EFT_IIB_ltau} and below -- introducing a gauge two-form $\calb_2$ in the EFT. We should then include an axionic string to be coupled to $\calb_2$, in a similar manner as in \eqref{Sugra_Sum_SStr}. The ten-dimensional picture in fact predicts the existence of such a string, by considering a D7-brane wrapped over the full internal space $X$. Its tension is easily obtained from \eqref{Sugra_ExtObj_TStrgf} as
\be
\label{SL_IIB_stringT}
\calt_{\rm string}= |e|M^2_{\rm P} \,\ell\simeq\, \frac{|e|M^2_{\rm P}}{\Im\tau}\,.
\ee

However, we also stated that the gauge two-form is gauged as in \eqref{EFT_IIB_tildeH3}, if the gauge two-form is supposed to encapsulate the superpotential axio-dilaton/complex structure moduli coupling $W \sim \tau h_A \Pi^A(\Phi)$ \eqref{EFT_IIB_hatWgf}. The gauging  \eqref{EFT_IIB_tildeH3} indeed couples to a string which serves as a boundary of membranes, with the string charge $e$ related to the membrane R-R charges $q_A^{\text{R-R}}$ as
\be
q_A^{\text{R-R}} = e h_A\,.
\ee
After crossing a membrane so determined, with a `hole' bounded by a string, both the axio-dilaton and the R-R fluxes shift as
\be
\tau \rightarrow \tau + e \,,\qquad m_A \to m_A + e h_A\,,
\ee
identifying the vacua on the two sides, owing to the fact that the potential is equal.

In order include such a string, this has to be treated semiclassically, satisfying \eqref{SL_HG_SCstr}. Here, the condition \eqref{SL_HG_SCstr} reads 
\be
\label{SL_IIB_SCstr}
\frac{\calt_{\rm string}}{\Lambda_{\rm UV}^2} \simeq \frac{|e|}{\Im\tau} \left(\frac{M_{\rm P}}{\Lambda_{\rm UV}}\right)^2\lesssim 1\;,
\ee
which is clearly satisfied in the weak string coupling limit $\Im \tau \gg 1$. 

The picture just described precisely matches the prediction of the Freed-Witten anomalies \cite{BerasaluceGonzalez:2012zn}. As briefly summarized in Appendix~\ref{app:FW_AM}, a string of charge $e$ develops an anomaly which is cured by attaching $q$ membranes of charge $e$. Furthermore, the condition \eqref{SL_HG_SLstr}, as can be easily seen from \eqref{SL_IIB_stringT}, exactly matches with the membrane condition \eqref{SL_IIB_RRcond2} for $q = e_\Lambda h^\Lambda$. This is an important consistency check of the dual formulation of Type IIB EFTs that we proposed in Section~\ref{sec:EFT_IIB_4d}: the two-form gauging \eqref{EFT_IIB_tildeH3} (allowed by the choice of the lattice \eqref{EFT_IIB_4dGammadyn}) naturally leads to the introduction of BPS--string/membrane junctions, which are in turn predicted from the ten-dimensional perspective and the Freed-Witten anomaly cancellation mechanism; additionally, in the chosen perturbative regime  \eqref{SL_IIB_RRcond2}, in the EFT both the string and the membranes constituting the junction can coexist.

So far, we have never discussed about the possible origin of the four-form term as it appears in the second line of \eqref{EFT_IIA_SLeg}:
\be
\label{SL_IIB_Stad}
\calq^{\rm bg} \int_\Sigma C_4\,,
\ee
where $C_4$ is trivially closed. Its ten-dimensional origin may be understood as follows. Let us consider the GKP Type IIB background \cite{GKP}: there, the five-form field strength is given by
\be
\tilde F_5 = (1+ *)  \d e^{4 A(y)} \wedge \d^4 x
\ee
where, we recall, $e^{2A(y)}$ is the warp factor that appears in the background metric \eqref{EFT_IIB_ds2}. Then, one may think of the four-form potential $C_4$, related to $\tilde F_5$ as in \eqref{EFT_IIB_Fs}, as composed by a closed part $\d_{10} C_4^{(0)} = 0$ in the full ten-dimensional space and a non-closed part $C_4$. Then, $C_4 |_{\rm ext} = e^{4A(y)} \d^4 x$ and the action of a D3-brane in the Einstein-frame simplifies as
\be
\label{SL_IIB_SD3}
S_{\rm D3} = - \frac{2\pi}{\ell_{\rm s}^4} \int_{\Sigma_{4}} \d^{4}\sigma \sqrt{-\det\left(P[g_{mn}]\right)} + \frac{2\pi}{\ell_{\rm s}^4}  \int_{\Sigma_{4}} (C_4 + C_4^{(0)}) =  \frac{2\pi}{\ell_{\rm s}^4} \int_{\Sigma_{4}} C_4^{(0)} \,,
\ee
where the Nambu-Goto term has been canceled by the Chern-Simons contribution. Hence, the `topological'  term \eqref{SL_IIB_SD3} may be understood as coming from spacetime filling D3-branes.

\section{A hierarchy of objects in Type IIA and M-theory EFTs}
\label{sec:LandSwamp_Hierarchy_IIA}

We now move the case of Type IIA and M-theory EFTs, which we will treat together. The perturbative regimes that we consider are distinguished by the string coupling constant $g_{\rm s}$: in weak coupling limit, the Type IIA effective description is valid and therethe requirement $g_{s}\ll 1$ determines a hierarchy of membranes, as for the Type IIB case; moving towards strongly coupled regions, the hierachy of membranes changes and, for $g_{\rm s} \gg 1$ the Type IIA effective description has to be abandoned, in favor of an M-theory description.

\subsection{Weak coupling}

The first case that we examine is the weak coupling limit, $g_s \ll 1$, where the Type IIA effective description is expected to be valid. We focus our attention of flux lattice \eqref{EFT_IIA_4dGamma} of Section~\ref{sec:EFT_IIA_4d}, by turning off the geometric and non-geometric fluxes as in \eqref{EFT_IIA_CYcond} and further considering, as a simplifying assumption, that no open string moduli are present, turning off also the D6 flux quanta $n_a^\alpha = 0$ in \eqref{EFT_IIA_D6coupling}. As for the Type IIB case, we would like to have an understanding of the mass scales entering the effective theory. It is then convenient to follow the simplification introduced in \cite{Hertzberg:2007wc}, introducing the moduli
\be
\label{SL_IIA_rhosigma}
\rho\equiv V^{\frac13}_{\rm s}\, ,\quad\quad
\sigma\equiv \frac1{g_{\rm s}}V^{\frac12}_{\rm s}\,,
\ee
which have diagonal kinetic terms
\be 
-\frac34 M^2_{\rm P}\frac{\del_\mu\rho\del^\mu\rho}{\rho^2}- M^2_{\rm P}\frac{\del_\mu\sigma\del^\mu\sigma}{\sigma^2}\, ,
\ee
and generically denoting the remaining moduli, associated with the K\"ahler and complex structure deformations, with $\chi^\alpha$. 

The background is characterized by picking up background fluxes into the lattice \eqref{EFT_IIA_4dGamma} and let us assume that, for the moment, the only local sources which enter the effective description are D6-branes and O6-planes. These contribute to the ten-dimensional effective action \eqref{EFT_IIA_S10d} as
\be
S_{\rm loc} = - \mu_6 \int_{\text{D6}} \d^7 \xi \sqrt{-\det {\bf h}}\, e^{\frac34 \phi} + 2 \mu_6 \int_{\text{O6}} \d^7 \xi \sqrt{-\det {\bf h}}\, e^{\frac34 \phi} 
\ee
where we have just restricted to their Nambu-Goto term, expressed in the Einstein frame. Recalling the identification \eqref{SL_EO_MP}, we can rewrite the effective potential  in the generic form \cite{Hertzberg:2007wc}
\be
\label{SL_IIA_Veff}
V=M^4_{\rm P}\left[\frac{ A_3(\chi)}{\rho^3\sigma^2}+ \sum_{n=0,2,4,6}\frac{ A_{n}(\chi)}{\rho^{n-3}\sigma^4}+\frac{ A_{\rm D6}(\chi)-A_{\rm O6}(\chi)}{\rho^3\sigma^3}\right]\, .
\ee
Here $A_3(\chi)$  comes from the NS-NS internal flux and then scales as $|h|^2$, while the $A_{n}$s quadratically depend on the R-R fluxes $e_a$, $m^a$.  We also notice that in the weak-coupling limit $g_{\rm s}\ll 1$, it is the first term of \eqref{SL_IIA_Veff} to dominate and thus we can get the following estimation for the mass of the moduli
\be
\label{SL_IIA_mphi}
m_\chi \simeq \sqrt{\frac{ A_3(\chi)}{\rho^3\sigma^2}} M_{\rm P}=\frac{g_{\rm s}}{\rho^3}\sqrt{A_3(\chi)} M_{\rm P}\,,
\ee
which agrees with its IIB counterpart given in \eqref{SL_IIB_Lflux}. The Kaluza-Klein mass scale $m_{\rm KK}$ can again be estimated by \eqref{SL_IIB_LKK}. Hence, the arguments leading to \eqref{SL_IIB_taurange} (with $\Im\tau=\frac1{g_{\rm s}}$) can be applied to the Type IIA case as well.

Let us now pass to investigate the membrane content that the four-dimensional theory may be endowed with. First, there can be included membranes that originate from compactifying D$p$-branes, with $p=2,4,6,8$, over an internal $(p-2)$-cycle. These wrap, respectively, zero-, two-, four- and six-cycles in the internal manifold $X$. The volume of such cycles are
\be
{\rm vol} (\Sigma_2) = \int_{\Sigma_2} J_{\rm c}\;, \qquad  {\rm vol}(\Sigma_4) = \frac12 \int_{\Sigma_4} J_{\rm c}^2\;, \qquad {\rm vol}(X) = \frac{1}{3!} \int_{X} J_{\rm c}^3\;.
\ee
By then recalling that $J_c = \varphi^i \omega_i$, the tensions of such membranes have the following schematic behaviors in terms of the K\"ahler moduli 
{\small\be
\calt^{(2)}_{\rm memb} \sim q_0 \,,\qquad \calt^{(4)}_{\rm memb} \sim q_i \varphi^i \,,\qquad\calt^{(6)}_{\rm memb} \sim \kappa_{ijk} p^i \varphi^j \varphi^k \,,\qquad \calt^{(8)}_{\rm memb} \sim \kappa_{ijk} p^0 \varphi^i \varphi^j \varphi^k \,,
\ee}
with $(q_a, p^a)$ a set of quantized constants, ultimately identified with the charges of the membranes. Therefore, given a generic four-dimensional membrane originating from a set of $p$-cycles chosen as above, this gives rise to a tension of the form
\be
\label{SL_IIA_TRR}
\calt^{\text{R-R}}_{\rm memb} = 2 \Mp^3 |q_A \calv^A(\varphi)|
\ee
where $q_A = (q_a, p^a)$ are the quantized charges and we introduced the periods as in \eqref{EFT_IIA_PiA} with the prepotential chosen as in \eqref{EFT_IIA_calg}, with $a_{ab}=0$. The relation \eqref{SL_IIA_TRR} matches with the pure four-dimensional counterpart \eqref{Sugra_ExtObj_Memb_Tmgf}.

A D$p$-brane generates a \emph{jump} for the the $(8-p)$-form background fluxes $N_A = (e_a, m^a)$ in \eqref{EFT_IIA_em} -- see also Table~\ref{tab:SL_IIA_charges}.\footnote{This can be easily inferred by the Bianchi identities associated to the $F_{p+1} = \d C_{p}$, which read
\be
\label{SL_IIA_BIF}
\d F_{p+1} = q \delta_{p+2} (\Sigma_{8-p})
\ee
where $\delta_{p+1}$ is a $(p+2)$-current form representing the $(8-p)$ cycle $\Sigma_{8-p}$. One may think of $\delta_{p+1}$ as the Poincar\'e dual representative of the cycle $\Sigma_{8-p}$. Integrating the previous relation along the directions transverse to $\Sigma_{8-p}$, we get the cohomological relation
\be
[F_{p+1}] |_{\rm right} - [F_{p+1}]|_{\rm left} = q {\rm PD}[\Gamma_{7-p} \subset \Sigma_{8-p}]
\ee
It is then clear that a D$p$-brane makes a $(8-p)$--flux jump.} Then, following the general expression for the tension \eqref{SL_EO_SNG4D}, we get the following estimation for the membrane tensions originating from D$p$-branes
\be
\label{SL_IIA_TDp}
\calt^{(p)}_{\rm memb}\simeq \frac{g^2_{\rm s}\, \cala^{(8-p)}(\chi)}{\rho^{\frac12(11-p)}}
\,M^3_{\rm P}\, .
\ee
where we have used the relation
\be
\label{SL_IIA_eK}
e^{K_{\rm k}} = \frac{1}{8 V_6}\ \,,\qquad e^{K_{\rm cs}} = \frac{e^{4\phi}}{V_6^2} \,,
\ee
where  $V_6 \equiv \frac{1}{3!} \int_X J^3$ is the internal volume measured in string units. The moduli-dependent quantity $\cala^{(8-p)}(\chi)$  linearly depends on the charges making the $(8-p)$--fluxes jump.

However, also ten-dimensional NS5-branes can make four-dimensional membranes be born, once wrapped over internal three-cycles $\Sigma_\Lambda$. Their tensions, again, can be inferred from the general expression \eqref{SL_EO_SNG4D}. Given that the volume of an internal three-cycle, proportional to $|\int_{\Sigma_\Lambda} e^{\tilde\phi} \calc \Omega|$, carry a linear dependence on the complex structure moduli
\be
{\rm vol}(\Sigma_\Lambda)  \sim T_\Lambda\,,
\ee
the tension of a NS-NS membrane is
\be
\label{SL_IIA_TNSb}
\calt^{\text{NS-NS}}_{\rm memb} = 2 \Mp^3 g_s^{-1} e^{\frac K2}  |r^\Lambda T_\Lambda|\,,
\ee
as a particular case of \eqref{Sugra_ExtObj_Memb_Tmgf}.

They induce jumps for the NS-NS background fluxes $h^\Lambda = (h_K, h^Q)$ in \eqref{EFT_IIA_hbg} and we can estimate their tensions as
\be
\label{SL_IIA_TNS}
\calt^{\text{NS-NS}}_{\rm memb} \simeq \frac{g_{\rm s}\, \cala^{\text{NS-NS}}(\chi)}{\rho^3} \,M^3_{\rm P}\, .
\ee

An immediate comparison of \eqref{SL_IIA_TDp} with \eqref{SL_IIA_TNS} shows off that, in the small string coupling limit $g_{\rm s} \to 0$, while keeping the volume $V_{\rm s}$ fixed
\be
\label{SL_IIA_Tcomp}
\calt^{\text{NS-NS}}_{\rm memb} \gg \calt^{(p)}_{\rm memb}
\ee

\begin{table}[!h]
	\begin{center}
		\begin{tabular}{ |c c c c c c |}
			\hline
			\rowcolor{ochre!30} Brane & wrapped over... & Charge & Flux & Behavior of $\calt_{\rm memb}/M^3_{\rm P}$ & Dynamical
			\\\hline\hline
			D2 & trivial cycle & $q_0$ & $e_0 = \int_X \overline{G}_6$  & $g_{\rm s}^2 \rho^{-\frac92}$ & \cmark \TBstrut
			\\
			D4 &  two-cycles $\tilde\Sigma_i$ &$q_i$ &	$e_i = \int_{\tilde \Sigma^i}  \overline{G}_4$ &$g_{\rm s}^2 \rho^{-\frac72}$ & \cmark \TBstrut
			\\
			D6 &  four-cycles $\Sigma^i$ & $p^i$ & $m^i = -\int_{\Sigma_i}  \overline{G}_2$ & $g_{\rm s}^2 \rho^{-\frac52}$ & \cmark \TBstrut
			\\
			D8 &  six-cycle $X$ & $p^0$ &	$m^0 =  \overline{G}_0$ & $g_{\rm s}^2 \rho^{-\frac32}$ & \cmark \TBstrut
			\\
			\hline
			\rowcolor{darkred!20}  NS5 & three-cycles $\Sigma_\Lambda$ &$r^\Lambda$ &	$h^\Lambda = \int_{\Sigma_\Lambda}  \overline{H}_3$ & $g_{\rm s} \rho^{-3}$ & \xmark \TBstrut
			\\
			\hline
		\end{tabular}
	\end{center}
	\caption{The branes originating the membranes which give the jumps over the R-R fluxes $e_a$, $m^a$ and the NS-NS fluxes $h_p$.}
	\label{tab:SL_IIA_charges}
\end{table}

We also notice that
\be
\label{SL_IIA_const}
\frac{\calt_{\rm mem}^{(p)}}{M^2_{\rm P}}\simeq \frac{g_s}{\rho^{\frac12(11-p)}} \frac{|q_{(p)}|}{|h|}\,  m_\chi\,, \quad \quad \frac{\calt_{\rm mem}^{\text{NS-NS}}}{M^2_{\rm P}}\simeq  \frac{|q^\text{NS-NS}|}{|h|}\,  m_\chi\,,
\ee
and so the condition \eqref{SL_IIA_Tcomp} is satisfied for a parametrically large fraction of R-R membranes at sufficiently weak coupling, while it essentialy excludes the NS-NS membranes. Therefore, once again the criterion  \eqref{SL_IIA_Tcomp} matches the choice of EFT flux lattice \eqref{EFT_IIA_4dGammadyn}.

\subsection{Moderately strong coupling}

The hierarchy of membrane tensions illustrated above changes as we change the perturbative regime that we are scanning. As an example , let us examine the case of moderately strong coupling $g_{\rm s}\simeq 1$. The estimate \eqref{SL_IIA_TDp} for D8-branes, which make the Romans mass $m^0$ jump, reads
\be 
\frac{\calt_{\rm memb}^{(8)}}{M^2_{\rm P}} \simeq \frac{\cala^{(0)}(\chi)}{\rho^{\frac32}}\, M_{\rm P} \simeq  \cala^{(0)}(\chi) \rho^\frac12 \, m_{\rm KK} > m_{\rm KK}  \,, 
\ee
which violates the EFT condition \eqref{SL_HG_SCmemb}, with $\Lambda_{\rm UV} \simeq m_{\rm KK}$. This fact is just a manifestation of the usual strong coupling obstruction for massive IIA, as explained, for instance, in the recent \cite{Aharony:2010af}. In our framework, we can interpret this result as stating that those membranes for which $\frac{\calt_{\rm memb}}{M^2_{\rm P}} > m_{\rm KK}$ not only do not correspond to elements of $\Gamma_{\rm EFT}$, but in fact must be excluded from the larger flux lattice $\Gamma$. This is to be expected, in the sense that if one works in the ten-dimensional supergravity approximation such flux lattice is defined at the compactification scale, being different for each compact manifold.

Notably, with the choice \eqref{EFT_IIA_CYcond}, turning off the open string fluxes and excluding the Romans mass $m^0$ from the full lattice $\Gamma$ \eqref{EFT_IIA_4dGamma}, the flux contributions to the tadpole conditions \eqref{EFT_IIA_Tad} disappear, and so does the obstruction to dualize all the remaining fluxes to three-form potentials. We are then led to an `unconstrained' dual three-form theory, with the fluxes $(e_i,m^j)$ regarded as generated by three-form potentials $(A^i_3,\tilde A_{3j})$ which are accommodated into $b_2^-$ double three-form multiplets of the kind \eqref{Sugra_MTF_Ex_Za}, and the remaining fluxes $e_0$ and $h^\Lambda$ generated by the three-forms $A^0_3$ and $\hat A_{3\Lambda}$ which are part of single three-form multiplets as \eqref{Sugra_MTF_Ex_STF}. 

We also notice that, in general, the strong coupling regime does not allow for identifying the K\"ahler potential of the effective theory, unless the theory is interpreted as stemming from M-theory (see below or the previous Section~\eqref{sec:EFT_Mth_G2}). Nevertheless, the structure of the three-form multiplets is dictated just by the superpotential, thus by holomorphic periods, and so it is expected to enjoy some protection mechanism against perturbative corrections.

\subsection{M-theory regime}

To conclude, we consider the very strong coupling regime $g_{\rm s}\gg 1$, in which the Type IIA description is no longer suitable and one must rather formulate the setup in terms of M-theory compactifications. The eleven-dimensional M-theory metric $\d s^2_{11}$ is related to the IIA string frame metric $\d s^2_{10}$  by 
\be
\label{SL_Mth_ds11}
\d s^2_{11}=e^{-\frac23\phi}\d s^2_{\rm 10}+\ell^2_{\rm M} e^{\frac43\phi}(\d y+C_1)^2\, ,
\ee
where $y$ is the $S^1$ coordinate $y\simeq y+1$ nontrivially fibered over the ten-dimensional base manifold, with connection $C_1$. For simplicity, we have chosen a parametrization such that the M-theory Planck length $\ell_{\rm M}$ coincides with the string length, $\ell_{\rm M}\equiv\ell_{\rm s}$. We can  consider a limit in which the internal seven-dimensional space $\hat X$ is large in natural $\ell_{\rm M}$-units, $ V_{\rm M}\gg 1$. Furthermore, as will be particularly useful to get energy estimates, we assume that $\hat X$ is approximately isotropic and homogeneous. Then, from  \eqref{SL_Mth_ds11} we get the relations $\rho= V_{\rm M}^{\frac37}$ and $g_{\rm s}=\langle e^\phi\rangle = V_{\rm M}^{\frac3{14}}=\rho^\frac12$. With such assumption, the estimates \eqref{SL_IIA_Tcomp}-\eqref{SL_IIA_TNS} can be rephrased exclusively in terms of $\rho$:
\be
\label{IIAmemTb}
\calt^{(p)}_{\rm memb}\simeq \frac{\cala^{(8-p)}}{\rho^{\frac12(9-p)}}\, M^3_{\rm P}\, , \quad\quad \calt^{\text{NS-NS}}_{\rm memb}\simeq \frac{\cala^\text{NS-NS}}{\rho^\frac52}M^3_{\rm P}\,.
\ee
In the M-theory regime, with metric \eqref{SL_Mth_ds11}, the KK-scale becomes
\be
m^{\rm M}_{\rm KK}=\frac{M_{\rm P}}{\rho^\frac32}\,.
\ee
and therefore we obtain that
\be
\frac{\calt^{(p)}_{\rm memb}}{M^2_{\rm P}}\simeq \cala^{(n)}\rho^{\frac12(p-6)} \, m_{\rm KK}^{\rm M}\, , \quad\quad  \frac{\calt^{\text{NS-NS}}_{\rm memb}}{M^2_{\rm P}}\simeq \cala^{(p)}\rho^{-1} \, m_{\rm KK}^{\rm M}\, .
\ee
Hence, in the geometric regime $\rho\gg 1$, both $\calt^{(8)}_{\rm memb}$ and $\calt^{(6)}_{\rm memb}$ violate the KK scale condition $\frac{\calt_{\rm memb}}{M^2_{\rm P}}  \lesssim m_{\rm KK}$. According to our criterion above the corresponding fluxes, namely the Romans mass and the IIA R-R two-form fluxes, must not be included. From the M-theory perspective, to geometric fluxes that vanish on $G_2$-holonomy spaces.

The flux lattice $\Gamma$ in this regime is parametrized by the former type IIA fluxes $e_0,e_i$ and $h^\Lambda$. In M-theory language, $e_0$ is identified with the internal $G_7$-flux $n_0$ over the entire $\hat X$, while $e_a,h^\Lambda$ recombine into the flux quanta $n_{\check\imath}$ of  $G_4\in H^4(\hat X;\mathbb{Z})$ as in \eqref{EFT_Mth_Wb} . The associated membranes correspond to M2-branes, which are just points in the internal manifold, and M5-branes, wrapped over internal three-cycles, respectively. 

\begin{figure}[t]
	\centering
	\includegraphics[width=15cm]{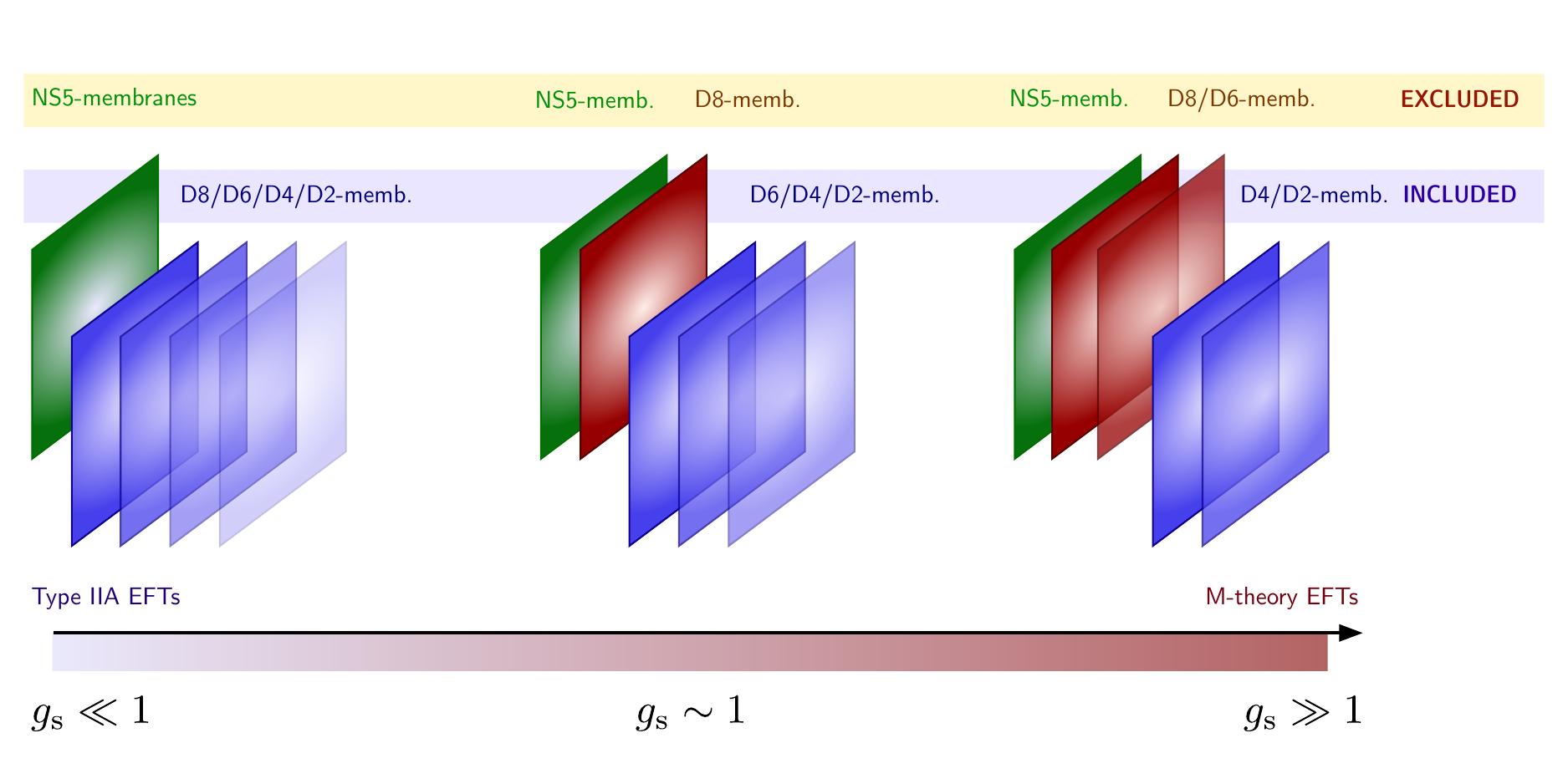}
	\caption{\footnotesize{How the membrane content changes across Type IIA and M-theory EFTs according to the chosen value of $g_{\rm s}$. Darker colors correspond to `heavier' membranes.} }\label{fig:MembMthTypeIIA}
\end{figure}

Interestingly, with this restricted choice of $\Gamma$ there are no tadpole conditions, and so in principle one may take $\Gamma_{\rm EFT} = \Gamma$. More precisely, one can see that $(n_0,n_{\check\imath})$ can be dualized to three-form potentials $A^0_3,A^{\check\imath}_3$ which can be incorporated, together with $\check s^0$ and $\check s^{\check \imath}$, into single three-form multiplets as in \eqref{EFT_Mth_STF}. One may then see if this choice is compatible with the hierarchy of corresponding membrane tensions.  By using  the K\"ahler potential of \cite{Beasley:2002db} in the one-modulus case, one can obtain a simple estimate $m_\phi\sim M_{\rm P}\rho^{-\frac52}$ of the scaling behaviour of the moduli masses. We then find that
\be
\frac{\calt^{(2)}_{\rm memb}}{M^2_{\rm P}} \simeq \rho^{-1}\, m_\phi\, , \quad \quad  \frac{\calt^{(4)}_{\rm memb}}{M^2_{\rm P}}\simeq \frac{\calt^{\text{NS-NS}}_{\rm memb}}{M^2_{\rm P}}\simeq \, m_\phi\, .
\ee
Therefore, to set $\Gamma_{\rm EFT} = \Gamma$ one must take a cut-off scale  such that $m_\phi \ll \Lambda_{\rm UV} \lesssim m^{\rm M}_{\rm KK}$.  As a result, the corresponding EFT has the attractive feature of including flux transitions that change significantly the masses of the would-be moduli, unlike in previous examples. The full picture that has emerged from Type IIA and M-theory EFTs is depicted in Fig.~\ref{fig:MembMthTypeIIA}.

While this is a perfectly consistent low-energy effective action, it does not incorporate anomalous axionic strings  in its spectrum of  fundamental extended objects, as well as the associated gauging of the two-forms dual to the corresponding axions. 
Indeed, such strings are given by M5-branes wrapping four-cycles of the compactification manifold. If the integral of the internal four-form flux does not vanish over the M5-brane, a Freed-Witten anomaly will be generated on its worldvolume \cite{Witten:1999vg}, which will then be cured by M2-branes ending on the corresponding 4d axionic string. In terms of the EFT Lagrangian, we will have a series of non-trivial gaugings of the 4d two-forms dual to the $C_3$-axions by the three-form potential $A_3^0$ coupling to the M2-branes. On the other hand, the gauging coefficients are nothing but the four-form fluxes $n_I$ and, by the reasoning of section \ref{sec:Sugra_GL}, we could adopt the dual description in terms of two-forms, in which the four-form fluxes are considered as part of the background fluxes $\caln^{\rm  bg}_\cala$ and specify the two-form gaugings.  In this case, we identify $\Gamma_{\rm EFT}$ with the one-dimensional lattice parametrized by $e_0$.

It seems that this class of compactifications allow for two different, complementary descriptions in terms of the three-form Lagrangians. Either we describe a 4d EFT containing $b_3 + 1$ classes of membranes, or we have an EFT with one class of membranes and one anomalous strings. It would be interesting to consider further examples of compactifications of this sort, and to understand whether the obstruction we find in incorporating all of these ingredients simultaneously is fundamental or can be overcome by adopting some so far unknown alternative scheme.

\chapter{Conclusions and future outlook}

In this work we have presented a full reformulation of generic $\caln=1$ four-dimensional supersymmetric theories including elements predicted from higher-dimensional string theory.  In both global and local supersymmetric theories we have introduced a hierarchy of gauge forms, summarized in Table~\ref{tab:Concl_pforms}. In order to construct manifestly $\caln=1$ supersymmetric Lagrangians, such gauge forms have been properly included among the components of supersymmetric multiplets: gauge two-forms find their place in linear multiplets, while gauge three- and four-forms are accommodated into special three- and four-form multiplets, the construction of some of which was only recently performed in \cite{Farakos:2017jme,Bandos:2018gjp,Lanza:2019xxg}. 

\begin{table}[!h]
	\begin{center}
		\begin{tabular}{ |c c c c  | }
			\hline
			\rowcolor{ochre!30}{ ${\bm p}$\bf-form} &  & {\bf Data} & {\bf Multiplet}
			\\
			\hline
			\hline
			Two-form & $\calb_2^\Lambda$ & \small{axion} & \small{Linear multiplet} $L^\Lambda$
			\\
			Three-form & $A_3^A$ & \small{dynamical background fluxes} & \small{Master three-form multiplet} $S^A$
			\\
			Four-form & $C_4^I$ & \small{tadpole cancellation conditions} & \small{Four-form multiplet} $\Gamma^I$
			\\
			\hline
		\end{tabular}
	\end{center}
	\caption{The role of the $p$-forms in four-dimensional effective theories of string and M-theory}
	\label{tab:Concl_pforms}
\end{table}

The introduction of such forms within a four-dimensional theory is related to the spectrum of the theory. Gauge two-forms may be regarded as `dual' of axionic fields, for which they provide an alternative representation. The rather peculiar gauge three-forms, although not carrying any propagating degrees of freedom, dynamically generate sets of constants which enter quadratically and linearly the potential. Such constants are arbitrary, but they may be constrained by means of gauge four-forms. In short, the newly introduced hierarchy delivers an \emph{alternative}, possibly more complete formulation of known $\caln=1$ supersymmetric theories. 

The proposed reformulation is necessary when it comes to coupling extended objects, summarized in Table~\ref{tab:Concl_ExtObj}. By introducing a set of \emph{super $p$-forms}, built out of the aforementioned supersymmetric multiplets, we were able to construct manifestly supersymmetric actions for BPS-strings, membranes and 3-branes. The action for strings and membranes are composed by a Nambu-Goto part, which includes the field-dependent tension of the object, and a Wess-Zumino part, which expresses the coupling of the object to the gauge $p$-forms and depends on its charges. The two terms are tightened together by the requirement that the action is $\kappa$-symmetric, allowing strings and membranes to preserve a $\caln=1$ worldvolume supersymmetry once a ground state is chosen. In turn, this implies that the tensions of the BPS-objects are completely determined by their charges. Instead, 3-branes appear in our framework only with a Wess-Zumino term, which depends on the charge, and preserve the full four-dimensional $\caln=1$ supersymmetry. 

\begin{table}[!h]
	\begin{center}
		\begin{tabular}{ |c c c  | }
			\hline
			\rowcolor{ochre!30}{\bf BPS--object} &  {\bf coupled to} & {\bf Data} 
			\\
			\hline
			\hline
			String & $\calb_2^\Lambda$ & axion monodromy
			\\
			Membrane & $A_3^A$ & shifts in background fluxes
			\\
			Topological 3-brane & $C_4^I$ & tadpole cancellation conditions
			\\
			\hline
		\end{tabular}
	\end{center}
	\caption{The extended BPS--objects included in the 4D EFTs.}
	\label{tab:Concl_ExtObj}
\end{table}

The inclusion of such objects has important physical consequences. The presence of a string signals a monodromy transformation which makes the axions transverse a certain number of periods. Membranes make the constants which are dual to the gauge three-forms jump. As a result, \emph{within the same theory}, one can describe a multi-branched potential. Membranes then open the possibility to get new vacua for a potential and new nonperturbative transitions among them can be triggered by the very existence of the membrane. Particular attention in this work has been devoted to the study of stable, BPS-domain wall solutions, which spontaneously break half of the bulk supersymmetry.

More complex BPS-objects were also built and are here summarized in Table~\ref{tab:Concl_Gaugings}. Membranes can be regarded as boundaries of 3-branes, identifying a `BPS--region', whose naturally defined gauge-invariant object is the gauged three-form $\d A_3^A + Q_I^A C_4^I$. Such a gauging serves to take into account how the tadpole cancellation conditions change across the various spacetime regions separated by the membranes. At the same time, strings may serve as boundaries for membranes and identify the gauged two-form $\d \calb_2^\Lambda + c_A^\Lambda A_3^A$. The discrete data which is encoded in this configuration is the cancellation of the Freed-Witten anomalies, which make the membrane `unstable', forming the hole bounded by the string.

\begin{table}[!h]
	\begin{center}
		\begin{tabular}{ |c c c | }
			\hline
			\rowcolor{ochre!30}{\bf Gauged ${\bm p}$-form} & {\bf BPS--object} & {\bf Data}
			\\
			\hline
			\hline
			$\d \calb_2^\Lambda + c_A^\Lambda A_3^A$ & \small{Membranes ending on a string} & \small{Freed-Witten anomaly cancellation}
			\\
			\hline
			$\d A_3^A + Q_I^A C_4^I$ & \small{3-branes ending on a membrane}  &  \small{changing tadpole cancellation}
			\\
			\hline
		\end{tabular}
	\end{center}
	\caption{The mutual gaugings of the $p$-forms of Table~\ref{tab:Concl_ExtObj} and the composite objects to which they couple.}
	\label{tab:Concl_Gaugings}
\end{table}

The results mentioned so far hold for \emph{any} $\caln=1$ global and local supersymmetric theory, namely with arbitrary K\"ahler potential and superpotential. They find fertile ground in the contexts of effective field theories stemming from compactifications of string and M-theory. In this work we focused on Type IIA and Type IIB string theory compactified over an orientifolded Calabi-Yau three-fold and on M-theory compactified over a seven-manifold with either $G_2$-holonomy or $G_2$-structure and briefly mentioned a possible F-theory uplift. In compactification scenarios, a potential for the otherwise \emph{moduli} of the compactification is introduced by threading the internal manifold with background fluxes and filling it with sources. Allowing them to vary, the background fluxes determine the maximal flux lattice
\be
\Gamma = \{ \caln^A \; | \; \caln^A \in \mathbb{Z} \}\,,
\ee
which enter linearly into the superpotential. In a three-form formulation, where membranes are also coupled, one may then think the lattice $\Gamma$ to be \emph{dynamically} spanned via membrane transitions.

However, we found some obstructions in interpreting the \emph{full} lattice $\Gamma$ dynamically: only a portion thereof $\Gamma_{\rm EFT} \subset \Gamma$ can be truly generated by gauge three-forms and undertake membrane transitions. The obstructions to identify $\Gamma_{\rm EFT}$ with $\Gamma$ are a priori all different, and arise independently from the EFT ingredients mentioned above: 
\begin{description}
	
	\item[Supersymmetry] The number of dynamical fluxes must be such that $n+1 \leq {\rm dim}\, \Gamma_{\rm EFT} \leq 2n+2$, where $n$ is the number of chiral fields entering the superpotential terms generated by the dynamical fluxes. This comes from the upper and lower bound on the number of three-forms per scalar in master multiplets.
	
	\item[Tadpoles] Fluxes typically contribute to tadpole conditions quadratically, by means of symmetric bilinear forms $\cali_I$. To implement tadpoles as three-form gaugings, $ \Gamma_{\rm EFT}$ must be an isotropic sublattice of $\Gamma$ with respect to each of these parings. As a result, at the level of the EFT tadople cancellation appears as a set of linear conditions on the dynamical fluxes. 
	
	\item[Axion-monodromy] In certain regimes of the compactification anomalous axionic strings appear at low energies, which means that membranes can nucleate holes in their worldvolume \cite{Evslin:2007ti,BerasaluceGonzalez:2012zn}. At the EFT level it makes sense to include the two-forms coupled electrically to the strings, together with a gauging encoding the anomaly. The gauging parameters will be fluxes that cannot belong to $ \Gamma_{\rm EFT}$.
\end{description}

As a result, the lattice is somehow `perceived differently' from the four-dimensional EFT perspective with respect to the higher-dimensional viewpoint. The most vivid argument is in the correct implementation of the tadpole cancellation condition. In higher dimensions, the tadpole cancellation involves quadratic combination of the fluxes in $\Gamma$; in four dimensions, a proper dynamical implementation of the tadpole require it to be reduced to a \emph{linear} constraint for $\Gamma_{\rm EFT}$ -- see Table~\ref{tab:Concl_stEFT}. 

\begin{table}[!ht]
	\begin{center}
		\begin{tabular}{|c||c|c|}
			\hline
			\rowcolor{ochre!30} & {\bf String Theory} & {\bf EFT}  \\
			\hline\hline
			Flux lattice & $\Gamma$ & $\Gamma_{\rm EFT}$\\
			\hline
			Tadpole condition & Quadratic & Linear\\
			\hline
		\end{tabular}
		\caption{\footnotesize Main difference between the string theory and the 4d EFT perspectives. For each region of field space in the  string theory construction a different sublattice $\Gamma_{\rm EFT} \subset \Gamma$ may be selected, but the linear dependence of the tadpole condition on the EFT fluxes applies to all of them. \label{tab:Concl_stEFT}}
	\end{center}
\end{table}

Then, while $\Gamma_{\rm EFT} \subset \Gamma$ is what can be rendered dynamical, the quotient $\Gamma/\Gamma_{\rm EFT}$, in one-to-one correspondence with the background, `frozen' fluxes, is what determines the gauging parameters in Table~\ref{tab:Concl_Gaugings}.

The choice of the dynamical sublattice is crucial, for it tells what is the explorable Landscape within a single effective theory and which nonperturbative effects may connect the various vacua of the Ladscape. However, such a choice is not arbitrary and we gave some physical arguments in Chapter~\ref{chapter:LandSwamp} to discriminate the well--motivated choices. For a given perturbative regime, certain kinds of membranes may be lighter, signaling that the associated flux transitions require less energy to take place. Choosing $\Gamma_{\rm EFT}$ then amounts in singling out which objects can be included within the EFT and which others may not, being too `heavy'. The picture that emerged draws effective field theories with a mutable spectrum of objects, according to the perturbative regime that is scanned, and with different amounts of fields which may render a potential multi-branched.

This is not the end of the story and the results presented in this work can be generalized and extended along diverse directions. A first, interesting technical development would be to include worldvolume matter supported on the  effective strings, membranes and 3-branes here considered. Such a matter content is indeed expected from higher-dimensional constructions, predicting, for example, gauge bosons and adjoint chiral fields living on 3-branes. Furthermore, it would be interesting to study the gauge theories living on top of membranes, better exploring the coupling of the Goldstone sector with the worldvolume fields and how the EFT treats this sector upon membrane-mediated transitions that change the number of 3-branes. Moreover, the presence of world-volume matter would allow for the study of possible novel swampland criteria, along the lines of \cite{Kim:2019vuc}.  

The theories here presented could be a starting point to extract models useful for phenomenology and cosmology. It would be interesting to question which Physics one should expect from the sub-ensemble of vacua dynamically connected through membrane nucleations within $\Gamma_{\rm EFT}$ and how physical observables like the cosmological constant, Yukawa couplings, soft terms vary significantly sailing through  the EFT landscape. Then, for those couplings that are effectively scanned over by the EFT, one may attempt to see if any physically relevant information can be extracted from a statistical analysis \cite{Douglas:2003um,Denef:2007pq}. 

The models here presented could well provide a compelling framework to test the recent Swampland conjectures \cite{Brennan:2017rbf,Palti:2019pca}. The simultaneous presence of gravity, gauge fields and extended objects in a controlled setup could indeed be crucial to inquire how the conjectures intertwine among each other.

\part{Appendices}

\begin{appendix}

	\chapter{Conventions on differential forms}
\label{app:Diff_Conv}

In generic $D$ dimensions, we define the totally antisymmetric Levi-Civita symbol such that
\be
\label{AppDiff_De}
\varepsilon^{0 1 \ldots (D-1)} = - \varepsilon_{0 1 \ldots (D-1)}= 1\;.
\ee
In the differential basis $\{\d x^\mu \}$, with $\mu = 0,\ldots , D-1$, a generic bosonic $p$--form $\omega_p$ has the following expansion
\be
\label{AppDiff_wp}
\omega_p = \frac{1}{p!} \omega_{\mu_1 \ldots \mu_p} \d x^{\mu_1} \wedge \ldots \wedge  \d x^{\mu_p}
\ee
The differential $\d$ acts on $\omega_p$ as
\be
\label{AppDiff_dwp}
\d \omega_p = \frac{1}{p!} \del_{[\sigma} \omega_{\mu_1 \ldots \mu_p]} \d x^\sigma \wedge \d x^{\mu_1} \wedge \ldots \wedge  \d x^{\mu_p}
\ee
The Hodge-dual of a $p$--form $\omega_p$ is the $(D-p)$--form defined as
\be
\label{AppDiff_*wp}	
*\omega = - \frac{e}{(D-p)! p!} \varepsilon_{\mu_1 \ldots \mu_D} g^{\mu_1 \nu_1} \ldots g^{\mu_p \nu_p} \omega_{\nu_1 \ldots \nu_p} \d x^{\mu_{p+1}} \wedge \ldots \wedge \d x^{\mu_{D}}
\ee
where the components given by
\be
\label{AppDiff_*wpc}
(*\omega)_{\mu_{p+1} \ldots \mu_D} = - \frac{e}{p!} \varepsilon_{\mu_1 \ldots \mu_D} g^{\mu_1 \nu_1} \ldots g^{\mu_p \nu_p} \omega_{\nu_1 \ldots \nu_p}\,.
\ee
The components $\omega^{\mu_1 \ldots \mu_p}$ are defined by raising the indices of the components of $\omega_p$ with the inverse $g^{\mu\nu}$ of the metric as usual. 

In any dimension $D$, we get the following useful identity
\be
\label{AppDiff_ww}
\begin{split}
	\omega \wedge * \omega &= - \frac{e}{p! (D-p)!} \omega_{\nu_1 \ldots \nu_p}  \varepsilon_{\mu_1 \ldots \mu_D} \omega^{\mu_1 \ldots \mu_p}  \d x^{\nu_1} \wedge \ldots \wedge  \d x^{\nu_p} \wedge  \d x^{\mu_{p+1}} \wedge \ldots \wedge \d x^{\mu_{D}}
	\\
	&= \frac{e}{p!}\, \omega^{\mu_1 \ldots \mu_p} \omega_{\mu_1 \ldots \mu_p} \d x^0 \wedge  \ldots \wedge  \d x^D \equiv e \, \omega \lrcorner\, \omega \d x^0 \wedge  \ldots \wedge  \d x^D \,,
\end{split}
\ee
and we have
\be
* (*\omega _p) = - (-)^{p(D-p)} \omega_p\;.
\ee

Finally, given the embedding of a $(p+1)$-dimensional bosonic manifold $\Sigma_{p+1}$, we use orientation conventions such that Stokes' theorem is
\be
\label{AppDiff_Stokes}
\int_{\Sigma_{p+1}}\d \omega_p=\int_{\del\Sigma_{p+1}} \omega_p\;.
\ee

\subsubsection*{Four-dimensional conventions}

In four dimensions, we define the Levi-Civita totally anti-symmetric symbol $\varepsilon^{mnpq}$ such that
\be
\label{AppDiff_4e}
\varepsilon^{0123} = - \varepsilon_{0123} = 1\;.
\ee
Then, the four-dimensional volume element is
\be
\label{AppDiff_41}
*1 = e\, \d x^0 \wedge  \d x^1 \wedge \d x^2 \wedge \d x^3 \equiv e\, \d^4 x\,.
\ee
Other useful identities that we used throughout this work include
\be
\label{AppDiff_4F4}
\begin{aligned}
	* F_4 &= -\frac{1}{e}\del^m \left( e (* A_3)_m \right) \;,
	\\ 
	F_4 &= - e\, (* F_4)\,\d^4x\;.
\end{aligned}
\ee

\subsubsection*{World-volume conventions}

The 3-brane worldvolume differentials enjoy the same conventions as the four-dimensional ones enlisted above.

For the membrane worldvolume we define the Levi-Civita antisymmetric symbol $\varepsilon^{ijk}$ such that
\be
\label{AppDiff_3e}
\varepsilon^{012} = - \varepsilon_{012} = 1\;,
\ee
and, given the local coordinates $\xi^i$, $i=0,1,2$, the worldvolume integration is performed by
\be
\label{AppDiff_3v}
\d^3 \xi \equiv \d \xi^0 \wedge \d \xi^1 \wedge \d \xi^2 = -\frac{1}{3!} \varepsilon_{ijk} \d \xi^i \wedge \d \xi^j \wedge \d \xi^k\;.
\ee

Equivalently, for the string worldvolume, the two-dimensional Levi-Civita symbol $\varepsilon^{ij}$ is defined such that
\be
\label{AppDiff_2e}
\varepsilon^{01} = - \varepsilon_{01} = 1\;,
\ee
and, denoting with $\zeta^i$, $i=0,1$ the local coordinates, the worlvolume integration is performed by
\be
\label{AppDiff_2v}
\d^2 \zeta \equiv \d \zeta^0 \wedge \d \zeta^1 = -\frac{1}{2!} \varepsilon_{ij} \d \zeta^i \wedge \d \zeta^j \;.
\ee

	\chapter{Conventions on  ${\cal N}=1$ superspace}
\label{app:Superspace_Conv}

The superspace conventions that we have used throughout this work coincide with those of \cite{Wess:1992cp}. In this appendix we collects  definitions and computations of the superfield components used in the main text. Furthermore, we also briefly recall how differential forms are defined in superspace.

\section{Component structure of ${\cal N}=1$ superfields}

In this section we provide all the components of the $\mathcal{N}=1$ multiplets considered throughout this work. We will be very schematic.

\begin{centerbox}
	Chiral multiplets
\end{centerbox}

The chiral superfield $\Phi$ satisfies the constraint
\begin{equation}
\bar{D}_{\dot\alpha} \Phi = 0
\label{Conv_ChiralA}
\end{equation}
and its expansion in components is 
\begin{equation}
\Phi = \varphi + \sqrt{2} \theta \psi + \theta^2 f + \ii \theta \sigma^m \bar\theta \partial_m \varphi -\frac{\ii}{\sqrt{2}} \theta^2 \partial_m \psi \sigma^m \bar\theta + \frac 14 \theta^2\bar\theta^2 \Box \varphi\;.
\label{Conv_ChiralB}
\end{equation}
Here $\varphi$ and $f$ are complex scalar fields, while $\psi$ is a Weyl spinor. 
The independent components of $\Phi$ can be defined by the projections
\begin{equation}\label{Conv_ChiralComp}
\begin{aligned}
\Phi | &= \varphi \,, \\
D_\alpha \Phi &|= \sqrt 2 \psi_\alpha \,,\\
-\frac14 D^2 \Phi| &= f \,,
\end{aligned}
\end{equation}
where the vertical line means that the quantity is evaluated at $\theta=\bar\theta=0$.

\begin{centerbox}
	Real multiplets
\end{centerbox}

A real scalar superfield $V$, beside being real $V = \bar V$, is unconstrained. It has the following component expansion 
\begin{equation}
\label{Conv_V}
\begin{split}
V =& \, u + \ii \theta \chi  - \ii \bar\theta \bar\chi + {\ii} \theta^2 \bar\varphi -  {i} \bar\theta^2 {\varphi} - \theta \sigma^m \bar\theta B_m  \\
& +\ii\theta^2\bar{\theta} \left( \bar{\lambda} 
+\frac{\ii}{2}\bar{\sigma}^m\partial_m \chi\right)-i\bar{\theta}^2 \theta \left( \lambda+\frac{\ii}{2}{\sigma}^m\partial_m \bar{\chi}\right)+\frac12\theta^2\bar\theta^2 \left({\rm D}-\frac12\Box u \right) \, , 
\end{split}
\end{equation}
where $u$ and $D$ are real scalar fields, $\varphi$ is a complex scalar field, $v_m$ is a real vector field and $\chi$ and $\lambda$ are Weyl spinors. 
The independent components of $V$ can be defined by projections
\begin{equation}
\label{Conv_Vprojections}
\begin{aligned}
V | &= u \,,\\
D_\alpha V| &= \ii \chi_\alpha\,,\\
\frac 14 \bar{\sigma}^{\dot\alpha \alpha}_m [D_\alpha, \bar{D}_{\dot\alpha}] V| &= B_m\,,\\ 
\frac{\ii}{4} D^2 V| &= \bar\varphi\,,\\
-\frac14 \bar D^2 D_\alpha V| &= -\ii\lambda_\alpha\,,\\
\frac{1}{16} D^2 \bar{D}^2 V| &= \frac12\left({\rm D} - \ii \partial^m v_m \right) \, .
\end{aligned}
\end{equation}

\begin{centerbox}
	Real linear multiplets
\end{centerbox}
A real linear superfield $L$ is a real superfield that obeys the constraint
\begin{equation}
D^2 L = 0\,, \qquad \bar{D}^2 L = 0 \,.
\label{Conv_RealLMdef}
\end{equation}
Its component expansion is 
\begin{equation}
\begin{split}
L &= l + \ii \theta \eta -\ii \bar\theta \bar\eta - \frac12 \theta \sigma_m \bar\theta \varepsilon^{mnpq} \partial_{n}\Lambda_{pq} \\
&\quad\,+ \frac12 \theta^2\bar\theta \bar\sigma^m \partial_m \eta + \frac12  \bar\theta^2  \theta \sigma^m \partial_m \bar\eta -\frac14 \theta^2\bar\theta^2 \Box l \, , 
\end{split}
\label{Conv_RealLM}
\end{equation}
where $l$ is a real scalar,  $\Lambda_{mn}$ is a rank 2 antisimmetric tensor  and $\eta$ is a Weyl spinor. The independent components of $L$ can be defined by projections 
\begin{equation}
\begin{aligned}
\label{Conv_RealLc}
L | &= l \,, \\
D_\alpha L| &= \ii \eta_\alpha\,,\\
-\frac12\bar{\sigma}^{m\,\dot\alpha \alpha}\left[D_\alpha,\bar{D}_{\dot{\alpha}}\right] L| &= \varepsilon^{mnpq} \partial_{n}\Lambda_{pq} \,.
\end{aligned}
\end{equation}

\begin{centerbox}
	Complex linear multiplets
\end{centerbox}
A complex linear multiplet $\Sigma$ satisfies the condition
\begin{equation}
\bar{D}^2 \Sigma = 0 \,.
\label{Conv_ComplexLMdef}
\end{equation}
Its component expansion is 
\begin{equation}
\begin{split}
\Sigma &= \sigma + \sqrt 2\theta \psi + \sqrt{2} \bar\theta \bar\rho - \theta \sigma_m \bar\theta \mathcal{B}^{m} + \theta^2 \bar s+ \sqrt 2\theta^2\bar\theta \bar\zeta \\
&\quad\,-\frac{\ii}{\sqrt{2}} \bar\theta^2 \theta \sigma^m \partial_m \bar\rho + \theta^2\bar\theta^2 \left(\frac{\ii}{2} \partial_m \mathcal{B}^{m} -\frac14 \Box \sigma \right)\,,
\end{split}
\label{Conv_ComplexLM}
\end{equation}
where $\sigma$ and $\bar s$ are complex scalars, $\rho$, $\psi$ and $\xi$ are Weyl spinors and $\mathcal{B}^{m}$ is a complex vector. The components of $\Sigma$ can be defined  by the projections
\begin{equation}
\label{Conv_SigmaCom}
\begin{aligned}
\Sigma | &= \sigma \,,\\
D_\alpha \Sigma| &= \sqrt 2 \psi_\alpha\,,\\
\bar D_{\dot \alpha} \Sigma| &= \sqrt 2 \bar \rho_{\dot\alpha}\,,\\
\frac14 \bar{\sigma}^{m\,\dot\alpha \alpha}
\left[D_\alpha,\bar{D}_{\dot{\alpha}}\right] \Sigma| &= \mathcal{B}^{m}\,,\\
-\frac14 D^2 \Sigma| &= \bar s\,,\\
\bar D_{\dot \alpha} D^2 \Sigma| & = -4\sqrt 2 \bar \zeta_{\dot \alpha} +2\sqrt 2 \ii \, \partial_m \psi^\alpha \sigma^m_{\alpha \dot \alpha}\,,\\
\frac{1}{16} \bar{D}^2 D^2 \Sigma| &= \ii \partial_m \mathcal{B}^m \,.
\end{aligned}
\end{equation}

\section{Differential forms in superspace}
\label{app:Superspace_Conv_Diff}

Also in the whole superspace $\mathbb{R}^{1,3|4}$ differential forms can be introduced. The most natural basis of differentials $\d {\frak z}^M = \{\d x^m, \d \theta^\alpha, \d \bar\theta_{\dot\alpha}\}$ is however not so useful in order to build supersymmetric objects. It is more convenient to introduce the differential basis
\be
\begin{aligned}
E^a &\equiv \d x^a - \ii \d \theta \sigma^a \bar \theta + \ii \theta \sigma^a \d \bar \theta\,,
\\
E^\alpha &\equiv \d \theta^\alpha\,, \qquad \quad\; E_{\dot \alpha} \equiv \d \bar \theta_{\dot\alpha}\,.
\end{aligned}
\ee
which we collectively denote as $E^A = \{E^m, E^\alpha, E_{\dot\alpha}\}$. As can be easily checked, this basis in invariant under the supersymmetry transformations \eqref{ExtObj_FrM_susy}.

The vierbein matrices which allow to pass from the basis $\d {\frak z}^M$ to $E^A$ are then defined as
\be
\label{SDiff_EA}
\begin{aligned}
	E^a &= \d x^m E_m^a + \d \theta^\mu E_\mu^a + \d \bar{\theta}_{\dot\mu} E^{\dot\mu a}\,,
	\\
	E^\alpha &= \d x^m E_m^\alpha + \d \theta^\mu E_\mu^\alpha + \d \bar{\theta}_{\dot\mu} E^{\dot\mu \alpha}\,,
	\\
	E_{\dot\alpha} &= \d x^m E_{m\dot\alpha} + \d \theta^\mu E_{\mu\dot\alpha} + \d \bar{\theta}_{\dot\mu} E^{\dot\mu}_{\dot\alpha}\,,
\end{aligned}
\ee
with
\begin{alignat}{5}
	E_m^a &= \delta^a_m\,,  &\qquad  &E_\mu^a = -\ii \sigma_{\mu\dot\mu}^a \bar\theta^{\dot\mu}\,,  &\qquad  &E^{\dot \mu a} = -\ii \bar\sigma^{a \dot\mu \mu} \theta_{\mu}   \,,
	\\
	E_m^\alpha &= 0\,,  &\qquad  &E_\mu^\alpha =\delta_\mu^\alpha\,,  &\qquad  &E^{\dot \mu \alpha} = 0   \,,
	\\
	 E_{m\dot\alpha} &= 0\,,  &\qquad  &E_{\mu\dot\alpha} = 0\,,  &\qquad  &E^{\dot\mu}_{\dot\alpha} = \delta^{\dot\mu}_{\dot\alpha}   \,.
\end{alignat}
We define the \emph{superspace differential} $\dwb$
\be
\label{SDiff_d}
\dwb = E^A D_A = E^a D_a + E^\alpha D_\alpha + E_{\dot \alpha} \bar D^{\bar \alpha}\;.
\ee
Given generic super $p$-form ${\bm \Omega}_p$, with the superspace differential expansion
\be
{\bm \Omega}_p =  E^{A_1} \wedge \ldots E^{A_p} \Omega_{A_p \ldots A_1}\;,
\ee
the action of the superspace differential \eqref{SDiff_d} is defined as
\be
\label{SDiff_dO}
\dwb {\bm \Omega}_p =  E^{A_1} \wedge \ldots E^{A_p} \wedge E^A D_{A}\Omega_{A_p \ldots A_1}\
\ee

We have also introduced the more common differential operator $\d$, acting on the left on a form as in \eqref{AppDiff_dwp}. If we focus on the bosonic components, the relation between the action of $\d$ and $\dwb$ on a generic super $p$-form ${\bm \Omega }_p$ is given by
\be
\d {\bm \Omega }_p|_{\rm bos} = \d \Omega_p = (-)^p\dwb {\bm \Omega}_p|_{\rm bos}\;.
\ee
Furthermore, in superspace, given the embedding of a $(p+1)$-dimensional bosonic manifold $\Sigma_{p+1}$ into superspace, we use orientation conventions such that Stokes' theorem reads $\int_{\Sigma_{p+1}}\d {\bf A}_p=\int_{\del\Sigma_{p+1}}{\bf A}_p$, compatibly with its purely bosonic counterpart \eqref{AppDiff_Stokes}.

Whenever the superspace coordinates ${\frak z}^M$ define a super-embedding of the form
\be
\xi^i \quad \mapsto \quad  {\frak z}^M = (x^m(\xi), \theta^\alpha(\xi),\bar\theta_{\dot\alpha}(\xi))\;,
\ee
with $\xi^i$ generic worldvolume local coordinates, the basis $E^A$ may be re-expressed in terms of a basis of worldvolume differentials $\{\d \xi^i\}$ as
\be
\label{SDiff_EAemb}
\begin{aligned}
	E^a &= \d \xi^i \left(\del_i x^m E_m^a +  \del_i\theta^\mu E_\mu^a + \del_i\bar{\theta}_{\dot\mu} E^{\dot\mu a} \right) \equiv \d \xi^i  E^a_i\;,
	\\
	E^\alpha &= \d \xi^i \left(\del_i x^m E_m^\alpha + \del_i\theta^\mu E_\mu^\alpha +  \del_i\bar{\theta}_{\dot\mu} E^{\dot\mu \alpha}\right)  \equiv \d \xi^i E^\alpha_i\;,
	\\
	E_{\dot\alpha} &=  \d \xi^i \left(\del_i x^m E_{m\dot\alpha} + \del_i\theta^\mu E_{\mu\dot\alpha} + \del_i\bar{\theta}_{\dot\mu} E^{\dot\mu}_{\dot\alpha}\right)  \equiv \d \xi^i  E_{\dot\alpha i}\;.
\end{aligned}
\ee
with the vierbein $E_i^A$ defined by the pullback of $E^A_M$ to the worldvolume as
\be
\label{SDiff_EAi}
\begin{aligned}
	E^a_i & = \del_i x^m- \ii \del_i\theta \sigma^m \bar\theta - \ii \del_i \bar\theta \bar\sigma^m \theta  \;,
	\\
	E^\alpha_i &= \del_i \theta^\alpha  \;, \qquad \qquad\;
	E_{\dot\alpha i} = \del_i \bar\theta_{\dot\alpha} \;.
\end{aligned}
\ee

	\chapter{Projections of ${\cal N}=1$ Supergravity superfields}
\label{app:Sugra_Conv}

We enlist the component structure of the multiplets in Supergravity that we have used throughout this work. We focus on the bosonic components only. The notation is that of \cite{Wess:1992cp}, except for the scalar curvature, here redefined as $R \to -R$.

\begin{centerbox}
	Chiral multiplets
\end{centerbox}

In supergravity, a chiral superfield $\Phi$ is a constrained superfield satisfying the covariantized chirality condition
\begin{equation}
\overline{\cal D}_{\dot \alpha} \Phi = 0 \, . \end{equation} 
Its component fields are defined as 
\begin{subequations}\label{ConvSG_Phi}
\begin{align}
\Phi |& = \varphi \, , 
\\
-\frac14 {\cal D}^2 \Phi | &  = F_{(\Phi)} \,  . 
\end{align}
\end{subequations}
with $\varphi$ and $F$ complex scalar fields. The vertical line means that the quantity is evaluated at $\theta=\bar\theta=0$.

\begin{centerbox}
	Supergravity multiplet
\end{centerbox}

The bosonic components of the off-shell Supergravity multiplet $\calr$ are the graviton $g_{mn}$, an auxiliary real vector $G^a$ and a complex scalar auxiliary field $M$. These components may be defined as projections of a chiral superfield $\calr$ as
\begin{subequations} \label{ConvSG_R}
	\begin{align}
	\calr |& = - \frac16 M \, , 
	\\
	-\frac14 {\cal D}^2 \calr | &  =  -\frac{1}{12} R  + \frac\ii6 \cald_a G^a - \frac1{18} G_a G^a -\frac19 M \bar M \,  . 
	\end{align}
\end{subequations}
\if{}

\begin{centerbox}
	Superspace chiral measure
\end{centerbox}

In defining the superspace chiral measure $\d^2 \Theta\, 2\cale$ -- for instance, in \eqref{Sugra_SW_L} -- the chiral density $2\cale$ appears, whose chiral projections are

\begin{subequations} 
	\label{ConvSG_Ec}
	\begin{align}
	2\cale |& = e \, , 
	\\
	-\frac14 {\cal D}^2 (2\cale)| &  =  - e \bar M\,  . 
	\end{align}
\end{subequations}

\fi{}

\begin{centerbox}
	Real multiplets
\end{centerbox}

An unconstrained, real multiplet $P$, in a Wess-Zumino gauge \cite{Wess:1992cp}, are
\begin{subequations} 
	\label{ConvSG_P}
	\begin{align}
	 P | &= 0  \;,
	 \\
	 - \frac\ii 4 \bar \cald^2  P |  & = z
	 \\
	 -\frac14 \bar\sigma_a^{\dot \alpha \alpha} [{\cal D}_\alpha , \bar{\cal D}_{\dot \alpha} ] P |& = A_{a} \, ,
	 \\
	  \frac\ii{16} \cald^2 (\bar \cald^2 - 8 \calr)  P |  & = \frac\ii2 d - \frac12 \cald^a A_a + \Re (\bar M z ) 
	\end{align}
\end{subequations}
where $z$ is a complex, propagating scalar field, $d$ a real, auxiliary scalar field and $A_a$ a real vector field, which can although be related, via Hodge--duality, to a three-form as in \eqref{GSS_In_CL_HD}.

\begin{centerbox}
	Real linear multiplets
\end{centerbox}

A real linear superfield $L$ satisfies the constraint
\be
(\bar\cald^2 - 8 \calr) L = 0\,,\qquad (\cald^2 - 8 \bar \calr) L = 0
\ee
which can be solved in terms of an unconstrained spinorial superfield $\Psi^\alpha$ as \cite{Buchbinder:1995uq}
\be
L =  \cald^\alpha (\bar \cald^2 - 8 \calr) \Psi_\alpha^\Lambda + \bar \cald_{\dot \alpha} (\cald^2 - 8\bar\calr) \bar \Psi^{\Lambda\dot\alpha}\,.
\ee

The bosonic components of $L$ are a real scalar $l$ and a gauge two-form $\calb_2$, which appear through its field strength $\calh_3 =\d\calb_2$. These are defined via:
\begin{subequations} 
	\label{ConvSG_L}
	\begin{align}
L | &= l  \;,
\\
- \frac14 \bar \cald^2  L |  & = \frac13 M l  \;,
\\
-\frac14 \bar\sigma_a^{\dot \alpha \alpha} [{\cal D}_\alpha , \bar{\cal D}_{\dot \alpha} ] L |& = \calh_a- \frac23 b_a l   \, ,
	\end{align}
\end{subequations}
where $\calh_a = \frac{1}{3!} \varepsilon_{abcd} \calh^{bcd} = \frac12 \varepsilon_{abcd} \del^{[b} \calb_2^{cd]}$.

\begin{centerbox}
	Complex linear multiplets
\end{centerbox}

A complex linear superfield $\Sigma$ satisfies the constraint \cite{Buchbinder:1995uq}
\be
\label{ConvSG_Sigmaconstr}
(\bar\cald^2 - 8 \calr) \Sigma = 0
\ee
which can be solved in terms of an unconstrained spinorial superfield $\Psi^\alpha$ as
\be
\Sigma = \bar \cald_{\dot \alpha} \bar \Psi^{\dot \alpha}
\ee
In the Wess-Zumino gauge, the bosonic components of $\Sigma$ read
\begin{subequations} 
\label{ConvSG_Sigma}
\begin{align}
\Sigma| &=0\,
\\
-\frac14 {\cal D}^2 \Sigma | &  = \bar s \, ,
\\
-\frac12 \bar\sigma_m^{\dot \alpha \alpha} [{\cal D}_\alpha , \bar{\cal D}_{\dot \alpha} ] \Sigma |&
= C_{m} \, ,
\\
\frac1{16}{\cal D}^2 \bar{\cal D}^2 \bar \Sigma |& = \frac\ii2 \,  {\cal D}_m C^m
+ \bar M s \, ,
	\end{align}
\end{subequations}
where $s$ is a complex scalar field and $C_{a}$ are component of a complex vector which can be related to those of a three-form as in \eqref{GSS_In_CL_HD}.

\if{}
\begin{centerbox}
	Single three-form multiplets
\end{centerbox}


\begin{centerbox}
	Nonlinear double three-form multiplets
\end{centerbox}

\fi{}
	\chapter{Super-Weyl invariant Supergravity}
\label{app:SW}

In Chapter~\ref{chapter:Sugra} we have extensively used the super-Weyl invariant formalism. This approach was firstly formulated in \cite{Kaplunovsky:1994fg} and we refer to \cite{Buchbinder:1995uq} for a more detailed treatment. Here we provide a review on how to construct generic super-Weyl invariant Lagrangians and how to compute their gauge--fixed components. For the sake of clarity, in this appendix we set $M_{\rm P}=1$, which can be re-installed with simple field redefinitions.

\section{K\"ahler and super-Weyl invariance in Supergravity}
\label{app:SW_KI}

Let us consider a set of $n$ chiral multiplets $\Phi^i$, whose bosonic components are
\be
\label{SW_Phimult}
\Phi^i = \{\varphi^i, F^i_\Phi\}\,,\quad \text{with}\quad i=1,\ldots,n\,,
\ee
where, as in \eqref{ConvSG_Phi}, $\varphi^i$  are the lowest component, complex scalar fields and $F^i_\Phi$ are the highest component, auxiliary complex scalar fields. These are `physical' fields, containing all the propagating degrees of freedom associated to the matter fields. 

The locally supersymmetric coupling of the matter multiplets $\Phi^i$ to gravity requires the introduction of another chiral multiplet $\calr$, the supergravity multiplet:
\be
\label{SW_Sugra_SW_calr}
\calr = \{e_a^m,\psi^m_\alpha, b^a, M\}
\ee
which accommodates the veilbein $e_a^m$, the gravitino $\psi^m$  and the nonpropgating fields, $b^a$ and $M$ which are a real vector and a complex scalar field. Up to two derivatives, the most general $\caln=1$ locally supersymmetric Lagrangian that we can build in superspace with only these ingredients is
\be\label{SW_SugraA}
\call= \int \d^4\theta\, E\,\Omega(\Phi,\bar \Phi)
+ \left(\int\d^2\Theta\,2\cale\,W(\Phi) +{\rm c.c }\right)\, ,
\ee
where $\Omega(\Phi,\bar \Phi) = -3e^{-\frac13 K(\Phi,\bar \Phi)}$, with $K(\Phi,\bar \Phi)$ being the ordinary K\"ahler potential and $W(\Phi)$, holomorphic in $\Phi^i$, is the superpotential. A direct computation of the bosonic components of the Lagrangian \eqref{SW_SugraA} leads to
\be
\label{SW_SugraALag}
\begin{split}
	e^{-1} \call_{\rm bos} &= -\frac16\Omega\, R - \Omega_{i \bar \jmath} \del_\mu \varphi^i \del^\mu \bar \varphi^{\bar\jmath} - \frac{1}{4\Omega} \left( \Omega_i \del_\mu \varphi^i - \bar \Omega_{\bar \jmath} \del_\mu \bar \varphi^{\bar\jmath} \right)^2
	\\
	&\quad\,+ \frac{\Omega}{9} \left| M - \frac{3 \Omega_{\bar \imath} \bar F^{\bar \jmath}_\Phi }{\Omega}\right|^2 + \left(\Omega_{i\bar \jmath} - \frac{\Omega_{i}\Omega_{\bar \jmath}}{\Omega}\right) F^i_\Phi \bar F^{\bar \jmath}_\Phi + \left(W_i F^i_\Phi + \bar W_{\bar\jmath} \bar F^{\bar \jmath}_\Phi  \right)\;.
\end{split}
\ee

The Lagrangian \eqref{SW_SugraA} exhibits a neat geometrical interpretation. The fields $\varphi^i$ parametrize a Hodge--K\"ahler manifold. This is a K\"ahler manifold $\scrm_{\rm scalar}$ of restricted type, namely it is defined by the existence of a line bundle $\scrl \rightarrow \scrm_{\rm scalar}$, whose first Chern class equals the cohomology class of the K\"ahler form $J$: $c_1(\scrl) = [J]$. Practically, this implies the following. Given a holomorphic section $\xi(z)$ over the line bundle $\scrl$, the first Chern class is determined by 
\be
\label{SW_c1}
c_1 (\scrl) = -\frac{\ii}{2\pi} \del \bar\del \log | \xi(z)|^2
\ee
where $|\xi(z)|^2 \equiv h(z,\bar z) \xi(z) \bar \xi(\bar z)$ is with $h(z,\bar z)$an hermitian fiber metric over $\scrl$. Then, recalling that $J = \frac\ii{2\pi} K_{i \bar \jmath} \d z^i \wedge \d \bar z^{\bar \jmath}$ is closed, locally, by the Poincar\'e-Lelong theorem, we may rewrite
\be
\label{SW_J2f}
J =  -\frac{\ii}{2\pi} \del \bar\del \rho
\ee
in terms of a \emph{potential} $\rho(z,\bar z)$. The comparison of \eqref{SW_c1} with \eqref{SW_J2f} shows that, when $\scrm_{\rm scalar}$ is Hodge--K\"ahler, then we may regard $\rho = e^K$ as the metric over the line bundle $\scrl$.

A K\"ahler transformations
\be
\label{SW_Ktransf}
K (\Phi, \bar \Phi) \rightarrow K (\Phi, \bar \Phi) + h(\Phi) + \bar h (\bar \Phi)
\ee
with $h(\Phi)$ an arbitrary holomorphic function of $\Phi^i$, leaves \eqref{SW_J2f} invariant. However, the Lagrangian \eqref{SW_SugraALag} is not invariant under the sole  \eqref{SW_Ktransf}.  The reason is that the superpotential and the auxiliary fields are sections of the $\scrl$-line bundle and, whenever \eqref{SW_Ktransf} acts over the K\"ahler potential, the following transformations have also to be performed:
\begin{subequations}
	\label{SW_WFtransf}
	\begin{align}
	\label{SW_Wtransf}
	&W(\Phi) \rightarrow e^{-h(\Phi)} W(\Phi)\,
	\\
	\label{SW_Ftransf}
	&M \rightarrow e^{-\frac23 (h(\Phi)+ \bar h(\Phi)) } M\,,\qquad F^i_\Phi \rightarrow F^i_\Phi e^{-\frac23 (h(\Phi)+ \bar h(\Phi)) }\,.
	\end{align}
\end{subequations}
A proper K\"ahler transformation then combines the three relations \eqref{SW_Ktransf}, \eqref{SW_WFtransf} and \eqref{SW_Ftransf}.

Still, the combined action of  \eqref{SW_Ktransf}, \eqref{SW_WFtransf} and \eqref{SW_Ftransf} does not leave the Lagrangian \eqref{SW_SugraALag} invariant. In fact, they have to be further combined with the \emph{super-Weyl transformation}, which acts on the super-vielbeins as \cite{Howe:1978km}
\be
\label{SW_EW}
E^a_M\rightarrow e^{\Upsilon+\bar\Upsilon}E^a_M\,,\quad E^\alpha_M\rightarrow e^{2\bar\Upsilon-\Upsilon}
\left(E^\alpha_M-\frac{\ii}{4}E^a_M\sigma^{\alpha\dot\alpha}_a\bar\cald_{\dot\alpha}\bar\Upsilon\right).
\ee
where $(a,\alpha)$ are flat superspace indices, $M=(m,\mu)$ are curved indices and $\Upsilon$ is an arbitrary chiral superfield parameterizing the super-Weyl transformation. In particular, \eqref{SW_EW} implies that
\be
E \rightarrow e^{2 \Upsilon + 2 \bar \Upsilon} E\,, \quad  \d^2 \Theta\, 2 \cale \rightarrow  e^{6\Upsilon} \d^2 \Theta\, 2 \cale
\ee
as well as, from the lower bosonic components of \eqref{SW_EW}, we recover the usual Weyl-rescaling \mbox{$e_m^a \to e_m^a e^{\Upsilon+\bar\Upsilon}$}. It can be immediately show that the superspace Lagrangian \eqref{SW_SugraA} is invariant if \eqref{SW_Ktransf} and \eqref{SW_WFtransf} are supplemented by \eqref{SW_EW} with the choice $6 \Upsilon = h(\Phi)$.

However, the Lagrangian \eqref{SW_SugraALag} is not expressed in the Einstein frame. It is then the lack of invariance of \eqref{SW_SugraALag} under the sole super-Weyl rescaling \eqref{SW_EW} that can be used to perform just a super-Weyl rescaling \eqref{SW_EW}, choosing $\Upsilon + \bar \Upsilon = \frac16 K(\Phi,\bar \Phi)$, which recasts the Lagrangian in the form
\be
\label{SW_SugraBLag}
\begin{split}
	e^{-1} \call_{\rm bos} &= \frac{1}2 R - K_{i\bar \jmath} \del_\mu \varphi^i \bar \del_\mu \bar \varphi^{\bar \jmath}
	\\
	&\quad\,+ e^{-K} K_{i\bar \jmath} F^i_\Phi \bar F^{\bar \jmath}_\Phi - \frac{e^{-K}}3 \left| \bar M + K_i F^i_\Phi \right|^2+ \left(W_i F^i_\Phi -  W \bar M + {\rm c.c.}\right)\;.
\end{split}
\ee
with a canonically normalized Einstein-Hilbert term. Further integrating out the auxiliary fields $F^i_\Phi$ and $M$, we arrive at the Lagrangian
\be
\label{SW_SugraCLag}
\begin{split}
	e^{-1} \call_{\rm bos} &= \frac{1}2 R - K_{i\bar \jmath} \del_\mu \varphi^i \del_\mu \bar \varphi^{\bar \jmath}- e^{K} \left( K^{\bar\jmath i} D_i W \bar D_{\bar \jmath} \bar W - 3 |W|^2 \right) \;,
\end{split}
\ee
where the last term is nothing but the well-known Cremmer et al. potential \cite{Cremmer:1978iv}. Indeed, this Lagrangian is invariant under \eqref{SW_Ktransf} and \eqref{SW_Wtransf}.

\section{Super-Weyl invariant Lagrangians with chiral multiplets}
\label{app:SW_C}

The super-Weyl invariant formalism takes somewhat another perspective. The Lagrangian \eqref{SW_SugraA} is rendered invariant under K\"ahler transformations \eqref{SW_Ktransf}-\eqref{SW_WFtransf} and super-Weyl transformations \eqref{SW_EW} \emph{separately}, while still arriving at the same component Lagrangian \eqref{SW_SugraBLag} or \eqref{SW_SugraCLag} when the the super-Weyl transformations are properly gauge-fixed. At the core of the formalism is the introduction of an unphysical, chiral compensator $U$, which transforms as
\be
\label{SW_Ytransfb}
U  \rightarrow  e^{-6 \Upsilon} U
\ee
under super-Weyl transformations. In turn, we introduce new chiral superfields $Z^a$
\be
\label{SW_Zmult}
Z^a = \{z^a,\, F^a_Z\}\,\quad \text{with}\quad a=1,\ldots,n+1\,,
\ee
where $z^a$ and $f^a$ are understood to be functions of the components of $\Phi^i$ and those of $U$, and we assume that, from $Z^a$, we can single out the compensator $U$ as
\be
\label{SW_homZ}
Z^a = U g^a (\Phi)
\ee
where $g^a$ are functions of the physical fields only and are inert under super-Weyl transformations.

The most general supergravity Lagrangian that we can build out the $Z^a$ multiplets has still the form \eqref{SW_SugraA} 
\be
\label{SW_SugraSWa}
\begin{split}
	\call = \int \d^4\theta\, E\, \calk (Z,\bar Z) + \left(\int \d^2 \Theta\, 2\cale\, \calw (Z) +{\rm c.c.}\right)
\end{split}
\ee
where $\calk({Z},\bar {Z})$ is the kinetic potential and $\calw(Z)$ the superpotential. Additionally, however, they satisfy the following homogeneity conditions
\be\label{SW_homKW}
\calk(\lambda Z,\bar\lambda\bar {Z})=|\lambda|^{\frac 23}\calk({Z},\bar {Z})\,,\qquad
\calw(\lambda Z)=\lambda\calw(Z).
\ee
with $\lambda$ an arbitrary chiral superfield. 

The ordinary K\"ahler potential $K(\Phi, \bar \Phi)$ and superpotential $W(\Phi)$ are then recovered by isolating the compensator $U$ as
\be
\label{SW_homKWb}
\calk({Z},\bar {Z})=-3|U|^{\frac23}e^{-\frac13 K(\Phi,\bar\Phi)}\,, \qquad 	\calw(Z)= U\,W(\Phi)
\ee
where we have set $K(\Phi,\bar\Phi)\equiv K(g(\Phi),\bar g (\bar\Phi))$ and $W(\Phi)\equiv W(g(\Phi))$. Such homogeneity properties, along with \eqref{SW_Ytransfb}, make the Lagrangian  \eqref{SW_SugraSWa} manifestly invariant under \eqref{SW_EW}. Indeed, the Lagrangian \eqref{SW_homKW} is also invariant under K\"ahler transformations. In fact, the split \eqref{SW_homZ} is clearly not unique, since  we may redefine
\be\label{SW_transfarb}
U\rightarrow e^{h(\Phi)}U\,,\quad g^a(\Phi)\rightarrow e^{-h(\Phi)} g^a(\Phi)\,.
\ee
Such a residual symmetry may be used to re-absorb a K\"ahler transformation \eqref{SW_Ktransf}-\eqref{SW_WFtransf}.

The bosonic components of the Lagrangian \eqref{SW_SugraSWa} acquire a very simple form
\be
\label{SW_SugraSWacomp}
\begin{split}
	e^{-1} \call_{\rm bos} &= -\frac16\calk\, R - \calk_{a\bar b} D_\mu z^a \bar D^\mu \bar z^b + \calk_{a\bar b}  f^a \bar f^b + \left( \hat\calw_a f^a + {\rm c.c.}\right)\,,
\end{split}
\ee
which is quite similar to the rigid case, as for example \eqref{GSS_DTF_LchirComp}. However, here we have redefined	$f^a \equiv  \bar M z^a - F^a_Z$ and introduced the $U(1)$-covariant derivatives
\be
D_\mu z^a = \del_\mu z^a + \ii A_\mu z^a\,\qquad {\rm with}\quad	A_\mu = \frac{3\ii}{2\calk} (\calk_a \del_\mu z^b - \calk_{\bar b} \del_\mu \bar z^b)\;.
\ee
In order to pass to the Einstein frame, we isolate the compensator $U$ and gauge-fix the super-Weyl invariance by setting \cite{Kaplunovsky:1994fg} 
\be
\label{SW_SWgf}
|U|^\frac23 = e^\frac{{K(\Phi, \bar \Phi)}}{3}\,.
\ee
In order to arrive at the bosonic components \eqref{SW_SugraBLag}, one has to isolate the compensator $u$ and split the index $a = (0,i)$, with $0$ associated to the compensator $u$ and $i$ to the physical fields $\varphi^i$, namely we may set $z^a = (u,u\varphi^i)$. Then, the kinetic matrix $\calk_{a\bar b}$ splits as
\be
\label{SW_GF_calkab}
\calk_{a \bar b} = e^{-K} \begin{pmatrix}
	-\frac13 & \frac{K_{\bar \jmath}}{3} \\ \frac{K_i}{3} & K_{i \bar \jmath} - \frac13 K_i K_{\bar \jmath}
\end{pmatrix}\;.
\ee
The Cremmer et al. potential instead stems after integrating out the auxiliary fields $f^a$. This requires the inverse of $\calk_{\bar b a}$, which is
\be
\label{SW_GF_Finv}
\calk^{\bar b a} =  e^{K} \begin{pmatrix}
	-3 + K^{\bar m n}K_n K_{\bar m} & K^{\bar m i} K_{\bar m} \\  K^{\bar\jmath n} K_{n}  & K^{\bar\jmath i}
\end{pmatrix} \;,
\ee

From the discussion above it is clear that we may assign a couple of \emph{super-Weyl weights} $(w_1, \bar w_2)$ to each physical quantity according to how they transform under \eqref{SW_Ytransfb} and its anti-holomorphic counterpart.  In Table~\ref{SW_weights}, we have collected the super-Weyl weights of all the relevant quantities which enter the supergravity actions considered in this Appendix and in Chapter~\ref{chapter:Sugra}. Super-Weyl invariant physical quantities are built simply by identifying combinations of objects in Table~\ref{SW_weights} so that the sum of their super-Weyl weights is zero.

\begin{table}[t]
	\begin{center}
		\begin{tabular}{|c c c|}
			\hline
			\rowcolor{ochre!30} & {\bf Supergravity quantity} & {\bf super-Weyl weights} \\
			\hline\hline
			\cellcolor{darkblue!20} $U$ & chiral compensator & $(1,0)$\\
			\hline
			\cellcolor{darkblue!20} $Z^a$ & chiral superfields & $(1,0)$\\
			\hline
			\cellcolor{darkblue!20} $L^\Lambda$ & linear multiplets & $(\frac13, \frac13)$\\
			\hline
			\cellcolor{darkblue!20} $\calk(Z, \bar Z, T)$ & kinetic function & $(\frac13, \frac13)$\\
			\hline
			\cellcolor{darkblue!20} $\calf(Z, \bar Z,L)$ & Legendre transform  & $(\frac13, \frac13)$ \\
			\hline
			\cellcolor{darkblue!20} $E$ & Berezinian super-determinant   & $(-\frac13, -\frac13)$ \\
			\hline
			\cellcolor{darkblue!20} $\d^2 \Theta\, 2 \cale$ & chiral superspace measure  & $(-1,0)$ \\
			\hline
		\end{tabular}
		\caption{Super-Weyl weights $(w_1, \bar w_2)$ of the relevant quantities which enter the Supergravity Lagrangian. The super-Weyl weight $w_1$ ($\bar w_2$) refers to how the sections transform in the holomorphic (anti-holomorphic) line bundle. Quantities which only depend on the physical fields $\Phi^i$ and $\ell^\Lambda$ carry zero super-Weyl weights.}
		\label{SW_weights}
	\end{center}
\end{table}

\section{Super-Weyl invariant Lagrangians with chiral and linear multiplets}
\label{sec:SW_L}

Let us now couple linear multiplets to the Lagrangians considered in the previous section. The linear multiplets are endowed with the components
\be
\label{SW_Lmult}
L^\Lambda = \{ l^\Lambda,  \calh_3^\Lambda\}\,,\quad \text{with}\quad \Lambda=1,\ldots,M\,,
\ee
with $l^\Lambda$ real scalar fields and $ H_3^\Lambda$ real field strengths of gauge two-forms $B_2^\Lambda$.  These transform under the super-Weyl transformations as $L^\Lambda \rightarrow e^{-2 \Upsilon -2 \bar \Upsilon} L^\Lambda$. The most general Lagrangian that we can build out of the chiral multiplets \eqref{SW_Zmult} and linear multiplets \eqref{SW_Lmult} is
\be
\label{SW_SWLina}
\begin{split}
	\call &= \int \d^4\theta\, E\, \calf (Z,\bar Z; L) + \left(\int \d^2 \theta\, 2\cale\, \calw (Z) +{\rm c.c.}\right)\,.
\end{split}
\ee
As in \eqref{Sugra_AL_Dual2Lag}, the kinetic function $ \calf (Z,\bar Z; L)$ can be seen as Legendre transform of a kinetic function $ \calk (Z,\bar Z; \Im T)$
\be
\label{SW_calkcalf}
 \calk (Z,\bar Z; \Im T) = \calf (Z,\bar Z; L) - \calf_\Lambda L^\Lambda\,,
\ee
with $\Im T_\Lambda$ dual to $L^\Lambda$ as in \eqref{Sugra_AL_Dual2L}. We require that $\calf$ satisfies the same homogeneity condition as $\calk$ in \eqref{SW_homKWb}, namely
\be
\label{SW_homKWc}
\calf(\lambda Z, \bar \lambda \bar Z; |\lambda|^\frac23 L) = |\lambda|^\frac23 \calf (Z,\bar Z; L)
\ee

The bosonic components of \eqref{SW_SWLina} are
\be
\label{SW_GF_Lag_LC}
\begin{split}
	e^{-1} \call_{\rm bos} &= -\frac16\tilde \calf\, R - \calf_{a\bar b} D_\mu z^a \bar D^\mu \bar z^b  + \frac1{4} \calf_{\Lambda \Sigma} \del_\mu l^\Lambda \del^\mu l^\Sigma + \frac1{4\cdot 3!} \calf_{\Lambda \Sigma} \calh_{\mu\nu\rho}^\Lambda \calh^{\Sigma\mu\nu\rho}
	\\
	&\quad\,+ \left(\frac\ii{2\cdot 3 !} \calf_{\bar a \Sigma} \varepsilon^{\mu\nu\rho\sigma}\calh^{\Sigma}_{\nu\rho\sigma} D_\mu \bar z^a+ {\rm c.c.}\right) + \calf_{a\bar b} f^a \bar f^b +\left(\calw_a f^a+{\rm c.c.}\right) 
\end{split}
\ee
where we have defined $\tilde \calf = \calf - l^\Lambda \calf_\Lambda$ and introduced the $U(1)$--covariant derivative
\be
\label{SW_LC_cov}
\begin{split}
	D_\mu z^a = \del_\mu z^a + \ii A_\mu z^a\;,
\end{split}
\ee
with
\be
\label{SW_LC_covb}
\begin{split}
	A_\mu &= \frac{3}{2(\tilde \calf -\tilde \calf_\Lambda l^\Lambda)} \left[\ii(\tilde\calf_{a}  \del_\mu z^a -  \bar{\tilde \calf}_{\bar a} \del_\mu \bar z^a ) + \frac{1}{3!} \tilde \calf_\Lambda \varepsilon_{\mu\nu\rho\sigma}\calh^{\Sigma\nu\rho\sigma} \right]\;.
\end{split}
\ee

In order to gauge fix the super-Weyl invariance, we need to isolate the compensator from the chiral multiplets $Z^a$ and linear multiplets $L^\Lambda$ as
\be
\label{SW_gfUL}
Z^a = U g^a(\Phi)\,,\qquad L^\Lambda =|U|^{\frac32} \tilde L^\Lambda
\ee
with $\tilde\calf = \calf - \calf_\Lambda L^\Lambda$ and, afterwards, set
\be
\label{SW_gfb}
|U|^\frac23 = e^\frac{{\tilde F(\Phi, \bar \Phi; L)}}{3}\,,
\ee
where $\tilde F = F - \ell^\Lambda F_\Lambda$, with $F(\Phi, \bar \Phi; L)$ the Legendre transform of the K\"ahler potential $K(\Phi, \bar \Phi, \Im T)$.

We now explain, with further details, the gauge-fixing procedure adopted in Section~\ref{sec:Sugra_LMDual} in order to pass from \eqref{Sugra_AL_Lbos} to \eqref{Sugra_AL_Lgf}. In principle, implementing the gauge-fixing in \eqref{SW_gfb} allows one to compute the Lagrangian in the Einstein frame. The gauge-fixing relations \eqref{SW_gfUL} may suggest that the proper linear multiplets to consider, after the gauge-fixing, are the $\tilde L^\Lambda$ obtained `extracting' the compensator form $L^\Lambda$. This is a necessary first step, but this is not the end of the story. In order to get kinetic terms in a compact form, it is then convenient to redefine the lowest components $\tilde l^\Lambda$ of $\tilde L^\Lambda$ as $\tilde l^\Lambda = e^{-\frac13 \tilde F} \ell^\Lambda$. Comparing this change of variables with \eqref{SW_gfUL}-\eqref{SW_gfb}, it is clear that we are somehow `undoing' the super-Weyl rescaling for the lowest component of $L$. Restoring the Planck mass, passing from our starting $l^\Lambda$ to $\ell^\Lambda$ simply results in \eqref{Sugra_AL_lgf}. In order to express the kinetic matrices in terms of the Legendre transform of the K\"ahler potential \eqref{Sugra_AL_F}, we preliminary notice that
\be
K (\varphi, \bar \varphi; \Im t ) = F (\varphi, \bar \varphi; \ell )  - \ell^\Lambda F_\Lambda(\varphi, \bar \varphi; \ell ) \equiv \tilde F (\varphi, \bar \varphi; \ell ) \;.
\ee
We then get the following identifications between the derivatives of the K\"ahler potential and its Legendre transform\footnote{We notice that partial derivatives of $K$ with respect to $\varphi^i$ are computed for constant $\Im t_\Lambda$, while those of $F$ for constant $\ell^\Lambda$.}
\be
\label{SW_GF_K}
\begin{aligned}
	K_i = \frac{\del K}{\del \varphi^i} \Big|_{\Im t_\Lambda} &= \frac{\del F}{\del \varphi^i} \Big|_{\ell^\Lambda} = F_i\;,
	\\
	K_{i \bar \jmath} + F^{\Lambda\Sigma} F_{\Lambda_i} F_{\Sigma \bar\jmath} &= \frac{\del^2 F}{\del \varphi^i \del \bar \varphi^{\bar\jmath}}  \Big|_{\ell^\Lambda}  = F_{i \bar \jmath}\;.
\end{aligned}
\ee
Now, from
\be
\tilde \calf \equiv \calf - \calf_\Lambda l^\Lambda = - 3 e^{-\frac{\tilde F}{3}} \equiv - 3 e^{-\frac{F-F_\Lambda \ell^\Lambda}{3}}
\ee
we get the following identities
\be
\label{SW_Fident}
\begin{aligned}
	-\frac{3}{\tilde \calf}\left(\calf_{a\bar b} -\frac{\tilde\calf_a \tilde\calf_{\bar b}}{\tilde\calf - \tilde\calf_\Pi l^\Pi} \right) &= F_{a\bar b}\;,
	\\
	-\frac{\tilde\calf}{3}\left(\calf_{\Lambda \Sigma}  - \frac{\tilde\calf_\Lambda \tilde\calf_\Sigma}{\tilde\calf - \tilde\calf_\Pi l^\Pi} \right) &= F_{\Lambda\Sigma}\;,
	\\
	\calf_{a \Sigma} - \frac{{\tilde \calf}_{a} \tilde\calf_\Sigma }{\tilde\calf - \tilde\calf_\Pi l^\Pi}&= F_{a\Lambda}\;.
\end{aligned}
\ee
which relate the super-Weyl invariant quantities with the gauge-fixed ones. These relations can be rewritten so as to get, more explicitly
\be
\label{SW_GF_calfLS}
\calf_{\Lambda \Sigma} =  -\frac{3}{\tilde\calf}\left(F_{\Lambda\Sigma} - \frac{\tilde F_\Lambda \tilde F_\Sigma}{3- \ell^\Pi \tilde F_\Pi}\right)
\ee
whence we can compute
\be
\label{SW_calfLSi}
\calf^{\Lambda \Sigma} =  - \frac{\tilde\calf}{3} \left(F^{\Lambda\Sigma} + \frac13 \ell^\Lambda \ell^\Sigma \right)
\ee
and
\be
\label{SW_calfLa}
\begin{aligned}
	\calf_{\Lambda i} &= \frac{1}{u} F_{\Lambda i} -\frac{1}{u} \frac{\tilde F_\Lambda \tilde F_i}{3- \ell^\Pi \tilde F_\Pi}\;,
	\\
	\calf_{\Lambda 0} &= \frac{1}{u} \frac{\tilde F_\Lambda}{3- \ell^\Pi \tilde F_\Pi}\;,
\end{aligned}
\ee
where the $0$th-index is associated to the compensator $u$, while the $i$s to the physical fields $\varphi^i$. Moreover, from \eqref{SW_calkcalf}, we may further relate the derivatives of $\calk$ with those of the Legendre transform $\calf$ as
\be
\label{LegA}
\begin{aligned}
	\calf_a |_{l^\Lambda} &= \calk_a |_{\Im t_\Lambda}\,, 
	\\
	\calf_{a \bar b} |_{l^\Lambda}  &= \calk_{a \bar b}|_{\Im t_\Lambda} - \calk^{\Lambda \Sigma}  \frac{\del \Im t_\Lambda}{\del z^a}  \frac{\del \Im t_\Sigma}{\del \bar z^b} = \calk_{a \bar b}|_{\Im t_\Lambda} +  \calf^{\Lambda \Sigma} \calf_{\Lambda a} \calf_{\Sigma \bar b}
\end{aligned}
\ee
where we have used 
\be
\begin{aligned}
	\calf_{\Lambda\Sigma} \calk^{\Sigma\Pi} &= -4 \delta_\Lambda{}^\Pi \quad \Rightarrow \quad  \calk^{\Lambda\Sigma} |_{l^\Pi} = -4 \calf^{\Lambda\Sigma}|_{\Im t_\Pi} 
	\\
	\calk^\Lambda_{a} &= - \calk^{\Lambda\Sigma} \frac{\del \Im t_\Sigma}{\del z^a}
	\\
	\calf_{\Lambda a} &= - 2 \frac{\del \Im t_\Lambda}{\del z^a}
\end{aligned}
\ee
These relations, combined with \eqref{SW_GF_K}, allow us to rewrite
\be
\calf_{i \bar \jmath} = \frac{e^{-\frac{\tilde F}{3}}}{|u|^\frac43} \left(  F_{i\bar \jmath} -\frac{\tilde F_i \tilde F_{\bar\jmath}}{3- \ell^\Pi \tilde F_\Pi}\right)
\ee	
whence
\be 
\label{SW_GF_calfab}
\calf_{a \bar b} = \frac{e^{-\frac{\tilde F}{3}}}{|u|^\frac43}  \begin{pmatrix}
	-\frac1{3- \ell^\Pi \tilde F_\Pi} & \frac{\tilde F_{\bar \jmath}}{3- \ell^\Pi \tilde F_\Pi} \\  \frac{\tilde F_{i}}{3- \ell^\Pi \tilde F_\Pi} & F_{i\bar \jmath} -\frac{\tilde F_i \tilde F_{\bar\jmath}}{3- \ell^\Pi \tilde F_\Pi}
\end{pmatrix}
\ee

We can now get the Lagrangian in the gauge-fixed Einstein frame. Integrating out the fields $f^a$, extracting the compensator $u$ and using \eqref{SW_GF_calfab}, \eqref{SW_calfLa} and \eqref{SW_GF_calfLS}, allows us to finally rewrite the Lagrangian \eqref{SW_GF_Lag_LC} as
\be
\label{SW_GF_Lag_LChb}
\begin{split}
	e^{-1} \call_{\rm bos} &= \frac{1}2 R -F_{i\bar \jmath}  \del \varphi^i \bar \del \bar \varphi^{\bar\jmath} + \frac14 F_{\Lambda\Sigma}  \left( \del_\mu \ell^\Lambda \del^\mu \ell^\Sigma + \frac1{ 3!}  \calh_{\mu\nu\rho}^\Lambda \calh^{\Sigma\mu\nu\rho}\right)
	\\
	&\quad\,+ \left\{\frac\ii{2\cdot 3 !} F_{\bar \imath \Sigma} \varepsilon^{\mu\nu\rho\sigma}\calh^{\Sigma}_{\nu\rho\sigma} \del_\mu \bar \varphi^{\bar\imath} + {\rm c.c.}\right\} 
	\\
	&\quad\,- e^{\tilde F} \left[ F^{\bar\jmath i} D_i W \bar D_{\bar \jmath} \bar W- (3 - \ell^\Lambda \tilde F_\Lambda) | W|^2 \right]
\end{split}
\ee
where the K\"ahler covariant derivatives are now given by
\be
D_i = \del_i + \tilde F_i\;.
\ee	

\section{Super-Weyl invariant Lagrangians with three-form and gauged linear multiplets}
\label{app:SW_GL}

Finally, let us consider the more general case where some chiral superfields are endowed with gauge three-forms, namely they are constrained as in \eqref{Sugra_MTF}, and also linear multiplets are present. The linear multiplets can also be gauged by the three-form potentials as in \eqref{Sugra_GL}. The most general Lagrangian including these ingredients (plus other hidden multiplets) is given by \eqref{Sugra_GL_L3f}.
However, in order to obtain its expression in bosonic components, it is convenient to start from the master Lagrangian \eqref{Sugra_GL_LM}. After integrating out $T_\Sigma$ from \eqref{Sugra_GL_LM}, we arrive at a master Lagrangian of the form
\be\label{SW_32formlagr}
\begin{split}
	\call&=\int\d^4\theta\, E \calf(Z,\bar Z,\hat L)+\left(\int\d^2\Theta\,2\cale\,\hat\calw(Z)+\text{c.c.}\right)
	\\
	&\quad\,+ \left[\int \d^2\Theta\, 2\cale\, X_A\calv^A(Z)+\frac\ii8\int \d^2\Theta\, 2\cale\, (\bar \cald^2 - 8 \calr)  (X_A-\bar X_A)P^A  +\text{c.c.}\right]\,.
\end{split}
\ee
We recall that $X_A$ is a chiral superfield, with bosonic components $\{x_A, F^{(X)}_{A}\}$ and $\calv^A$ are homogeneous of degree one as in \eqref{Sugra_MTF_WcalvA}. The homogeneity properties for $\calf$ and $\hat W$ are the same as in \eqref{SW_homKWb}-\eqref{SW_homKWc}. 

The bosonic components of the Lagrangian \eqref{SW_32formlagr} are 
\be
\small{
\label{SW_CLgauged}
\begin{split}
	e^{-1} \call_{\rm bos} &= -\frac16\tilde \calf\,R - \calf_{a\bar b} D z^a \bar D \bar z^b  + \frac1{4} \calf_{\Lambda \Sigma} \del_\mu l^\Lambda \del^\mu l^\Sigma + \frac1{4\cdot 3!} \calf_{\Lambda \Sigma} \hat\calh_{\mu\nu\rho}^\Lambda \hat\calh^{\Sigma\mu\nu\rho}
	\\
	&\quad\,+ \left(\frac\ii{2\cdot 3 !} \calf_{\bar a \Sigma} \varepsilon^{\mu\nu\rho\sigma}\hat\calh^{\Sigma}_{\nu\rho\sigma} D_\mu \bar z^{\bar a}+ {\rm c.c.}\right) +
	\\
	&\quad\,+\calf_{a\bar b} f^a \bar f^b+\calf_{\Lambda \Sigma} c_A^\Lambda c_B^\Sigma \calv^A \bar \calv^B
	\\
	&\quad\,+\left\{ -\frac{\ii}{2} \calf_\Lambda c_A^\Lambda \calv^A_a f^a - \ii \calf_{\Lambda b}f^b c_A^\Lambda \calv^A+\hat\calw_a f^a+{\rm c.c.}\right\}
	\\
	&\quad\, + \Big[\left(F^{(X)}_{A}- \bar M x_A\right) \left(- s^A + \calv^A_a z^a\right) \\
	&\quad\quad\,+ x_A \calv^A_a F^a_Z - \frac\ii2 x _A d^A 
	- \frac1{2 \cdot 3! e} \varepsilon^{\mu\nu\rho\sigma} \del_\mu x_A  A^A_{\nu\rho\sigma}  - x_A \Re (\bar M s^A) + {\rm c.c.} \Big]\,,
\end{split}}
\ee
where now $\hat \calh^\Lambda_{\mu\nu\rho} = \calh^\Lambda_{\mu\nu\rho} + c^\Lambda_A A_{\mu\nu\rho}^A$ and with the same covariant derivative as \eqref{SW_LC_cov}, modulo the replacement $\calh^\Lambda \to \hat \calh^\Lambda$. 

We now proceed as follows. First, we integrate out the real auxiliary fields $d^A$, which constrain $x_A$ to be real, and $F^{(X)}_{A}$, identifying $\calv^A(z) = s^A$ as in \eqref{Sugra_MTF_Compa}. Then, \eqref{SW_CLgauged} becomes
\be
\label{SW_CLgaugeda}
\small{\begin{split}
	e^{-1} \call_{\rm bos} &=- \frac16\tilde \calf\,R - \calf_{a\bar b} D z^a \bar D \bar z^b  + \frac1{4} \calf_{\Lambda \Sigma} \del_\mu l^\Lambda \del^\mu l^\Sigma + \frac1{4\cdot 3!} \calf_{\Lambda \Sigma} \hat\calh_{\mu\nu\rho}^\Lambda \hat\calh^{\Sigma\mu\nu\rho}
	\\
	&\quad\,+ \left(\frac\ii{2\cdot 3 !} \calf_{\bar a \Sigma} \varepsilon^{\mu\nu\rho\sigma}\hat\calh^{\Sigma}_{\nu\rho\sigma} D_\mu \bar z^{\bar a}+ {\rm c.c.}\right)
	\\
	&\quad\,+\calf_{a\bar b} f^a \bar f^b+\calf_{\Lambda \Sigma} c_A^\Lambda c_B^\Sigma \calv^A \bar \calv^B
	\\
	&\quad\,+\left\{ -\frac{\ii}{2} \calf_\Lambda c_A^\Lambda \calv^A_a f^a - \ii \calf_{\Lambda b}f^b c_A^\Lambda \calv^A+\hat\calw_a f^a+{\rm c.c.}\right\}
	\\
	&\quad\, + \Big[ x_A \calv^A_a f^a 
	- \frac1{2 \cdot 3! e} \varepsilon^{\mu\nu\rho\sigma} \del_\mu x_A\,  A^A_{\nu\rho\sigma} + {\rm c.c.} \Big]\,.
\end{split}}
\ee
Then, we further integrate out the auxiliary fields $f^a$ of the chiral multiplets $Z^a$ via
\be
\bar f^b = - \calf^{a\bar b} \left(x_A \calv^A_a - \frac{\ii}{2} \calf_\Lambda c_A^\Lambda \calv^A_a - \ii \calf_{a\Lambda} c_E^\Lambda \calv^E+\hat\calw_a  \right)
\ee
and, subsequently, the real Lagrange multipliers $x_A$. We then arrive at the super-Weyl invariant Lagrangian
\be
\label{SW_CLgaugedb}
\begin{split}
	e^{-1} \call_{\rm bos} &= -\frac16\tilde \calf\,R - \calf_{a\bar b} D z^a \bar D \bar z^b  + \frac1{4} \calf_{\Lambda \Sigma} \del_\mu l^\Lambda \del^\mu l^\Sigma + \frac1{4\cdot 3!} \calf_{\Lambda \Sigma} \hat\calh_{\mu\nu\rho}^\Lambda \hat\calh^{\Sigma\mu\nu\rho}
	\\
	&\quad\,+ \left(\frac\ii{2\cdot 3 !} \calf_{\bar a \Sigma} \varepsilon^{\mu\nu\rho\sigma}\hat\calh^{\Sigma}_{\nu\rho\sigma} D_\mu \bar z^{\bar a}+ {\rm c.c.}\right)+ e^{-1} \call_{\text{3-forms}}\,,
\end{split}
\ee
where the three-form Lagrangian $\call_{\text{three-forms}}$ can be recast as in \eqref{Sugra_MTF_L3fComp}, with
\begin{subequations}\label{ThVdefgl}
	\begin{align}
	T^{AB}(z,\bar z)&\equiv 2\,\Re\left(\calf^{\bar b a}\,\calv_a^A\bar\calv_{\bar b}^B\right)\,,\label{ThVdef1gl}
	\\
	h^A(z,\bar z)&\equiv 2\Re \left[\calf^{\bar b a} \left(\bar{\hat\calw}_{\bar b} + \frac{\ii}{2} \calf_\Lambda c_{B}^\Lambda \bar\calv^B_{\bar b} + \ii \calf_{\Lambda \bar b} c_F^\Lambda \bar \calv^{\bar F} \right) \calv_{a}^A \right]\,,\label{ThVdef2gl}
	\\
	\begin{split}
	\hat V(z,\bar z)&\equiv \calf^{\bar b a} \left(\hat\calw_a - \frac{\ii}{2} \calf_\Lambda c_A^\Lambda  \calv^A_a - \ii \calf_{\Lambda a} c_E^\Lambda \calv^E \right) \\ &\quad\quad \times \left(\bar{\hat\calw}_{\bar b} + \frac{\ii}{2} \calf_\Lambda c_{B}^\Lambda \bar\calv^B_{\bar b} + \ii \calf_{\Lambda \bar b} c_F^\Lambda \bar \calv^{\bar F} \right)
	\\
	&\quad\, - \calf_{\Lambda \Sigma} c_A^\Lambda c_B^\Sigma \calv^A \bar \calv^B\,.\label{ThVdef3gl}
	\end{split}
	\end{align}
\end{subequations}
We can now proceed to gauge-fixing the super-Weyl invariance by following the same procedure described in the previous subsection.
After having imposed the Einstein frame condition  \eqref{SW_gfb}, we arrive at
\be
\label{SW_GF_Lag_LChg}
\begin{split}
	e^{-1} \call_{\rm bos} &= \frac{1}2 R  -F_{i\bar \jmath}  \del \phi^i \bar \del \bar \phi^{\bar\jmath} + \frac14 F_{\Lambda\Sigma}  \left( \del_\mu \ell^\Lambda \del^\mu \ell^\Sigma + \frac1{ 3!}  \hat\calh_{\mu\nu\rho}^\Lambda \hat\calh^{\Sigma\mu\nu\rho}\right)
	\\
	&\quad\,+ \left\{ \frac\ii{2\cdot 3 !} F_{\bar \imath \Sigma} \varepsilon^{\mu\nu\rho\sigma}\hat\calh^{\Sigma}_{\nu\rho\sigma} \del_\mu \bar \phi^{\bar\imath} + {\rm c.c.}\right\} + e^{-1} \call_{\text{3-forms}}\,,
\end{split}
\ee
where the three-form Lagrangian has the same form as  \eqref{Sugra_MTF_L3fComp} with
\begin{subequations}\label{SW_ThVdefgl_wf}
	\small{\begin{align}
	T^{AB}&\equiv 2e^{\tilde F}\,\Re\left({F}^{i \bar \jmath}\,D_i  \Pi^A \bar D_{\bar \jmath} \bar\Pi^B-(3 - \ell^\Lambda \tilde F_\Lambda) \Pi^A\bar\Pi^B\right)\,,\label{ThVdef1gl_wf}
	\\
	\begin{split}
	h^A&\equiv 2e^{\tilde F}\Re \Bigg[{F}^{i \bar \jmath}\,\left( D_{\bar \jmath}  \bar{\hat W} + \frac{\ii}{2} F_\Lambda c_{B}^\Lambda D_{\bar \jmath} \bar \Pi^B + \ii F_{\Lambda \bar \jmath} c_F^\Lambda \bar \Pi^{\bar F}\right) D_i\Pi^A 
	\\
	&\qquad\qquad\, - (3 - \ell^\Lambda \tilde F_\Lambda) \left(\overline{\hat W} + \frac\ii2 F_\Lambda c_D^\Lambda \bar\Pi^D\right) \Pi^A - \ii \tilde F_{\Lambda} c_F^\Lambda \bar \Pi^{\bar F} \Pi^A\Bigg]\,,\label{ThVdef2gl_wf}
	\end{split}
	\\
	\begin{split}
	\hat V&\equiv e^{\tilde F} {F}^{\bar \jmath i} \left( D_{i}  {\hat W} - \frac{\ii}{2} F_\Lambda c_{A}^\Lambda D_{i} \Pi^A - \ii F_{\Lambda i} c_F^\Lambda \Pi^{F} \right) \left(D_{\bar \jmath}  \bar{\hat W} + \frac{\ii}{2} F_\Lambda c_{B}^\Lambda D_{\bar \jmath} \bar \Pi^B + \ii F_{\Lambda \bar \jmath} c_F^\Lambda \bar \Pi^{\bar F} \right)
	\\
	&\quad\,- (3 - \ell^\Lambda \tilde F_\Lambda) e^{\tilde F} \left|\hat W - \frac\ii2 F_\Lambda c_A^\Lambda \Pi^A \right|^2 - e^{\tilde F} F_{\Lambda \Sigma} c_A^\Lambda c_B^\Sigma \Pi^A \bar \Pi^B
	\\
	&\quad\,+e^{\tilde F}\left[ \ii c_F^\Lambda \tilde F_\Lambda \Pi^F \left( \bar{\hat W} + \frac\ii2 F_\Lambda c_A^\Lambda \bar\Pi^A \right)+{\rm c.c.}\right] .\label{ThVdef3gl_wf}
	\end{split}
	\end{align}}
\end{subequations}
In addition to the various identities among the kinetic matrices enlisted in the previous section, we have also employed the following relations
\be
\begin{aligned}
	\calf^{\Lambda \Sigma} \calf_{\Lambda 0} \calf_{\Sigma \bar 0} &= - \frac{e^{-\frac{\tilde F}{3}}}{3|u|^\frac43}  \frac{\tilde F_{\Lambda} \ell^\Lambda}{3- \ell^\Pi \tilde F_\Pi}\;,
	\\
	\calf^{\Lambda \Sigma} \calf_{\Lambda i} \calf_{\Sigma \bar 0} &=  -\frac{e^{-\frac{\tilde F}{3}}}{3|u|^\frac43} \frac{3 F_{\Sigma i} \ell^\Sigma- (\ell^\Pi \tilde F_\Pi)\tilde F_{i}}{3- \ell^\Pi \tilde F_\Pi}\;,
	\\
	\calf^{\Lambda \Sigma} \calf_{\Lambda i} \calf_{\Sigma \bar \jmath} &= \frac{e^{-\frac{\tilde F}{3}}}{|u|^\frac43} \left( F^{\Lambda \Sigma} F_{i\Lambda} F_{\bar\jmath \Sigma} + \frac13 F_i F_{\bar \jmath}  - \frac{\tilde F_i \tilde F_{\bar\jmath}}{3- \ell^\Pi \tilde F_\Pi}\right)\;,
\end{aligned}
\ee
which can be obtained from \eqref{SW_calfLSi} and \eqref{SW_calfLa}.

	\chapter{Freed-Witten anomalies and axion monodromies}
\label{app:FW_AM}

The configurations that we introduced in Sections~\ref{sec:Intro_3branes} and~\ref{sec:Intro_AnomStr}, namely of membranes ending on strings or of 3-branes ending on membranes are not unusual in string theory constructions. In fact, `attaching' 3-branes or membranes to a string provides a mechanism to cure the anomalies of the gauge theories defined over the worldvoumes of such objects. This effect, known as Freed-Witten anomaly cancellation mechanism \cite{Freed:1999vc,Maldacena:2001xj}, has a higher-dimensional origin in string theory contexts. The descriptions that we gave in Sections~\ref{sec:Intro_3branes} and~\ref{sec:Intro_AnomStr}, as well as in supergravity settings in Sections~\ref{sec:Sugra_ExtObj_3bMemb} and~\ref{sec:Sugra_ExtObj_MembString}, provides effective four-dimensional descriptions of such effects inherited from higher dimensions.

\section{Freed-Witten anomaly cancellation and Hanany-Witten brane creation effects}

From the ten-dimensional string theory description, configurations of branes with nontrivial background fluxes may lead to anomalies in the gauge theories living on their worldvolumes. The Freed-Witten anomaly cancellation mechanisms are as follows:
\begin{itemize}
	\item consider a D$p$-brane whose worldvolume $S_{p+1}$ is threaded by a nontrivial background NS-NS $\overline{H}_3$ flux. Then, the D$p$-brane has to emit a D$(p-2)$-brane spanning the directions orthogonal to $\overline{H}_3$ \cite{Maldacena:2001xj};
	\item consider an NS5-brane whose worldvolume $S_{6}$ is threaded by nontrivial background R-R $\overline{F}_p$ fluxes. Then, the NS5-brane has to emit a D$(6-p)$-brane spanning the directions orthogonal to $\overline{F}_p$;
	\item consider a D$p$-brane whose worldvolume $S_{p+1}$ is threaded by a nontrivial background $\overline{F}_p$. Then, an F1-string is emitted spanning the direction orthogonal to $\overline{F}_p$.
\end{itemize}

The Freed-Witten anomaly cancellation conditions, as explained in \cite{BerasaluceGonzalez:2012zn}, are in one-to-one correspondence with the Hanany-Witten brane creation effects \cite{Hanany:1996ie}. To exemplify the discussion, let us consider two examples:
\begin{enumerate}
	\item consider an NS5-brane stretching along all the external directions $x^0,\ldots,x^3$ and the internal $x^4$ and $x^5$ and a D8-brane covering all the directions but $x^9$, as summarized in Table~\ref{tab:app_HW1}. The NS5-brane may transverse the D8-brane along the $x^9$-direction and this results in creating a D6-brane covering the directions enlisted in Table~\ref{tab:app_HW1}.
	\begin{table}[!ht]
		\begin{center}
			\begin{tabular}{| c || c | c c c c c c |}
				\hline
				\rowcolor{ochre!30}  & $x^{0, \ldots, 3}$ & $x^4$ & $x^5$ & $x^6$ & $x^7$ &  $x^8$ &  $x^9$  \\ \hline
				{\bf NS5} & $\times$ &  $\times$ & $\times$ & $-$ & $-$ & $-$ & $-$  \\ \hline
				{\bf D8} & $\times$ &  $\times$ & $\times$ & $\times$ & $\times$ & $\times$ & $-$  \\ \hline\hline
				\rowcolor{bluegreen!20} $[\overline{H}_3]$ & $-$ &  $-$ & $-$ & $\bullet$ & $\bullet$ & $\bullet$ & $-$  \\ \hline\hline
				\rowcolor{darkred!20}  {\bf D6} & $\times$ &  $\times$ & $\times$ & $-$ & $-$ & $-$ & $\times$  \\ \hline
			\end{tabular}
			\caption{\footnotesize The brane/flux-configuration considered in the first example of Hanany-Witten effect, where $\times$ denotes the spanned direction, while $-$ a direction orthogonal to the brane. Analogously, $\bullet$ denotes the cycle spannes, in homology, by the background flux. \label{tab:app_HW1}}
		\end{center}
	\end{table}
	In terms of Freed-Witten anomaly cancellation mechanism, the NS5-brane may be understood as a source for an $H_3$ flux over the D8-brane worldvolume. This anomalous configuration is cured by emitting a D6-brane in the directions orthogonal to $H_3$.
	
	\item consider a D0-brane and a D8-brane covering all the directions but $x^9$, as in Table~\ref{tab:app_HW2}. The D0-brane may transverse the D8-brane along the $x^9$-direction and this results in creating an F1-string stretching along $x^0$ and $x^9$. 
	\begin{table}[!ht]
		\begin{center}
			\begin{tabular}{| c || c c c c | c c c c c c |}
				\hline
				\rowcolor{ochre!30}  & $x^0$ & $x^1$ & $x^2$ & $x^3$ & $x^4$ & $x^5$ & $x^6$ & $x^7$ &  $x^8$ &  $x^9$  \\ \hline
				{\bf D0} & $\times$ & $-$ & $-$ & $-$ &  $\times$ & $\times$ & $-$ & $-$ & $-$ & $-$  \\ \hline
				{\bf D8} & $\times$ & $\times$ & $\times$ & $\times$ &  $\times$ & $\times$ & $\times$ & $\times$ & $\times$ & $-$  \\ \hline\hline
				\rowcolor{bluegreen!20} $[\overline{F}_8]$ & $-$ & $\bullet$ & $\bullet$ & $\bullet$ & $\bullet$ & $\bullet$ & $\bullet$ & $\bullet$ & $\bullet$ &  $-$  \\ \hline\hline
				\rowcolor{darkred!20}  {\bf F1} & $\times$ & $-$ & $-$ & $-$ & $-$ & $-$ & $-$ & $-$ & $-$ & $\times$  \\ \hline
			\end{tabular}
			\caption{\footnotesize The brane/flux-configuration considered in the second example of Hanany-Witten effect. \label{tab:app_HW2}}
		\end{center}
	\end{table}
	Again, this effect may be understood in terms of Freed-Witten anomaly cancellations, as the D0-brane generate an $F_8$ fluxes over the D8-brane worldvolume. According the third point enlisted above, an F1-string cures such an anomalous configuration.
\end{enumerate}

\section{Freed-Witten anomalies and four-dimensional EFTs}

\begin{figure}[h]
	\centering
	\includegraphics[width=8cm]{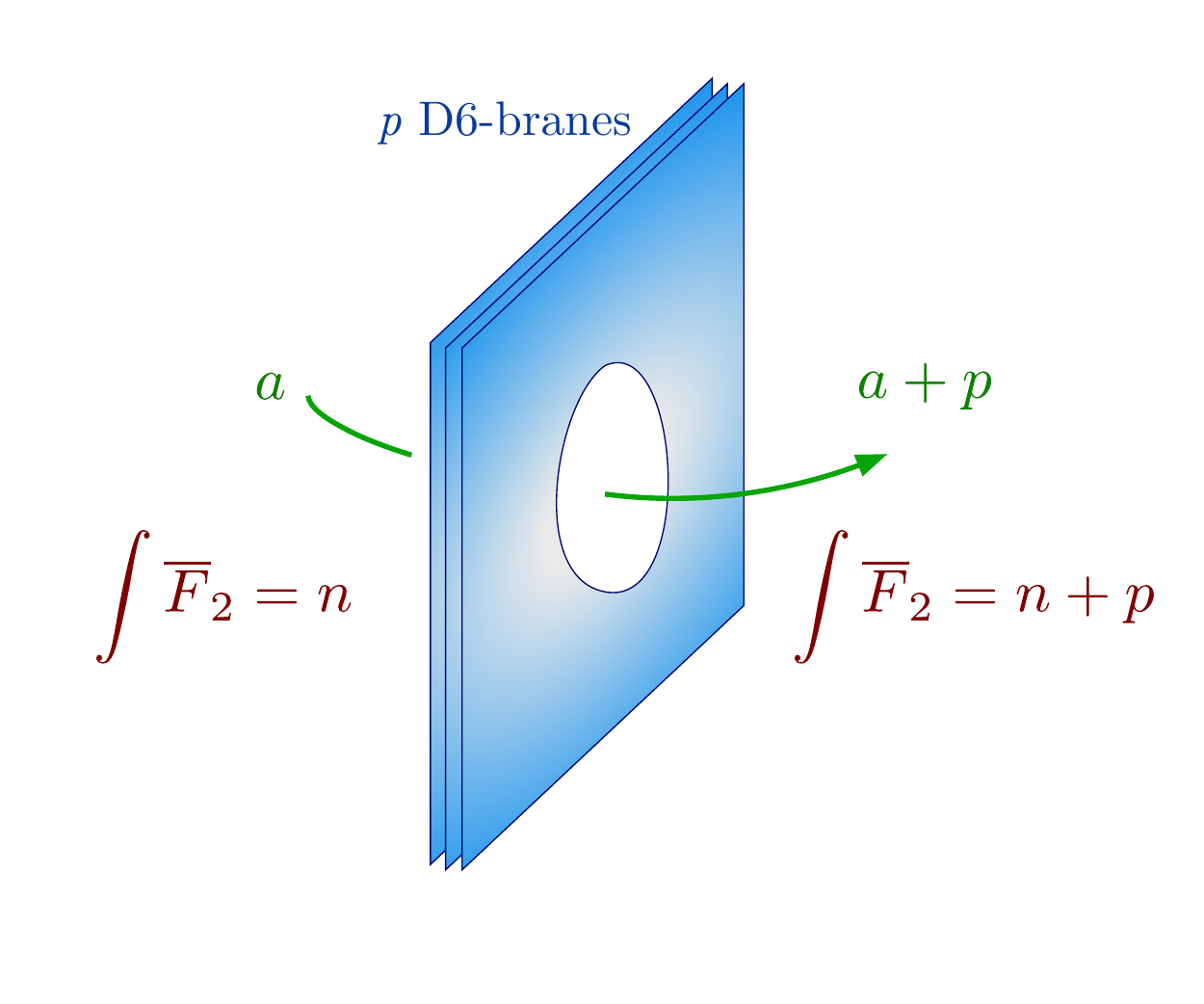}
	\caption{\footnotesize{An example of Freed-Witten anomaly cancellation mechanism.}}\label{fig:app_FWsb}
\end{figure}

Four-dimensional effective theories inevitably inherit such higher-dimensional phenomena: what in higher dimensions were branes emitting other branes, in four dimensions we get extended objects, whose boundary is nontrivial, being constituted by another object of one dimension less. To exemplify the discussion, let us again consider a simple example. An NS5-brane wrapping a special Lagrangian four cycle in the internal space gives rise to a supersymmetric string in four dimensions. Assume that over the NS5-brane a nontrivial zero-form flux $\overline{F}_0 = p$ is turned on. Now, according to the Freed-Witten anomaly cancellation mechanism, such a configuration is cured by attaching $p$ D6-branes to the string, whose number is counted by the zero-form flux, as depicted in Fig.~\ref{fig:app_FWsb}. These D6-branes look like membranes in four dimensions.

A D6-brane generates a jump for the background two-form flux $\overline{F}_2$, from $n$ units on the left to $n+p$ units on the right. At the level of four-dimensional theories, the superpotential induced by such a flux, schematically $\calw_{\text{R-R}} = n Z$ (see \eqref{EFT_IIA_calwRRsw} for the complete expression), is different on the two sides of the membrane. However, also a string is present, which couples to a gauge two-form in such a way that, after transversing the string, the dual axion is subjected to a monodromy transformation that changes $a \to a + p$. The superpotential associated to such a configuration, as explained in Section~\ref{sec:Sugra_GL}, is of the form $\calw_{\text{string}} = - T Z  $, with $a = \Re T$. It is then clear that the dull superpotential $\calw = \calw_{\text{R-R}} + \calw_{\text{string}} $ is unaltered if both a monodromy transformation and a flux jump are performed.

This simple example can be further generalized to more complicated monodromy transformations as in \cite{BerasaluceGonzalez:2012zn,Montero:2017yja,Herraez:2018vae}. In particular, it can be also generalized to the case where the four-dimensional `anomalous' object is a membrane, whose anomaly is cured by attaching 3-branes to it, in a similar fashion that we did for a string.

In other words, the Freed-Witten anomaly cancellation mechanism is inherited in the four-dimensional theory with the appearance of three elements, although connected among each other: the composite extended objects that constitute the spectrum of the theory; the gauging of forms that are coupled to the objects; the symmetries of the superpotential and, thus, the potential. 

\end{appendix}

\printbibliography

\end{document}